# Comparative genomic analysis of the human gut microbiome reveals a broad distribution of metabolic pathways for the degradation of host-synthetized mucin glycans

**Dmitry A. Ravcheev[1], Ines Thiele[1,*]**

[1]Luxembourg Centre for Systems Biomedicine, University of Luxembourg, Esch-sur-Alzette, Luxembourg

**\* Correspondence:**
Corresponding Author: University of Luxembourg, Campus Belval, 7, Avenue des Hauts-Fourneaux, L-4362, Esch-sur-Alzette
ines.thiele@uni.lu

**Keywords: Human gut microbiome, Comparative genomics, Mucin glycans, Carbohydrate utilization, Metabolism reconstruction**

**Abstract**

The colonic mucus layer is a dynamic and complex structure formed by secreted and transmembrane mucins, which are high-molecular-weight and heavily glycosylated proteins. Colonic mucus consists of a loose outer layer and a dense epithelium-attached layer. The outer layer is inhabited by various representatives of the human gut microbiota (HGM). Glycans of the colonic mucus can be used by the HGM as a source of carbon and energy when dietary fibers are not sufficiently available. Both commensals and pathogens can utilize mucin glycans. Commensals are mostly involved in the cleavage of glycans, while pathogens mostly utilize monosaccharides released by commensals. This HGM-derived degradation of the mucus layer increases pathogen susceptibility and causes many other health disorders. Here, we analyzed 397 individual HGM genomes to identify pathways for the cleavage of host-synthetized mucin glycans to monosaccharides as well as for the catabolism of the derived monosaccharides. Our key results are as follows: (i) Genes for the cleavage of mucin glycans were found in 86% of the analyzed genomes, whereas genes for the catabolism of derived monosaccharides were found in 89% of the analyzed genomes. (ii) Comparative genomic analysis identified four alternative forms of the monosaccharide-catabolizing enzymes and four alternative forms of monosaccharide transporters. (iii) Eighty-five percent of the analyzed genomes may be involved in exchange pathways for the monosaccharides derived from cleaved mucin glycans. (iv) The analyzed genomes demonstrated different abilities to degrade known mucin glycans. Generally, the ability to degrade at least one type of mucin glycan was predicted for 81% of the analyzed genomes. (v) Eighty-two percent of the analyzed genomes can form mutualistic pairs that are able to degrade mucin glycans and are not degradable by any of the paired organisms alone. Taken together, these findings provide further insight into the inter-microbial communications of the HGM as well as into host-HGM interactions.

## 1. Introduction

The colonic mucus layer is a dynamic and complex structure that is mainly composed of the glycoprotein mucin-2 (MUC2) (Johansson et al., 2008). MUC2 is characterized by abundant glycosylation. Highly variable glycan structures are O-linked to serine or threonine residues that are concentrated in so-called PTS (proline, threonine, and serine) domains. The mass of the glycosylated MUC2 protein is approximately 2.5 MDa, and more than 80% of this mass comes from carbohydrates (Lang et al., 2007). Glycosylated MUC2 can form a gel-like structure due to its N- and C-terminal domains that form numerous cross-links between cysteine residues (Johansson et al., 2013). Due to these cross-links, MUC2 forms a mucus structure. The colonic mucus consists of two layers, a loose outer layer and a dense epithelium-attached layer (Johansson et al., 2011). The inner layer acts as a physical barrier preventing bacteria from accessing the epithelium, whereas the outer layer is densely populated by various commensal microbes (Johansson et al., 2011;Johansson et al., 2015;Li et al., 2015).

Interactions between the human gut microbiota (HMG) and colonic mucus are not limited to the HGM members inhabiting the outer layer of mucus. Thus, once in the outer mucus layer, gut microbes not only avoid washout by the contents flowing through the colon but also are able to access mucin glycans that can be used as sources of carbon and energy. Both commensal and pathogenic microbes are able to degrade mucin glycans. The commensal microbes are usually able to cleave glycans with secreted glycosyl hydrolases (GHs) and lyases and further catabolize the derived monosaccharides, whereas most intestinal pathogens use monosaccharides released by commensal-secreted enzymes (Martens et al., 2008;Koropatkin et al., 2012;Marcobal et al., 2013;Cameron and Sperandio, 2015;Pacheco and Sperandio, 2015). Being both an environmental niche and a food source, the mucus layer plays a key role in shaping the HGM composition (Koropatkin et al., 2012;Johansson et al., 2015). In turn, the HGM is able to modulate mucus chemical composition via the degradation of glycans and peptides and the local release of bioactive factors that can change the expression patterns of the mucin-producing host cells (Deplancke and Gaskins, 2001). The degradation of mucin glycans is highly dependent on the host diet. Thus, decreased levels of fibers in the diet force the HGM to degrade more mucin glycans, resulting in the thinning and depletion of the mucus layer, which subsequently enhances host susceptibility to pathogens (Marcobal et al., 2013;Tailford et al., 2015;Desai et al., 2016). Additionally, depletion of the colonic mucus is associated with such disorders as Crohn's disease, celiac disease, colonic ischemia, compound exocytosis, and ulcerative colitis (Png et al., 2010;Joossens et al., 2011;Parmar et al., 2012;Johansson et al., 2013;Arike and Hansson, 2016;Cockburn and Koropatkin, 2016).

Because of the significant mucus-microbiota interaction, multiple HGM organisms have been analyzed for their abilities to degrade mucin glycans. To date, more than 50 mucus-degrading bacterial strains, generally belonging to *Akkermansia muciniphila*, *Bacteroides* spp., *Barnesiella intestinihominis*, *Bifidobacterium* spp., *Eubacterium* spp., and *Ruminococcus* spp. (Derrien et al., 2004;Martens et al., 2008;Png et al., 2010;Kiyohara et al., 2012;Pudlo et al., 2015;Tailford et al., 2015;Desai et al., 2016), have been identified. In this study, we analyzed the degradation of mucin glycans by the HGM using comparative genomic analysis. The comparative genomic analysis of sugar utilization by microbes is a recent but actively developing research area. There are several approaches to the genome-based computational analysis of sugar metabolism. The first approach is a pathway-centric one. In this approach, a certain pathway, for example, utilization of N-acetylgalactosamine, is analyzed in a large number of related genomes (Rodionov et al., 2000;Yang et al., 2006;Suvorova et al., 2011;Leyn et al., 2012;Zhang et al., 2015). The second approach is a regulator-centric one in which the family of proteins that are transcriptional regulators of sugar metabolism is analyzed in multiple genomes, and regulated metabolic pathways are reconstructed (Ravcheev et al., 2012;Rodionova et al., 2012;Kazanov et al., 2013;Ravcheev et al., 2014;Suvorova et al., 2015). The third approach is a taxonomy-centric one. In this approach, a set of genomes belonging to a certain taxon is selected, and all possible pathways for carbohydrate metabolism are reconstructed (Rodionov et al., 2010;Rodionov et al., 2011;Ravcheev et al., 2013;Rodionov et al., 2013;Khoroshkin et al., 2016).

Another approach for the computational analysis of metabolic pathways is a community-centric one. In this approach, metabolic pathways are reconstructed for microorganisms cohabiting a certain environment such as the human intestine. Unlike functional analysis of metagenomes, community-centric analysis is based on a reconstruction of metabolic pathways in individual genomes of microbes found in the studied environmental community and further prediction of interactions between different microorganisms. However, this approach has not previously been used for the detailed analysis of sugar utilization. It has been repeatedly used to reconstruct other metabolic pathways, including respiration (Ravcheev and Thiele, 2014), biosynthesis of B-vitamins (Magnusdottir et al., 2015) and quinones (Ravcheev and Thiele, 2016), as well central carbon metabolism and biosynthesis of amino acids and nucleotides (Magnusdottir et al., 2017), in multiple HGM genomes.

Here, we used a microbiome-centric approach to analyze the degradation of mucin glycans in HGM genomes. This analysis includes metabolic pathways for cleavage of mucin glycans to monosaccharides as well for utilization of these monosaccharides as carbon and energy sources. Additionally, we predicted exchange pathways for mucin glycan-derived monosaccharides, the specificity of various HGM strains for different types of mucin glycans, and mutualistic relationships for the cleavage of mucin glycans by different HGM organisms.

## 2. Materials and methods

### 2.1. Analyzed genomes

The analyzed genomes were selected using the following steps. (1) All genomes listed in the Human Microbiome Project (HMP, http://www.hmpdacc.org/HMRGD/) as of 17.09.2015 and 459 genomes with the body site "Gastrointestinal tract" (i.e., isolated from the intestine) were selected. (2) All the genomes absent from the PubSEED (Overbeek et al., 2005;Disz et al., 2010) and Integrated Microbial Genomes (IMG) databases (Markowitz et al., 2014) were excluded. The remaining 397 genomes (Supplementary Table S1) represent 288 microbial species, 89 genera, 45 families, 19 orders, 14 classes, and 8 phyla. All the selected genomes, except 2 Archaea, are bacterial. The phyletic distribution of the analyzed genomes is in good agreement with that observed in various HGM (Eckburg et al., 2005;Goodman et al., 2011;Walker et al., 2011;Graf et al., 2015). Thus, the most represented phyla are Actinobacteria (37 genomes, 9.3% of the analyzed genomes), Bacteroidetes (69 genomes, 17.3%), Firmicutes (197 genomes, 49.6%), and Proteobacteria (71 genomes, 17.9%).

### 2.2. Tools and databases

The PubSEED platform was used to annotate the genes responsible for the degradation of mucin glycans. To avoid misannotation, all of the proteins with the same function were checked for orthology. Orthologs were defined as the best bidirectional hits that have a similar genomic context. To search for the best bidirectional hits, a BLAST algorithm (Altschul et al., 1997) was implemented in PubSEED and the IMG platform (cutoff = $e^{-20}$). Additionally, in the search for orthologs, the GenomeExplorer program package (Mironov et al., 2000) was used, and orthologs were determined as the best bidirectional hits with protein identity no less than 20%. To analyze genomic context and gene occurrence, we used PubSEED and STRING v9.1 (Franceschini et al., 2013) along with phylogenetic trees for protein domains in MicrobesOnline (Dehal et al., 2010). To analyze protein domain structure, we searched the Pfam (Finn et al., 2014) and CDD (Marchler-Bauer et al., 2013) databases and the "Domains & Families" option of the MicrobesOnline platform. Additionally, functional annotations of the analyzed genes were performed using the UniProt (Magrane and Consortium, 2011), KEGG (Kanehisa et al., 2012), and MetaCyc (Caspi et al., 2014) databases.

Multiple protein alignments were performed using the MUSCLE v. 3.8.31 tool (Edgar, 2004a;b). Phylogenetic trees were constructed using the maximum-likelihood method with the default parameters implemented in PhyML-3.0 (Guindon et al., 2010). The obtained trees were visualized and midpoint-rooted using the interactive viewer Dendroscope, version 3.2.10, build 19 (Huson et al., 2007). To clarify the taxonomic affiliations of the analyzed genomes, the NCBI Taxonomy database (http://www.ncbi.nlm.nih.gov/taxonomy) was used. Analysis of gene occurrence was performed using the "Phylogenetic Profiler for Single Genes" option of the IMG platform.

All of the annotated genes are represented as a subsystem in PubSEED (http://pubseed.theseed.org/SubsysEditor.cgi; subsystem names are "Galactose utilization HGM", "L-fucose utilization HGM", "N-Acetylgalactosamine utilization HGM", "N-Acetylglucosamine utilization HGM", and "N-Acetylneuraminic acid utilization HGM") and in Supplementary Tables S2-S7. The protein sequences for the annotated genes in FASTA format are represented in the file Sequences S1 in the Supplementary Materials.

## 3. Results

### 3.1. Repertoire of enzymes for the degradation of mucin glycans

Mucin glycans are complex polysaccharides that contain five different monosaccharides, L-fucose (Fuc), D-galactose (Gal), N-acetyl-D-galactosamine (GalNAc), N-acetyl-D-glucosamine (GlcNAc), and N-acetylneuraminic acid (Neu5Ac), and can form various glycosidic bonds (Podolsky, 1985;Tailford et al., 2015). Evidently, degradation of such complex structures requires a large number of bacterial proteins interacting with the mucin glycans. Because of this complexity and the variability of mucin glycans, systematization of bacterial proteins for the degradation of mucin glycans became the first goal of this study.

All bacterial proteins involved in the degradation of mucin glycans were divided into two groups: (1) GHs, which split glycans to oligo- and monosaccharides as well separating glycans from mucin proteins, and (2) enzymes required for the catabolism of the derived monosaccharides. This division was made based on the following aspects of mucin glycan degradation by the HGM. First, splitting of glycans and catabolism of the derived monosaccharides are spatially separated; the first process occurs outside of the microbial cell, while the second process occurs in the cell cytoplasm. Such spatial separation is important for the metabolic modeling of HGM-host metabolism. Second, some HGM organisms have only glycan-cleaving hydrolases or only monosaccharide-catabolizing enzymes. For example, *Bacteroides thetaiotaomicron* encodes only sialidases that can release Neu5Ac from mucin glycans but not genes for the catabolism of this compound (Marcobal et al., 2011). On the other hand, *Clostridium difficile* encodes only genes for the catabolism of N-acetylneuraminic acid but not sialidases (Sebaihia et al., 2006). Such differentiation of the enzymes provides for the cross-talk of the HGM organisms, as was shown for *B. thetaiotaomicron* releasing Neu5Ac and *Salmonella typhimurium* consuming it (Ng et al., 2013). Thus, the distribution of the monosaccharide-releasing enzymes and monosaccharide catabolic pathways (CPs) should be analyzed separately.

To identify all the GHs for mucin glycan degradation, we first identified all the glycosyl bonds previously detected in mucin glycans of the human intestine (Podolsky, 1985;Larsson et al., 2009;Tailford et al., 2015), which resulted in a collection of 21 different glycosyl bonds (Figure 1A and Supplementary Table S8). Then, we searched for all the enzymes able to hydrolase such bonds using the KEGG (Kanehisa et al., 2012) and MetaCyc (Caspi et al., 2014) databases. Briefly, we searched the databases for both of the monosaccharides that form the bond after the reactions connected to this monosaccharide were filtered by EC number to identify all the glycoside hydrolases (i.e., enzymes with EC 3.2.-.-) for which this monosaccharide can be a substrate or product. After that, all the identified glycoside hydrolases were manually checked for the corresponding analyzed bond. Finally, we collected 9 types of GHs (Supplementary Table S8). For further information on these enzymes, such as protein families and experimentally analyzed representatives, we carried out a search by EC number in the CAZy database (Cantarel et al., 2012).

Pathways for catabolism of the derived monosaccharides were identified here as sets of reactions necessary to convert the monosaccharides into any intermediates of central metabolism. The pathway data were extracted from the KEGG (Kanehisa et al., 2012) and MetaCyc (Caspi et al., 2014) databases as well as previous publications. For the two monosaccharides GalNAc and GlcNAc, only one pathway per monosaccharide has been described (Figure 1B). GalNAc is catabolized through tagatose 6-phosphate to glyceraldehyde 3-phosphate and dihydroxyacetone phosphate (Leyn et al., 2012;Bidart et al., 2014), whereas GlcNAc is converted into fructose 6-phosphate (Afzal et al., 2015a;Plumbridge, 2015;Uhde et al., 2016). For Fuc, Gal, and Neu5Ac, two alternative pathways for the catabolism of each monosaccharide have been described. Thus, Fuc may be catabolized through fuculose 1-phosphate to lactaldehyde and dihydroxyacetone phosphate or through fucolactone to lactate and pyruvate (Yew et al., 2006;Hobbs et al., 2013). For the genes encoding enzymes of the latter pathway, no four-letter abbreviations have been designated. Thus, for these genes, we introduced the designation *fclABCDE* (from fucolactone; the last letter corresponds to the order of catalyzed reactions in the pathway). Gal catabolism can also occur through two alternative pathways: galactose 1-phosphate and UDP galactose (the Leloir pathway) (Bettenbrock and Alpert, 1998;Afzal et al., 2015b) and galactose 6-phosphate and tagatose 6-phosphate (Zeng et al., 2010). The last two steps of the second pathway, phosphorylation and aldol splitting, are shared with the pathway for GalNAc catabolism. Neu5Ac is converted to fructose 6-phosphate by two pathways, through GlcNAc or GlcNAc 6-phosphate (Vimr et al., 2004;Brigham et al., 2009). Therefore, these pathways overlap with GlcNAc by two or three reactions, respectively (Figure 1B).

### 3.2. Utilization of L-fucose

Intestinal mucin glycans contain Fuc moieties connected to Gal by α1,2-linkage as well to GlcNAc by α1,2-, α1,3-, or α1,4- linkages (Podolsky, 1985;Tailford et al., 2015). During the degradation of mucin glycans, these moieties can be removed by α-L-fucosidases (Katayama et al., 2004;Nagae et al., 2007;Ashida et al., 2009). α-L-fucosidases found in the analyzed genomes belong to three families: GH29, GH42, and GH95. At least one α-L-fucosidase was found in 131 genomes (Supplementary Table S3). Surprisingly, all of these genomes belong to only four phyla: Actinobacteria, Bacteroidetes, Firmicutes, and Verrucomicrobia. The largest number of genes encoding α-L-fucosidases were found in representatives of Firmicutes (*Lachnospiraceae bacterium* 3_1_57FAA_CT1, 16 genes) and Bacteroidetes (*Bacteroides coprophilus* DSM 18228, 14 genes).

Genes for both the alternative pathways for Fuc catabolism were found in the analyzed genomes. Thus, genes for the pathway through fuculose 1-phosphate (*fucIKA* genes) were found in the 125 genomes belonging to all studied bacterial phyla except Synergistetes and Tenericutes. The main cause of misannotation for the fuc genes is their similarity to the genes for rhamnose utilization. Both the FucK and RhaB proteins belong to the FGGY family of carbohydrate kinases (Pfam: PF02782), whereas the FucA and RhaD proteins belong to the Aldolase_II family (Pfam: PF00596). We used the following steps to distinguish *fuc* and *rha* genes. (1) The *fucI* gene was used as a signature gene for the Fuc utilization pathway because this gene has no homologs among rhamnose-catabolizing enzymes. (2) Phylogenetic maximum-likelihood trees were built for the FucK/RhaB- (Supplementary Figure S1) and FucA/RhaD-like proteins (Supplementary Figure S2) found in the analyzed genomes. (3) Genes that chromosomally clustered with *fucI* were annotated as *fucK* and *fucA*. (4) Single-copy genes found in genomes having *fucI* but lacking any *rha* genes were annotated as *fucK* and *fucA*. (5) The remaining non-annotated genes were annotated by their positions on the phylogenetic trees relative to the previously annotated genes. Thus, using both phylogeny and genomic context, Fuc catabolism through fuculose 1-phosphate was reconstructed in the analysis. The genes for the pathway through fucolactone (*fclABCDE*) were found in only 8 genomes, all belonging to Actinobacteria. Nevertheless, orthologs for all 6 previously described genes (Yew et al., 2006) were found in only 2 *Streptomyces* spp. genomes, whereas another 6 genomes had only *fclBC* genes, which were clustered together with each other and with genes for Fuc transporters (see below). A detailed analysis of this gene cluster revealed non-orthologous replacements for the *fclADE* genes. First, the *fcl* gene clusters contain the protein homologous to 4-hydroxy-tetrahydrodipicolinate synthase and belonging to the DHDPS protein family (Pfam: PF00701). This family includes enzymes such as N-acetylneuraminate lyase and *trans*-o-hydroxybenzylidenepyruvate hydratase-aldolase, which produces pyruvate as one of its products. Because pyruvate is also a product of the FclE-catalyzed reaction (Figure 1B), we concluded that this DHDPS family protein is an alternative form of FclE and named it FclE2.

The Fcl pathway contains two dehydrogenases, L-fucose dehydrogenase FclA and 2-keto-3-deoxy-L-fuconate dehydrogenase FclD. These dehydrogenases were previously described in *Xanthomonas campestris* (Yew et al., 2006) and *Burkholderia multivorans* (Hobbs et al., 2013). In *X. campestris*, FucA belongs to the aldo/keto reductase family (Pfam: PF00248), whereas FucD belongs to the short-chain dehydrogenase family (Pfam: PF00106). On the other hand, in *B. multivorans*, both FucA and FucD belong to the short-chain dehydrogenase family. The proteins belonging to the short-chain dehydrogenase family were also encoded in the *fcl* gene cluster in genomes lacking the *fclAD* genes. To predict their functions, we constructed phylogenetic trees for these proteins and experimentally analyzed the FclA and FclD proteins belonging to the short-chain dehydrogenase family (Supplementary Figure S3). As expected, in the resulting phylogenetic tree, the FclD proteins from *X. campestris*, *B. multivorans* and *Streptomyces* spp. were clustered together, forming a monophyletic branch. The proteins from *C. ammoniagenes* and *Bifidobacterium* spp. formed a monophyletic branch separated from both FclD proteins and the branch for *B. multivorans* FclA protein. Thus, we proposed that the proteins from *C. ammoniagenes* and *Bifidobacterium* spp. are bifunctional enzymes having both FclA and FclD activities. This assumption indirectly corroborated that only one copy of these proteins was found in the *fcl* gene clusters. Thus, proteins from *C. ammoniagenes* and *Bifidobacterium* spp. were designated FclA2/FclD2.

In total, three different Fuc-specific transport systems were found in the analyzed genomes, including two different permeases and one ABC transporter. The first permease, herein referred to as FucP1, was previously analyzed in *Escherichia coli* (Gunn et al., 1994) and predicted in the genomes of *Bacteroides* spp. (Hooper et al., 1999;Ravcheev et al., 2013). This transporter is highly distributed in the analyzed genomes, co-clustering with the *fuc* genes in 88 genomes and with the *fcl* genes in 4 genomes. An alternative Fuc permease named FucP2 belongs to the Sugar_tr family (Pfam: PF00083). The gene encoding this transporter is located inside the *fuc* gene cluster in the genomes of *Pediococcus acidilactici* and *Lactobacillus rhamnosus*. A third transport system, the ABC-type system, was predicted in this study. The closest experimentally analyzed homolog of the substrate-binding subunit of this system is the substrate-binding protein AraF of the arabinose-specific ABC system *from E. coli* (Johnson and Schleif, 2000). Predicted Fuc-specific transport systems have been predicted by chromosomal clustering with the *fuc* genes in 25 analyzed genomes, including *Actinomyces* spp.,

*Clostridium* spp., *Coprococcus* spp., *Lachnospiraceae bacterium*, *Ruminococcus* spp., and *Enterobacter* spp. Additionally, genes of this system co-cluster together with the *fcl* genes in the genomes of *Streptomyces* sp. HGB0020, *Bifidobacterium longum* ATCC 15697, and *Bifidobacterium pseudocatenulatum* DSM 20438.

Generally, α-L-fucosidases were found in 131 analyzed genomes, whereas Fuc CPs were found in 133 genomes. Both α-fucosidases and CPs were found together in 72 analyzed genomes (Figure 2A, Supplementary Table S3).

### 3.3. Utilization of N-acetyl-D-galactosamine

GalNAc plays a crucial role in mucin glycans, forming links with side-chain oxygen atoms of Ser/Thr residues in mucin peptide chains. This linkage between the glycan and peptide parts of mucin is mediated by various endo-α-N-acetylgalactosaminidases (Ashida et al., 2008;Koutsioulis et al., 2008;Kiyohara et al., 2012). Additionally, in intestinal mucin glycans, GalNAc moieties can be connected to Gal by α1,3- or β1,4-linkage and to GalNAc by α1,6- or β1,3-linkage (Podolsky, 1985;Tailford et al., 2015). Release of GalNAc from mucin glycans can be mediated by β-N-hexosaminidases (Cabezas, 1989;Zwierz et al., 1999) and exo-α-N-acetylgalactosaminidases (Hoskins et al., 1997).

At least one GalNAc-releasing GH (GalNAc-GH) was found in 218 genomes (Supplementary Table S4). No GalNAc-GHs were found in the genomes of Archaea or in the bacterial phyla Fusobacteria and Synergistetes. The maximal number of genes for the GalNAc-GHs was found in *Bacteroides* sp. 1_1_6 and *Bacteroides* sp. D22. Both of these organisms have 17 GalNAc-GH genes per genome. All exo-α-N-acetylgalactosaminidases in the analyzed genomes belong to the GH27 family. This type of GalNAc-GHs were found in only 13 genomes belonging to only four phyla: Actinobacteria, Bacteroidetes, Firmicutes, and Proteobacteria. Endo-α-N-acetylgalactosaminidases are also rarely represented in the analyzed genomes. These enzymes, belonging to the GH101 or GH129 families, were found in only 15 genomes from only two phyla: Actinobacteria and Firmicutes. In contrast, β-N-hexosaminidases were found in 217 genomes and in all bacterial phyla except Fusobacteria and Synergistetes. All the β-N-hexosaminidases found in the analyzed genomes belong to the GH3 or GH20 families.

It was previously found that at least some Firmicutes and Proteobacteria can utilize only galactosamine (GalN) but not GalNAc (Leyn et al., 2012;Zhang et al., 2015). The crucial feature of the GalNAc-utilizing microorganisms is the presence of the *agaA* gene for N-acetylgalactosamine-6-phosphate deacetylase. For proper pathway reconstruction, AgaA should be distinguished from the N-acetylglucosamine-6-phosphate deacetylase NagA, which is involved in GlcNAc utilization. For this purpose, we used the following pipeline. (1) The *agaS/agaI* genes were used as signature genes for the GalNAc utilization pathway, and *nagB* was used as a signature gene for the GlcNAc utilization pathway. These genes were selected because AgaS and AgaI have no homologs in the GlcNAc utilization pathway and NagB has no homologs among the GalNAc catabolic enzymes. (2) A phylogenetic maximum-likelihood tree was built for AgaA/NagA-like proteins (Supplementary Figure S4) found in the analyzed genomes. (3) Genes chromosomally clustered with *agaS* or *agaI* were annotated as *agaA*, while genes chromosomally clustered with *nagB* were annotated as *nagA*. (4) Genes that co-occurred in the genomes with only *agaS* or *agaI* were annotated as *agaA*, and genes that co-occurred in the genomes with only *nagB* were annotated as *nagA*. (5) The remaining non-annotated genes were annotated by their positions on the phylogenetic trees relative to the previously annotated genes.

GalNAc can be phosphorylated by the kinase AgaK and by the phosphotransferase transport system (PTS). No AgaK genes were found in the analyzed genomes with the GalNAc CPs. Three PTSs associated with *aga* genes were previously identified, including a GalNAc-specific (AgaPTS), a GalN-specific (GamPTS), and a GnbPTS with multiple specificities (Leyn et al., 2012;Bidart et al., 2014;Zhang et al., 2015). The GnbPTS can transport and phosphorylate three different compounds: GalNAc, lacto-N-biose (Galβ-1,3-GlcNAc), and galacto-N-biose (Galβ-1,3-GalNAc). Both lacto-N-biose and galacto-N-biose contain the glycosyl bonds found in intestinal mucin glycans (Supplementary Table S1). Thus, GnbPTS was included in the three analyzed pathways as GalNAc, GlcNAc, and Gal catabolism. Lacto- and galacto-N-biose then are hydrolyzed by intracellular GH GnbG. Formally, GnbG is a GH, but because of its intracellular localization here, it is considered to be a part of the GalNAc, GlcNAc and Gal CPs.

Thus, AgaPTS and GnbPTS but not GamPTS are involved in GalNAc catabolism. We used the following steps to distinguish the various types of PTSs associated with *aga* genes. (1) A phylogenetic tree for the PTS component EIIC was constructed (Supplementary Figure S5). (2) The systems that co-clustered with the *gnbG* gene were annotated as GnbPTSs. (3) The systems that co-clustered with the *agaS/agaI* and *agaA* genes were annotated as AgaPTS. (4) The systems that co-clustered with the *agaS/agaI* genes but not the *agaA* genes were annotated as GamPTS.

In addition to GamPTS, lacto- and galacto-N-biose can be imported into the cell by the ABC transport system (LnbABC in Supplementary Tables S2, S4-5, and S7) and then hydrolyzed by intracellular phosphorylase LnbP (Nishimoto and Kitaoka, 2007). As with GamPTS-GamB, the LnbABC-LnbP system was considered to be a part of the GalNAc, GlcNAc and Gal CPs.

Generally, the GalNAc CP was found in the 76 analyzed genomes belonging to the phyla Actinobacteria, Firmicutes, and Proteobacteria (Figure 2A, Supplementary Table S4). Distribution of the GalNAc-releasing GHs is much broader; these enzymes were found in 218 genomes. Both GHs and CPs were found in only 27 of analyzed genomes.

### 3.4. Utilization of N-acetyl-D-glucosamine

Intestinal mucin glycans contain GlcNAc moieties that form various glycosyl bonds, such as α1,4- and β1,3-linkages with Gal and β1,3- and β1,6-linkages with GalNAc (Podolsky, 1985;Tailford et al., 2015). The α-linkages are hydrolyzed by α-N-acetylglucosaminidases (Shimada et al., 2015), whereas the β-linkages with Gal or GalNAc are hydrolyzed by β-N-hexosaminidases (see subsection **3.3**).

At least one GlcNAc-releasing GH (GlcNAc-GH) was found in 257 genomes (Supplementary Table S5). The maximal numbers of the genes encoding GlcNAc-GHs were found in *Bacteroides* sp. 1_1_14 (23 genes) and *B. thetaiotaomicron* (22 genes). In addition to β-N-hexosaminidases (see subsection **3.3**), α-N-acetylglucosaminidases (GH89 family) were found in 60 genomes (Supplementary Table S5). All genomes in which α-N-acetylglucosaminidase genes were found belong to the phyla Actinobacteria, Bacteroidetes, Firmicutes, Proteobacteria, and Verrucomicrobia.

The GlcNAc CP (Yang et al., 2006;Plumbridge, 2015) was found in 218 of the analyzed genomes (Supplementary Table S5). This pathway is broadly distributed among analyzed taxa and is absent only in Archaea and the bacterial phyla Synergistetes and Verrucomicrobia. Various GlcNAc-specific transport systems were identified in the analyzed genomes. These are the PTSs NagE (Plumbridge et al., 1993;Plumbridge, 2015) and GnbPTS (see subsection **3.3**), the ABC transport systems NgcEFG (Xiao et al., 2002) and LnbABC (Nishimoto and Kitaoka, 2007), and the permease NagP (Ravcheev et al., 2013). Here, in addition to these previously known transporters, we predicted an alternative ABC transporter for GlcNAc, which was named NgcABCD. Genes for this system co-cluster together with the *nagKAB* genes in the genomes of *Bifidobacterium* spp. The following differences exist between NgcEFG and NgcABCD. First, the *ngcEFG* operon encodes only the substrate-binding protein and two intermembrane proteins, whereas the *ngcABCD* operon encodes an additional protein, an ATP-binding protein. Second, the substrate-binding subunits of these systems belong to different protein families; NgcE is a member of the SBP_bac_8 family (Pfam: PF13416), whereas NgcA is a member of the SBP_bac_5 family (Pfam: PF00496).

Generally, GlcNAc CP was found in 218 analyzed genomes, whereas GlcNAc-GHs were found in 266 genomes. Both the pathway and the GHs were found in 155 analyzed genomes (Figure 2, Supplementary Table S5).

### 3.5. Utilization of N-acetyl-D-neuraminic acid

Neu5Ac is commonly found in the terminal location of intestinal mucin glycans (Johansson et al., 2011;McGuckin et al., 2011), forming α2,3-linkages with Gal and α2,6-linkages with Gal or GalNAc (Tailford et al., 2015). These bonds can be hydrolyzed by sialidases (Juge et al., 2016). Sialidases, all belonging to the GH33 family, were found in 112 analyzed genomes. Sialidases were found in the genomes of all bacterial phyla except Fusobacteria and Synergistetes. The maximal number of sialidase-encoding genes was found in *Bacteroides fragilis* 638R and *Bacteroides* sp. D22 genomes (7 genes per genome).

Two CPs for Neu5Ac have been described previously (Vimr et al., 2004;Brigham et al., 2009). These pathways can be distinguished by the sugar epimerases; the first pathway is characterized by the presence of N-acetylmannosamine-6-phosphate 2-epimerase (NanE), whereas the second pathway is characterized by the presence of N-acetylglucosamine 2-epimerase (NanE-II) (Figure 1B). These epimerases are homologous to each other, and to distinguish them, we constructed a maximum-likelihood phylogenetic tree (Supplementary Figure S6). Nonetheless, NanE and NanE-II are homologs; they form clearly distinguishable monophyletic branches on the tree. Thus, the first Neu5Ac CP, with the NanE enzyme, was found in 131 analyzed genomes, belonging to the phyla Actinobacteria, Firmicutes, Fusobacteria, and Proteobacteria. The second pathway, with the NanE-II enzyme, was found in 56 genomes. Most of these genomes are Bacteroidetes, only 3 of them belong to Firmicutes, and *Akkermansia muciniphila* is a representative of Verrucomicrobia.

The CPs were found in 187 analyzed genomes, belonging to all bacterial phyla with the exception of Synergistetes and Tenericutes. For Neu5Ac, various types of transporters have been previously described (Thomas, 2016). In the analyzed genomes, we identified the following Neu5Ac transporters: an MFS-type transporter (NanT), a sodium solute symporter (NanX), two ABC transport systems (NanABC and NanABC2), and a TRAP transport system (NeuT) (Supplementary Table S7). No novel genes for Neu5Ac transport or catabolism were predicted.

Generally, the Neu5Ac CPs were found in 189 analyzed genomes, while both the pathways and the sialidases were found in only 80 analyzed genomes.

### 3.6. Utilization of D-galactose

In human intestinal mucin glycans, Gal can form α1,3-linkages with other Gal, β1,3-linkages with GalNAc and GlcNAc, and β1,4-linkages with GlcNAc (Podolsky, 1985;Tailford et al., 2015). These linkages can be hydrolyzed by various α- (Wakinaka et al., 2013;Han et al., 2014;Reddy et al., 2016) and β-galactosidases (Husain, 2010;Park and Oh, 2010;Michlmayr and Kneifel, 2014;Solomon et al., 2015). At least one galactosidase gene was found in 310 analyzed genomes, belonging to all bacterial phyla except Synergistetes. The maximal numbers of galactosidase genes were found in *Bacteroides cellulosilyticus* DSM 14838 (56 genes) and *Bacteroides* sp. D2 (50 genes). The α-galactosidases found in the analyzed genomes belong to families GH4, GH27, GH36, and GH43, whereas the β-galactosidases belong to families GH2, GH35, and GH42.

For Gal utilization, two alternative pathways are possible. The Leloir pathway, in which Gal is utilized through UDP galactose (Bettenbrock and Alpert, 1998;Afzal et al., 2015b), was found in 335 analyzed genomes belonging to all bacterial phyla except Synergistetes. The alternative pathway, the utilization of Gal through tagatose 6-phosphate (Zeng et al., 2010), was found in only 29 analyzed genomes, belonging to Actinobacteria and Firmicutes. Surprisingly, genes for the Leloir pathway were also found in all 29 of these genomes.

Analysis of the Leloir pathway revealed that 79 of the analyzed genomes have *galK* genes for galactokinase but lack *galU* and *galT* genes for uridylyltransferases. To predict non-orthologous displacement of these genes, we used an analysis of gene occurrence also known as "Phyletic patterns" (Osterman and Overbeek, 2003;Tatusov et al., 2003). For prediction of the non-orthologous displacement of *galT*, two sets of genomes were selected. The first set included finished genomes having *galK* but not *galT*. The second set included finished genomes having both *galK* and *galT* (Supplementary Table S9). Finished genomes were selected because draft genomes do not allow us to distinguish the actual absence of the gene in the genome or the location of the gene in an unsequenced part of the genome. Candidate functional analogs of GalT were identified as genes present in all genomes having only *galK* but absent in genomes having *galKT* genes. This gene search resulted in 62 candidates (Supplementary Table S9). All of these candidates, with the exception of membrane, regulatory, and secreted proteins, were then checked for additional confirmation of their role in *galT* non-orthologous displacement (Supplementary Figure S7). This check included an analysis of sequence similarity of these proteins to known proteins and protein families as well as an analysis of their chromosomal co-clustering with previously annotated genes. The best candidate was the gene *Amuc*_0031 in the *A. muciniphila* genome. First, the encoded protein belongs to the nucleotidyl transferase (NTP_transferase, Pfam: PF00483) family. Members of the NTP_transferase family are able to transform phosphosugars onto nucleotide sugars (Jensen and Reeves, 1998). Second, in *A. muciniphila,* this gene was co-clustered with *galE*. Homologs of this gene were found in all analyzed genomes, and a phylogenetic maximum-likelihood tree was constructed (Supplementary Figure S7). Analysis of the tree together with the gene occurrence patterns showed that all homologs found in genomes lacking *galT* form a monophyletic branch. Additionally, proteins belonging to this branch are often co-clustered with genes for Gal metabolism, such as the *galE* gene, the *galMP* gene, or the a- and b-galactosidase genes. Thus, the identified genes were predicted to be functional analogs of *galT* and were designated *galY*. The *galY* genes were found in 70 genomes (Supplementary Table S7), including 68 genomes of Bacteroidetes as well as single representatives of Verrucomicrobia (*A. muciniphila*) and Firmicutes (*Holdemania filiformis* DSM 12042). Unfortunately, analysis of gene occurrence did not provide any results for non-orthologous displacement of *galU*.

A number of previously known Gal transporters were found in the analyzed genomes. These are galactose (Essenberg et al., 1997) and galactose/lactose (Luesink et al., 1998) permeases, here referred to as GalP1 and GalP2, respectively; the galactose/methyl galactoside ABC transport system Mgl (Weickert and Adhya, 1993); and the galactose-specific PTS (Zeng et al., 2012). Additionally, Gal can be imported into the cell as part of lacto-N-biose and galacto-N-biose by the GnbPTS and/or LnbABC systems (see subsections **3.3** and **3.4**). Additionally, we predicted a new Gal-specific transporter, here referred to as GalP3. This transporter is a member of the SSS family (sodium solute symporter, Pfam: PF00474). It was initially predicted due to co-clustering of its gene together with *galK* in the genomes of *Propionibacterium* spp. and with *galKT* in the genomes of *Streptomyces* spp. Further analysis revealed the presence of this transporter in 43 analyzed genomes. In most of these genomes, the gene for the novel transporter is co-clustered with the genes for the Leloir pathway or galactosidases (Supplementary Table S7). Generally, Gal CPs were found in 355 analyzed genomes, whereas both Gal-GHs and the pathways were found in 296 genomes.

## 4. Discussion

### 4.1. Distribution of GHs and CPs in the analyzed HGM genomes

In this study, we analyzed the distribution of genes required for utilization of mucin glycans in 397 genomes of microbes found in the human gastrointestinal tract. The analyzed genes encode extracellular enzymes for cleavage of mucin glycans to monosaccharides as well as transport proteins and enzymes for subsequent utilization of the derived monosaccharides. These genes were conditionally divided into five groups, representing the five monosaccharides found in mucin glycans. Each of these monosaccharide-specific groups of genes can, in turn, be subdivided into extracellular GHs (Figure 1A) and genes for monosaccharide-utilizing CPs (Figure 1B). Analyzing the distribution of glycan-utilizing genes, we observed two general trends: (1) numerous genomes contain only CPs or only GHs and (2) pathways for different monosaccharides are distributed very differently from each other (Figure 2).

The presence of only CPs or only GHs has been previously described for the metabolism of Fuc (Pacheco et al., 2012;Conway and Cohen, 2015) or Neu5Ac (Ng et al., 2013) in the HGM. Nonetheless, such partial pathways have been previously described for only some HGM genomes. Here, our results demonstrated that such partial pathways are characteristic for the degradation of mucin glycans by the HGM. Thus, 339 (85%) of the analyzed genomes have a partial pathway for at least one of the mucin-derived monosaccharides. Generally, 726 partial pathways were found, including 312 cases with the presence of only CPs and 414 cases with the presence of only GHs (Supplementary Table S10). Such a wide distribution of partial pathways indicates the existence of multiple exchange pathways for mucin glycan-derived monosaccharides in the HGM (see subsection **4.2**).

A distinct distribution of pathways for utilization of the analyzed monosaccharides has not been described previously at the level of large microbial communities, such as the HGM. At the level of monosaccharide-specific CPs, sorting by the number of genomes where they were found produces the following sequence: Gal > GlcNAc > Neu5Ac > Fuc > GalNAc. A similar sequence was found for the GHs: Gal > GlcNAc > GalNAc > Fuc > Neu5Ac. The differences between these two sequences can be explained as follows. (1) GHs specific for Fuc and Neu5Ac were found in 72 and 80 genomes, respectively. Therefore, their switching places may be the result of a minor bias caused by the selection of the analyzed genomes. Because only genomes available in the HMP, KEGG and PubSEED databases were analyzed (see subsection **2.2**), the set of selected genomes is slightly biased. Thus, the set of analyzed genomes is enriched by genomes of Proteobacteria, especially by Enterobacteriales. (2) This difference also indicates that gastric mucin glycans are mainly neutral and that sialylation of them is quite rare (Rossez et al., 2012). (3) The higher number of genomes with GalNAc-GHs in comparison with Fuc- and Neu5Ac-GHs can be explained by the fact that the majority of GalNAc-GHs are β-hexosaminidases, which are enzymes specific to GalNAc and GlcNAc. This explanation is in good agreement with the large number of genomes having GlcNAc-specific CPs and/or GHs.

The different distributions of pathways for different monosaccharides may be due to their dissemination in nature, particularly in the human intestine. Therefore, in animals, Fuc and Neu5Ac are found mostly on terminal units of carbohydrate chains linked to proteins or lipids (Staudacher et al., 1999;Vimr, 2013;Pickard and Chervonsky, 2015). GalNAc is a slightly more disseminated since it is not only a terminal monosaccharide but also acts as a connecting link between the glycan and protein portions of N- or O-linked proteoglycans (Ashida et al., 2008;Koutsioulis et al., 2008;Kilcoyne et al., 2012;Kiyohara et al., 2012). Thus, in the human intestine, these three monosaccharides are parts of host-synthesized glycans or are derived from dietary components of animal origin. GlcNAc is much more broadly disseminated and, in addition to being found in glycoproteins, it can be found in heparin, chondroitin sulfate, hyaluronan, and various human milk oligosaccharides (HMOs) (Lamberg and Stoolmiller, 1974;Garrido et al., 2011). Gal is even more broadly disseminated than GlcNAc and can be found in both animal- or plant-synthesized polysaccharides. Thus, Gal is a building block of arabinogalactan, pectic galactan, and HMOs. As a component of side chains, Gal may be found in type II rhamnogalacturonan and as a terminal unit in α-mannans, galactomannan, xyloglucan, and xylan (Kunz et al., 2000;Mohnen, 2008;Garrido et al., 2011;Marcobal et al., 2011).

To confirm that the distribution of monosaccharide-utilizing pathways and monosaccharide-specific GHs reflects dissemination of these monosaccharides in nature, we analyzed data for these monosaccharides from the KEGG database. For each of the monosaccharides, two parameters were analyzed: (1) the number of reactions that the monosaccharide is involved in as a product or a substrate and (2) the number of glycans containing this monosaccharide (Supplementary Figure S8). The numbers of reactions and glycans for the monosaccharides are in line with the distribution of CPs and GHs in the analyzed genomes. Thus, analysis of the numbers of reactions for each monosaccharide resulted in the following sequence: Gal > GlcNAc > Neu5Ac > Fuc > GalNAc, which is identical to the sequence for genomes with monosaccharide-specific CPs. For the number of glycans, the sequence is as follows: Gal > GlcNAc > Fuc > GalNAc > Neu5Ac, which is similar to the sequence for genomes with monosaccharide-specific GHs.

Examining the variation of the monosaccharide-specific pathways, we wondered how the combinations of the utilized monosaccharides varied across the HGM genomes. Because only CPs or only GHs were found in 85% of the genomes, combinations of CPs and GHs were analyzed separately. We used binary information regarding the distribution of pathways in 397 HGM genomes, i.e., the presence or absence of a CP or GH in a genome (Supplementary Table S10). We investigated the $2^5=32$ possible patterns of the eight studied pathways. Only 22 (69%) and 19 (60%) of the 32 possible patterns were found for CPs and GHs, respectively. The most frequent pattern for CPs represents catabolism of only Gal and GlcNAc. This pattern was found in 59 analyzed genomes, including Bifidobacteriacea, various Firmicutes, and Fusobacteria. Other frequently observed CP patterns are as follows: (1) absence of CPs for any analyzed monosaccharides, (2) utilization of all monosaccharides except GalNAc, (3) utilization of Gal only, and (4) utilization of Gal and Neu5Ac only. The most frequent pattern for GHs is the presence of GHs specific to all five monosaccharides. This pattern was found in 81 genomes, including some Actinobacteria, multiple Bacteroidetes, Firmicutes (mostly belonging to the Lachnospiraceae family), and *A. muciniphila*. Other frequently observed GH patterns are as follows: (1) presence of GHs specific to GalNAc, GlcNAc, and Gal; (2) absence of GHs for any analyzed monosaccharides; (3) presence of only Gal-specific GHs; and (4) presence of only GlcNAc- and Gal-specific GHs.

Theoretically, the combination of the observed CP and GH patterns should result in 22*19=418 combined patterns; however, only 102 (24%) of them were actually observed, which indicates interdependence of GHs and CPs. This interdependence appears to be rather trivial because only non-digestible carbohydrates are available for HGM organisms, especially in the large intestine (Walker et al., 2011), so CPs are highly dependent on GHs. On the other hand, no significant correlations were observed between CP and GH patterns. Together with CP dependence on GH repertoire, this absence of correlations again indicates intensive exchange by mucin-derived monosaccharides in the HGM.

Surprisingly, the combined pattern corresponding to the most frequent GH and CP patterns was found in only three genomes: *Clostridium nexile* DSM 1787, *Lachnospiraceae bacterium* 2_1_46FAA, and *Ruminococcus lactaris* ATCC 29176. The combined pattern we observed most frequently was the absence of all analyzed GHs and CPs. This pattern was found in 28 genomes, belonging to Archaea, some Firmicutes, and Beta- and Epsilonproteobacteria. Other frequently observed combined patterns were as follows: (1) presence of GHs specific for GalNAc, GlcNAc, and Gal together with CPs for GlcNAc and Gal; (2) presence of GHs for all five monosaccharides together with utilization of all these monosaccharides except GalNAc; and (3) presence of GHs for all five monosaccharides together with utilization of Neu5Ac and Gal.

Taken together, an optimal strategy for glycan-utilizing HGM microorganisms includes (1) the presence of CPs specific to GlcNAc and Gal as the components of multiple host- and dietary-derived carbohydrates and (2) the presence of GHs specific to as large as possible a number of glycan-building monosaccharides.

## 4.2. Exchange pathways for the analyzed HGM genomes

The presence of only CPs for a certain monosaccharide in one HGM organism and only GHs for this monosaccharide in another organism allows us to predict possible exchange pathways. Previously, exchange pathways in the HGM have been found for Fuc (Pacheco et al., 2012;Conway and Cohen, 2015) and Neu5Ac (Ng et al., 2013). Here, we predicted multiple exchange pathways for all five monosaccharides forming mucin glycans. The 339 (85%) analyzed genomes demonstrated the presence of only CPs or only GHs for at least one monosaccharide; therefore, the majority of HGM organisms are involved in these exchange pathways.

Based on the presence or absence of CPs and GHs, each organism can be classified as a "donor" (having GHs but not CPs) or "acceptor" (having CPs but not GHs) in relation to a certain monosaccharide. Among the analyzed genomes, 181 (46%) organisms can only be "donors" while 103 (33%) organisms can only be "acceptors" for the studied monosaccharides (Supplementary Table S10). Additionally, 55 (14%) organisms were classified as "mixed", being "donors" for some monosaccharides and "acceptors" for others.

In summary (Figure 3 and Supplementary Table S11), data on possible exchange pathways demonstrate the following features of exchange of mucin-derived monosaccharides in the HGM: (1) larger number of "donors" relative to "acceptors" and (2) taxonomy-specific distribution of donors and acceptors.

Generally, 414 "donor" roles were found to be distributed among 236 genomes, whereas 312 "acceptor" roles were found to be distributed among 158 genomes. The significant predominance of "donors" reflects an adaptation of the analyzed organisms to the environment of the human intestine. No free monosaccharides are available in the large intestine (Walker et al., 2011), so the only way for HGM organisms to obtain monosaccharides is from the cleavage of polysaccharides that are non-digestible for the host. Thus, HGM organisms should have GHs not only for monosaccharides that they can catabolize but also for all monosaccharides present in the available polysaccharides. Additionally, the appearance of a new "donor" role during evolution is much more probable than the appearance of a new "acceptor" role. With the appearance of a "donor", only one new function should be acquired through horizontal gene transfer or duplication and neofunctionalization. Indeed, horizontal transfer has been previously demonstrated for genes encoding sialidase (Roggentin et al., 1993). In contrast, the appearance of a new "acceptor" requires horizontal transfer of multiple genes for enzymes and transporters. Because genes for CPs are not always encoded as a single locus (Leyn et al., 2012;Ravcheev et al., 2013;Rodionov et al., 2013;Khoroshkin et al., 2016), horizontal transfer of all genes required for the catabolism of a certain monosaccharide would be nearly impossible. Similarly, the appearance of a new CP through duplication and neofunctionalization requires the simultaneous change of specificities for multiple genes, which is also very unlikely.

The distribution of "donor" and "acceptor" roles among the analyzed genomes is taxon-specific (Figure 3). Thus, no "donors" were found among Fusobacteria, and no "acceptors" were found among Tenericutes and Verrumicrobia. Among Bacteroidetes, only 2 strains of *Bacteroides eggerthii* were identified as "acceptors", whereas 62 strains from this group were identified as "donors". The opposite is observed for Proteobacteria, for which only 9 genomes can be "donors" and 51 genomes can be "acceptors". At the phylum level, the most intensive exchange of monosaccharides should be between "donor" Bacteroidetes and "acceptor" Proteobacteria. In some cases, up to 4 different monosaccharides are able to be exchanged. Thereby, exchange pathways in the HGM are formed by the interaction of two factors, (1) evolutionary history of microbial taxa and (2) adaptation of HGM organisms to the intestinal environment.

## 4.3. Specificity of the HGM organism to mucin glycans

Mucin glycans are complex polysaccharides characterized by a variety of monosaccharide building blocks and bonds between these monosaccharides (Martens et al., 2009;Koropatkin et al., 2012;Rossez et al., 2012;Johansson et al., 2015;Tailford et al., 2015). Tens of mucin glycan motifs and structures have been described to date. These structures vary in terms of both monosaccharides and glycosidic bonds (Podolsky, 1985;Rossez et al., 2012;Tailford et al., 2015). Furthermore, the analyzed genomes vary in the patterns of GHs able to cleave mucin glycans (Supplementary Table S10). Thus, we proposed that the analyzed organisms should demonstrate some preferences for cleaved mucin glycans.

Data on the known structures of mucin glycans found in the human intestine were collected from the literature (Podolsky, 1985;Rossez et al., 2012), resulting in 56 different glycan structures (Supplementary Figure S9 and Supplementary Table S12). For each analyzed genome, the ability to cleave each glycan structure was predicted. Glycan was considered able to be cleaved by a certain organism if the GHs for all glycoside bonds in the glycan were found in the genome of the organism. Bonds between GalNAc and Ser/Thr residues of the mucin peptide were excluded from the analysis because this bond is cleaved by endo-α-N-acetylgalactosaminidases found in only 15 of the analyzed genomes (Supplementary Table S4). On the basis of this prediction for each genome, the pattern of likely cleaved glycans was determined (Supplementary Table S13). Among 397 analyzed genomes, 321 (81%) were able to cleave at least one of the glycans; generally, 20 different glycan-cleavage patterns were defined. It has previously been estimated that approximately 40% of bacteria have glycan-degrading enzymes (Arike and Hansson, 2016). Here, we demonstrated that, at least for the HGM microbes, this figure is actually at least two-fold higher.

Based on the glycan-cleavage patterns, only 8 analyzed organisms are able to cleave all 56 glycan structures belonging to the phyla Bacteroidetes (*Bacteroides ovatus* SD CMC 3f, *Bacteroides* sp. 2_2_4, and *Bacteroides* sp. 3_1_23) and Firmicutes (*Clostridium perfringens* WAL-14572, 3 strains of *Lachnospiraceae bacterium*, and *Ruminococcus torques* ATCC 27756). The three most frequently observed patterns were the following. (1) Only Core 1 (Tailford et al., 2015) structures can be cleaved, i.e., no GHs except β-galactosidases are present. This pattern was found in 77 genomes belonging mostly to Lactobacillaceae, Ruminococcaceae and Enterobacteriales. (2) Glycans having only poly-lacto-N-biose but lacking any specific groups (Supplementary Table S12) can be cleaved. These genomes have only GHs for the hydrolysis of β-Gal and β-GlcNAc bonds. This pattern was found in 56 genomes belonging mostly to Actinobacteria, Firmicutes and Enterobacteriales. (3) All glycans lacking α-GalNAc groups can be cleaved. These genomes have all GHs but not α-N-acetylgalactosaminidases. This pattern was found in 41 genomes belonging mostly to Bacteroidaceae.

The predicted glycan-cleavage patterns demonstrate some taxon-specific features. For example, for 66 genomes of glycan-cleaving Bacteroidetes, only 8 patterns were identified, whereas for 46 genomes of glycan-cleaving Proteobacteria, only 4 patterns were identified. Thus, the glycan-cleavage abilities of HGM organisms also depend on the evolutionary history of the taxon, whereas the correlation between these abilities and taxonomy is not so strong. On the other hand, mucin glycans can also be cleaved by multistrain communities in which each strain hydrolyzes a part of the glycosyl bonds (Figure 5).

A minimal multistrain community should include two different strains. Here, we proposed the existence of mutualistic pairs of HGM organisms. We defined a mutualistic pair as a pair of organisms that can cleave more mucin glycans than a union of the glycan-cleavage patterns of these two organisms. To identify such mutualistic pairs, we used the following procedure. (1) All possible pairs of analyzed genomes were analyzed with the exception of pairs in which at least one of the organisms can cleave all 56 glycan structures (Supplementary Figure S9). (2) For each of the analyzed pairs, two glycan-cleavage patterns were predicted. (3) The so-called "summary pattern" was defined as a union of the patterns for each organism in the pair. (4) The so-called "mutualistic pattern" was defined as a pattern predicted on the basis of the GHs present in both genomes. (5) For a pair of genomes, the common set of GHs was considered to be the sum of GHs found in each genome. (6) For the common set of GHs, the ability to cleave each glycan structure was predicted. (7) All the glycans able to be cleaved by the common set of GHs formed the "mutualistic pattern" for this pair of organisms. (8) If the "mutualistic pattern" contained glycans absent in the "summary pattern", this pair of organisms was considered mutualistic. In total, 325 (82%) analyzed genomes were able to form mutualistic pairs (Figure 5 and Supplementary Table S14).

The identified mutualistic pairs able to cleave all 56 glycan structures were considered to be highly beneficial. Thus, in agreement with predicted glycan-cleavage patterns, *Collinsella stercoris* DSM 13279 and *Clostridium spiroforme* DSM 1552 can cleave 27 different glycans each, *Eubacterium dolichum* DSM 3991 can cleave 9 glycans, and *Cedecea davisae* DSM 4568 cannot cleave any mucin glycans. Each of these genomes can form highly beneficial pairs with various *Bacteroides* spp. that alone can cleave 40 to 41 glycans. *Clostridium celatum* DSM 1785 has a glycan-cleavage pattern that includes 50 glycans. This genome can form highly beneficial pairs with *Bifidobacterium* spp. that can cleave 2 to 22 glycans, with *Lactobacillus* spp. that can cleave 0 to 22 glycans, and with Enterobacteriaceae that can cleave 0 to 9 glycans. All highly beneficial pairs described are organized in a similar manner. The five strains listed above are distantly related to each other, and three of them, *C. celatum*, *C. davisae*, and *C. spiroforme*, are pathogens (Akinosoglou et al., 2012;Papatheodorou et al., 2012;Agergaard et al., 2016), whereas no data about the pathogenicity of *C. stercoris* and *E. dolichum* were found. Each of these five strains forms highly beneficial pairs with a large number of organisms closely related to each other and highly represented in healthy HGM (Eckburg et al., 2005;Goodman et al., 2011;Walker et al., 2011;Graf et al., 2015). Based on these features of highly beneficial pairs, we proposed that these five organisms can be harmful to human health not only due to pathogenicity itself but also because they can greatly increase the ability of the HGM to forage the host mucus layer.

## 4.4. Unresolved problems and possible solutions

This study resulted in the prediction of a number of novel genes involved in utilization of human mucin glycans. Nonetheless, some problems related to monosaccharide utilization remain unresolved (Supplementary Tables S3-S8). These problems are the absence of one or two steps of certain CPs as well as the absence of known transporters in the presence of a corresponding CP. At least one such problem was detected for 90 (23%) of the analyzed genomes. The most frequently observed problems are as follows: (1) the absence of known Gal transporters in 30 genomes, mostly Firmicutes; (2) the absence of L-fuculose phosphate aldolase in 23 genomes belonging to Bacteroidetes, Clostridia, and some Actinobacteria; (3) the absence of L-fuculokinase in 19 genomes belonging to Bacteroidetes, Clostridia, and some Actinobacteria; and (4) the absence of galactose kinase, which was observed in 7 genomes of Firmicutes. Other problems have been observed in 1 to 6 analyzed genomes.

These unresolved problems can be explained by three non-exclusive hypotheses: (1) the incompleteness of genome sequences, (2) non-orthologous replacements for enzymes and transporters, and (3) the existence of alternative reactions and pathways. A total of 326 (82%) of the analyzed genomes have draft status, and some genes for the transport and utilization of monosaccharides may thus be absent from the current version of the genome. Indeed, 77 genomes with absent genes have draft status. Therefore, obtaining the finished genomes for the studied organisms will help us to fill the gaps in the incomplete pathways.

The problem of pathway incompleteness is only partially resolved by the finished versions of the analyzed genomes because incomplete pathways were also found in 13 finished genomes. For example, the finished genome of *Clostridium difficile* NAP07 lacks genes for Gal- and GalNAc-specific transporters, as well as for galactose kinase. These gaps may be filled by prediction or experimental identification of non-orthologous replacements, namely, genes that are not orthologs of the previously known genes but have the same functions. Such replacements have been previously described for the analyzed monosaccharide-utilization pathways. Thus, pairs of non-orthologous proteins were known for galactosamine-6-phosphate isomerase (AgaS and AgaI), glucosamine-6-phosphate deaminase (NagB1 and NagB2), and N-acetylglucosamine kinase (NagK1 and NagK2). Moreover, in this study, we predicted 4 non-orthologous replacements for enzymes (FclA2, FclD2, FclE2, and GalY) and 4 non-orthologous replacements for transporters (FucABC, FucP2, GalP3, and NgcABCD). The idea of non-orthologous replacement is very promising because these replacements can be found with computational methods alone.

Another possible way to resolve problems with incomplete pathways is the prediction of alternative reactions or pathways. Unlike the case of non-orthologous displacements, here, we should predict genes with previously unknown functions, but involved into metabolism of analyzed compounds. For example, catabolism of Fuc, Neu5Ac, and Gal is possible via two different pathways for each of these compounds (Figure 1B). The prediction of novel reactions in pathways is usually more difficult than the detection of non-orthologous displacements but is also possible using only computational methods. For example, an alternative pathway for the biosynthesis of menaquinone was discovered using comparative genomics techniques (Hiratsuka et al., 2008), and gaps in this pathway have subsequently been filled by computational analysis (Ravcheev and Thiele, 2016).

Taken together, all the remaining problems may be resolved using comparative genomics-based analysis. The availability of an increasing number of microbial genomes as well as completion of existing genome sequences will provide significant opportunities for the computational analyses of these microbes and the resolution of the described problems.

## 4.5. Concluding remarks and future plans

This study included a comprehensive computational analysis of the degradation of mucin glycans by HGM microorganisms. In addition to novel functional annotations, which are standard for comparative genomics studies, this analysis also predicted commensal exchange pathways between different microbes, specificity of the HGM strains to various types of glycans, and mutualistic inter-strain interactions at the level of mucin glycan degradation. These results demonstrate the efficiency of the selected approach: the analysis of individual genomes for members of the microbial community.

Although this study significantly improved our understanding of the HGM and its interactions with the host, there is still much to be discovered in this field. The areas lacking evidence indicate the future directions for the analysis of the HGM as it relates to mucin glycan degradation. The first two future directions, which are defined by unresolved problems (see subsection **4.4**), are to update the metabolic reconstructions with full versions of all analyzed genomes and to fill in the remaining gaps in the studied metabolic pathways. The third future direction is to expand the reconstructed metabolic, transport, and exchange pathways to novel genomes; this is aimed at the growing number of microbial genomes, including members of the HGM community. Mucin glycans are not the only glycans that can be degraded by HGM organisms. Thus, the fourth future direction is pathway reconstruction for the degradation of dietary-derived glycans.

The HGM has been intensively studied in relation to its impact on human health, and more than 50 human diseases have been shown to be associated with HGM alterations (Potgieter et al., 2015). Nonetheless, the use of HGM taxonomical composition as a diagnostic tool is currently hampered by the high variability of the HGM depending on various factors (Kurokawa et al., 2007;Clemente et al., 2012;Yatsunenko et al., 2012;Suzuki and Worobey, 2014;Allais et al., 2015) and by functional redundancy of the HGM. As it relates to the HGM, functional redundancy is defined as the ability of multiple bacteria, both closely and distantly related to each other, to implement the same metabolic functions (Moya and Ferrer, 2016). On the other hand, more than a dozen HGM-produced metabolites are associated with various human diseases (Potgieter et al., 2015), indicating possible associations between human health and the presence or absence of certain metabolic pathways in the HGM. Therefore, functional redundancy may provide a possible approach for the use of the HGM to distinguish health states of an individual. Indeed, a comparison of metagenomics data from healthy or diseased subjects may highlight HGM genes that are associated with disease. Because accurate functional annotation of all the genes in each metagenome is extremely time-consuming and costly, it makes sense to compare different microbiomes for the presence or absence of particular metabolic pathways that are already annotated and possibly associated with a health state. Because the state of the intestinal mucus layer is closely associated with human health (Png et al., 2010;Johansson et al., 2013;Cockburn and Koropatkin, 2016;Desai et al., 2016), HGM genes involved in the degradation of mucin glycans are perfect candidates to be tested for associations with health and disease. Thus, the fifth future direction is an analysis of the presence or absence of the analyzed genes in health and disease HGM metagenomes.

Computational modeling of metabolism (Palsson, 2006;Orth et al., 2010) may be used for elucidation of mucin glycan degradation by HGM organisms. Previously, genome-based models have been published for single representatives of the HGM (Thiele et al., 2005;Orth et al., 2011;Thiele et al., 2011;Thiele et al., 2012;Heinken et al., 2014) as well as for whole HGM communities (Levy and Borenstein, 2013;Bauer et al., 2015;Heinken and Thiele, 2015;Noecker et al., 2016;Shashkova et al., 2016;Magnusdottir et al., 2017). Additionally, computational models for human metabolism are also available (Thiele et al., 2013b;Mardinoglu et al., 2014), and host-microbial metabolic interactions have been modeled (Heinken et al., 2013;Thiele et al., 2013a;Shoaie and Nielsen, 2014;Levy et al., 2015;Heinken et al., 2016). Thus, the final future direction is to update the existing HGM and host-HGM metabolic models of reactions for all the pathways reconstructed in this study. Such an update would improve the existing models and would improve our understanding of the interaction between humans and their microbiome.

## 5. Conflict of Interest

The authors declare that the research was conducted in the absence of any commercial or financial relationships that could be construed as potential conflicts of interest.

## 6. Author Contributions

DAR and IT conceived of and designed the research project and wrote the manuscript. DAR performed the genomic analysis of the pathways for utilization of mucin glycans. All authors read and approved the final manuscript.

## 7. Funding

The presented project is supported by the National Research Fund (#6847110), Luxembourg, and co-funded under the Marie Curie Actions of the European Commission (FP7-COFUND).

## 8. References

Afzal, M., Shafeeq, S., Ahmed, H., and Kuipers, O.P. (2015a). Sialic acid-mediated gene expression in *Streptococcus pneumoniae* and the role of NanR as a transcriptional activator of the nan gene cluster. *Appl Environ Microbiol*.

Afzal, M., Shafeeq, S., Manzoor, I., and Kuipers, O.P. (2015b). GalR Acts as a Transcriptional Activator of *galKT* in the Presence of Galactose in *Streptococcus pneumoniae*. *J Mol Microbiol Biotechnol* 25**,** 363-371.

Agergaard, C.N., Hoegh, S.V., Holt, H.M., and Justesen, U.S. (2016). Two Serious Cases of Infection with *Clostridium celatum* after 40 Years in Hiding? *J Clin Microbiol* 54**,** 236-238.

Akinosoglou, K., Perperis, A., Siagris, D., Goutou, P., Spiliopoulou, I., Gogos, C.A., and Marangos, M. (2012). Bacteraemia due to *Cedecea davisae* in a patient with sigmoid colon cancer: a case report and brief review of the literature. *Diagn Microbiol Infect Dis* 74**,** 303-306.

Allais, L., Kerckhof, F.M., Verschuere, S., Bracke, K.R., De Smet, R., Laukens, D., Van Den Abbeele, P., De Vos, M., Boon, N., Brusselle, G.G., Cuvelier, C.A., and Van De Wiele, T. (2015). Chronic cigarette smoke exposure induces microbial and inflammatory shifts and mucin changes in the murine gut. *Environ Microbiol*.

Altschul, S.F., Madden, T.L., Schaffer, A.A., Zhang, J., Zhang, Z., Miller, W., and Lipman, D.J. (1997). Gapped BLAST and PSI-BLAST: a new generation of protein database search programs. *Nucleic Acids Research* 25**,** 3389-3402.

Arike, L., and Hansson, G.C. (2016). The Densely O-Glycosylated MUC2 Mucin Protects the Intestine and Provides Food for the Commensal Bacteria. *J Mol Biol* 428**,** 3221-3229.

Ashida, H., Maki, R., Ozawa, H., Tani, Y., Kiyohara, M., Fujita, M., Imamura, A., Ishida, H., Kiso, M., and Yamamoto, K. (2008). Characterization of two different endo-alpha-N-acetylgalactosaminidases from probiotic and pathogenic enterobacteria, *Bifidobacterium longum* and *Clostridium perfringen*s. *Glycobiology* 18**,** 727-734.

Ashida, H., Miyake, A., Kiyohara, M., Wada, J., Yoshida, E., Kumagai, H., Katayama, T., and Yamamoto, K. (2009). Two distinct alpha-L-fucosidases from *Bifidobacterium bifidum* are essential for the utilization of fucosylated milk oligosaccharides and glycoconjugates. *Glycobiology* 19, 1010-1017.

Bauer, E., Laczny, C.C., Magnusdottir, S., Wilmes, P., and Thiele, I. (2015). Phenotypic differentiation of gastrointestinal microbes is reflected in their encoded metabolic repertoires. *Microbiome* 3, 55.

Bettenbrock, K., and Alpert, C.A. (1998). The gal genes for the Leloir pathway of *Lactobacillus casei* 64H. *Applied and Environmental Microbiology* 64, 2013-2019.

Bidart, G.N., Rodriguez-Diaz, J., Monedero, V., and Yebra, M.J. (2014). A unique gene cluster for the utilization of the mucosal and human milk-associated glycans galacto-N-biose and lacto-N-biose in *Lactobacillus casei*. *Mol Microbiol* 93, 521-538.

Brigham, C., Caughlan, R., Gallegos, R., Dallas, M.B., Godoy, V.G., and Malamy, M.H. (2009). Sialic acid (N-acetyl neuraminic acid) utilization by *Bacteroides fragilis* requires a novel N-acetyl mannosamine epimerase. *J Bacteriol* 191, 3629-3638.

Cabezas, J.A. (1989). Some comments on the type references of the official nomenclature (IUB) for beta-N-acetylglucosaminidase, beta-N-acetylhexosaminidase and beta-N-acetylgalactosaminidase. *Biochem J* 261, 1059-1060.

Cameron, E.A., and Sperandio, V. (2015). Frenemies: Signaling and Nutritional Integration in Pathogen-Microbiota-Host Interactions. *Cell Host Microbe* 18, 275-284.

Cantarel, B.L., Lombard, V., and Henrissat, B. (2012). Complex carbohydrate utilization by the healthy human microbiome. *PLoS One* 7, e28742.

Caspi, R., Altman, T., Billington, R., Dreher, K., Foerster, H., Fulcher, C.A., Holland, T.A., Keseler, I.M., Kothari, A., Kubo, A., Krummenacker, M., Latendresse, M., Mueller, L.A., Ong, Q., Paley, S., Subhraveti, P., Weaver, D.S., Weerasinghe, D., Zhang, P., and Karp, P.D. (2014). The MetaCyc database of metabolic pathways and enzymes and the BioCyc collection of Pathway/Genome Databases. *Nucleic Acids Res* 42, D459-471.

Clemente, J.C., Ursell, L.K., Parfrey, L.W., and Knight, R. (2012). The impact of the gut microbiota on human health: an integrative view. *Cell* 148, 1258-1270.

Cockburn, D.W., and Koropatkin, N.M. (2016). Polysaccharide Degradation by the Intestinal Microbiota and Its Influence on Human Health and Disease. *J Mol Biol* 428, 3230-3252.

Conway, T., and Cohen, P.S. (2015). Commensal and Pathogenic *Escherichia coli* Metabolism in the Gut. *Microbiol Spectr* 3.

Dehal, P.S., Joachimiak, M.P., Price, M.N., Bates, J.T., Baumohl, J.K., Chivian, D., Friedland, G.D., Huang, K.H., Keller, K., Novichkov, P.S., Dubchak, I.L., Alm, E.J., and Arkin, A.P. (2010). MicrobesOnline: an integrated portal for comparative and functional genomics. *Nucleic Acids Res* 38, D396-D400.

Deplancke, B., and Gaskins, H.R. (2001). Microbial modulation of innate defense: goblet cells and the intestinal mucus layer. *Am J Clin Nutr* 73, 1131S-1141S.

Derrien, M., Vaughan, E.E., Plugge, C.M., and De Vos, W.M. (2004). *Akkermansia muciniphila* gen. nov., sp. nov., a human intestinal mucin-degrading bacterium. *International Journal of Systematic and Evolutionary Microbiology* 54, 1469-1476.

Desai, Mahesh s., Seekatz, Anna m., Koropatkin, Nicole m., Kamada, N., Hickey, Christina a., Wolter, M., Pudlo, Nicholas a., Kitamoto, S., Terrapon, N., Muller, A., Young, Vincent b., Henrissat, B., Wilmes, P., Stappenbeck, Thaddeus s., Núñez, G., and Martens, Eric c. (2016). A Dietary Fiber-Deprived Gut

Microbiota Degrades the Colonic Mucus Barrier and Enhances Pathogen Susceptibility. *Cell* 167, 1339-1353.e1321.

Disz, T., Akhter, S., Cuevas, D., Olson, R., Overbeek, R., Vonstein, V., Stevens, R., and Edwards, R.A. (2010). Accessing the SEED genome databases via Web services API: tools for programmers. *BMC Bioinformatics* 11, 319.

Eckburg, P.B., Bik, E.M., Bernstein, C.N., Purdom, E., Dethlefsen, L., Sargent, M., Gill, S.R., Nelson, K.E., and Relman, D.A. (2005). Diversity of the human intestinal microbial flora. *Science* 308, 1635-1638.

Edgar, R.C. (2004a). MUSCLE: a multiple sequence alignment method with reduced time and space complexity. *BMC Bioinformatics* 5, 113.

Edgar, R.C. (2004b). MUSCLE: multiple sequence alignment with high accuracy and high throughput. *Nucleic Acids Research* 32, 1792-1797.

Essenberg, R.C., Candler, C., and Nida, S.K. (1997). *Brucella abortus* strain 2308 putative glucose and galactose transporter gene: cloning and characterization. *Microbiology* 143 ( Pt 5), 1549-1555.

Finn, R.D., Bateman, A., Clements, J., Coggill, P., Eberhardt, R.Y., Eddy, S.R., Heger, A., Hetherington, K., Holm, L., Mistry, J., Sonnhammer, E.L., Tate, J., and Punta, M. (2014). Pfam: the protein families database. *Nucleic Acids Res* 42, D222-230.

Franceschini, A., Szklarczyk, D., Frankild, S., Kuhn, M., Simonovic, M., Roth, A., Lin, J., Minguez, P., Bork, P., Von Mering, C., and Jensen, L.J. (2013). STRING v9.1: protein-protein interaction networks, with increased coverage and integration. *Nucleic Acids Res* 41, D808-815.

Garrido, D., Kim, J.H., German, J.B., Raybould, H.E., and Mills, D.A. (2011). Oligosaccharide binding proteins from *Bifidobacterium longum* subsp. infantis reveal a preference for host glycans. *PLoS One* 6, e17315.

Goodman, A.L., Kallstrom, G., Faith, J.J., Reyes, A., Moore, A., Dantas, G., and Gordon, J.I. (2011). Extensive personal human gut microbiota culture collections characterized and manipulated in gnotobiotic mice. *Proc Natl Acad Sci U S A* 108, 6252-6257.

Graf, D., Di Cagno, R., Fak, F., Flint, H.J., Nyman, M., Saarela, M., and Watzl, B. (2015). Contribution of diet to the composition of the human gut microbiota. *Microb Ecol Health Dis* 26, 26164.

Guindon, S., Dufayard, J.F., Lefort, V., Anisimova, M., Hordijk, W., and Gascuel, O. (2010). New algorithms and methods to estimate maximum-likelihood phylogenies: assessing the performance of PhyML 3.0. *Syst Biol* 59, 307-321.

Gunn, F.J., Tate, C.G., and Henderson, P.J. (1994). Identification of a novel sugar-H+ symport protein, FucP, for transport of L-fucose into *Escherichia coli*. *Mol Microbiol* 12, 799-809.

Han, Y.R., Youn, S.Y., Ji, G.E., and Park, M.S. (2014). Production of alpha- and beta-galactosidases from *Bifidobacterium longum* subsp. *longum* RD47. *J Microbiol Biotechnol* 24, 675-682.

Heinken, A., Khan, M.T., Paglia, G., Rodionov, D.A., Harmsen, H.J., and Thiele, I. (2014). Functional metabolic map of *Faecalibacterium prausnitzii*, a beneficial human gut microbe. *J Bacteriol* 196, 3289-3302.

Heinken, A., Ravcheev, D.A., and Thiele, I. (2016). "Systems biology of bacteria-host interactions," in *The Human Microbiota and Chronic Disease*. John Wiley & Sons, Inc.), 113-137.

Heinken, A., Sahoo, S., Fleming, R.M., and Thiele, I. (2013). Systems-level characterization of a host-microbe metabolic symbiosis in the mammalian gut. *Gut Microbes* 4, 28-40.

Heinken, A., and Thiele, I. (2015). Anoxic conditions promote species-specific mutualism between gut microbes *in silico*. *Appl Environ Microbiol*.

Hiratsuka, T., Furihata, K., Ishikawa, J., Yamashita, H., Itoh, N., Seto, H., and Dairi, T. (2008). An alternative menaquinone biosynthetic pathway operating in microorganisms. *Science* 321**,** 1670-1673.

Hobbs, M.E., Vetting, M., Williams, H.J., Narindoshvili, T., Kebodeaux, D.M., Hillerich, B., Seidel, R.D., Almo, S.C., and Raushel, F.M. (2013). Discovery of an L-fucono-1,5-lactonase from cog3618 of the amidohydrolase superfamily. *Biochemistry* 52**,** 239-253.

Hooper, L.V., Xu, J., Falk, P.G., Midtvedt, T., and Gordon, J.I. (1999). A molecular sensor that allows a gut commensal to control its nutrient foundation in a competitive ecosystem. *Proceedings of the National Academy of Sciences of the United States of America* 96**,** 9833-9838.

Hoskins, L.C., Boulding, E.T., and Larson, G. (1997). Purification and characterization of blood group A-degrading isoforms of alpha-N-acetylgalactosaminidase from *Ruminococcus torques strain* IX-70. *J Biol Chem* 272**,** 7932-7939.

Husain, Q. (2010). Beta galactosidases and their potential applications: a review. *Crit Rev Biotechnol* 30**,** 41-62.

Huson, D.H., Richter, D.C., Rausch, C., Dezulian, T., Franz, M., and Rupp, R. (2007). Dendroscope: An interactive viewer for large phylogenetic trees. *BMC Bioinformatics* 8**,** 460.

Jensen, S.O., and Reeves, P.R. (1998). Domain organisation in phosphomannose isomerases (types I and II). *Biochimica et Biophysica Acta (BBA) - Protein Structure and Molecular Enzymology* 1382**,** 5-7.

Johansson, M.E., Jakobsson, H.E., Holmen-Larsson, J., Schutte, A., Ermund, A., Rodriguez-Pineiro, A.M., Arike, L., Wising, C., Svensson, F., Backhed, F., and Hansson, G.C. (2015). Normalization of Host Intestinal Mucus Layers Requires Long-Term Microbial Colonization. *Cell Host Microbe* 18**,** 582-592.

Johansson, M.E., Larsson, J.M., and Hansson, G.C. (2011). The two mucus layers of colon are organized by the MUC2 mucin, whereas the outer layer is a legislator of host-microbial interactions. *Proc Natl Acad Sci U S A* 108 Suppl 1**,** 4659-4665.

Johansson, M.E., Phillipson, M., Petersson, J., Velcich, A., Holm, L., and Hansson, G.C. (2008). The inner of the two Muc2 mucin-dependent mucus layers in colon is devoid of bacteria. *Proc Natl Acad Sci U S A* 105**,** 15064-15069.

Johansson, M.E., Sjovall, H., and Hansson, G.C. (2013). The gastrointestinal mucus system in health and disease. *Nat Rev Gastroenterol Hepatol* 10**,** 352-361.

Johnson, C.M., and Schleif, R.F. (2000). Cooperative action of the catabolite activator protein and AraC in vitro at the *araFGH* promoter. *J Bacteriol* 182**,** 1995-2000.

Joossens, M., Huys, G., Cnockaert, M., De Preter, V., Verbeke, K., Rutgeerts, P., Vandamme, P., and Vermeire, S. (2011). Dysbiosis of the faecal microbiota in patients with Crohn's disease and their unaffected relatives. *Gut* 60**,** 631-637.

Juge, N., Tailford, L., and Owen, C.D. (2016). Sialidases from gut bacteria: a mini-review. *Biochem Soc Trans* 44**,** 166-175.

Kanehisa, M., Goto, S., Sato, Y., Furumichi, M., and Tanabe, M. (2012). KEGG for integration and interpretation of large-scale molecular data sets. *Nucleic Acids Res* 40**,** D109-114.

Katayama, T., Sakuma, A., Kimura, T., Makimura, Y., Hiratake, J., Sakata, K., Yamanoi, T., Kumagai, H., and Yamamoto, K. (2004). Molecular cloning and characterization of *Bifidobacterium bifidum* 1,2-alpha-L-fucosidase (AfcA), a novel inverting glycosidase (glycoside hydrolase family 95). *J Bacteriol* 186**,** 4885-4893.

Kazanov, M.D., Li, X., Gelfand, M.S., Osterman, A.L., and Rodionov, D.A. (2013). Functional diversification of ROK-family transcriptional regulators of sugar catabolism in the Thermotogae phylum. *Nucleic Acids Res* 41**,** 790-803.

Khoroshkin, M.S., Leyn, S.A., Van Sinderen, D., and Rodionov, D.A. (2016). Transcriptional Regulation of Carbohydrate Utilization Pathways in the *Bifidobacterium* Genus. *Front Microbiol* 7**,** 120.

Kilcoyne, M., Gerlach, J.Q., Gough, R., Gallagher, M.E., Kane, M., Carrington, S.D., and Joshi, L. (2012). Construction of a natural mucin microarray and interrogation for biologically relevant glyco-epitopes. *Anal Chem* 84**,** 3330-3338.

Kiyohara, M., Nakatomi, T., Kurihara, S., Fushinobu, S., Suzuki, H., Tanaka, T., Shoda, S., Kitaoka, M., Katayama, T., Yamamoto, K., and Ashida, H. (2012). alpha-N-acetylgalactosaminidase from infant-associated bifidobacteria belonging to novel glycoside hydrolase family 129 is implicated in alternative mucin degradation pathway. *J Biol Chem* 287**,** 693-700.

Koropatkin, N.M., Cameron, E.A., and Martens, E.C. (2012). How glycan metabolism shapes the human gut microbiota. *Nature Reviews Microbiology* 10**,** 323-335.

Koutsioulis, D., Landry, D., and Guthrie, E.P. (2008). Novel endo-alpha-N-acetylgalactosaminidases with broader substrate specificity. *Glycobiology* 18**,** 799-805.

Kunz, C., Rudloff, S., Baier, W., Klein, N., and Strobel, S. (2000). Oligosaccharides in human milk: structural, functional, and metabolic aspects. *Annu Rev Nutr* 20**,** 699-722.

Kurokawa, K., Itoh, T., Kuwahara, T., Oshima, K., Toh, H., Toyoda, A., Takami, H., Morita, H., Sharma, V.K., Srivastava, T.P., Taylor, T.D., Noguchi, H., Mori, H., Ogura, Y., Ehrlich, D.S., Itoh, K., Takagi, T., Sakaki, Y., Hayashi, T., and Hattori, M. (2007). Comparative metagenomics revealed commonly enriched gene sets in human gut microbiomes. *DNA Research* 14**,** 169-181.

Lamberg, S.I., and Stoolmiller, A.C. (1974). Glycosaminoglycans. A biochemical and clinical review. *J Invest Dermatol* 63**,** 433-449.

Lang, T., Hansson, G.C., and Samuelsson, T. (2007). Gel-forming mucins appeared early in metazoan evolution. *Proc Natl Acad Sci U S A* 104**,** 16209-16214.

Larsson, J.M., Karlsson, H., Sjovall, H., and Hansson, G.C. (2009). A complex, but uniform O-glycosylation of the human MUC2 mucin from colonic biopsies analyzed by nanoLC/MSn. *Glycobiology* 19**,** 756-766.

Levy, R., and Borenstein, E. (2013). Metabolic modeling of species interaction in the human microbiome elucidates community-level assembly rules. *Proc Natl Acad Sci U S A* 110**,** 12804-12809.

Levy, R., Carr, R., Kreimer, A., Freilich, S., and Borenstein, E. (2015). NetCooperate: a network-based tool for inferring host-microbe and microbe-microbe cooperation. *BMC Bioinformatics* 16**,** 164.

Leyn, S.A., Gao, F., Yang, C., and Rodionov, D.A. (2012). N-Acetylgalactosamine utilization pathway and regulon in proteobacteria: Genomic reconstruction and experimental characterization in *Shewanella Journal of Biological Chemistry* 287**,** 28047-28056.

Li, H., Limenitakis, J.P., Fuhrer, T., Geuking, M.B., Lawson, M.A., Wyss, M., Brugiroux, S., Keller, I., Macpherson, J.A., Rupp, S., Stolp, B., Stein, J.V., Stecher, B., Sauer, U., Mccoy, K.D., and Macpherson, A.J. (2015). The outer mucus layer hosts a distinct intestinal microbial niche. *Nat Commun* 6**,** 8292.

Luesink, E.J., Van Herpen, R.E., Grossiord, B.P., Kuipers, O.P., and De Vos, W.M. (1998). Transcriptional activation of the glycolytic las operon and catabolite repression of the gal operon in *Lactococcus lactis* are mediated by the catabolite control protein CcpA. *Mol Microbiol* 30**,** 789-798.

Magnusdottir, S., Heinken, A., Kutt, L., Ravcheev, D.A., Bauer, E., Noronha, A., Greenhalgh, K., Jager, C., Baginska, J., Wilmes, P., Fleming, R.M., and Thiele, I. (2017). Generation of genome-scale metabolic reconstructions for 773 members of the human gut microbiota. *Nat Biotechnol* 35**,** 81-89.

Magnusdottir, S., Ravcheev, D.A., De Crecy-Lagard, V., and Thiele, I. (2015). Systematic genome assessment of B-vitamin biosynthesis suggests co-operation among gut microbes. *Frontiers in Genetics* 6**,** 148.

Magrane, M., and Consortium, U. (2011). UniProt Knowledgebase: a hub of integrated protein data. *Database (Oxford)* 2011**,** bar009.

Marchler-Bauer, A., Zheng, C., Chitsaz, F., Derbyshire, M.K., Geer, L.Y., Geer, R.C., Gonzales, N.R., Gwadz, M., Hurwitz, D.I., Lanczycki, C.J., Lu, F., Lu, S., Marchler, G.H., Song, J.S., Thanki, N., Yamashita, R.A., Zhang, D., and Bryant, S.H. (2013). CDD: conserved domains and protein three-dimensional structure. *Nucleic Acids Research* 41**,** D348-D352.

Marcobal, A., Barboza, M., Sonnenburg, E.D., Pudlo, N., Martens, E.C., Desai, P., Lebrilla, C.B., Weimer, B.C., Mills, D.A., German, J.B., and Sonnenburg, J.L. (2011). *Bacteroides* in the infant gut consume milk oligosaccharides via mucus-utilization pathways. *Cell Host Microbe* 10**,** 507-514.

Marcobal, A., Southwick, A.M., Earle, K.A., and Sonnenburg, J.L. (2013). A refined palate: bacterial consumption of host glycans in the gut. *Glycobiology* 23**,** 1038-1046.

Mardinoglu, A., Agren, R., Kampf, C., Asplund, A., Uhlen, M., and Nielsen, J. (2014). Genome-scale metabolic modelling of hepatocytes reveals serine deficiency in patients with non-alcoholic fatty liver disease. *Nat Commun* 5**,** 3083.

Markowitz, V.M., Chen, I.M., Palaniappan, K., Chu, K., Szeto, E., Pillay, M., Ratner, A., Huang, J., Woyke, T., Huntemann, M., Anderson, I., Billis, K., Varghese, N., Mavromatis, K., Pati, A., Ivanova, N.N., and Kyrpides, N.C. (2014). IMG 4 version of the integrated microbial genomes comparative analysis system. *Nucleic Acids Res* 42**,** D560-567.

Martens, E.C., Chiang, H.C., and Gordon, J.I. (2008). Mucosal glycan foraging enhances fitness and transmission of a saccharolytic human gut bacterial symbiont. *Cell Host Microbe* 4**,** 447-457.

Martens, E.C., Koropatkin, N.M., Smith, T.J., and Gordon, J.I. (2009). Complex glycan catabolism by the human gut microbiota: the *Bacteroidetes* Sus-like paradigm. *Journal of Biological Chemistry* 284**,** 24673-24677.

Mcguckin, M.A., Linden, S.K., Sutton, P., and Florin, T.H. (2011). Mucin dynamics and enteric pathogens. *Nat Rev Microbiol* 9**,** 265-278.

Michlmayr, H., and Kneifel, W. (2014). beta-Glucosidase activities of lactic acid bacteria: mechanisms, impact on fermented food and human health. *FEMS Microbiol Lett* 352**,** 1-10.

Mironov, A.A., Vinokurova, N.P., and Gel'fand, M.S. (2000). Software for analyzing bacterial genomes. *Mol Biol (Mosk)* 34**,** 253-262.

Mohnen, D. (2008). Pectin structure and biosynthesis. *Curr Opin Plant Biol* 11**,** 266-277.

Moya, A., and Ferrer, M. (2016). Functional Redundancy-Induced Stability of Gut Microbiota Subjected to Disturbance. *Trends in Microbiology* 24**,** 402–413.

Nagae, M., Tsuchiya, A., Katayama, T., Yamamoto, K., Wakatsuki, S., and Kato, R. (2007). Structural basis of the catalytic reaction mechanism of novel 1,2-alpha-L-fucosidase from *Bifidobacterium bifidum*. *J Biol Chem* 282**,** 18497-18509.

Ng, K.M., Ferreyra, J.A., Higginbottom, S.K., Lynch, J.B., Kashyap, P.C., Gopinath, S., Naidu, N., Choudhury, B., Weimer, B.C., Monack, D.M., and Sonnenburg, J.L. (2013). Microbiota-liberated host sugars facilitate post-antibiotic expansion of enteric pathogens. *Nature* 502**,** 96-99.

Nishimoto, M., and Kitaoka, M. (2007). Identification of N-acetylhexosamine 1-kinase in the complete lacto-N-biose I/galacto-N-biose metabolic pathway in *Bifidobacterium longum*. *Appl Environ Microbiol* 73**,** 6444-6449.

Noecker, C., Eng, A., Srinivasan, S., Theriot, C.M., Young, V.B., Jansson, J.K., Fredricks, D.N., and Borenstein, E. (2016). Metabolic Model-Based Integration of Microbiome Taxonomic and Metabolomic Profiles Elucidates Mechanistic Links between Ecological and Metabolic Variation. *mSystems* 1.

Orth, J.D., Conrad, T.M., Na, J., Lerman, J.A., Nam, H., Feist, A.M., and Palsson, B.O. (2011). A comprehensive genome-scale reconstruction of Escherichia coli metabolism--2011. *Mol Syst Biol* 7**,** 535.

Orth, J.D., Thiele, I., and Palsson, B.O. (2010). What is flux balance analysis? *Nat Biotechnol* 28**,** 245-248.

Osterman, A., and Overbeek, R. (2003). Missing genes in metabolic pathways: a comparative genomics approach. *Curr Opin Chem Biol* 7**,** 238-251.

Overbeek, R., Begley, T., Butler, R.M., Choudhuri, J.V., Chuang, H.Y., Cohoon, M., De Crecy-Lagard, V., Diaz, N., Disz, T., Edwards, R., Fonstein, M., Frank, E.D., Gerdes, S., Glass, E.M., Goesmann, A., Hanson, A., Iwata-Reuyl, D., Jensen, R., Jamshidi, N., Krause, L., Kubal, M., Larsen, N., Linke, B., Mchardy, A.C., Meyer, F., Neuweger, H., Olsen, G., Olson, R., Osterman, A., Portnoy, V., Pusch, G.D., Rodionov, D.A., Ruckert, C., Steiner, J., Stevens, R., Thiele, I., Vassieva, O., Ye, Y., Zagnitko, O., and Vonstein, V. (2005). The subsystems approach to genome annotation and its use in the project to annotate 1000 genomes. *Nucleic Acids Res* 33**,** 5691-5702.

Pacheco, A.R., Curtis, M.M., Ritchie, J.M., Munera, D., Waldor, M.K., Moreira, C.G., and Sperandio, V. (2012). Fucose sensing regulates bacterial intestinal colonization. *Nature* 492**,** 113-117.

Pacheco, A.R., and Sperandio, V. (2015). Enteric Pathogens Exploit the Microbiota-generated Nutritional Environment of the Gut. *Microbiol Spectr* 3.

Palsson, B. (2006). *Systems biology : properties of reconstructed networks.* Cambridge: Cambridge University Press.

Papatheodorou, P., Wilczek, C., Nolke, T., Guttenberg, G., Hornuss, D., Schwan, C., and Aktories, K. (2012). Identification of the cellular receptor of *Clostridium spiroforme* toxin. *Infect Immun* 80**,** 1418-1423.

Park, A.R., and Oh, D.K. (2010). Galacto-oligosaccharide production using microbial beta-galactosidase: current state and perspectives. *Appl Microbiol Biotechnol* 85**,** 1279-1286.

Parmar, A.S., Alakulppi, N., Paavola-Sakki, P., Kurppa, K., Halme, L., Farkkila, M., Turunen, U., Lappalainen, M., Kontula, K., Kaukinen, K., Maki, M., Lindfors, K., Partanen, J., Sistonen, P., Matto, J., Wacklin, P., Saavalainen, P., and Einarsdottir, E. (2012). Association study of FUT2 (rs601338) with celiac disease and inflammatory bowel disease in the Finnish population. *Tissue Antigens* 80**,** 488-493.

Pickard, J.M., and Chervonsky, A.V. (2015). Intestinal fucose as a mediator of host-microbe symbiosis. *J Immunol* 194**,** 5588-5593.

Plumbridge, J. (2015). Regulation of the Utilization of Amino Sugars by *Escherichia coli* and *Bacillus subtilis* : Same Genes, Different Control. *J Mol Microbiol Biotechnol* 25**,** 154-167.

Plumbridge, J.A., Cochet, O., Souza, J.M., Altamirano, M.M., Calcagno, M.L., and Badet, B. (1993). Coordinated regulation of amino sugar-synthesizing and -degrading enzymes in *Escherichia coli* K-12. *Journal of Bacteriology* 175**,** 4951-4956.

Png, C.W., Linden, S.K., Gilshenan, K.S., Zoetendal, E.G., Mcsweeney, C.S., Sly, L.I., Mcguckin, M.A., and Florin, T.H. (2010). Mucolytic bacteria with increased prevalence in IBD mucosa augment in vitro utilization of mucin by other bacteria. *Am J Gastroenterol* 105**,** 2420-2428.

Podolsky, D.K. (1985). Oligosaccharide structures of human colonic mucin. *J Biol Chem* 260**,** 8262-8271.

Potgieter, M., Bester, J., Kell, D.B., and Pretorius, E. (2015). The dormant blood microbiome in chronic, inflammatory diseases. *FEMS Microbiol Rev* 39, 567-591.

Pudlo, N.A., Urs, K., Kumar, S.S., German, J.B., Mills, D.A., and Martens, E.C. (2015). Symbiotic Human Gut Bacteria with Variable Metabolic Priorities for Host Mucosal Glycans. *MBio* 6, e01282-01215.

Ravcheev, D.A., Godzik, A., Osterman, A.L., and Rodionov, D.A. (2013). Polysaccharides utilization in human gut bacterium *Bacteroides thetaiotaomicron*: comparative genomics reconstruction of metabolic and regulatory networks. *BMC Genomics* 14, 873.

Ravcheev, D.A., Khoroshkin, M.S., Laikova, O.N., Tsoy, O.V., Sernova, N.V., Petrova, S.A., Rakhmaninova, A.B., Novichkov, P.S., Gelfand, M.S., and Rodionov, D.A. (2014). Comparative genomics and evolution of regulons of the LacI-family transcription factors. *Frontiers in Microbiology* 5.

Ravcheev, D.A., Li, X., Latif, H., Zengler, K., Leyn, S.A., Korostelev, Y.D., Kazakov, A.E., Novichkov, P.S., Osterman, A.L., and Rodionov, D.A. (2012). Transcriptional regulation of central carbon and energy metabolism in bacteria by redox-responsive repressor Rex. *J Bacteriol* 194, 1145-1157.

Ravcheev, D.A., and Thiele, I. (2014). Systematic genomic analysis reveals the complementary aerobic and anaerobic respiration capacities of the human gut microbiota. *Frontiers in Microbiology* 5.

Ravcheev, D.A., and Thiele, I. (2016). Genomic analysis of the human gut microbiome suggests novel enzymes involved in quinone biosynthesis. *Frontiers in Microbiology* 7, 128.

Reddy, S.K., Bagenholm, V., Pudlo, N.A., Bouraoui, H., Koropatkin, N.M., Martens, E.C., and Stalbrand, H. (2016). A beta-mannan utilization locus in Bacteroides ovatus involves a GH36 alpha-galactosidase active on galactomannans. *FEBS Lett* 590, 2106-2118.

Rodionov, D.A., Mironov, A.A., Rakhmaninova, A.B., and Gelfand, M.S. (2000). Transcriptional regulation of transport and utilization systems for hexuronides, hexuronates and hexonates in gamma purple bacteria. *Molecular Microbiology* 38, 673-683.

Rodionov, D.A., Novichkov, P.S., Stavrovskaya, E.D., Rodionova, I.A., Li, X., Kazanov, M.D., Ravcheev, D.A., Gerasimova, A.V., Kazakov, A.E., Kovaleva, G.Y., Permina, E.A., Laikova, O.N., Overbeek, R., Romine, M.F., Fredrickson, J.K., Arkin, A.P., Dubchak, I., Osterman, A.L., and Gelfand, M.S. (2011). Comparative genomic reconstruction of transcriptional networks controlling central metabolism in the *Shewanella* genus. *BMC Genomics* 12 Suppl 1, S3.

Rodionov, D.A., Rodionova, I.A., Li, X., Ravcheev, D.A., Tarasova, Y., Portnoy, V.A., Zengler, K., and Osterman, A.L. (2013). Transcriptional regulation of the carbohydrate utilization network in *Thermotoga maritima*. *Front Microbiol* 4, 244.

Rodionov, D.A., Yang, C., Li, X., Rodionova, I.A., Wang, Y., Obraztsova, A.Y., Zagnitko, O.P., Overbeek, R., Romine, M.F., Reed, S., Fredrickson, J.K., Nealson, K.H., and Osterman, A.L. (2010). Genomic encyclopedia of sugar utilization pathways in the *Shewanella* genus. *BMC Genomics* 11, 494.

Rodionova, I.A., Yang, C., Li, X., Kurnasov, O.V., Best, A.A., Osterman, A.L., and Rodionov, D.A. (2012). Diversity and versatility of the *Thermotoga maritima* sugar kinome. *J Bacteriol* 194, 5552-5563.

Roggentin, P., Schauer, R., Hoyer, L.L., and Vimr, E.R. (1993). The sialidase superfamily and its spread by horizontal gene transfer. *Mol Microbiol* 9, 915-921.

Rossez, Y., Maes, E., Lefebvre Darroman, T., Gosset, P., Ecobichon, C., Joncquel Chevalier Curt, M., Boneca, I.G., Michalski, J.C., and Robbe-Masselot, C. (2012). Almost all human gastric mucin O-glycans harbor blood group A, B or H antigens and are potential binding sites for *Helicobacter pylori*. *Glycobiology* 22, 1193-1206.

Sebaihia, M., Wren, B.W., Mullany, P., Fairweather, N.F., Minton, N., Stabler, R., Thomson, N.R., Roberts, A.P., Cerdeno-Tarraga, A.M., Wang, H., Holden, M.T., Wright, A., Churcher, C., Quail, M.A., Baker, S., Bason, N., Brooks, K., Chillingworth, T., Cronin, A., Davis, P., Dowd, L., Fraser, A., Feltwell, T., Hance, Z., Holroyd, S., Jagels, K., Moule, S., Mungall, K., Price, C., Rabbinowitsch, E., Sharp, S., Simmonds, M., Stevens, K., Unwin, L., Whitehead, S., Dupuy, B., Dougan, G., Barrell, B., and Parkhill, J. (2006). The multidrug-resistant human pathogen *Clostridium difficile* has a highly mobile, mosaic genome. *Nat Genet* 38, 779-786.

Shashkova, T., Popenko, A., Tyakht, A., Peskov, K., Kosinsky, Y., Bogolubsky, L., Raigorodskii, A., Ischenko, D., Alexeev, D., and Govorun, V. (2016). Agent Based Modeling of Human Gut Microbiome Interactions and Perturbations. *PLoS One* 11, e0148386.

Shimada, Y., Watanabe, Y., Wakinaka, T., Funeno, Y., Kubota, M., Chaiwangsri, T., Kurihara, S., Yamamoto, K., Katayama, T., and Ashida, H. (2015). alpha-N-Acetylglucosaminidase from *Bifidobacterium bifidum* specifically hydrolyzes alpha-linked N-acetylglucosamine at nonreducing terminus of O-glycan on gastric mucin. *Appl Microbiol Biotechnol* 99, 3941-3948.

Shoaie, S., and Nielsen, J. (2014). Elucidating the interactions between the human gut microbiota and its host through metabolic modeling. *Front Genet* 5, 86.

Solomon, H.V., Tabachnikov, O., Lansky, S., Salama, R., Feinberg, H., Shoham, Y., and Shoham, G. (2015). Structure-function relationships in Gan42B, an intracellular GH42 beta-galactosidase from *Geobacillus stearothermophilus*. *Acta Crystallogr D Biol Crystallogr* 71, 2433-2448.

Staudacher, E., Altmann, F., Wilson, I.B., and Marz, L. (1999). Fucose in N-glycans: from plant to man. *Biochim Biophys Acta* 1473, 216-236.

Suvorova, I.A., Korostelev, Y.D., and Gelfand, M.S. (2015). GntR Family of Bacterial Transcription Factors and Their DNA Binding Motifs: Structure, Positioning and Co-Evolution. *PLoS One* 10, e0132618.

Suvorova, I.A., Tutukina, M.N., Ravcheev, D.A., Rodionov, D.A., Ozoline, O.N., and Gelfand, M.S. (2011). Comparative genomic analysis of the hexuronate metabolism genes and their regulation in gammaproteobacteria. *J Bacteriol* 193, 3956-3963.

Suzuki, T.A., and Worobey, M. (2014). Geographical variation of human gut microbial composition. *Biol Lett* 10, 20131037.

Tailford, L., Crost, E., Kavanaugh, D., and Juge, N. (2015). Mucin glycan foraging in the human gut microbiome. *Frontiers in Genetics* 6, 81.

Tatusov, R.L., Fedorova, N.D., Jackson, J.D., Jacobs, A.R., Kiryutin, B., Koonin, E.V., Krylov, D.M., Mazumder, R., Mekhedov, S.L., Nikolskaya, A.N., Rao, B.S., Smirnov, S., Sverdlov, A.V., Vasudevan, S., Wolf, Y.I., Yin, J.J., and Natale, D.A. (2003). The COG database: an updated version includes eukaryotes. *BMC Bioinformatics* 4, 41.

Thiele, I., Fleming, R.M., Que, R., Bordbar, A., Diep, D., and Palsson, B.O. (2012). Multiscale modeling of metabolism and macromolecular synthesis in *E. coli* and its application to the evolution of codon usage. *PLoS One* 7, e45635.

Thiele, I., Heinken, A., and Fleming, R.M. (2013a). A systems biology approach to studying the role of microbes in human health. *Current Opinion in Biotechnology* 24, 4-12.

Thiele, I., Hyduke, D.R., Steeb, B., Fankam, G., Allen, D.K., Bazzani, S., Charusanti, P., Chen, F.C., Fleming, R.M., Hsiung, C.A., De Keersmaecker, S.C., Liao, Y.C., Marchal, K., Mo, M.L., Ozdemir, E., Raghunathan, A., Reed, J.L., Shin, S.I., Sigurbjornsdottir, S., Steinmann, J., Sudarsan, S., Swainston, N., Thijs, I.M., Zengler, K., Palsson, B.O., Adkins, J.N., and Bumann, D. (2011). A community effort towards a knowledge-base and mathematical model of the human pathogen *Salmonella Typhimurium* LT2. *BMC Systems Biology* 5**,** 8.

Thiele, I., Swainston, N., Fleming, R.M., Hoppe, A., Sahoo, S., Aurich, M.K., Haraldsdottir, H., Mo, M.L., Rolfsson, O., Stobbe, M.D., Thorleifsson, S.G., Agren, R., Bolling, C., Bordel, S., Chavali, A.K., Dobson, P., Dunn, W.B., Endler, L., Hala, D., Hucka, M., Hull, D., Jameson, D., Jamshidi, N., Jonsson, J.J., Juty, N., Keating, S., Nookaew, I., Le Novere, N., Malys, N., Mazein, A., Papin, J.A., Price, N.D., Selkov, E., Sr., Sigurdsson, M.I., Simeonidis, E., Sonnenschein, N., Smallbone, K., Sorokin, A., Van Beek, J.H., Weichart, D., Goryanin, I., Nielsen, J., Westerhoff, H.V., Kell, D.B., Mendes, P., and Palsson, B.O. (2013b). A community-driven global reconstruction of human metabolism. *Nat Biotechnol* 31**,** 419-425.

Thiele, I., Vo, T.D., Price, N.D., and Palsson, B.O. (2005). Expanded metabolic reconstruction of *Helicobacter pylori* (iIT341 GSM/GPR): an in silico genome-scale characterization of single- and double-deletion mutants. *J Bacteriol* 187**,** 5818-5830.

Thomas, G.H. (2016). Sialic acid acquisition in bacteria-one substrate, many transporters. *Biochem Soc Trans* 44**,** 760-765.

Uhde, A., Bruhl, N., Goldbeck, O., Matano, C., Gurow, O., Ruckert, C., Marin, K., Wendisch, V.F., Kramer, R., and Seibold, G.M. (2016). Transcription of Sialic Acid Catabolism Genes in *Corynebacterium glutamicum* Is Subject to Catabolite Repression and Control by the Transcriptional Repressor NanR. *J Bacteriol* 198**,** 2204-2218.

Vimr, E.R. (2013). Unified theory of bacterial sialometabolism: how and why bacteria metabolize host sialic acids. *ISRN Microbiol* 2013**,** 816713.

Vimr, E.R., Kalivoda, K.A., Deszo, E.L., and Steenbergen, S.M. (2004). Diversity of microbial sialic acid metabolism. *Microbiol Mol Biol Rev* 68**,** 132-153.

Wakinaka, T., Kiyohara, M., Kurihara, S., Hirata, A., Chaiwangsri, T., Ohnuma, T., Fukamizo, T., Katayama, T., Ashida, H., and Yamamoto, K. (2013). Bifidobacterial alpha-galactosidase with unique carbohydrate-binding module specifically acts on blood group B antigen. *Glycobiology* 23**,** 232-240.

Walker, A.W., Ince, J., Duncan, S.H., Webster, L.M., Holtrop, G., Ze, X., Brown, D., Stares, M.D., Scott, P., Bergerat, A., Louis, P., Mcintosh, F., Johnstone, A.M., Lobley, G.E., Parkhill, J., and Flint, H.J. (2011). Dominant and diet-responsive groups of bacteria within the human colonic microbiota. *ISME J* 5**,** 220-230.

Weickert, M.J., and Adhya, S. (1993). The galactose regulon of Escherichia coli. *Mol Microbiol* 10**,** 245-251.

Xiao, X., Wang, F., Saito, A., Majka, J., Schlosser, A., and Schrempf, H. (2002). The novel S*treptomyces olivaceoviridis* ABC transporter Ngc mediates uptake of N-acetylglucosamine and N,N'-diacetylchitobiose. *Mol Genet Genomics* 267**,** 429-439.

Yang, C., Rodionov, D.A., Li, X., Laikova, O.N., Gelfand, M.S., Zagnitko, O.P., Romine, M.F., Obraztsova, A.Y., Nealson, K.H., and Osterman, A.L. (2006). Comparative genomics and experimental characterization of N-acetylglucosamine utilization pathway of *Shewanella oneidensis*. *Journal of Biological Chemistry* 281**,** 29872-29885.

Yatsunenko, T., Rey, F.E., Manary, M.J., Trehan, I., Dominguez-Bello, M.G., Contreras, M., Magris, M., Hidalgo, G., Baldassano, R.N., Anokhin, A.P., Heath, A.C., Warner, B., Reeder, J., Kuczynski, J., Caporaso, J.G., Lozupone, C.A., Lauber, C., Clemente, J.C., Knights, D., Knight, R., and Gordon, J.I. (2012). Human gut microbiome viewed across age and geography. *Nature* 486**,** 222-227.

Yew, W.S., Fedorov, A.A., Fedorov, E.V., Rakus, J.F., Pierce, R.W., Almo, S.C., and Gerlt, J.A. (2006). Evolution of enzymatic activities in the enolase superfamily: L-fuconate dehydratase from *Xanthomonas campestris*. *Biochemistry* 45**,** 14582-14597.

Zeng, L., Das, S., and Burne, R.A. (2010). Utilization of lactose and galactose by S*treptococcus mutans*: transport, toxicity, and carbon catabolite repression. *J Bacteriol* 192**,** 2434-2444.

Zeng, L., Martino, N.C., and Burne, R.A. (2012). Two gene clusters coordinate galactose and lactose metabolism in *Streptococcus gordonii*. *Appl Environ Microbiol* 78**,** 5597-5605.

Zhang, H., Ravcheev, D.A., Hu, D., Zhang, F., Gong, X., Hao, L., Cao, M., Rodionov, D.A., Wang, C., and Feng, Y. (2015). Two novel regulators of N-acetyl-galactosamine utilization pathway and distinct roles in bacterial infections. *Microbiologyopen* 4**,** 983-1000.

Zwierz, K., Zalewska, A., and Zoch-Zwierz, A. (1999). Isoenzymes of N-acetyl-beta-hexosaminidase. *Acta Biochim Pol* 46**,** 739-751.

## 9. Figure Legends

**Figure 1. Enzymes involved in the degradation of mucin glycans.** (A) Splitting of a hypothetical mucin glycan by GHs. The following enzymes are shown: EC 3.2.1.14, Chitinase; EC 3.2.1.18, Neuraminidase; EC 3.2.1.22, α-galactosidase; EC 3.2.1.23, Beta-galactosidase; EC 3.2.1.49, α-N-acetylgalactosaminidase; EC 3.2.1.50, α-N-acetylglucosaminidase; EC 3.2.1.51, α-L-fucosidase; EC 3.2.1.52, β-N-hexosaminidase; EC 3.2.1.97, and Endo-α-N-acetylgalactosaminidase. (B) Known pathways for utilization of the derived monosaccharides. Protein abbreviations are shown; for the corresponding functions see Table S2 in the Supplementary Materials.

**Figure 2. Distribution of the CPs and GHs for mucin glycan-derived monosaccharides in the analyzed genomes.** (A) Pathways and GHs are grouped according to the utilized monosaccharide. The bar "Catabolic pathways (CPs)" includes all the identified pathways, independent of the presence of GHs. The same is true for the bar "Glycosyl hydrolases (GHs)". The bar "Both CPs and GHs" corresponds to the presence of GHs and pathways for this monosaccharide in the same genomes. (B) The numbers of monosaccharides, cleaved and/or catabolized in the analyzed genomes. The "Both CPs and GHs" bars correspond to the presence in the same genome of both the pathway and the GH for a certain monosaccharide.

**Figure 3. Distribution of monosaccharide donors and acceptors by taxa.** Only genomes classified as "donors" (the horizontal axis) and "acceptors" (the vertical axis) are shown. The numbers of exchanged monosaccharides are shown in agreement with the color scale. For details on each pair of organisms and exchanged monosaccharides, see Supplementary Table S11.

**Figure 4. Cleavage of a hypothetical mucin glycan by a community of HGM inhabitants.** For each reaction, the taxa with the corresponding specific GHs are shown.

**Figure 5. Mutualistic pairs for cleavage of mucin glycans.** The numbers of cleaved glycans are shown in agreement with the color scale. The numbers of glycans cleaved by single organisms are shown for all genomes. For the pairs of organisms, the numbers of cleaved glycans are shown only for mutualistic pairs. For details on each mutualistic pair of organisms and cleaved glycans, see Supplementary Table S14.

**Figure 1**

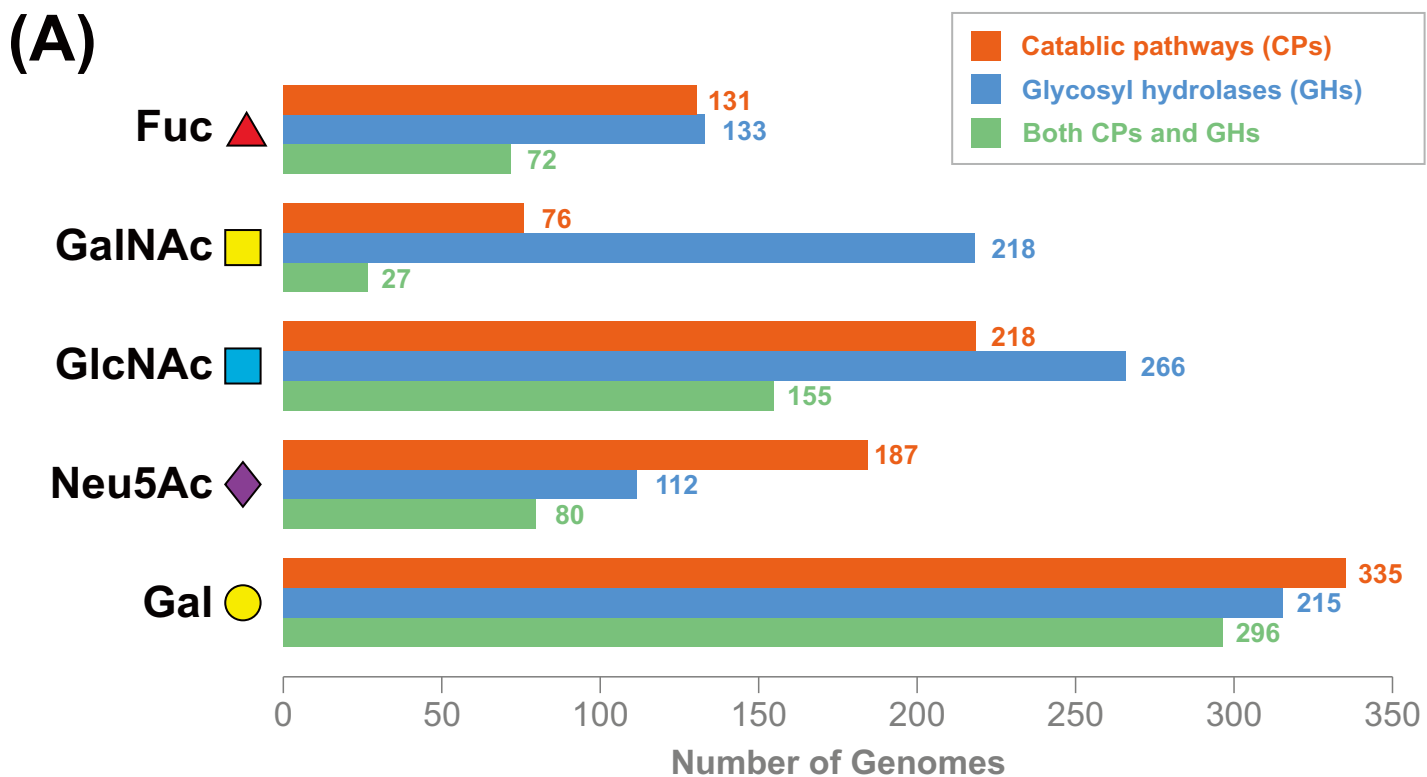

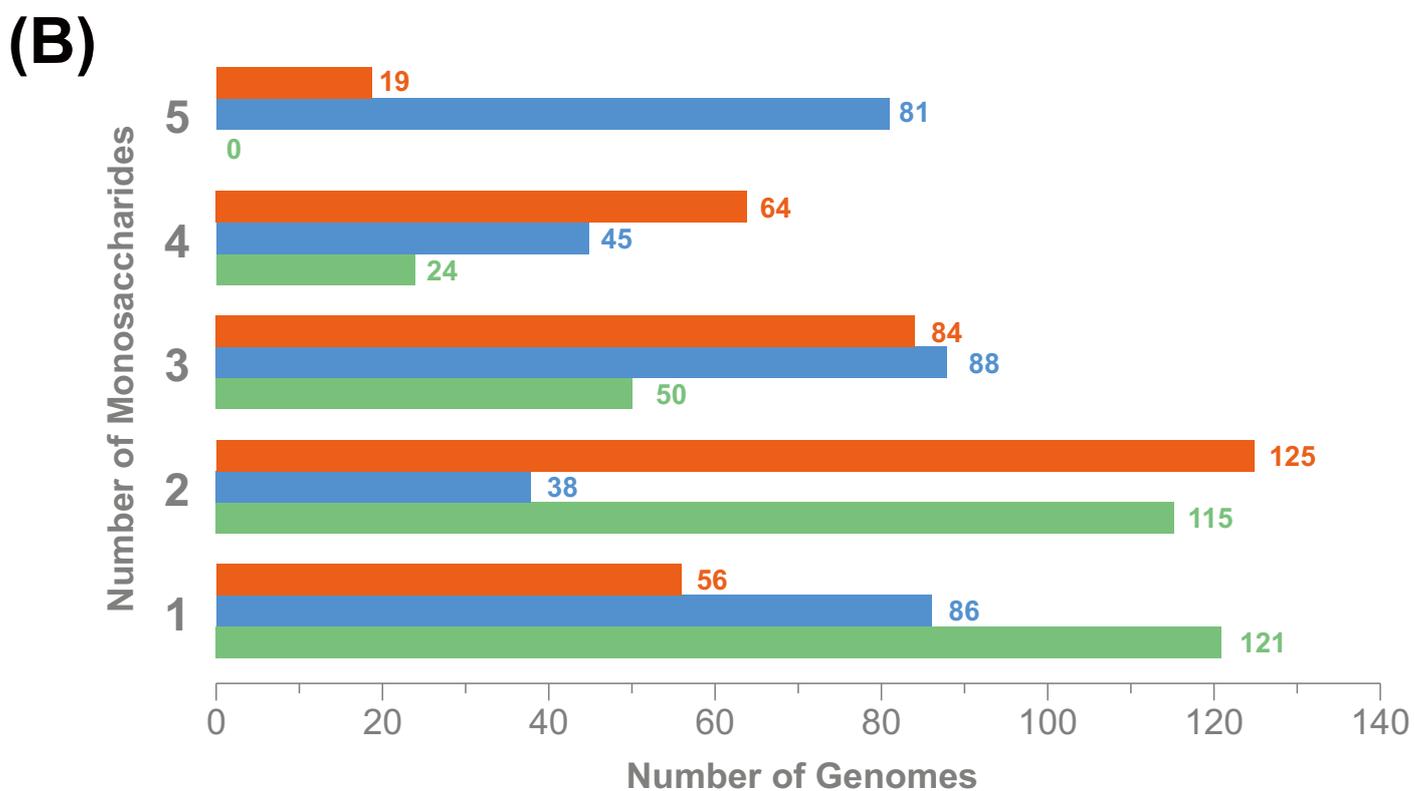

**Figure 2**

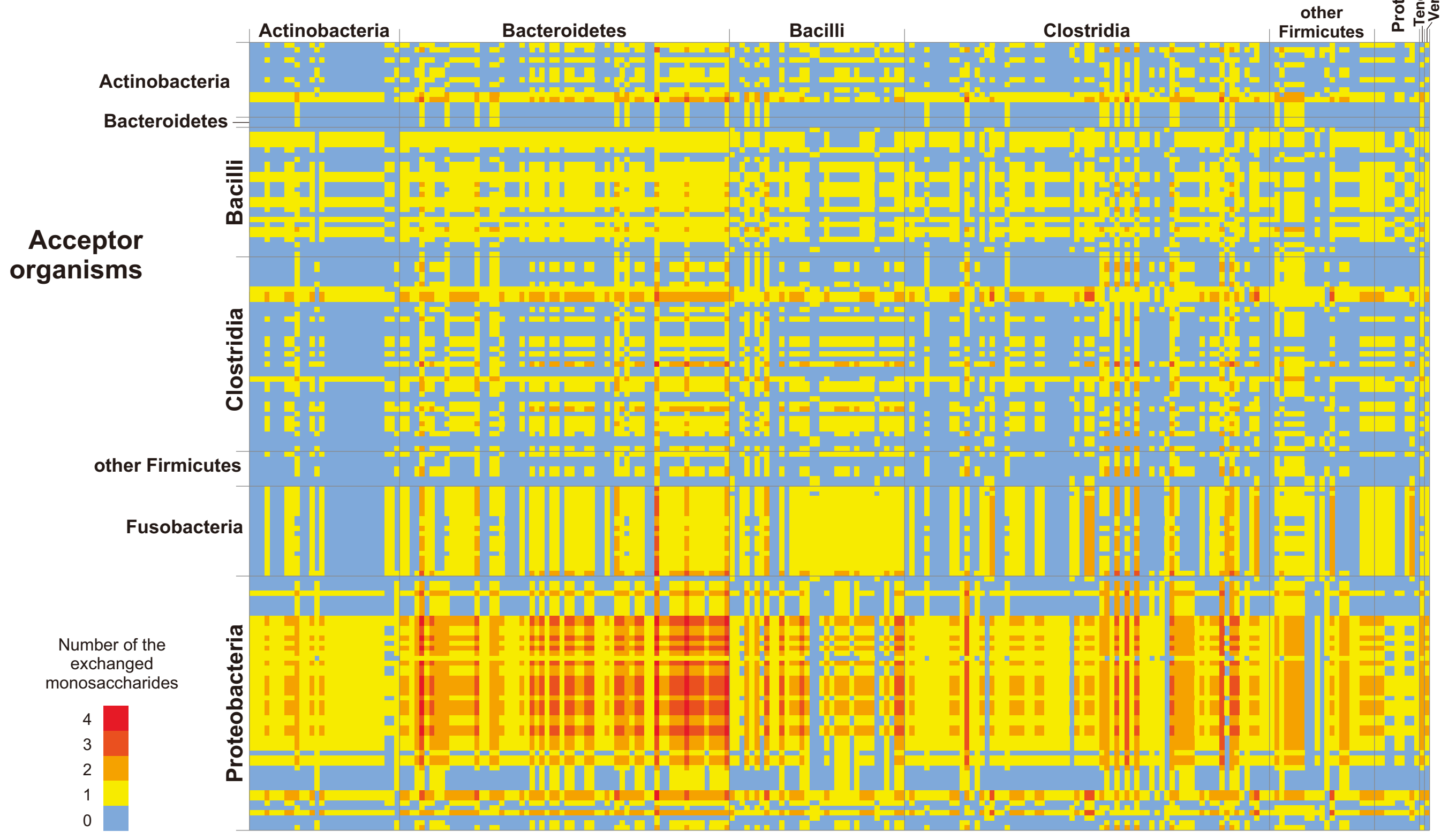

**Figure 3**

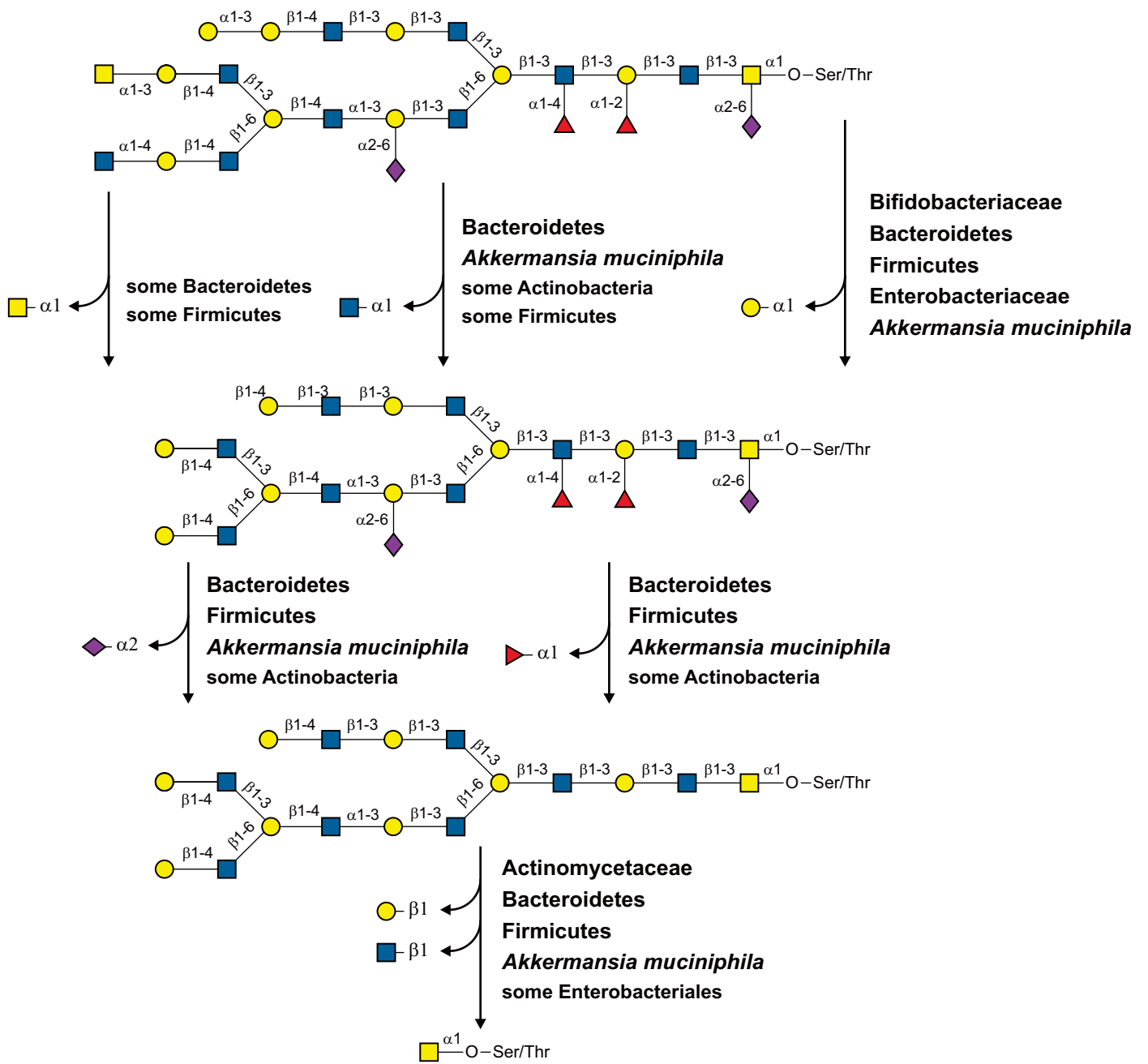

**Figure 4**

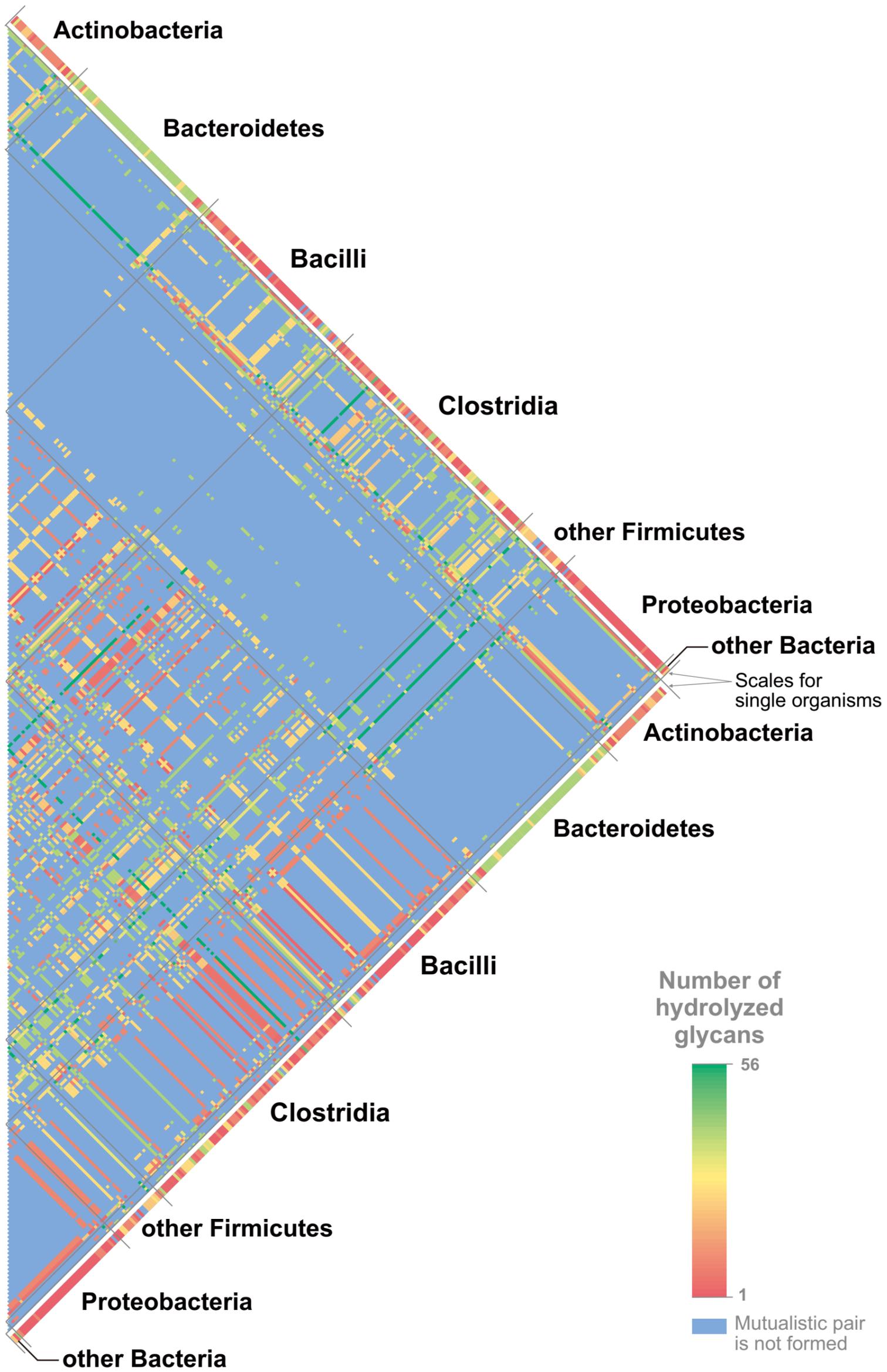

Actinobacteria

Bacteroidetes

Bacilli

Clostridia

other Firmicutes

Proteobacteria

other Bacteria

Scales for single organisms

Actinobacteria

Bacteroidetes

Bacilli

Clostridia

other Firmicutes

Proteobacteria

other Bacteria

Number of hydrolyzed glycans

Mutualistic pair is not formed

**Figure 5**

**Figure S1.** Maximum-likelihood tree for the FucK and RhaB proteins. Functions are shown by colors. The SEED identifiers for proteins are shown; their sequences, see the file Sequences S1 in the Supplementary materials. Genome names are shown in brackets.

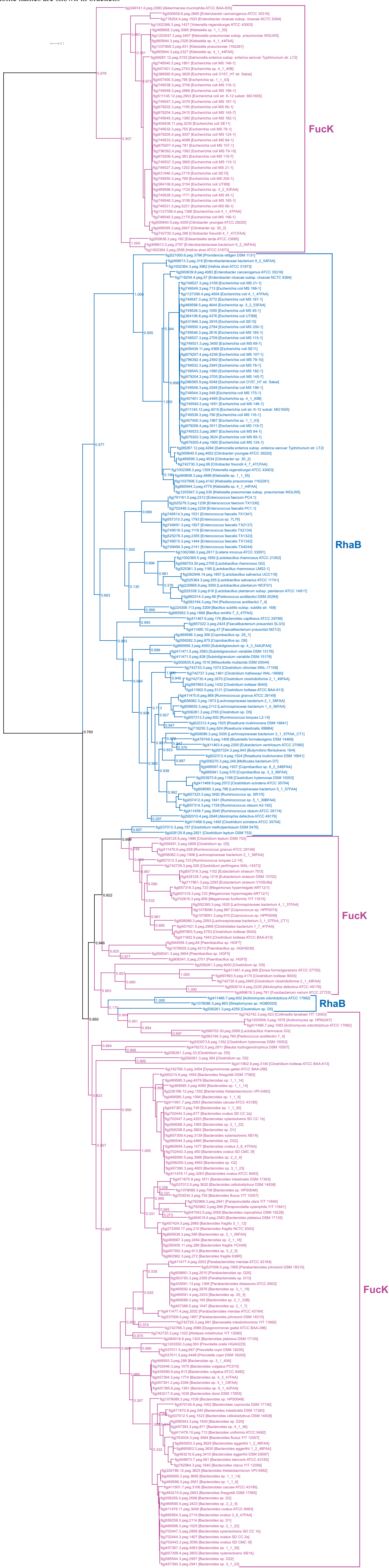

**Figure S1.** Maximum-likelihood tree for the FucA and RhaD proteins. Functions are shown by colors. The SEED identifiers for proteins are shown; for their sequences, see the file Sequences S1 in the Supplementary materials. Genome names are shown in brackets.

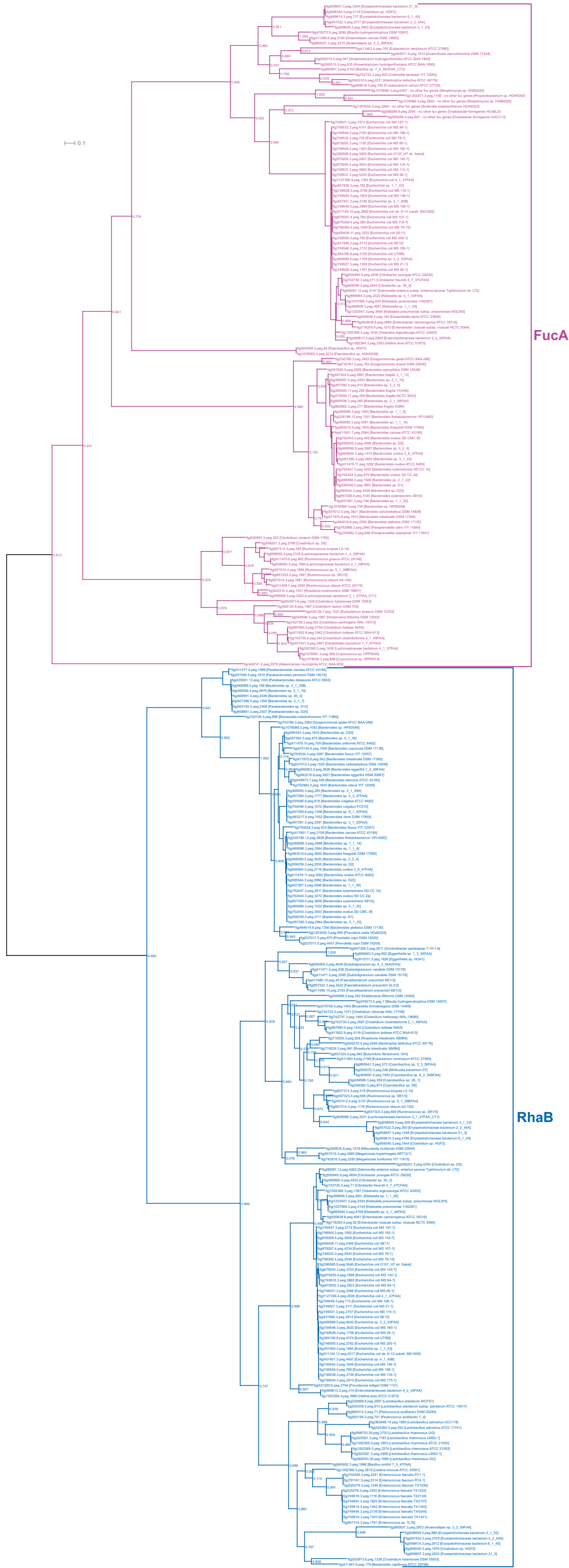

**Figure S3.** Maximum-likelihood tree for the FclA and FclD proteins. Functions are shown by colors. The SEED identifiers for proteins are shown; for their sequences, see the file Sequences S1 in the Supplementary materials. For the experimentally analyzed proteins (shown in bold) locus_tag identifiers are shown. Genome names are shown in brackets.

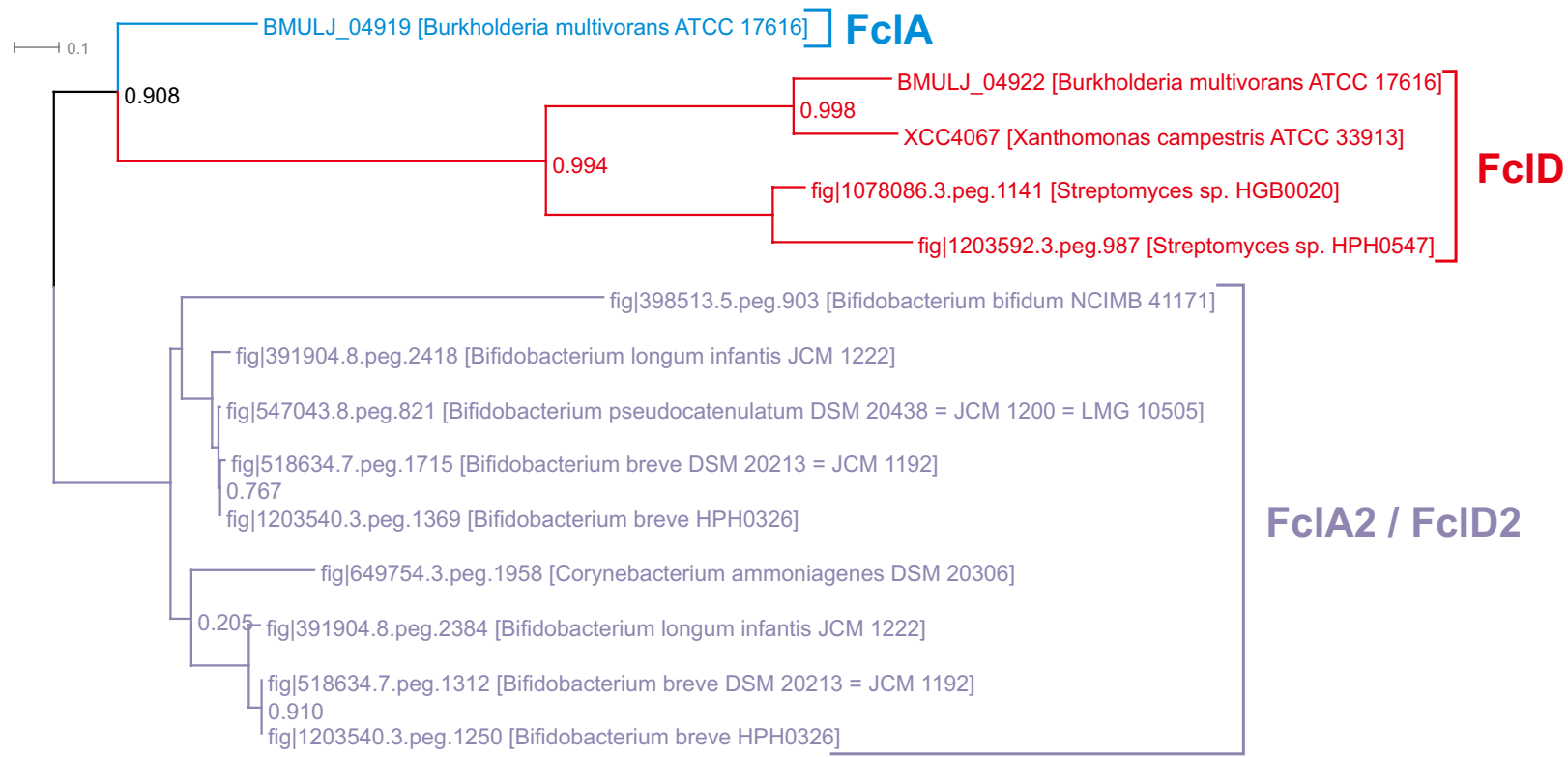

**Figure S4.** Maximum-likelihood tree for the AgaA and NagA proteins. Functions are shown by colors. The SEED identifiers for proteins are shown; for their sequences, see the file Sequences S1 in the Supplementary materials. Genome names are shown in brackets.

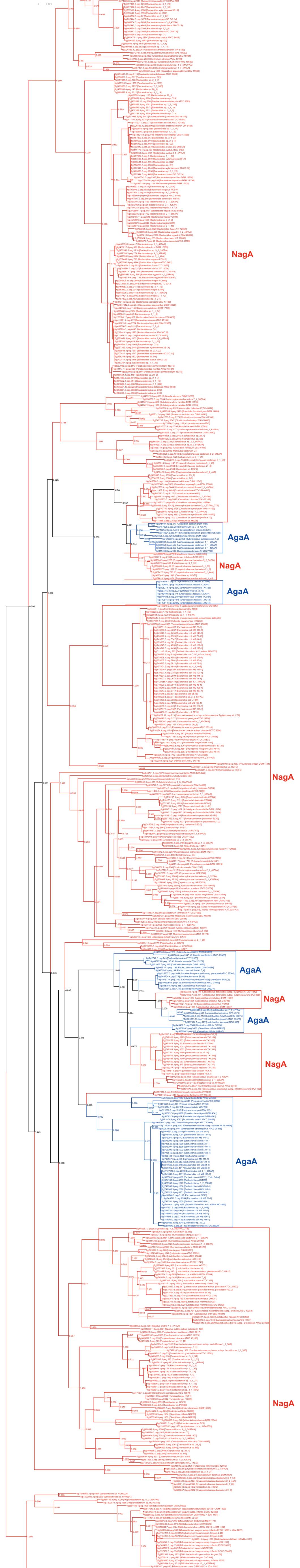

**Figure S5.** Maximum-likelihood tree for EIIC subunits of the AgaPTS, GamPTS, and GnbPTS. Functions are shown by colors. The SEED identifiers for proteins are shown; for their sequences, see the file Sequences S1 in the Supplementary materials. Genome names are shown in brackets.

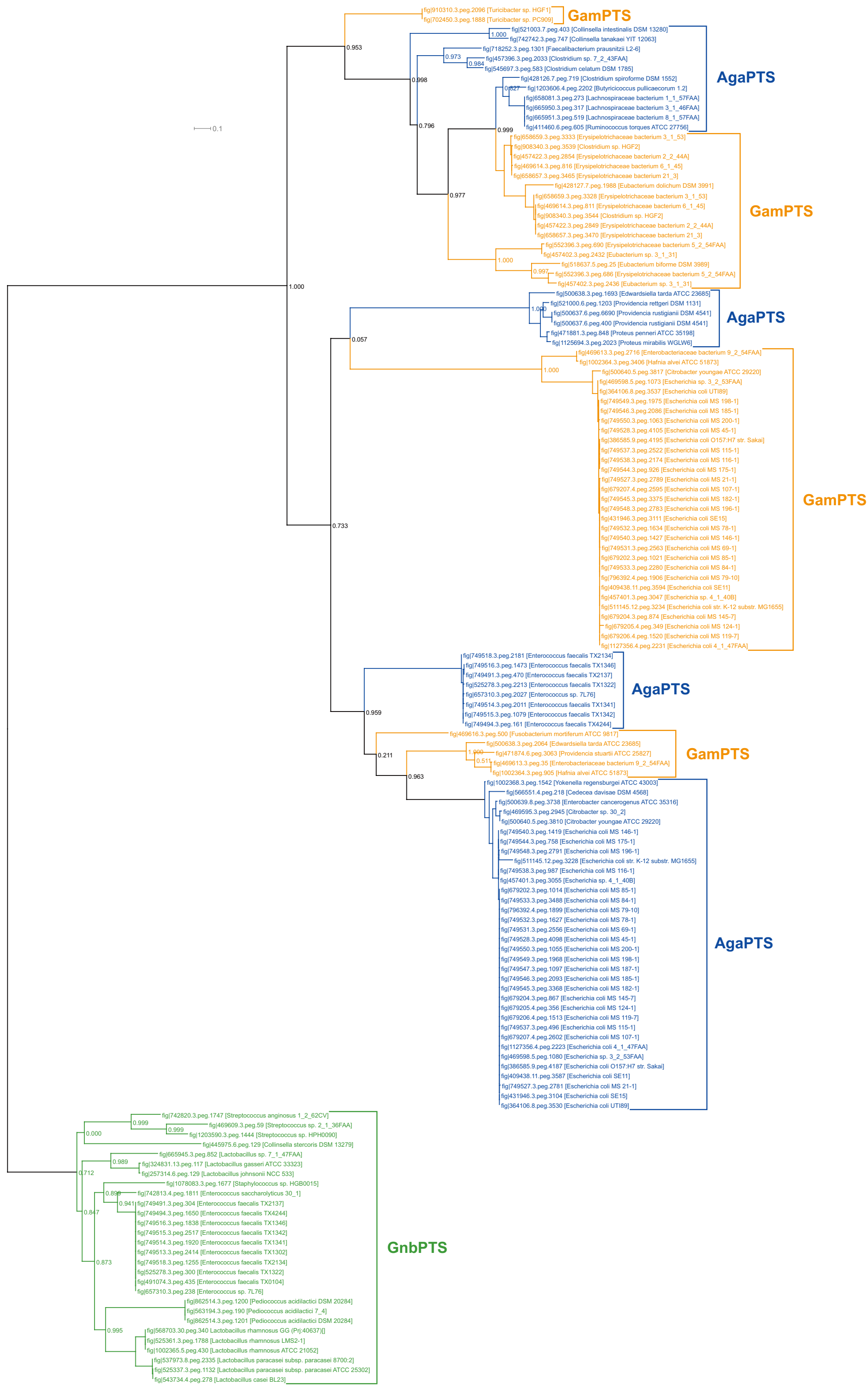

**Figure S6.** Maximum-likelihood tree for NanE and NanE-II proteins. Functions are shown by colors. The SEED identifiers for proteins are shown; for their sequences, see the file Sequences S1 in the Supplementary materials. Genome names are shown in brackets.

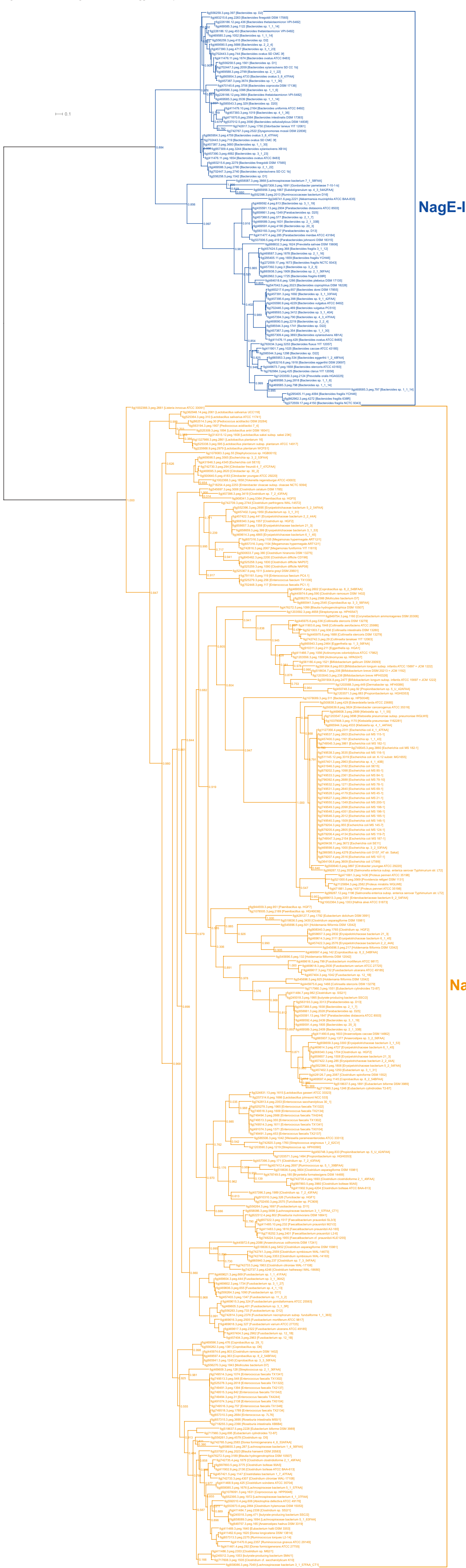

**Figure S7.** Maximum-likelihood tree for the GalY proteins and their homologs. GalY proteins are shown in blue. The SEED identifiers for proteins are shown; for their sequences, see the file Sequences S1 in the Supplementary materials. Genome names are shown in brackets.

**Figure S8.** Abundance of mucin glycan-forming monosaccharides in the KEGG database. Number of reactions associated with each monosaccharides is shown in agreement with the KEGG COMPOUND. Number of glycans containing each monosaccharide is shown in agreement with KEGG GLYCAN.

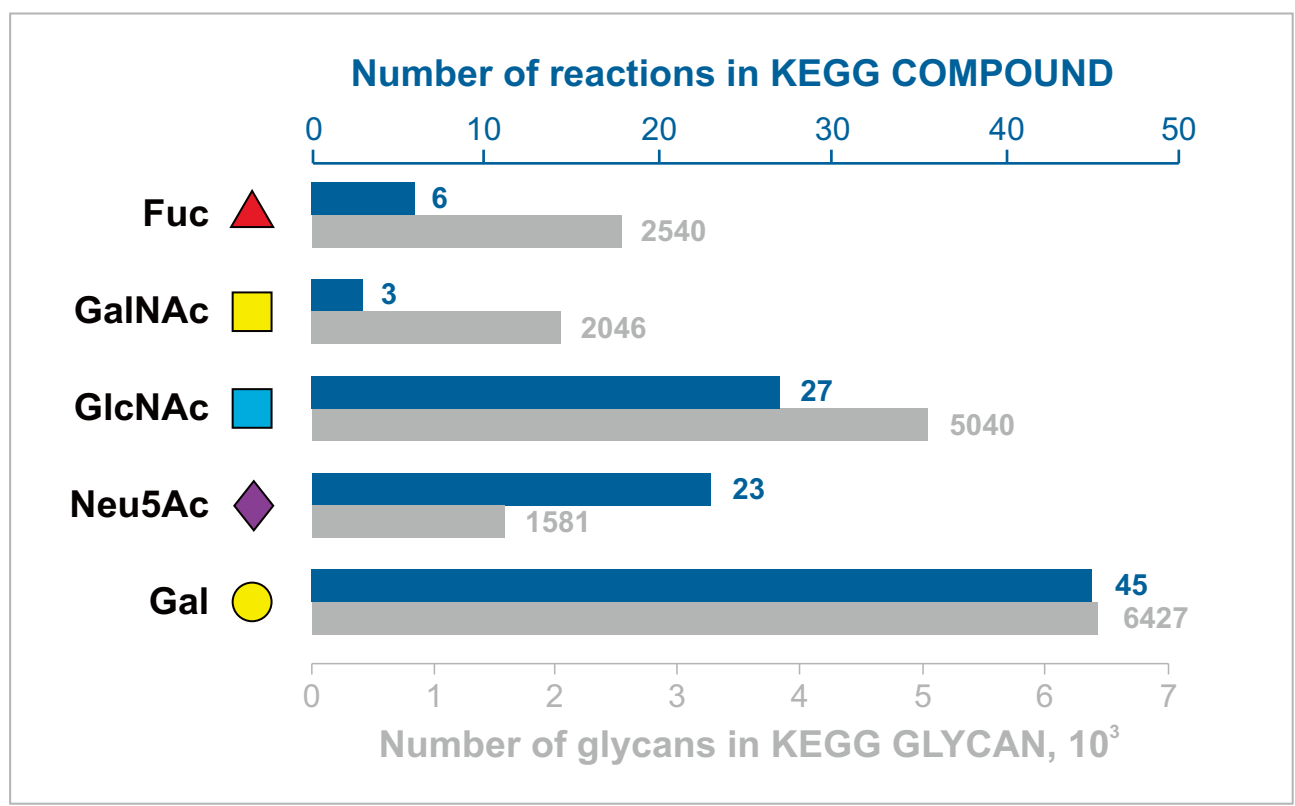

**Figure S9.** Known structures of mucin glycans detected in human intestine.

## 1. Structures listed by Podolsky, 1985

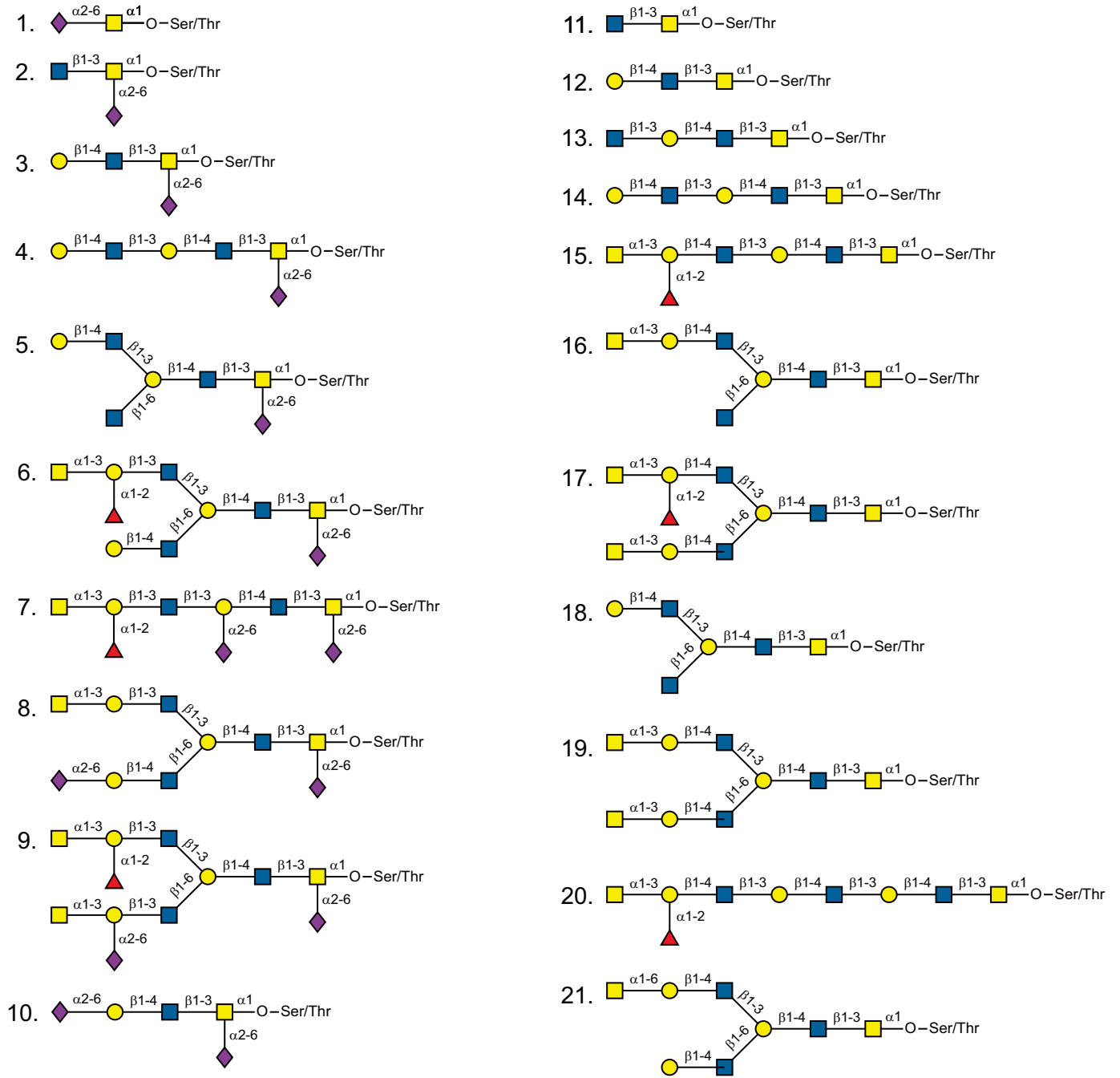

## 2. Structures listed by Rossez *et al.*, 2012

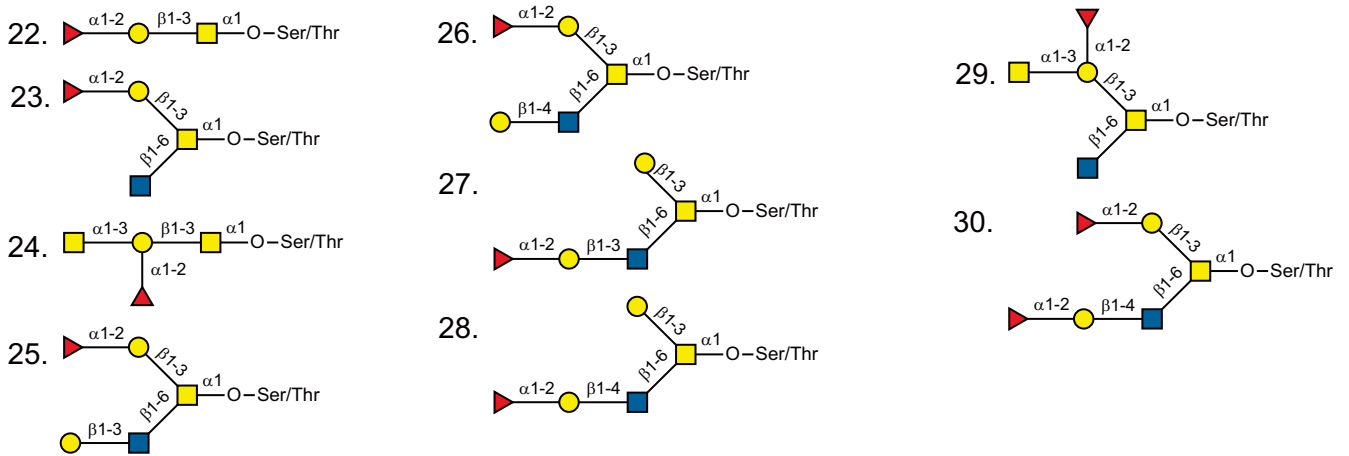

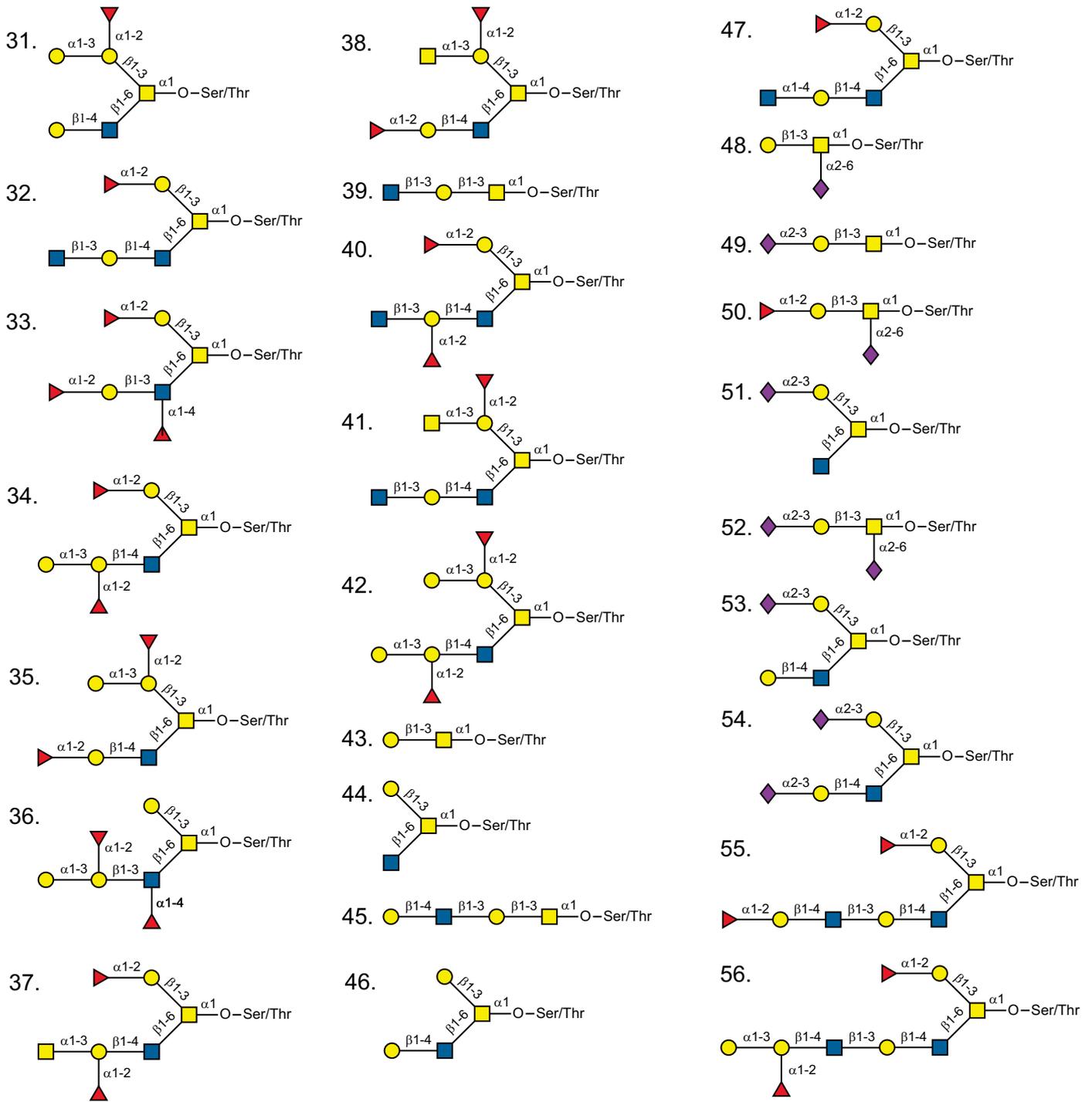

**Table S1.** Analyzed HGM genomes.

| Organism name | IDs | | | Taxonomy | | | | | Genome |
|---|---|---|---|---|---|---|---|---|---|
| | PubSEED ID | IMG ID | NCBI TaxID | Domain | Phylum | Class | Order | Family | status |
| Abiotrophia defectiva ATCC 49176 | 592010.4 | 2562617177 | 592010 | Bacteria | Firmicutes | Bacilli | Lactobacillales | Aerococcaceae | Draft |
| Acidaminococcus sp. D21 | 563191.3 | 643886056 | 563191 | Bacteria | Firmicutes | Negativicutes | Selenomonadales | Veillonellaceae | Draft |
| Acidaminococcus sp. HPA0509 | 1203555.3 | 2541047012 | 1203555 | Bacteria | Firmicutes | Negativicutes | Selenomonadales | Acidaminococcaceae | Draft |
| Acinetobacter junii SH205 | 575587.3 | 646206269 | 575587 | Bacteria | Proteobacteria | Gammaproteobacteria | Pseudomonadales | Moraxellaceae | Draft |
| Actinomyces odontolyticus ATCC 17982 | 411466.7 | 2623620210 | 411466 | Bacteria | Actinobacteria | Actinobacteria | Actinomycetales | Actinomycetaceae | Draft |
| Actinomyces sp. HPA0247 | 1203556.3 | 2541047011 | 1203556 | Bacteria | Actinobacteria | Actinobacteria | Actinomycetales | Actinomycetaceae | Draft |
| Akkermansia muciniphila ATCC BAA-835 | 349741.6 | 642555104 | 349741 | Bacteria | Verrucomicrobia | Verrucomicrobiae | Verrucomicrobiales | Verrucomicrobiaceae | Finished |
| Alistipes indistinctus YIT 12060 | 742725.3 | 2513237277 | 742725 | Bacteria | Bacteroidetes | Bacteroidia | Bacteroidales | Rikenellaceae | Draft |
| Alistipes shahii WAL 8301 | 717959.3 | 650377904 | 717959 | Bacteria | Bacteroidetes | Bacteroidia | Bacteroidales | Rikenellaceae | Draft |
| Anaerobaculum hydrogeniformans ATCC BAA-1850 | 592015.5 | 647000209 | 592015 | Bacteria | Synergistetes | Synergistia | Synergistales | Synergistaceae | Draft |
| Anaerococcus hydrogenalis DSM 7454 | 561177.4 | 642979323 | 561177 | Bacteria | Firmicutes | Clostridia | Clostridiales | Clostridiales Family XI. Incertae Sedis | Draft |
| Anaerofustis stercorihominis DSM 17244 | 445971.6 | 641736193 | 445971 | Bacteria | Firmicutes | Clostridia | Clostridiales | Eubacteriaceae | Draft |
| Anaerostipes caccae DSM 14662 | 411490.6 | 641736227 | 411490 | Bacteria | Firmicutes | Clostridia | Clostridiales | Lachnospiraceae | Draft |
| Anaerostipes hadrus DSM 3319 | 649757.3 | 2534681714 | 649757 | Bacteria | Firmicutes | Clostridia | Clostridiales | Lachnospiraceae | Draft |
| Anaerostipes sp. 3_2_56FAA | 665937.3 | 649989907 | 665937 | Bacteria | Firmicutes | Clostridia | Clostridiales | Lachnospiraceae | Draft |
| Anaerotruncus colihominis DSM 17241 | 445972.6 | 641736271 | 445972 | Bacteria | Firmicutes | Clostridia | Clostridiales | Ruminococcaceae | Draft |
| Bacillus smithii 7_3_47FAA | 665952.3 | 2513237385 | 665952 | Bacteria | Firmicutes | Bacilli | Bacillales | Bacillaceae | Draft |
| Bacillus sp. 7_6_55CFAA_CT2 | 665957.3 | 2513237325 | 665957 | Bacteria | Firmicutes | Bacilli | Bacillales | Bacillaceae | Draft |
| Bacillus subtilis subsp. subtilis str. 168 | 224308.113 | 646311909 | 224308 | Bacteria | Firmicutes | Bacilli | Bacillales | Bacillaceae | Finished |
| Bacteroides caccae ATCC 43185 | 411901.7 | 2623620314 | 411901 | Bacteria | Bacteroidetes | Bacteroidia | Bacteroidales | Bacteroidaceae | Draft |
| Bacteroides capillosus ATCC 29799 | 411467.6 | 2623620352 | 411467 | Bacteria | Bacteroidetes | Bacteroidia | Bacteroidales | Bacteroidaceae | Draft |
| Bacteroides cellulosilyticus DSM 14838 | 537012.5 | 643886111 | 537012 | Bacteria | Bacteroidetes | Bacteroidia | Bacteroidales | Bacteroidaceae | Draft |
| Bacteroides clarus YIT 12056 | 762984.3 | 651324010 | 762984 | Bacteria | Bacteroidetes | Bacteroidia | Bacteroidales | Bacteroidaceae | Draft |
| Bacteroides coprocola DSM 17136 | 470145.6 | 642791613 | 470145 | Bacteria | Bacteroidetes | Bacteroidia | Bacteroidales | Bacteroidaceae | Draft |
| Bacteroides coprophilus DSM 18228 | 547042.5 | 643886197 | 547042 | Bacteria | Bacteroidetes | Bacteroidia | Bacteroidales | Bacteroidaceae | Draft |
| Bacteroides dorei DSM 17855 | 483217.6 | 642979370 | 483217 | Bacteria | Bacteroidetes | Bacteroidia | Bacteroidales | Bacteroidaceae | Draft |
| Bacteroides eggerthii 1_2_48FAA | 665953.3 | 649989912 | 665953 | Bacteria | Bacteroidetes | Bacteroidia | Bacteroidales | Bacteroidaceae | Draft |
| Bacteroides eggerthii DSM 20697 | 483216.6 | 642979334 | 483216 | Bacteria | Bacteroidetes | Bacteroidia | Bacteroidales | Bacteroidaceae | Draft |
| Bacteroides finegoldii DSM 17565 | 483215.6 | 2562617142 | 483215 | Bacteria | Bacteroidetes | Bacteroidia | Bacteroidales | Bacteroidaceae | Draft |
| Bacteroides fluxus YIT 12057 | 763034.3 | 651324011 | 763034 | Bacteria | Bacteroidetes | Bacteroidia | Bacteroidales | Bacteroidaceae | Draft |
| Bacteroides fragilis 3_1_12 | 457424.5 | 645058788 | 457424 | Bacteria | Bacteroidetes | Bacteroidia | Bacteroidales | Bacteroidaceae | Draft |
| Bacteroides fragilis 638R | 862962.3 | 650377910 | 862962 | Bacteria | Bacteroidetes | Bacteroidia | Bacteroidales | Bacteroidaceae | Finished |
| Bacteroides fragilis NCTC 9343 | 272559.17 | 2623620861 | 272559 | Bacteria | Bacteroidetes | Bacteroidia | Bacteroidales | Bacteroidaceae | Finished |
| Bacteroides fragilis YCH46 | 295405.11 | 2623620898 | 295405 | Bacteria | Bacteroidetes | Bacteroidia | Bacteroidales | Bacteroidaceae | Finished |
| Bacteroides intestinalis DSM 17393 | 471870.8 | 642791621 | 471870 | Bacteria | Bacteroidetes | Bacteroidia | Bacteroidales | Bacteroidaceae | Draft |
| Bacteroides ovatus 3_8_47FAA | 665954.3 | 651324012 | 665954 | Bacteria | Bacteroidetes | Bacteroidia | Bacteroidales | Bacteroidaceae | Draft |
| Bacteroides ovatus ATCC 8483 | 411476.11 | 641380449 | 411476 | Bacteria | Bacteroidetes | Bacteroidia | Bacteroidales | Bacteroidaceae | Draft |
| Bacteroides ovatus SD CC 2a | 702444.3 | 647000212 | 702444 | Bacteria | Bacteroidetes | Bacteroidia | Bacteroidales | Bacteroidaceae | Draft |
| Bacteroides ovatus SD CMC 3f | 702443.3 | 647000213 | 702443 | Bacteria | Bacteroidetes | Bacteroidia | Bacteroidales | Bacteroidaceae | Draft |
| Bacteroides pectinophilus ATCC 43243 | 483218.5 | 642979337 | 483218 | Bacteria | Bacteroidetes | Bacteroidia | Bacteroidales | Bacteroidaceae | Draft |
| Bacteroides plebeius DSM 17135 | 484018.6 | 642979351 | 484018 | Bacteria | Bacteroidetes | Bacteroidia | Bacteroidales | Bacteroidaceae | Draft |
| Bacteroides sp. 1_1_14 | 469585.3 | 648861000 | 469585 | Bacteria | Bacteroidetes | Bacteroidia | Bacteroidales | Bacteroidaceae | Draft |
| Bacteroides sp. 1_1_30 | 457387.3 | 651324013 | 457387 | Bacteria | Bacteroidetes | Bacteroidia | Bacteroidales | Bacteroidaceae | Draft |
| Bacteroides sp. 1_1_6 | 469586.3 | 646206272 | 469586 | Bacteria | Bacteroidetes | Bacteroidia | Bacteroidales | Bacteroidaceae | Draft |
| Bacteroides sp. 2_1_16 | 469587.3 | 647533110 | 469587 | Bacteria | Bacteroidetes | Bacteroidia | Bacteroidales | Bacteroidaceae | Draft |
| Bacteroides sp. 2_1_22 | 469588.3 | 647533111 | 469588 | Bacteria | Bacteroidetes | Bacteroidia | Bacteroidales | Bacteroidaceae | Draft |
| Bacteroides sp. 2_1_33B | 469589.3 | 647533112 | 469589 | Bacteria | Bacteroidetes | Bacteroidia | Bacteroidales | Bacteroidaceae | Draft |
| Bacteroides sp. 2_1_56FAA | 665938.3 | 651324014 | 665938 | Bacteria | Bacteroidetes | Bacteroidia | Bacteroidales | Bacteroidaceae | Draft |
| Bacteroides sp. 2_1_7 | 457388.5 | 645058782 | 457388 | Bacteria | Bacteroidetes | Bacteroidia | Bacteroidales | Bacteroidaceae | Draft |
| Bacteroides sp. 2_2_4 | 469590.5 | 646206266 | 469590 | Bacteria | Bacteroidetes | Bacteroidia | Bacteroidales | Bacteroidaceae | Draft |
| Bacteroides sp. 20_3 | 469591.4 | 648861001 | 469591 | Bacteria | Bacteroidetes | Bacteroidia | Bacteroidales | Bacteroidaceae | Draft |
| Bacteroides sp. 3_1_19 | 469592.4 | 648861002 | 469592 | Bacteria | Bacteroidetes | Bacteroidia | Bacteroidales | Bacteroidaceae | Draft |
| Bacteroides sp. 3_1_23 | 457390.3 | 648861003 | 457390 | Bacteria | Bacteroidetes | Bacteroidia | Bacteroidales | Bacteroidaceae | Draft |
| Bacteroides sp. 3_1_33FAA | 457391.3 | 647533113 | 457391 | Bacteria | Bacteroidetes | Bacteroidia | Bacteroidales | Bacteroidaceae | Draft |

| Organism | | | | | | | | | |
|---|---|---|---|---|---|---|---|---|---|
| Bacteroides sp. 3_1_40A | 469593.3 | 649989913 | 469593 | Bacteria | Bacteroidetes | Bacteroidia | Bacteroidales | Bacteroidaceae | Draft |
| Bacteroides sp. 3_2_5 | 457392.3 | 2558309115 | 457392 | Bacteria | Bacteroidetes | Bacteroidia | Bacteroidales | Bacteroidaceae | Draft |
| Bacteroides sp. 4_1_36 | 457393.3 | 649989914 | 457393 | Bacteria | Bacteroidetes | Bacteroidia | Bacteroidales | Bacteroidaceae | Draft |
| Bacteroides sp. 4_3_47FAA | 457394.3 | 646206274 | 457394 | Bacteria | Bacteroidetes | Bacteroidia | Bacteroidales | Bacteroidaceae | Draft |
| Bacteroides sp. 9_1_42FAA | 457395.6 | 646206263 | 457395 | Bacteria | Bacteroidetes | Bacteroidia | Bacteroidales | Bacteroidaceae | Draft |
| Bacteroides sp. D1 | 556258.5 | 646206264 | 556258 | Bacteria | Bacteroidetes | Bacteroidia | Bacteroidales | Bacteroidaceae | Draft |
| Bacteroides sp. D2 | 556259.3 | 2562617173 | 556259 | Bacteria | Bacteroidetes | Bacteroidia | Bacteroidales | Bacteroidaceae | Draft |
| Bacteroides sp. D20 | 585543.3 | 647533114 | 585543 | Bacteria | Bacteroidetes | Bacteroidia | Bacteroidales | Bacteroidaceae | Draft |
| Bacteroides sp. D22 | 585544.3 | 648861004 | 585544 | Bacteria | Bacteroidetes | Bacteroidia | Bacteroidales | Bacteroidaceae | Draft |
| Bacteroides sp. HPS0048 | 1078089.3 | 2529293266 | 1078089 | Bacteria | Bacteroidetes | Bacteroidia | Bacteroidales | Bacteroidaceae | Draft |
| Bacteroides stercoris ATCC 43183 | 449673.7 | 641736196 | 449673 | Bacteria | Bacteroidetes | Bacteroidia | Bacteroidales | Bacteroidaceae | Draft |
| Bacteroides thetaiotaomicron VPI-5482 | 226186.12 | 2623620899 | 226186 | Bacteria | Bacteroidetes | Bacteroidia | Bacteroidales | Bacteroidaceae | Finished |
| Bacteroides uniformis ATCC 8492 | 411479.1 | 641380447 | 411479 | Bacteria | Bacteroidetes | Bacteroidia | Bacteroidales | Bacteroidaceae | Draft |
| Bacteroides vulgatus ATCC 8482 | 435590.9 | 2623620366 | 435590 | Bacteria | Bacteroidetes | Bacteroidia | Bacteroidales | Bacteroidaceae | Finished |
| Bacteroides vulgatus PC510 | 702446.3 | 647000214 | 702446 | Bacteria | Bacteroidetes | Bacteroidia | Bacteroidales | Bacteroidaceae | Draft |
| Bacteroides xylanisolvens SD CC 1b | 702447.3 | 2617271081 | 702447 | Bacteria | Bacteroidetes | Bacteroidia | Bacteroidales | Bacteroidaceae | Draft |
| Bacteroides xylanisolvens XB1A | 657309.4 | 650377911 | 657309 | Bacteria | Bacteroidetes | Bacteroidia | Bacteroidales | Bacteroidaceae | Finished |
| Barnesiella intestinihominis YIT 11860 | 742726.3 | 2529292932 | 742726 | Bacteria | Bacteroidetes | Bacteroidia | Bacteroidales | Porphyromonadaceae | Draft |
| Bifidobacterium adolescentis L2-32 | 411481.5 | 640963015 | 411481 | Bacteria | Actinobacteria | Actinobacteria | Bifidobacteriales | Bifidobacteriaceae | Draft |
| Bifidobacterium angulatum DSM 20098 = JCM 7096 | 518635.7 | 2562617143 | 518635 | Bacteria | Actinobacteria | Actinobacteria | Bifidobacteriales | Bifidobacteriaceae | Draft |
| Bifidobacterium animalis subsp. lactis AD011 | 442563.4 | 643348515 | 442563 | Bacteria | Actinobacteria | Actinobacteria | Bifidobacteriales | Bifidobacteriaceae | Finished |
| Bifidobacterium bifidum NCIMB 41171 | 398513.5 | 643886040 | 398513 | Bacteria | Actinobacteria | Actinobacteria | Bifidobacteriales | Bifidobacteriaceae | Draft |
| Bifidobacterium breve DSM 20213 = JCM 1192 | 518634.7 | 643886102 | 518634 | Bacteria | Actinobacteria | Actinobacteria | Bifidobacteriales | Bifidobacteriaceae | Draft |
| Bifidobacterium breve HPH0326 | 1203540.3 | 2541047014 | 1203540 | Bacteria | Actinobacteria | Actinobacteria | Bifidobacteriales | Bifidobacteriaceae | Draft |
| Bifidobacterium catenulatum DSM 16992 = JCM 1194 | 566552.6 | 642979312 | 566552 | Bacteria | Actinobacteria | Actinobacteria | Bifidobacteriales | Bifidobacteriaceae | Draft |
| Bifidobacterium dentium ATCC 27678 | 473819.7 | 641736189 | 473819 | Bacteria | Actinobacteria | Actinobacteria | Bifidobacteriales | Bifidobacteriaceae | Draft |
| Bifidobacterium gallicum DSM 20093 | 561180.4 | 2562617144 | 561180 | Bacteria | Actinobacteria | Actinobacteria | Bifidobacteriales | Bifidobacteriaceae | Draft |
| Bifidobacterium longum DJO10A (Prj:321) | 205913.11 | 642555107 | 205913 | Bacteria | Actinobacteria | Actinobacteria | Bifidobacteriales | Bifidobacteriaceae | Finished |
| Bifidobacterium longum NCC2705 | 206672.9 | 637000031 | 206672 | Bacteria | Actinobacteria | Actinobacteria | Bifidobacteriales | Bifidobacteriaceae | Finished |
| Bifidobacterium longum subsp. infantis 157F | 565040.3 | 649633017 | 565040 | Bacteria | Actinobacteria | Actinobacteria | Bifidobacteriales | Bifidobacteriaceae | Finished |
| Bifidobacterium longum subsp. infantis ATCC 15697 = JCM | 391904.8 | 651053007 | 391904 | Bacteria | Actinobacteria | Actinobacteria | Bifidobacteriales | Bifidobacteriaceae | Finished |
| Bifidobacterium longum subsp. infantis ATCC 55813 | 548480.3 | 643886140 | 548480 | Bacteria | Actinobacteria | Actinobacteria | Bifidobacteriales | Bifidobacteriaceae | Draft |
| Bifidobacterium longum subsp. infantis CCUG 52486 | 537937.5 | 643886216 | 537937 | Bacteria | Actinobacteria | Actinobacteria | Bifidobacteriales | Bifidobacteriaceae | Draft |
| Bifidobacterium longum subsp. longum 2-2B | 1161745.3 | 2531839495 | 1161745 | Bacteria | Actinobacteria | Actinobacteria | Bifidobacteriales | Bifidobacteriaceae | Draft |
| Bifidobacterium longum subsp. longum 44B | 1161743.3 | 2529293221 | 1161743 | Bacteria | Actinobacteria | Actinobacteria | Bifidobacteriales | Bifidobacteriaceae | Draft |
| Bifidobacterium longum subsp. longum F8 | 722911.3 | 650377912 | 722911 | Bacteria | Actinobacteria | Actinobacteria | Bifidobacteriales | Bifidobacteriaceae | Finished |
| Bifidobacterium longum subsp. longum JCM 1217 | 565042.3 | 649633019 | 565042 | Bacteria | Actinobacteria | Actinobacteria | Bifidobacteriales | Bifidobacteriaceae | Finished |
| Bifidobacterium pseudocatenulatum DSM 20438 = JCM 120 | 547043.8 | 2597490216 | 547043 | Bacteria | Actinobacteria | Actinobacteria | Bifidobacteriales | Bifidobacteriaceae | Draft |
| Bifidobacterium sp. 12_1_47BFAA | 469594.3 | 649989916 | 469594 | Bacteria | Actinobacteria | Actinobacteria | Bifidobacteriales | Bifidobacteriaceae | Draft |
| Bilophila wadsworthia 3_1_6 | 563192.3 | 2562617176 | 563192 | Bacteria | Proteobacteria | Deltaproteobacteria | Desulfovibrionales | Desulfovibrionaceae | Draft |
| Blautia hansenii DSM 20583 | 537007.6 | 2562617096 | 537007 | Bacteria | Firmicutes | Clostridia | Clostridiales | unclassified Clostridiales | Draft |
| Blautia hydrogenotrophica DSM 10507 | 476272.5 | 643886199 | 476272 | Bacteria | Firmicutes | Clostridia | Clostridiales | unclassified Clostridiales | Draft |
| Bryantella formatexigens DSM 14469 | 478749.5 | 2562617090 | 478749 | Bacteria | Firmicutes | Clostridia | Clostridiales | Lachnospiraceae | Draft |
| Burkholderiales bacterium 1_1_47 | 469610.4 | 648861006 | 469610 | Bacteria | Proteobacteria | Betaproteobacteria | Burkholderiales | unclassified Burkholderiales | Draft |
| butyrate-producing bacterium SM4/1 | 245012.3 | 2622736406 | 245012 | Bacteria | Firmicutes | Clostridia | Clostridiales | unclassified Clostridiales | Finished |
| butyrate-producing bacterium SS3/4 | 245014.3 | 650377989 | 245014 | Bacteria | Firmicutes | Clostridia | Clostridiales | unclassified Clostridiales | Finished |
| butyrate-producing bacterium SSC/2 | 245018.3 | 650377990 | 245018 | Bacteria | Firmicutes | Clostridia | Clostridiales | unclassified Clostridiales | Finished |
| Butyricicoccus pullicaecorum 1.2 | 1203606.4 | 2537561731 | 1203606 | Bacteria | Firmicutes | Clostridia | Clostridiales | Clostridiaceae | Draft |
| Butyrivibrio crossotus DSM 2876 | 511680.4 | 645951834 | 511680 | Bacteria | Firmicutes | Clostridia | Clostridiales | Lachnospiraceae | Draft |
| Butyrivibrio fibrisolvens 16/4 | 657324.3 | 650377919 | 657324 | Bacteria | Firmicutes | Clostridia | Clostridiales | Lachnospiraceae | Finished |
| Campylobacter coli JV20 | 864566.3 | 648276627 | 864566 | Bacteria | Proteobacteria | Epsilonproteobacteria | Campylobacterales | Campylobacteraceae | Draft |
| Campylobacter upsaliensis JV21 | 888826.3 | 649989919 | 888826 | Bacteria | Proteobacteria | Epsilonproteobacteria | Campylobacterales | Campylobacteraceae | Draft |
| Catenibacterium mitsuokai DSM 15897 | 451640.5 | 643886110 | 451640 | Bacteria | Firmicutes | Erysipelotrichi | Erysipelotrichales | Erysipelotrichaceae | Draft |
| Cedecea davisae DSM 4568 | 566551.4 | 2541046972 | 566551 | Bacteria | Proteobacteria | Gammaproteobacteria | Enterobacteriales | Enterobacteriaceae | Draft |
| Citrobacter freundii 4_7_47CFAA | 742730.3 | 2513237266 | 742730 | Bacteria | Proteobacteria | Gammaproteobacteria | Enterobacteriales | Enterobacteriaceae | Draft |
| Citrobacter sp. 30_2 | 469595.3 | 646206256 | 469595 | Bacteria | Proteobacteria | Gammaproteobacteria | Enterobacteriales | Enterobacteriaceae | Draft |
| Citrobacter youngae ATCC 29220 | 500640.5 | 2562617093 | 500640 | Bacteria | Proteobacteria | Gammaproteobacteria | Enterobacteriales | Enterobacteriaceae | Draft |

| Name | ID1 | ID2 | ID3 | Domain | Phylum | Class | Order | Family | Status |
|---|---|---|---|---|---|---|---|---|---|
| Clostridiales bacterium 1_7_47FAA | 457421.5 | 643886007 | 457421 | Bacteria | Firmicutes | Clostridia | Clostridiales | unclassified Clostridiales | Draft |
| Clostridium asparagiforme DSM 15981 | 518636.5 | 643886112 | 518636 | Bacteria | Firmicutes | Clostridia | Clostridiales | Clostridiaceae | Draft |
| Clostridium bartlettii DSM 16795 | 445973.7 | 641736113 | 445973 | Bacteria | Firmicutes | Clostridia | Clostridiales | Clostridiaceae | Draft |
| Clostridium boltae 90A5 | 997893.5 | 2537561659 | 997893 | Bacteria | Firmicutes | Clostridia | Clostridiales | Clostridiaceae | Draft |
| Clostridium boltae ATCC BAA-613 | 411902.9 | 641380428 | 411902 | Bacteria | Firmicutes | Clostridia | Clostridiales | Clostridiaceae | Draft |
| Clostridium celatum DSM 1785 | 545697.3 | 2524023202 | 545697 | Bacteria | Firmicutes | Clostridia | Clostridiales | Clostridiaceae | Draft |
| Clostridium cf. saccharolyticum K10 | 717608.3 | 650377923 | 717608 | Bacteria | Firmicutes | Clostridia | Clostridiales | Clostridiaceae | Draft |
| Clostridium citroniae WAL-17108 | 742733.3 | 2513237360 | 742733 | Bacteria | Firmicutes | Clostridia | Clostridiales | Clostridiaceae | Draft |
| Clostridium clostridioforme 2_1_49FAA | 742735.4 | 2513237315 | 742735 | Bacteria | Firmicutes | Clostridia | Clostridiales | Clostridiaceae | Draft |
| Clostridium difficile CD196 | 645462.3 | 646311914 | 645462 | Bacteria | Firmicutes | Clostridia | Clostridiales | Clostridiaceae | Draft |
| Clostridium difficile NAP07 | 525258.3 | 647000227 | 525258 | Bacteria | Firmicutes | Clostridia | Clostridiales | Clostridiaceae | Finished |
| Clostridium difficile NAP08 | 525259.3 | 647000228 | 525259 | Bacteria | Firmicutes | Clostridia | Clostridiales | Clostridiaceae | Draft |
| Clostridium hathewayi WAL-18680 | 742737.3 | 2513237275 | 742737 | Bacteria | Firmicutes | Clostridia | Clostridiales | Clostridiaceae | Draft |
| Clostridium hiranonis DSM 13275 | 500633.7 | 642979371 | 500633 | Bacteria | Firmicutes | Clostridia | Clostridiales | Clostridiaceae | Draft |
| Clostridium hylemonae DSM 15053 | 553973.6 | 642979359 | 553973 | Bacteria | Firmicutes | Clostridia | Clostridiales | Clostridiaceae | Draft |
| Clostridium leptum DSM 753 | 428125.8 | 641380427 | 428125 | Bacteria | Firmicutes | Clostridia | Clostridiales | Clostridiaceae | Draft |
| Clostridium methylpentosum DSM 5476 | 537013.3 | 643886206 | 537013 | Bacteria | Firmicutes | Clostridia | Clostridiales | Clostridiaceae | Draft |
| Clostridium nexile DSM 1787 | 500632.7 | 642979369 | 500632 | Bacteria | Firmicutes | Clostridia | Clostridiales | Clostridiaceae | Draft |
| Clostridium perfringens WAL-14572 | 742739.3 | 2537561919 | 742739 | Bacteria | Firmicutes | Clostridia | Clostridiales | Clostridiaceae | Draft |
| Clostridium ramosum DSM 1402 | 445974.6 | 641736198 | 445974 | Bacteria | Firmicutes | Erysipelotrichi | Erysipelotrichales | Erysipelotrichaceae | Draft |
| Clostridium scindens ATCC 35704 | 411468.9 | 641736197 | 411468 | Bacteria | Firmicutes | Clostridia | Clostridiales | Clostridiaceae | Draft |
| Clostridium sp. 7_2_43FAA | 457396.3 | 646206260 | 457396 | Bacteria | Firmicutes | Clostridia | Clostridiales | Clostridiaceae | Draft |
| Clostridium sp. 7_3_54FAA | 665940.3 | 2513237367 | 665940 | Bacteria | Firmicutes | Clostridia | Clostridiales | Clostridiaceae | Draft |
| Clostridium sp. D5 | 556261.3 | 651285002 | 556261 | Bacteria | Firmicutes | Clostridia | Clostridiales | Clostridiaceae | Draft |
| Clostridium sp. HGF2 | 908340.3 | 649989921 | 908340 | Bacteria | Firmicutes | Clostridia | Clostridiales | Clostridiaceae | Draft |
| Clostridium sp. L2-50 | 411489.7 | 641380448 | 411489 | Bacteria | Firmicutes | Clostridia | Clostridiales | Clostridiaceae | Draft |
| Clostridium sp. M62/1 | 411486.3 | 2562617079 | 411486 | Bacteria | Firmicutes | Clostridia | Clostridiales | Clostridiaceae | Draft |
| Clostridium sp. SS2/1 | 411484.7 | 641736270 | 411484 | Bacteria | Firmicutes | Clostridia | Clostridiales | Clostridiaceae | Draft |
| Clostridium spiroforme DSM 1552 | 428126.7 | 641736169 | 428126 | Bacteria | Firmicutes | Erysipelotrichi | Erysipelotrichales | Erysipelotrichaceae | Draft |
| Clostridium sporogenes ATCC 15579 | 471871.7 | 642791615 | 471871 | Bacteria | Firmicutes | Clostridia | Clostridiales | Clostridiaceae | Draft |
| Clostridium symbiosum WAL-14163 | 742740.3 | 649989922 | 742740 | Bacteria | Firmicutes | Clostridia | Clostridiales | Clostridiaceae | Draft |
| Clostridium symbiosum WAL-14673 | 742741.3 | 649989923 | 742741 | Bacteria | Firmicutes | Clostridia | Clostridiales | Clostridiaceae | Draft |
| Collinsella aerofaciens ATCC 25986 | 411903.6 | 640612206 | 411903 | Bacteria | Actinobacteria | Actinobacteria | Coriobacteriales | Coriobacteriaceae | Draft |
| Collinsella intestinalis DSM 13280 | 521003.7 | 2562617146 | 521003 | Bacteria | Actinobacteria | Actinobacteria | Coriobacteriales | Coriobacteriaceae | Draft |
| Collinsella stercoris DSM 13279 | 445975.6 | 642979321 | 445975 | Bacteria | Actinobacteria | Actinobacteria | Coriobacteriales | Coriobacteriaceae | Draft |
| Collinsella tanakaei YIT 12063 | 742742.3 | 2513237344 | 742742 | Bacteria | Actinobacteria | Actinobacteria | Coriobacteriales | Coriobacteriaceae | Draft |
| Coprobacillus sp. 29_1 | 469596.3 | 649989924 | 469596 | Bacteria | Firmicutes | Erysipelotrichi | Erysipelotrichales | Erysipelotrichaceae | Draft |
| Coprobacillus sp. 3_3_56FAA | 665941.3 | 2513237376 | 665941 | Bacteria | Firmicutes | Erysipelotrichi | Erysipelotrichales | Erysipelotrichaceae | Draft |
| Coprobacillus sp. 8_2_54BFAA | 469597.4 | 2513237364 | 469597 | Bacteria | Firmicutes | Erysipelotrichi | Erysipelotrichales | Erysipelotrichaceae | Draft |
| Coprobacillus sp. D6 | 556262.3 | 2547132035 | 556262 | Bacteria | Firmicutes | Erysipelotrichi | Erysipelotrichales | Erysipelotrichaceae | Draft |
| Coprococcus catus GD/7 | 717962.3 | 650377925 | 717962 | Bacteria | Firmicutes | Clostridia | Clostridiales | Lachnospiraceae | Finished |
| Coprococcus comes ATCC 27758 | 470146.3 | 643886116 | 470146 | Bacteria | Firmicutes | Clostridia | Clostridiales | Lachnospiraceae | Draft |
| Coprococcus eutactus ATCC 27759 | 411474.6 | 641380422 | 411474 | Bacteria | Firmicutes | Clostridia | Clostridiales | Lachnospiraceae | Draft |
| Coprococcus sp. ART55/1 | 751585.3 | 2565956545 | 751585 | Bacteria | Firmicutes | Clostridia | Clostridiales | Lachnospiraceae | Finished |
| Coprococcus sp. HPP0048 | 1078091.3 | 2541046996 | 1078091 | Bacteria | Firmicutes | Clostridia | Clostridiales | Lachnospiraceae | Draft |
| Coprococcus sp. HPP0074 | 1078090.3 | 2541046997 | 1078090 | Bacteria | Firmicutes | Clostridia | Clostridiales | Lachnospiraceae | Draft |
| Corynebacterium ammoniagenes DSM 20306 | 649754.3 | 2623620528 | 649754 | Bacteria | Actinobacteria | Actinobacteria | Actinomycetales | Corynebacteriaceae | Draft |
| Corynebacterium sp. HFH0082 | 1078764.3 | 2541046993 | 1078764 | Bacteria | Actinobacteria | Actinobacteria | Actinomycetales | Corynebacteriaceae | Draft |
| Dermabacter sp. HFH0086 | 1203568.3 | 2541047015 | 1203568 | Bacteria | Actinobacteria | Actinobacteria | Actinomycetales | Dermabacteraceae | Draft |
| Desulfitobacterium hafniense DP7 | 537010.4 | 2522572064 | 537010 | Bacteria | Firmicutes | Clostridia | Clostridiales | Peptococcaceae | Draft |
| Desulfovibrio piger ATCC 29098 | 411464.8 | 642979316 | 411464 | Bacteria | Proteobacteria | Deltaproteobacteria | Desulfovibrionales | Desulfovibrionaceae | Draft |
| Desulfovibrio sp. 3_1_syn3 | 457398.5 | 2562617147 | 457398 | Bacteria | Proteobacteria | Deltaproteobacteria | Desulfovibrionales | Desulfovibrionaceae | Draft |
| Dorea formicigenerans 4_6_53AFAA | 742765.5 | 2513237343 | 742765 | Bacteria | Firmicutes | Clostridia | Clostridiales | Lachnospiraceae | Draft |
| Dorea formicigenerans ATCC 27755 | 411461.4 | 641736133 | 411461 | Bacteria | Firmicutes | Clostridia | Clostridiales | Lachnospiraceae | Draft |
| Dorea longicatena DSM 13814 | 411462.6 | 2623620323 | 411462 | Bacteria | Firmicutes | Clostridia | Clostridiales | Lachnospiraceae | Draft |
| Dysgonomonas gadei ATCC BAA-286 | 742766.3 | 651324026 | 742766 | Bacteria | Bacteroidetes | Bacteroidia | Bacteroidales | Porphyromonadaceae | Draft |
| Dysgonomonas mossii DSM 22836 | 742767.3 | 651324027 | 742767 | Bacteria | Bacteroidetes | Bacteroidia | Bacteroidales | Porphyromonadaceae | Draft |

| Organism | ID1 | ID2 | ID3 | Domain | Phylum | Class | Order | Family | Status |
|---|---|---|---|---|---|---|---|---|---|
| Edwardsiella tarda ATCC 23685 | 500638.3 | 647000237 | 500638 | Bacteria | Proteobacteria | Gammaproteobacteria | Enterobacteriales | Enterobacteriaceae | Draft |
| Eggerthella sp. 1_3_56FAA | 665943.3 | 649989930 | 665943 | Bacteria | Actinobacteria | Actinobacteria | Coriobacteriales | Coriobacteriaceae | Draft |
| Eggerthella sp. HGA1 | 910311.3 | 651324028 | 910311 | Bacteria | Actinobacteria | Actinobacteria | Coriobacteriales | Coriobacteriaceae | Draft |
| Enterobacter cancerogenus ATCC 35316 | 500639.8 | 2562617148 | 500639 | Bacteria | Proteobacteria | Gammaproteobacteria | Enterobacteriales | Enterobacteriaceae | Draft |
| Enterobacter cloacae subsp. cloacae NCTC 9394 | 718254.4 | 650377928 | 718254 | Bacteria | Proteobacteria | Gammaproteobacteria | Enterobacteriales | Enterobacteriaceae | Draft |
| Enterobacteriaceae bacterium 9_2_54FAA | 469613.3 | 2562617162 | 469613 | Bacteria | Proteobacteria | Gammaproteobacteria | Enterobacteriales | Enterobacteriaceae | Draft |
| Enterococcus faecalis PC1.1 | 702448.3 | 2513237319 | 791166 | Bacteria | Firmicutes | Bacilli | Lactobacillales | Enterococcaceae | Draft |
| Enterococcus faecalis TX0104 | 491074.3 | 643886065 | 491074 | Bacteria | Firmicutes | Bacilli | Lactobacillales | Enterococcaceae | Draft |
| Enterococcus faecalis TX1302 | 749513.3 | 2531839256 | 749513 | Bacteria | Firmicutes | Bacilli | Lactobacillales | Enterococcaceae | Draft |
| Enterococcus faecalis TX1322 | 525278.3 | 643886064 | 525278 | Bacteria | Firmicutes | Bacilli | Lactobacillales | Enterococcaceae | Draft |
| Enterococcus faecalis TX1341 | 749514.3 | 2531839255 | 749514 | Bacteria | Firmicutes | Bacilli | Lactobacillales | Enterococcaceae | Draft |
| Enterococcus faecalis TX1342 | 749515.3 | 2534681822 | 749515 | Bacteria | Firmicutes | Bacilli | Lactobacillales | Enterococcaceae | Draft |
| Enterococcus faecalis TX1346 | 749516.3 | 2534681821 | 749516 | Bacteria | Firmicutes | Bacilli | Lactobacillales | Enterococcaceae | Draft |
| Enterococcus faecalis TX2134 | 749518.3 | 648276641 | 749518 | Bacteria | Firmicutes | Bacilli | Lactobacillales | Enterococcaceae | Draft |
| Enterococcus faecalis TX2137 | 749491.3 | 2529292855 | 749491 | Bacteria | Firmicutes | Bacilli | Lactobacillales | Enterococcaceae | Draft |
| Enterococcus faecalis TX4244 | 749494.3 | 2529292856 | 749494 | Bacteria | Firmicutes | Bacilli | Lactobacillales | Enterococcaceae | Draft |
| Enterococcus faecium PC4.1 | 791161.5 | 647000238 | 791161 | Bacteria | Firmicutes | Bacilli | Lactobacillales | Enterococcaceae | Draft |
| Enterococcus faecium TX1330 | 525279.3 | 643886120 | 525279 | Bacteria | Firmicutes | Bacilli | Lactobacillales | Enterococcaceae | Draft |
| Enterococcus saccharolyticus 30_1 | 742813.4 | 2513237314 | 742813 | Bacteria | Firmicutes | Bacilli | Lactobacillales | Enterococcaceae | Draft |
| Enterococcus sp. 7L76 | 657310.3 | 650377930 | 657310 | Bacteria | Firmicutes | Bacilli | Lactobacillales | Enterococcaceae | Finished |
| Erysipelotrichaceae bacterium 2_2_44A | 457422.3 | 2513237345 | 457422 | Bacteria | Firmicutes | Erysipelotrichi | Erysipelotrichales | Erysipelotrichaceae | Draft |
| Erysipelotrichaceae bacterium 21_3 | 658657.3 | 2513237332 | 658657 | Bacteria | Firmicutes | Erysipelotrichi | Erysipelotrichales | Erysipelotrichaceae | Draft |
| Erysipelotrichaceae bacterium 3_1_53 | 658659.3 | 649989945 | 658659 | Bacteria | Firmicutes | Erysipelotrichi | Erysipelotrichales | Erysipelotrichaceae | Draft |
| Erysipelotrichaceae bacterium 5_2_54FAA | 552396.3 | 2562617161 | 552396 | Bacteria | Firmicutes | Erysipelotrichi | Erysipelotrichales | Erysipelotrichaceae | Draft |
| Erysipelotrichaceae bacterium 6_1_45 | 469614.3 | 2531839268 | 469614 | Bacteria | Firmicutes | Erysipelotrichi | Erysipelotrichales | Erysipelotrichaceae | Draft |
| Escherichia coli 4_1_47FAA | 1127356.4 | 2531839265 | 562 | Bacteria | Proteobacteria | Gammaproteobacteria | Enterobacteriales | Enterobacteriaceae | Draft |
| Escherichia coli MS 107-1 | 679207.4 | 648276649 | 679207 | Bacteria | Proteobacteria | Gammaproteobacteria | Enterobacteriales | Enterobacteriaceae | Draft |
| Escherichia coli MS 115-1 | 749537.3 | 648276650 | 749537 | Bacteria | Proteobacteria | Gammaproteobacteria | Enterobacteriales | Enterobacteriaceae | Draft |
| Escherichia coli MS 116-1 | 749538.3 | 648276651 | 749538 | Bacteria | Proteobacteria | Gammaproteobacteria | Enterobacteriales | Enterobacteriaceae | Draft |
| Escherichia coli MS 119-7 | 679206.4 | 648276652 | 679206 | Bacteria | Proteobacteria | Gammaproteobacteria | Enterobacteriales | Enterobacteriaceae | Draft |
| Escherichia coli MS 124-1 | 679205.4 | 648276653 | 679205 | Bacteria | Proteobacteria | Gammaproteobacteria | Enterobacteriales | Enterobacteriaceae | Draft |
| Escherichia coli MS 145-7 | 679204.3 | 649989948 | 679204 | Bacteria | Proteobacteria | Gammaproteobacteria | Enterobacteriales | Enterobacteriaceae | Draft |
| Escherichia coli MS 146-1 | 749540.3 | 648276654 | 749540 | Bacteria | Proteobacteria | Gammaproteobacteria | Enterobacteriales | Enterobacteriaceae | Draft |
| Escherichia coli MS 175-1 | 749544.3 | 648276655 | 749544 | Bacteria | Proteobacteria | Gammaproteobacteria | Enterobacteriales | Enterobacteriaceae | Draft |
| Escherichia coli MS 182-1 | 749545.3 | 648276656 | 749545 | Bacteria | Proteobacteria | Gammaproteobacteria | Enterobacteriales | Enterobacteriaceae | Draft |
| Escherichia coli MS 185-1 | 749546.3 | 648276657 | 749546 | Bacteria | Proteobacteria | Gammaproteobacteria | Enterobacteriales | Enterobacteriaceae | Draft |
| Escherichia coli MS 187-1 | 749547.3 | 648276658 | 749547 | Bacteria | Proteobacteria | Gammaproteobacteria | Enterobacteriales | Enterobacteriaceae | Draft |
| Escherichia coli MS 196-1 | 749548.3 | 648276659 | 749548 | Bacteria | Proteobacteria | Gammaproteobacteria | Enterobacteriales | Enterobacteriaceae | Draft |
| Escherichia coli MS 198-1 | 749549.3 | 648276660 | 749549 | Bacteria | Proteobacteria | Gammaproteobacteria | Enterobacteriales | Enterobacteriaceae | Draft |
| Escherichia coli MS 200-1 | 749550.3 | 648276661 | 749550 | Bacteria | Proteobacteria | Gammaproteobacteria | Enterobacteriales | Enterobacteriaceae | Draft |
| Escherichia coli MS 21-1 | 749527.3 | 648276662 | 749527 | Bacteria | Proteobacteria | Gammaproteobacteria | Enterobacteriales | Enterobacteriaceae | Draft |
| Escherichia coli MS 45-1 | 749528.3 | 648276663 | 749528 | Bacteria | Proteobacteria | Gammaproteobacteria | Enterobacteriales | Enterobacteriaceae | Draft |
| Escherichia coli MS 69-1 | 749531.3 | 648276664 | 749531 | Bacteria | Proteobacteria | Gammaproteobacteria | Enterobacteriales | Enterobacteriaceae | Draft |
| Escherichia coli MS 78-1 | 749532.3 | 648276665 | 749532 | Bacteria | Proteobacteria | Gammaproteobacteria | Enterobacteriales | Enterobacteriaceae | Draft |
| Escherichia coli MS 79-10 | 796392.4 | 2531839147 | 796392 | Bacteria | Proteobacteria | Gammaproteobacteria | Enterobacteriales | Enterobacteriaceae | Draft |
| Escherichia coli MS 84-1 | 749533.3 | 648276666 | 749533 | Bacteria | Proteobacteria | Gammaproteobacteria | Enterobacteriales | Enterobacteriaceae | Draft |
| Escherichia coli MS 85-1 | 679202.3 | 2531839146 | 679202 | Bacteria | Proteobacteria | Gammaproteobacteria | Enterobacteriales | Enterobacteriaceae | Draft |
| Escherichia coli O157:H7 str. Sakai | 386585.9 | 2623620697 | 386585 | Bacteria | Proteobacteria | Gammaproteobacteria | Enterobacteriales | Enterobacteriaceae | Finished |
| Escherichia coli SE11 | 409438.11 | 643348551 | 409438 | Bacteria | Proteobacteria | Gammaproteobacteria | Enterobacteriales | Enterobacteriaceae | Finished |
| Escherichia coli SE15 | 431946.3 | 646862327 | 562 | Bacteria | Proteobacteria | Gammaproteobacteria | Enterobacteriales | Enterobacteriaceae | Finished |
| Escherichia coli str. K-12 substr. MG1655 | 511145.12 | 646311926 | 511145 | Bacteria | Proteobacteria | Gammaproteobacteria | Enterobacteriales | Enterobacteriaceae | Finished |
| Escherichia coli UTI89 | 364106.8 | 2623620781 | 364106 | Bacteria | Proteobacteria | Gammaproteobacteria | Enterobacteriales | Enterobacteriaceae | Finished |
| Escherichia sp. 1_1_43 | 457400.3 | 646206261 | 457400 | Bacteria | Proteobacteria | Gammaproteobacteria | Enterobacteriales | Enterobacteriaceae | Draft |
| Escherichia sp. 3_2_53FAA | 469598.5 | 646206259 | 469598 | Bacteria | Proteobacteria | Gammaproteobacteria | Enterobacteriales | Enterobacteriaceae | Draft |
| Escherichia sp. 4_1_40B | 457401.3 | 2562617151 | 457401 | Bacteria | Proteobacteria | Gammaproteobacteria | Enterobacteriales | Enterobacteriaceae | Draft |
| Eubacterium biforme DSM 3989 | 518637.5 | 642979360 | 518637 | Bacteria | Firmicutes | Erysipelotrichi | Erysipelotrichales | Erysipelotrichaceae | Draft |
| Eubacterium cylindroides T2-87 | 717960.3 | 650377935 | 717960 | Bacteria | Firmicutes | Erysipelotrichi | Erysipelotrichales | Erysipelotrichaceae | Finished |
| Eubacterium dolichum DSM 3991 | 428127.7 | 641380446 | 428127 | Bacteria | Firmicutes | Erysipelotrichi | Erysipelotrichales | Erysipelotrichaceae | Draft |
| Eubacterium hallii DSM 3353 | 411469.3 | 643886203 | 411469 | Bacteria | Firmicutes | Clostridia | Clostridiales | Eubacteriaceae | Draft |
| Eubacterium rectale DSM 17629 | 657318.4 | 650377936 | 657318 | Bacteria | Firmicutes | Clostridia | Clostridiales | Eubacteriaceae | Finished |
| Eubacterium rectale M104/1 | 657317.3 | 650377937 | 657317 | Bacteria | Firmicutes | Clostridia | Clostridiales | Eubacteriaceae | Finished |
| Eubacterium siraeum 70/3 | 657319.3 | 650377938 | 657319 | Bacteria | Firmicutes | Clostridia | Clostridiales | Eubacteriaceae | Finished |
| Eubacterium siraeum DSM 15702 | 428128.7 | 2519899684 | 428128 | Bacteria | Firmicutes | Clostridia | Clostridiales | Eubacteriaceae | Draft |
| Eubacterium siraeum V10Sc8a | 717961.3 | 2565956546 | 717961 | Bacteria | Firmicutes | Clostridia | Clostridiales | Eubacteriaceae | Finished |
| Eubacterium sp. 3_1_31 | 457402.3 | 2513237348 | 457402 | Bacteria | Firmicutes | Clostridia | Clostridiales | Eubacteriaceae | Draft |
| Eubacterium ventriosum ATCC 27560 | 411463.4 | 2623620332 | 411463 | Bacteria | Firmicutes | Clostridia | Clostridiales | Eubacteriaceae | Draft |
| Faecalibacterium cf. prausnitzii KLE1255 | 748224.3 | 649989951 | 748224 | Bacteria | Firmicutes | Clostridia | Clostridiales | Ruminococcaceae | Draft |
| Faecalibacterium prausnitzii A2-165 | 411483.3 | 645951831 | 411483 | Bacteria | Firmicutes | Clostridia | Clostridiales | Ruminococcaceae | Draft |
| Faecalibacterium prausnitzii L2-6 | 718252.3 | 650377941 | 718252 | Bacteria | Firmicutes | Clostridia | Clostridiales | Ruminococcaceae | Finished |
| Faecalibacterium prausnitzii M21/2 | 411485.1 | 641380420 | 411485 | Bacteria | Firmicutes | Clostridia | Clostridiales | Ruminococcaceae | Draft |
| Faecalibacterium prausnitzii SL3/3 | 657322.3 | 650377940 | 657322 | Bacteria | Firmicutes | Clostridia | Clostridiales | Ruminococcaceae | Finished |
| Fusobacterium gonidiaformans ATCC 25563 | 469615.3 | 645951804 | 469615 | Bacteria | Fusobacteria | Fusobacteriia | Fusobacteriales | Fusobacteriaceae | Draft |
| Fusobacterium mortiferum ATCC 9817 | 469616.3 | 646206254 | 469616 | Bacteria | Fusobacteria | Fusobacteriia | Fusobacteriales | Fusobacteriaceae | Draft |
| Fusobacterium necrophorum subsp. funduliforme 1_1_36S | 742814.3 | 2513237330 | 742814 | Bacteria | Fusobacteria | Fusobacteriia | Fusobacteriales | Fusobacteriaceae | Draft |
| Fusobacterium sp. 1_1_41FAA | 469621.3 | 647533168 | 469621 | Bacteria | Fusobacteria | Fusobacteriia | Fusobacteriales | Fusobacteriaceae | Draft |
| Fusobacterium sp. 11_3_2 | 457403.3 | 651324032 | 457403 | Bacteria | Fusobacteria | Fusobacteriia | Fusobacteriales | Fusobacteriaceae | Draft |
| Fusobacterium sp. 12_1B | 457404.3 | 2562617095 | 457404 | Bacteria | Fusobacteria | Fusobacteriia | Fusobacteriales | Fusobacteriaceae | Draft |
| Fusobacterium sp. 2_1_31 | 469599.3 | 646206262 | 469599 | Bacteria | Fusobacteria | Fusobacteriia | Fusobacteriales | Fusobacteriaceae | Draft |
| Fusobacterium sp. 21_1A | 469601.3 | 2562617152 | 469601 | Bacteria | Fusobacteria | Fusobacteriia | Fusobacteriales | Fusobacteriaceae | Draft |
| Fusobacterium sp. 3_1_27 | 469602.3 | 2563366552 | 469602 | Bacteria | Fusobacteria | Fusobacteriia | Fusobacteriales | Fusobacteriaceae | Finished |
| Fusobacterium sp. 3_1_33 | 469603.3 | 2562617153 | 469603 | Bacteria | Fusobacteria | Fusobacteriia | Fusobacteriales | Fusobacteriaceae | Draft |
| Fusobacterium sp. 3_1_36A2 | 469604.3 | 2554235746 | 469604 | Bacteria | Fusobacteria | Fusobacteriia | Fusobacteriales | Fusobacteriaceae | Finished |
| Fusobacterium sp. 3_1_5R | 469605.3 | 645951866 | 469605 | Bacteria | Fusobacteria | Fusobacteriia | Fusobacteriales | Fusobacteriaceae | Draft |
| Fusobacterium sp. 4_1_13 | 469606.3 | 646206253 | 469606 | Bacteria | Fusobacteria | Fusobacteriia | Fusobacteriales | Fusobacteriaceae | Draft |
| Fusobacterium sp. 7_1 | 457405.3 | 646206252 | 457405 | Bacteria | Fusobacteria | Fusobacteriia | Fusobacteriales | Fusobacteriaceae | Finished |
| Fusobacterium sp. D11 | 556264.3 | 645058833 | 556264 | Bacteria | Fusobacteria | Fusobacteriia | Fusobacteriales | Fusobacteriaceae | Draft |
| Fusobacterium sp. D12 | 556263.3 | 2562617180 | 556263 | Bacteria | Fusobacteria | Fusobacteriia | Fusobacteriales | Fusobacteriaceae | Draft |
| Fusobacterium ulcerans ATCC 49185 | 469617.3 | 2562617154 | 469617 | Bacteria | Fusobacteria | Fusobacteriia | Fusobacteriales | Fusobacteriaceae | Draft |
| Fusobacterium varium ATCC 27725 | 469618.3 | 646206275 | 469618 | Bacteria | Fusobacteria | Fusobacteriia | Fusobacteriales | Fusobacteriaceae | Draft |
| Gordonibacter pamelaeae 7-10-1-b | 657308.3 | 650377943 | 657308 | Bacteria | Actinobacteria | Actinobacteria | Coriobacteriales | Coriobacteriaceae | Finished |
| Hafnia alvei ATCC 51873 | 1002364.3 | 2513237377 | 1002364 | Bacteria | Proteobacteria | Gammaproteobacteria | Enterobacteriales | Enterobacteriaceae | Draft |
| Helicobacter bilis ATCC 43879 | 613026.4 | 2562617155 | 613026 | Bacteria | Proteobacteria | Epsilonproteobacteria | Campylobacterales | Helicobacteraceae | Draft |
| Helicobacter canadensis MIT 98-5491 (Prj:30719) | 537970.9 | 647533173 | 537970 | Bacteria | Proteobacteria | Epsilonproteobacteria | Campylobacterales | Helicobacteraceae | Draft |
| Helicobacter cinaedi ATCC BAA-847 | 1206745.3 | 2554235363 | 1206745 | Bacteria | Proteobacteria | Epsilonproteobacteria | Campylobacterales | Helicobacteraceae | Finished |
| Helicobacter cinaedi CCUG 18818 | 537971.5 | 643886141 | 537971 | Bacteria | Proteobacteria | Epsilonproteobacteria | Campylobacterales | Helicobacteraceae | Draft |
| Helicobacter pullorum MIT 98-5489 | 537972.5 | 643886218 | 537972 | Bacteria | Proteobacteria | Epsilonproteobacteria | Campylobacterales | Helicobacteraceae | Draft |
| Helicobacter winghamensis ATCC BAA-430 | 556267.4 | 646206255 | 556267 | Bacteria | Proteobacteria | Epsilonproteobacteria | Campylobacterales | Helicobacteraceae | Draft |
| Holdemania filiformis DSM 12042 | 545696.5 | 643886103 | 545696 | Bacteria | Firmicutes | Erysipelotrichi | Erysipelotrichales | Erysipelotrichaceae | Draft |
| Klebsiella pneumoniae 1162281 | 1037908.3 | 2563366764 | 1037908 | Bacteria | Proteobacteria | Gammaproteobacteria | Enterobacteriales | Enterobacteriaceae | Draft |
| Klebsiella pneumoniae subsp. pneumoniae WGLW5 | 1203547.3 | 2537561874 | 1203547 | Bacteria | Proteobacteria | Gammaproteobacteria | Enterobacteriales | Enterobacteriaceae | Draft |
| Klebsiella sp. 1_1_55 | 469608.3 | 647533174 | 469608 | Bacteria | Proteobacteria | Gammaproteobacteria | Enterobacteriales | Enterobacteriaceae | Draft |
| Klebsiella sp. 4_1_44FAA | 665944.3 | 2513237265 | 665944 | Bacteria | Proteobacteria | Gammaproteobacteria | Enterobacteriales | Enterobacteriaceae | Draft |
| Lachnospiraceae bacterium 1_1_57FAA | 658081.3 | 651324046 | 658081 | Bacteria | Firmicutes | Clostridia | Clostridiales | Lachnospiraceae | Draft |
| Lachnospiraceae bacterium 1_4_56FAA | 658655.3 | 651324047 | 658655 | Bacteria | Firmicutes | Clostridia | Clostridiales | Lachnospiraceae | Draft |
| Lachnospiraceae bacterium 2_1_46FAA | 742723.3 | 2562617190 | 742723 | Bacteria | Firmicutes | Clostridia | Clostridiales | Lachnospiraceae | Draft |
| Lachnospiraceae bacterium 2_1_58FAA | 658082.3 | 651324049 | 658082 | Bacteria | Firmicutes | Clostridia | Clostridiales | Lachnospiraceae | Draft |
| Lachnospiraceae bacterium 3_1_46FAA | 665950.3 | 651324050 | 665950 | Bacteria | Firmicutes | Clostridia | Clostridiales | Lachnospiraceae | Draft |
| Lachnospiraceae bacterium 3_1_57FAA_CT1 | 658086.3 | 2562617188 | 658086 | Bacteria | Firmicutes | Clostridia | Clostridiales | Lachnospiraceae | Draft |
| Lachnospiraceae bacterium 4_1_37FAA | 552395.3 | 651324052 | 552395 | Bacteria | Firmicutes | Clostridia | Clostridiales | Lachnospiraceae | Draft |
| Lachnospiraceae bacterium 5_1_57FAA | 658085.3 | 651324053 | 658085 | Bacteria | Firmicutes | Clostridia | Clostridiales | Lachnospiraceae | Draft |
| Lachnospiraceae bacterium 5_1_63FAA | 658089.3 | 649989957 | 658089 | Bacteria | Firmicutes | Clostridia | Clostridiales | Lachnospiraceae | Draft |
| Lachnospiraceae bacterium 6_1_63FAA | 658083.3 | 651324054 | 658083 | Bacteria | Firmicutes | Clostridia | Clostridiales | Lachnospiraceae | Draft |
| Lachnospiraceae bacterium 7_1_58FAA | 658087.3 | 2513237329 | 658087 | Bacteria | Firmicutes | Clostridia | Clostridiales | Lachnospiraceae | Draft |
| Lachnospiraceae bacterium 8_1_57FAA | 665951.3 | 649989958 | 665951 | Bacteria | Firmicutes | Clostridia | Clostridiales | Lachnospiraceae | Draft |

| Organism | ID1 | ID2 | ID3 | Domain | Phylum | Class | Order | Family | Status |
|---|---|---|---|---|---|---|---|---|---|
| Lachnospiraceae bacterium 9_1_43BFAA | 658088.3 | 651324055 | 658088 | Bacteria | Firmicutes | Clostridia | Clostridiales | Lachnospiraceae | Draft |
| Lactobacillus acidophilus ATCC 4796 | 525306.3 | 643886014 | 525306 | Bacteria | Firmicutes | Bacilli | Lactobacillales | Lactobacillaceae | Draft |
| Lactobacillus acidophilus NCFM | 272621.13 | 637000138 | 272621 | Bacteria | Firmicutes | Bacilli | Lactobacillales | Lactobacillaceae | Draft |
| Lactobacillus amylolyticus DSM 11664 | 585524.3 | 647000263 | 585524 | Bacteria | Firmicutes | Bacilli | Lactobacillales | Lactobacillaceae | Draft |
| Lactobacillus antri DSM 16041 | 525309.3 | 645951838 | 525309 | Bacteria | Firmicutes | Bacilli | Lactobacillales | Lactobacillaceae | Draft |
| Lactobacillus brevis ATCC 367 | 387344.15 | 639633027 | 387344 | Bacteria | Firmicutes | Bacilli | Lactobacillales | Lactobacillaceae | Finished |
| Lactobacillus brevis subsp. gravesensis ATCC 27305 | 525310.3 | 643886059 | 525310 | Bacteria | Firmicutes | Bacilli | Lactobacillales | Lactobacillaceae | Draft |
| Lactobacillus buchneri ATCC 11577 | 525318.3 | 643886062 | 525318 | Bacteria | Firmicutes | Bacilli | Lactobacillales | Lactobacillaceae | Draft |
| Lactobacillus casei ATCC 334 | 321967.11 | 639633028 | 321967 | Bacteria | Firmicutes | Bacilli | Lactobacillales | Lactobacillaceae | Finished |
| Lactobacillus casei BL23 | 543734.4 | 642555134 | 543734 | Bacteria | Firmicutes | Bacilli | Lactobacillales | Lactobacillaceae | Finished |
| Lactobacillus crispatus 125-2-CHN | 575595.3 | 647533176 | 575595 | Bacteria | Firmicutes | Bacilli | Lactobacillales | Lactobacillaceae | Draft |
| Lactobacillus delbrueckii subsp. bulgaricus ATCC 11842 | 390333.7 | 2623620931 | 390333 | Bacteria | Firmicutes | Bacilli | Lactobacillales | Lactobacillaceae | Finished |
| Lactobacillus delbrueckii subsp. bulgaricus ATCC BAA-365 | 321956.7 | 639633029 | 321956 | Bacteria | Firmicutes | Bacilli | Lactobacillales | Lactobacillaceae | Finished |
| Lactobacillus fermentum ATCC 14931 | 525325.3 | 643886061 | 525325 | Bacteria | Firmicutes | Bacilli | Lactobacillales | Lactobacillaceae | Draft |
| Lactobacillus fermentum IFO 3956 | 334390.5 | 641522633 | 334390 | Bacteria | Firmicutes | Bacilli | Lactobacillales | Lactobacillaceae | Finished |
| Lactobacillus gasseri ATCC 33323 | 324831.13 | 639633030 | 324831 | Bacteria | Firmicutes | Bacilli | Lactobacillales | Lactobacillaceae | Finished |
| Lactobacillus helveticus DPC 4571 | 405566.6 | 641228495 | 405566 | Bacteria | Firmicutes | Bacilli | Lactobacillales | Lactobacillaceae | Finished |
| Lactobacillus helveticus DSM 20075 | 585520.4 | 645951865 | 585520 | Bacteria | Firmicutes | Bacilli | Lactobacillales | Lactobacillaceae | Draft |
| Lactobacillus hilgardii ATCC 8290 | 525327.3 | 643886044 | 525327 | Bacteria | Firmicutes | Bacilli | Lactobacillales | Lactobacillaceae | Draft |
| Lactobacillus johnsonii NCC 533 | 257314.6 | 2623620932 | 257314 | Bacteria | Firmicutes | Bacilli | Lactobacillales | Lactobacillaceae | Finished |
| Lactobacillus paracasei subsp. paracasei 8700:2 | 537973.8 | 643886139 | 537973 | Bacteria | Firmicutes | Bacilli | Lactobacillales | Lactobacillaceae | Finished |
| Lactobacillus paracasei subsp. paracasei ATCC 25302 | 525337.3 | 643886051 | 525337 | Bacteria | Firmicutes | Bacilli | Lactobacillales | Lactobacillaceae | Draft |
| Lactobacillus plantarum 16 | 1327988.3 | 2563366583 | 1327988 | Bacteria | Firmicutes | Bacilli | Lactobacillales | Lactobacillaceae | Finished |
| Lactobacillus plantarum subsp. plantarum ATCC 14917 | 525338.3 | 2562617181 | 525338 | Bacteria | Firmicutes | Bacilli | Lactobacillales | Lactobacillaceae | Draft |
| Lactobacillus plantarum WCFS1 | 220668.9 | 2623620707 | 220668 | Bacteria | Firmicutes | Bacilli | Lactobacillales | Lactobacillaceae | Finished |
| Lactobacillus reuteri CF48-3A | 525341.3 | 2502171173 | 525341 | Bacteria | Firmicutes | Bacilli | Lactobacillales | Lactobacillaceae | Draft |
| Lactobacillus reuteri DSM 20016 | 557436.4 | not in IMG | 557436 | Bacteria | Firmicutes | Bacilli | Lactobacillales | Lactobacillaceae | Finished |
| Lactobacillus reuteri F275 | 299033.6 | 640427118 | 299033 | Bacteria | Firmicutes | Bacilli | Lactobacillales | Lactobacillaceae | Finished |
| Lactobacillus reuteri MM2-3 | 585517.3 | 2502171171 | 585517 | Bacteria | Firmicutes | Bacilli | Lactobacillales | Lactobacillaceae | Draft |
| Lactobacillus reuteri MM4-1A | 548485.3 | 2502171170 | 548485 | Bacteria | Firmicutes | Bacilli | Lactobacillales | Lactobacillaceae | Draft |
| Lactobacillus reuteri SD2112 | 491077.3 | 650716048 | 491077 | Bacteria | Firmicutes | Bacilli | Lactobacillales | Lactobacillaceae | Finished |
| Lactobacillus rhamnosus ATCC 21052 | 1002365.5 | 2534681855 | 1002365 | Bacteria | Firmicutes | Bacilli | Lactobacillales | Lactobacillaceae | Draft |
| Lactobacillus rhamnosus GG (Prj:40637) | 568703.3 | 644736382 | 568703 | Bacteria | Firmicutes | Bacilli | Lactobacillales | Lactobacillaceae | Finished |
| Lactobacillus rhamnosus LMS2-1 | 525361.3 | 643886127 | 525361 | Bacteria | Firmicutes | Bacilli | Lactobacillales | Lactobacillaceae | Draft |
| Lactobacillus ruminis ATCC 25644 | 525362.3 | 2562617182 | 525362 | Bacteria | Firmicutes | Bacilli | Lactobacillales | Lactobacillaceae | Draft |
| Lactobacillus sakei subsp. sakei 23K | 314315.12 | 637000142 | 314315 | Bacteria | Firmicutes | Bacilli | Lactobacillales | Lactobacillaceae | Finished |
| Lactobacillus salivarius ATCC 11741 | 525364.3 | 643886048 | 525364 | Bacteria | Firmicutes | Bacilli | Lactobacillales | Lactobacillaceae | Draft |
| Lactobacillus salivarius UCC118 | 362948.14 | 2623620709 | 362948 | Bacteria | Firmicutes | Bacilli | Lactobacillales | Lactobacillaceae | Finished |
| Lactobacillus sp. 7_1_47FAA | 665945.3 | 2513237395 | 665945 | Bacteria | Firmicutes | Bacilli | Lactobacillales | Lactobacillaceae | Draft |
| Lactobacillus ultunensis DSM 16047 | 525365.3 | 643886047 | 525365 | Bacteria | Firmicutes | Bacilli | Lactobacillales | Lactobacillaceae | Draft |
| Leuconostoc mesenteroides subsp. cremoris ATCC 19254 | 586220.3 | 643886109 | 586220 | Bacteria | Firmicutes | Bacilli | Lactobacillales | Leuconostocaceae | Draft |
| Listeria grayi DSM 20601 | 525367.9 | 2562617157 | 525367 | Bacteria | Firmicutes | Bacilli | Bacillales | Listeriaceae | Draft |
| Listeria innocua ATCC 33091 | 1002366.3 | 2537561540 | 1002366 | Bacteria | Firmicutes | Bacilli | Bacillales | Listeriaceae | Draft |
| Megamonas funiformis YIT 11815 | 742816.3 | 2513237399 | 742816 | Bacteria | Firmicutes | Negativicutes | Selenomonadales | Veillonellaceae | Draft |
| Megamonas hypermegale ART12/1 | 657316.3 | 650377958 | 158847 | Bacteria | Firmicutes | Negativicutes | Selenomonadales | Veillonellaceae | Finished |
| Methanobrevibacter smithii ATCC 35061 | 420247.6 | 640427121 | 420247 | Archaea | Euryarchaeota | Methanobacteria | Methanobacteriales | Methanobacteriaceae | Finished |
| Methanosphaera stadtmanae DSM 3091 | 339860.6 | 2623620943 | 339860 | Archaea | Euryarchaeota | Methanobacteria | Methanobacteriales | Methanobacteriaceae | Finished |
| Mitsuokella multacida DSM 20544 | 500635.8 | 2562617158 | 500635 | Bacteria | Firmicutes | Negativicutes | Selenomonadales | Veillonellaceae | Draft |
| Mollicutes bacterium D7 | 556270.3 | 646206251 | 556270 | Bacteria | Tenericutes | Mollicutes | unclassified Mollicutes | unclassified Mollicutes | Draft |
| Odoribacter laneus YIT 12061 | 742817.3 | 2513237280 | 742817 | Bacteria | Bacteroidetes | Bacteroidia | Bacteroidales | Porphyromonadaceae | Draft |
| Oxalobacter formigenes HOxBLS | 556268.6 | 646206250 | 556268 | Bacteria | Proteobacteria | Betaproteobacteria | Burkholderiales | Oxalobacteraceae | Draft |
| Oxalobacter formigenes OXCC13 | 556269.4 | 646206265 | 556269 | Bacteria | Proteobacteria | Betaproteobacteria | Burkholderiales | Oxalobacteraceae | Draft |
| Paenibacillus sp. HGF5 | 908341.3 | 651324080 | 908341 | Bacteria | Firmicutes | Bacilli | Bacillales | Paenibacillaceae | Draft |
| Paenibacillus sp. HGF7 | 944559.3 | 651324081 | 944559 | Bacteria | Firmicutes | Bacilli | Bacillales | Paenibacillaceae | Draft |
| Paenibacillus sp. HGH0039 | 1078505.3 | 2541046994 | 1078505 | Bacteria | Firmicutes | Bacilli | Bacillales | Paenibacillaceae | Draft |
| Parabacteroides distasonis ATCC 8503 | 435591.13 | 2623620331 | 435591 | Bacteria | Bacteroidetes | Bacteroidia | Bacteroidales | Porphyromonadaceae | Finished |
| Parabacteroides johnsonii DSM 18315 | 537006.5 | 642979358 | 537006 | Bacteria | Bacteroidetes | Bacteroidia | Bacteroidales | Porphyromonadaceae | Draft |

| Organism | Col2 | Col3 | Col4 | Domain | Phylum | Class | Order | Family | Status |
|---|---|---|---|---|---|---|---|---|---|
| Parabacteroides merdae ATCC 43184 | 411477.4 | 640963016 | 411477 | Bacteria | Bacteroidetes | Bacteroidia | Bacteroidales | Porphyromonadaceae | Draft |
| Parabacteroides sp. D13 | 563193.3 | 646206279 | 563193 | Bacteria | Bacteroidetes | Bacteroidia | Bacteroidales | Porphyromonadaceae | Draft |
| Parabacteroides sp. D25 | 658661.3 | not in IMG | 658661 | Bacteria | Bacteroidetes | Bacteroidia | Bacteroidales | Porphyromonadaceae | Draft |
| Paraprevotella clara YIT 11840 | 762968.3 | 2513237390 | 762968 | Bacteria | Bacteroidetes | Bacteroidia | Bacteroidales | Prevotellaceae | Draft |
| Paraprevotella xylaniphila YIT 11841 | 762982.3 | 651324083 | 762982 | Bacteria | Bacteroidetes | Bacteroidia | Bacteroidales | Prevotellaceae | Draft |
| Parvimonas micra ATCC 33270 | 411465.1 | 641380421 | 411465 | Bacteria | Firmicutes | Clostridia | Clostridiales | Clostridiales Family XI. Incertae Sedis | Draft |
| Pediococcus acidilactici 7_4 | 563194.3 | 647533194 | 563194 | Bacteria | Firmicutes | Bacilli | Lactobacillales | Lactobacillaceae | Draft |
| Pediococcus acidilactici DSM 20284 | 862514.3 | 649989981 | 862514 | Bacteria | Firmicutes | Bacilli | Lactobacillales | Lactobacillaceae | Draft |
| Phascolarctobacterium sp. YIT 12067 | 626939.3 | 649989983 | 626939 | Bacteria | Firmicutes | Negativicutes | Selenomonadales | Acidaminococcaceae | Draft |
| Prevotella copri DSM 18205 | 537011.5 | 2562617166 | 537011 | Bacteria | Bacteroidetes | Bacteroidia | Bacteroidales | Prevotellaceae | Draft |
| Prevotella oralis HGA0225 | 1203550.3 | 2541047008 | 1203550 | Bacteria | Bacteroidetes | Bacteroidia | Bacteroidales | Prevotellaceae | Draft |
| Prevotella salivae DSM 15606 | 888832.3 | 649989989 | 888832 | Bacteria | Bacteroidetes | Bacteroidia | Bacteroidales | Prevotellaceae | Draft |
| Propionibacterium sp. 5_U_42AFAA | 450748.3 | 2513237388 | 450748 | Bacteria | Actinobacteria | Actinobacteria | Actinomycetales | Propionibacteriaceae | Draft |
| Propionibacterium sp. HGH0353 | 1203571.3 | 2541047016 | 1203571 | Bacteria | Actinobacteria | Actinobacteria | Actinomycetales | Propionibacteriaceae | Draft |
| Proteus mirabilis WGLW6 | 1125694.3 | 2537561878 | 1125694 | Bacteria | Proteobacteria | Gammaproteobacteria | Enterobacteriales | Enterobacteriaceae | Draft |
| Proteus penneri ATCC 35198 | 471881.3 | 643886117 | 471881 | Bacteria | Proteobacteria | Gammaproteobacteria | Enterobacteriales | Enterobacteriaceae | Draft |
| Providencia alcalifaciens DSM 30120 | 520999.6 | 642979317 | 520999 | Bacteria | Proteobacteria | Gammaproteobacteria | Enterobacteriales | Enterobacteriaceae | Draft |
| Providencia rettgeri DSM 1131 | 521000.6 | 2562617097 | 521000 | Bacteria | Proteobacteria | Gammaproteobacteria | Enterobacteriales | Enterobacteriaceae | Draft |
| Providencia rustigianii DSM 4541 | 500637.6 | 642979318 | 500637 | Bacteria | Proteobacteria | Gammaproteobacteria | Enterobacteriales | Enterobacteriaceae | Draft |
| Providencia stuartii ATCC 25827 | 471874.6 | 641736172 | 471874 | Bacteria | Proteobacteria | Gammaproteobacteria | Enterobacteriales | Enterobacteriaceae | Draft |
| Pseudomonas sp. 2_1_26 | 665948.3 | 2513237259 | 665948 | Bacteria | Proteobacteria | Gammaproteobacteria | Pseudomonadales | Pseudomonadaceae | Draft |
| Ralstonia sp. 5_7_47FAA | 658664.3 | 649989994 | 658664 | Bacteria | Proteobacteria | Betaproteobacteria | Burkholderiales | Burkholderiaceae | Draft |
| Roseburia intestinalis L1-82 | 536231.5 | 2562617159 | 536231 | Bacteria | Firmicutes | Clostridia | Clostridiales | Lachnospiraceae | Draft |
| Roseburia intestinalis M50/1 | 657315.3 | 650377964 | 166486 | Bacteria | Firmicutes | Clostridia | Clostridiales | Lachnospiraceae | Finished |
| Roseburia intestinalis XB6B4 | 718255.3 | 650377965 | 718255 | Bacteria | Firmicutes | Clostridia | Clostridiales | Lachnospiraceae | Finished |
| Roseburia inulinivorans DSM 16841 | 622312.4 | 643886006 | 360807 | Bacteria | Firmicutes | Clostridia | Clostridiales | Lachnospiraceae | Draft |
| Ruminococcaceae bacterium D16 | 552398.3 | 2562617183 | 552398 | Bacteria | Firmicutes | Clostridia | Clostridiales | Ruminococcaceae | Draft |
| Ruminococcus bromii L2-63 | 657321.5 | 650377966 | 657321 | Bacteria | Firmicutes | Clostridia | Clostridiales | Ruminococcaceae | Finished |
| Ruminococcus gnavus ATCC 29149 | 411470.6 | 640963057 | 411470 | Bacteria | Firmicutes | Clostridia | Clostridiales | Ruminococcaceae | Draft |
| Ruminococcus lactaris ATCC 29176 | 471875.6 | 642791604 | 471875 | Bacteria | Firmicutes | Clostridia | Clostridiales | Ruminococcaceae | Draft |
| Ruminococcus obeum A2-162 | 657314.3 | 650377967 | 657314 | Bacteria | Firmicutes | Clostridia | Clostridiales | Ruminococcaceae | Finished |
| Ruminococcus obeum ATCC 29174 | 411459.7 | 640963024 | 411459 | Bacteria | Firmicutes | Clostridia | Clostridiales | Ruminococcaceae | Draft |
| Ruminococcus sp. 18P13 | 213810.4 | 650377969 | 213810 | Bacteria | Firmicutes | Clostridia | Clostridiales | Ruminococcaceae | Finished |
| Ruminococcus sp. 5_1_39BFAA | 457412.4 | 2562617160 | 457412 | Bacteria | Firmicutes | Clostridia | Clostridiales | Ruminococcaceae | Draft |
| Ruminococcus sp. SR1/5 | 657323.3 | 650377968 | 657323 | Bacteria | Firmicutes | Clostridia | Clostridiales | Ruminococcaceae | Finished |
| Ruminococcus torques ATCC 27756 | 411460.6 | 640963025 | 411460 | Bacteria | Firmicutes | Clostridia | Clostridiales | Ruminococcaceae | Draft |
| Ruminococcus torques L2-14 | 657313.3 | 650377970 | 657313 | Bacteria | Firmicutes | Clostridia | Clostridiales | Ruminococcaceae | Finished |
| Salmonella enterica subsp. enterica serovar Typhimurium st | 99287.12 | 2623620649 | 99287 | Bacteria | Proteobacteria | Gammaproteobacteria | Enterobacteriales | Enterobacteriaceae | Finished |
| Staphylococcus sp. HGB0015 | 1078083.3 | 2541046995 | 1078083 | Bacteria | Firmicutes | Bacilli | Bacillales | Staphylococcaceae | Draft |
| Streptococcus anginosus 1_2_62CV | 742820.3 | 649990001 | 742820 | Bacteria | Firmicutes | Bacilli | Lactobacillales | Streptococcaceae | Draft |
| Streptococcus equinus ATCC 9812 | 525379.3 | 649990006 | 525379 | Bacteria | Firmicutes | Bacilli | Lactobacillales | Streptococcaceae | Draft |
| Streptococcus infantarius subsp. infantarius ATCC BAA-102 | 471872.6 | 641736171 | 471872 | Bacteria | Firmicutes | Bacilli | Lactobacillales | Streptococcaceae | Draft |
| Streptococcus sp. 2_1_36FAA | 469609.3 | 647533231 | 469609 | Bacteria | Firmicutes | Bacilli | Lactobacillales | Streptococcaceae | Draft |
| Streptococcus sp. HPH0090 | 1203590.3 | 2541047013 | 1203590 | Bacteria | Firmicutes | Bacilli | Lactobacillales | Streptococcaceae | Draft |
| Streptomyces sp. HGB0020 | 1078086.3 | 2541046998 | 1078086 | Bacteria | Actinobacteria | Actinobacteria | Actinomycetales | Streptomycetaceae | Draft |
| Streptomyces sp. HPH0547 | 1203592.3 | 2541047017 | 1203592 | Bacteria | Actinobacteria | Actinobacteria | Actinomycetales | Streptomycetaceae | Draft |
| Subdoligranulum sp. 4_3_54A2FAA | 665956.3 | 2513237311 | 665956 | Bacteria | Firmicutes | Clostridia | Clostridiales | Ruminococcaceae | Draft |
| Subdoligranulum variabile DSM 15176 | 411471.5 | 2562617078 | 411471 | Bacteria | Firmicutes | Clostridia | Clostridiales | Ruminococcaceae | Draft |
| Succinatimonas hippei YIT 12066 | 762983.3 | 649990019 | 762983 | Bacteria | Proteobacteria | Gammaproteobacteria | Aeromonadales | Succinivibrionaceae | Draft |
| Sutterella wadsworthensis 3_1_45B | 742821.3 | 649990021 | 742821 | Bacteria | Proteobacteria | Betaproteobacteria | Burkholderiales | Sutterellaceae | Draft |
| Sutterella wadsworthensis HGA0223 | 1203554.3 | 2541047009 | 1203554 | Bacteria | Proteobacteria | Betaproteobacteria | Burkholderiales | Sutterellaceae | Draft |
| Tannerella sp. 6_1_58FAA_CT1 | 665949.3 | 2513237267 | 665949 | Bacteria | Bacteroidetes | Bacteroidia | Bacteroidales | Porphyromonadaceae | Draft |
| Turicibacter sp. HGF1 | 910310.3 | 651324107 | 910310 | Bacteria | Firmicutes | Erysipelotrichi | Erysipelotrichales | Erysipelotrichaceae | Draft |
| Turicibacter sp. PC909 | 702450.3 | 647000330 | 702450 | Bacteria | Firmicutes | Erysipelotrichi | Erysipelotrichales | Erysipelotrichaceae | Draft |
| Veillonella sp. 3_1_44 | 457416.3 | 647533241 | 457416 | Bacteria | Firmicutes | Negativicutes | Selenomonadales | Veillonellaceae | Draft |
| Veillonella sp. 6_1_27 | 450749.3 | 647533242 | 450749 | Bacteria | Firmicutes | Negativicutes | Selenomonadales | Veillonellaceae | Draft |
| Weissella paramesenteroides ATCC 33313 | 585506.3 | 645058793 | 585506 | Bacteria | Firmicutes | Bacilli | Lactobacillales | Leuconostocaceae | Draft |

| Yokenella regensburgei ATCC 43003 | 1002368.3 | 2534682241 | 1002368 | Bacteria | Proteobacteria | Gammaproteobacteria | Enterobacteriales | Enterobacteriaceae | Draft |

**Table S2.** Functions of all the genes for utilization of mucin glycans, analyzed in this work. Functions predicted in this work are marked by bold blue font. For the protein sequences see the file Sequences S1 in the Supplementary materials.

| Utilized monosaccharide (No of Supplementary table) | Abbreviation | Function |
|---|---|---|
| N-Acetylgalactosamine (S4) | AaeL | Exported apha-N-acetylgalactosaminidase (EC 3.2.1.49) |
| Galactose (S7) | Aga27 | Alpha-galactosidase (EC 3.2.1.22), GH27 family |
| Galactose (S7) | Aga36 | Alpha-galactosidase (EC 3.2.1.22), GH36 family |
| Galactose (S7) | Aga4 | Alpha-galactosidase (EC 3.2.1.22), GH4 family |
| Galactose (S7) | Aga43 | Alpha-galactosidase (EC 3.2.1.22), GH43 family |
| N-Acetylgalactosamine (S4) | AgaA | N-acetylgalactosamine-6-phosphate deacetylase (EC 3.5.1.80) |
| N-Acetylgalactosamine (S4) | AgaI | Galactosamine-6-phosphate isomerase (galactosamine-6-phosphate deaminase) (EC 5.3.1.-) |
| N-Acetylgalactosamine (S4) | AgaK | N-acetylgalactosamine kinase, ROK-type (EC 2.7.1.157) |
| N-Acetylgalactosamine (S4) | AgaPTS | PTS system, N-acetylgalactosamine-specific |
| N-Acetylgalactosamine (S4) | AgaS | Galactosamine-6-phosphate isomerase (EC 5.3.1.-) |
| N-Acetylgalactosamine (S4), Galactose (S7) | AgaY / LacD | Tagatose 1,6-bisphosphate aldolase (EC 4.1.2.40) |
| N-Acetylgalactosamine (S4), Galactose (S7) | AgaZ / LacC | Tagatose-6-phosphate kinase (EC 2.7.1.144) |
| Fucose (S3) | AlfA29 | Alpha-L-fucosidase (EC 3.2.1.51), GH29 family |
| Fucose (S3) | AlfA42 | Alpha-L-fucosidase (EC 3.2.1.51), GH42 family |
| Fucose (S3) | AlfA95 | Alpha-L-fucosidase (EC 3.2.1.51), GH95 family |
| N-Acetylglucosamine (S5) | Ang | Alpha-N-acetylglucosaminidase (EC 3.2.1.50) |
| N-Acetylgalactosamine (S4) | Apr101 | Endo-alpha-N-acetylgalactosaminidase (EC 3.2.1.97), GH101 family |
| N-Acetylgalactosamine (S4) | Apr129 | Endo-alpha-N-acetylgalactosaminidase (EC 3.2.1.97), GH129 family |
| Galactose (S7) | Bga2 | Beta-galactosidase (EC 3.2.1.23), GH2 family |
| Galactose (S7) | Bga35 | Beta-galactosidase (EC 3.2.1.23), GH35 family |
| Galactose (S7) | Bga42 | Beta-galactosidase (EC 3.2.1.23), GH42 family |
| N-Acetylgalactosamine (S4), N-Acetylglucosamine (S5) | Bha20 | Beta-hexosaminidase (EC 3.2.1.52), GH20 family |
| N-Acetylgalactosamine (S4), N-Acetylglucosamine (S5) | Bha3 | Beta-hexosaminidase (EC 3.2.1.52), GH3 family |
| Fucose (S3) | FclA1 | L-fuco-beta-pyranose dehydrogenase (EC 1.1.1.122) |
| **Fucose (S3)** | **FclA2** | **L-fuco-beta-pyranose dehydrogenase, type 2 (EC 1.1.1.122)** |
| Fucose (S3) | FclB | L-fuconolactone hydrolase |
| Fucose (S3) | FclC | L-fuconate dehydratase (EC 4.2.1.68) |
| Fucose (S3) | FclD1 | 2-keto-3-deoxy-L-fuconate dehydrogenase |
| **Fucose (S3)** | **FclD2** | **2-keto-3-deoxy-L-fuconate dehydrogenase, type 2, putative** |
| Fucose (S3) | FclE1 | 2,4-diketo-3-deoxy-L-fuconate hydrolase |
| **Fucose (S3)** | **FclE2** | **Predicted 2,4-diketo-3-deoxy-L-fuconate hydrolase, DHDPS family** |
| Fucose (S3) | FucA1 | L-fuculose phosphate aldolase (EC 4.1.2.17) |
| **Fucose (S3)** | **FucABC** | **Fucose ABC transport system** |
| Fucose (S3) | FucI | L-fucose isomerase (EC 5.3.1.25) |
| Fucose (S3) | FucK | L-fuculokinase (EC 2.7.1.51) |
| Fucose (S3) | FucP1 | Fucose permease |
| **Fucose (S3)** | **FucP2** | **Fucose transporter, Sugar_tr family, predicted** |
| Fucose (S3) | FucU1 | L-fucose mutarotase (EC 5.1.3.29) |
| Fucose (S3) | FucU2 | L-fucose mutarotase, type 2 (EC 5.1.3.n2) |
| Galactose (S7) | GalE | UDP-glucose 4-epimerase (EC 5.1.3.2) |
| Galactose (S7) | GalK | Galactokinase (EC 2.7.1.6) |
| Galactose (S7) | GalM | Aldose 1-epimerase (EC 5.1.3.3) |
| Galactose (S7) | GalP1 | Predicted galactose transporter, LysE superfamily |
| Galactose (S7) | GalP2 | Lactose and galactose permease, GPH translocator family |
| **Galactose (S7)** | **GalP3** | **Predicted sodium-dependent galactose transporter** |
| Galactose (S7) | GalPTS | PTS system, galactose-specific |
| Galactose (S7) | GalT | Galactose-1-phosphate uridylyltransferase (EC 2.7.7.10) |
| Galactose (S7) | GalU | UTP--glucose-1-phosphate uridylyltransferase (EC 2.7.7.9) |
| **Galactose (S7)** | **GalY** | **Predicted galactose-1-phosphate uridylyltransferase (EC 2.7.7.10)** |
| N-Acetylgalactosamine (S4) | GamPTS | PTS system, galactosamine-specific |
| N-Acetylgalactosamine (S4), N-Acetylglucosamine (S5), Galactose (S7) | GnbG | Phospho-beta-Galactosidase, specific for lacto-N-biose and galacto-N-biose, GH35 family |
| N-Acetylgalactosamine (S4), N-Acetylglucsoamine (S5), Galactose (S7) | GnbPTS | PTS system, galacto-N-biose, lacto-N-biose, and N-acetylgalactosamine-specific |
| Galactose (S7) | LacAB | Galactose-6-phosphate isomerase, (EC 5.3.1.26) |
| N-Acetylgalactosamine (S4), N-Acetylglucsoamine (S5), Galactose (S7) | LnbABC | Lacto-N-biose / Galacto-N-biose ABC transporter |
| N-Acetylgalactosamine (S4), N-Acetylglucsoamine (S5), Galactose (S7) | LnbP | Lacto-N-biose / Galacto-N-biose phosphorylase (EC 2.4.1.211) |
| Galactose (S7) | MglABC | Galactose/methyl galactoside ABC transport system |
| N-Acetylglucsoamine (S5), N-Acetylneuraminic acid (S6) | NagA | N-acetylglucosamine-6-phosphate deacetylase (EC 3.5.1.25) |
| N-Acetylglucsoamine (S5), N-Acetylneuraminic acid (S6) | NagB1 | Glucosamine-6-phosphate deaminase [isomerizing], alternative (EC 3.5.99.6) |
| N-Acetylglucsoamine (S5), N-Acetylneuraminic acid (S6) | NagB2 | Glucosamine-6-phosphate deaminase (EC 3.5.99.6) |
| N-Acetylglucsoamine (S5) | NagE | PTS system, N-acetylglucosamine-specific |
| N-Acetylglucsoamine (S5), N-Acetylneuraminic acid (S6) | NagK1 | N-acetylglucosamine kinase of eukaryotic type (EC 2.7.1.59) |
| N-Acetylglucsoamine (S5), N-Acetylneuraminic acid (S6) | NagK2 | Predicted N-acetyl-glucosamine kinase 2, ROK family (EC 2.7.1.59) |
| N-Acetylglucsoamine (S5) | NagP | N-acetylglucosamine permease, MFS family |
| N-Acetylneuraminic acid (S6) | NanA | N-acetylneuraminate lyase (EC 4.1.3.3) |
| N-Acetylneuraminic acid (S6) | NanABC | Predicted N-acetylneuraminate ABC transporter |
| N-Acetylneuraminic acid (S6) | NanABC2 | Predicted N-acetylneuraminate ABC transporter |
| N-Acetylneuraminic acid (S6) | NanE | N-acetylmannosamine-6-phosphate 2-epimerase (EC 5.1.3.9) |
| N-Acetylneuraminic acid (S6) | NanE-II | N-acetylglucosamine 2-epimerase (EC 5.1.3.8) |
| N-Acetylneuraminic acid (S6) | NanK | N-acetylmannosamine kinase (EC 2.7.1.60) |
| N-Acetylneuraminic acid (S6) | NanP | Sialidase precursor (EC 3.2.1.18) |
| N-Acetylneuraminic acid (S6) | NanT | Sialic acid transporter (permease) |
| N-Acetylneuraminic acid (S6) | NanX | Predicted sialic acid transporter |
| N-Acetylneuraminic acid (S6) | NeuT | Predicted N-acetylneuraminate TRAP-type transport system |
| **N-Acetylglucsoamine (S5)** | **NgcABCD** | **Predicted N-acetylglucosamine ABC transporter** |
| N-Acetylglucsoamine (S5) | NgcEFG | N-acetyl-D-glucosamine ABC transporter |

**Table S3.** Genes involved in depletion and utilization of fucose. The PubSEED identifiers are shown. For details on analyzed organisms see Table S1, for details on gene functions see Table S2.

| Organism (PubSEED ID) | Catabolic Pahtway I | | | | | | Catabolic Pathway II | | | | | Mutarotase | | Transport | | | Hydrolases Alfa-L-fucosidase | | | Problems (if any) |
|---|---|---|---|---|---|---|---|---|---|---|---|---|---|---|---|---|---|---|---|---|
| | Isomerase | Kinase | Aldolase | Dehydrogenase I | | Hydrolase I | Dehydratase | Dehydrogenase II | | Hydrolase II | | | | Permease | | ABC | | | | |
| | FucI | FucK | FucA | FclA1 | FclA2 | FclB | FclC | FclD1 | FclD2 | FclE1 | FclE2 | FucU1 | FucU2 | FucP1 | FucP2 | transporter | AlfA29 | AlfA42 | AlfA95 | |
| Abiotrophia defectiva ATCC 49176 (592010.4) | 2230 | 2229 | 2231 | | | | | | | | | 2235 | | | | 2232, 2233, 2234 | 1484, 1611 | | | |
| Acidaminococcus sp. D21 (563191.3) | | | | | | | | | | | | | | | | | | | | |
| Acidaminococcus sp. HPA0509 (1203555.3) | | | | | | | | | | | | | | | | | | | | |
| Acinetobacter junii SH205 (575587.3) | | | | | | | | | | | | | | | | | | | | |
| Actinomyces odontolyticus ATCC 17982 (411466.7) | 1082 | 1083 | | | | | | | | | | 1085 | | | | 1079, 1080, 1081 | 396, 1383, 1505, 1897 | | | L-fuculose phosphate aldolase is not found |
| Actinomyces sp. HPA0247 (1203556.3) | 1579 | 1578 | | | | | | | | | | 1576 | | | | 1580, 1581, 1582 | 424, 1050, 1900 | | | L-fuculose phosphate aldolase is not found |
| Akkermansia muciniphila ATCC BAA-835 (349741.6) | 2082 | 2080 | 2079 | | | | | | | | | 2403 | | 2083 | | | 13, 172, 446, 962 | | | |
| Alistipes indistinctus YIT 12060 (742725.3) | | | | | | | | | | | | | | | | | 339, 458, 861, 1048, 1997 | | | |
| Alistipes shahii WAL 8301 (717959.3) | | | | | | | | | | | | | | | | | 2434 | | | |
| Anaerobaculum hydrogeniformans ATCC BAA-1850 (592015.5) | | | | | | | | | | | | | | | | | | | | |
| Anaerococcus hydrogenalis DSM 7454 (561177.4) | | | | | | | | | | | | | | | | | | | | |
| Anaerofustis stercorihominis DSM 17244 (445971.6) | | | | | | | | | | | | | | | | | | | | |
| Anaerostipes caccae DSM 14662 (411490.6) | | | | | | | | | | | | | | | | | | | | |
| Anaerostipes hadrus DSM 3319 (649757.3) | | | | | | | | | | | | | | | | | | | | |
| Anaerostipes sp. 3_2_56FAA (665937.3) | | | | | | | | | | | | | | | | | | | | |
| Anaerotruncus colihominis DSM 17241 (445972.6) | 2663 | 2662, 2666 | | | | | | | | | | 2664 | | | | | | | | L-fuculose phosphate aldolase is not found |
| Bacillus smithii 7_3_47FAA (665952.3) | | | | | | | | | | | | | | | | | | | | |
| Bacillus sp. 7_6_55CFAA_CT2 (665957.3) | | | | | | | | | | | | | | | | | | | | |
| Bacillus subtilis subsp. subtilis str. 168 (224308.113) | | | | | | | | | | | | | | | | | | | | |
| Bacteroides caccae ATCC 43185 (411901.7) | 2065 | 2063 | 2064 | | | | | | | | | | 2062 | 2061 | | | 82, 547, 1156, 1241, 1249, 2489, 2536, 2669, 3008, 3119, 3380, 3585 | | | |
| Bacteroides capillosus ATCC 29799 (411467.6) | | | | | | | | | | | | | | | | | | | | |
| Bacteroides cellulosilyticus DSM 14838 (537012.5) | 3622, 3623 | 3620 | 3621 | | | | | | | | | | 3619 | 3618 | | | 1104, 1113, 1116, 1186, 1942, 1945, 2120, 2214, 4949 | | | |
| Bacteroides clarus YIT 12056 (762984.3) | | | | | | | | | | | | | | | | | 2867 | | | |
| Bacteroides coprocola DSM 17136 (470145.6) | | | | | | | | | | | | | 3113 | 2411 | | | 985, 1296, 1610 | | | |
| Bacteroides coprophilus DSM 18228 (547042.5) | 3061 | 3058 | 3059 | | | | | | | | | | 960 | 486, 3057 | | | 603, 1104, 1751, 1769, 1843, 1846, 1909, 2018, 2225, 2427, 2618, 3016, 3050, 3179 | | | |

| | | | | | | | | | | | | | |
|---|---|---|---|---|---|---|---|---|---|---|---|---|---|
| Bacteroides dorei DSM 17855 (483217.6) | 1641 | | | | | | | | | | 1640 | 295, 460, 737, 1090, 1389, 1396, 1399, 1496, 1914, 2938 | |
| Bacteroides eggerthii 1_2_48FAA (665953.3) | 1089 | | | | | | | | | | | 543, 1869, 1871, 1872, 1908 | |
| Bacteroides eggerthii DSM 20697 (483216.6) | 2986 | | | | | | | | | 135 | | 129, 134, 202, 239, 241, 1925 | |
| Bacteroides finegoldii DSM 17565 (483215.6) | 1656 | 1654 | 1655 | | | | | | | | 1653 | 808, 3068 | |
| Bacteroides fluxus YIT 12057 (763034.3) | 758 | 755 | 756 | | | | | | | | 754 | 100, 1084, 2884, 3259 | |
| Bacteroides fragilis 3_1_12 (457424.5) | 2683 | 2680 | 2681 | | | | | | | 1557 | 2679 | 946, 1785, 1791, 3290, 3430, 3477, 3577, 4522 | |
| Bacteroides fragilis 638R (862962.3) | 269 | 272 | 271 | | | | | | | 2669 | 273 | 29, 733, 882, 888, 3196, 3332, 3386, 3749 | 4535 |
| Bacteroides fragilis NCTC 9343 (272559.17) | 207 | 210 | 209 | | | | | | | 2619 | 211 | 29, 651, 792, 798, 1768, 3117, 3246, 3294, 3662 | 4376 |
| Bacteroides fragilis YCH46 (295405.11) | 283 | 286 | 285 | | | | | | | 2497 | 287 | 70, 747, 885, 891, 1671, 3113, 3240, 3287, 3648 | 4297 |
| Bacteroides intestinalis DSM 17393 (471870.8) | 1809 | 1811 | 1810 | | | | | | | 872, 1812 | 1813 | 223, 225, 787, 869, 2604, 3055 | |
| Bacteroides ovatus 3_8_47FAA (665954.3) | 1475 | 1477 | 1476 | | | | | | | 1478 | 1479 | 2122, 4662, 4701 | |
| Bacteroides ovatus ATCC 8483 (411476.11) | 3281 | 3283 | 3282 | | | | | | | 3284 | 3285 | 473, 1693, 3089, 3716, 3823, 3868, 3913 | |
| Bacteroides ovatus SD CC 2a (702444.3) | 675 | 677 | 676 | | | | | | | 678 | 679 | 1611, 1861, 3887, 4703 | |
| Bacteroides ovatus SD CMC 3f (702443.3) | 448 | 450 | 449 | | | | | | | 451 | 452 | 4198, 4578, 4613 | |
| Bacteroides pectinophilus ATCC 43243 (483218.5) | | | | | | | | | | | | | |
| Bacteroides plebeius DSM 17135 (484018.6) | 2585 | 2583 | 2584 | | | | | | | 642, 1522, 1648 | 2582 | 35, 554, 1508, 1523, 1559, 1666, 1803, 2180, 3304 | |

| Organism | | | | | | | | | |
|---|---|---|---|---|---|---|---|---|---|
| Bacteroides sp. 1_1_14 (469585.3) | 4082 | 4079, 4080 | 4081 | 3598, 4078 | 4077 | 645, 1387, 1732, 1902, 2593, 3202, 3597, 3730, 4289 | | |
| Bacteroides sp. 1_1_30 (457387.3) | 747 | 749 | 748 | 750 | 751 | 2347, 2348, 4204, 4241, 5185 | | |
| Bacteroides sp. 1_1_6 (469586.3) | 1092 | 1094 | 1093 | 1095 | 1096 | 1323, 2469, 3466, 3597, 3883, 4088, 4248, 4803 | | |
| Bacteroides sp. 2_1_16 (469587.3) | 2451 | 2454 | 2453 | 1110 | 2455 | 1994, 2136, 2142, 2280, 2841, 2888, 3014, 3716 | 3941 | |
| Bacteroides sp. 2_1_22 (469588.3) | 1567 | 1565 | 1566 | 1564 | 1563 | 754, 4210, 4244, 4549 | | |
| Bacteroides sp. 2_1_33B (469589.3) | | | | 3830 | 3829 | 3790 | | |
| Bacteroides sp. 2_1_56FAA (665938.3) | 263 | 266 | 265 | 2813 | 267 | 86, 803, 954, 960, 1919, 3228, 3779, 3992, 4041 | 4349 | |
| Bacteroides sp. 2_1_7 (457388.5) | | | | 4130 | 4129 | 3810 | | |
| Bacteroides sp. 2_2_4 (469590.5) | 3688 | 3686 | 3687 | 3685 | 3684 | 145, 146, 147, 281, 317, 1316, 1318, 4737 | | |
| Bacteroides sp. 20_3 (469591.4) | | | | 1471 | 1472 | 4663 | | |
| Bacteroides sp. 3_1_19 (469592.4) | | | | 3940 | | 4084 | | |
| Bacteroides sp. 3_1_23 (457390.3) | 4801 | 4803 | 4802 | 4804 | 4805 | 1587, 2978, 4916, 4951 | | |
| Bacteroides sp. 3_1_33FAA (457391.3) | 3286 | | | 1294, 1314 | 3285 | 1204, 1317, 1629, 1813, 2059, 3530, 3670, 3739, 3741 | | |
| Bacteroides sp. 3_1_40A (469593.3) | 635 | | | 3600, 3671 | 636 | 13, 403, 779, 1926, 1942, 3475, 3597, 3772, 4743 | | L-fuculokinase and L-fuculose phosphate aldolase are not found |
| Bacteroides sp. 3_2_5 (457392.3) | 916 | 913 | 914 | 4025 | 912 | 303, 309, 460, 1088, 1965, 2096, 2142, 3080 | 3345 | |
| Bacteroides sp. 4_1_36 (457393.3) | 253 | 250, 256 | | | 251 | 770, 1511 | | L-fuculose phosphate aldolase is not found |

| Organism | | | | | | | | | | | | Notes |
|---|---|---|---|---|---|---|---|---|---|---|---|---|
| Bacteroides sp. 4_3_47FAA (457394.3) | 70 | | | | | | | | 1598, 1620 | 71 | 156, 498, 852, 1505, 1623, 1708, 2741, 2787, 2800, 3236, 4184 | L-fuculokinase and L-fuculose phosphate aldolase are not found |
| Bacteroides sp. 9_1_42FAA (457395.6) | 3147 | | | | | | | | 53, 73 | 3146 | 76, 331, 586, 850, 3379, 3414, 3718, 3722, 3993 | L-fuculokinase and L-fuculose phosphate aldolase are not found |
| Bacteroides sp. D1 (556258.5) | 3800 | 3802 | 3801 | | | | | | 3803 | 3804 | 605, 3644, 3678, 4452 | |
| Bacteroides sp. D2 (556259.3) | 4995 | 4993 | 4994 | | | | | | 4992 | 4991 | 3077, 3113, 3267 | |
| Bacteroides sp. D20 (585543.3) | 2023 | 2020 | | | | | | | 2021 | | 15, 1520 | L-fuculose phosphate aldolase is not found |
| Bacteroides sp. D22 (585544.3) | 4458 | 4460 | 4459 | | | | | | 4461 | 4462 | 1170, 1205, 3485 | |
| Bacteroides sp. HPS0048 (1078089.3) | 707 | 709 | 708 | | | | | | 710 | 711 | 3184 | |
| Bacteroides stercoris ATCC 43183 (449673.7) | | | | | | | | | | | 1321, 1462, 2247, 2811 | |
| Bacteroides thetaiotaomicron VPI-5482 (226186.12) | 1300 | 1302 | 1301 | | | | | | 1303, 3726 | 1304 | 1663, 1891, 2256, 3020, 3725, 3861, 4021, 4203, 4792 | |
| Bacteroides uniformis ATCC 8492 (411479.10) | 517 | 514, 521 | | | | | | | | 515 | 1323, 3291 | L-fuculose phosphate aldolase is not found |
| Bacteroides vulgatus ATCC 8482 (435590.9) | 1430 | | | | | | | | 197, 220 | 1429 | 106, 223, 550, 573, 574, 1080, 1296, 1812, 2515, 4161 | L-fuculokinase and L-fuculose phosphate aldolase are not found |
| Bacteroides vulgatus PC510 (702446.3) | 138 | | | | | | | | 2269, 2292 | 137 | 707, 1759, 1901, 1975, 2265, 2377, 2594, 3857 | L-fuculokinase and L-fuculose phosphate aldolase are not found |
| Bacteroides xylanisolvens SD CC 1b (702447.3) | 4201 | 4203 | 4202 | | | | | | 4204 | 4205 | 244, 279, 1159, 2102 | |
| Bacteroides xylanisolvens XB1A (657309.4) | 3141 | 3139 | 3140 | | | | | | 3138 | 3137 | 225, 261, 784, 3616 | |
| Barnesiella intestinihominis YIT 11860 (742726.3) | | | | | | | | | | | 53, 315 | |
| Bifidobacterium adolescentis L2-32 (411481.5) | | | | | | | | | | | | |
| Bifidobacterium angulatum DSM 20098 = JCM 7096 (518635.7) | | | | | | | | | | | | |
| Bifidobacterium animalis subsp. lactis AD011 (442563.4) | | | | | | | | | | | | |
| Bifidobacterium bifidum NCIMB 41171 (398513.5) | | | | 903 | | 901 | | | | | | L-fuconolactone hydrolase, 2-keto-3-deoxy-L-fuconate dehydrogenase, 2,4-diketo-3-deoxy-L-fuconate hydrolase, and transporters are not found |
| Bifidobacterium breve DSM 20213 = JCM 1192 (518634.7) | | | | 1312, 1715 | 1314, 1716 | 1310, 1714 | 1312, 1715 | 1315, 1717 | | 1313 | | |
| Bifidobacterium breve HPH0326 (1203540.3) | | | | 1250, 1369 | 1248, 1370 | 1252, 1368 | 1250, 1369 | 1247, 1371 | | 1249 | | |

| Organism | | | | | | | | | | | | | | |
|---|---|---|---|---|---|---|---|---|---|---|---|---|---|---|
| Bifidobacterium catenulatum DSM 16992 = JCM 1194 (566552.6) | | | | | | | | | | | | | | |
| Bifidobacterium dentium ATCC 27678 (473819.7) | | | | | | | | | | | | 802 | | |
| Bifidobacterium gallicum DSM 20093 (561180.4) | | | | | | | | | | | | | | |
| Bifidobacterium longum DJO10A (Prj:321) (205913.11) | | | | | | | | | | | | | | |
| Bifidobacterium longum NCC2705 (206672.9) | | | | | | | | | | | | | | |
| Bifidobacterium longum subsp. infantis 157F (565040.3) | | | | | | | | | | | | | | |
| Bifidobacterium longum subsp. infantis ATCC 15697 = JCM 1222 (391904.8) | 2384, 2418 | 2382 | | 354, 2385, 2419 | 2384, 2418 | 2417, 2425 | 2381, 2416 | | 355, 2383 | 351, 352, 353, 2277, 2278, 2279, 2421, 2422, 2423 | 253, 439, 2415 | 295, 356 | | |
| Bifidobacterium longum subsp. infantis ATCC 55813 (548480.3) | | | | | | | | | | | | | | |
| Bifidobacterium longum subsp. infantis CCUG 52486 (537937.5) | | | | | | | | | | | | | | |
| Bifidobacterium longum subsp. longum 2-2B (1161745.3) | | | | | | | | | | | | | | |
| Bifidobacterium longum subsp. longum 44B (1161743.3) | | | | | | | | | | | | | | |
| Bifidobacterium longum subsp. longum F8 (722911.3) | | | | | | | | | | | | | | |
| Bifidobacterium longum subsp. longum JCM 1217 (565042.3) | | | | | | | | | | | | | | |
| Bifidobacterium pseudocatenulatum DSM 20438 = JCM 1200 (547043.8) | 821 | 820 | | 822 | 821 | 819 | | | | 823, 824, 825 | | | | |
| Bifidobacterium sp. 12_1_47BFAA (469594.3) | | | | | | | | | | | | | | |
| Bilophila wadsworthia 3_1_6 (563192.3) | | | | | | | | 3386 | | | | | | |
| Blautia hansenii DSM 20583 (537007.6) | | | | | | | | | | | 2138 | | 2574 | |
| Blautia hydrogenotrophica DSM 10507 (476272.5) | 335, 2914 | 2911, 2915, 2916 | 2912 | | | | 2913 | | | | | | | |
| Bryantella formatexigens DSM 14469 (478749.5) | | | | | | | | | | | | | | |
| Burkholderiales bacterium 1_1_47 (469610.4) | | | | | | | | | | | | | | |
| butyrate-producing bacterium SM4/1 (245012.3) | | | | | | | | | | | | | | |
| butyrate-producing bacterium SS3/4 (245014.3) | | | | | | | | | | | | | | |
| butyrate-producing bacterium SSC/2 (245018.3) | | | | | | | | | | | | | | |
| Butyricicoccus pullicaecorum 1.2 (1203606.4) | | | | | | | | | | | | | | |
| Butyrivibrio crossotus DSM 2876 (511680.4) | | | | | | | | | | | | | | |
| Butyrivibrio fibrisolvens 16/4 (657324.3) | | | | | | | | | | | | | | |
| Campylobacter coli JV20 (864566.3) | | | | | | | | | | | | | | |
| Campylobacter upsaliensis JV21 (888826.3) | | | | | | | | | | | | | | |
| Catenibacterium mitsuokai DSM 15897 (451640.5) | | | | | | | | | | | | | | |
| Cedecea davisae DSM 4568 (566551.4) | | | | | | | | | | | | | | |
| Citrobacter freundii 4_7_47CFAA (742730.3) | 269 | 268 | 271 | | | | 267 | | 270 | | | | | |
| Citrobacter sp. 30_2 (469595.3) | 2646 | 2647 | 2644 | | | | 2648 | | 2645 | | | | | |
| Citrobacter youngae ATCC 29220 (500640.5) | 4208 | 4209 | 4206 | | | | 4210 | | | | | | | |
| Clostridiales bacterium 1_7_47FAA (457421.5) | 2905 | 2906 | 2907 | | | | 2908 | | | 2909, 2910, 2911 | | | | |
| Clostridium asparagiforme DSM 15981 (518636.5) | | | | | | | | | | | | | | |
| Clostridium bartlettii DSM 16795 (445973.7) | | | | | | | | | | | | | | |
| Clostridium bolteae 90A5 (997893.5) | 5752 | 1153, 5753 | 5754 | | | | 5755 | | | 5756, 5757, 5758 | | | | |
| Clostridium bolteae ATCC BAA-613 (411902.9) | 1944 | 1943, 3140 | 1942 | | | | 1941 | | | 1938, 1939, 1940 | | | | |
| Clostridium oleatum DSM 1785 (545697.3) | 299 | 300, 301 | 303 | | | | | 570 | | 567, 568, 569 | | | 97, 2000 | |
| Clostridium cf. saccharolyticum K10 (717608.3) | | | | | | | | | | | | | | |
| Clostridium citroniae WAL-17108 (742733.3) | | | | | | | | | | | | | | |
| Clostridium clostridioforme 2_1_49FAA (742735.4) | 342 | 343 | 344 | | | | 345 | | | 346, 347, 348 | | | | |
| Clostridium difficile CD196 (645462.3) | | | | | | | | | | | | | | |
| Clostridium difficile NAP07 (525258.3) | | | | | | | | | | | | | | |
| Clostridium difficile NAP08 (525259.3) | | | | | | | | | | | | | | |
| Clostridium hathewayi WAL-18680 (742737.3) | | | | | | | | | | | | | | |
| Clostridium hiranonis DSM 13275 (500633.7) | | | | | | | | | | | | | | |
| Clostridium hylemonae DSM 15053 (553973.6) | 1347, 1356 | 1352 | 1348, 1358 | | | | | | | | | | | |
| Clostridium leptum DSM 753 (428125.8) | 869 | 1986 | 1987 | | | | | | | 749, 750, 751 | 371, 2627, 2637 | | 745, 2447 | |
| Clostridium methylpentosum DSM 5476 (537013.3) | 1311, 2438 | | | | | | 1310 | | | 1307, 1308, 1309 | 874, 1258, 1261, 2299, 2382, 2383 | 422, 423, 452, 454, 1356, 1361 | | L-fuculokinase and L-fuculose phosphate aldolase are not found |

| Organism | | | | | | | | | | | | | | | | Notes |
|---|---|---|---|---|---|---|---|---|---|---|---|---|---|---|---|---|
| Clostridium nexile DSM 1787 (500632.7) | | | | | | | | | | | | 3649, 3652 | | 2103, 2104, 2171 | | |
| Clostridium perfringens WAL-14572 (742739.3) | 499 | 500 | 502 | | | | | | 501 | | 503 | 1542 | | 1541 | | |
| Clostridium ramosum DSM 1402 (445974.6) | | | | | | | | | | | | | | | | |
| Clostridium scindens ATCC 35704 (411468.9) | 166 | | | | | | | | 199 | | | 156, 158, 159, 196, 197, 198 | 155 | | | L-fuculokinase and L-fuculose phosphate aldolase are not found |
| Clostridium sp. 7_2_43FAA (457396.3) | | | | | | | | | | | | | | | | |
| Clostridium sp. 7_3_54FAA (665940.3) | | | | | | | | | | | | | | | | |
| Clostridium sp. D5 (556261.3) | 2859 | 33, 394, 2858 | 30, 2796 | | | | | | 1248, 2860 | | | 1693, 1694, 1695, 1696, 2848, 2850, 2851, 2852, 2853, 3603, 3604, 3605, 3709, 3710, 3711 | 129, 961, 2609, 3511, 3636, 3714, 3874 | 665, 3606 | 175 | |
| Clostridium sp. HGF2 (908340.3) | | | | | | | | | | | | | | | | |
| Clostridium sp. L2-50 (411489.7) | | | | | | | | | | | | | | | | |
| Clostridium sp. M62/1 (411486.3) | | | | | | | | | | | | | | | | |
| Clostridium sp. SS2/1 (411484.7) | | | | | | | | | | | | | | | | |
| Clostridium spiroforme DSM 1552 (428126.7) | | | | | | | | | | | | 1021 | | | | |
| Clostridium sporogenes ATCC 15579 (471871.7) | | | | | | | | | | | | | | | | |
| Clostridium symbiosum WAL-14163 (742740.3) | | | | | | | | | | | | | | | | |
| Clostridium symbiosum WAL-14673 (742741.3) | | | | | | | | | | | | | | | | |
| Collinsella aerofaciens ATCC 25986 (411903.6) | | | | | | | | | | | | | | | | |
| Collinsella intestinalis DSM 13280 (521003.7) | | | | | | | | | | | | | | | | |
| Collinsella stercoris DSM 13279 (445975.6) | | | | | | | | | | | | | | 933 | | |
| Collinsella tanakaei YIT 12063 (742742.3) | 921 | 923 | 920 | | | | | | 922 | | | | | | | |
| Coprobacillus sp. 29_1 (469596.3) | | | | | | | | | | | | | | | | |
| Coprobacillus sp. 3_3_56FAA (665941.3) | | | | | | | | | | | | | | | | |
| Coprobacillus sp. 8_2_54BFAA (469597.4) | | | | | | | | | | | | | | | | |
| Coprobacillus sp. D6 (556262.3) | | | | | | | | | | | | | | | | |
| Coprococcus catus GD/7 (717962.3) | | | | | | | | | | | | | | | | |
| Coprococcus comes ATCC 27758 (470146.3) | | | | | | | | | | | | | | | | |
| Coprococcus eutactus ATCC 27759 (411474.6) | | | | | | | | | | | | | | | | |
| Coprococcus sp. ART55/1 (751585.3) | | | | | | | | | | | | | | | | |
| Coprococcus sp. HPP0048 (1078091.3) | 911 | 910 | 909 | | | | | | | | | 914, 915, 916, 917 | 913, 1543 | | | |
| Coprococcus sp. HPP0074 (1078090.3) | 886 | 887 | 888 | | | | | | | | | 880, 881, 882, 883 | 38, 884 | | | |
| Corynebacterium ammoniagenes DSM 20306 (649754.3) | 1958 | 2488 | 1956 | | 1958 | | 1960 | | 1954 | | 1955 | | | | | |
| Corynebacterium sp. HFH0082 (1078764.3) | | | | | | | | | | | | | | | | |
| Dermabacter sp. HFH0086 (1203568.3) | | | | | | | | | | | | | | | | |
| Desulfitobacterium hafniense DP7 (537010.4) | | | | | | | | | | | | | | | | |
| Desulfovibrio piger ATCC 29098 (411464.8) | | | | | | | | | | | | | | | | |
| Desulfovibrio sp. 3_1_syn3 (457398.5) | | | | | | | | | | | | | | | | |
| Dorea formicigenerans 4_6_53AFAA (742765.5) | | | | | | | | | | | | 3265 | | | | |
| Dorea formicigenerans ATCC 27755 (411464.4) | | | | | | | | | | | | | | | | |
| Dorea longicatena DSM 13814 (411462.6) | | | | | | | | | | | | | | | | |
| Dysgonomonas gadei ATCC BAA-286 (742766.3) | 3451 | 3454 | 3453 | | | | | | | 3452 | | 549, 3282, 3804 | | | | |
| Dysgonomonas mossii DSM 22836 (742767.3) | 792 | 794 | 793 | | | | | | | 795 | | 337, 342, 1513, 1867 | | | | |
| Edwardsiella tarda ATCC 23685 (500638.3) | 183 | 182 | 185 | | | | | | 181 | | 184 | | | | | |
| Eggerthella sp. 1_3_56FAA (665943.3) | | | | | | | | | | | | | | | | |
| Eggerthella sp. HGA1 (910311.3) | | | | | | | | | | | | | | | | |
| Enterobacter canerogenus ATCC 35316 (500639.8) | 2894 | 2890, 2895 | 2888 | | | | | | 2896 | | | 2891, 2892, 2893 | | | | |
| Enterobacter cloacae subsp. cloacae NCTC 9394 (718254.4) | 1504 | 1503, 1508 | 1510 | | | | | | 1502 | | | 1505, 1506, 1507 | | | | |
| Enterobacteriaceae bacterium 9_2_54FAA (469613.3) | 2798 | 2797 | 2800 | | | | | | 2796 | | 2799 | | | | | |
| Enterococcus faecalis PC1.1 (702448.3) | | | | | | | | | | | | | | | | |
| Enterococcus faecalis TX0104 (491074.3) | | | | | | | | | | | | | | | | |
| Enterococcus faecalis TX1302 (749513.3) | | | | | | | | | | | | | | | | |
| Enterococcus faecalis TX1322 (525278.3) | | | | | | | | | | | | | | | | |
| Enterococcus faecalis TX1341 (749514.3) | | | | | | | | | | | | | | | | |
| Enterococcus faecalis TX1342 (749515.3) | | | | | | | | | | | | | | | | |
| Enterococcus faecalis TX1346 (749516.3) | | | | | | | | | | | | | | | | |
| Enterococcus faecalis TX2134 (749518.3) | | | | | | | | | | | | | | | | |
| Enterococcus faecalis TX2137 (749491.3) | | | | | | | | | | | | | | | | |
| Enterococcus faecalis TX4244 (749494.3) | | | | | | | | | | | | | | | | |
| Enterococcus faecium PC4.1 (791161.5) | | | | | | | | | | | | | | | | |
| Enterococcus faecium TX1330 (525279.3) | | | | | | | | | | | | | | | | |

| Organism | 1 | 2 | 3 | 4 | 5 | 6 | 7 | 8 | 9 | 10 | 11 | 12 | 13 | 14 | 15 | 16 | 17 | 18 | 19 | 20 |
|---|---|---|---|---|---|---|---|---|---|---|---|---|---|---|---|---|---|---|---|---|
| Enterococcus saccharolyticus 30_1 (742813.4) | | | | | | | | | | | | | | 386, 2522, 2526 | | | | | | |
| Enterococcus sp. 7L76 (657310.3) | | | | | | | | | | | | | | | | | | | | |
| Erysipelotrichaceae bacterium 2_2_44A (457422.3) | | | | | | | | | | | | | | | | | | | | |
| Erysipelotrichaceae bacterium 21_3 (658657.3) | | | | | | | | | | | | | | | | | | | | |
| Erysipelotrichaceae bacterium 3_1_53 (658659.3) | | | | | | | | | | | | | | | | | | | | |
| Erysipelotrichaceae bacterium 5_2_54FAA (552396.3) | | | | | | | | | | | | | | | | | | | | |
| Erysipelotrichaceae bacterium 6_1_45 (469614.3) | | | | | | | | | | | | | | | | | | | | |
| Escherichia coli 4_1_47FAA (1127356.4) | 1365 | 1366 | 1362 | | | | | | | 1367 | 1364 | | | | | | | | | |
| Escherichia coli MS 107-1 (679207.4) | 782 | 781 | 784 | | | | | | | 780 | 783 | | | | | | | | | |
| Escherichia coli MS 115-1 (749537.3) | 3801 | 3800 | 3804 | | | | | | | 3799 | 3802 | | | | | | | | | |
| Escherichia coli MS 116-1 (749538.3) | 3708 | 3709 | 3706 | | | | | | | 3710 | 3707 | | | | | | | | | |
| Escherichia coli MS 119-7 (679206.4) | 382 | 383 | 380 | | | | | | | 384 | 381 | | | | | | | | | |
| Escherichia coli MS 124-1 (679205.4) | 3056 | 3057 | 3054 | | | | | | | 3058 | 3055 | | | | | | | | | |
| Escherichia coli MS 145-7 (679204.3) | 2409 | 2410 | 2407 | | | | | | | 2411 | 2408 | | | | | | | | | |
| Escherichia coli MS 146-1 (749540.3) | 1802 | 1801 | 1804 | | | | | | | 1800 | 1803 | | | | | | | | | |
| Escherichia coli MS 175-1 (749544.3) | | | | | | | | | | | | | | | | | | | | |
| Escherichia coli MS 182-1 (749545.3) | 1361 | 1360 | 1363 | | | | | | | 1359 | 1362 | | | | | | | | | |
| Escherichia coli MS 185-1 (749546.3) | 3109 | 3108 | 3112 | | | | | | | 3107 | 3110 | | | | | | | | | |
| Escherichia coli MS 187-1 (749547.3) | 3378 | 3374, 3379 | 3372 | | | | | | | 3380 | | | | 3375, 3376, 3377 | | | | | | |
| Escherichia coli MS 196-1 (749548.3) | 2867 | 2866 | 2869 | | | | | | | 2865 | 2868 | | | | | | | | | |
| Escherichia coli MS 198-1 (749549.3) | 2180 | 2179 | 2183 | | | | | | | 2178 | 2181 | | | | | | | | | |
| Escherichia coli MS 200-1 (749550.3) | 768 | 769 | 765 | | | | | | | 770 | 767 | | | | | | | | | |
| Escherichia coli MS 21-1 (749527.3) | 1203 | 1202 | 1206 | | | | | | | 1201 | 1204 | | | | | | | | | |
| Escherichia coli MS 45-1 (749528.3) | 1770 | 1771 | 1767 | | | | | | | 1772 | 1769 | | | | | | | | | |
| Escherichia coli MS 69-1 (749531.3) | 5232 | 5231 | 5235 | | | | | | | 5230 | 5233 | | | | | | | | | |
| Escherichia coli MS 78-1 (749532.3) | 754 | 755 | 752 | | | | | | | 756 | 753 | | | | | | | | | |
| Escherichia coli MS 79-10 (796392.4) | 1561 | 1562 | 1559 | | | | | | | 1563 | 1560 | | | | | | | | | |
| Escherichia coli MS 84-1 (749533.3) | 4099 | 4098 | 4101 | | | | | | | 4097 | 4100 | | | | | | | | | |
| Escherichia coli MS 85-1 (679202.3) | 1194 | 1195 | 1192 | | | | | | | 1196 | 1193 | | | | | | | | | |
| Escherichia coli O157:H7 str. Sakai (386585.9) | 3828 | 3829 | 3826 | | | | | | | 3830 | 3827 | | | | | | | | | |
| Escherichia coli SE11 (409438.11) | 3234 | 3235 | 3232 | | | | | | | 3236 | 3233 | | | | | | | | | |
| Escherichia coli SE15 (431946.3) | 2718 | 2719 | 2715, 4404 | | | | | | | 2720 | 2717 | | | | | | | | | |
| Escherichia coli str. K-12 substr. MG1655 (511145.12) | 2902 | 2903 | 2900 | | | | | | | 2904 | 2901 | | | | | | | | | |
| Escherichia coli UTI89 (364106.8) | 3153 | 3154 | 3150 | | | | | | | 3155 | 3152 | | | | | | | | | |
| Escherichia sp. 1_1_43 (457400.3) | 794 | 795 | 792 | | | | | | | 796 | 793 | | | | | | | | | |
| Escherichia sp. 3_2_53FAA (469598.5) | 1725 | 1724 | 1728, 3622 | | | | | | | 1723 | 1726 | | | | | | | | | |
| Escherichia sp. 4_1_40B (457401.3) | 2744 | 2743 | 2746 | | | | | | | 2742 | 2745 | | | | | | | | | |
| Eubacterium biforme DSM 3989 (518637.5) | | | | | | | | | | | | | | | | | | 1701 | | |
| Eubacterium cylindroides T2-87 (717960.3) | | | | | | | | | | | | | | | | | | | | |
| Eubacterium dolichum DSM 3991 (428127.7) | | | | | | | | | | | | | | | | | | | | |
| Eubacterium hallii DSM 3353 (411469.3) | | | | | | | | | | | | | | | | | | | | |
| Eubacterium rectale DSM 17629 (657318.4) | | | | | | | | | | | | | | | | | | | | |
| Eubacterium rectale M104/1 (657317.3) | | | | | | | | | | | | | | | | | | | | |
| Eubacterium siraeum 70/3 (657319.3) | 1103 | 1102 | 1100 | | | | | | | 1101 | | | | | | | | | 1111 | |
| Eubacterium siraeum DSM 15702 (428128.7) | 1218 | 1219 | 1221 | | | | | | | 1220 | | | | | | | | 2078 | 1212 | |
| Eubacterium siraeum V10Sc8a (717961.3) | 2292 | 2293 | 2295 | | | | | | | 2294 | | | | | | | | | 2286 | |
| Eubacterium sp. 3_1_31 (457402.3) | | | | | | | | | | | | | | | | | | | | |
| Eubacterium ventriosum ATCC 27560 (411463.4) | | | | | | | | | | | | | | | | | | | | |
| Faecalibacterium cf. prausnitzii KLE1255 (748224.3) | | | | | | | | | | | | | | | | | | | | |
| Faecalibacterium prausnitzii A2-165 (411483.3) | | | | | | | | | | | | | | | | | | | | |
| Faecalibacterium prausnitzii L2-6 (718252.3) | | | | | | | | | | | | | | | | | | | | |
| Faecalibacterium prausnitzii M21/2 (411485.10) | | | | | | | | | | | | | | | | | | | | |
| Faecalibacterium prausnitzii SL3/3 (657322.3) | | | | | | | | | | | | | | | | | | | | |
| Fusobacterium gonidiaformans ATCC 25563 (469615.3) | | | | | | | | | | | | | | | | | | | | |
| Fusobacterium mortiferum ATCC 9817 (469616.3) | | | | | | | | | | | | | | | | | | | | |
| Fusobacterium necrophorum subsp. funduliforme 1_1_36S (742814.3) | | | | | | | | | | | | | | | | | | | | |
| Fusobacterium sp. 1_1_41FAA (469621.3) | | | | | | | | | | | | | | | | | | | | |
| Fusobacterium sp. 11_3_2 (457403.3) | | | | | | | | | | | | | | | | | | | | |
| Fusobacterium sp. 12_1B (457404.3) | | | | | | | | | | | | | | | | | | | | |
| Fusobacterium sp. 2_1_31 (469599.3) | | | | | | | | | | | | | | | | | | | | |
| Fusobacterium sp. 21_1A (469601.3) | | | | | | | | | | | | | | | | | | | | |
| Fusobacterium sp. 3_1_27 (469602.3) | | | | | | | | | | | | | | | | | | | | |
| Fusobacterium sp. 3_1_33 (469603.3) | | | | | | | | | | | | | | | | | | | | |
| Fusobacterium sp. 3_1_36A2 (469604.3) | | | | | | | | | | | | | | | | | | | | |
| Fusobacterium sp. 3_1_5R (469605.3) | | | | | | | | | | | | | | | | | | | | |
| Fusobacterium sp. 4_1_13 (469606.3) | | | | | | | | | | | | | | | | | | | | |
| Fusobacterium sp. 7_1 (457405.3) | | | | | | | | | | | | | | | | | | | | |
| Fusobacterium sp. D11 (556264.3) | | | | | | | | | | | | | | | | | | | | |
| Fusobacterium sp. D12 (556263.3) | | | | | | | | | | | | | | | | | | | | |
| Fusobacterium ulcerans ATCC 49185 (469617.3) | | | | | | | | | | | | | | | | | | | | |
| Fusobacterium varium ATCC 27725 (469618.3) | 792 | 791 | 793 | | | | | | | 788 | 789 | | | | | | | | | |

| Name | | | | | | | | | | | | | | Notes |
|---|---|---|---|---|---|---|---|---|---|---|---|---|---|---|
| Gordonibacter pamelaeae 7-10-1-b (657308.3) | | | | | | | | | | | | | | |
| Hafnia alvei ATCC 51873 (1002364.3) | 2555 | 2556 | 2553 | | | | | | 2557 | 2554 | | | | |
| Helicobacter bilis ATCC 43879 (613026.4) | | | | | | | | | | | | | | |
| Helicobacter canadensis MIT 98-5491 (Prj:30719) (537970.9) | | | | | | | | | | | | | | |
| Helicobacter cinaedi ATCC BAA-847 (1206745.3) | | | | | | | | | | | | | | |
| Helicobacter cinaedi CCUG 18818 (537971.5) | | | | | | | | | | | | | | |
| Helicobacter pullorum MIT 98-5489 (537972.5) | | | | | | | | | | | | | | |
| Helicobacter winghamensis ATCC BAA-430 (556267.4) | | | | | | | | | | | | | | |
| Holdemania filiformis DSM 12042 (545696.5) | 1989 | 1988 | 1987 | | | | | | | | | 749, 2038 | | |
| Klebsiella pneumoniae 1162281 (1037908.3) | 632 | 631 | 634 | | | | | | 630 | 633 | | | | |
| Klebsiella pneumoniae subsp. pneumoniae WGLW5 (1203547.3) | 3406 | 3407 | 3404 | | | | | | 3408 | 3405 | | | | |
| Klebsiella sp. 1_1_55 (469608.3) | 3083 | 3085 | 3081 | | | | | | 3086 | 3082 | | | | |
| Klebsiella sp. 4_1_44FAA (665944.3) | 2325 | 2326, 2327 | 2323 | | | | | | 2328 | 2324 | | | | |
| Lachnospiraceae bacterium 1_1_57FAA (658081.3) | | | | | | | | | | | | | 1329, 1582 | |
| Lachnospiraceae bacterium 1_4_56FAA (658655.3) | 2064 | 2065 | | | | | | | 2063 | | 2069, 2070, 2071, 2072 | 1116, 2725 | | L-fuculose phosphate aldolase is not found |
| Lachnospiraceae bacterium 2_1_46FAA (742723.3) | | | | | | | | | | | | 378 | 1785, 1909, 1910 | |
| Lachnospiraceae bacterium 2_1_58FAA (658082.3) | 1907 | 1908 | | | | | | | 1906 | | 532, 533, 534, 816, 817, 818, 819, 1911, 1912, 1913, 1914, 1916 | 814, 1149 | 531 | L-fuculose phosphate aldolase is not found |
| Lachnospiraceae bacterium 3_1_46FAA (665950.3) | | | | | | | | | | | | 1265 | 1171, 1642 | |
| Lachnospiraceae bacterium 3_1_57FAA_CT1 (658086.3) | 2094, 2095 | 2093 | | | | | | | 2092, 4203 | | 733, 734, 735, 885, 886, 887, 888 | 144, 719, 720, 721, 971, 976, 1188, 1239, 2928, 2941, 3811, 3832, 4202, 5393, 5779, 7036 | | L-fuculose phosphate aldolase is not found |
| Lachnospiraceae bacterium 4_1_37FAA (552395.3) | 1624 | 1625 | 1626 | | | | | | | | 1619, 1620, 1621, 1622 | 741, 1623 | | |
| Lachnospiraceae bacterium 5_1_57FAA (658085.3) | 146 | | | | | | | | | | 137, 138, 139 | 136 | | L-fuculokinase and L-fuculose phosphate aldolase are not found |
| Lachnospiraceae bacterium 5_1_63FAA (658089.3) | | | | | | | | | | | | | | |
| Lachnospiraceae bacterium 6_1_63FAA (658090.3) | | | | | | | | | | | | 1461 | | |
| Lachnospiraceae bacterium 7_1_58FAA (658087.3) | | | | | | | | | | | | | | |
| Lachnospiraceae bacterium 8_1_57FAA (665951.3) | | | | | | | | | | | | 1946 | 1188, 2069 | |
| Lachnospiraceae bacterium 9_1_43BFAA (658088.3) | 331 | 330 | 329 | | | | | | | | 334, 335, 336, 337 | 333, 1006 | | |
| Lactobacillus acidophilus ATCC 4796 (525306.3) | | | | | | | | | | | | | | |
| Lactobacillus acidophilus NCFM (272621.13) | | | | | | | | | | | | | | |
| Lactobacillus amylolyticus DSM 11664 (585524.3) | | | | | | | | | | | | | | |
| Lactobacillus antri DSM 16041 (525309.3) | | | | | | | | | | | | | | |
| Lactobacillus brevis ATCC 367 (387344.15) | | | | | | | | | | | | | | |
| Lactobacillus brevis subsp. gravesensis ATCC 27305 (525310.3) | | | | | | | | | | | | | | |
| Lactobacillus buchneri ATCC 11577 (525318.3) | | | | | | | | | | | | | | |
| Lactobacillus casei ATCC 334 (321967.11) | | | | | | | | | | | | 1797, 2695 | | |
| Lactobacillus casei BL23 (543734.4) | | | | | | | | | | | | 1938, 2699, 2806 | | |
| Lactobacillus crispatus 125-2-CHN (575595.3) | | | | | | | | | | | | | | |
| Lactobacillus delbrueckii subsp. bulgaricus ATCC 11842 (390333.7) | | | | | | | | | | | | | | |
| Lactobacillus delbrueckii subsp. bulgaricus ATCC BAA-365 (321956.7) | | | | | | | | | | | | | | |
| Lactobacillus fermentum ATCC 14931 (525325.3) | | | | | | | | | | | | | | |
| Lactobacillus fermentum IFO 3956 (334390.5) | | | | | | | | | | | | | | |
| Lactobacillus gasseri ATCC 33323 (324831.13) | | | | | | | | | | | | | | |
| Lactobacillus helveticus DPC 4571 (405566.6) | | | | | | | | | | | | | | |
| Lactobacillus helveticus DSM 20075 (585520.4) | | | | | | | | | | | | | | |
| Lactobacillus hilgardii ATCC 8290 (525327.3) | | | | | | | | | | | | | | |

| Organism | | | | | | | | | | | |
|---|---|---|---|---|---|---|---|---|---|---|---|
| Lactobacillus johnsonii NCC 533 (257314.6) | | | | | | | | | | | |
| Lactobacillus paracasei subsp. paracasei 8700:2 (537973.8) | | | | | | | | | | 844, 1584 | |
| Lactobacillus paracasei subsp. paracasei ATCC 25302 (525337.3) | | | | | | | | | | 678 | |
| Lactobacillus plantarum 16 (1327988.3) | | | | | | | | | | | |
| Lactobacillus plantarum subsp. plantarum ATCC 14917 (525338.3) | | | | | | | | | | | |
| Lactobacillus plantarum WCFS1 (220668.9) | | | | | | | | | | | |
| Lactobacillus reuteri CF48-3A (525341.3) | | | | | | | | | | | |
| Lactobacillus reuteri DSM 20016 (557436.4) | | | | | | | | | | | |
| Lactobacillus reuteri F275 (299033.6) | | | | | | | | | | | |
| Lactobacillus reuteri MM2-3 (585517.3) | | | | | | | | | | | |
| Lactobacillus reuteri MM4-1A (548485.3) | | | | | | | | | | | |
| Lactobacillus reuteri SD2112 (491077.3) | | | | | | | | | | | |
| Lactobacillus rhamnosus ATCC 21052 (1002365.5) | | | | | | | | | | 1776, 1907 | |
| Lactobacillus rhamnosus GG (Prj:40637) (568703.30) | 2700 | 2696, 2699 | | | | | 2697 | 2698 | | 1882, 2667, 2757 | |
| Lactobacillus rhamnosus LMS2-1 (525361.3) | | | | | | | | | | 810, 1104 | |
| Lactobacillus ruminis ATCC 25644 (525362.3) | | | | | | | | | | | |
| Lactobacillus sakei subsp. sakei 23K (314315.12) | | | | | | | | | | | |
| Lactobacillus salivarius ATCC 11741 (525364.3) | | | | | | | | | | | |
| Lactobacillus salivarius UCC118 (362948.14) | | | | | | | | | | | |
| Lactobacillus sp. 7_1_47FAA (665945.3) | | | | | | | | | | | |
| Lactobacillus ultunensis DSM 16047 (525365.3) | | | | | | | | | | | |
| Leuconostoc mesenteroides subsp. cremoris ATCC 19254 (586220.3) | | | | | | | | | | | |
| Listeria grayi DSM 20601 (525367.9) | | | | | | | | | | | |
| Listeria innocua ATCC 33091 (1002366.3) | | | | | | | | | | | |
| Megamonas funiformis YIT 11815 (742816.3) | 2445 | 608 | 2444 | | | | 609 | 610 | | | |
| Megamonas hypermegale ART12/1 (657316.3) | 725, 726 | 722, 723 | 724 | | | | 721 | 719, 720 | | | |
| Methanobrevibacter smithii ATCC 35061 (420247.6) | | | | | | | | | | | |
| Methanosphaera stadtmanae DSM 3091 (339860.6) | | | | | | | | | | | |
| Mitsuokella multacida DSM 20544 (500635.8) | | | | | | | | | | | |
| Mollicutes bacterium D7 (556270.3) | | | | | | | | | | | |
| Odoribacter laneus YIT 12061 (742817.3) | | | | | | | | | | 1543, 1594, 1703, 2078 | |
| Oxalobacter formigenes HOxBLS (556268.6) | | | | | | | | | | | |
| Oxalobacter formigenes OXCC13 (556269.4) | | | | | | | | | | | |
| Paenibacillus sp. HGF5 (908341.3) | 2703 | 2701 | 2702 | | | | 2700 | | 2819, 2820, 2821 | 740, 1967, 3557, 3558, 3943, 3994, 4098, 4856 | 5680 |
| Paenibacillus sp. HGF7 (944559.3) | 83 | 84 | 85 | | | | 3873 | | | 2441, 4089, 4594, 4601 | |
| Paenibacillus sp. HGH0039 (1078505.3) | 4212 | 4213 | 4214 | | | | 4211 | | | 3884, 3891, 4803, 5330 | |
| Parabacteroides distasonis ATCC 8503 (435591.13) | | | | | | | 98 | | | 2711 | |
| Parabacteroides johnsonii DSM 18315 (537006.5) | | | | | | | | 1589 | | 280, 401 | |
| Parabacteroides merdae ATCC 43184 (411477.4) | | | | | | | | | | 125, 263, 1035 | |
| Parabacteroides sp. D13 (563193.3) | | | | | | | 4086 | 4087 | | 4126 | |
| Parabacteroides sp. D25 (658661.3) | | | | | | | 4209 | 4210 | | 3962 | |
| Paraprevotella clara YIT 11840 (762968.3) | 2939 | 2941 | 2940 | | | | 2942 | 2943 | | 1622, 1993, 2945, 2973 | |
| Paraprevotella xylaniphila YIT 11841 (762982.3) | 897 | 895 | 896 | | | | 894 | 893 | | 863, 890, 891, 2026, 2122 | |
| Parvimonas micra ATCC 33270 (411465.10) | | | | | | | | | | | |
| Pediococcus acidilactici 7_4 (563194.3) | 761 | 760, 763 | | | | | 757 | 758, 759 | | 130, 766 | |
| Pediococcus acidilactici DSM 20284 (862514.3) | | | | | | | | | | 1263 | |
| Phascolarctobacterium sp. YIT 12067 (626939.3) | | | | | | | | | | | |
| Prevotella copri DSM 18205 (537011.5) | | | | | | | | | | | |

| Organism | | | | | | | | | | | | | | Notes |
|---|---|---|---|---|---|---|---|---|---|---|---|---|---|---|
| Prevotella oralis HGA0225 (1203550.3) | | | | | | | | | | | 910, 2398 | | | |
| Prevotella salivae DSM 15606 (888832.3) | | | | | | | | | | | 1319, 1561, 2450 | | | |
| Propionibacterium sp. 5_U_42AFAA (450748.3) | 1158, 1159 | | 1160 | | | | | | | 1149, 1154, 1156 | 1157 | | | L-fuculokinase is not found |
| Propionibacterium sp. HGH0353 (1203571.3) | | | | | | | | | | | | | | |
| Proteus mirabilis WGLW6 (1125694.3) | | | | | | | | | | | | | | |
| Proteus penneri ATCC 35198 (471881.3) | | | | | | | | | | | | | | |
| Providencia alcalifaciens DSM 30120 (520999.6) | | | | | | | | | | | | | | |
| Providencia rettgeri DSM 1131 (521000.6) | | | | | | | | | | | | | | |
| Providencia rustigianii DSM 4541 (500632.6) | | | | | | | | | | | | | | |
| Providencia stuartii ATCC 25827 (471874.6) | | | | | | | | | | | | | | |
| Pseudomonas sp. 2_1_26 (665948.3) | | | | | | | | | | | | | | |
| Ralstonia sp. 5_7_47FAA (658664.3) | | | | | | | | | | | | | | |
| Roseburia intestinalis L1-82 (536231.5) | | | | | | | | | | 2317 | | | | |
| Roseburia intestinalis M50/1 (657315.3) | | | | | | | | | | 3601 | | | | |
| Roseburia intestinalis XB6B4 (718255.3) | | | | | | | | | | 2479 | | | | |
| Roseburia inulinivorans DSM 16841 (622312.4) | 1515 | | 1531 | | | | | | 1522 | 1517, 1518, 1519, 1520, 1521, 1977, 1978, 1979, 1980 | 1693, 1976 | | | L-fuculokinase is not found |
| Ruminococcaceae bacterium D16 (552398.3) | | | | | | | | | | | | | | |
| Ruminococcus bromii L2-63 (657321.5) | | | | | | | | | | | | | | |
| Ruminococcus gnavus ATCC 29149 (411470.6) | 3426 | 929 | | | | | | | 931 | 922, 923, 924, 925, 926, 1098, 1099, 1100, 3314, 3315, 3316, 3317 | 2956, 3319 | 1101 | 717 | L-fuculose phosphate aldolase is not found |
| Ruminococcus lactaris ATCC 29176 (471875.6) | | | | | | | | | | | 1096 | | 1849 | |
| Ruminococcus obeum A2-162 (657314.3) | 1696 | | 1681 | | | | | | 1714 | | | | | L-fuculokinase is not found |
| Ruminococcus obeum ATCC 29174 (411459.7) | 3072 | | 3092 | | | | | | 3067 | 342, 343, 344 | 345 | | | L-fuculokinase is not found |
| Ruminococcus sp. 18P13 (213810.4) | | | | | | | | | | | | | | |
| Ruminococcus sp. 5_1_39BFAA (457412.4) | 1849 | | 1864 | | | | | | 3466 | 2466 | 2200, 2462 | | | L-fuculokinase is not found |
| Ruminococcus sp. SR1/5 (657323.3) | 1662 | | 1681 | | | | | | 3509 | 2568, 2569 | | | | L-fuculokinase is not found |
| Ruminococcus torques ATCC 27756 (411460.6) | | | | | | | | | | 1986 | | 1649, 2451 | | |
| Ruminococcus torques L2-14 (657313.3) | 724 | 723 | 583 | | | | | | 717 | 718, 719, 720, 721 | | | 678 | |
| Salmonella enterica subsp. enterica serovar Typhimurium str. LT2 (99287.12) | 3152 | 3153 | 3147 | | | | | | 3154 | 3150, 3151 | | | | |
| Staphylococcus sp. HGB0015 (1078083.3) | | | | | | | | | | | | | | |
| Streptococcus anginosus 1_2_62CV (742820.3) | | | | | | | | | | | | | | |
| Streptococcus equinus ATCC 9812 (525379.3) | | | | | | | | | | | | | | |
| Streptococcus infantarius subsp. infantarius ATCC BAA-102 (471872.6) | | | | | | | | | | | | | | |
| Streptococcus sp. 2_1_36FAA (469609.3) | | | | | | | | | | | 153, 1790 | | | |
| Streptococcus sp. HPH0090 (1203590.3) | | | | | | | | | | | 1211 | 1062 | | |
| Streptomyces sp. HGB0020 (1078086.3) | | | | 1140 | 1138 | 1142 | 1141 | 5070 | 1139 | 1143, 1144, 1145 | 1166, 1168, 1586 | | | |
| Streptomyces sp. HPH0547 (1203582.3) | | | | 986 | 985 | 988 | 987 | 2325 | | | 2155 | | | Transporters are not found |
| Subdoligranulum sp. 4_3_54A2FAA (665956.3) | | | | | | | | | | | | | | |
| Subdoligranulum variabile DSM 15176 (411471.5) | | | | | | | | | | | 2979 | | | |
| Succinatimonas hippei YIT 12066 (762983.3) | 1625 | | | | | | | | 1626 | | | | | |
| Sutterella wadsworthensis 3_1_45B (742821.3) | | | | | | | | | | | | | | |
| Sutterella wadsworthensis HGA0223 (1203554.3) | | | | | | | | | | | | | | |
| Tannerella sp. 6_1_58FAA_CT1 (665949.3) | | | | | | | | | | | 973 | | | |
| Turicibacter sp. HGF1 (910310.3) | | | | | | | | | | | 79, 1295 | | | |
| Turicibacter sp. PC909 (702450.3) | | | | | | | | | | | 516, 1320 | | | |
| Veillonella sp. 3_1_44 (457416.3) | | | | | | | | | | | | | | |
| Veillonella sp. 6_1_27 (450749.3) | | | | | | | | | | | | | | |
| Weissella paramesenteroides ATCC 33313 (585506.3) | | | | | | | | | | | | | | |
| Yokenella regensburgei ATCC 43003 (1002368.3) | 1436 | 1437 | 1434 | | | | | | 1438 | 1435 | | | | |

Table S4. Genes involved in depletion and utilization of N-acetylgalactosamine. The PubSEED identifiers are shown. For details on analyzed organisms see Table S1, for details on gene functions see Table S2.

| Organism (PubSEED ID) | Kinase I | Deacetylase | Isomerase | | Kinase II | Aldolase | Phospho-beta-galactosidase | Phosphorylase | PTS systems | | ABC system | Beta-hexosaminidases | | Endo-alpha-N-acetylgalactosaminidases | | Exo-alpha-N-acetylgalactosaminidase | Problems (if any) |
|---|---|---|---|---|---|---|---|---|---|---|---|---|---|---|---|---|---|
| | AgaK | AgaA | AgaS | AgaI | AgaZ | AgaY | GnbG | LnbP | AgaPTS | GnbPTS | LnbABC | Bha3 | Bha20 | Apr101 | Apr129 | AaeL | |
| Abiotrophia defectiva ATCC 49176 (592010.4) | | | | | | | 3045 | | | | 758, 759, 760, 3041, 3042, 3043 | | 2689 | 3047 | | | |
| Acidaminococcus sp. D21 (563191.3) | | | | | | | | | | | | | | | | | |
| Acidaminococcus sp. HPA0509 (1203555.3) | | | | | | | | | | | | | | | | | |
| Acinetobacter junii SH205 (575587.3) | | | | | | | | | | | | | | | | | |
| Actinomyces odontolyticus ATCC 17982 (411466.7) | | | | | 1018 | 1019 | | | | | | | | | | | |
| Actinomyces sp. HPA0247 (1203556.3) | | | | | | | | | | | | | 130 | | | | |
| Akkermansia muciniphila ATCC BAA-835 (349741.6) | | | | | | | | | | | | 2416 | 421, 451, 986, 1169, 1892, 2063, 2191, 2301, 2302, 2446, 2460 | | | | |
| Alistipes indistinctus YIT 12060 (742725.3) | | | | | | | | | | | | | 913, 2149, 2200 | | | | |
| Alistipes shahii WAL 8301 (717959.3) | | | | | | | | | | | | 3053, 3056, 3057 | 383, 2285, 3055, 3058, 3073 | | | | |
| Anaerobaculum hydrogeniformans ATCC BAA-1850 (592015.5) | | | | | 898 | | | | | | | | | | | | |
| Anaerococcus hydrogenalis DSM 7454 (561177.4) | | | | | | | | | | | | | | | | | |
| Anaerofustis stercorihominis DSM 17244 (445971.6) | | | | | | 413 | | | | | | | | | | | |
| Anaerostipes caccae DSM 14662 (411490.6) | | | | | 1040, 3509 | 746, 1039, 1941, 3511 | | | | | | 1081 | | | | | |
| Anaerostipes hadrus DSM 3319 (649757.3) | | | | | 1244 | 1242 | | | | | | 1225 | | | | | |
| Anaerostipes sp. 3_2_56FAA (665937.3) | | | | | 146, 2545, 2546 | 144, 429, 2543 | | | | | | 188 | | | | | |
| Anaerotruncus colihominis DSM 17241 (445972.6) | | | | | | | | | | | | | | | | | |
| Bacillus smithii 7_3_47FAA (665952.3) | | | | | 318 | 317 | | | | | | 2068 | | | | | |
| Bacillus sp. 7_6_55CFAA_CT2 (665957.3) | | | | | 1168 | | | | | | | | | | | | |
| Bacillus subtilis subsp. subtilis str. 168 (224308.113) | | | | | | | | | | 3350, 3351 | 167 | | | | | | |
| Bacteroides caccae ATCC 43185 (411901.7) | | | | | | | | | | | | 325, 2648 | 1016, 1017, 1021, 2633, 2643, 2667, 3165, 3362, 3410, 3423, 3502, 3508 | | | | |
| Bacteroides capillosus ATCC 29799 (411467.6) | | | | | | | | 848 | | | 850, 851, 852 | 1352 | | | | | |
| Bacteroides cellulosilyticus DSM 14838 (537012.5) | | | | | | | | | | | | 1267, 2062, 2066, 2067, 2104 | 37, 950, 1188, 2126, 2422, 3129 | | | | |
| Bacteroides clarus YIT 12056 (762984.3) | | | | | | | | | | | | 2102 | 1623, 2048, 3000 | | | | |
| Bacteroides coprocola DSM 17136 (470145.6) | | | | | | | | | | | | 46, 1649, 1996 | 463, 908, 914, 915, 2401, 3707 | | | | |
| Bacteroides coprophilus DSM 18228 (547042.5) | | | | | | | | | | | | 547, 1770, 2949 | 578, 782, 783, 896, 1099, 1246, 1558, 2137, 2288, 2913, 2914, 3123 | | | | |
| Bacteroides dorei DSM 17855 (483217.6) | | | | | | | | | | | | 415, 575 | 474, 630, 634, 636, 1087, 1401, 1441, 2459, 3166, 3335 | | | | |
| Bacteroides eggerthii 1_2_48FAA (665953.3) | | | | | | | | | | | | 3587, 3588 | 670, 1693 | | | | |
| Bacteroides eggerthii DSM 20697 (483216.6) | | | | | | | | | | | | 2645 | 352 | | | | |
| Bacteroides finegoldii DSM 17565 (483215.6) | | | | | | | | | | | | 3350, 3645 | 532, 1700, 2282, 2283, 3069 | | | | |
| Bacteroides fluxus YIT 12057 (763034.3) | | | | | | | | | | | | 3105, 3106 | 1188, 2029, 2231, 2698, 2717, 3256, 3264 | | | | |
| Bacteroides fragilis 3_1_12 (457424.5) | | | | | | | | | | | | 2412, 2413 | 387, 391, 393, 888, 952, 1609, 1929, 2578, 3292, 3295, 3302, 3599 | | | | |

| Organism | | | | | | | | | | | | | | | | | |
|---|---|---|---|---|---|---|---|---|---|---|---|---|---|---|---|---|---|
| Bacteroides fragilis 638R (862962.3) | | | | | | | | | | | 576, 577 | 391, 738, 1038, 1739, 1743, 1745, 2289, 3199, 3202, 3210, 3823, 4278 | | | | | |
| Bacteroides fragilis NCTC 9343 (272559.17) | | | | | | | | | | | 502, 503 | 321, 656, 943, 1778, 1782, 1784, 2181, 3119, 3122, 3130, 3685, 4157 | | | | | |
| Bacteroides fragilis YCH46 (295405.11) | | | | | | | | | | | 581, 582 | 405, 752, 1030, 1679, 1683, 1685, 2074, 3116, 3119, 3127, 3668, 4100 | | | | | |
| Bacteroides intestinalis DSM 17393 (471870.8) | | | | | | | | | | | 507, 512, 513, 1330 | 561, 789, 796, 797, 1715, 3500, 3952 | | | | | |
| Bacteroides ovatus 3_8_47FAA (665954.3) | | | | | | | | | | | 4523 | 1381, 2124, 2395, 2808, 3273, 4373, 4374, 4576, 4727, 4728 | | | | | |
| Bacteroides ovatus ATCC 8483 (411476.11) | | | | | | | | | | | 537, 2236, 2315 | 993, 994, 1351, 1728, 1872, 1873, 2004, 2490, 2619, 2959, 3245, 3718, 4371 | | | | | |
| Bacteroides ovatus SD CC 2a (702444.3) | | | | | | | | | | | 2617, 4819 | 17, 351, 2781, 3077, 3387, 3388, 3417, 3889, 4736, 4737 | | | | | |
| Bacteroides ovatus SD CMC 3f (702443.3) | | | | | | | | | | | 85, 4858 | 992, 1159, 1160, 2371, 2619, 2620, 3221, 3235, 4196, 4730, 4966 | | 4964 | | | |
| Bacteroides pectinophilus ATCC 43243 (483218.5) | | | | | | | | | | | | | | | | | |
| Bacteroides plebeius DSM 17135 (484018.6) | | | | | | | | | | | 38 | 130, 896, 897, 975, 1279, 2458, 2642 | | | | | |
| Bacteroides sp. 1_1_14 (469585.3) | | | | | | | | | | | 4779 | 199, 257, 792, 1057, 1142, 1145, 1146, 1358, 1402, 1640, 3200, 3206, 3532, 4136, 4666 | | | | | |
| Bacteroides sp. 1_1_30 (457387.3) | | | | | | | | | | | 4426 | 708, 1297, 1298, 1543, 3697, 3699, 4488, 4508, 5025, 5183 | | | | | |
| Bacteroides sp. 1_1_6 (469586.3) | | | | | | | | | | | 4596 | 134, 137, 138, 411, 1045, 2012, 2068, 2556, 2810, 2823, 3389, 3851, 4611, 4797, 4798, 4805 | | | | | |
| Bacteroides sp. 2_1_16 (469587.3) | | | | | | | | | | | 2732, 2733 | 591, 1325, 1888, 1892, 1894, 1999, 2563, 3001, 3009, 3012, 3739 | | | | | |
| Bacteroides sp. 2_1_22 (469588.3) | | | | | | | | | | | 577, 4138 | 466, 1602, 1680, 1681, 2803, 2804, 3590, 4190, 4551 | | | | | |
| Bacteroides sp. 2_1_33B (469589.3) | | | | | | | | | | | 930, 931, 2907 | 225, 808, 1625, 3100, 3882 | | | | | |

| Organism | | | | | | | | | | | | | | | A | B | | C | |
|---|---|---|---|---|---|---|---|---|---|---|---|---|---|---|---|---|---|---|---|
| Bacteroides sp. 2_1_56FAA (665938.3) | | | | | | | | | | | | | | | 551, 552 | 381, 808, 1112, 1929, 1933, 1935, 2408, 3206, 3781, 3784, 3793 | | | |
| Bacteroides sp. 2_1_7 (457388.5) | | | | | | | | | | | | | | | 2424 | 572, 1099, 1720, 3165, 3379, 4182 | | | |
| Bacteroides sp. 2_2_4 (469590.5) | | | | | | | | | | | | | | | 378, 1260 | 14, 149, 1391, 1392, 1963, 3277, 3278, 3733, 3966, 5483, 5484 | | 1394 | |
| Bacteroides sp. 20_3 (469591.4) | | | | | | | | | | | | | | | 1686, 4443 | 1415, 2485, 3124, 3326, 4195 | | | |
| Bacteroides sp. 3_1_19 (469592.4) | | | | | | | | | | | | | | | 2013 | 818, 1473, 2927, 3315, 3888, 4241 | | | |
| Bacteroides sp. 3_1_23 (457390.3) | | | | | | | | | | | | | | | 1840, 1913, 3075 | 1589, 2061, 2417, 2840, 3135, 3642, 4446, 4721, 4722, 4762 | | 2063 | |
| Bacteroides sp. 3_1_33FAA (457391.3) | | | | | | | | | | | | | | | 1242 | 912, 1190, 1713, 1715, 1719, 1868, 2842, 3673, 3737 | | | |
| Bacteroides sp. 3_1_40A (469593.3) | | | | | | | | | | | | | | | 3727, 3735 | 234, 401, 832, 1025, 1946, 2532, 2838, 3385, 3389, 3391, 3790, 3791 | | | |
| Bacteroides sp. 3_2_5 (457392.3) | | | | | | | | | | | | | | | 604, 605 | 455, 797, 1514, 1967, 1970, 1978, 2336, 3149, 4354, 4356, 4360 | | | |
| Bacteroides sp. 4_1_36 (457393.3) | | | | | | | | | | | | | | | 33, 34, 38, 894, 895 | 268, 1837, 2096, 3003, 3266 | | | |
| Bacteroides sp. 4_3_47FAA (457394.3) | | | | | | | | | | | | | | | 1545 | 550, 765, 769, 771, 1365, 1489, 1710, 2458, 2952, 4574 | | | |
| Bacteroides sp. 9_1_42FAA (457395.6) | | | | | | | | | | | | | | | 636 | 420, 422, 426, 572, 2484, 2735, 3470, 3724, 3995 | | | |
| Bacteroides sp. D1 (556258.5) | | | | | | | | | | | | | | | 429, 3517 | 304, 1565, 1566, 1810, 3291, 3292, 3465, 3765, 4450 | | | |
| Bacteroides sp. D2 (556259.3) | | | | | | | | | | | | | | | 2130, 4854 | 392, 393, 1206, 1647, 2103, 2592, 3265, 3716, 3717, 3971, 4792, 5047 | | | |
| Bacteroides sp. D20 (585543.3) | | | | | | | | | | | | | | | 1651, 1652, 1802, 1803, 1807 | 62, 292, 1194, 2336, 2558 | | | |
| Bacteroides sp. D22 (585544.3) | | | | | | | | | | | | | | | 1093, 2778, 3070 | 145, 146, 1092, 1094, 1095, 1315, 1319, 1320, 1907, 2835, 3302, 3483, 3778, 4367 | | | |
| Bacteroides sp. HPS0048 (1078089.3) | | | | | | | | | | | | | | | 1104, 1105 | 5, 729, 730, 1199, 1200, 1558, 2876, 2942, 2943, 2953, 3179, 3943, 3948, 3967 | | | |
| Bacteroides stercoris ATCC 43183 (449673.7) | | | | | | | | | | | | | | | 379 | 548, 1455, 1461, 2245, 2379, 2694, 3236 | | | |

| Organism | | | | | | | | | | | | | |
|---|---|---|---|---|---|---|---|---|---|---|---|---|---|
| Bacteroides thetaiotaomicron VPI-5482 (226186.12) | | | | | | | | 2501 | 191, 456, 459, 460, 507, 1068, 1659, 1665, 2517, 3240, 3657, 3930, 4413, 4472, 4761 | | | | |
| Bacteroides uniformis ATCC 8492 (411479.10) | | | | | | | | 689, 690, 3474, 3475, 3478 | 43, 999, 1376, 1724, 3172 | | | | |
| Bacteroides vulgatus ATCC 8482 (435590.9) | | | | | | | | 150 | 89, 404, 552, 1244, 2282, 3342, 4247, 4249, 4253 | | | | |
| Bacteroides vulgatus PC510 (702446.3) | | | | | | | | 1436 | 216, 649, 709, 763, 765, 769, 1958, 2250, 2425, 2896 | | | | |
| Bacteroides xylanisolvens SD CC 1b (702447.3) | | | | | | | | 2034, 4399 | 129, 130, 350, 1720, 1721, 2100, 3926, 3927, 4163, 4692 | | | | |
| Bacteroides xylanisolvens XB1A (657309.4) | | | | | | | | 520, 984 | 73, 74, 573, 1113, 2296, 3256, 3618, 3931, 4504, 4505 | | | | |
| Barnesiella intestinihominis YIT 11860 (742726.3) | | 942 | 941 | | | | | 2312 | 101, 308, 379, 380, 513, 2614, 2835, 2904, 2964, 2975 | | | | |
| Bifidobacterium adolescentis L2-32 (411481.5) | | | | | | | | 1012 | | | | | |
| Bifidobacterium angulatum DSM 20098 = JCM 7096 (518635.7) | | | | | | | | 1101 | | | | | |
| Bifidobacterium animalis subsp. lactis AD011 (442563.4) | | | | | | | 490, 492 | 1067 | | | | | |
| Bifidobacterium bifidum NCIMB 41171 (398513.5) | | | | | 481, 905 | | 482, 483, 484 | 133 | 441, 1540, 1769 | 1379 | | | |
| Bifidobacterium breve DSM 20213 = JCM 1192 (518634.7) | | | | | 1473 | | 524, 527, 1469, 1471, 1472 | 744, 745 | 1506 | 977 | | | |
| Bifidobacterium breve HPH0326 (1203540.3) | | | | | 1077 | | 507, 509, 510, 612, 613, 615, 616, 1078, 1079, 1081 | 1986, 1987 | | | | | |
| Bifidobacterium catenulatum DSM 16992 = JCM 1194 (566552.6) | | | | | | | | 948 | | | | | |
| Bifidobacterium dentium ATCC 27678 (473819.7) | | | | | | | | 75 | | | | | |
| Bifidobacterium gallicum DSM 20093 (561180.4) | | | | | | | | | | | | | |
| Bifidobacterium longum DJO10A (Prj:321) (205913.11) | | | | | 1835 | | 918, 923, 1022, 1023, 1837, 1838, 1839 | 359 | 1253 | 125 | | | |
| Bifidobacterium longum NCC2705 (206672.9) | | | | | 1696 | | 875, 880, 995, 997, 1692, 1693, 1694 | 735 | 59 | 520 | | | |
| Bifidobacterium longum subsp. infantis 157F (565040.3) | | | | | 1729 | | 359, 360, 361, 1731, 1732, 1733 | 625, 1342 | 1471 | 938 | | | |
| Bifidobacterium longum subsp. infantis ATCC 15697 = JCM 1222 (391904.8) | | | | | 2251 | | 909, 910, 911, 2252, 2253, 2254 | 1915 | 471, 748, 2431 | 1541 | | | |
| Bifidobacterium longum subsp. infantis ATCC 55813 (548480.3) | | | | | 648 | | 645, 646, 647 | 104 | 892 | 12 | | | |
| Bifidobacterium longum subsp. infantis CCUG 52486 (537937.5) | | | | | 1177 | | 1173, 1174, 1175 | 12 | 2008 | 1133 | | | |
| Bifidobacterium longum subsp. longum 2-2B (1161745.3) | | | | | 514 | | 515, 516, 517, 838, 839 | 2072 | 1617 | | | | |
| Bifidobacterium longum subsp. longum 44B (1161743.3) | | | | | 1601 | | 103, 104, 105, 1598, 1599, 1600 | 1807 | 1528 | 189 | | | |

| | | | | | | | | | | | | | | |
|---|---|---|---|---|---|---|---|---|---|---|---|---|---|---|
| Bifidobacterium longum subsp. longum F8 (722911.3) | | | | | | 1599 | | 858, 859, 860, 1600, 1601, 1602 | 602 | 91 | | | 390 | |
| Bifidobacterium longum subsp. longum JCM 1217 (565042.3) | | | | | | 1679 | | 458, 1680, 1681, 1682 | 618 | 1443 | 822 | | | |
| Bifidobacterium pseudocatenulatum DSM 20438 = JCM 1200 (547043.8) | | | | | | | | | 1008 | 1742 | | | | |
| Bifidobacterium sp. 12_1_47BFAA (469594.3) | | | | | | 583 | | 475, 476, 585, 586, 587 | 277 | 78 | 1871 | | | |
| Bilophila wadsworthia 3_1_6 (563192.3) | | | | | | | | | 1035 | | | | | |
| Blautia hansenii DSM 20583 (537007.6) | | | | | | 1753 | | 811, 812, 813, 1757, 1758 | 1035 | | | | | |
| Blautia hydrogenotrophica DSM 10507 (476272.5) | | | | 1033 | | 583 | | 585, 586, 587 | 646 | | | | | |
| Bryantella formatexigens DSM 14469 (478749.5) | | | | | | 554 | | 550, 555, 556, 557 | 1695, 1696 | 279 | | | | |
| Burkholderiales bacterium 1_1_47 (469610.4) | | | | | | | | | | | | | | |
| butyrate-producing bacterium SM4/1 (245012.3) | | | | | | | | | 2299 | | | | | |
| butyrate-producing bacterium SS3/4 (245014.3) | | | | | | | | | | | | | | |
| butyrate-producing bacterium SSC/2 (245018.3) | | | | 181 | 179, 2271 | | | | 131 | | | | | |
| Butyricicoccus pullicaecorum 1.2 (1203606.4) | 2195 | 2198 | | 2192 | 2193 | | | 2199, 2201, 2202, 2203 | 2205 | | | | | |
| Butyrivibrio crossotus DSM 2876 (511680.4) | | | | | | | | | | | | | | |
| Butyrivibrio fibrisolvens 16/4 (657324.3) | | | | | | | | 2729, 2730, 2731 | 1716 | | | | | |
| Campylobacter coli JV20 (864566.3) | | | | | | | | | | | | | | |
| Campylobacter upsaliensis JV21 (888826.3) | | | | | | | | | | | | | | |
| Catenibacterium mitsuokai DSM 15897 (451640.5) | | | | | | | | | 1730 | 704, 705, 706, 1150, 1856 | | | | |
| Cedecea davisae DSM 4568 (566551.4) | | 215 | | 220 | 214 | | | | | | | 213 | | |
| Citrobacter freundii 4_7_47CFAA (742730.3) | | | | 1446 | 934, 935, 1445 | | | | | | | | | |
| Citrobacter sp. 30_2 (469595.3) | 2948 | 2949 | | 2943 | 2950, 3862, 3863 | | | 2944, 2945, 2946, 2947 | 4080 | | | | | |
| Citrobacter youngae ATCC 29220 (500640.5) | 3813 | 3814 | 3819 | 3808 | 141, 142, 3815 | | | 3809, 3810, 3811, 3812 | | | | | | |
| Clostridiales bacterium 1_7_47FAA (457421.5) | | | | | 2378 | 4154 | | 4156, 4157, 4158 | 1584, 2883, 3168 | | | | | |
| Clostridium asparagiforme DSM 15981 (518636.5) | | | | | | | | | 1805, 2939, 3111 | | | | | |
| Clostridium bartlettii DSM 16795 (445973.7) | | | | | 2253 | | | | 139, 323 | | | | | |
| Clostridium bolteae 90A5 (997893.5) | | | | | | 4411 | | 4408, 4409, 4410 | 3453, 3548 | | | | | |
| Clostridium bolteae ATCC BAA-613 (411902.9) | | | | | 140, 147 | 1483 | | 1480, 1481, 1482 | 5814, 5902 | | | | | |
| Clostridium celatum DSM 1785 (545697.3) | 577 | 578 | | 585 | 574 | 2573, 2574 | 582, 583, 584 | 2577, 2578, 2579 | 163, 1196 | 2632 | 1671, 3216 | | | |
| Clostridium cf. saccharolyticum K10 (717608.3) | | | | | | | | | 3073 | | | | | |
| Clostridium citroniae WAL-17108 (742733.3) | | | | | | 2512 | | | 132, 5679 | | | | | |
| Clostridium clostridioforme 2_1_49FAA (742735.4) | | | | | 4527 | 2652 | | 2653, 2655, 2656 | 2087 | | | | | No transporters found |
| Clostridium difficile CD196 (645462.3) | 3389 | 3385 | | 3387 | 2994, 3384 | | | | | | | | | No transporters found |
| Clostridium difficile NAP07 (525258.3) | 513 | 509 | | 511 | 508, 1052 | | | | | | | | | No transporters found |
| Clostridium difficile NAP08 (525259.3) | 3486 | 3482 | | 3484 | 1166, 3481 | | | | | | | | | |
| Clostridium hathewayi WAL-18680 (742737.3) | | | | | | 4106, 4107 | | 4108, 4109, 4110 | 960, 1448, 1593, 3978, 4273 | | | | | |
| Clostridium hiranonis DSM 13275 (500633.7) | | | | 1060, 1113 | 1058 | | | | 1544, 1546, 1549 | | | | | |
| Clostridium hylemonae DSM 15053 (553973.6) | | | | 1189 | 1192 | 3210 | | | 3213, 3214, 3215 | | | | | |
| Clostridium leptum DSM 753 (428125.8) | | | | 5 | 9 | | | | 744 | | | | | |
| Clostridium methylpentosum DSM 5476 (537013.3) | | | | | | | | | | | | | | |

| Organism | | | | | | | | | | | | | | |
|---|---|---|---|---|---|---|---|---|---|---|---|---|---|---|
| Clostridium nexile DSM 1787 (500632.7) | | | | | | 1721 | | 1712, 1713, 1714, 3026, 3027, 3028 | 2441 | 613, 614, 1955 | | | | |
| Clostridium perfringens WAL-14572 (742739.3) | | 2682 | | 1913 | 2683 | 1927 | 2689 | 1921, 1922, 1923 | 2776 | 679, 1027 | | 930 | | |
| Clostridium ramosum DSM 1402 (445974.6) | | | | | | 1039 | | 1030, 1034, 1035, 1036, 1038 | 741, 1205, 1472, 1631, 2275 | 1501, 1517, 1518 | | | | |
| Clostridium scindens ATCC 35704 (411468.9) | | | | | 2580 | 2577 | 2966 | 117, 118, 119, 2960, 2961, 2962 | | | | | | |
| Clostridium sp. 7_2_43FAA (457396.3) | 2038 | 2040 | | 2043 | 1396, 2042 | 3457 | 2032, 2033, 2034, 2037 | 3455, 3456, 3459 | 87, 117, 2545 | 69, 139, 2261 | | | | |
| Clostridium sp. 7_3_54FAA (665940.3) | | | | | | 3618 | | | 4282 | | | | | |
| Clostridium sp. D5 (556261.3) | | | | 2836, 4279 | 2308, 2832, 4277 | 3342, 819 | | 820, 821, 822, 2033, 2034, 2037, 2038, 2039 | | | | | | |
| Clostridium sp. HGF2 (908340.3) | | 3550 | 3118 | 2391 | | 1474, 2964 | | 2962, 2965, 2966, 2967 | | | | | | |
| Clostridium sp. L2-50 (411489.7) | | | | | | | | | 1103, 2279 | | | | | |
| Clostridium sp. M62/1 (411486.3) | | | | | | | | | 883 | | | | | |
| Clostridium sp. SS2/1 (411484.7) | | | | 1976 | 1178, 1974 | | | | 1928 | | | | | |
| Clostridium spiroforme DSM 1552 (428126.7) | 726 | 725 | | 728 | 729 | | 718, 719, 720, 724 | | 110, 632, 1502, 1503 | 436, 1806, 1927, 2020, 2021 | | 471 | | |
| Clostridium sporogenes ATCC 15579 (471871.7) | | | | | 1607 | | | | | | | | | |
| Clostridium symbiosum WAL-14163 (742740.3) | | | | | | 366 | | | 2861 | | | | | |
| Clostridium symbiosum WAL-14673 (742741.3) | | | | | | 150 | | | 3939 | | | | | |
| Collinsella aerofaciens ATCC 25986 (411903.6) | 2045, 2053 | 2052 | | 791, 2056 | 799 | 2047 | 2048, 2049, 2050, 2059 | | | | | | | |
| Collinsella intestinalis DSM 13280 (521003.7) | 399 | 400 | | 384 | 381 | | 393, 394, 402, 403, 404, 406 | | | 769 | | | | |
| Collinsella stercoris DSM 13279 (445975.6) | 135 | 134 | | 159 | 161, 776 | 127 | 150 | 128, 129, 130, 132 | 363, 1473 | 284, 945, 1661 | | 672 | | |
| Collinsella tanakaei YIT 12063 (742742.3) | 743 | 744 | | 741 | 704 | | 746, 747, 748, 750 | | | | | | | |
| Coprobacillus sp. 29_1 (469596.3) | | | | 704 | 272, 705, 1430 | 1527 | | 1528, 1530, 1531, 1532 | | 242 | | | | |
| Coprobacillus sp. 3_3_56FAA (665941.3) | | | | | | 1492 | | 1483, 1487, 1488, 1489, 1491 | 2379, 2856, 3052 | 940, 941, 946, 3081, 3097, 3098 | | | | |
| Coprobacillus sp. 8_2_54BFAA (469597.4) | | | | | | 596 | | 586, 591, 592, 593, 595 | 1455, 2774 | 3167, 3174, 3175, 3518, 3519, 3535, 3563 | | | | |
| Coprobacillus sp. D6 (556262.3) | | | | 1159 | 1158, 1407, 3455 | 3551 | | 3552, 3554, 3555, 3556 | | 1437 | | | | |
| Coprococcus catus GD/7 (717962.3) | | | | | | | | | 894 | | | | | |
| Coprococcus comes ATCC 27758 (470146.3) | | | | | | | | 1105, 1106, 1107, 1108 | | | | | | |
| Coprococcus eutactus ATCC 27759 (411474.6) | | | | | | | | | 1610 | | | | | |
| Coprococcus sp. ART55/1 (751585.3) | | | | | | | | | 1312 | | | | | |
| Coprococcus sp. HPP0048 (1078091.3) | | | | | | 2284 | | 2290, 2291, 2292 | 2618 | | | | | |
| Coprococcus sp. HPP0074 (1078090.3) | | | | | | 2064 | | 2070, 2071, 2072 | 2765, 2846 | | | | | |
| Corynebacterium ammoniagenes DSM 20306 (649754.3) | | | | | | | | | 1483 | | | | | |
| Corynebacterium sp. HFH0082 (1078764.3) | | | | | | | | | | | | | | |
| Dermabacter sp. HFH0086 (1203568.3) | | | | | | | | | | | | | | |

| | | | | | | | | | | | |
|---|---|---|---|---|---|---|---|---|---|---|---|
| Desulfitobacterium hafniense DP7 (537010.4) | | | | 3905, 4278 | | | | | 2987 | | |
| Desulfovibrio piger ATCC 29098 (411464.8) | | | | | | | | | 2651 | | |
| Desulfovibrio sp. 3_1_syn3 (457398.5) | | | | | | | | | | | |
| Dorea formicigenerans 4_6_53AFAA (742765.5) | | | | | | | | | | | |
| Dorea formicigenerans ATCC 27755 (411461.4) | | | | | | | | | 351 | | |
| Dorea longicatena DSM 13814 (411462.6) | | | | | | | | | | | |
| Dysgonomonas gadei ATCC BAA-286 (742766.3) | | | | | | | | | | 2324, 3483 | 1924, 4134 |
| Dysgonomonas mossii DSM 22836 (742767.3) | | | | | | | | | | 48, 1500 | 1263, 1693 |
| Edwardsiella tarda ATCC 23685 (500638.3) | 1690 | 1695, 2062 | 1696 | 1685, 2067 | | | 1691, 1692, 1693, 1694 | | | 1775 | |
| Eggerthella sp. 1_3_56FAA (665943.3) | | | | | | | | | | | |
| Eggerthella sp. HGA1 (910311.3) | | | | | | | | | | | |
| Enterobacter cancerogenus ATCC 35316 (500639.8) | 3741 | 3742 | 3736 | 3743 | | | 3737, 3738, 3739 | | | 4025 | |
| Enterobacter cloacae subsp. cloacae NCTC 9394 (718254.4) | 2633 | 2634 | 2631 | 2635 | | | 2632 | | | 3658 | |
| Enterobacteriaceae bacterium 9_2_54FAA (469613.3) | | 33, 2719 | 32 | 38, 2788 | | | | | | 364 | |
| Enterococcus faecalis PC1.1 (702448.3) | | | 655, 1367, 2507 | 2508 | | | | | | | |
| Enterococcus faecalis TX0104 (491074.3) | | 440 | 438, 1218, 1241 | 439, 1221 | 437 | | 433, 434, 435, 436 | | 2801 | | |
| Enterococcus faecalis TX1302 (749513.3) | | 2419 | 1698, 1722, 2417 | 2418 | 2416 | | 2412, 2413, 2414, 2415 | | 506 | | |
| Enterococcus faecalis TX1322 (525278.3) | 2212 | 294 | 297, 1533, 1567, 1931 | 295, 296, 1536, 1932 | 298 | | 2210, 2211, 2213, 2214 | 299, 300, 301, 302 | 2575 | | |
| Enterococcus faecalis TX1341 (749514.3) | 2012 | 1915 | 1917, 2357 | 1916, 2354 | 1918 | | 2010, 2011, 2013, 2014 | 1919, 1920, 1921, 1922 | 1112 | 1923 | |
| Enterococcus faecalis TX1342 (749515.3) | 1078 | 2522 | 1315, 1339, 2520 | 2521 | 2519 | | 1076, 1077, 1079, 1080 | 2515, 2516, 2517, 2518 | | | |
| Enterococcus faecalis TX1346 (749516.3) | 1472 | 1833 | 950, 1835 | 947, 1834 | 1836 | | 1470, 1471, 1473, 1474 | 1837, 1838, 1839, 1840 | 2280 | | |
| Enterococcus faecalis TX2134 (749518.3) | 2180 | 1250 | 290, 315, 1252 | 312, 1251 | 1253 | | 2178, 2179, 2181, 2182 | 1254, 1255, 1256, 1257 | 1827 | | |
| Enterococcus faecalis TX2137 (749491.3) | 471 | 309 | 307, 2280, 2304 | 308, 2301 | 306 | | 469, 470, 472, 473 | 302, 303, 304, 305 | 1356 | | |
| Enterococcus faecalis TX4244 (749494.3) | 160 | 1655 | 1653, 2031 | 1654 | 1652 | | 158, 159, 161, 162 | 1648, 1649, 1650, 1651 | 1917 | 1647 | |
| Enterococcus faecium PC4.1 (791161.5) | | | 671, 1416, 2587 | 2588 | | | | | | | |
| Enterococcus faecium TX1330 (525279.3) | | | 391, 1613, 1732 | 1614, 1733 | | | | | | | |
| Enterococcus saccharolyticus 30_1 (742813.4) | | 1807 | 869, 1366, 1808, 1850 | 872, 1809 | | | 1810, 1811, 1812, 1813 | | 981 | | |
| Enterococcus sp. 7L76 (657310.3) | 2028 | 244 | 241, 242, 1664 | 243, 1661 | 240 | | 2026, 2027, 2029, 2030 | 236, 237, 238, 239 | | | |
| Erysipelotrichaceae bacterium 2_2_44A (457422.3) | | 2843 | 4656 | 4296 | | 1268, 4021 | 3495, 3496, 3497, 3500 | | | | |
| Erysipelotrichaceae bacterium 21_3 (658657.3) | | 3476 | 1871 | 4471 | | 3498, 531 | 4019, 4022, 4023, 4024 | | | | |
| Erysipelotrichaceae bacterium 3_1_53 (658659.3) | | 3323 | 2681 | 4692 | | 1540, 305 | 1537, 1538, 1539, 1542 | | | | |
| Erysipelotrichaceae bacterium 5_2_54FAA (552396.3) | | 682 | | 1672 | | 2260, 2266 | 2257, 2258, 2261, 2262, 2263 | | 2652 | | |
| Erysipelotrichaceae bacterium 6_1_45 (469614.3) | | 805 | 3243 | 2178 | | 1202, 4306 | 1200, 1203, 1204, 1205 | | | | |

| Strain | | | | | | | | |
|---|---|---|---|---|---|---|---|---|
| Escherichia coli 4_1_47FAA (1127356.4) | 2226 | 2227 | 2233 | 2221, 3583 | 2228, 2229, 3582 | | | 2222, 2223, 2224, 2225 |
| Escherichia coli MS 107-1 (679207.4) | 2599 | 2598 | 2593 | 2604, 3851 | 2597, 3850 | | | 2600, 2601, 2602, 2603 |
| Escherichia coli MS 115-1 (749537.3) | 493 | 492 | 2524 | 498, 874, 875 | 491, 873, 1412 | | | 494, 495, 496, 497 |
| Escherichia coli MS 116-1 (749538.3) | 990 | 992 | 2176 | 985, 1312, 1313 | 993, 1041, 1314 | | | 986, 987, 988, 989 |
| Escherichia coli MS 119-7 (679206.4) | 1516 | 1517 | 1522 | 1510, 1576 | 1518, 1575 | | | 1511, 1512, 1513, 1514, 1515 |
| Escherichia coli MS 124-1 (679205.4) | 353 | 352 | 347 | 358, 2955 | 351, 2954 | | | 354, 355, 356, 357 |
| Escherichia coli MS 145-2 (679204.3) | 870 | 871 | 876 | 123, 865 | 124, 872 | | | 866, 867, 868, 869 |
| Escherichia coli MS 146-1 (749540.3) | 1422 | 1424 | 1429 | 17, 1417 | 16, 1425 | | | 1418, 1419, 1420, 1421 |
| Escherichia coli MS 175-1 (749544.3) | 761 | 763 | 924 | 756 | 764 | | | 757, 758, 759, 760 |
| Escherichia coli MS 182-1 (749545.3) | 3371 | 3372 | 3377 | 1678, 3366 | 1679, 3373 | | | 3367, 3368, 3369, 3370 |
| Escherichia coli MS 185-1 (749546.3) | 2090 | 2089 | 2084 | 2095, 2853 | 1418, 1419, 1830, 2088, 2854, 3978 | | | 2091, 2092, 2093, 2094 |
| Escherichia coli MS 187-1 (749547.3) | 1094 | 1093 | | 1099, 3102 | 1092, 1100, 3101 | | | 1095, 1096, 1097, 1098 |
| Escherichia coli MS 196-1 (749548.3) | 2788 | 2786 | 2781 | 2793, 2946, 2947, 2948 | 2785, 2949 | | | 2789, 2790, 2791, 2792 |
| Escherichia coli MS 198-1 (749549.3) | 1971 | 1972 | 2024 | 1966, 5003 | 1973, 5002 | | | 1967, 1968, 1969, 1970 |
| Escherichia coli MS 200-1 (749550.3) | 1058 | 1060 | 1065 | 1053, 1569 | 301, 1061, 1312, 1568, 2297, 2298 | | | 1054, 1055, 1056, 1057 |
| Escherichia coli MS 21-1 (749527.3) | 2784, 2785 | 2786 | 2791 | 1495, 2779 | 1494, 2178, 2787, 3902, 3903 | | | 2780, 2781, 2782, 2783 |
| Escherichia coli MS 45-1 (749528.3) | 4101 | 4102, 4315 | 4107 | 1587, 4096 | 1586, 2935, 2936, 3037, 4103, 4217, 4313, 4314 | | | 4097, 4098, 4099, 4100 |
| Escherichia coli MS 69-1 (749531.3) | 2559 | 2560 | 2565 | 2055, 2554 | 2054, 2561 | | | 2555, 2556, 2557, 2558 |
| Escherichia coli MS 78-1 (749532.3) | 1630 | 1631 | 1636 | 1625, 4086 | 1632, 4085 | | | 1626, 1627, 1628, 1629 |
| Escherichia coli MS 79-10 (796392.4) | 1902 | 1903 | 1908 | 1421, 1897 | 1422, 1904 | | | 1898, 1899, 1900, 1901 |
| Escherichia coli MS 84-1 (749533.3) | 3485 | 3484 | 2282 | 419, 3490 | 418, 3483 | | | 3486, 3487, 3488, 3489 |

| Organism | | | | | | | | | | | | | |
|---|---|---|---|---|---|---|---|---|---|---|---|---|---|
| Escherichia coli MS 85-1 (679202.3) | 1017 | 1018 | 1023 | 1012, 2581 | 1019, 2580 | | | 1013, 1014, 1015, 1016 | | | | | |
| Escherichia coli O157:H7 str. Sakai (386585.9) | 4190 | 4192 | 4197, 4198 | 3030, 4185 | 3031, 4193, 4548 | | | 4186, 4187, 4188, 4189 | | | | | |
| Escherichia coli SE11 (409438.11) | 3590 | 3591 | 3596 | 2516, 3585 | 2517, 3592 | | | 3586, 3587, 3588, 3589 | | | | | |
| Escherichia coli SE15 (431946.3) | 3107 | 3108 | 3113 | 2063, 3102 | 2064, 3109, 3221, 3641, 3642 | | | 3103, 3104, 3105, 3106 | | | | | |
| Escherichia coli str. K-12 substr. MG1655 (511145.12) | 3229 | 3231 | 3236 | 2172, 3226 | 2173, 3232 | | | 3227, 3228 | | | | | |
| Escherichia coli UTI89 (364106.8) | 3533 | 3534 | 3539 | 1176, 2397, 3528 | 1174, 1175, 2398, 3535, 3647, 3921, 4110, 4111 | | | 3529, 3530, 3531, 3532 | | | | | |
| Escherichia sp. 1_1_43 (457400.3) | | | | | | | | | | | | | |
| Escherichia sp. 3_2_53FAA (469598.5) | 1077 | 1076 | 1071 | 1082, 2678 | 962, 1075, 2679, 4279, 4486, 4487 | | | 1078, 1079, 1080, 1081 | | | | | |
| Escherichia sp. 4_1_40B (457401.3) | 3052 | 3050 | 3045 | 2059, 3057 | 2058, 2554, 3049 | | | 3053, 3054, 3055, 3056 | | | | | |
| Eubacterium biforme DSM 3989 (518637.5) | | 21 | | 29 | | | 1074 | | 1075, 1077, 1078, 1079 | | | | | |
| Eubacterium cylindroides T2-87 (717960.3) | | | | | | | 1442, 1443 | | 1438, 1439, 1441 | | | | | |
| Eubacterium dolichum DSM 3991 (428127.7) | | 825, 1984 | | 1786 | | | | | | 44, 1823, 1958 | 721, 1589 | | 42 | |
| Eubacterium hallii DSM 3353 (411469.3) | | | | | | | | | | 1337 | | | | |
| Eubacterium rectale DSM 17629 (657318.4) | | | | | | | 2888 | | 2884, 2885, 2886 | | 2338 | | | |
| Eubacterium rectale M104/1 (657317.3) | | | | | | | 959 | | 955, 956, 957, 2443, 2444, 2445 | | 1569 | | | |
| Eubacterium siraeum 70/3 (657319.3) | | | | | | | | | | | 2364 | | | |
| Eubacterium siraeum DSM 15702 (428128.7) | | | | | | | | | | | 2382 | | | |
| Eubacterium siraeum V10Sc8a (717961.3) | | | | | | | | | | | | | | |
| Eubacterium sp. 3_1_31 (457402.3) | | 2428 | | 1386 | | | 231, 238 | | 235, 236, 237, 240 | | 1992 | | | |
| Eubacterium ventriosum ATCC 27560 (411463.4) | | | | | | | | | | | | | | |
| Faecalibacterium cf. prausnitzii KLE1255 (748224.3) | | 1442 | | | | | 2586 | | 2582, 2583, 2584 | | | | | |
| Faecalibacterium prausnitzii A2-165 (411483.3) | | | | | | | 1155 | | 1151, 1152, 1153 | | 152 | | | |
| Faecalibacterium prausnitzii L2-6 (718252.3) | 1293 | 1294 | 1292 | 1297 | | | 113 | | 1295, 1300, 1301, 1302 | 115, 116, 117, 3087, 3088, 3089 | | | | |
| Faecalibacterium prausnitzii M21/2 (411485.10) | | | | | | | 2375 | | 2377, 2378, 2379 | | | | | |
| Faecalibacterium prausnitzii SL3/3 (657322.3) | | | | | | | 2621 | | 2617, 2618, 2619 | | | | | |
| Fusobacterium gondiiformans ATCC 25563 (469615.3) | | | | | | | | | | | | | | |
| Fusobacterium mortiferum ATCC 9817 (469616.3) | 506 | | 507 | 509 | 508, 741 | | | | | | | | | |
| Fusobacterium necrophorum subsp. funduliforme 1_1_36S (742814.3) | | | | | | | | | | | | | | |
| Fusobacterium sp. 1_1_41FAA (469621.3) | | | | | | | | | | | | | | |
| Fusobacterium sp. 11_3_2 (457403.3) | | | | | | | | | | | | | | |
| Fusobacterium sp. 12_1B (457404.3) | 3263 | | 3262 | 3260 | 3261 | | | | | | | | | |
| Fusobacterium sp. 2_1_31 (469599.3) | | | | | | | | | | | | | | |
| Fusobacterium sp. 21_1A (469601.3) | | | | | | | | | | | | | | |
| Fusobacterium sp. 3_1_27 (469602.3) | | | | | | | | | | | | | | |
| Fusobacterium sp. 3_1_33 (469603.3) | | | | | | | | | | | | | | |
| Fusobacterium sp. 3_1_36A2 (469604.3) | | | | | | | | | | | | | | |
| Fusobacterium sp. 3_1_5R (469605.3) | | | | | | | | | | | | | | |
| Fusobacterium sp. 4_1_13 (469606.3) | | | | | | | | | | | | | | |
| Fusobacterium sp. 7_1 (457405.3) | | | | | | | | | | | | | | |
| Fusobacterium sp. D11 (556264.3) | | | | | | | | | | | | | | |
| Fusobacterium sp. D12 (556263.3) | | | | | | | | | | | | | | |
| Fusobacterium ulcerans ATCC 49185 (469617.3) | 3036 | | 3037 | 722, 3039 | 721, 3038 | | | | | | | | | |
| Fusobacterium varium ATCC 27725 (469618.3) | 1210 | | 1211 | 1213 | 1212 | | | | | | | | | |
| Gordonibacter pamelaeae 7-10-1-b (657308.3) | | | | | | | | | | | | | | |
| Hafnia alvei ATCC 51873 (1002364.3) | | 907, 3409 | | 908 | 902, 2566 | | | | | | | 4030 | | |
| Helicobacter bilis ATCC 43879 (613026.4) | | | | | | | | | | | | | | |
| Helicobacter canadensis MIT 98-5491 (Pr;30719) (537970.9) | | | | | | | | | | | | | | |
| Helicobacter cinaedi ATCC BAA-847 (1206745.3) | | | | | | | | | | | | | | |
| Helicobacter cinaedi CCUG 18818 (537971.5) | | | | | | | | | | | | | | |
| Helicobacter pullorum MIT 98-5489 (537972.5) | | | | | | | | | | | | | | |
| Helicobacter winghamensis ATCC BAA-430 (556267.4) | | | | | | | | | | | | | | |
| Holdemania filiformis DSM 12042 (545696.5) | | | | | | | | | | 1238 | | | | |
| Klebsiella pneumoniae 1162281 (1037908.3) | | | | | 4163 | | | | | | | | | |
| Klebsiella pneumoniae subsp. pneumoniae WGLW5 (1203547.3) | | | | | | | | | | | | | | |
| Klebsiella sp. 1_1_55 (469608.3) | | | | 2810 | 2806, 4917 | | | | | | | | | |
| Klebsiella sp. 4_1_44FAA (665944.3) | | | | | 4752 | | | | | | | | | |
| Lachnospiraceae bacterium 1_1_57FAA (658081.3) | | 265 | 269 | 262 | 263 | | 613 | 270, 272, 273, 274 | 605, 606, 615, 616, 617, 620 | | 471, 480, 1560 | | | 706, 719, 1762 |
| Lachnospiraceae bacterium 1_4_56FAA (658655.3) | | | | | | | 359 | | 53, 54, 55, 56 | 1305, 1436, 2820 | | | | |
| Lachnospiraceae bacterium 2_1_46FAA (742723.3) | | | | | | | 1802 | | | | 1388, 1390 | 1316, 1389 | | |
| Lachnospiraceae bacterium 2_1_58FAA (658082.3) | | | | 1925 | 1929 | | 1677 | | 1671, 1683, 2212, 2213, 2214 | 507 | | | | |
| Lachnospiraceae bacterium 3_1_46FAA (665950.3) | | 309 | 313 | 306 | 307 | | 2413 | 314, 316, 317, 318 | 2406, 2407, 2415, 2416, 2417, 2420 | | 1149, 2283 | | | 525 |
| Lachnospiraceae bacterium 3_1_57FAA_CT1 (658086.3) | | | | 96 | | | 2098, 6940 | | 2099, 2100, 2101 | | 998, 3499 | 4301 | | |
| Lachnospiraceae bacterium 4_1_37FAA (552395.3) | | | | | | | 2536 | | 2527, 2528, 2529 | | 1742, 2355, 2915 | | | |
| Lachnospiraceae bacterium 5_1_57FAA (658085.3) | | | | 786 | 789 | | 243 | | 237, 238, 239, 1688, 1689, 1690 | | | | | |
| Lachnospiraceae bacterium 5_1_63FAA (658089.3) | | | | | 1308 | | | | | 2486 | | | | |
| Lachnospiraceae bacterium 6_1_63FAA (658083.3) | | | | | | | 642 | | 646, 647, 1608, 1609, 1610 | | | | | |
| Lachnospiraceae bacterium 7_1_58FAA (658087.3) | | | | 5075 | | | | | | | | | | |
| Lachnospiraceae bacterium 8_1_57FAA (665951.3) | | 527 | 523 | 530 | 529 | | 851 | 518, 519, 520, 522 | 844, 847, 848, 849, 857, 858 | | 980, 2091 | | | 2579, 2591, 2654 |
| Lachnospiraceae bacterium 9_1_43BFAA (658088.3) | | | | | | | 2228 | | 2220, 2221, 2222 | 31, 186, 2068, 2545, 2634, 2639 | | | | |
| Lactobacillus acidophilus ATCC 4796 (525306.3) | | | | | | | | | | | | | | |
| Lactobacillus acidophilus NCFM (272621.13) | | | | | | | | | | | | | | |
| Lactobacillus amylolyticus DSM 11664 (585524.3) | | | | | | | | | | | | 1618 | | |
| Lactobacillus antri DSM 16041 (525309.3) | | | | | | | | | | | | | | |
| Lactobacillus brevis ATCC 367 (387344.15) | | | | | | | | | | | | | | |
| Lactobacillus brevis subsp. gravesensis ATCC 27305 (525310.3) | | | | | 1784 | | | | | | 1041 | | | |
| Lactobacillus buchneri ATCC 11577 (525318.3) | | | | | 1293 | | | | | | 778 | | | |
| Lactobacillus casei ATCC 334 (321967.11) | | | | 680, 2598 | 681, 2540 | | | | | | 315 | | | |
| Lactobacillus casei BL23 (543734.4) | | 275 | 274 | 705, 2694 | 386, 706, 2632 | 276 | | | 277, 278, 279, 280 | | 271 | | | |
| Lactobacillus crispatus 125-2-CHN (575595.3) | | | | 1902 | | | | | | | | | | |
| Lactobacillus delbrueckii subsp. bulgaricus ATCC 11842 (390333.7) | | | | | | | | | | | | | | |
| Lactobacillus delbrueckii subsp. bulgaricus ATCC BAA 365 (321956.7) | | | | | | | | | | | | | | |
| Lactobacillus fermentum ATCC 14931 (525325.3) | | | | | | | | | | | | | | |
| Lactobacillus fermentum IFO 3956 (334390.5) | | | | | | | | | | | | | | |
| Lactobacillus gasseri ATCC 33323 (324831.13) | | 115 | 114 | 147, 160 | 144 | 17 | | | 116, 117, 118, 119, 124 | | | | | |

| Organism | | | | | | | | | | | | |
|---|---|---|---|---|---|---|---|---|---|---|---|---|
| Lactobacillus helveticus DPC 4571 (405566.6) | 531 | | | | | | | | | | | |
| Lactobacillus helveticus DSM 20075 (585520.4) | 1139 | | | | | | | | | | | |
| Lactobacillus hilgardii ATCC 8290 (525327.3) | | | | 1543 | | | | 653 | | | | |
| Lactobacillus johnsonii NCC 533 (257314.6) | 127 | 126 | 155, 170 | 152 | 16 | | 128, 129, 130, 131, 135 | | | | | |
| Lactobacillus paracasei subsp. paracasei 8700:2 (537973.8) | 2332 | 2331 | 1460, 2338, 2612 | 1099, 2613 | 2333 | | 2334, 2335, 2336, 2337 | 2328, 2784 | | | | |
| Lactobacillus paracasei subsp. paracasei ATCC 25302 (525337.3) | 1054 | 1053 | 1849, 2392 | 1904, 2393 | | | 1132, 1133, 1134 | 1050 | | | | |
| Lactobacillus plantarum 16 (1327988.3) | | | 1691 | | | | | | | | | |
| Lactobacillus plantarum subsp. plantarum ATCC 14917 (525338.3) | | | 2027 | | | | | | | | | |
| Lactobacillus plantarum WCFS1 (220668.9) | | | 1773 | | | | | | | | | |
| Lactobacillus reuteri CF48-3A (525341.3) | | | | | | | | | | | | |
| Lactobacillus reuteri DSM 20016 (557436.4) | | | | | | | | | | | | |
| Lactobacillus reuteri F275 (299033.6) | | | | | | | | | | | | |
| Lactobacillus reuteri MM2-3 (585517.3) | | | | | | | | | | | | |
| Lactobacillus reuteri MM4-1A (548485.3) | | | | | | | | | | | | |
| Lactobacillus reuteri SD2112 (491077.3) | | | | | | | | | | | | |
| Lactobacillus rhamnosus ATCC 21052 (1002365.5) | 426 | 425 | 433, 523, 2045, 2742 | 500, 2046, 2671 | 427, 428 | | 429, 430, 431, 432 | | | | | |
| Lactobacillus rhamnosus GG (Prj:40637) (568703.30) | 335 | 334 | 343, 667, 2662 | 415, 668, 2584 | 336, 337 | | 338, 340, 341, 342 | 435 | | | | |
| Lactobacillus rhamnosus LMS2-1 (525361.3) | 1785 | 1784 | 1200, 1980, 2182 | 1264, 1953, 2183 | 1786 | | 1787, 1788, 1789, 1791 | | | | | |
| Lactobacillus ruminis ATCC 25644 (525362.3) | | | 520 | | | | | 602 | | | | |
| Lactobacillus sakei subsp. sakei 23K (314315.12) | | | | | | | | | | | | |
| Lactobacillus salivarius ATCC 11741 (525364.3) | | | | | | | | 1471, 1472 | | | | |
| Lactobacillus salivarius UCC118 (362948.14) | | | 241 | | | | | 1663 | | | | |
| Lactobacillus sp. 7_1_47FAA (665945.3) | 854 | 855 | 1111 | 843 | 849 | | 850, 851, 852, 853 | | | | | |
| Lactobacillus ultunensis DSM 16047 (525365.3) | | | | | | | | | | | | |
| Leuconostoc mesenteroides subsp. cremoris ATCC 19254 (586220.3) | | | | | | | | | | | | |
| Listeria grayi DSM 20601 (525367.9) | | | | 645 | | | | | | | | |
| Listeria innocua ATCC 33091 (1002366.3) | | | 527, 778 | 1141 | | | | | | | | |
| Megamonas funiformis YIT 11815 (742816.3) | | | | | | | | 1387, 1388 | | | | |
| Megamonas hypermegale ART12/1 (657316.3) | | | | | | | | 1826 | | | | |
| Methanobrevibacter smithii ATCC 35061 (420247.6) | | | | | | | | | | | | |
| Methanosphaera stadtmanae DSM 3091 (339860.6) | | | | | | | | | | | | |
| Mitsuokella multacida DSM 20544 (500635.8) | | | | | | | | 637 | | | | |
| Mollicutes bacterium D7 (556270.3) | | | | | | 1624 | 1615, 1619, 1620, 1621, 1623 | 835, 1175, 1469, 1743, 1881 | 1910, 1927, 1928, 2482, 2483 | | | |
| Odoribacter laneus YIT 12061 (742817.3) | | | | | | | | 2659 | 783, 1217, 1614, 1720, 3233 | | | |
| Oxalobacter formigenes HOxBLS (556268.6) | | | | | | | | | | | | |
| Oxalobacter formigenes OXCC13 (556269.4) | | | | | | | | | | | | |
| Paenibacillus sp. HGF5 (908341.3) | | | 6203 | 5909 | | | | 3687, 5435 | | | | |
| Paenibacillus sp. HGF7 (944559.3) | | | 3326 | 3325 | | | | 412, 1976, 3722 | 2816 | | | |
| Paenibacillus sp. HGH0039 (1078505.3) | | | 4398 | 4399 | | | | 1240, 2100, 3354 | 330 | | | |
| Parabacteroides distasonis ATCC 8503 (435591.13) | | | | | | | | 2321 | 46, 652, 1252, 2504, 2909 | | | |
| Parabacteroides johnsonii DSM 18315 (537006.5) | | | | | | | | 2863 | 398, 608, 651, 792, 1061, 1113, 1922, 3536 | | | |
| Parabacteroides merdae ATCC 43184 (411477.4) | | | | | | | | 728 | 248, 288, 370, 401, 433, 862, 1720, 1922, 2721 | | | |
| Parabacteroides sp. D13 (563193.3) | | | | | | | | 2866 | 743, 1422, 1983, 3485, 4031 | | | |
| Parabacteroides sp. D25 (658661.3) | | | | | | | | | 144, 416, 1344, 2458, 3505, 4154 | | | |
| Paraprevotella clara YIT 11840 (762968.3) | | | | | | | | | | | | |
| Paraprevotella xylaniphila YIT 11841 (762982.3) | | | | | | | | | | | | |
| Parvimonas micra ATCC 33270 (411465.10) | | | | | | | | | | | | |
| Pediococcus acidilactici 7_4 (563194.3) | 194 | 193 | 1853 | 1858 | 188 | | 187, 189, 190, 191 | | | | | |

| Organism | | | | | | | | | | | | |
|---|---|---|---|---|---|---|---|---|---|---|---|---|
| Pediococcus acidilactici DSM 20284 (862514.3) | 1196 | 1197 | | 1410 | 1203 | | 1199, 1200, 1201, 1202, 1204 | | | | | |
| Phascolarctobacterium sp. YIT 12067 (626939.3) | | | | | | | | | | | | |
| Prevotella copri DSM 18205 (537011.5) | | | | | | | | 298, 4027 | | | | |
| Prevotella oralis HGA0225 (1203550.3) | | | | | | | | | 1139, 2050, 2182 | | | |
| Prevotella salivae DSM 15606 (888832.3) | | | | | | | | | 518, 1151, 1565, 1571 | | | |
| Propionibacterium sp. 5_U_42AFAA (450748.3) | | | | | 572 | | 575, 576, 577 | 252, 2246 | 553, 2413 | | | |
| Propionibacterium sp. HGH0353 (1203571.3) | | | | | 2212 | | | | 1217 | | | |
| Proteus mirabilis WGLW6 (1125694.3) | 2020 | 2025 | 2026 | 2031 | | | 2021, 2022, 2023, 2024 | | | | | |
| Proteus penneri ATCC 35198 (471881.3) | 843, 844 | 851, 852, 853 | 854, 855 | 863 | | | 845, 846, 847, 848, 849, 850 | | | | | |
| Providencia alcalifaciens DSM 30120 (520999.6) | | | | | | | | | | | | |
| Providencia rettgeri DSM 1131 (521000.6) | 1206 | 1201 | 1200 | 1195 | | | 1202, 1203, 1204, 1205 | 1693 | | | | |
| Providencia rustigianii DSM 4541 (500637.6) | 404, 6686 | 398, 6692 | 397, 6693 | 391, 6699 | | | 399, 400, 401, 6688, 6689, 6690, 6691 | | | | | |
| Providencia stuartii ATCC 25827 (471874.6) | 3067 | 3061 | 3060 | 3066 | | | 3062, 3063, 3064, 3065 | 3256 | 966, 2757 | | | |
| Pseudomonas sp. 2_1_26 (665948.3) | | | | | | | | | | | | |
| Ralstonia sp. 5_7_47FAA (658664.3) | | | 4146 | | | | | | | | | |
| Roseburia intestinalis L1-82 (536231.5) | | | | | 3250 | | 3251, 3252, 3253, 3255 | 423 | 1506 | | | |
| Roseburia intestinalis M50/1 (657315.3) | | | | | 3559 | | 3553, 3554, 3556, 3557, 3558 | 796 | | | | |
| Roseburia intestinalis XB6B4 (718255.3) | | | | | 2521 | | 2522, 2523, 2524, 2526 | 2929 | 1473 | | | |
| Roseburia inulinivorans DSM 16841 (622312.4) | | | | | 1688 | | 1694, 1695, 1696 | 2587 | | | | |
| Ruminococcaceae bacterium D16 (552398.3) | | | | | 347 | | 343, 344, 345 | 2512 | | | | |
| Ruminococcus bromii L2-63 (657321.5) | | | | | | | | 848 | | | | |
| Ruminococcus gnavus ATCC 29149 (411470.6) | | | 915 | 911 | 1054 | | 1048, 1393, 1394, 1395 | | | | | |
| Ruminococcus lactaris ATCC 29176 (471875.6) | | | | | 1843 | | 1845, 1846, 1847 | 1628, 1662, 2484 | 1632, 1871 | | | |
| Ruminococcus obeum A2-162 (657314.3) | | | 1719 | 1715 | 2912 | | 3540, 3541, 3542 | | | | | |
| Ruminococcus obeum ATCC 29174 (411459.7) | | | 3062 | 3053, 3066 | 339 | | 342, 343, 344, 347 | | | | | |
| Ruminococcus sp. 18P13 (213810.4) | | | | | | | | | | | | |
| Ruminococcus sp. 5_1_39BFAA (457412.4) | | | | | 611 | | 1422, 1423, 1424 | | | | | |
| Ruminococcus sp. SR1/5 (657323.3) | | | 3503 | 3507 | 1418 | | 1413, 1414, 1415 | | | | | |
| Ruminococcus torques ATCC 27756 (411460.6) | 613 | 609 | 616 | 615 | 1844 | 604, 605, 606, 608 | 1837, 1840, 1841, 1842, 1850 | | 22, 1627, 2034 | | 442, 984, 994 | |
| Ruminococcus torques L2-14 (657313.3) | | | 554 | 558, 2221 | | | | | | | | |
| Salmonella enterica subsp. enterica serovar Typhimurium str. LT2 (99287.12) | | | 3456 | 3452, 4000 | | | | | | | | |
| Staphylococcus sp. HGB0015 (1078083.3) | 1680 | 1672 | 1673 | 784, 1674 | 1675 | | 1676, 1677, 1678, 1679 | | | | | |
| Streptococcus anginosus 1_2_62CV (742820.3) | | 1750 | | 118, 1751 | 1745 | | 1746, 1747, 1748, 1749 | | | | | |
| Streptococcus equinus ATCC 9812 (525379.3) | | | 297 | | | | | | | | | |
| Streptococcus infantarius subsp. infantarius ATCC BAA-102 (471872.6) | | | 266, 654 | 653 | | | | | | | | |
| Streptococcus sp. 2_1_36FAA (469609.3) | | 62 | 703, 1476, 1483 | 1475, 1482 | 57 | | 58, 59, 60, 61 | 419 | | | | |
| Streptococcus sp. HPH0090 (1203590.3) | | 1441 | 193, 397 | 398 | 1446 | | 1442, 1443, 1444, 1445 | 1002, 1048, 1448 | | | | |
| Streptomyces sp. HGB0020 (1078086.3) | | | 3644 | | | | 2898, 3737, 5165, 8190 | 1167, 2955, 4922, 5130, 8394 | | | | |
| Streptomyces sp. HPH0547 (1203592.3) | | | 4042 | | | | 43, 2838, 5766 | 389, 2269, 2386, 2503 | | | | |
| Subdoligranulum sp. 4_3_54A2FAA (665956.3) | | | | | | 447 | 443, 444, 445 | 204, 1927, 2746, 3106, 4057 | | | | |
| Subdoligranulum variable DSM 15176 (411471.5) | | | | | | 2375, 3110 | 2377, 2378, 2379, 3112, 3113, 3114 | 1056, 4229 | | | | |
| Succinatimonas hippei YIT 12066 (762983.3) | | | | | | | | | | | | |
| Sutterella wadsworthensis 3_1_45B (742821.3) | | | | | | | | | | | | |
| Sutterella wadsworthensis HGA0223 (1203554.3) | | | | | | | | | | | | |
| Tannerella sp. 6_1_58FAA_CT1 (665949.3) | | | | | | | | 1827 | 2544, 2545 | | | |
| Turicibacter sp. HGF1 (910310.3) | | 1114 | 35 | 1113 | | 1099 | 1100, 1101, 1110, 1116 | | 302 | | | |
| Turicibacter sp. PC909 (702450.3) | | 369 | 1430 | 370 | | 384 | 367, 373, 382, 383 | | 806 | | | |
| Veillonella sp. 3_1_44 (457416.3) | | | | | | | | 839, 889 | | | | |
| Veillonella sp. 6_1_27 (450749.3) | | | | | | | | 879, 940 | | | | |
| Weissella paramesenteroides ATCC 33313 (585506.3) | | | | | | | | | | | | |
| Yokenella regensburgei ATCC 43003 (1002368.3) | 1539 | 1538 | 1544 | 1537 | | | 1541, 1542, 1543 | | 316 | | | |

**Table S5.** Genes involved in depletion and utilization of N-acetylglucosamine. The PubSEED identifiers are shown. For details on analyzed organisms see Table S1, for details on gene functions see Table S2.

| Organism (PubSEED ID) | Kinase | | Deacetylase | Deaminase | | Phospho-beta-galactosidase | Phosphorylase | PTS systems | | ABC systems | | | Permease | Beta-hexosaminidase | | Alpha-N-acetylglucosaminidase | Problems (if any) |
|---|---|---|---|---|---|---|---|---|---|---|---|---|---|---|---|---|---|
| | NagK1 | NagK2 | NagA | NagB1 | NagB2 | GnbG | LnbP | NagE | GnbPTS | NgcEFG | NgcABCD | LnbABC | NagP | Bha3 | Bha20 | Ang | |
| Abiotrophia defectiva ATCC 49176 (592010.4) | | | 1693, 2925 | | 1694, 2924 | | 3045 | | | 1673, 1990, 1991, 2311, 2312, 2313, 2506, 2514, 2515, 2823, 2824 | | 758, 759, 760, 3041, 3042, 3043 | | 2689 | | | |
| Acidaminococcus sp. D21 (563191.3) | | | 918 | | | | | | | | | | | | | | |
| Acidaminococcus sp. HPA0509 (1203555.3) | | | 1878 | | | | | | | | | | | | | | |
| Acinetobacter junii SH205 (575587.3) | | | | | | | | | | | | | | | | | |
| Actinomyces odontolyticus ATCC 17982 (411466.7) | | 61, 1017 | 1628 | 1016 | 59 | | | | | 1022, 1023, 1024 | | | | | | | |
| Actinomyces sp. HPA0247 (1203556.3) | | 52 | 1177 | | 50 | | | | | | | | | 130 | | | Transporter is not found |
| Akkermansia muciniphila ATCC BAA-835 (349741.6) | | | 1075 | | 2070 | | | | | | | | | 421, 451, 986, 1169, 1892, 2063, 2191, 2301, 2302, 2446, 2460 | 2416 | 72, 1384 | |
| Alistipes indistinctus YIT 12060 (742725.3) | | | 1108 | | 1109, 1793, 1795 | | | | | | | | | 913, 2149, 2200 | | | |
| Alistipes shahii WAL 8301 (717959.3) | | 2772 | | | 2404, 2405 | | | | | | | | | 383, 2285, 3055, 3058, 3073 | 3053, 3056, 3057 | 1230 | |
| Anaerobaculum hydrogeniformans ATCC BAA-1850 (592015.5) | | | | | | | | | | | | | | | | | |
| Anaerococcus hydrogenalis DSM 7454 (561177.4) | | | | | | | | | | | | | | | | | |
| Anaerofustis stercorihominis DSM 17244 (445971.6) | | | | | | | | | | | | | | | | | |
| Anaerostipes caccae DSM 14662 (411490.6) | | | 46 | | 47, 2358, 3068 | | | 48, 48 | | | | | | | | 1081 | |
| Anaerostipes hadrus DSM 3319 (649757.3) | | | 1969 | | 163 | | | 1970, 1970, 1971 | | | | | | | | 1225 | |
| Anaerostipes sp. 3_2_56FAA (665937.3) | | | 3347 | | 269, 3346 | | | 3345, 3345 | | | | | | | | 188 | |
| Anaerotruncus colihominis DSM 17241 (445972.6) | | | 2267 | | 2258, 2268, 2817 | | | 2270, 2270 | | | | | | | | | |
| Bacillus smithii 7_3_47FAA (665952.3) | | | 1254 | | 78 | | | 1752, 1752 | | | | | | | | 2068 | |
| Bacillus sp. 7_6_55CFAA_CT2 (665957.3) | | | 821 | | 822 | | | 5203, 5203 | | | | | | | | | |
| Bacillus subtilis subsp. subtilis str. 168 (224308.113) | | | 3601 | | 235, 3602 | | | 801, 801 | | | | 3350, 3351 | | | | 167 | |
| Bacteroides caccae ATCC 43185 (411901.7) | | 321 | 771, 772 | | 1643, 2122 | | | | | | | | | 1016, 1017, 1021, 2633, 2643, 2667, 3165, 3362, 3410, 3423, 3502, 3508 | 325, 2648 | 245, 1846, 2376 | |
| Bacteroides capillosus ATCC 29799 (411467.6) | | | 2734 | | 1155, 2735 | | 848 | | | | | 850, 851, 852 | | | | 1352 | |
| Bacteroides cellulosilyticus DSM 14838 (537012.5) | | 976, 2271 | 2425 | 3600 | 2424, 2830, 4912, 4996 | | | | | | | | 975 | 37, 950, 1188, 2126, 2422, 3129 | 1267, 2062, 2066, 2067, 2104 | 1796, 2025 | |
| Bacteroides clarus YIT 12056 (762984.3) | | 2106, 2501 | 203, 337 | | 336, 2802 | | | | | | | | 2502 | 1623, 2048, 3000 | 2102 | | |
| Bacteroides coprocola DSM 17136 (470145.6) | | 1566 | 538, 539 | | 41, 42 | | | | | | | | | 463, 908, 914, 915, 2401, 3707 | 46, 1649, 1996 | | |
| Bacteroides coprophilus DSM 18228 (547042.5) | | 2945 | 2344, 2345 | | 788, 789 | | | | | | | | | 578, 782, 783, 896, 1099, 1246, 1558, 2137, 2288, 2913, 2914, 3123 | 547, 1770, 2949 | 2066, 3031, 3260 | |
| Bacteroides dorei DSM 17855 (483217.6) | | 364 | 505, 639 | | 654, 655 | | | | | | | | | 474, 630, 634, 636, 1087, 1401, 1441, 2459, 3166, 3335 | 415, 575 | 2803, 4255 | |

| Species | | | | | | | | | | | | | | | |
|---|---|---|---|---|---|---|---|---|---|---|---|---|---|---|---|
| Bacteroides eggerthii 1_2_48FAA (665953.3) | 2039, 3592 | 298, 626 | 297, 1611, 2158 | | | | | | | | 2040 | 670, 1693 | 3587, 3588 | | |
| Bacteroides eggerthii DSM 20697 (483216.6) | 54, 2641 | 1738, 2008 | 1074, 1579, 1737 | | | | | | | | 53 | 352 | 2645 | | |
| Bacteroides finegoldii DSM 17565 (483215.6) | 3013, 3346 | 2724, 2725 | 1318, 2171 | | | | | | | | 3014 | 532, 1700, 2282, 2283, 3069 | 3350, 3645 | 2273 | |
| Bacteroides fluxus YIT 12057 (763034.3) | 2480, 3111 | 692, 2829 | 693, 1162, 1850 | | | | | | | | | 1188, 2029, 2231, 2698, 2717, 3256, 3264 | 3105, 3106 | 1695, 2256 | |
| Bacteroides fragilis 3_1_12 (457424.5) | 2418 | 2892, 3090 | 1880, 3091 | | | | | | | | | 387, 391, 393, 888, 952, 1609, 1929, 2578, 3292, 3295, 3302, 3599 | 2412, 2413 | 9 | |
| Bacteroides fragilis 638R (862962.3) | 572 | 2840, 3043 | 987, 3044 | | | | | | | | | 391, 738, 1038, 1739, 1743, 1745, 2289, 3199, 3202, 3210, 3823, 4278 | 576, 577 | 670 | |
| Bacteroides fragilis NCTC 9343 (272559.17) | 498 | 2771, 2978 | 894, 2979 | | | | | | | | | 321, 656, 943, 1778, 1782, 1784, 2181, 3119, 3122, 3130, 3685, 4157 | 502, 503 | 593 | |
| Bacteroides fragilis YCH46 (295405.11) | 577 | 2648, 2983 | 982, 2984 | | | | | | | | | 405, 752, 1030, 1679, 1683, 1685, 2074, 3116, 3119, 3127, 3668, 4100 | 581, 582 | 689 | |
| Bacteroides intestinalis DSM 17393 (471870.8) | 168, 528 | | 2678, 4191, 4686 | | | | | | | | | 561, 789, 796, 797, 1715, 3500, 3952 | 507, 512, 513, 1330 | 1420 | |
| Bacteroides ovatus 3_8_47FAA (665954.3) | 3368, 4519 | 1101, 1102, 2899 | 2900, 3004, 3613 | | | | | | | | 3367 | 1381, 2124, 2395, 2808, 3273, 4373, 4374, 4576, 4727, 4728 | 4523 | 69, 73, 4744 | |
| Bacteroides ovatus ATCC 8483 (411476.11) | 541, 4656 | 126, 127, 2864 | 1505, 2863, 4190 | | | | | | | | 4655 | 993, 994, 1351, 1728, 1872, 1873, 2004, 2490, 2619, 2959, 3245, 3718, 4371 | 537, 2236, 2315 | 751, 755, 1396, 4409 | |
| Bacteroides ovatus SD CC 2a (702444.3) | 434, 2613 | 1675, 4485, 4486 | 80, 301, 1676 | | | | | | | | 435 | 17, 351, 2781, 3077, 3387, 3388, 3417, 3889, 4736, 4737 | 2617, 4819 | 4024, 4836, 4841, 4842 | |
| Bacteroides ovatus SD CMC 3f (702443.3) | 4854, 5435 | 2084, 2579, 2580 | 1903, 2083, 4919 | | | | | | | | 5436 | 992, 1159, 1160, 2371, 2619, 2620, 3221, 3235, 4196, 4730, 4966 | 85, 4858 | 731, 732, 2332, 3299, 3300, 3304 | |
| Bacteroides pectinophilus ATCC 43243 (483218.5) | | | | | | | | | | | | | | | |
| Bacteroides plebeius DSM 17135 (484018.6) | 2792 | 1144, 1145 | 1772, 1773 | | | | | | | | | 130, 896, 897, 975, 1279, 2458, 2642 | 38 | 3368 | |
| Bacteroides sp. 1_1_14 (469585.3) | 1329, 4768 | 2384, 2385, 3522 | 884, 1890, 3521 | | | | | | | | 1328 | 199, 257, 792, 1057, 1142, 1145, 1146, 1358, 1402, 1640, 3200, 3206, 3532, 4136, 4666 | 4779 | 222, 223, 1120, 1121, 2642, 3524 | |

| | | | | | | | | | | | | | |
|---|---|---|---|---|---|---|---|---|---|---|---|---|---|
| Bacteroides sp. 1_1_30 (457387.3) | 2726, 4422 | 2, 3, 2647 | 2082, 2646, 2976 | | | | | | | 2725 | 708, 1297, 1298, 1543, 3697, 3699, 4488, 4508, 5025, 5183 | 4426 | 2392, 2396 |
| Bacteroides sp. 1_1_6 (469586.3) | 3762, 4640 | 851, 852, 3379 | 387, 2281, 3378 | | | | | | | 3761 | 134, 137, 138, 411, 1045, 2012, 2068, 2556, 2810, 2823, 3389, 3851, 4611, 4797, 4798, 4805 | 4596 | 106, 2045, 3381 |
| Bacteroides sp. 2_1_16 (469587.3) | 658, 4134 | 3151, 3353 | 1274, 3150 | | | | | | | | 591, 1325, 1888, 1892, 1894, 1999, 2563, 3001, 3009, 3012, 3739 | 2732, 2733 | 1943 |
| Bacteroides sp. 2_1_22 (469588.3) | 2728 | 22, 1957, 1958 | 23, 2634 | | | | | | | 2253 | 466, 1602, 1680, 1681, 2803, 2804, 3590, 4190, 4551 | 577, 4138 | 799, 803, 2786 |
| Bacteroides sp. 2_1_33B (469589.3) | 2254, 4134 | 2237, 3381, 3382 | 1848, 3380 | | | | | | | | 225, 808, 1625, 3100, 3882 | 930, 931, 2907 | |
| Bacteroides sp. 2_1_56FAA (665938.3) | 1691, 1827 | 4456, 4702 | 1060, 4455 | | | | | | | | 381, 808, 1112, 1929, 1933, 1935, 2408, 3206, 3781, 3784, 3793 | 551, 552 | 733 |
| Bacteroides sp. 2_1_7 (457388.5) | 547 | 378, 3711, 3712 | 899, 3710 | | | | | | | | 572, 1099, 1720, 3165, 3379, 4182 | 2424 | |
| Bacteroides sp. 2_2_4 (469590.5) | 634, 878 | 3593, 5110, 5111 | 944, 1649, 3594 | | | | | | | 640 | 14, 149, 1391, 1392, 1963, 3277, 3278, 3733, 3966, 5483, 5484 | 378, 1260 | 2002, 4017, 4181, 4185, 5507 |
| Bacteroides sp. 20_3 (469591.4) | 383, 639 | 140, 1154, 1155 | 635, 1156 | | | | | | | | 1415, 2485, 3124, 3326, 4195 | 1686, 4443 | |
| Bacteroides sp. 3_1_19 (469592.4) | 11, 468 | 1012, 3516, 3517 | 448, 3518 | | | | | | | | 818, 1473, 2927, 3315, 3888, 4241 | 2013 | |
| Bacteroides sp. 3_1_23 (457390.3) | 3071, 3882 | 613, 614, 2739 | 953, 2738, 4269 | | | | | | | 3881 | 1589, 2061, 2417, 2840, 3135, 3642, 4446, 4721, 4722, 4762 | 1840, 1913, 3075 | 1372, 1376, 4498, 4705 |
| Bacteroides sp. 3_1_33FAA (457391.3) | 1295 | 1153, 1710 | 1694, 1695 | | | | | | | | 912, 1190, 1713, 1715, 1719, 1868, 2842, 3673, 3737 | 1242 | 2250, 3916 |
| Bacteroides sp. 3_1_40A (469593.3) | 3670 | 3394, 3823 | 3409, 3410 | | | | | | | | 234, 401, 832, 1025, 1946, 2532, 2838, 3385, 3389, 3391, 3790, 3791 | 3727, 3735 | 147, 3929 |
| Bacteroides sp. 3_2_5 (457392.3) | 609 | 1606, 1826 | 1565, 1827 | | | | | | | | 455, 797, 1514, 1967, 1970, 1978, 2336, 3149, 4354, 4356, 4360 | 604, 605 | 518 |
| Bacteroides sp. 4_1_36 (457393.3) | 889 | | 1614, 2968, 3374 | | | | | | | | 268, 1837, 2096, 3003, 3266 | 33, 34, 38, 894, 895 | |
| Bacteroides sp. 4_3_47FAA (457394.3) | 1599 | 774, 1459 | 787, 788 | | | | | | | | 550, 765, 769, 771, 1365, 1489, 1710, 2458, 2952, 4574 | 1545 | 296, 2181 |

| Organism | | | | | | | | | | | | | |
|---|---|---|---|---|---|---|---|---|---|---|---|---|---|
| Bacteroides sp. 9_1_42FAA (457395.6) | 54 | 417, 524 | 400, 401 | | | | | | | 420, 422, 426, 572, 2484, 2735, 3470, 3724, 3995 | 636 | 1032, 3912 | |
| Bacteroides sp. D1 (556258.5) | 3521, 3924 | 2034, 2802, 2803 | 1395, 2035, 3157 | | | | | | 3923 | 304, 1565, 1566, 1810, 3291, 3292, 3465, 3765, 4450 | 429, 3517 | 650, 654, 1548 | |
| Bacteroides sp. D2 (556259.3) | 923, 4858 | 2681, 4490, 4491 | 617, 1394, 2682 | | | | | | 924 | 392, 393, 1206, 1647, 2103, 2592, 3265, 3716, 3717, 3971, 4792, 5047 | 2130, 4854 | 1169, 1597, 2384, 2388 | |
| Bacteroides sp. D20 (585543.3) | 1646 | | 411, 936, 2501 | | | | | | | 62, 292, 1194, 2336, 2558 | 1651, 1652, 1802, 1803, 1807 | | |
| Bacteroides sp. D22 (585544.3) | 2774, 3855 | 1563, 1564, 2093 | 2092, 2582, 4263 | | | | | | 3854 | 145, 146, 1092, 1094, 1095, 1315, 1319, 1320, 1907, 2835, 3302, 3483, 3778, 4367 | 1093, 2778, 3070 | 542, 546 | |
| Bacteroides sp. HPS0048 (1078089.3) | | | 1377, 3765 | | | | | | | 5, 729, 730, 1199, 1200, 1558, 2876, 2942, 2943, 2953, 3179, 3943, 3948, 3967 | 1104, 1105 | 409, 413, 1204, 2093, 2951 | |
| Bacteroides stercoris ATCC 43183 (449673.7) | 383, 3289 | 97, 1575 | 1574, 2872 | | | | | | | 548, 1455, 1461, 2245, 2379, 2694, 3236 | 379 | | |
| Bacteroides thetaiotaomicron VPI-5482 (226186.12) | 2545, 4734 | 689, 690, 3647 | 259, 3646, 4194 | | | | | | 4733 | 191, 456, 459, 460, 507, 1068, 1659, 1665, 2517, 3240, 3657, 3930, 4413, 4472, 4761 | 2501 | 437, 3649, 4436 | |
| Bacteroides uniformis ATCC 8492 (411479.10) | 696 | | 181, 1567, 1758 | | | | | | | 43, 999, 1376, 1724, 3172 | 689, 690, 3474, 3475, 3478 | | |
| Bacteroides vulgatus ATCC 8482 (435590.9) | 198 | 60, 4244 | 4231, 4232 | | | | | | | 89, 404, 552, 1244, 2282, 3342, 4247, 4249, 4253 | 150 | 1927 | |
| Bacteroides vulgatus PC510 (702446.3) | 2291 | 760, 1928 | 466, 467 | | | | | | | 216, 649, 709, 763, 765, 769, 1958, 2250, 2425, 2896 | 1436 | 517, 2736 | |
| Bacteroides xylanisolvens SD CC 1b (702447.3) | 3634, 4395 | 2747, 2748, 4648 | 2400, 2508, 4647 | | | | | | 3635 | 129, 130, 350, 1720, 1721, 2100, 3926, 3927, 4163, 4692 | 2034, 4399 | 1110, 1114, 2046 | |
| Bacteroides xylanisolvens XB1A (657309.4) | 515, 2695 | 1696, 2448, 2449 | 1695, 2032, 3239 | | | | | | 2694 | 73, 74, 573, 1113, 2296, 3256, 3618, 3931, 4504, 4505 | 520, 984 | 731, 735, 1850, 3263 | |
| Barnesiella intestinihominis YIT 11860 (742726.3) | | | 2704, 2705 | | | | | | | 101, 308, 379, 380, 513, 2614, 2835, 2904, 2964, 2975 | 2312 | 2781, 2785, 2966 | |
| Bifidobacterium adolescentis L2-32 (411481.5) | 1801 | 59 | | | | | 1797, 1798, 1799, 1800 | | | | | 1012 | Deaminase is not found |
| Bifidobacterium angulatum DSM 20098 = JCM 7096 (518635.7) | 565 | | | | | | 566, 567, 568, 569 | | | | | 1101 | |
| Bifidobacterium animalis subsp. lactis AD011 (442563.4) | | | | | | | | 490, 492 | | | | 1067 | |

| Organism | | | | | | | | | | | |
|---|---|---|---|---|---|---|---|---|---|---|---|
| Bifidobacterium bifidum NCIMB 41171 (398513.5) | 479, 1224, 1226, 1507 | 1228, 1610 | 1227 | 481, 905 | 912, 912, 912 | 1229, 1231, 1232, 1233 | 482, 483, 484 | 441, 1540, 1769 | 133 | 1567 | |
| Bifidobacterium breve DSM 20213 = JCM 1192 (518634.7) | 882, 883, 1821 | 880, 1824 | 881, 1823 | 1473 | 18, 19, 19 | 1394 | 524, 527, 1469, 1471, 1472 | 1506 | 744, 745 | | |
| Bifidobacterium breve HPH0326 (1203540.3) | 1043, 1412 | 1415 | 1414 | 1077 | 19, 20, 20 | 1159 | 507, 509, 510, 612, 613, 615, 616, 1078, 1079, 1081 | | 1986, 1987 | | |
| Bifidobacterium catenulatum DSM 16992 = JCM 1194 (566552.6) | | 190 | | | | | | | 948 | | |
| Bifidobacterium dentium ATCC 27678 (473819.7) | 909, 1413 | 907 | | | | 1414, 1415, 1416, 1417 | | | | | Deaminase is not found |
| Bifidobacterium gallicum DSM 20093 (561180.4) | | 1408 | 1559 | | | | | | 75 | | |
| Bifidobacterium longum DJO10A (Prj:321) (205913.11) | 493, 494, 495 | 491 | 492 | 1835 | | 488, 489, 490 | 918, 923, 1022, 1023, 1837, 1838, 1839 | 1253 | 359 | | |
| Bifidobacterium longum NCC2705 (206672.9) | 198, 199, 200, 1746 | 202 | 201 | 1696 | | 203, 204, 205, 206 | 875, 880, 995, 997, 1692, 1693, 1694 | 59 | 735 | | |
| Bifidobacterium longum subsp. infantis 157F (565040.3) | 1332, 1333, 1334 | 1330 | 1331 | 1729 | | 1326, 1327, 1328, 1329 | 359, 360, 361, 1731, 1732, 1733 | 1471 | 625, 1342 | | |
| Bifidobacterium longum subsp. infantis ATCC 15697 = JCM 1222 (391904.8) | 905 | 908 | 907 | 2251 | 2545, 2546, 2546 | | 909, 910, 911, 2252, 2253, 2254 | 471, 748, 2431 | 1915 | | |
| Bifidobacterium longum subsp. infantis ATCC 55813 (548480.3) | 1366, 1367 | 1369 | 1368 | 648 | | 1371, 1372, 1373 | 645, 646, 647 | 892 | 104 | | |
| Bifidobacterium longum subsp. infantis CCUG 52486 (537937.5) | 1377, 1379, 1380 | 641, 1382 | 1381 | 1177 | | 1383, 1384, 1385, 1386 | 1173, 1174, 1175 | 2008 | 12 | | |
| Bifidobacterium longum subsp. longum 2-2B (1161745.3) | 341, 342, 343 | 339 | 340 | 514 | | 338, 1977, 1978, 1979 | 515, 516, 517, 838, 839 | 1617 | 2072 | | |
| Bifidobacterium longum subsp. longum 44B (1161743.3) | 584, 585, 586 | 582 | 583 | 1601 | | 581, 1033, 1034, 1035 | 103, 104, 105, 1598, 1599, 1600 | 1528 | 1807 | | |
| Bifidobacterium longum subsp. longum F8 (722911.3) | 1929, 1930, 1931 | 1933 | 1932 | 1599 | | 1934, 1935, 1936, 1937 | 858, 859, 860, 1601, 1602 | 91 | 602 | | |
| Bifidobacterium longum subsp. longum JCM 1217 (565042.3) | 1308, 1309, 1310 | 1306 | 1307 | 1679 | | 1302, 1303, 1304, 1305 | 458, 1680, 1681, 1682 | 1443 | 618 | | |
| Bifidobacterium pseudocatenulatum DSM 20438 = JCM 1200 (547043.8) | 630, 1743 | 1740 | 1741 | | | 625, 626, 627, 628 | | 1742 | 1008 | | |
| Bifidobacterium sp. 12_1_47BFAA (469594.3) | 1513, 1514, 1515 | 1518 | 1517 | 583 | | 1519, 1913, 1914, 1915 | 475, 476, 585, 586, 587 | 78 | 277 | | |
| Bilophila wadsworthia 3_1_6 (563192.3) | | | | | | | | | 1035 | | |
| Blautia hansenii DSM 20583 (537007.6) | | 2031, 2798 | 817, 2799 | 1753 | 2600, 2600, 2601 | | 811, 812, 813, 1757, 1758 | | | | |
| Blautia hydrogenotrophica DSM 10507 (476272.5) | | 3234 | 1236, 3311 | 583 | | | 585, 586, 587 | | 646 | | |
| Bryantella formatexigens DSM 14469 (478749.5) | 4208 | 1279, 2475 | 1280, 2474 | 554 | | 635, 756, 757, 771, 809, 810, 2012, 2013, 2014, 2497, 2499, 3680, 3841, 3842, 3994, 3995, 4151, 4152, 4153 | 550, 555, 556, 557 | 279 | 1695, 1696 | | |
| Burkholderiales bacterium 1_1_47 (469610.4) | | | | | | | | | | | |
| butyrate-producing bacterium SM4/1 (245012.3) | | | 1055 | | | | | | 2299 | | |
| butyrate-producing bacterium SS3/4 (245014.3) | | 646 | 647 | | | | | | | | |
| butyrate-producing bacterium SSC/2 (245018.3) | | 1999 | 473 | | 1992, 1993, 1993 | | | | 131 | | |

| | | | | | | | | | | | | |
|---|---|---|---|---|---|---|---|---|---|---|---|---|
| Butyricicoccus pullicaecorum 1.2 (1203606.4) | | | | 641, 2843 | | 640, 640, 640, 1200, 1200, 1200 | | | | 2205 | | Deacetylase is not found |
| Butyrivibrio crossotus DSM 2876 (511680.4) | | | | | | | | | | | | |
| Butyrivibrio fibrisolvens 16/4 (657324.3) | | | | | | | | 2729, 2730, 2731 | | 1716 | | |
| Campylobacter coli JV20 (864566.3) | | | | | | | | | | | | |
| Campylobacter upsaliensis JV21 (888826.3) | | | | | | | | | | | | |
| Catenibacterium mitsuokai DSM 15897 (451640.5) | | 1825 | | 429, 989 | | 281, 281, 281 | | | 704, 705, 706, 1150, 1856 | 1730 | | |
| Cedecea davisae DSM 4568 (566551.4) | 1464 | 959 | | 960 | | 961, 961, 961 | | | | | | |
| Citrobacter freundii 4_7_47CFAA (742730.3) | | 2811 | | 2812 | | 2813, 2813, 2813 | | | | | 2064 | |
| Citrobacter sp. 30_2 (469595.3) | 1778 | 1321 | | 1322 | | 1323, 1323, 1323 | | | | 4080 | | |
| Citrobacter youngae ATCC 29220 (500640.5) | 1202 | 3177 | | 3178 | | 3179, 3179, 3179 | | | | | 659 | |
| Clostridiales bacterium 1_7_47FAA (457421.5) | | 3312, 4354 | | 4352, 4353, 5717 | 4154 | 1669, 1669, 1671 | | 4156, 4157, 4158 | | 1584, 2883, 3168 | | |
| Clostridium asparagiforme DSM 15981 (518636.5) | | 1833, 2622, 3333 | | 94, 465, 906, 1839, 1840, 3334 | | 694, 696, 696, 2125, 2125 | | | | 1805, 2939, 3111 | | |
| Clostridium bartlettii DSM 16795 (445973.7) | | | | | | | | | | 139, 323 | | |
| Clostridium bolteae 90A5 (997893.5) | 5207 | 6127 | | 4812, 4894 | 4411 | 1740, 1742, 1742 | | 4408, 4409, 4410 | | 3453, 3548 | | |
| Clostridium bolteae ATCC BAA-613 (411902.9) | | 4493 | | 936, 1657 | 1483 | 3975, 3975, 3977 | | 1480, 1481, 1482 | | 5814, 5902 | | |
| Clostridium celatum DSM 1785 (545697.3) | | 3277 | | 2387 | 2573, 2574 | 297, 298, 298 | | 2577, 2578, 2579 | 2632 | 163, 1196 | 2671 | |
| Clostridium cf. saccharolyticum K10 (717608.3) | | 1503 | | 1504 | | 436, 438, 438 | | | | 3073 | | |
| Clostridium citroniae WAL-17108 (742733.3) | | 2681, 5553, 6173 | | 2682, 5759, 6182 | 2512 | 205, 205, 205 | | | | 132, 5679 | | |
| Clostridium clostridioforme 2_1_49FAA (742735.4) | | 5004 | | 1363, 4630 | 2652 | 1259, 1259, 1259 | | 2653, 2655, 2656 | | 2087 | | |
| Clostridium difficile CD196 (645462.3) | | 925 | | 926 | | 3021, 3021, 3021 | 2509 | | | | | |
| Clostridium difficile NAP07 (525258.3) | | 1900 | | 1899 | | 1034, 1034, 1034, 1046, 1046, 1046 | | | | | | |
| Clostridium difficile NAP08 (525259.3) | | 1964 | | 1965 | | 1160, 1160, 1160, 1706, 1706, 1706 | | | | | | |
| Clostridium hathewayi WAL-18680 (742737.3) | | 627, 3547, 4459, 4554 | 352 | 621, 3525, 4460, 5054 | 4106, 4107 | 2326, 2326, 2326, 5033, 5033, 5033 | | 4108, 4109, 4110 | | 960, 1448, 1593, 3978, 4273 | | |
| Clostridium hiranonis DSM 13275 (500633.7) | | 1746 | | 1747 | | | | | | 1544, 1546, 1549 | | |
| Clostridium hylemonae DSM 15053 (553973.6) | | 2858 | | 1168, 2865, 3353 | 3210 | | | 3213, 3214, 3215 | | | | |
| Clostridium leptum DSM 753 (428125.8) | | 602 | | 755 | | | | | | 744 | | |
| Clostridium methylpentosum DSM 5476 (537013.3) | | | | | | | | | | | | |
| Clostridium nexile DSM 1787 (500632.7) | | 656 | | 655 | 1721 | 3724, 3724, 3725 | | 1712, 1713, 1714, 3026, 3027, 3028 | 613, 614, 1955 | 2441 | | |
| Clostridium perfringens WAL-14572 (742739.3) | | 1842 | | 2487 | 1927 | 515, 515, 515, 3169, 3170, 3170 | | 1921, 1922, 1923 | 679, 1027 | 2776 | 305, 307 | |
| Clostridium ramosum DSM 1402 (445974.6) | 861 | 2050, 2352 | | 575, 1785, 1946, 2908 | 1039 | 501, 501, 501, 2453, 2454 | | 1030, 1034, 1035, 1036, 1038 | 1501, 1517, 1518 | 741, 1205, 1472, 1631, 2275 | | |
| Clostridium scindens ATCC 35704 (411468.9) | | 432 | | 2964 | 2966 | | | 117, 118, 119, 2960, 2961, 2962 | | | | |

| | | | | | | | | | | | | | | | |
|---|---|---|---|---|---|---|---|---|---|---|---|---|---|---|---|
| Clostridium sp. 7_2_43FAA (457396.3) | | | 2860 | 21 | | 3457 | 1466, 1466, 1466, 3463, 3464, 3464 | | | | 3455, 3456, 3459 | 69, 139, 2261 | 87, 117, 2545 | | |
| Clostridium sp. 7_3_54FAA (665940.3) | | | 2695 | 1895, 4893 | | 3618 | | | | | | | 4282 | | |
| Clostridium sp. D5 (556261.3) | | 748 | 967, 4582 | 1045, 4581 | | 3342, 819 | 4585, 4586, 4586 | | | | 820, 821, 822, 2033, 2034, 2037, 2038, 2039 | | | | |
| Clostridium sp. HGF2 (908340.3) | | | 1346, 1992, 2904 | 1341, 2679, 2681, 2682 | | 1474, 2964 | | | | | 2962, 2965, 2966, 2967 | | | | |
| Clostridium sp. L2-50 (411489.7) | | | | | | | | | | | | | 1103, 2279 | | |
| Clostridium sp. M62/1 (411486.3) | | | 2355 | 2354 | | | 2007, 2007, 2009 | | | | | | 883 | | |
| Clostridium sp. SS2/1 (411484.7) | | | 986 | 2340 | | | 979, 980, 980 | | | | | | 1928 | | |
| Clostridium spiroforme DSM 1552 (428126.7) | | | | 1540 | | | 366, 367, 367 | | | | | 436, 1806, 1927, 2020, 2021 | 110, 632, 1502, 1503 | 2029 | Deacetylase is not found |
| Clostridium sporogenes ATCC 15579 (471871.7) | | | 804 | 805, 1535 | | | 799, 799, 800, 895, 895, 895 | | 2051, 2052 | | | | | | |
| Clostridium symbiosum WAL-14163 (742740.3) | | | 2736 | 2149, 4897 | | 366 | | | | | | | 2861 | | |
| Clostridium symbiosum WAL-14673 (742741.3) | | | 3393 | 2975, 3644 | | 150 | | | | | | | 3939 | | |
| Collinsella aerofaciens ATCC 25986 (411903.6) | | | | 658 | 2047 | | | | 2048, 2049, 2050, 2059 | | | 769 | | | |
| Collinsella intestinalis DSM 13280 (521003.7) | | | | 314 | | | | | | | | | | | |
| Collinsella stercoris DSM 13279 (445975.6) | | | 635 | 294, 636 | 127 | | | | 128, 129, 130, 132 | | | 284, 945, 1661 | 363, 1473 | 581 | Kinase is not found |
| Collinsella tanakaei YIT 12063 (742742.3) | | | | 1224 | | | | | | | | | | | |
| Coprobacillus sp. 29_1 (469596.3) | | | 1355, 1361, 2040, 2663 | 2041, 3024, 3096, 3832 | | 1527 | 249, 250, 250 | | | | 1528, 1530, 1531, 1532 | 242 | | | |
| Coprobacillus sp. 3_3_56FAA (665941.3) | 2598 | 1314 | 2932, 3323 | 1994, 2330, 2561, 3427 | | 1492 | 669, 670, 670, 1891, 1891, 1891, 2636, 2636, 2636 | | | | 1483, 1487, 1488, 1489, 1491 | 940, 941, 946, 3081, 3097, 3098 | 2379, 2856, 3052 | | |
| Coprobacillus sp. 8_2_54BFAA (469597.4) | 2541 | 414 | 1380, 3382 | 1035, 2587, 2822, 3276 | | 596 | 933, 933, 933, 1837, 1837, 1838, 2502, 2502, 2502 | | | | 586, 591, 592, 593, 595 | 3167, 3174, 3175, 3518, 3519, 3535, 3563 | 1455, 2774 | | |
| Coprobacillus sp. D6 (556262.3) | | | 285, 2085, 3380, 3386 | 2086, 2441, 2512, 3558 | | 3551 | 1429, 1429, 1430 | | | | 3552, 3554, 3555, 3556 | 1437 | | | |
| Coprococcus catus GD/7 (717962.3) | | | 1169 | 1168 | | | | | | | | | 894 | | Kinase is not found |
| Coprococcus comes ATCC 27758 (470146.3) | | | 921 | 920 | | | 922, 923, 924, 925 | | | | 1105, 1106, 1107, 1108 | | | | |
| Coprococcus eutactus ATCC 27759 (411474.6) | | | | | | | | | | | | | 1610 | | |
| Coprococcus sp. ART55/1 (751585.3) | | | | | | | | | | | | | 1312 | | |
| Coprococcus sp. HPP0048 (1078091.3) | | | 1628 | 2398 | | 2284 | 2658, 2658, 2659, 2662 | | | | 2290, 2291, 2292 | | 2618 | | |
| Coprococcus sp. HPP0074 (1078090.3) | | | 2570 | 2160 | | 2064 | 2799, 2799, 2801 | | | | 2070, 2071, 2072 | | 2765, 2846 | | |
| Corynebacterium ammoniagenes DSM 20306 (649754.3) | | | 1162 | 1163, 1365 | | | 1366, 1366, 1366 | | | | | | 1483 | | |
| Corynebacterium sp. HFH0082 (1078764.3) | | | | 1058 | | | | | | | | | | | |
| Dermabacter sp. HFH0086 (1203568.3) | | | 1885 | 1448 | | | 1303, 1303, 1303 | | | | | | | | |
| Desulfitobacterium hafniense DP7 (537010.4) | | | | 2421 | | | | | 3307 | | | | 2987 | | |
| Desulfovibrio piger ATCC 29098 (411464.8) | | | | | | | | | | | | | | | |
| Desulfovibrio sp. 3_1_syn3 (457398.5) | | | | | | | | | | | | | 2651 | | |
| Dorea formicigenerans 4_6_53AFAA (742765.5) | | | 2580 | 446, 2581 | | | | | 703 | | | | | | Kinase is not found |
| Dorea formicigenerans ATCC 27755 (411461.4) | | | 298 | 291, 2352 | | | | | 2626, 2627 | | | | 351 | | Kinase is not found |

| Organism | | | | | | | | | | | | |
|---|---|---|---|---|---|---|---|---|---|---|---|---|
| Dorea longicatena DSM 13814 (411462.6) | | 1626 | 1532, 1619 | | | | | | | | | |
| Dysgonomonas gadei ATCC BAA-286 (742766.3) | 692 | 2316 | 2317, 2594, 2595 | | | | | | | | 1924, 4134 | 2324, 3483 |
| Dysgonomonas mossi DSM 22836 (742767.3) | 2584 | | 660, 661 | | | | | | | | 1263, 1693 | 48, 1500 |
| Edwardsiella tarda ATCC 23685 (500638.3) | 1136 | 1782 | 1781 | | | 1780, 1780, 1780 | | | | | 1775 | |
| Eggerthella sp. 1_3_56FAA (665943.3) | | 2466 | 2467 | | | | | | | | | |
| Eggerthella sp. HGA1 (910311.3) | | 209 | 208 | | | | | | | | | |
| Enterobacter cancerogenus ATCC 35316 (500639.8) | 942 | 2210 | 2211 | | | 2212, 2212, 2212 | | | | 4025 | | |
| Enterobacter cloacae subsp. cloacae NCTC 9394 (718254.4) | | 1109 | 1518 | | | 1108, 1108, 1108 | | | | 3658 | | |
| Enterobacteriaceae bacterium 9_2_54FAA (469613.3) | | 358 | 359 | | | 360, 360, 360 | | | | 364 | | |
| Enterococcus faecalis PC1.1 (702448.3) | | 45 | 1967 | | | 1848, 1848, 1848 | | | | | | |
| Enterococcus faecalis TX0104 (491074.3) | | 15 | 2461, 2476 | 437 | | 139, 139, 139 | 433, 434, 435, 436 | 2298, 2299, 2300 | | | 2801 | |
| Enterococcus faecalis TX1302 (749513.3) | | 2003 | 133 | 2416 | | 2138, 2138, 2138, 2138, 2489, 2489, 2489 | 2412, 2413, 2414, 2415 | 1499, 1500 | | | 506 | |
| Enterococcus faecalis TX1322 (525278.3) | | 703 | 2324 | 298 | | 583, 583, 583 | 299, 300, 301, 302 | 1350, 1351 | | | 2575 | |
| Enterococcus faecalis TX1341 (749514.3) | | 2552 | 1562 | 1918 | | 498, 498, 498 | 1919, 1920, 1921, 1922 | 2836, 2837 | | | 1112 | |
| Enterococcus faecalis TX1342 (749515.3) | | 2106 | 1412, 1427 | 2519 | | 2233, 2233, 2233 | 2515, 2516, 2517, 2518 | 1957, 1958 | | | | |
| Enterococcus faecalis TX1346 (749516.3) | | 537 | 1072, 1088 | 1836 | | 2628, 2628, 2628 | 1837, 1838, 1839, 1840 | 1257, 1258 | | | 2280 | |
| Enterococcus faecalis TX2134 (749518.3) | | 2963 | 1085 | 1253 | | 579, 579, 579 | 1254, 1255, 1256, 1257 | 2660 | | | 1827 | |
| Enterococcus faecalis TX2137 (749491.3) | | 877 | 1795 | 306 | | 752, 752, 752 | 302, 303, 304, 305 | 2801, 2802 | | | 1356 | |
| Enterococcus faecalis TX4244 (749494.3) | | 1240 | 2109 | 1652 | | 1381, 1381, 1381 | 1648, 1649, 1650, 1651 | 2253, 2254 | | | 1917 | |
| Enterococcus faecium PC4.1 (791161.5) | | 46 | 2035 | | | 1910, 1910, 1910 | | | | | | |
| Enterococcus faecium TX1330 (525279.3) | | 1138 | 1885 | | | 2166, 2166, 2166 | | | | | | |
| Enterococcus saccharolyticus 30_1 (742813.4) | | 2285 | 1520 | 1809 | | | 1810, 1811, 1812, 1813 | | | | 981 | |
| Enterococcus sp. 7L76 (657310.3) | | 2584 | 1762, 1776 | 240 | | 2773, 2773, 2773 | 236, 237, 238, 239 | 545 | | | | |
| Erysipelotrichaceae bacterium 2_2_44A (457422.3) | | 785, 2394, 3564 | 791, 1901, 1903, 1904, 2307 | | 1268, 4021 | | | | | 3495, 3496, 3497, 3500 | | |
| Erysipelotrichaceae bacterium 21_3 (658657.3) | | 1077, 2652, 3964 | 1072, 2200, 2202, 2203, 2568 | | 3498, 531 | | | | | 4019, 4022, 4023, 4024 | | |
| Erysipelotrichaceae bacterium 3_1_53 (658659.3) | | 53, 902, 1588 | 47, 3873 | | 1540, 305 | | | | | 1537, 1538, 1539, 1542 | | |
| Erysipelotrichaceae bacterium 5_2_54FAA (552396.3) | | 56, 1943, 2205 | 2201 | | 2260, 2266 | 73, 73 | | | | 2257, 2258, 2261, 2262, 2263 | 2652 | 2413, 2593 |
| Erysipelotrichaceae bacterium 6_1_45 (469614.3) | | 1143, 2928, 4186 | 399, 400, 402, 2847, 4181 | | 1202, 4306 | | | | | 1200, 1203, 1204, 1205 | | |
| Escherichia coli 4_1_47FAA (1127356.4) | | 479 | 480 | | | 482, 482, 482 | | | | | | |
| Escherichia coli MS 107-1 (679207.4) | 1836 | 2785 | 2784 | | | 2782, 2782, 2782 | | | | | | |

| Organism | | | | | | | | | | |
|---|---|---|---|---|---|---|---|---|---|---|
| Escherichia coli MS 115-1 (749537.3) | | 3985 | 3984 | | 3982, 3982, 3982 | | | | | |
| Escherichia coli MS 116-1 (749538.3) | 2112 | 4287 | 4288 | | 4290, 4290, 4290 | | | | | |
| Escherichia coli MS 119-7 (679206.4) | 871 | 4582, 5234 | 4583 | | 4585, 4585, 4585 | | | | | |
| Escherichia coli MS 124-1 (679205.4) | 4108 | 682 | 683 | | 685, 685, 685 | | | | | |
| Escherichia coli MS 145-7 (679204.3) | 3623 | 2594 | 2593 | | 2591, 2591, 2591 | | | | | |
| Escherichia coli MS 146-1 (749540.3) | 1147 | 2459 | 2458 | | 2456, 2456, 2456 | | | | | |
| Escherichia coli MS 175-1 (749544.3) | 1972 | | | | | | | | | |
| Escherichia coli MS 182-1 (749545.3) | 839 | 4959 | 4958 | | 4956, 4956, 4956 | | | | | |
| Escherichia coli MS 185-1 (749546.3) | 4731 | 129 | 130 | | 132, 132, 132 | | | | | |
| Escherichia coli MS 187-1 (749547.3) | 666 | 3717 | 3718 | | 3720, 3720, 3720 | | | | | |
| Escherichia coli MS 196-1 (749548.3) | 5030 | 4456 | 4455 | | 4453, 4453, 4453 | | | | | |
| Escherichia coli MS 198-1 (749549.3) | 1124 | 4821 | 4822 | | 4824, 4824, 4824 | | | | | |
| Escherichia coli MS 200-1 (749550.3) | 2046 | 4782 | 4781 | | 4779, 4779, 4779 | | | | | |
| Escherichia coli MS 21-1 (749527.3) | 3367 | 2619 | 2618 | | 2616, 2616, 2616 | | | | | |
| Escherichia coli MS 45-1 (749528.3) | 764 | 3971 | 3970 | | 3968, 3968, 3968 | | | | | |
| Escherichia coli MS 69-1 (749531.3) | 3045 | 4037 | 4036 | | 4034, 4034, 4034 | | | | | |
| Escherichia coli MS 78-1 (749532.3) | | 3612 | 3613 | | 3615, 3615, 3615 | | | | | |
| Escherichia coli MS 79-10 (796392.4) | 4513 | 2328 | 2329 | | 2331, 2331, 2331 | | | | | |
| Escherichia coli MS 84-1 (749533.3) | 650 | 5347 | 5348 | | 5350, 5350, 5350 | | | | | |
| Escherichia coli MS 85-1 (679202.3) | 4732 | 2801 | 2800 | | 2798, 2798, 2798 | | | | | |
| Escherichia coli O157:H7 str. Sakai (386585.9) | 1599 | 819 | 820 | | 822, 822, 822 | | | | | |
| Escherichia coli SE11 (409438.11) | 1320 | 861 | 862 | | 864, 864, 864 | | | | | |
| Escherichia coli SE15 (431946.3) | 1068 | 631 | 632 | | 634, 634, 634 | | | | | |
| Escherichia coli str. K-12 substr. MG1655 (511145.12) | 1165 | 702 | 703 | | 705, 705, 705 | | | | | |
| Escherichia coli UTI89 (364106.8) | 1322 | 766 | 767 | | 769, 769, 769 | | | | | |
| Escherichia sp. 1_1_43 (457400.3) | | | | | | | | | | |
| Escherichia sp. 3_2_53FAA (469598.5) | 819 | 341 | 340 | | 338, 338, 338 | | | | | |
| Escherichia sp. 4_1_40B (457401.3) | 1071 | 1840 | 1839 | | 1837, 1837, 1837 | | | | | |
| Eubacterium biforme DSM 3989 (518637.5) | | 272 | 30 | 1074 | 2113, 2113, 2113 | | 1075, 1077, 1078, 1079 | | | |
| Eubacterium cylindroides T2-87 (717960.3) | | | 1444 | 1442, 1443 | | | 1438, 1439, 1441 | | | |
| Eubacterium dolichum DSM 3991 (428127.7) | | 276, 845 | 282 | | | | | 721, 1589 | 44, 1823, 1958 | 31 |
| Eubacterium hallii DSM 3353 (411469.3) | | 1642 | 1639 | | | | | | 1337 | |
| Eubacterium rectale DSM 17629 (657318.4) | | 602 | 601 | 2888 | 603, 604, 604 | | 2884, 2885, 2886 | | 2338 | |
| Eubacterium rectale M104/1 (657317.3) | | 1724 | 1725 | 959 | 1722, 1722, 1723 | | 955, 956, 957, 2443, 2444, 2445 | | 1569 | |
| Eubacterium siraeum 70/3 (657319.3) | | | | | | | | | | |
| Eubacterium siraeum DSM 15702 (428128.7) | | | | | | | | | 2364 | |
| Eubacterium siraeum V10Sc8a (717961.3) | | | | | | | | | 2382 | |
| Eubacterium sp. 3_1_31 (457402.3) | | 303, 1628, 2852 | 307 | 231, 238 | 2834, 2834, 2834 | | 235, 236, 237, 240 | | 1992 | |
| Eubacterium ventriosum ATCC 27560 (411463.4) | | 865 | 864, 1748, 2197 | | 861, 862, 862 | | | | | |
| Faecalibacterium cf. prausnitzii KLE1255 (748224.3) | | | 7 | 2586 | 1728, 1728, 1728 | | 2582, 2583, 2584 | | | |
| Faecalibacterium prausnitzii A2-165 (411483.3) | | 104 | 105 | 1155 | 106, 106 | 137 | 1151, 1152, 1153 | | 152 | |

| Organism | | | | | | | | | | | | | |
|---|---|---|---|---|---|---|---|---|---|---|---|---|---|
| Faecalibacterium prausnitzii L2-6 (718252.3) | | | | 1289 | 113 | | 115, 116, 117, 3087, 3088, 3089 | | | | | | |
| Faecalibacterium prausnitzii M21/2 (411485.10) | | | 1937 | 498, 1059 | 2375 | 1936, 1936, 1936, 1938 | 2377, 2378, 2379 | | | | | | |
| Faecalibacterium prausnitzii SL3/3 (657322.3) | | | 1777 | 847, 901 | 2621 | 1776, 1776, 1778 | 2617, 2618, 2619 | | | | | | |
| Fusobacterium gonidiaformans ATCC 25563 (469615.3) | | | 53 | 1472 | | 374, 374, 374 | | | | | | | |
| Fusobacterium mortiferum ATCC 9817 (469616.3) | | | 1073, 1669 | 1668 | | 2480, 2480, 2480 | | | | | | | |
| Fusobacterium necrophorum subsp. funduliforme 1_1_36S (742814.3) | | | 1309, 1310 | 1481, 1482 | | 1332, 1332, 1332, 2260, 2260, 2260 | | | | | | | |
| Fusobacterium sp. 1_1_41FAA (469621.3) | | | 980 | 1079 | | 1513, 1513, 1513 | | | | | | | |
| Fusobacterium sp. 11_3_2 (457403.3) | | | 178 | 442 | | 1444, 1444, 1444 | | | | | | | |
| Fusobacterium sp. 12_1B (457404.3) | | | 625 | 1423 | | 47, 47, 47, 1116, 1116, 1116, 1267, 1267, 1267, 2580, 2580, 2580 | | | | | | | |
| Fusobacterium sp. 2_1_31 (469599.3) | | | 525 | 398 | | 1457, 1457, 1457 | | | | | | | |
| Fusobacterium sp. 21_1A (469601.3) | | | 733 | 933 | | 198, 198, 198 | | | | | | | |
| Fusobacterium sp. 3_1_27 (469602.3) | | | 828 | 818 | | 1665, 1665, 1665 | | | | | | | |
| Fusobacterium sp. 3_1_33 (469603.3) | | | 768 | 550 | | 1391, 1391, 1391 | | | | | | | |
| Fusobacterium sp. 3_1_36A2 (469604.3) | | | 1169 | 1179 | | 509, 509, 509 | | | | | | | |
| Fusobacterium sp. 3_1_5R (469605.3) | | | 150 | 72 | | 353, 353, 353 | | | | | | | |
| Fusobacterium sp. 4_1_13 (469606.3) | | | 1527 | 1537 | | 722, 722, 722 | | | | | | | |
| Fusobacterium sp. 7_1 (457405.3) | | | 780 | 470 | | 1659, 1659, 1659 | | | | | | | |
| Fusobacterium sp. D11 (556264.3) | | | 1986 | 239 | | 1019, 1019, 1019 | | | | | | | |
| Fusobacterium sp. D12 (556263.3) | | | 1492 | 1065 | | 787, 787, 787 | | | | | | | |
| Fusobacterium ulcerans ATCC 49185 (469617.3) | | | 195 | 2144 | | 654, 654, 654, 1200, 1200, 1200, 2284, 2284, 2284 | | | | | | | |
| Fusobacterium varium ATCC 27725 (469618.3) | | | 2033 | 2614 | | 1725, 1725, 1725, 2388, 2388, 2388, 2433, 2433, 2849, 2849, 2849 | | | | | | | |
| Gordonibacter pamelaeae 7-10-1-b (657308.3) | | | | 1694 | | | | | | | | | |
| Hafnia alvei ATCC 51873 (1002364.3) | | | 4024 | 2230, 4025 | | 4026, 4026, 4026 | | | 4030 | | | | |
| Helicobacter bilis ATCC 43879 (613026.4) | | | | | | | | | | | | | |
| Helicobacter canadensis MIT 98-5491 (Prj:30719) (537970.9) | | | | | | | | | | | | | |
| Helicobacter cinaedi ATCC BAA-847 (1206745.3) | | | | | | | | | | | | | |
| Helicobacter cinaedi CCUG 18818 (537971.5) | | | | | | | | | | | | | |
| Helicobacter pullorum MIT 98-5489 (537972.5) | | | | | | | | | | | | | |
| Helicobacter winghamensis ATCC BAA-430 (556267.4) | | | | | | | | | | | | | |
| Holdemania filiformis DSM 12042 (545696.5) | | | 1184, 1185, 2175 | 1188, 1784, 2076 | | 2165, 2165, 2165 | | | 1238 | | | | |
| Klebsiella pneumoniae 1162281 (1037908.3) | | | 2760 | 2761 | | 2762, 2762, 2762 | | | | | | | |

| | | | | | | | | | | | | |
|---|---|---|---|---|---|---|---|---|---|---|---|---|
| Klebsiella pneumoniae subsp. pneumoniae WGLW5 (1203547.3) | | | 849 | 850 | | 851, 851, 851 | | | | | | |
| Klebsiella sp. 1_1_55 (469608.3) | | 2261 | 1792 | 1793 | | 1794, 1794, 1794 | | | | | | |
| Klebsiella sp. 4_1_44FAA (665944.3) | | | 1579 | 1580 | | 1581, 1581, 1581 | | | | | | |
| Lachnospiraceae bacterium 1_1_57FAA (658081.3) | | | | 1668 | 613 | | 605, 606, 615, 616, 617, 620 | | 471, 480, 1560 | | 1077 | |
| Lachnospiraceae bacterium 1_4_56FAA (658655.3) | | | 2720 | 941 | 359 | 938, 938, 939 | 53, 54, 55, 56 | | | 1305, 1436, 2820 | | |
| Lachnospiraceae bacterium 2_1_46FAA (742723.3) | | | 1212 | 1213 | 1802 | 481, 482, 482 | | | 1316, 1389 | 1388, 1390 | 1922 | |
| Lachnospiraceae bacterium 2_1_58FAA (658082.3) | | | 2454 | 618 | 1677 | 615, 615, 615 | 1671, 1683, 2212, 2213, 2214 | | | 507 | | |
| Lachnospiraceae bacterium 3_1_46FAA (665950.3) | | | | 1250 | 2413 | | 2406, 2407, 2415, 2416, 2417, 2420 | | 1149, 2283 | | 2534 | |
| Lachnospiraceae bacterium 3_1_57FAA_CT1 (658086.3) | | | 7342 | 7343 | 2098, 6940 | 7344, 7344, 7347 | 2099, 2100, 2101 | | 4301 | 998, 3499 | | |
| Lachnospiraceae bacterium 4_1_37FAA (552395.3) | | | 1969 | 2419 | 2536 | 2948, 2948, 2949, 2952 | 2527, 2528, 2529 | | | 1742, 2355, 2915 | | |
| Lachnospiraceae bacterium 5_1_57FAA (658085.3) | | | 1669 | 241 | 243 | | 237, 238, 239, 1688, 1689, 1690 | | | | | |
| Lachnospiraceae bacterium 5_1_63FAA (658089.3) | | | 982 | 1692 | | 983, 983, 984 | | | | 2486 | | |
| Lachnospiraceae bacterium 6_1_63FAA (658083.3) | | | 1271, 2492 | 1272, 1614 | 642 | 1066, 1066, 1067 | 646, 647, 1608, 1609, 1610 | | | | | |
| Lachnospiraceae bacterium 7_1_58FAA (658087.3) | | | 3534, 3860 | 3535, 3861 | | | | | | | | |
| Lachnospiraceae bacterium 8_1_57FAA (665951.3) | | | | 1984 | 851 | | 844, 847, 848, 849, 857, 858 | | 980, 2091 | | 1478 | |
| Lachnospiraceae bacterium 9_1_43BFAA (658088.3) | | | 1719 | 2131 | 2228 | 2582, 2582, 2583, 2586 | 2220, 2221, 2222 | | | 31, 186, 2068, 2545, 2634, 2639 | | |
| Lactobacillus acidophilus ATCC 4796 (525306.3) | 39 | | | 1612 | | | | | | | | |
| Lactobacillus acidophilus NCFM (272621.13) | 1433 | | 140 | 1852 | | | | 1564 | | | | |
| Lactobacillus amylolyticus DSM 11664 (585524.3) | 417 | | 1310 | 909 | | | | | | 1618 | | Transporter is not found |
| Lactobacillus antri DSM 16041 (525309.3) | | | 1891 | 1892 | | | | | | | | |
| Lactobacillus brevis ATCC 367 (387344.15) | | | 520 | 666 | | | | | | | | |
| Lactobacillus brevis subsp. gravesensis ATCC 27305 (525310.3) | | | 2620 | | | | | | | 1041 | | |
| Lactobacillus buchneri ATCC 11577 (525318.3) | | | 1815 | | | | | | | 778 | | |
| Lactobacillus casei ATCC 334 (321967.11) | | | 1787 | 2831 | | | | | | 315 | | |
| Lactobacillus casei BL23 (543734.4) | | | 1929 | 2971 | 276 | 277, 278, 279, 280 | | | | 271 | | |
| Lactobacillus crispatus 125-2-CHN (575595.3) | 764 | | 1861 | 1510 | | | | | | | | Transporter is not found |
| Lactobacillus delbrueckii subsp. bulgaricus ATCC 11842 (390333.7) | | | 177 | 1918 | | | | | | | | |
| Lactobacillus delbrueckii subsp. bulgaricus ATCC BAA-365 (321956.7) | | | 171 | 1913 | | | | | | | | Transporter is not found |
| Lactobacillus fermentum ATCC 14931 (525325.3) | | | | | | | | | | | | |
| Lactobacillus fermentum IFO 3956 (334390.5) | | | | | | | | | | | | |
| Lactobacillus gasseri ATCC 33323 (324831.13) | 1559 | | | 1776 | 17 | 116, 117, 118, 119, 124 | | | | | | |
| Lactobacillus helveticus DPC 4571 (405566.6) | 1656, 1657 | | | 2205 | | | | | | | | |
| Lactobacillus helveticus DSM 20075 (585520.4) | 1317 | | | 60 | | | | | | | | |
| Lactobacillus hilgardi ATCC 8290 (525327.3) | | | 2003 | | | | | | | 653 | | |
| Lactobacillus johnsonii NCC 533 (257314.6) | 1610 | | | 1824 | 16 | 128, 129, 130, 131, 135 | | | | | | |
| Lactobacillus paracasei subsp. paracasei 8700:2 (537973.8) | | | 835 | 1126 | 2333 | 2334, 2335, 2336, 2337 | | | | 2328, 2784 | | |
| Lactobacillus paracasei subsp. paracasei ATCC 25302 (525337.3) | | | 687 | 1689 | | 1132, 1133, 1134 | | | | 1050 | | |
| Lactobacillus plantarum 16 (1327988.3) | | | 501 | 207 | | 2068, 2068, 2068, 2408, 2408, 2408 | | | | | | |

|  |  |  |  |  |  |  |  |  |  |  |  |
| --- | --- | --- | --- | --- | --- | --- | --- | --- | --- | --- | --- |
| Lactobacillus plantarum subsp. plantarum ATCC 14917 (525338.3) |  | 1151 | 160 |  |  | 990, 990, 990, 991, 991, 991, 2851, 2851, 2851 |  |  |  |  |  |
| Lactobacillus plantarum WCFS1 (220668.9) |  | 466 | 188 |  |  | 2129, 2129, 2129, 2478, 2478, 2478 |  |  |  |  |  |
| Lactobacillus reuteri CF48-3A (525341.3) |  |  | 1764 |  |  |  |  |  |  |  |  |
| Lactobacillus reuteri DSM 20016 (557436.4) |  |  | 1035 |  |  |  |  |  |  |  |  |
| Lactobacillus reuteri F275 (299033.6) |  |  | 1010 |  |  |  |  |  |  |  |  |
| Lactobacillus reuteri MM2-3 (585517.3) |  |  | 1454 |  |  |  |  |  |  |  |  |
| Lactobacillus reuteri MM4-1A (548485.3) |  |  | 967 |  |  |  |  |  |  |  |  |
| Lactobacillus reuteri SD2112 (491077.3) |  |  | 1846 |  |  |  |  |  |  |  |  |
| Lactobacillus rhamnosus ATCC 21052 (1002365.5) |  | 1920 | 1648 | 427, 428 |  | 429, 430, 431, 432 |  |  |  |  |  |
| Lactobacillus rhamnosus GG (Prj:40637) (568703.30) |  | 1869 | 2936 | 336, 337 |  | 338, 340, 341, 342 |  |  | 435 |  |  |
| Lactobacillus rhamnosus LMS2-1 (525361.3) |  | 799 | 1698 | 1786 |  | 1787, 1788, 1789, 1791 |  |  |  |  |  |
| Lactobacillus ruminis ATCC 25644 (525362.3) |  | 2024 | 234 |  |  |  |  |  | 602 |  |  |
| Lactobacillus sakei subsp. sakei 23K (314315.12) |  | 1555 | 409 |  |  |  |  |  |  |  |  |
| Lactobacillus salivarius ATCC 11741 (525364.3) |  | 1583 | 1584 |  |  | 58, 58, 58 |  |  | 1471, 1472 |  |  |
| Lactobacillus salivarius UCC118 (362948.14) |  | 1549 | 1548 |  |  |  |  |  | 1663 |  |  |
| Lactobacillus sp. 7_1_47FAA (665945.3) | 1083 | 617 | 718 | 849 |  | 850, 851, 852, 853 | 1052 |  |  |  |  |
| Lactobacillus ultunensis DSM 16047 (525365.3) | 1503 | 1964 | 1115 |  |  |  |  |  |  |  | Transporter is not found |
| Leuconostoc mesenteroides subsp. cremoris ATCC 19254 (586220.3) |  | 791 | 850 |  |  |  |  |  |  |  |  |
| Listeria grayi DSM 20601 (525367.9) |  | 963 | 754, 964, 2263 |  |  |  |  |  |  |  |  |
| Listeria innocua ATCC 33091 (1002366.3) |  | 1546 | 800, 1468, 1547, 2734 |  |  |  | 49, 50, 51, 2452, 2453 |  |  |  | Kinase is not found |
| Megamonas funiformis YIT 11815 (742816.3) |  | 1644 | 2074 |  |  | 1536, 1536 |  |  | 1387, 1388 |  |  |
| Megamonas hypermegale ART12/1 (657316.3) |  | 440 | 1094 |  |  | 2021, 2021 |  |  | 1826 |  |  |
| Methanobrevibacter smithii ATCC 35061 (420247.6) |  |  |  |  |  |  |  |  |  |  |  |
| Methanosphaera stadtmanae DSM 3091 (339860.6) |  |  |  |  |  |  |  |  |  |  |  |
| Mitsuokella multacida DSM 20544 (500635.8) |  | 348 | 449 |  |  |  |  |  | 637 |  |  |
| Mollicutes bacterium D7 (556270.3) |  | 2290 | 1547, 2043 | 882, 2580, 3191, 3284 | 1624 | 46, 46, 46, 404, 405, 405 | 1615, 1619, 1620, 1621, 1623 | 1910, 1927, 1928, 2482, 2483 | 835, 1175, 1469, 1743, 1881 |  |  |
| Odoribacter splanchnicus DSM 20712 (709991.3) |  |  | 2356 |  |  |  |  | 247, 348, 2552, 2553 | 3376 |  |  |
| Oxalobacter formigenes HOxBLS (556268.6) |  |  |  |  |  |  |  |  |  |  |  |
| Oxalobacter formigenes OXCC13 (556269.4) |  |  |  |  |  |  |  |  |  |  |  |
| Paenibacillus sp. HGF5 (908341.3) | 3284, 6169 | 2275, 2485, 5378 | 2274, 2486, 5379 |  |  | 390, 390, 390, 1379, 1379, 1379 |  |  | 3687, 5435 |  |  |
| Paenibacillus sp. HGF7 (944559.3) |  | 3153 | 3152 |  |  | 1305, 1305, 1305, 3133, 3133, 3133 |  | 2816 | 412, 1976, 3722 |  |  |
| Paenibacillus sp. HGH0039 (1078505.3) | 2034 | 4055 | 4400 | 4056 |  | 896, 896, 896, 4076, 4076 |  | 330 | 1240, 2100, 3354 |  |  |
| Parabacteroides distasonis ATCC 8503 (435591.13) |  | 2842, 3552 | 225, 226, 3115 | 227, 3530 |  |  |  | 46, 652, 1252, 2504, 2909 | 2321 |  |  |
| Parabacteroides johnsonii DSM 18315 (537006.5) |  | 3317 | 2442, 2443, 2444 | 2441, 3494 |  |  |  | 398, 608, 651, 792, 1061, 1113, 1922, 3536 | 2863 | 353 |  |
| Parabacteroides merdae ATCC 43184 (411477.4) | 1578 | 3334, 3335 | 1695, 3333 |  |  |  |  | 248, 288, 370, 401, 433, 862, 1720, 1922, 2721 | 728 | 200, 1856 |  |
| Parabacteroides sp. D13 (563193.3) | 11, 432 | 1058, 3904, 3905 | 411, 3903 |  |  |  |  | 743, 1422, 1983, 3485, 4031 | 2866 |  |  |

| Organism | | | | | | | | | | | | |
|---|---|---|---|---|---|---|---|---|---|---|---|---|
| Parabacteroides sp. D25 (658661.3) | | 2922 | 807, 3683, 3684 | | 2944, 3685 | | | | | 144, 416, 1344, 2458, 3505, 4154 | | |
| Paraprevotella clara YIT 11840 (762968.3) | | | | | | | | | | | | |
| Paraprevotella xylaniphila YIT 11841 (762982.3) | | | | | | | | | | | | |
| Parvimonas micra ATCC 33270 (411465.10) | | 682 | | | | | | | | | | |
| Pediococcus acidilactici 7_4 (563194.3) | | | 1492 | | 1057 | 188 | 187, 189, 190, 191 | | | | | |
| Pediococcus acidilactici DSM 20284 (862514.3) | | | 965 | | 1780 | 1203 | 1199, 1200, 1201, 1202, 1204 | | | | | |
| Phascolarctobacterium sp. YIT 12067 (626939.3) | | | | | | | | | | | | |
| Prevotella copri DSM 18205 (537011.5) | | 384, 2765 | 642, 771, 4375, 4510 | | | | | | | | 298, 4027 | |
| Prevotella oralis HGA0225 (1203550.3) | | 208 | | | 2046, 2049 | | | | | 1139, 2050, 2182 | 1976 | |
| Prevotella salivae DSM 15606 (888832.3) | | 2547 | 1558, 1559 | | | | | | | 518, 1151, 1565, 1571 | | |
| Propionibacterium sp. 5_U_42AFAA (450748.3) | 1698 | | 1020 | | 1012 | 572 | 1236 | | 575, 576, 577 | 553, 2413 | 252, 2246 | |
| Propionibacterium sp. HGH0353 (1203571.3) | | | 1668 | | 1660 | 2212 | | | | 1217 | | |
| Proteus mirabilis WGLW6 (1125694.3) | | | 387 | | 388 | | 389, 389, 389 | | | | | |
| Proteus penneri ATCC 35198 (471881.3) | | 1305 | 4620 | | 4619 | | 4617, 4617, 4617, 4618, 4618, 4618 | | | | | |
| Providencia alcalifaciens DSM 30120 (520999.6) | | 3402 | 3264 | | 3263 | | 3262, 3262, 3262 | | | | | |
| Providencia rettgeri DSM 1131 (521000.6) | | 1547 | 3087, 3112 | | 3086, 3113 | | 3114, 3114, 3114 | | | | 1693 | |
| Providencia rustigianii DSM 4541 (500637.6) | | 90, 6992 | 2653, 4581 | | 2654, 4580 | | 2655, 2655, 2655, 4579, 4579, 4579 | | | | | |
| Providencia stuartii ATCC 25827 (471874.6) | | 3144 | 756 | | 757 | | 758, 758, 758 | | | 966, 2757 | 3256 | |
| Pseudomonas sp. 2_1_26 (665948.3) | | | 480 | 479 | | | 477, 477, 477 | | | | | |
| Ralstonia sp. 5_7_47FAA (658664.3) | | | | | | | | | | | | |
| Roseburia intestinalis L1-82 (536231.5) | | | 2027 | | 2028 | 3250 | 2535, 2535, 2536 | 644 | 3251, 3252, 3253, 3255 | 1506 | 423 | |
| Roseburia intestinalis M50/1 (657315.3) | | | 1783 | | 1782 | 3559 | 3031, 3032, 3032 | 2089, 3161 | 3553, 3554, 3556, 3557, 3558 | | 796 | |
| Roseburia intestinalis XB6B4 (718255.3) | | | 3120, 3121 | | 3119 | 2521 | 1973, 1974, 1974, 1975, 1975, 1975 | 1188, 1308, 1315, 2076, 3368, 3369 | 2522, 2523, 2526 | 1473 | 2929 | |
| Roseburia inulinivorans DSM 16841 (622312.4) | | | 2448, 3995 | | 795, 2449 | 1688 | 3682, 3682, 3683 | | 1694, 1695, 1696 | | 2587 | |
| Ruminococcaceae bacterium D16 (552398.3) | | | 52 | | 51 | 347 | | | 343, 344, 345 | | 2512 | |
| Ruminococcus bromii L2-63 (657321.5) | | | | | | | | | | | 848 | |
| Ruminococcus gnavus ATCC 29149 (411470.6) | | | 3204 | | 1071, 2646 | 1054 | 2649, 2649, 2649 | | 1048, 1393, 1394, 1395 | | | |
| Ruminococcus lactaris ATCC 29176 (471875.6) | | | 2329 | | 1814 | 1843 | 1817, 1817, 1817 | | 1845, 1846, 1847 | 1632, 1871 | 1628, 1662, 2484 | 1299 |
| Ruminococcus obeum A2-162 (657314.3) | | | 1118 | | 3354 | 2912 | 3112, 3112, 3112 | | 3540, 3541, 3542 | | | |
| Ruminococcus obeum ATCC 29174 (411459.7) | | | 2341 | | 200 | 339 | | | 342, 343, 344, 347 | | | |
| Ruminococcus sp. 18P13 (213810.4) | | | | | | | | | | | | |
| Ruminococcus sp. 5_1_39FAA (457412.4) | | | 2698 | | 2699 | 611 | | | 1422, 1423, 1424 | | | |
| Ruminococcus sp. SR1/5 (657323.3) | | 1960 | 1214 | | 972 | 1418 | 2355, 2355, 2356 | | 1413, 1414, 1415 | | | |
| Ruminococcus torques ATCC 27756 (411460.6) | | | | | 1971 | 1844 | | | 1837, 1840, 1841, 1842, 1850 | | 22, 1627, 2034 | 1404 |
| Ruminococcus torques L2-14 (657313.3) | | | 808, 2281 | | 809, 2273 | | 776, 776 | | | | | |

| Organism | | | | | | | | | | | | | | |
|---|---|---|---|---|---|---|---|---|---|---|---|---|---|---|
| Salmonella enterica subsp. enterica serovar Typhimurium str. LT2 (99287.12) | 1289 | | 713 | | 714 | | 715, 715, 715 | | | | | | | |
| Staphylococcus sp. HGB0015 (1078083.3) | | | | | 56 | 1675 | 1676, 1677, 1678, 1679 | | | | | | | |
| Streptococcus anginosus 1_2_62CV (742820.3) | | | 1108 | | 858 | 1745 | 1746, 1747, 1748, 1749 | | | | | | | |
| Streptococcus equinus ATCC 9812 (525379.3) | | | 1465 | | 1070 | | | | | | | | | |
| Streptococcus infantarius subsp. infantarius ATCC BAA-102 (471872.6) | | | 176 | | 497, 1165 | | | | | | | | | |
| Streptococcus sp. 2_1_36FAA (469609.3) | | | 556 | | 1537 | 57 | 58, 59, 60, 61 | | | | | 419 | | |
| Streptococcus sp. HPH0090 (1203590.3) | | | 1120 | | 696 | 1446 | 1442, 1443, 1444, 1445 | | | | | 1002, 1048, 1448 | | |
| Streptomyces sp. HGB0020 (1078086.3) | 3801, 5253 | 3295, 6873 | 6874 | 3645 | 8192 | | | 5255, 5256, 5257 | | | 1167, 2955, 4922, 5130, 8394 | 2898, 3737, 5165, 8190 | | |
| Streptomyces sp. HPH0547 (1203592.3) | 4620, 4638 | 5571 | 5570 | 2839, 4043 | | | | | | | 389, 2269, 2386, 2503 | 43, 2838, 5766 | 1769 | Transporter is not found |
| Subdoligranulum sp. 4_3_54A2FAA (665956.3) | | | 303, 518, 4373 | | 297, 476, 4374 | 447 | | | | 443, 444, 445 | | 204, 1927, 2746, 3106, 4057 | | |
| Subdoligranulum variabile DSM 15176 (411471.5) | | | 1467, 1900, 4520, 4957 | | 1466, 1899, 4519, 4956 | 2375, 3110 | 1465, 1465, 1465, 4518, 4518, 4518 | | | 2377, 2378, 2379, 3112, 3113, 3114 | | 1056, 4229 | | |
| Succinatimonas hippei YIT 12066 (762983.3) | | | 1293 | | | | | | | | | | | |
| Sutterella wadsworthensis 3_1_45B (742821.3) | | | | | | | | | | | | | | |
| Sutterella wadsworthensis HGA0223 (1203554.3) | | | | | | | | | | | | | | |
| Tannerella sp. 6_1_58FAA_CT1 (665949.3) | | | | | 2008, 2009 | | | | | | 2544, 2545 | | 1827 | |
| Turicibacter sp. HGF1 (910310.3) | | | 2230, 2553 | | 1326, 2552 | 1099 | 2551, 2551 | | | 1100, 1101, 1110, 1116 | 302 | | | |
| Turicibacter sp. PC909 (702450.3) | | | 2042, 2454 | | 2286, 2455 | 384 | 2456, 2456 | | | 367, 373, 382, 383 | 806 | | | |
| Veillonella sp. 3_1_44 (457416.3) | | | | | | | | | | | | 839, 889 | | |
| Veillonella sp. 6_1_27 (450749.3) | | | | | | | | | | | | 879, 940 | | |
| Weissella paramesenteroides ATCC 33313 (585506.3) | | | 1380 | | 578 | | | 1538 | | | | | | Kinase is not found |
| Yokenella regensburgei ATCC 43003 (1002368.3) | | | 2353 | | 2354 | | 2355, 2355, 2355 | | | | 316 | | | |

**Table S6.** Genes involved in depletion and utilization of N-acetylglneuramic acid. The PubSEED identifiers are shown. For details on analyzed organisms see Table S1, for details on gene functions see Table S2.

| Organism (PubSEED ID) | Catabolic pathway I | Catabolic pathway II | | | | Common part of the pathways | | | | Transporters | | | | | Hydrolases | Problems (if any) |
| | Kinase I | Epimerase I | Epimerase II | Kinase II | | Lyase | Deacetylase | Deaminase | | Permease | Na$^+$ symporter | ABC Transporters | | TRAP transproter | Sialydase | |
| | NanK | NanE | NanE-II | NagK1 | NagK2 | NanA | NagA | NagB1 | NagB2 | NanT | NanX | NanABC | NanABC2 | NeuT | NanP | |
|---|---|---|---|---|---|---|---|---|---|---|---|---|---|---|---|---|
| Abiotrophia defectiva ATCC 49176 (592010.4) | | 606 | | | | | 1693, 2925 | | 1694, 2924 | | | | | | 193 | |
| Acidaminococcus sp. D21 (563191.3) | 1146 | | | | | | 918 | | | | | | | | | |
| Acidaminococcus sp. HPA0509 (1203555.3) | 1254 | | | | | | 1878 | | | | | | | | | |
| Acinetobacter junii SH205 (575587.3) | | | | | | | | | | | | | | | | |
| Actinomyces odontolyticus ATCC 17982 (411466.7) | 1055 | 1056 | | | 61, 1017 | 1054 | 1628 | 1016 | 59 | 276 | | | 1052 | | 277, 1482 | |
| Actinomyces sp. HPA0247 (1203556.3) | 1600 | 1599 | | | 52 | 1601 | 1177 | | 50 | 273 | | | 1604, 1605 | | 276, 1996 | |
| Akkermansia muciniphila ATCC BAA-835 (349741.6) | | | 2221 | | | 2220 | 1075 | | 2070 | | | | | | 707, 913, 2085 | |
| Alistipes indistinctus YIT 12060 (742725.3) | | | | | | 203 | 1108 | | 1109, 1793, 1795 | | | | | | | |
| Alistipes shahii WAL 8301 (717959.3) | | | | 2772 | | | | | 2404, 2405 | | | | | | 1555, 1556, 3051 | |
| Anaerobaculum hydrogeniformans ATCC BAA-1850 (592015.5) | | | | | | | | | | | | | | | | |
| Anaerococcus hydrogenalis DSM 7454 (561177.4) | | | | | | | | | | | | | | | | |
| Anaerofustis stercorihominis DSM 17244 (445971.6) | | | | | | | | | | | 795 | | | | | |
| Anaerostipes caccae DSM 14662 (411490.6) | | 1603 | | | | | 46 | | 47, 2358, 3068 | | | | | 1736 | | |
| Anaerostipes hadrus DSM 3319 (649757.3) | | 165 | | | | | 1969 | | 163 | | | | | 589 | | |
| Anaerostipes sp. 3_2_56FAA (665937.3) | | 1377 | | | | | 3347 | | 269, 3346 | | | | | 1546 | | |
| Anaerotruncus colihominis DSM 17241 (445972.6) | | 2088 | | | | 2004, 2082 | 2267 | | 2258, 2268, 2817 | | | 2083, 2084, 2085 | | | | |
| Bacillus smithii 7_3_47FAA (665952.3) | | | | | | | 1254 | | 78 | | | | | | | |
| Bacillus sp. 7_6_55CFAA_CT2 (665957.3) | 5162 | | | | | | 821 | | 822 | | | | | | | |
| Bacillus subtilis subsp. subtilis str. 168 (224308.113) | | | | | | | 3601 | | 235, 3602 | | | | | | | |
| Bacteroides caccae ATCC 43185 (411901.7) | | | 1025 | | 321 | 1026, 2256 | 771, 772 | | 1643, 2122 | 1027, 2258 | | | | | 166, 1022, 1985, 2257, 2262 | |
| Bacteroides capillosus ATCC 29799 (411467.6) | | | | | | | 2734 | | 1155, 2735 | | | | | | | |
| Bacteroides cellulosilyticus DSM 14838 (537012.5) | | | 3590 | 976, 2271 | | | 2425 | 3600 | 2424, 2830, 4912, 4996 | | | | | | | |
| Bacteroides clarus YIT 12056 (762984.3) | | | 425 | 2106, 2501 | 424 | 203, 337 | | | 336, 2802 | 423 | | | | | 426, 427, 431, 529 | |
| Bacteroides coprocola DSM 17136 (470145.6) | | 3706 | | | 1566 | | 538, 539 | | 41, 42 | | | | | | 1294 | |
| Bacteroides coprophilus DSM 18228 (547042.5) | | | 2023 | | 2945 | 2022 | 2344, 2345 | | 788, 789 | 2024 | | | | | 736, 787, 2026 | |
| Bacteroides dorei DSM 17855 (483217.6) | | | 657 | | 364 | 658 | 505, 639 | | 654, 655 | 656 | | | | | 629, 3327 | |
| Bacteroides eggerthii 1_2_48FAA (665953.3) | | | 534 | 2039, 3592 | 533 | 298, 626 | | | 297, 1611, 2158 | 532 | 1104 | | | | | |
| Bacteroides eggerthii DSM 20697 (483216.6) | | | 1918 | 54, 2641 | 1917 | 1738, 2008 | | | 1074, 1579, 1737 | 1916 | 2972, 2982 | | | | | |
| Bacteroides finegoldii DSM 17565 (483215.6) | | | 2263, 2279 | 3013, 3346 | | | 2724, 2725 | | 1318, 2171 | | | | | | 718, 2281, 4174 | |

| | | | | | | | | | | | | |
|---|---|---|---|---|---|---|---|---|---|---|---|---|
| Bacteroides fluxus YIT 12057 (763034.3) | | 3253 | 2480, 3111 | 3252 | 692, 2829 | 693, 1162, 1850 | 3254 | | | | 1561, 2077, 3269 | |
| Bacteroides fragilis 3_1_12 (457424.5) | | 368 | 2418 | 367 | 2892, 3090 | 1880, 3091 | 369 | | | | 386, 2166, 4179 | |
| Bacteroides fragilis 638R (862962.3) | | 1725, 4272 | 572 | 1724, 3935 | 2840, 3043 | 987, 3044 | 1726, 3930, 4273 | | | | 732, 1347, 1738, 3931, 4240, 4269, 4270 | |
| Bacteroides fragilis NCTC 9343 (272559.17) | | 1673, 4150 | 498 | 1672, 3796 | 2771, 2978 | 894, 2979 | 1674, 3791, 4151 | | | | 1290, 1777, 3792, 4126, 4148 | |
| Bacteroides fragilis YCH46 (295405.11) | | 1659, 4094 | 577 | 1658, 3784 | 2648, 2983 | 982, 2984 | 1660, 3779, 4095 | | | | 1311, 1678, 3780, 4064, 4158 | |
| Bacteroides intestinalis DSM 17393 (471870.8) | | 2564 | 168, 528 | | | 2678, 4191, 4686 | | | | | 3956, 4189 | |
| Bacteroides ovatus 3_8_47FAA (665954.3) | | 4733, 4759 | 3368, 4519 | 1898 | 1101, 1102, 2899 | 2900, 3004, 3613 | 1896 | | | | 1893, 1897, 2892, 4731, 5126 | |
| Bacteroides ovatus ATCC 8483 (411476.11) | | 429, 1654, 1674 | 541, 4656 | 430, 3571 | 126, 127, 2864 | 1505, 2863, 4190 | 427, 3569 | | | | 696, 1875, 2871, 3566, 3570 | |
| Bacteroides ovatus SD CC 2a (702444.3) | | | 434, 2613 | 2455 | 1675, 4485, 4486 | 80, 301, 1676 | 2453 | | | | 949, 1667, 2450, 2454, 3390 | |
| Bacteroides ovatus SD CMC 3f (702443.3) | | 719, 744 | 4854, 5435 | | 2084, 2579, 2580 | 1903, 2083, 4919 | | | | | 1162, 2091, 3529 | |
| Bacteroides pectinophilus ATCC 43243 (483218.5) | | | | | | | | | | | | |
| Bacteroides plebeius DSM 17135 (484018.6) | | 1286 | 2792 | 841, 1287 | 1144, 1145 | 1772, 1773 | 1285 | 842 | | | 837, 1283, 1870 | |
| Bacteroides sp. 1_1_14 (469585.3) | | 797, 798, 1052, 1122, 3539 | 1329, 4768 | 796 | 2384, 2385, 3522 | 884, 1890, 3521 | 799 | | | | 791, 1141, 1401, 2809, 3542 | |
| Bacteroides sp. 1_1_30 (457387.3) | | 354, 3674, 3693 | 2726, 4422 | 355, 4148 | 2, 3, 2647 | 2082, 2646, 2976 | 352, 4150 | | | | 2655, 3695, 4149, 4153, 5080 | |
| Bacteroides sp. 1_1_6 (469586.3) | | 2818, 3396 | 3762, 4640 | 2819 | 851, 852, 3379 | 387, 2281, 3378 | 2817 | 1482 | | | 133, 1655, 2824, 3399 | |
| Bacteroides sp. 20_3 (469591.4) | 1805 | 4190 | 658, 4134 | 4189 | 3151, 3353 | 1274, 3150 | 4191 | | | 3740 | 1608, 4194, 4547 | |
| Bacteroides sp. 2_1_16 (469587.3) | | 1878 | 2728 | 1877, 3843 | 22, 1957, 1958 | 23, 2634 | 1879, 3838 | | | | 1887, 3839, 4137 | |
| Bacteroides sp. 2_1_22 (469588.3) | | 2780, 2799 | 2254, 4134 | 3926 | 2237, 3381, 3382 | 1848, 3380 | 3924 | | | | 15, 886, 2801, 3921, 3925 | |
| Bacteroides sp. 2_1_33B (469589.3) | 2409 | 1631 | 1691, 1827 | 1632 | 4456, 4702 | 1060, 4455 | 1629, 1630 | | | 946 | 1626, 3009, 3596 | |
| Bacteroides sp. 2_1_56FAA (665938.3) | | 1909 | 547 | 1908, 3059 | 378, 3711, 3712 | 899, 3710 | 1910, 3064 | | | | 1928, 3063, 4158 | |
| Bacteroides sp. 2_1_7 (457388.5) | 1938 | 577 | 634, 878 | 578 | 3593, 5110, 5111 | 944, 1649, 3594 | 576 | | | | 573, 2518, 3113 | |
| Bacteroides sp. 2_2_4 (469590.5) | | 2219, 5686 | 383, 639 | 2220 | 140, 1154, 1155 | 635, 1156 | 2217 | | | | 3586, 4408, 5486 | |
| Bacteroides sp. 3_1_19 (469592.4) | 2439 | 813 | 11, 468 | 812 | 1012, 3516, 3517 | 448, 3518 | 814 | | | 3649 | 817, 1265, 1918 | |
| Bacteroides sp. 3_1_23 (457390.3) | | 4692, 4717 | 3071, 3882 | 1444 | 613, 614, 2739 | 953, 2738, 4269 | 1442 | | | | 40, 690, 1439, 1443, 2747, 4719 | |

| | | | | | | | | | | | |
|---|---|---|---|---|---|---|---|---|---|---|---|
| Bacteroides sp. 3_1_33FAA (457391.3) | | | 1692 | 1295 | 1691 | 1153, 1710 | 1694, 1695 | 1693 | | | 920, 1720 |
| Bacteroides sp. 3_1_40A (469593.3) | | | 3412 | 3670 | 3413 | 3394, 3823 | 3409, 3410 | 3411 | 4072 | | 2263, 3384, 4178 |
| Bacteroides sp. 3_2_5 (457392.3) | | | 3 | 609 | 4, 3258 | 1606, 1826 | 1565, 1827 | 2, 3253 | | | 1232, 3254, 3540, 4361 |
| Bacteroides sp. 4_1_36 (457393.3) | | | 1019 | 889 | | | 1614, 2968, 3374 | | | | |
| Bacteroides sp. 4_3_47FAA (457394.3) | | | 790 | 1599 | 791 | 774, 1459 | 787, 788 | 789 | 3113 | | 764, 2404, 2447 |
| Bacteroides sp. 9_1_42FAA (457395.6) | | | 398 | 54 | 397 | 417, 524 | 400, 401 | 399 | | | 427, 2498 |
| Bacteroides sp. D1 (556258.5) | | | 1542, 1561 | 3521, 3924 | 2332 | 2034, 2802, 2803 | 1395, 2035, 3157 | 2334 | | | 738, 1219, 1563, 2027, 2333, 2337 |
| Bacteroides sp. D2 (556259.3) | | | 397, 415 | 923, 4858 | 3430, 3431 | 2681, 4490, 4491 | 617, 1394, 2682 | 3433 | | | 395, 2673, 3432, 3436, 5119 |
| Bacteroides sp. D20 (585543.3) | | | 329 | 1646 | 3523 | | 411, 936, 2501 | | 3524 | | 3520 |
| Bacteroides sp. D22 (585544.3) | | | 1298, 1741 | 2774, 3855 | 1297, 1742, 2327 | 1563, 1564, 2093 | 2092, 2582, 4263 | 1296, 1739, 2325 | | | 1299, 1300, 1314, 1802, 2100, 2322, 2326, 3551 |
| Bacteroides sp. HPS0048 (1078089.3) | | 511 | | | 2880 | | 1377, 3765 | | | | 91, 2877, 3949 |
| Bacteroides stercoris ATCC 43183 (449673.7) | | | 1658 | 383, 3289 | 1657 | 97, 1575 | 1574, 2872 | 1656 | | | 1659, 1660, 1729 |
| Bacteroides thetaiotaomicron VPI-5482 (226186.12) | | | 436, 453, 3664 | 2545, 4734 | | 689, 690, 3647 | 259, 3646, 4194 | | | | 455, 2798, 3667 |
| Bacteroides uniformis ATCC 8492 (411479.10) | | | 2164 | 696 | 2787 | | 181, 1567, 1758 | | 2789 | | 2784 |
| Bacteroides vulgatus ATCC 8482 (435590.9) | | | 4229 | 198 | 4228 | 60, 4244 | 4231, 4232 | 4230 | | | 635, 2271, 4254 |
| Bacteroides vulgatus PC510 (702446.3) | | | 469 | 2291 | 470 | 760, 1928 | 466, 467 | 468 | | | 205, 770 |
| Bacteroides xylanisolvens SD CC 1b (702447.3) | | | 2059, 2740 | 3634, 4395 | 4031 | 2747, 2748, 4648 | 2400, 2508, 4647 | 4033 | | | 1718, 2312, 4032, 4036, 4278, 4657 |
| Bacteroides xylanisolvens XB1A (657309.4) | | | 3244, 3265, 3893 | 515, 2695 | 3461, 3892 | 1696, 2448, 2449 | 1695, 2032, 3239 | 3459, 3895 | | | 668, 1704, 3456, 3460, 4507 |
| Barnesiella intestinihominis YIT 11860 (742726.3) | | | | | | | 2704, 2705 | | | | 342, 732 |
| Bifidobacterium adolescentis L2-32 (411481.5) | | | | 1801 | | 59 | | | | | |
| Bifidobacterium angulatum DSM 20098 = JCM 7096 (518635.7) | | | | 565 | | | | | | | |
| Bifidobacterium animalis subsp. lactis AD011 (442563.4) | | | | | | | | | | | |
| Bifidobacterium bifidum NCIMB 41171 (398513.5) | | | | 479, 1224, 1226, 1507 | | 1228, 1610 | 1227 | | | | 1282, 1283 |
| Bifidobacterium breve DSM 20213 = JCM 1192 (518634.7) | 1536 | 206 | | 882, 883, 1821 | 212 | 880, 1824 | 881, 1823 | | | 208, 210, 211 | 214 |
| Bifidobacterium breve HPH0326 (1203540.3) | 237 | 238 | | 1043, 1412 | 245 | 1415 | 1414 | | | 244 | 248, 1934 |
| Bifidobacterium catenulatum DSM 16992 = JCM 1194 (566552.6) | | | | | | 190 | | | | | |

| Organism | | | | | | | | | | | | | |
|---|---|---|---|---|---|---|---|---|---|---|---|---|---|
| Bifidobacterium dentium ATCC 27678 (473819.7) | | | | | 909, 1413 | | 907 | | | | | | |
| Bifidobacterium gallicum DSM 20093 (561180.4) | | 1521 | | | | 1560 | 1408 | | 1559 | | 1561, 1562, 1563, 1564 | | |
| Bifidobacterium longum DJO10A (Prj:321) (205913.11) | | | | | 493, 494, 495 | | 491 | | 492 | | | | |
| Bifidobacterium longum NCC2705 (206672.9) | | | | | 198, 199, 200, 1746 | | 202 | | 201 | | | | |
| Bifidobacterium longum subsp. infantis 157F (565040.3) | | | | | 1332, 1333, 1334 | | 1330 | | 1331 | | | | |
| Bifidobacterium longum subsp. infantis ATCC 15697 = JCM 1222 (391904.8) | 652 | 653, 2477 | | | 905 | 660 | 908 | | 907 | | | | 654, 2424 |
| Bifidobacterium longum subsp. infantis ATCC 55813 (548480.3) | | | | | 1366, 1367 | | 1369 | | 1368 | | | | |
| Bifidobacterium longum subsp. infantis CCUG 52486 (537937.5) | | | | | 1377, 1379, 1380 | | 641, 1382 | | 1381 | | | | |
| Bifidobacterium longum subsp. longum 2-2B (1161745.3) | 33 | | | | 341, 342, 343 | | 339 | | 340 | | | | |
| Bifidobacterium longum subsp. longum 44B (1161743.3) | 1649 | | | | 584, 585, 586 | | 582 | | 583 | | | | |
| Bifidobacterium longum subsp. longum F8 (722911.3) | | | | | 1929, 1930, 1931 | | 1933 | | 1932 | | | | |
| Bifidobacterium longum subsp. longum JCM 1217 (565042.3) | | | | | 1308, 1309, 1310 | | 1306 | | 1307 | | | | |
| Bifidobacterium pseudocatenulatum DSM 20438 = JCM 1200 (547043.8) | | | | | 630, 1743 | | 1740 | | 1741 | | | | |
| Bifidobacterium sp. 12_1_47BFAA (469594.3) | | | | | 1513, 1514, 1515 | | 1518 | | 1517 | | | | |
| Bilophila wadsworthia 3_1_6 (563192.3) | | | | | | | | | | | | 2980, 3193 | |
| Blautia hansenii DSM 20583 (537007.6) | 2021 | 2023 | | | | 2022 | 2031, 2798 | | 817, 2799 | | | 2025 | |
| Blautia hydrogenotrophica DSM 10507 (476272.5) | 3219 | 1099, 3189 | | | | 2917, 3190 | 3234 | | 1236, 3311 | 3191 | | | |
| Bryantella formatexigens DSM 14469 (478749.5) | 184 | 185 | | | 4208 | 3, 183 | 1279, 2475 | | 1280, 2474 | | 4, 5, 187, 188, 189 | 769 | |
| Burkholderiales bacterium 1_1_47 (469610.4) | | | | | | | | | | | | | |
| butyrate-producing bacterium SM4/1 (245012.3) | 1049, 1050 | 1053 | | | | | | | 1055 | | 84, 85 | | |
| butyrate-producing bacterium SS3/4 (245014.3) | | | | | | | 646 | | 647 | | | | |
| butyrate-producing bacterium SSC/2 (245018.3) | | 471, 1995 | | | | | 1999 | | 473 | | | | |
| Butyricicoccus pullicaecorum 1.2 (1203606.4) | | | | | | | | | 641, 2843 | | | | |
| Butyrivibrio crossotus DSM 2876 (511680.4) | | | | | | | | | | | | | |
| Butyrivibrio fibrisolvens 16/4 (657324.3) | | | | | | | | | | | | | |
| Campylobacter coli JV20 (864566.3) | | | | | | | | | | | | | |
| Campylobacter upsaliensis JV21 (888826.3) | | | | | | | | | | | | | |
| Catenibacterium mitsuokai DSM 15897 (451640.5) | 500 | | | | | | 1825 | | 429, 989 | 960 | | | |
| Cedecea davisae DSM 4568 (566551.4) | 346 | | | | | 1464 | 959 | | 960 | | | | |
| Citrobacter freundii 4_7_47CFAA (742730.3) | 1520 | 294 | | | | 1522 | 2811 | | 2812 | 1521 | | 2100 | |
| Citrobacter sp. 30_2 (469595.3) | 3019 | 2620 | | | 1778 | 3021 | 1321 | | 1322 | 3020 | | 3187 | |
| Citrobacter youngae ATCC 29220 (500640.5) | 3896 | 3897, 4183 | | | | 1202 | 3899 | | 3177 | 3178 | 3898 | 613 | |

| Organism | | | | | | | | | | | | | |
|---|---|---|---|---|---|---|---|---|---|---|---|---|---|
| Clostridiales bacterium 1_7_47FAA (457421.5) | 1141 | 1147 | | | 1145 | 3312, 4354 | | 4352, 4353, 5717 | | 1143 | | 890, 3112, 4350 | |
| Clostridium asparagiforme DSM 15981 (518636.5) | 3603 | 3430, 3604, 5452 | | | 3594, 3909 | 1833, 2622, 3333 | | 94, 465, 906, 1839, 1840, 3334 | | 2141, 2143 | 5177 | | |
| Clostridium bartlettii DSM 16795 (445973.7) | | | | | | | | | | | | | |
| Clostridium bolteae 90A5 (997893.5) | 3770 | 3775, 3960 | | 5207 | 273, 2663, 3774, 3962 | 6127 | | 4812, 4894 | | 3772 | 4050 | 4302, 5479 | |
| Clostridium bolteae ATCC BAA-613 (411902.9) | 2131 | 2136, 4204 | | | 595, 2135, 4202, 6013 | 4493 | | 936, 1657 | | 2133 | 4119 | 83 | |
| Clostridium celatum DSM 1785 (545697.3) | 976 | 3006 | | | | 3277 | | 2387 | | | | | 143 |
| Clostridium cf. saccharolyticum K10 (717608.3) | 1509 | 1505 | | | 1506 | 1503 | | 1504 | | 1507 | | 2059, 2060, 3207 | |
| Clostridium citroniae WAL-17108 (742733.3) | 4302, 5276 | 1963, 4307 | | | 1964, 4306, 4353 | 2681, 5553, 6173 | | 2682, 5759, 6182 | | 4304, 4305 | 11, 1199, 1200, 1959, 1960, 1961, 4639, 4640, 4641 | 2322, 4847 | |
| Clostridium clostridioforme 2_1_49FAA (742735.4) | 1884 | 1693, 1879 | | | 1691, 1880, 4223 | 5004 | | 1363, 4630 | | 1882 | 476, 1604, 4225, 4226, 4227 | 1613, 1613, 2769, 3193 | |
| Clostridium difficile CD196 (645462.3) | 2195 | 2200 | | | 2199 | 925 | | 926 | 2198 | | | | 2507 |
| Clostridium difficile NAP07 (525258.3) | 1835 | 1830 | | | 1831 | 1900 | | 1899 | 1832 | | | | 1490 |
| Clostridium difficile NAP08 (525259.3) | 1095 | 1090 | | | 1091 | 1964 | | 1965 | 1092 | | | | 778 |
| Clostridium hathewayi WAL-18680 (742737.3) | 1003, 4245 | 4246 | | | 4896, 5127 | 627, 3547, 4459, 4554 | 352 | 621, 3525, 4460, 5054 | | 143, 4240, 4241, 4242 | 5013 | | 854 |
| Clostridium hiranonis DSM 13275 (500633.7) | 384 | 380 | | | 381 | 1746 | | 1747 | | 382 | | | |
| Clostridium hylemonae DSM 15053 (553973.6) | 2860 | 2864 | | | 2863 | 2858 | | 1168, 2865, 3353 | | 2862 | | | 3500 |
| Clostridium leptum DSM 753 (428125.8) | | | | | | 602 | | 755 | | | | | |
| Clostridium methylpentosum DSM 5476 (537013.3) | | | | | | | | | | | | | |
| Clostridium nexile DSM 1787 (500632.7) | | | | | | 656 | | 655 | | | | 261 | 489, 2775 |
| Clostridium perfringens WAL-14572 (742739.3) | 1742, 2740 | 2744 | | | 2743 | 1842 | | 2487 | | 2742 | | | 152, 1947, 2737 |
| Clostridium ramosum DSM 1402 (445974.6) | 2081 | 590, 803 | | 861 | | 2050, 2352 | | 575, 1785, 1946, 2908 | | | | | 143, 1584 |
| Clostridium scindens ATCC 35704 (411468.9) | 430 | 425 | | | 426 | 432 | | 2964 | | 427 | | | 2284 |
| Clostridium sp. 7_2_43FAA (457396.3) | 55 | 171, 1989, 3419 | | | 180, 1993 | 2860 | | 21 | | 175, 176, 177 | | | 179 |
| Clostridium sp. 7_3_54FAA (665940.3) | 238 | 237 | | | 246 | 2695 | | 1895, 4893 | | | | 1241, 2299, 2300 | |
| Clostridium sp. D5 (556261.3) | 4580 | 4579 | | 748 | 38, 1421 | 967, 4582 | | 1045, 4581 | | | | | |

| Organism | | | | | | | | | | | | |
|---|---|---|---|---|---|---|---|---|---|---|---|---|
| Clostridium sp. HGF2 (908340.3) | 1586, 2540 | 1557, 1704, 1765 | | | | 1346, 1992, 2904 | 1341, 2679, 2681, 2682 | | | | 4037 | |
| Clostridium sp. L2-50 (411489.7) | | | | | | | | | | | | |
| Clostridium sp. M62/1 (411486.3) | 2348 | 2353 | | | 2352 | 2355 | 2354 | 2350, 2351 | | | 1916, 1917 | |
| Clostridium sp. SS2/1 (411484.7) | | 982, 2338 | | | | 986 | 2340 | | | | 506 | |
| Clostridium spiroforme DSM 1552 (428126.7) | | 2087 | | | | | 1540 | | | | | |
| Clostridium sporogenes ATCC 15579 (471871.7) | | | | | | 804 | 805, 1535 | | | | | 214, 2570 |
| Clostridium symbiosum WAL-14163 (742740.3) | 3352 | 3353 | | | 3344 | 2736 | 2149, 4897 | | | | 1824, 3839, 3840, 4346 | |
| Clostridium symbiosum WAL-14673 (742741.3) | 2560 | 2559 | | | 2568 | 3393 | 2975, 3644 | | | | 4358, 4482, 4483 | |
| Collinsella aerofaciens ATCC 25986 (411903.6) | 1949 | 1948 | | | 1950 | | 658 | | | 1953, 1954, 1955, 1956 | | |
| Collinsella intestinalis DSM 13280 (521003.7) | 935 | 936 | | | 934 | | 314 | | | 931, 932 | | 733 |
| Collinsella stercoris DSM 13279 (445975.6) | | 536, 1466, 1688 | | | | 635 | 294, 636 | | | | | |
| Collinsella tanakaei YIT 12063 (742742.3) | 28 | 29 | | | 27 | | 1224 | | | | | |
| Coprobacillus sp. 29_1 (469596.3) | 212, 1614, 1737 | 476 | | | | 1355, 1361, 2040, 2663 | 2041, 3024, 3096, 3832 | | | | | 248, 518, 519, 1064 |
| Coprobacillus sp. 3_3_56FAA (665941.3) | 3290 | 1245, 2545 | 2598 | 1314 | | 2932, 3323 | 1994, 2330, 2561, 3427 | | | | | 672, 801, 3166 |
| Coprobacillus sp. 8_2_54BFAA (469597.4) | 3413 | 142, 363, 1145, 2602 | 2541 | 414 | | 1380, 3382 | 1035, 2587, 2822, 3276 | | | | | 45, 149, 152, 1639, 1835 |
| Coprobacillus sp. D6 (556262.3) | 1467, 1674, 2903 | 1381 | | | | 285, 2085, 3380, 3386 | 2086, 2441, 2512, 3558 | | | | | 1341, 1342, 1343, 1431, 3093 |
| Coprococcus catus GD/7 (717962.3) | | | | | | 1169 | 1168 | | | | | |
| Coprococcus comes ATCC 27758 (470146.3) | | | | | | 921 | 920 | | | | 2991 | |
| Coprococcus eutactus ATCC 27759 (411474.6) | 2177 | | | | | | | | | | | |
| Coprococcus sp. ART55/1 (751585.3) | 707 | | | | | | | | | | | |
| Coprococcus sp. HPP0048 (1078091.3) | 1629 | 1631 | | | 1630 | 1628 | 2398 | | | 1634, 1635, 1636 | | 412, 2182 |
| Coprococcus sp. HPP0074 (1078090.3) | | | | | | 2570 | 2160 | | | | | 1447, 1942 |
| Corynebacterium ammoniagenes DSM 20306 (649754.3) | 1161 | 1160 | | | 1164 | 1162 | 1163, 1365 | | | 1166, 1167, 1168, 1169 | | |
| Corynebacterium sp. HFH0082 (1078764.3) | | | | | | | 1058 | | | | | |
| Dermabacter sp. HFH0086 (1203568.3) | 448 | 449 | | | 447 | 1885 | 1448 | | | | | 446, 451 |
| Desulfitobacterium hafniense DP7 (537010.4) | | | | | | | 2421 | | | 733, 1247, 1248, 2966 | | |
| Desulfovibrio piger ATCC 29098 (411464.8) | | | | | | | | | | | | |
| Desulfovibrio sp. 3_1_syn3 (457398.5) | | | | | | | | | | | 1676 | |
| Dorea formicigenerans 4_6_53AFAA (742765.5) | 2582 | 2583 | | | | 2580 | 446, 2581 | | | | | 701 |
| Dorea formicigenerans ATCC 27755 (411461.4) | 296 | 292 | | | 293 | 298 | 291, 2352 | 294 | | | | 2628 |
| Dorea longicatena DSM 13814 (411462.6) | 1624 | 1620 | | | 1621 | 1626 | 1532, 1619 | 1622 | | | | |

| Organism | | | | | | | | | | | | | |
|---|---|---|---|---|---|---|---|---|---|---|---|---|---|
| Dysgonomonas gadei ATCC BAA-286 (742766.3) | | | | | 692 | | 2316 | 2317, 2594, 2595 | | | | | |
| Dysgonomonas mossii DSM 22836 (742767.3) | | | 2522 | | 2584 | | | 660, 661 | | | | | |
| Edwardsiella tarda ATCC 23685 (500638.3) | 428, 999 | 429 | | | 1136 | 1001, 1002, 1030 | 1782 | 1781 | 1000 | 1032 | | | |
| Eggerthella sp. 1_3_56FAA (665943.3) | 2463 | 2464 | | | | 2462 | 2466 | | 2467 | | | 2457, 2459, 2460, 2461 | |
| Eggerthella sp. HGA1 (910311.3) | 212 | 211 | | | | 213 | 209 | | 208 | | | 214, 215, 216, 218 | |
| Enterobacter cancerogenus 35316 (500639.8) | 3823 | 3824 | | | 942 | 3826 | 2210 | 2211 | 3825 | | | | 3087 |
| Enterobacter cloacae subsp. cloacae NCTC 9394 (718254.4) | 2724 | 2253 | | | | 2726 | 1109 | 1518 | 2725 | | | | 1878 |
| Enterobacteriaceae bacterium 9_2_54FAA (469613.3) | 3352 | 3351 | | | 1738 | 358 | 359 | 1737 | | | | | |
| Enterococcus faecalis PC1.1 (702448.3) | 1197, 2176 | 117 | | | | | 45 | 1967 | | | | | |
| Enterococcus faecalis TX0104 (491074.3) | 1617 | 1371, 2138 | | | | | 15 | 2461, 2476 | | | | | |
| Enterococcus faecalis TX1302 (749513.3) | 2523 | 355, 548 | | | | | 2003 | 133 | | | | | |
| Enterococcus faecalis TX1322 (525278.3) | 219 | 1965, 2618 | | | | | 703 | 2324 | | | | | |
| Enterococcus faecalis TX1341 (749514.3) | 1855 | 1074, 1611 | | | | | 2552 | 1562 | | | | | |
| Enterococcus faecalis TX1342 (749515.3) | 2596 | 842 | | | | | 2106 | 1412, 1427 | | | | | |
| Enterococcus faecalis TX1346 (749516.3) | 654 | 757 | | | | | 537 | 1072, 1088 | | | | | |
| Enterococcus faecalis TX2134 (749518.3) | 1167 | 1609, 1789 | | | | | 2963 | 1085 | | | | | |
| Enterococcus faecalis TX2137 (749491.3) | 1986 | 453, 1394 | | | | | 877 | 1795 | | | | | |
| Enterococcus faecalis TX4244 (749494.3) | 1726 | 31, 2666 | | | | | 1240 | 2109 | | | | | |
| Enterococcus faecium PC4.1 (791161.5) | 1238, 2249 | 119 | | | | | 46 | 2035 | | | | | |
| Enterococcus faecium TX1330 (525279.3) | 727 | 256 | | | | | 1138 | 1885 | | | | | |
| Enterococcus saccharolyticus 30_1 (742813.4) | 179, 1708, 2243, 2529 | 2353 | | | | | 2285 | 1520 | | | | | |
| Enterococcus sp. 7L76 (657310.3) | 276 | 2684 | | | | | 2584 | 1762, 1776 | | | | | |
| Erysipelotrichaceae bacterium 21_3 (658657.3) | 1389, 2061 | 1358, 1508, 2832 | | | | | 785, 2394, 3564 | 791, 1901, 1903, 1904, 2307 | | | | | 111 |
| Erysipelotrichaceae bacterium 2_2_44A (457422.3) | 403, 4061 | 285, 441, 2576 | | | | | 1077, 2652, 3964 | 1072, 2200, 2202, 2203, 2568 | | | | | 1677 |
| Erysipelotrichaceae bacterium 3_1_53 (658659.3) | 451, 4493 | 399, 3300 | | | | | 53, 902, 1588 | 47, 3873 | | | | | 1310 |
| Erysipelotrichaceae bacterium 5_2_54FAA (552396.3) | | 1806, 2695 | | | | | 56, 1943, 2205 | 2201 | | | | | 874 |
| Erysipelotrichaceae bacterium 6_1_45 (469614.3) | 539, 4836 | 3111, 4727, 4865 | | | | | 1143, 2928, 4186 | 399, 400, 402, 2847, 4181 | | | | | 1706 |
| Escherichia coli 4_1_47FAA (1127356.4) | 2310 | 2311 | | | 2313 | | 479 | 480 | 2312 | | | | 1551 |
| Escherichia coli MS 107-1 (679207.4) | 2517 | 2516 | | | 1836 | 2514 | 2785 | 2784 | 2515 | | | | |
| Escherichia coli MS 115-1 (749537.3) | 2602 | 2603 | | | | 2605 | 3985 | | 3984 | 2604 | | | |
| Escherichia coli MS 116-1 (749538.3) | 3534 | 3535 | | | 2112 | 3537 | 4287 | | 4288 | 3536 | | | |
| Escherichia coli MS 119-7 (679206.4) | 4133 | 4134 | | | 871 | 4136 | 4582, 5234 | | 4583 | 4135 | | | |
| Escherichia coli MS 124-1 (679205.4) | 2806 | 2805 | | | 4108 | 2803 | 682 | | 683 | 2804 | | | |
| Escherichia coli MS 145-7 (679204.3) | 954 | 955 | | | 3623 | 957 | 2594 | | 2593 | 65, 956 | | | |
| Escherichia coli MS 146-1 (749540.3) | 1508 | 1509 | | | 1147 | 1511 | 2459 | | 2458 | 1510 | | | |
| Escherichia coli MS 175-1 (749544.3) | | | | | | 1972 | | | | | | | |
| Escherichia coli MS 182-1 (749545.3) | 3862 | 3860, 3861 | | | 839 | 3858 | 4959 | | 4958 | 3859 | | | |
| Escherichia coli MS 185-1 (749546.3) | 2013 | 2012 | | | 4731 | 2010 | 129 | | 130 | 2011 | | 1920 | |
| Escherichia coli MS 187-1 (749547.3) | 2153 | 2154 | | | 666 | 2156 | 3717 | | 3718 | 2155 | | | |
| Escherichia coli MS 196-1 (749548.3) | 4352 | 4351 | | | 5030 | 4349 | 4456 | | 4455 | 4350 | | 732 | |
| Escherichia coli MS 198-1 (749549.3) | 7, 2097 | 2098 | | | 1124 | 6, 1976 | 4821 | | 4822 | 8, 2099 | | 4035 | |
| Escherichia coli MS 200-1 (749550.3) | 1350 | 1349 | | | 2046 | 1347 | 4782 | | 4781 | 136, 1348 | | 3916, 4065 | |
| Escherichia coli MS 21-1 (749527.3) | 2863 | 2864 | | | 3367 | 2866 | 2619 | | 2618 | 2865 | | 2224 | |
| Escherichia coli MS 45-1 (749528.3) | 343, 4178 | 4179 | | | 764 | 344, 4181 | 3971 | | 3970 | 342, 4180 | | 510 | |
| Escherichia coli MS 69-1 (749531.3) | 1041, 2639 | 2640 | | | 3045 | 1042, 2642 | 4037 | | 4036 | 1040, 2641 | | 1944 | |
| Escherichia coli MS 78-1 (749532.3) | 1272 | 1271 | | | | 1269 | 3612 | | 3613 | 1270 | | | |
| Escherichia coli MS 79-10 (796392.4) | 2687 | 2688 | | | 4513 | 2690 | 2328 | | 2329 | 2689 | | | |
| Escherichia coli MS 84-1 (749533.3) | 2360 | 2361 | | | 650 | 2363 | 5347 | | 5348 | 2362 | | | |
| Escherichia coli MS 85-1 (679202.3) | 1097 | 1098 | | | 4732 | 1100 | 2801 | | 2800 | 1099 | | | |
| Escherichia coli O157:H7 str. Sakai (386585.9) | 4275 | 4276 | | | 1599 | 4278 | 819 | | 820 | 4277 | | | |
| Escherichia coli SE11 (409438.11) | 3671 | 3672 | | | 1320 | 3674 | 861 | | 862 | 3673 | | | |
| Escherichia coli SE15 (431946.3) | 3181 | 3182, 4345 | | | 1068 | 3184 | 631 | | 632 | 3183 | | 2979 | |
| Escherichia coli str. K-12 substr. MG1655 (511145.12) | 3318 | 3319 | | | 1165 | 3321 | 702 | | 703 | 3320 | | | |
| Escherichia coli UTI89 (364106.8) | 3608 | 3609 | | | 1322 | 3611 | 766 | | 767 | 3610 | | 3410 | |
| Escherichia sp. 1_1_43 (457400.3) | 874, 1160 | 1161 | | | | 875, 1163 | | | 873, 1162 | | | | |
| Escherichia sp. 3_2_53FAA (469598.5) | 1001 | 1000, 3565 | | | 819 | 998 | 341 | | 340 | 999 | | 1200 | |
| Escherichia sp. 4_1_40B (457401.3) | 2964 | 2963 | | | 1071 | 2961 | 1840 | | 1839 | 2962 | | | |
| Eubacterium biforme DSM 3989 (518637.5) | | 1891, 2228 | | | | 272 | 30 | | | | 941 | | |
| Eubacterium cylindroides T2-87 (717960.3) | | 695, 1246, 1551 | | | | | 1444 | | | | | | Kinase I is not found |
| Eubacterium dolichum DSM 3991 (428127.7) | | 1792 | | | | 276, 845 | 282 | | | | | | |
| Eubacterium hallii DSM 3353 (411469.3) | | 1640 | | | | 1642 | 1639 | | | | 587 | | |
| Eubacterium rectale DSM 17629 (657318.4) | | | | | | 602 | 601 | | | | 2275 | | |
| Eubacterium rectale M104/1 (657317.3) | | | | | | 1724 | 1725 | | | | 1649 | | Transporter is not found |
| Eubacterium siraeum 70/3 (657319.3) | | | | | | | | 2128 | | | | | |
| Eubacterium siraeum DSM 15702 (428128.7) | | | | | | | | 2046 | | | | | |
| Eubacterium siraeum V10Sc8a (717961.3) | | | | | | | | 2523 | | | | | |
| Eubacterium sp. 3_1_31 (457402.3) | | 1250, 1950 | | | | 303, 1628, 2852 | 307 | | | | | | |
| Eubacterium ventriosum ATCC 27560 (411463.4) | | | | | | 865 | 864, 1748, 2197 | | | | | | |
| Faecalibacterium cf. prausnitzii KLE1255 (748224.3) | 1956 | 1955 | | | 1953 | | | 7 | | 1950, 1951, 1952 | | 77 | |
| Faecalibacterium prausnitzii A2-165 (411483.3) | 1817 | 1816 | | | 1814 | 104 | | 105 | | 1811, 1812, 1813 | | | |
| Faecalibacterium prausnitzii L2-6 (718252.3) | 2400 | 2401 | | | 2403 | | 1289 | | | 2405, 2406, 2407 | | 656 | |
| Faecalibacterium prausnitzii M21/2 (411485.10) | 231 | 232 | | | 234 | 1937 | | 498, 1059 | | 235, 236, 237 | | 1412 | |
| Faecalibacterium prausnitzii SL3/3 (657322.3) | | 1517 | | | 1515 | 1777 | 847, 901 | | | 1512, 1513, 1514 | | 2058 | Transporter is not found |
| Fusobacterium gonidiaformans ATCC 25563 (469615.3) | 322 | 324 | | | 323 | 53 | 1472 | | | | 320, 321, 321 | | |
| Fusobacterium mortiferum ATCC 9817 (469616.3) | 2503 | 799, 2505 | | | 2361 | 1073, 1669 | 1668 | 2499 | | 2501, 2502, 2502 | | | |

| Organism | C1 | C2 | C3 | C4 | C5 | C6 | C7 | C8 | C9 | C10 | C11 | C12 | C13 |
|---|---|---|---|---|---|---|---|---|---|---|---|---|---|
| Fusobacterium necrophorum subsp. funduliforme 1_1_36S (742814.3) | 2378, 2379, 2380 | 2376 | | | | 2377 | 1309, 1310 | 1481, 1482 | | | | 2381, 2381, 2382 | |
| Fusobacterium sp. 11_3_2 (457403.3) | 1345 | 1347 | | | | 1346 | 980 | 1079 | | | | 930, 1343, 1344, 1344 | |
| Fusobacterium sp. 12_1B (457404.3) | 2966 | 1042, 2962, 2963 | | | | 2964 | 178 | 442 | | | | 2968, 2968, 2969 | |
| Fusobacterium sp. 1_1_41FAA (469621.3) | 671 | 669 | | | | 670 | 625 | 1423 | | | | 672, 672, 673, 800 | |
| Fusobacterium sp. 21_1A (469601.3) | | | | | | | 525 | 398 | | | | 1343 | |
| Fusobacterium sp. 2_1_31 (469599.3) | | | | | | | 733 | 933 | | | | 50 | |
| Fusobacterium sp. 3_1_27 (469602.3) | 1736 | 1734 | | | | 1735 | 828 | 818 | | | | 687, 1737, 1737, 1738 | |
| Fusobacterium sp. 3_1_33 (469603.3) | | | | | | | 768 | 550 | | | | 250 | |
| Fusobacterium sp. 3_1_36A2 (469604.3) | 442 | 444 | | | | 443 | 1169 | 1179 | | | | 440, 441, 441, 1302 | |
| Fusobacterium sp. 3_1_5R (469605.3) | 403 | 401 | | | | 402 | 150 | 72 | | | | 404, 404, 405 | |
| Fusobacterium sp. 4_1_13 (469606.3) | 653 | 655 | | | | 654 | 1527 | 1537 | | | | 318, 651, 652, 652 | |
| Fusobacterium sp. 7_1 (457405.3) | | | | | | | 780 | 470 | | | | 1329 | |
| Fusobacterium sp. D11 (556264.3) | 1088 | 1090, 1697 | | | | 1089 | 1986 | 239 | | | | 1086, 1087, 1087, 1341 | |
| Fusobacterium sp. D12 (556263.3) | 731 | 733 | | | | 732 | 1492 | 1065 | | | | 729, 730, 730 | |
| Fusobacterium ulcerans ATCC 49185 (469617.3) | 2319 | 732, 2322 | | | | 2321 | 195 | 2144 | | | | 1276, 2316, 2317, 2317 | |
| Fusobacterium varium ATCC 27725 (469618.3) | 324 | 327, 2930 | | | | 326 | 2033 | 2614 | | | | 321, 322, 322, 645 | |
| Gordonibacter pamelaeae 7-10-1-b (657308.3) | | | 1691 | | | 1690 | | 1694 | | | 1686, 1687, 1688, 1689 | | |
| Hafnia alvei ATCC 51873 (1002364.3) | 1352 | 1353 | | | | 134 | 4024 | 2230, 4025 | 133 | | | | |
| Helicobacter bilis ATCC 43879 (613026.4) | | | | | | | | | | | | | |
| Helicobacter canadensis MIT 98-5491 (Prj:30719) (537970.9) | | | | | | | | | | | | 941 | |
| Helicobacter cinaedi ATCC BAA-847 (1206745.3) | | | | | | | | | | | | 1116 | |
| Helicobacter cinaedi CCUG 18818 (537971.5) | | | | | | | | | | | | 1243 | |
| Helicobacter pullorum MIT 98-5489 (537972.5) | | | | | | | | | | | | 362 | |
| Helicobacter winghamensis ATCC BAA-430 (556267.4) | | | | | | | | | | | | 941 | |
| Holdemania filiformis DSM 12042 (545696.5) | | 132, 217, 501, 925 | | | | | 1184, 1185, 2175 | 1188, 1784, 2076 | | | | | |
| Klebsiella pneumoniae 1162281 (1037908.3) | 4821 | 1170 | | | | | 2760 | 2761 | | | | | |
| Klebsiella pneumoniae subsp. pneumoniae WGLW5 (1203547.3) | 3084 | 3896 | | | | | 849 | 850 | | | | | |
| Klebsiella sp. 1_1_55 (469608.3) | 86 | 2889 | | | 2261 | | 1792 | 1793 | | | | | |
| Klebsiella sp. 4_1_44FAA (665944.3) | 4348 | 4533 | | | | | 1579 | 1580 | | | | | |
| Lachnospiraceae bacterium 1_1_57FAA (658081.3) | | | | | | | | 1668 | | | | | 1719, 1809, 2267, 2539, 2540 |
| Lachnospiraceae bacterium 1_4_56FAA (658655.3) | 286 | 287 | | | | | 2720 | 941 | | | | | |
| Lachnospiraceae bacterium 2_1_46FAA (742723.3) | | | | | | | 1212 | 1213 | | | | | 1454 |
| Lachnospiraceae bacterium 2_1_58FAA (658082.3) | | | | | | | 2454 | 618 | | | | | |
| Lachnospiraceae bacterium 3_1_46FAA (665950.3) | | | | | | | | 1250 | | | | | 571, 1075, 2928 |

| Organism | | | | | | | | | | | | | | |
|---|---|---|---|---|---|---|---|---|---|---|---|---|---|---|
| Lachnospiraceae bacterium 3_1_57FAA_CT1 (658086.3) | 5697, 7047 | 5699, 7049 | | | 5701 | 7342 | 7343 | | | 5702, 5703, 5704 | | | 2757 | |
| Lachnospiraceae bacterium 4_1_37FAA (552395.3) | 1970 | 1972 | | | 1971 | 1969 | 2419 | | | 1975, 1976, 1977 | | | 268, 2689 | |
| Lachnospiraceae bacterium 5_1_57FAA (658085.3) | 1671 | 1676 | | | 1675 | 1669 | 241 | | 1674 | | | | 2127 | |
| Lachnospiraceae bacterium 5_1_63FAA (658089.3) | | 1694 | | | 1124 | 982 | 1692 | | | | | 731 | | |
| Lachnospiraceae bacterium 6_1_63FAA (658083.3) | | | | | | 1271, 2492 | 1272, 1614 | | | | | | | |
| Lachnospiraceae bacterium 7_1_58FAA (658087.3) | | | 3868 | | 3862 | 3534, 3860 | 3535, 3861 | | | 3863, 3864, 3865 | | | 1408 | |
| Lachnospiraceae bacterium 8_1_57FAA (665951.3) | | | | | | | 1984 | | | | | | 1695, 1800, 1801, 1937, 2355 | |
| Lachnospiraceae bacterium 9_1_43BFAA (658088.3) | | | | | | 1719 | 2131 | | | | | | 842, 1278, 2348 | |
| Lactobacillus acidophilus ATCC 4796 (525306.3) | | | | 39 | | | 1612 | | | | | | | |
| Lactobacillus acidophilus NCFM (272621.13) | | | | 1433 | | 140 | 1852 | | | | | | | |
| Lactobacillus amylolyticus DSM 11664 (585524.3) | | | | 417 | | 1310 | 909 | | | | | | | |
| Lactobacillus antri DSM 16041 (525309.3) | 1895 | 1894 | | | 1896 | 1891 | 1892 | | 1898 | | | | | |
| Lactobacillus brevis 367 (387344.15) | | | | | | 520 | 666 | | | | | | | |
| Lactobacillus brevis subsp. gravesensis ATCC 27305 (525310.3) | | | | | | 2620 | | | | | | | | |
| Lactobacillus buchneri ATCC 11577 (525318.3) | | | | | | 1815 | | | | | | | | |
| Lactobacillus casei ATCC 334 (321967.11) | | | | | | 1787 | 2831 | | | | | | | |
| Lactobacillus casei BL23 (543734.4) | 2706 | | | | | 1929 | 2971 | | | | | | | |
| Lactobacillus crispatus 125-2-CHN (575595.3) | | | | 764 | | 1861 | 1510 | | | | | | | |
| Lactobacillus delbrueckii subsp. bulgaricus ATCC 11842 (390333.7) | | | | | | 177 | 1918 | | | | | | | |
| Lactobacillus delbrueckii subsp. bulgaricus ATCC BAA-365 (321956.7) | | | | | | 171 | 1913 | | | | | | | |
| Lactobacillus fermentum ATCC 14931 (525325.3) | | | | | | | | | | | | | | |
| Lactobacillus fermentum IFO 3956 (334390.5) | | | | | | | | | | | | | | |
| Lactobacillus gasseri ATCC 33323 (324831.13) | 195 | 1615 | | 1559 | | | 1776 | | | | | | | |
| Lactobacillus helveticus DPC 4571 (405566.6) | | | | 1656, 1657 | | | 2205 | | | | | | | |
| Lactobacillus helveticus DSM 20075 (585520.4) | | | | 1317 | | | 60 | | | | | | | |
| Lactobacillus hilgardii ATCC 8290 (525327.3) | | | | | | 2003 | | | | | | | | |
| Lactobacillus johnsonii NCC 533 (257314.6) | 207 | 1666 | | 1610 | | | 1824 | | | | | | | |
| Lactobacillus paracasei subsp. paracasei 8700:2 (537973.8) | 1220, 1475 | | | | | 835 | 1126 | | | | | | | |
| Lactobacillus paracasei subsp. paracasei ATCC 25302 (525337.3) | 1832 | | | | | 687 | 1689 | | | | | | | |
| Lactobacillus plantarum 16 (1327988.3) | 2400, 2859 | 2861 | | | 2860 | 501 | 207 | | 2855 | | | | | |
| Lactobacillus plantarum subsp. plantarum ATCC 14917 (525338.3) | 324, 646, 2855 | 595 | | | | 1151 | 160 | | | | | | | |
| Lactobacillus plantarum WCFS1 (220668.9) | 361, 2474, 2976, 3030 | 2979 | | | 2977 | 466 | 188 | | 2973 | | | | | |
| Lactobacillus reuteri CF48-3A (525341.3) | | | | | | | 1764 | | | | | | | |
| Lactobacillus reuteri DSM 20016 (557436.4) | | | | | | | 1035 | | | | | | | |
| Lactobacillus reuteri F275 (299033.6) | | | | | | | 1010 | | | | | | | |
| Lactobacillus reuteri MM2-3 (585517.3) | | | | | | | 1454 | | | | | | | |
| Lactobacillus reuteri MM4-1A (548485.3) | | | | | | | 967 | | | | | | | |
| Lactobacillus reuteri SD2112 (491077.3) | | | | | | | 1846 | | | | | | | |
| Lactobacillus rhamnosus ATCC 21052 (1002365.5) | | | | | | 1920 | 1648 | | | | | | | |

| Organism | | | | | | | | | | | | |
|---|---|---|---|---|---|---|---|---|---|---|---|---|
| Lactobacillus rhamnosus GG (Prj:40637) (568703.30) | 98, 2688 | | | | | | 1869 | 2936 | | | | |
| Lactobacillus rhamnosus LMS2-1 (525361.3) | 1581 | | | | | | 799 | 1698 | | | | |
| Lactobacillus ruminis ATCC 25644 (525362.3) | 677 | | | | | | 2024 | 234 | | | | |
| Lactobacillus sakei subsp. sakei 23K (314315.12) | 118, 1610 | 1608 | | | | 1607 | 1555 | 409 | | 1609 | | |
| Lactobacillus salivarius ATCC 11741 (525364.3) | 312 | 310 | | | | 311 | 1583 | 1584 | | 316 | | |
| Lactobacillus salivarius UCC118 (362948.14) | 2059 | 2061 | | | | 2060 | 1549 | 1548 | | | | 2052, 2053 |
| Lactobacillus sp. 7_1_47FAA (665945.3) | | | 1083 | | | | 617 | 718 | | | | |
| Lactobacillus ultunensis DSM 16047 (525365.3) | | | 1503 | | | | 1964 | 1115 | | | | |
| Leuconostoc mesenteroides subsp. cremoris ATCC 19254 (586220.3) | | | | | | | 791 | 850 | | | | |
| Listeria grayi DSM 20601 (525367.9) | 1178, 1406, 1443, 1911 | 1511 | | | | | 963 | 754, 964, 2263 | | | | |
| Listeria innocua ATCC 33091 (1002366.3) | 2666, 2688 | 2661 | | | | | 1546 | 800, 1468, 1547, 2734 | | | | |
| Megamonas funiformis YIT 11815 (742816.3) | 2072 | 2067 | | | | 2068 | 1644 | 2074 | | 2069 | | |
| Megamonas hypermegale ART12/1 (657316.3) | 1097, 1098 | 1104, 1105 | | | | 1102, 1103 | 440 | 1094 | | 1101 | | |
| Methanobrevibacter smithii ATCC 35061 (420247.6) | | | | | | | | | | | | |
| Methanosphaera stadtmanae DSM 3091 (339860.6) | | | | | | | | | | | | |
| Mitsuokella multacida DSM 20544 (500635.8) | 359, 1512 | | | | | | 348 | 449 | | | | |
| Mollicutes bacterium D7 (556270.3) | 2076 | 1843, 2566 | | | 2290 | | 1547, 2043 | 882, 2580, 3191, 3284 | | | | 1995, 2402 |
| Odoribacter laneus YIT 12061 (742817.3) | | 1750 | | | 1748 | | | 2356 | | 1749 | 1709 | 2156, 2674 |
| Oxalobacter formigenes HOxBLS (556268.6) | | | | | | | | | | | | |
| Oxalobacter formigenes OXCC13 (556269.4) | | | | | | | | | | | | |
| Paenibacillus sp. HGF5 (908341.3) | | 5364 | | 3284, 6169 | | | 2275, 2485, 5378 | 2274, 2486, 5379 | | 888 | | 4097 |
| Paenibacillus sp. HGF7 (944559.3) | | 951 | | | | | 3153 | 3152 | | | | |
| Paenibacillus sp. HGH0039 (1078505.3) | | 2189 | | | 2034 | | 4055 | 4400 | 4056 | 4868 | | 1485, 1486 |
| Parabacteroides distasonis ATCC 8503 (435591.13) | | 1847 | 2904 | | 2842, 3552 | 2903 | 225, 226, 3115 | 227, 3530 | 2905 | | | 1676, 2414, 2908 |
| Parabacteroides johnsonii DSM 18315 (537006.5) | | | 419 | | 3317 | 418 | 2442, 2443, 2444 | 2441, 3494 | 420 | | | 421, 2117 |
| Parabacteroides merdae ATCC 43184 (411477.4) | | | 285 | | 1578 | 284, 1788 | 3334, 3335 | 1695, 3333 | 286 | 1787 | | 287, 596, 1792, 1794 |
| Parabacteroides sp. D13 (563193.3) | | 3313 | 737 | | 11, 432 | 736 | 1058, 3904, 3905 | 411, 3903 | 738 | | 2021 | 742, 1205, 2768 |
| Parabacteroides sp. D25 (658661.3) | | 2028 | 1349 | | 2922 | 1350 | 807, 3683, 3684 | 2944, 3685 | 1348 | | | 199, 1345, 1438 |
| Paraprevotella clara YIT 11840 (762968.3) | | | | | | | | | | | | |
| Paraprevotella xylaniphila YIT 11841 (762982.3) | | | | | 682 | | | | | | | |
| Parvimonas micra ATCC 33270 (411465.10) | | | | | | | | | | | | |
| Pediococcus acidilactici 7_4 (563194.3) | 735, 793 | 1907 | | | | 789 | 1492 | 1057 | | 790, 791 | | |
| Pediococcus acidilactici DSM 20284 (862514.3) | 31, 77 | 30 | | | | 34, 1580 | 965 | 1780 | 33 | | | 41 |
| Phascolarctobacterium sp. YIT 12067 (626939.3) | | | | | | | | | | | 381 | |
| Prevotella copri DSM 18205 (537011.5) | | | | | 384, 2765 | | | 642, 771, 4375, 4510 | | | | |
| Prevotella oralis HGA0225 (1203550.3) | | | 1528, 2124 | | 208 | 2123 | | 2046, 2049 | 2122 | | | 834 |
| Prevotella salivae DSM 15606 (888832.3) | | | 1624 | | 2547 | 1625, 1842 | | 1558, 1559 | 1626 | | | 1564, 1839 |

| | | | | | | | | | | | | | | | |
|---|---|---|---|---|---|---|---|---|---|---|---|---|---|---|---|---|
| Propionibacterium sp. 5_U_42AFAA (450748.3) | 573 | 92, 833 | | 1698 | | 93 | 1020 | | 1012 | 95, 1242, 1428 | | | | | | 54, 1240, 1241, 2302 |
| Propionibacterium sp. HGH0353 (1203571.3) | | 683, 1494 | | | | 684 | 1668 | | 1660 | 686, 2058 | | | | | | |
| Proteus mirabilis WGLW6 (1125694.3) | 2583 | 2582 | | | 2587 | 387 | | 388 | | 2584 | | | | | | |
| Proteus penneri ATCC 35198 (471881.3) | 1438, 1439 | 1436, 1437 | | | 1305 | 1443, 1444 | 4620 | | 4619 | | 1440 | | | | | |
| Providencia alcalifaciens DSM 30120 (520999.6) | | | | | 3402 | | 3264 | | 3263 | | | | | | | |
| Providencia rettgeri DSM 1131 (521000.6) | 3568 | 3569 | | | 1547 | 3564 | 3087, 3112 | | 3086, 3113 | | 3567 | | | | | |
| Providencia rustigianii DSM 4541 (500637.6) | | | | | 90, 6992 | | 2653, 4581 | | 2654, 4580 | | | | | | | |
| Providencia stuartii ATCC 25827 (471874.6) | | | | | 3144 | | 756 | | 757 | | | | | | | |
| Pseudomonas sp. 2_1_26 (665948.3) | | | | | | | 480 | 479 | | | | | | | | 450, 2388, 4078 |
| Ralstonia sp. 5_7_47FAA (658664.3) | | | | | | | | | | | | | | | | |
| Roseburia intestinalis L1-82 (536231.5) | | | | | | | 2027 | | 2028 | | | | | 1146, 1550 | | |
| Roseburia intestinalis M50/1 (657315.3) | 3689 | 3695 | | | | | 1783 | | 1782 | | | | | 2807, 3086 | | |
| Roseburia intestinalis XB6B4 (718255.3) | 2396 | 2390 | | | | | 3120, 3121 | | 3119 | | | | | 1204, 2027 | | |
| Roseburia inulinivorans DSM 16841 (622312.4) | 803 | 802 | | | | 800 | 2448, 3995 | | 795, 2449 | | | | 797, 798, 799 | 1563 | | |
| Ruminococcaceae bacterium D16 (552398.3) | 2573 | | 2013 | | | 2014 | 52 | | 51 | 2015 | | | | | | |
| Ruminococcus bromii L2-63 (657321.5) | | | | | | | | | | | | | | | | |
| Ruminococcus gnavus ATCC 29149 (411470.6) | 2355 | 2357 | | | | 2356 | 3204 | | 1071, 2646 | | | | 2360, 2361, 2362 | | | 2358 |
| Ruminococcus lactaris ATCC 29176 (471875.6) | | | | | | | 2329 | | 1814 | | | | | 756 | | 1195, 2021 |
| Ruminococcus obeum A2-162 (657314.3) | | | | | | | 1118 | | 3354 | | | | | 1558 | | |
| Ruminococcus obeum ATCC 29174 (411459.7) | | | | | | 2080 | 2341 | | 200 | | | | | 3232 | | |
| Ruminococcus sp. 18P13 (213810.4) | | | | | | | | | | | | | | | | |
| Ruminococcus sp. 5_1_39BFAA (457412.4) | 2689 | 2697 | | | | 2691 | 2698 | | 2699 | | | | | | | |
| Ruminococcus sp. SR1/5 (657323.3) | | | | | | 1960 | 1214 | | 972 | | | | | | | |
| Ruminococcus torques ATCC 27756 (411460.6) | | | | | | | | | 1971 | | | | | | | 395, 1279 |
| Ruminococcus torques L2-14 (657313.3) | 2279 | 2275 | | | | 2276 | 808, 2281 | | 809, 2273 | 2277 | | | | | | |
| Salmonella enterica subsp. enterica serovar Typhimurium str. LT2 (99287.12) | 3537 | 1196, 3538 | | | 1289 | 3540 | 713 | | 714 | 1199, 3539 | 1195 | | | 3361 | | 978 |
| Staphylococcus sp. HGB0015 (1078083.3) | 2076 | 55 | | | | 573, 2074 | | 56 | 574, 2073 | | | | | | | 576, 1266, 1267 |
| Streptococcus anginosus 1_2_62CV (742820.3) | 1767 | 1760 | | | | 1766 | 1108 | | 858 | | | | 1761, 1762, 1763 | | | |
| Streptococcus equinus ATCC 9812 (525379.3) | 1614 | | | | | | 1465 | | 1070 | | | | | | | |
| Streptococcus infantarius subsp. infantarius ATCC BAA-102 (471872.6) | 662 | | | | | | 176 | | 497, 1165 | | | | | | | |
| Streptococcus sp. 2_1_36FAA (469609.3) | 133, 1786 | 128 | | | | 129 | 556 | | 1537 | | 130 | | | | | |
| Streptococcus sp. HPH0090 (1203590.3) | 1212, 1556 | 1219 | | | | 1213 | 1120 | | 696 | | | | 1216, 1217, 1218 | | | 197, 1220 |
| Streptomyces sp. HGB0020 (1078086.3) | 2157 | | | 3801, 5253 | 3295, 6873 | 5490 | 6874 | 3645 | 8192 | | | | 5493, 5494 | | | 1717, 5489, 5824 |
| Streptomyces sp. HPH0547 (1203592.3) | 3641 | 4658 | | 4620, 4638 | 5571 | 58 | 5570 | 2839, 4043 | 59 | | | | | | | 55, 1802 |
| Subdoligranulum sp. 4_3_54A2FAA (665956.3) | | | 1667 | | | 1664 | 303, 518, 4373 | | 297, 476, 4374 | | | | 1661, 1662, 1663, 3401 | | | 2382, 2467, 2970, 2978, 3483 |
| Subdoligranulum variabile DSM 15176 (411471.5) | | | | | | | 1467, 1900, 4520, 4957 | | 1466, 1899, 4519, 4956 | | | | 124, 959, 5083, 5651 | | | |
| Succinatimonas hippei YIT 12066 (762983.3) | | | | | | | 1293 | | | | | | | 1977, 1978 | | |
| Sutterella wadsworthensis 3_1_45B (742821.3) | | | | | | | | | | | | | | | | |
| Sutterella wadsworthensis HGA0223 (1203554.3) | | | | | | | | | | | | | | | | |

| | | | | | | | | | | |
|---|---|---|---|---|---|---|---|---|---|---|
| Tannerella sp. 6_1_58FAA_CT1 (665949.3) | | | | | 341 | | | 2008, 2009 | | |
| Turicibacter sp. HGF1 (910310.3) | 1327, 2390 | 326 | | | 1330 | 2230, 2553 | | 1326, 2552 | | 603, 604, 605 |
| Turicibacter sp. PC909 (702450.3) | 664, 808, 2285 | 2575 | | | 2282 | 2042, 2454 | | 2286, 2455 | | 2402, 2403, 2404 |
| Veillonella sp. 3_1_44 (457416.3) | | | | | | | | | | |
| Veillonella sp. 6_1_27 (450749.3) | | | | | | | | | | |
| Weissella paramesenteroides ATCC 33313 (585506.3) | | 1042 | | | | 1380 | | 578 | | |
| Yokenella regensburgei ATCC 43003 (1002368.3) | 1461 | 1858 | | | 1459 | 2353 | | 2354 | 1460 | |

**Table S7.** Genes involved in depletion and utilization of galactose. The PubSEED identifiers are shown. For details on analyzed organisms see Table S1, for details on gene functions see Table S2.

| Organism (PubSEED ID) | Catabolic pathway I | | | | | | Catabolic pathway II | | | | | Transport | | | | | | | Hydrolases | | | | | | | Problems (if any) |
|---|---|---|---|---|---|---|---|---|---|---|---|---|---|---|---|---|---|---|---|---|---|---|---|---|---|---|
| | Uridylyltransferase I | Epimerase I | Kinase | Uridylyltransferase II | Uridylyltransferase III | Epimerase II | Phosphorylase | Isomerase | Kinase II | Aldolase | Phospho-beta-galactosidase | ABC systems | | Permeases | | Na⁺ symporter | PTS systems | | Beta-galactosidases | | | | Alpha-galactosidases | | | |
| | GalU | GalM | GalK | GalT | GalY | GalE | LnbP | LacAB | LacC | LacD | GnbG | MglABC | LnbABC | GalP1 | GalP2 | GalP3 | GalPTS | GnbPTS | Bga2 | Bga35 | Bga42 | Aga4 | Aga27 | Aga36 | Aga43 | |
| Abiotrophia defectiva ATCC 49176 (592010.4) | | 1505 | 265 | 1211 | | 3049 | 3045 | | | | | 1913, 1914, 1915, 1916 | 758, 759, 760, 3041, 3042, 3043 | | | | | | 761 | 1670 | | | 2310 | 2310 | | |
| Acidaminococcus sp. D21 (563191.3) | 419, 1505 | | | | | 548, 1926, 1967 | | | | | | | | | | | | | | | | | | | | |
| Acidaminococcus sp. HPA0509 (1203555.3) | 622, 1197 | | | | | 486, 1612, 1835 | | | | | | | | | | | | | | | | | | | | |
| Acinetobacter junii SH205 (575587.3) | 2884 | | | | | 961, 1635, 2887 | | | | | | | | | | | | | | | | | | | | |
| Actinomyces odontolyticus ATCC 17982 (411466.7) | 327 | | 176 | | | 120, 1876 | | 1813 | 1018 | 1019 | | | | | | 1696 | | | | | | | 1025 | | | Galactose-1-phosphate uridylyltransferase is not found |
| Actinomyces sp. HPA0247 (1203556.3) | 353 | | 172 | | | 114, 1434 | | 1362 | | | | | | | | 1248 | | | | | | | | | | Galactose-1-phosphate uridylyltransferase is not found |
| Akkermansia muciniphila ATCC BAA-835 (349741.6) | 1975 | 1482 | 1100 | | 37, 38 | 36, 1274 | | | | | | | | | | 1101 | | | 336, 608, 936, 1885, 1887, 1888, 1890 | 874, 1912 | | | 971, 1346 | 971 | | |
| Alistipes indistinctus YIT 12060 (742725.3) | | 983, 1894 | 2233 | 982 | | 1028 | | | | | | 984 | | | | | | | 145, 1251, 1406, 1986, 2368 | | | | 422, 2225 | 2225 | | |
| Alistipes shahii WAL 8301 (717959.3) | | 1826, 2802 | 439 | 703 | | 296 | | | | | | 184 | | | | | | | 35, 38, 767, 932, 933, 1311, 2716 | | | | 34, 105, 1071, 1807, 2088 | 105, 1071, 2088 | | |
| Anaerobaculum hydrogeniformans ATCC BAA-1850 (592015.5) | 883 | | | | | 1437, 1671 | | | 898 | | | | | | | | | | | | | | | | | |
| Anaerococcus hydrogenalis DSM 7454 (561177.4) | | | | | | 1885, 1886 | | | | | | | | | | | | | | | | | | | | |
| Anaerofustis stercorihominis DSM 17244 (445971.6) | | | | | | | | | | | 413 | | | | | | | | | | | | | | | |
| Anaerostipes caccae DSM 14662 (411490.6) | 279 | 1024 | 1025 | 1023 | | 1021, 1022 | | | 1040, 3509 | 746, 1039, 1941, 3511 | | | | | | 1026 | | | | | | | 1879 | 1879 | | |
| Anaerostipes hadrus DSM 3319 (649757.3) | | 2309, 2320, 2319, 2680 | 2321 | | | 320, 1705, 2322 | | | 1244 | 1242 | | | | | | 2318 | | | 2143, 2144 | | | | | | | |
| Anaerostipes sp. 3_2_56FAA (665937.3) | 2945 | 129 | 130 | 128 | | 127 | | | 146, 2545, 2546 | 144, 429, 2543 | | | | | | 131 | | | | | | | 3397 | 3397 | | |
| Anaerotruncus colihominis DSM 17241 (445972.6) | 2937 | | 3172 | 57 | 58 | 893, 1163, 1167 | | | | | | 679, 681, 682, 686 | | | | | | | | | | | | | | |
| Bacillus smithii 7_3_47FAA (665952.3) | 1614, 1939, 3590 | | | | | 951, 1611, 3587 | | 315, 316 | 318 | 317 | | | | | | | 319, 320, 321 | | | | | | 1628 | 1628 | | |
| Bacillus sp. 7_6_55CFAA_CT2 (665957.3) | 4442, 4795 | | | 5468 | | 1471, 1624, 4445, 4460, 5844 | | 1168 | | | | | | | | | | | | | | | | | | Galactose kinase and transporter are not found |
| Bacillus subtilis subsp. subtilis str. 168 (224308.113) | 1878, 3174, 3670, 3675 | 1897 | 3929 | 3928 | | 760, 3107, 3997 | | | | | | | 3350, 3351 | | | | | | | 740, 3513 | 745, 3119 | | | | | Transporter is not found |
| Bacteroides caccae ATCC 43185 (411901.7) | | 2174, 3292 | 2172 | 3128 | | 665, 826, 842, 2903 | | | | | | | | 2173 | | | | | 10, 1150, 1273, 1424, 1598, 1715, 1960, 2446, 2668, 2714, 3365, 3629 | 247, 1659, 1845, 3023 | | | 260, 1271, 3244 | 1271, 3244 | | |
| Bacteroides capillosus ATCC 29799 (411467.6) | 1888 | 867 | 3502 | 3504 | | 855, 3503 | 948 | | | | | 733, 737, 738, 850, 851, 852, 739 | | | | | | | 1144, 2810, 2828, 2829, 3217, 3522 | | | | | 3081 | | |

| Species | | | | | | | | | | | | | | | | | | | | | |
|---|---|---|---|---|---|---|---|---|---|---|---|---|---|---|---|---|---|---|---|---|---|
| Bacteroides cellulosilyticus DSM 14838 (537012.5) | | 983, 1946, 4504, 4885 | 4506 | | 102, 103 | 377, 656, 3700, 3858, 3924 | | | | | | | 4505 | 2323 | | 178, 179, 205, 214, 406, 407, 408, 621, 1187, 1206, 1248, 1368, 1375, 1498, 1551, 1618, 1757, 1764, 1896, 1974, 2122, 2294, 2316, 2328, 2543, 2548, 2550, 2676, 2992, 3116, 3280, 4047 | 180, 637, 785, 2542, 3102 | 108, 785, 1893, 2999 | 1770, 1771, 1772, 3605, 4025, 4642, 4781, 5064 | 1772, 4025, 4642, 4781, 5064 | |
| Bacteroides clarus YIT 12056 (762984.3) | | 373 | 371 | | 2589 | 798 | | | | | | | 372 | | | 225, 587, 1659, 1664, 1769, 1777, 2168 | | | 2116, 2124, 2674 | 2116, 2674 | |
| Bacteroides coprocola DSM 17136 (470145.6) | 2943 | 2685, 3092, 3121 | 3094 | | 3155 | 573, 2177 | | | | | | | 3093 | | | 1295, 1650, 1685, 1744, 2155, 2186, 2554, 3214, 3495, 3660, 3661 | | | 377, 2763, 3565 | 377, 3565 | |
| Bacteroides coprophilus DSM 18228 (547042.5) | | 3047 | 3045 | | 2303 | 660 | | | | | | | 3046 | 2729 | | 80, 121, 299, 950, 1117, 2609, 2610, 2723, 2728, 2781, 2782, 2832, 3146, 3173, 3262 | 1773, 3141, 3144 | | 531, 2028 | 2028 | |
| Bacteroides dorei DSM 17855 (483217.6) | | 1375, 3350, 3466, 3469, 4170 | 3467 | | 59, 890 | 891, 2494, 2504 | | | | | | | 3468 | | | 44, 134, 328, 373, 395, 638, 1023, 1088, 1633, 1861, 2100, 2848, 2849, 2878, 2882, 2908, 3007, 3154, 3309, 3349, 3964, 3972, 3973, 4291, 4293 | 1136 | 374, 375 | 992, 4166 | 4166 | |
| Bacteroides eggerthii 1_2_48FAA (665953.3) | | 1890, 3240, 3253 | 3242 | | 1197 | 2383, 2669, 3214 | | | | | | | 3241 | | | 504, 505, 674, 675, 1018, 1026, 1155, 1165, 1166, 1167, 1170, 2146, 2342, 2897, 3387, 3546, 3776 | 668, 673, 2274 | | 2323 | 2323 | |

| Organism | | | | | | | | | | | | | | | | | | | |
|---|---|---|---|---|---|---|---|---|---|---|---|---|---|---|---|---|---|---|---|
| Bacterides eggerthi DSM 20697 (483216.6) | 216, 1162, 1174 | 1172 | | 612 | 82, 1197, 1202, 2162, 2386 | | | | | | 1173 | | | 570, 579, 580, 581, 584, 659, 1085, 1898, 1899, 2344, 2680, 3059, 3067 | 951 | | | | |
| Bacteroides finegoldii DSM 17565 (483215.6) | 2218, 2226 | 2224 | | 2955 | 1794, 1811, 2793, 3155 | | | | | | 2225 | | | 127, 342, 1297, 1310, 1951, 1956, 1957, 1959, 2992, 3068, 3116, 3818 | 2207, 2832 | 1309 | | 125, 3512, 4046 | 3512 |
| Bacteroides fluxus YIT 12057 (763034.3) | 517, 654 | 656 | | 1246 | 164, 1058, 2206, 2392, 3194 | | | | | | 655 | | | 193, 198, 226, 509, 554, 556, 1086, 1673, 2321, 3096, 3267 | 207, 225, 1563 | | | 119, 1662, 3492 | 119, 3492 |
| Bacteroides fragilis 3_1_12 (457424.5) | 320, 2600, 4068 | 318 | | 4636 | 219, 480, 1527, 1562, 1994, 3668 | | | | | | 319 | | | 328, 395, 398, 545, 864, 947, 1828, 1834, 2129, 2292, 2579, 2614, 2618, 2782, 3291, 3435, 3653, 4521 | 2291, 3652, 4334 | | | 152, 2660, 2665, 4062 | 2660 |
| Bacteroides fragilis 638R (862962.3) | 361, 1672, 4454 | 1670 | | 4122 | 809, 1463, 1464, 1574, 1903, 2653, 2662, 2674, 3896 | | | | | | 1671 | | | 30, 146, 342, 346, 390, 639, 734, 932, 937, 1681, 1747, 1941, 3197, 3336 | 640, 3880, 4015, 4282, 4283, 4319 | | | 287, 293, 561, 1512, 4461 | 293, 561 |
| Bacteroides fragilis NCTC 9343 (272559.17) | 293, 1622, 4301 | 1620 | | 4012 | 1011, 1527, 1889, 2603, 2612, 2624, 3727, 3728, 3765 | | | | | | 1621 | | | 30, 153, 273, 277, 320, 563, 652, 841, 846, 1631, 1786, 1923, 3118, 3250 | 564, 3748, 3876, 4162 | | | 223, 228, 486, 1386, 4307 | 228, 486 |
| Bacteroides fragilis YCH46 (295405.11) | 377, 1605, 4225 | 1603 | | 3958 | 1512, 1797, 2476, 2502, 3500 | | | | | | 1604 | | | 71, 224, 355, 359, 404, 640, 748, 932, 937, 1614, 1687, 1690, 1834, 3115, 3244 | 642, 3741, 3867, 4104 | | | 301, 311, 565, 1451, 4231 | 311, 565 |

| Name | | | | | | | | | | | | | | | | | | | | | | |
|---|---|---|---|---|---|---|---|---|---|---|---|---|---|---|---|---|---|---|---|---|---|---|
| Bacteroides intestinalis DSM 17393 (471870.8) | 175, 867, 1175, 2613, 2628 | 2615 | | 4579 | 2172, 2372, 2503, 2515, 2897, 2900 | | | | | | 2614 | | | | 323, 350, 539, 788, 799, 829, 865, 901, 1009, 1016, 1108, 1163, 1246, 1353, 1467, 1470, 1472, 1618, 2242, 2596, 2682, 2839, 2864, 2975, 4059, 4115, 4116, 4499, 4502, 4505, 4510, 4542 | 303, 3973 | 4049, 4574 | 1002, 2550, 2881, 3023, 4544 | 2881, 3023 | | |
| Bacteroides ovatus 3_8_47FAA (665954.3) | 2943, 4833, 4840 | 4835 | | 2667 | 1000, 1209, 1226, 1547, 1742, 3406 | | | | | | 4834 | | | | 127, 258, 308, 310, 315, 317, 376, 592, 976, 1291, 2123, 2340, 2704, 2800, 2958, 2981, 3385, 3709, 3738, 3739, 3743, 4093, 4563, 5006, 5020 | 2492, 2829, 4120, 5049, 5180 | 2979, 2980, 4592 | 129, 415, 3516, 4460, 4748, 4919 | 415, 3516, 4919 | | |
| Bacteroides ovatus ATCC 8483 (411476.11) | 1581, 1585, 2824 | 1583 | | 3102 | 24, 210, 3172, 4710 | | | | | | 1584 | | | | 232, 400, 419, 420, 422, 811, 1182, 1541, 1731, 1732, 1748, 1820, 1821, 1827, 2172, 2189, 2225, 2265, 2279, 2285, 2331, 2348, 2599, 2810, 2967, 3066, 3717, 4205, 4241, 4317, 4673 | 2940, 4408 | 4206, 4207 | 606, 800, 813, 1684, 2101, 2324 | 800, 2101 | | |

| Organism | | | | | | | | | | | | | | | | | | | |
|---|---|---|---|---|---|---|---|---|---|---|---|---|---|---|---|---|---|---|---|
| Bacteroides ovatus SD CC 2a (702444.3) | 306, 1735, 1741 | 1739 | | 2031 | 254, 775, 1727 | | | | | | 1740 | | | 54, 55, 152, 284, 285, 286, 321, 502, 1056, 1934, 2304, 2712, 2715, 3670, 3888, 3976, 4176, 4287, 4334, 4680, 4794 | 280, 1651, 2032 | 1450, 1651, 2032 | 3216, 3801, 4288, 4348, 4350 | 2713, 3049, 3797, 4149, 4678 | 2713, 3049, 3797 |
| Bacteroides ovatus SD CMC 3f (702443.3) | 1595, 1599, 2021 | 1597 | | 3011 | 1270, 1623, 1663, 1684, 1739, 1740, 4326 | | | | | | 1596 | 1611 | | 60, 256, 355, 815, 900, 1295, 1296, 1558, 2006, 2249, 2956, 3041, 3122, 3351, 3352, 3356, 3859, 3861, 3875, 4197, 5415 | 280, 2153, 2333, 2904 | 1559, 1560, 3227, 3229 | | 1941, 2942, 2943, 2945, 3858, 3873, 4795 | 1941 |
| Bacteroides pectinophilus ATCC 43243 (483218.5) | | | | 983 | 168, 251, 994, 2781 | | | | | | | | | | | | | | |
| Bacteroides plebeius DSM 17135 (484018.6) | 2011, 2124 | 2009 | | 86 | 814, 1890 | | | | | | 2010 | | | 45, 460, 471, 488, 895, 912, 979, 980, 1214, 1330, 1350, 1363, 1367, 1640, 1641, 1670, 2133, 2630, 3219, 3286, 3290, 3306, 3320 | 3341, 3415 | 984 | | 463, 493, 1240, 2195, 3374, 3475 | 3374, 3475 |
| Bacteroides sp. 1_1_14 (469585.3) | 974, 990, 2690, 3477, 3480 | 988 | | 3742 | 2271 | | | | | | 989 | | | 21, 85, 646, 758, 761, 766, 767, 1147, 1345, 1361, 1799, 2189, 2206, 2207, 2208, 2212, 2213, 2476, 2594, 2687, 2884, 3201, 3330, 3459, 3909, 3910, 3956 | 770, 913, 1059, 3589 | 753 | | 763, 2931, 3772, 4679, 4817 | 3772, 4817 |

| Species | | | | | | | | | | | | | | | | | | |
|---|---|---|---|---|---|---|---|---|---|---|---|---|---|---|---|---|---|---|
| Bacteroides sp. 1_1_30 (457387.3) | 2585, 3597, 3602 | 3600 | 4626 | 82, 2774, 3024, 3347, 3367 | | | | | | 3601 | | | 323, 344, 345, 348, 916, 1149, 1392, 1743, 1757, 2093, 2119, 2248, 2354, 2357, 2571, 2741, 4598, 4846, 5184 | 2717 | 2094, 2095, 2460, 4501 | 1729, 1730, 1732, 2246, 2355, 2455, 2509, 4367 | 2355, 2455, 2509 | |
| Bacteroides sp. 1_1_6 (469586.3) | 1017, 1143, 1446, 3319, 3322 | 1141 | 3609 | 913, 1407 | | | | | | 1142 | | | 139, 471, 475, 476, 485, 710, 1324, 1374, 1755, 1801, 2162, 2226, 2252, 2253, 2254, 2260, 2365, 3777, 3854, 4019, 4247, 4804, 5119, 5140 | 532, 2248, 2808, 3456 | 2259 | 1240, 1444, 1937, 2251, 3383, 4712 | 1444, 1937 | |
| Bacteroides sp. 20_3 (469591.4) | 845, 1633 | 1631 | 2110 | 3604 | | | | | | 1632 | | | 98, 120, 121, 1147, 1920, 2275, 2837, 3273, 3639, 4829, 4830 | 2164 | | 1456, 2113, 3670 | 1456, 3670 | |
| Bacteroides sp. 2_1_16 (469587.3) | 1827, 2534, 4010 | 1825 | 4486 | 119, 1091, 1115, 1392, 1622, 1727, 3572, 4464 | | | | | | 1826 | | | 155, 1836, 1896, 1899, 1995, 2185, 2190, 2281, 2397, 2517, 2521, 2562, 2792, 2884, 3013 | 2793, 3797, 4272 | | 1669, 2466, 2471, 2717, 4004 | 2471, 2717 | |
| Bacteroides sp. 2_1_22 (469588.3) | 127, 2705, 2711 | 2709 | 982 | 1915, 2416, 2875, 3786 | | | | | | 2710 | | | 142, 292, 552, 733, 760, 763, 1010, 1347, 1486, 2298, 2302, 2303, 2323, 2395, 2530, 3317, 3342, 3343, 3528, 4550 | 326, 1128, 3641 | 864, 3318, 4069, 4071 | 205, 731, 761, 860, 4058 | 205, 761, 860 | |
| Bacteroides sp. 2_1_33B (469589.3) | 1431, 3623, 3683 | 3621 | 1139 | 1074, 3797 | | | | | | 3622 | | | 551, 1040, 1279, 1470, 2131, 2270, 2271, 2525, 3388 | 1181 | | 1013, 3682, 3845 | 1013, 3682, 3845 | |

| Species | | | | | | | | | | | | | | | | | | | |
|---|---|---|---|---|---|---|---|---|---|---|---|---|---|---|---|---|---|---|---|
| Bacteroides sp. 2_1_56FAA (665938.3) | 350, 1855, 4275 | 1853 | | 4863 | 889, 1184, 1758, 2040, 2053, 2797, 2818, 3096, 3159 | | | | | | | | 1854 | | 87, 205, 329, 334, 380, 704, 804, 1004, 1009, 1866, 1937, 2106, 3780, 3996 | 705, 3112, 4755 | | 278, 283, 536, 1675, 4282 | 283, 536 |
| Bacteroides sp. 2_1_7 (457388.5) | 722, 3092 | 3094 | | 2800 | 2870 | | | | | | | | 3093 | | 403, 404, 421, 1509, 2049, 2904, 3718 | 2751 | | 2928, 4144 | 2928, 4144 |
| Bacteroides sp. 2_2_4 (469590.5) | 860, 872, 2667 | 862 | | 3381 | 3806, 3881, 4617, 4820 | | | | | | | | 861 | 884 | 148, 620, 1290, 1320, 1482, 1496, 1498, 1664, 1824, 1825, 1909, 2186, 2207, 2208, 2213, 2682, 2812, 2830, 3408, 4320, 4594, 4727, 4728, 4735, 4949 | 2001, 2879, 3527, 4973 | 4, 6, 1665, 1666, 1667, 2635 | 1056, 1484, 1499, 2740, 2841, 2843, 2844, 4318, 4742, 5656, 5657 | 1056, 2740, 4318 |
| Bacteroides sp. 3_1_19 (469592.4) | 645, 1286 | 1284 | | 2555 | 733, 3787, 3864 | | | | | | | | 1285 | | 163, 967, 989, 1754, 2328, 2772, 3510, 3752 | 2606 | | 2558, 3718, 3926 | 3718, 3926 |
| Bacteroides sp. 3_1_23 (457390.3) | 874, 880, 2675 | 876 | | 2992 | 501, 782, 856, 1226, 3926 | | | | | | | | 875 | 892 | 93, 97, 98, 118, 478, 920, 1588, 1693, 1707, 1709, 1831, 1875, 1877, 1882, 1884, 1947, 1948, 1967, 2472, 2490, 2649, 2783, 2849, 2956, 3281, 3898, 4283, 4309, 4310, 4392, 4881 | 2427, 2818, 3305, 4483 | 2703, 3141, 3143, 4284, 4285 | 1695, 1710, 2458, 2459, 2461, 2570, 3008 | 2570 |

| | | | | | | | | | | | | | | | | | | | | |
|---|---|---|---|---|---|---|---|---|---|---|---|---|---|---|---|---|---|---|---|---|
| Bacteroides sp. 3_1_33FAA (457391.3) | | 793, 796, 889, 2337, 3764 | 795 | | 1444, 3025 | 119, 1443 | | | | | | 794 | | | 164, 373, 374, 382, 890, 938, 1085, 1263, 1285, 1711, 1919, 2125, 2202, 2225, 2382, 3006, 3109, 3278, 3428, 3429, 3462, 3467, 3475, 3497, 3672 | 3568 | 1283, 1284 | 2341, 2348, 3472 | 2341 | |
| Bacteroides sp. 3_1A (469593.3) | | 1910, 2382, 2385 | 2383 | | 2857, 4273 | 1396, 4225, 4274, 4385 | | | | | | 2384 | | | 63, 276, 402, 643, 878, 1351, 2837, 2839, 2946, 3273, 3281, 3282, 3283, 3393, 3681, 3704, 3860, 3886, 4311, 4768 | 448 | 3682, 3683, 3684 | 227, 241, 902, 3380 | 227 | |
| Bacteroides sp. 3_2_5 (457392.3) | | 56, 826, 3419 | 58 | | 4308 | 149, 1445, 2725, 2948, 3192, 3190, 3238, 3696, 3708, 4020 | | | | | | 57 | | | 47, 256, 261, 459, 546, 798, 841, 845, 974, 1087, 1966, 2100, 3747, 4352 | 545, 3213, 4085 | | 208, 620, 893, 898, 3413 | 620, 893 | |
| Bacteroides sp. 4_1_36 (457393.3) | | 184, 1569, 1573, 2207 | 1571 | | 3694 | 1065, 1197, 1212, 1368 | | | | | | 1570 | | | 194, 643, 696, 699, 771, 773, 780, 902, 1002, 1070, 1813, 2053, 2141, 2142, 2213, 3026, 3511 | 2042 | 772 | 1312, 2138 | 1312 | |
| Bacteroides sp. 4_3_47FAA (457394.3) | | 1935, 1938, 1937, 2815 | | | 1016, 2935 | 966, 1017 | | | | | | 1936 | | | 78, 205, 596, 773, 1053, 1566, 1589, 1709, 1786, 2676, 2698, 2718, 2951, 2953, 3618, 3810, 4381, 4382, 4390 | 1663 | 1587, 1588 | 616, 759, 760, 4348 | | |

| Name | | | | | | | | | | | | | | | | | | | | | | | | |
|---|---|---|---|---|---|---|---|---|---|---|---|---|---|---|---|---|---|---|---|---|---|---|---|---|
| Bacteroides sp. 9_1_42FAA (457395.6) | | 1121, 2369, 2372, 2467, 3694 | 2371 | | 151, 2970 | 150, 2080 | | | | | | | | 2370 | | | | 20, 418, 616, 655, 678, 902, 996, 1008, 1349, 1570, 1571, 1579, 2121, 2468, 2469, 2516, 2950, 3053, 3139, 3285, 3286, 3317, 3321, 3347, 3523, 3994 | 3326, 3950 | 676, 677 | | 1125, 1133 | 1125 | |
| Bacteroides sp. D1 (556258.5) | | 357, 1466, 1473 | 1471 | | 2071 | 1051, 2471, 2760, 4268 | | | | | | | | 1472 | | | | 128, 372, 405, 584, 611, 614, 962, 1257, 1261, 1262, 1295, 1748, 2099, 2544, 2685, 3121, 3122, 3146, 3936, 4451 | 162, 1860, 1966 | 717, 3145, 3409, 3411 | | 582, 612, 713, 875, 3449 | 612, 713, 875 | |
| Bacteroides sp. D2 (556259.3) | | 495, 505, 2754 | 497 | | 1794 | 283, 878, 4349, 4377 | | | | | | | | 496 | | | | 66, 536, 590, 896, 1260, 1341, 1379, 1683, 1684, 1685, 1690, 1933, 1934, 1999, 2141, 2249, 2267, 2268, 2493, 2584, 2644, 2770, 2776, 2939, 2953, 3146, 3266, 3670, 3991, 4605, 4891, 5010, | 565, 1124, 1170, 2613, 3003 | 1377, 1378, 4769, 4771 | | 1939, 2247, 2856, 2964, 2966, 2967, 4955 | 2856 | |
| Bacteroides sp. D20 (585543.3) | | 977, 981, 1943, 3518 | 979 | | 641 | 200, 1135, 2907, 3088, 3207, 3471, 3485 | | | | | | | | 980 | | | | 84, 528, 1443, 1446, 1521, 1523, 1530, 1658, 1758, 1898, 1904, 1906, 1953, 1990, 1991, 2278, 2279, 2370, 2606, 3578 | 2379 | 1522 | | 315, 317, 1080, 2281 | 1080 | |

| Genome | | Data | | | Data | Data | | | | | | | | Data | | | | Data | | | | Data | | Data | Data | Data | Data | |
|---|---|---|---|---|---|---|---|---|---|---|---|---|---|---|---|---|---|---|---|---|---|---|---|---|---|---|---|---|
| Bacteroides sp. D22 (585544.3) | | 1042, 4651, 4655 | 4653 | | 2988 | 936, 1400, 1490, 1506, 1645, 2193, 3884 | | | | | | | | 4654 | | | | 286, 414, 882, 901, 1028, 1667, 1711, 1730, 1731, 1735, 2593, 2618, 2699, 2973, 3045, 3484, 3868, 4183 | 832, 2162 | 2594, 2595, 2845, 2847 | | | 412, 425, 867, 868, 871, 981, 2714 | 425, 981 | | | | |
| Bacteroides sp. HPS0048 (1078089.3) | | 265, 813, 2515, 3721 | 3723 | | 4790 | 1628, 2235, 2812, 2813, 3474, 3477, 3634, 4680 | | | | | | | | 3722 | | | | 23, 88, 235, 354, 636, 1055, 1419, 1963, 2090, 2518, 3180, 3183, 3947, 4402, 4404, 4412 | 466 | | | | 120, 151, 1949, 4049 | 151, 4049 | | | | |
| Bacteroides stercoris ATCC 43183 (449673.7) | | 1612, 2824 | 1610 | | 334 | 1097 | | | | | | | | 1611 | | | | 537, 809, 1435, 1454, 2246, 3212 | | | | | 2341 | 2341 | | | | |
| Bacteroides thetaiotaomicron VPI-5482 (226186.12) | | 355, 370, 2896, 3593, 3596 | 368 | | 3872 | 631 | | | | | | | | 369 | | | | 461, 773, 1000, 1009, 1010, 1013, 1664, 1815, 2254, 2255, 2721, 2969, 3019, 3153, 3177, 3241, 3359, 3360, 3407, 3577, 4112, 4218, 4223, 4248, 4312, 4748, 4765 | 289, 3714, 4227 | 4219 | | | 2706, 2898, 3123, 3194, 3651, 4224 | 2898, 3194 | | | | |
| Bacteroides uniformis ATCC 8492 (411479.10) | | 1277, 1526, 1529, 2781, 2994 | 1528 | | 2484 | 2346, 2727, 2745, 2892 | | | | | | | | 1527 | | | | 328, 424, 578, 682, 1223, 1289, 1395, 2888, 2988, 3061, 3062, 3218, 3280, 3287, 3290, 3382, 3385, 3436, 3904, 3912 | 3232 | 3289 | | | 2269, 3065 | 2269 | | | | |

| | | | | | | | | | | | | | | | |
|---|---|---|---|---|---|---|---|---|---|---|---|---|---|---|---|
| Bacteroides vulgatus ATCC 8482 (435590.9) | | 770, 2297, 2399, 2402, 3717, 3718 | 2400 | | 386, 3988 | 3047, 3987, 4127 | | | 2401 | 169, 189, 307, 403, 405, 551, 625, 1029, 1198, 1422, 1836, 1853, 1876, 1877, 2234, 2244, 2296, 2769, 2775, 2776, 3001, 3949, 4245 | 507 | 187, 188 | 660, 763 | 763 | |
| Bacteroides vulgatus PC510 (702446.3) | | 231, 323, 3430, 3433 | 3431 | | 1253, 1628 | 172, 1254 | | | 3432 | 130, 230, 648, 650, 708, 761, 1063, 1761, 1844, 1878, 2301, 2338, 2345, 2346, 2643, 2775, 3179, 3196, 3205 | 919 | 2302, 2303 | 1033, 2496, 2889 | 2889 | |
| Bacteroides xylanisolvens SD CC 1b (702447.3) | | 2585, 2591, 4565 | 2589 | | 2852 | 760, 1854, 4805 | | | 2590 | 31, 171, 400, 401, 1150, 1153, 1180, 1817, 2101, 2824, 3069, 3102, 3103, 3312, 3393, 3456, 3639, 3643, 3644, 3723, 4007, 4550 | 1252, 1551, 3690 | 158, 920, 2027, 2065, 2417 | 1152, 1182, 2031, 2142, 3241 | 1152, 2031, 3241 | |
| Bacteroides xylanisolvens XB1A (657309.4) | | 353, 364, 1616 | 362 | | 3848 | 1837, 2756 | | | 363 | 764, 777, 805, 1008, 1420, 1444, 1603, 2043, 2069, 2158, 2708, 3617, 3818, 3900, 3903, 3904, 3957, 4643 | 1076, 1766 | 584, 586, 689, 2044, 2046 | 453, 694, 776, 807, 1405, 1406, 1408, 1536 | 694, 776, 1536 | |
| Barnesiella intestinihominis YIT 11860 (742726.3) | | 1724, 2373 | 1723 | | 367 | 2491 | 941 | | | 511, 725, 738, 2376, 2871 | | | 108 | 2381 | |
| Bifidobacterium adolescentis L2-32 (411481.5) | 1142 | 1788 | 137 | 138 | 706 | | | | 1914, 63 | 25, 64, 1984 | 60, 749, 1375, 1938 | | 18, 318, 830, 1978 | 18, 318, 830, 1979 | 1977 |
| Bifidobacterium angulatum DSM 20098 = JCM 7096 (518635.7) | 849 | | 280 | 281 | 282 | | | | 285, 287 | 119, 283, 284 | 291, 284 | | 246, 728 | 246, 728 | |
| Bifidobacterium animalis subsp. lactis AD011 (442563.4) | 1392 | | 585 | 586 | 466 | | | 490, 492 | 481, 482 | 149, 271, 483 | 271 | 48, 491 | 921, 1598 | 921, 1598 | |
| Bifidobacterium bifidum NCIMB 41171 (398513.5) | 403 | | 1067 | 1066 | 474, 1614 | 481, 905 | | 482, 483, 484 | 1612 | 833, 1143, 1478, 1479, 1511 | 849, 1704 | | 1721 | 1721 | |
| Bifidobacterium breve DSM 20213 = JCM 1192 (518634.7) | 1055 | | 487 | 486 | 93, 1476 | 1473 | | 524, 527, 1469, 1471, 1472 | 1513, 80 | 81, 1512 | 430, 525 | | 8 | 8 | |

| Organism | C1 | C2 | C3 | C4 | C5 | C6 | C7 | C8 | C9 | C10 | C11 | C12 | C13 | C14 | C15 | C16 | C17 | C18 | C19 | C20 | C21 | C22 | C23 |
|---|---|---|---|---|---|---|---|---|---|---|---|---|---|---|---|---|---|---|---|---|---|---|---|
| Bifidobacterium breve HPH0326 (1203540.3) | 1736 | | 575 | 574 | | | 119, 1074 | 1077 | | | | | 507, 509, 510, 612, 613, 615, 616, 1078, 1079, 1081 | 1038, 80 | | | 81, 1039 | 511, 614 | | 8 | 8 | | |
| Bifidobacterium catenulatum DSM 16992 = JCM 1194 (566552.6) | 643 | | 1028 | 1029 | | | 1151, 1153 | | | | | | 1238 | | | | 1237, 1268, 1310 | 287, 1112, 1529 | | 201, 992, 1557 | 201, 992 | | |
| Bifidobacterium dentium ATCC 27678 (473819.7) | 1757 | | 1165 | 1164 | | | 205 | | | | | | | 1304, 1305 | | | 505, 744 | 82, 1059 | | 246, 500, 592, 849, 858, 895 | 246, 500, 592, 849, 858, 895 | 856 | |
| Bifidobacterium gallicum DSM 20093 (561180.4) | 227 | | 407 | 408 | | | 989 | | | | | | | | | | 809 | | | 1178 | 1178 | | Transporter is not found |
| Bifidobacterium longum DJO10A (Pc;321) (205913.11) | 314 | | 438 | 437 | | | 1832 | 1835 | | | | | 918, 923, 1022, 1023, 1837, 1838, 1839 | 1820 | | | 1819 | 921, 1025 | | 1053, 1308 | 1053 | | |
| Bifidobacterium longum NCC2705 (206672.9) | 436 | | 926 | 927 | | | 1699, 1726 | 1696 | | | | | 875, 880, 995, 997, 1692, 1693, 1694 | 679 | | | 680 | 878, 994 | | 1422, 1476 | 1422 | | |
| Bifidobacterium longum subsp. infantis 157F (565040.3) | 1039 | | 426 | 425 | | | 1698, 1726 | 1729 | | | | | 359, 360, 361, 1731, 1732, 1733 | 680 | | | 679 | 362, 472 | | 1934, 1985 | 1985 | | |
| Bifidobacterium longum subsp. infantis ATCC 15697 = JCM 1222 (391904.8) | 1193 | | 2146 | 2147 | | | 541, 633, 2248 | 2251 | | | | | 909, 910, 911, 2252, 2253, 2254 | 2199, 2410, 2411, 2448, 270, 523 | 1436 | | 2413 | 2098, 2202, 2490 | | 2534 | 2534 | | |
| Bifidobacterium longum subsp. infantis ATCC 55813 (548480.3) | 1185 | | 297 | 298 | | | 651, 678, 1050 | 648 | | | | | 645, 646, 647 | 1893, 452, 794 | | | 453, 793 | 250, 1891 | | 1009, 1598 | 1598 | | |
| Bifidobacterium longum subsp. infantis CCUG 52486 (537007.5) | 1657 | | 206 | 207 | | | 1180, 1207 | 1177 | | | | | 1173, 1174, 1175 | 246, 975 | | | 974 | 157, 248 | | 681, 1708 | 681 | | |
| Bifidobacterium longum subsp. longum 2-2B (1161745.3) | 1486 | | 1294 | 1293 | | | 484, 511 | 514 | | | | | 515, 516, 517, 838, 839 | 281 | | | 282 | 836, 1341 | | 138, 1514 | 1514 | | |
| Bifidobacterium longum subsp. longum 44B (1161743.3) | 766 | | 1263 | 1264 | | | 1604, 1631, 1798 | 1601 | | | | | 103, 104, 105, 1598, 1599, 1600 | 1859 | | | 1858 | 102, 535 | | 211, 375 | 375 | | |
| Bifidobacterium longum subsp. longum F8 (722911.3) | 299 | | 798 | 799 | | | 1189, 1596 | 1599 | | | | | 858, 859, 860, 1600, 1601, 1602 | 549 | | | 550 | 752, 856 | | 1549, 1837 | 1549 | | |
| Bifidobacterium longum subsp. longum JCM 1217 (565042.3) | 1119 | | 413 | 412 | | | 1648, 1676 | 1679 | | | | | 458, 1680, 1681, 1682 | 1541, 366 | | | 1542 | 362, 459 | | 1887, 1953 | 1953 | | |
| Bifidobacterium pseudocatenulatum DSM 20438 = JCM 1200 (547043.8) | 1258, 1259 | | 932 | 931 | | | 208, 210, 1612 | | | | | | | 1749, 734, 815 | | | 788, 816, 1750, 1753 | 110, 759, 835 | | 1722, 1878 | 1722, 1878 | | |
| Bifidobacterium sp. 12_1_47BFAA (469594.3) | 1664 | | 1013 | 1014 | | | 552, 580, 1894, 1899 | 583 | | | | | 475, 476, 585, 586, 587 | 331 | | | 330 | 478, 968 | | 744, 1085 | 744 | | |
| Bilophila wadsworthia 3_1_6 (563192.3) | 954 | 32 | | | | | 986, 1355 | | | | | | | | | | | | | | | | |
| Blautia hansenii DSM 20583 (537007.6) | | | 542 | 719, 1523 | 1524 | | 806 | 1753 | | | | | 653, 657, 658, 659 | 811, 812, 813, 1757, 1758 | | | 1683, 1812, 1930, 2045, 2332 | | | 88 | 88 | | |
| Blautia hydrogenotrophica DSM 10507 (476272.5) | | 2869 | 372 | 373 | | | 596 | 583 | 1033 | | | | 585, 586, 587 | | | | 1073 | | | | | | Transporter is not found |
| Bryantella formatexigens DSM 14469 (478749.5) | | | 3236 | 1433 | 2156 | | 2589 | 554 | | | | | 1634, 1637, 1638, 1639, 2753 | 550, 555, 556, 557 | | | 964, 965, 985, 1691, 1974, 2504, 2766, 2812, 2851, 3870 | 966 | 2871 | 674, 684, 4155 | 581, 681, 1952 | 581, 1952 | |
| Burkholderiales bacterium 1_1_47 (469610.4) | 108 | | | 1722 | | | 603, 1632 | | | | | | 1849 | | | | | | 1904, 1905 | | | | |
| butyrate-producing bacterium SM4/1 (245012.3) | | | | 1722 | | | 2237 | | | | | | 1849 | | | | | | 1904, 1905 | | | | |
| butyrate-producing bacterium SS3/4 (245014.3) | | 1361 | 1837 | 899 | | | 1013, 3086 | | | | | | 968, 969, 970, 971 | | | | | | | | | | |
| butyrate-producing bacterium SSC/2 (245018.3) | | 1126, 1138, 2276 | 1137 | 1139 | | | 1140, 2694 | | 181 | 179, 2271 | | | | | 1136 | | | | | | | | |
| Butyricicoccus pullicaecorum 1.2 (1203606.4) | | 1111, 1160 | 1163 | 1161 | | | 1162 | | 2192 | 2193 | | | 1157, 1158, 1159, 2294, 2300, 2301, 2302 | | | | 3040 | | | 2962 | 2962 | | |
| Butyrivibrio crossotus DSM 2876 (511680.4) | | | 2197 | 2195 | | | 453, 508, 2144, 2196 | | | | | | | | 2198 | | 2199 | | | | | | |
| Butyrivibrio fibrisolvens 16/4 (657324.3) | | 514 | 2726 | 1942 | | | 1856, 2547, 2562 | | | | | | 1516, 1517 | 2729, 2730, 2731 | | | 1385, 2728, 3236 | 613 | 220, 221 | 586, 587, 1389, 2056 | 586, 587, 1389 | | |
| Campylobacter coli JV20 (864566.3) | 693 | | | | | | 204 | | | | | | | | | | | | | | | | |
| Campylobacter upsaliensis JV21 (888826.3) | 1343 | | | | | | 750, 1037, 1409 | | | | | | | | | | | | | | | | |
| Catenibacterium mitsuokai DSM 15897 (451640.5) | | 202, 1928 | 175, 556, 1389, 1879, 1897, 1927 | 175, 1929 | | | 465, 546, 1926 | | | | | | | | 1541 | | 112 | 1268 | | 1019 | 1019 | | |
| Cedecea davisae DSM 4568 (566551.4) | 1301, 2239, 2677 | 1015 | 1016 | 1017 | | | 1018, 1018 | | 220 | 214 | | | 2774, 2775, 2776 | | | | | | | 3198 | 3198 | | |
| Citrobacter freundii 4_7_47CFAA (742730.3) | 3912, 4367 | 2888 | 2889 | 2890 | | | 449, 2891, 2891 | | 1446 | 934, 935, 1445 | | | 4474, 4475, 4476 | | | | 2462 | | | 766 | | | |
| Citrobacter sp. 30_2 (469595.3) | 138, 483 | 1407 | 1408 | 1409 | | | 484, 1410, 1410, 2476 | | 2943 | 2950, 3862, 3863 | | | 592, 593, 594 | | | | 982 | | 4479 | | | | |
| Citrobacter youngae ATCC 29220 (500640.5) | 1827, 2170 | 3256 | 3257 | 3258 | | | 3259, 3259, 4007 | | 3808 | 141, 142, 3815 | | | 2288, 2289, 2290 | | | | 1867, 2805, 2867, 3766 | | | 2801 | 2801 | | |

| Organism | | | | | | | | | | | | | | | | | | | | Notes |
|---|---|---|---|---|---|---|---|---|---|---|---|---|---|---|---|---|---|---|---|---|
| Clostridiales bacterium 1_7_47FAA (457421.5) | | 609 | 2132 | 3194 | 1583, 5582, 5585 | 4154 | | | 2378 | 4998, 4999, 5000, 5001 | 4156, 4157, 4158 | 1332 | | 325, 2539, 4566, 4895, 5089 | | | | 2185, 2659 | 2185, 2659 | |
| Clostridium asparagiforme DSM 15981 (518636.5) | | 2178, 3861 | 1958 | 1874 | 862, 4490, 5292 | | | | | 2786, 2787, 2788, 4849, 4850, 4851, 4852, 4854 | | | | 1702, 2459, 2562 | 3719 | | 2268 | 170 | 170 | |
| Clostridium bartlettii DSM 16795 (445973.7) | 2397 | | | | 2219 | | | | 2253 | | | | | | | | | | | |
| Clostridium bolteae 90A5 (997893.5) | | 2024 | 1518 | 6057 | 1821, 5447, 5449 | 4411 | | | | 2040, 2041, 2042, 2565, 2566, 2567, 2568 | 4408, 4409, 4410 | 5864 | | 69, 3740 | | | 1099, 1810, 4946 | 1130, 4831 | 1130, 4831 | |
| Clostridium bolteae ATCC BAA-613 (411902.9) | | 2646 | 5037 | 4426 | 3898, 4068, 4070 | 1483 | | | 140, 147 | 1313, 1314, 1315, 1316, 2628, 2629, 2630 | 1480, 1481, 1482 | 1864 | | 538, 2103 | | | 885, 3192, 3909 | 1640, 3163 | 1640, 3163 | |
| Clostridium celatum DSM 1785 (545697.3) | 1528, 2692 | 603 | 617 | 618 | 199 | 2573, 2574 | 585 | 574 | | 2466, 2472, 2473, 2474 | 2577, 2578, 2579 | 1197 | | 623, 2065, 2344, 2374, 3008 | | | | | | |
| Clostridium cf. saccharolyticum K10 (717608.3) | | 817 | 3091 | 3552 | 55 | | | | | 1407, 1408, 1409, 1411 | | | | | | 1341 | | | | |
| Clostridium citroniae WAL-17108 (742733.3) | | 45 | 506 | 5650 | 131, 5404, 5406 | 2512 | | | | 3498, 3499, 3500, 3501 | | | | 979, 3594, 5806 | | 573 | 142, 4585, 6067 | 815, 2890, 3231, 4589 | 2890, 4589 | |
| Clostridium clostridioforme 2_1_49FAA (742735.4) | | 3822 | 3497 | 4956 | 503, 3171, 3173, 3893 | 2652 | | | 4527 | 981, 983, 984, 3804, 3805, 3806, 4830, 4831, 4832 | 2653, 2655, 2656 | | | 4686, 4766 | | 137, 4513 | 3904 | 273, 4653, 5156 | 273, 4653, 5156 | |
| Clostridium difficile CD196 (645462.3) | 2677, 3445 | | | 1252 | 2676 | | | | 3387 | 2994, 3384 | | | | | | | | | | Galactose kinase and transporter are not found |
| Clostridium difficile NAP07 (525258.3) | 550, 3473 | | | 3479 | 3472 | | | | 511 | 508, 1052 | | | | | | | | | | Galactose kinase and transporter are not found |
| Clostridium difficile NAP08 (525259.3) | 3523, 3584 | | | 2724 | 3585 | | | | 3484 | 1166, 3481 | | | | | | | | | | Galactose kinase and transporter are not found |
| Clostridium hathewayi WAL-18680 (742737.3) | | 2395, 4478, 4917 | 1176 | 3179 | 699, 701, 714, 715, 812, 2327, 2713, 4774 | 4106, 4107 | | | | 1251, 4810, 4812, 4813 | 4108, 4109, 4110 | | | 569, 1206, 4423 | 1002, 2910, 3196 | | 4564 | 146, 4509 | 146, 4509 | |
| Clostridium hiranonis DSM 13275 (500633.7) | 2125 | | | 996 | 787 | | | | 1060, 1113 | 1058 | | | | | | | | | | Galactose kinase and transporter are not found |
| Clostridium hylemonae DSM 15053 (553973.6) | | 382, 1946, 3355 | 941 | 940 | 2618, 3218 | 3210 | 1189 | 1192 | | 216, 217, 218, 222 | 3213, 3214, 3215 | | | 383, 1518 | | | | 507 | 507 | |
| Clostridium leptum DSM 753 (428125.8) | 419 | 539, 972 | 1931 | 1925 | 1520 | | 5 | 9 | | | | | | 466, 497, 767, 2309 | 778 | 18 | 2455 | 489, 2128, 2635 | 489, 2128, 2635 | Transporter is not found |
| Clostridium methylpentosum DSM 5476 (537013.3) | 186 | | | | 414 | | | | | | | | | | | | | | | |
| Clostridium nexile DSM 1787 (500632.7) | | 3596 | 3598 | 3597 | 1686, 1707, 2694 | 1721 | | | | 792, 793, 794, 798 | 1712, 1713, 1714, 3026, 3027, 3028 | | | 278, 2119, 3063, 3643 | 3029 | | | 275 | 275 | |
| Clostridium perfringens WAL-14572 (742739.3) | 1992, 1996 | 1009 | 1010 | 1011 | 825, 1993, 2000, 2256 | 1927 | 1913 | 2683 | | 1006, 1007, 1008 | 1921, 1922, 1923 | | | 205, 272, 932, 2783 | 2761 | | | 1990 | 1990 | |
| Clostridium ramosum DSM 1402 (445974.6) | 2151 | 62 | 61 | 63 | 64, 223, 1696 | 1039 | | | | | 1030, 1034, 1035, 1036, 1038 | | | 55, 79, 1637 | 1047 | | | 47 | 47 | Transporter is not found |
| Clostridium scindens ATCC 35704 (411468.9) | | 35 | 3455 | 3454 | 124, 2957 | 2966 | 2580 | 2577 | | 1656, 1661, 1662, 1663, 1667, 1668 | 117, 118, 119, 2960, 2961, 2962 | | | 125 | | | | 123 | | |
| Clostridium sp. 7_2_43FAA (457396.3) | 269 | 425 | 1486 | 1484 | 488, 1485 | 3457 | 2043 | 1396, 2042 | | 883, 884, 2641, 2646, 2647, 2648 | 3455, 3456, 3459 | | | 198, 199, 214, 430 | 160 | | | 2240 | 2240 | |
| Clostridium sp. 7_3_54FAA (665940.3) | | 3877 | 4266 | 560 | 2255, 3788, 3807 | 3618 | | | | 4480, 4483, 4484, 4485 | | | | 2989, 3332 | | | | | | |
| Clostridium sp. D5 (556261.3) | | 4375 | 2938 | 2936 | 2937, 3669, 3342, 819 | 4678, 4767, 4915 | 2836, 4279 | 2308, 2832, 4277 | | 4226, 4227, 4228 | 820, 821, 822, 2033, 2034, 2037, 2038, 2039 | 161 | 1935 | 488, 1131, 2629, 2633, 3635, 4229, 4231 | 688, 689, 914 | 42, 691, 3437 | 4191 | 1807, 2325 | 1807, 2325 | |
| Clostridium sp. HGF2 (908340.3) | | | | 3232 | 3566, 4136 | 1474, 2964 | 2391 | | | | 2962, 2965, 2966, 2967 | | | | 2900 | | | | | |
| Clostridium sp. L2-50 (411489.7) | | | 1067 | 1066 | 1994, 2086 | | | | | 1959, 1960, 1961, 1965 | | 1416 | | 1417 | | | | 63, 64 | 63, 64 | |
| Clostridium sp. M62/1 (411486.3) | | 212 | 866 | 2512 | 320, 3363 | | | | | 2636, 2637, 2638 | | | | | | 524 | | | | |
| Clostridium sp. SS2/1 (411484.7) | 852, 863, 1183 | 862 | 864 | | 865, 2767 | | 1976 | 1178, 1974 | | | | | 861 | 1517, 1518 | | | | | | |
| Clostridium spiroforme DSM 1552 (428126.7) | 235 | | | | 838 | | 728 | 729 | | | | | | 1339, 2167, 2168, 2169, 2269 | | | | | | |
| Clostridium sporogenes ATCC 15579 (471871.7) | 463 | | | | 868, 1901 | | | | 1607 | | | | | | | | | | | |
| Clostridium symbiosum WAL-14163 (742740.3) | | 1568 | 2873 | 993 | 72, 1626, 3029 | 366 | | | | 1782, 1783, 1784, 1785 | | | | 587, 799 | | | | | | |
| Clostridium symbiosum WAL-14673 (742741.3) | | 4226 | 3951 | 670, 2301 | 519, 800, 4158 | 150 | | | | 2016, 2017, 2018, 2019 | | | 671 | 565, 672, 883, 884 | | | | | | |
| Collinsella aerofaciens ATCC 25986 (411903.6) | 948 | 768 | 794 | 444, 793 | 1712 | | 788, 789 | 791, 2056 | 799 | 2047 | | | | 2048, 2049, 2050, 2059 | | | | | | |
| Collinsella intestinalis DSM 13280 (521003.7) | 253 | | | | 487 | | 386, 387 | 384 | 381 | | | | | | | | | | | |
| Collinsella stercoris DSM 13279 (445975.6) | 359 | 138, 829 | 137, 828 | 136, 827 | 625 | | 155, 156 | 159 | 161, 776 | 127 | 831, 832, 833 | | | 128, 129, 130, 132 | 1635 | | | | | |
| Collinsella tanakaei YIT 12063 (742742.3) | 622, 1893 | 1362 | 1363 | 1364 | 981 | | 730, 731 | 741 | 704 | | | | | | | | | | | Transporter is not found |

| Organism | | | | | | | | | | | | | | | | | | | | | | Notes |
|---|---|---|---|---|---|---|---|---|---|---|---|---|---|---|---|---|---|---|---|---|---|---|
| Coprobacillus sp. 29_1 (469596.3) | 2740 | 2976 | 2975 | 2977 | | 584, 1142, 1146, 1193, 1194 | 1527 | | 704 | 272, 705, 1430 | | 1528, 1530, 1531, 1532 | | | | 663, 1743 | 81 | | | | | Transporter is not found |
| Coprobacillus sp. 3_3_56FAA (665941.3) | 3217 | 1085 | 1086 | 1084 | | 740, 1083, 1163 | 1492 | | | | | 1483, 1487, 1488, 1489, 1491 | | | | 1066, 1094, 3902 | 1500 | | 1102 | 1102 | | Transporter is not found |
| Coprobacillus sp. 8_2_54BFAA (469597.4) | 3485 | 3064 | 3063 | 3065 | | 283, 1761, 3066 | 596 | | | | | 586, 591, 592, 593, 595 | | | | 147, 217, 3055, 3083 | 604 | | 3047 | 3047 | | Transporter is not found |
| Coprobacillus sp. D6 (556262.3) | 363 | 2560 | 2561 | 2559 | | 1277, 3174, 3178, 3223, 3224 | 3551 | | 1159 | 1158, 1407, 3455 | | 3552, 3554, 3555, 3556 | | | | 1199, 2909 | 1597 | | | | | Transporter is not found |
| Coprococcus catus GD/7 (717962.3) | | 2369 | 22 | 2148 | | 23 | | | | | | | | | | 269 | | | | | | Transporter is not found |
| Coprococcus comes ATCC 27758 (470146.3) | | 2341 | 1070 | 1069 | | 199, 200 | | | | | 1169, 1170, 1171, 1172, 1173 | 1105, 1106, 1107, 1108 | 427 | | | 1166, 1167, 1659 | 676 | | 1126, 1280, 2555 | 1126, 1280, 2555 | | |
| Coprococcus eutactus ATCC 27759 (411474.6) | | | 1663 | 1662 | | 1661 | | | | | 1988, 1989, 1990, 1994 | | | 2263 | | 1668, 2262 | 915 | | 520, 1454, 1515, 1906, 2218 | 520, 1454, 1515, 1906, 2218 | | |
| Coprococcus sp. ART55/1 (751585.3) | | | 1269 | 1270 | | 1271 | | | | | 993, 998, 999, 1000 | | | 2584 | | 1242, 2583 | 2529 | | 1093, 1399, 1460, 1474, 2559 | 1093, 1399, 1460, 1474, 2559 | | |
| Coprococcus sp. HPP0048 (1078091.3) | | 1091 | 1602 | 1601 | | 2296 | 2284 | | | | 1593, 1594, 1596, 1597 | 2290, 2291, 2292 | 821, 958 | | | 158, 717, 940, 1014, 1022, 2369 | | | 2623 | 1335, 1918, 2047, 2623 | | |
| Coprococcus sp. HPP0074 (1078090.3) | | 488 | 2589 | 2590 | | 1779, 2076 | 2064 | | | | 2594, 2595, 2597, 2598 | 2070, 2071, 2072 | 1055, 1703, 623 | | | 560, 569, 641, 1182, 1702, 2135 | | | 257, 2769 | 257, 2769 | | |
| Corynebacterium ammoniagenes DSM 20306 (649754.3) | 730 | | 2090 | | | 238 | | 1185 | | | | | | | | | | | | | | Galactose-1-phosphate uridylyltransferase and transporter are not found |
| Corynebacterium sp. HFH0082 (1078764.3) | 2670 | | 1866 | | | 638, 1404 | | 1014 | | | | | | | | | | | | | | Galactose-1-phosphate uridylyltransferase and transporter are not found |
| Dermabacter sp. HFH0086 (1203568.3) | 1035, 1497 | | 1036, 1056 | | | 165, 1016 | | 1308 | | | | | | | | | | | 869 | 869 | | Galactose-1-phosphate uridylyltransferase and transporter are not found |
| Desulfotobacterium hafniense DP7 (537010.4) | 4837 | | | | | 284, 4780 | | | 3905, 4278 | | | | | | | | | | | | | |
| Desulfovibrio piger ATCC 29098 (411464.8) | 1264, 1285 | | | | | 1165, 1166, 1375, 1436, 1511 | | | | | | | | | | | | | | | | |
| Desulfovibrio sp. 3_1_syn3 (457398.5) | 2203 | | | | | 423, 520, 673, 735 | | | | | | | | | | | | | | | | |
| Dorea formicigenerans 4_6_53AFAA (742765.5) | | 1936 | 2199, 3234 | 2200, 3162, 3163, 3231 | | 2546, 3232, 3233 | | | | | 2192, 2193, 2197 | | | | | 2698, 3236, 3263, 3264 | | | | | | |
| Dorea formicigenerans ATCC 27755 (411461.4) | | 1602 | 1840 | 1841 | | 326 | | | | | 1832, 1833, 1834, 1838 | | | 5 | | 4, 2657 | | | | | | |
| Dorea longicatena DSM 13814 (411462.6) | | 2325 | 2298 | 2299 | | 1524, 1775, 2374 | | | | | 1499, 1503, 1504, 1505 | | | 2203 | | 165, 1573, 1574, 2400, 2401, 2402, 2403, 2525 | | | 1547 | | | |
| Dysgonomonas gadei ATCC BAA-286 (742766.3) | | 873, 1638, 3950 | 1879 | | 1795, 2422 | 3193 | | | | | | | | 2883 | | 88, 300, 407, 1488, 1566, 1593, 1594, 1692, 2392, 2394, 2533, 2535, 3414, 3520, 3901, 4196 | 1387, 4065 | | 1499, 1968, 2543, 2544, 2547, 4063 | 1968 | | |
| Dysgonomonas mossii DSM 22836 (742767.3) | | 484, 898 | 1312 | | 1650 | 355, 2111, 2123, 2654 | | | | | | | | 1384 | | 339, 1116, 1723, 3263 | 1023 | | 2511 | | | |
| Edwardsiella tarda ATCC 23685 (500638.3) | 2181, 2460 | 1723 | 1722 | 1721 | | 1720, 2179, 2182 | | | 1696 | 1685, 2067 | | | | | | 950, 951, 2782 | | | | | | Transporter is not found |
| Eggerthella sp. 1_3_56FAA (665943.3) | 2219 | | | | | 1050, 2020 | | | | | | | | | | | | | | | | |
| Eggerthella sp. HGA1 (910311.3) | 2802 | | | | | 1156, 2014 | | | | | | | | | | | | | | | | |
| Enterobacter cancerogenus ATCC 35316 (500639.8) | 1613, 1966 | 2268 | 2269 | 2270 | | 1956, 2271, 2271 | | 3736 | 3743 | | 21, 22, 23 | | | 625 | | 2930 | | | | | | |
| Enterobacter cloacae subsp. cloacae NCTC 9394 (718254.4) | 3407, 4153 | 1049 | 1048 | 1047 | | 1046, 1046 | | 2631 | 2635 | | 3292, 3293, 3294 | | | 3708 | | 1466, 2902 | | | 47 | 47 | | |
| Enterobacteriaceae bacterium 9_2_54FAA (469613.3) | 2420, 2623 | 427, 2789 | 428 | 429 | | 430 | | 32 | 38, 2788 | | 535, 536, 537 | | | | | 477 | | | | | | |
| Enterococcus faecalis PC1.1 (702448.3) | 1024, 1226 | 923, 1539, 1572 | 924 | 756 | | 315, 757, 1275 | | 2505, 2506 | 655, 1367, 2507 | 2508 | | | | 922 | | 634, 635, 636 | 1961 | | 637, 2029, 2174 | 637, 2029, 2174 | | |
| Enterococcus faecalis TX0104 (491074.3) | 356 | 181, 2001 | 182 | 392, 1048 | | 2546, 2954 | | 462, 463 | 438, 1218, 1241 | 439, 1221 | 437 | | | 464, 465, 466 | 433, 434, 435, 436 | 1113 | 475, 1333 | | | | | |

| Organism | | | | | | | | | | | | | | | | | | | | |
|---|---|---|---|---|---|---|---|---|---|---|---|---|---|---|---|---|---|---|---|---|
| Enterococcus faecalis TX1302 (749513.3) | 2361 | 904, 1114 | 903 | 651, 1015 | | 100, 1520 | | 2455, 2456 | 1698, 1722, 2417 | 2418 | 2416 | | 2457, 2458, 2459 | 2412, 2413, 2414, 2415 | | 1609 | | | | |
| Enterococcus faecalis TX1322 (525278.3) | 361 | 986, 1799 | 985 | 712, 982 | | 43, 711, 983, 984, 1484 | | 1929, 1930 | 297, 1533, 1567, 1931 | 295, 296, 1536, 1932 | 298 | | | 299, 300, 301, 302 | 1027 | 277 | | | | |
| Enterococcus faecalis TX1341 (749514.3) | 1973 | 224, 1315 | 223 | 658, 898 | | 1679, 2403 | | 1877, 1878 | 1917, 2357 | 1916, 2354 | 1918 | | 1874, 1875, 1876 | 1919, 1920, 1921, 1922 | | 2241 | | | | |
| Enterococcus faecalis TX1342 (749515.3) | 2459 | 145, 470 | 471 | 2, 1587 | | 1, 480, 1386, 1586 | | 2560, 2561 | 1315, 1339, 2520 | 2521 | 2519 | | 2562, 2563, 2564 | 2515, 2516, 2517, 2518 | 771 | 1227 | | | | |
| Enterococcus faecalis TX1346 (749516.3) | 150 | 398, 1515 | 399 | 1438, 1750 | | 864, 996 | | | 950, 1835 | 947, 1834 | 1836 | | | 1837, 1838, 1839, 1840 | | | | | | Galactose-6-phosphate isomerase and transporter is not found |
| Enterococcus faecalis TX2134 (749518.3) | 1306 | 734, 902 | 735 | 237, 2457 | | 238, 1593, 2458, 2502 | | | 290, 315, 1252 | 312, 1251 | 1253 | | | 1254, 1255, 1256, 1257 | | 1230 | | | | Galactose-6-phosphate isomerase and transporter is not found |
| Enterococcus faecalis TX2137 (749491.3) | 251 | 1058, 1188 | 1187 | 61, 2415 | | 2156, 2354 | | 346, 347 | 307, 2280, 2304 | 308, 2301 | 306 | | 348, 349, 350 | 302, 303, 304, 305 | 2481 | | | | | |
| Enterococcus faecalis TX4244 (749494.3) | 1596 | 1010, 2502 | 2501 | 557, 558 | | 2082, 2700 | | 1692, 1693 | 1653, 2031 | 1654 | 1652 | | 1694, 1695, 1696 | 1648, 1649, 1650, 1651 | | 710 | | | | |
| Enterococcus faecium PC4.1 (791161.5) | 1063, 1267 | 954, 1596, 1630 | 955 | 777 | | 324, 778, 1319 | | 2585, 2586 | 671, 1416, 2587 | 2588 | | 953 | 650, 651, 652 | 2030 | | | | 653, 2098, 2247 | 653, 2098, 2247 | |
| Enterococcus faecium TX1330 (525279.3) | 697, 1975 | 1067, 1399, 1433, 1778 | 1068 | 491 | | 492, 649, 2412 | | 1611, 1612, 1730, 1731 | 391, 1613, 1732 | 1614, 1733 | | 1066 | 371, 372 | 1880 | | | | 373, 1658 | 373, 1658 | |
| Enterococcus saccharolyticus 30_1 (742813.4) | 1192 | 1771 | 2005 | 2007 | | 2006, 2064 | | | 869, 1366 | 872, 1808, 1850 | 1809 | | 1810, 1811, 1812, 1813 | 692, 2441 | 1244 | 929, 1243, 1322, 1923 | 85 | 928, 992 | 928, 992 | Galactose-6-phosphate isomerase and transporter is not found |
| Enterococcus sp. 7L76 (657310.3) | 171 | 1342, 1799 | 1798 | 1090, 1641 | | 477, 1710 | | | 241, 242, 1664 | 243, 1661 | 240 | | 236, 237, 238, 239 | 1492 | | | | | | Galactose-6-phosphate isomerase and transporter is not found |
| Erysipelotrichaceae bacterium 21_3 (658657.3) | | | 3141 | | | 1926, 3561 | 1268, 4021 | | 4471 | | | 3495, 3496, 3497, 3500 | | | 3960 | | | | | |
| Erysipelotrichaceae bacterium 2_2_44A (457422.3) | | 4919 | | | | 2681, 2734 | 3498, 531 | | 4296 | | | 4019, 4022, 4023, 4024 | | | 3567 | | | | | |
| Erysipelotrichaceae bacterium 3_1_53 (658659.3) | | | | | | 4037, 4117 | 1540, 305 | | 4692 | | | 1537, 1538, 1539, 1542 | | | | | | | | |
| Erysipelotrichaceae bacterium 5_2_54FAA (552396.3) | | | | | | 625 | 2260, 2266 | | 1672 | | | 2257, 2258, 2261, 2262, 2263 | | | | | | | | |
| Erysipelotrichaceae bacterium 6_1_45 (469614.3) | | | 3589 | | | 674, 719 | 1202, 4306 | | 2178 | | | 1200, 1203, 1204, 1205 | | | 1139 | | | | | |
| Escherichia coli 4_1_47FAA (1127356.4) | 3683, 4221 | 563 | 564 | 565 | | 566, 566 | | 2221, 3583 | 2228, 2229, 3582 | | | 4032, 4033, 4034 | 118 | | 993 | | | | | |
| Escherichia coli MS 107-1 (679207.4) | 1675, 4610 | 1409 | 1410 | 1411 | | 1412, 1412 | | 2604, 3851 | 2597, 3850 | | | 4098, 4099, 4100 | 2129, 3792 | | 3007 | | | 4011 | 4011 | |
| Escherichia coli MS 115-1 (749537.3) | 2773, 3752 | 3445 | 3446 | 3447 | | 3448, 3448 | | 498, 874, 875 | 491, 873, 1412 | | | 2429, 2430, 2431, 2432 | 28 | | 4288 | | | | | |
| Escherichia coli MS 116-1 (749538.3) | 1280, 2705, 4974 | 779 | 778 | 777 | | 776, 776, 2706 | | 985, 1312, 1313 | 993, 1041, 1314 | | | 3289, 3290, 3291 | 1483, 3623, 3830 | 2015 | 1247 | | | | | |
| Escherichia coli MS 119-7 (679206.4) | 766, 1629 | 2735 | 2734 | 2733 | | 2732, 2732 | | 1510, 1576 | 1518, 1575 | | | 2368, 2369, 2370 | 209, 1283, 1442 | | 979 | | | 4956 | 4956 | |
| Escherichia coli MS 124-1 (679205.4) | 2991, 3286 | 2516 | 2517 | 2518 | | 2519, 2519 | | 358, 2955 | 351, 2954 | | | 4942, 4943, 4944 | 45 | | 4495 | | | | | |
| Escherichia coli MS 145-7 (679204.3) | 1472, 5238 | 4169 | 4168 | 4167 | | 4166, 4166, 5237 | | 123, 865 | 124, 872 | | | 177, 178, 179 | 802, 2063 | | 3960 | | | | | |
| Escherichia coli MS 146-1 (749540.3) | 3360, 4297 | 2665 | 2666 | 2667 | | 2668, 2668 | | 17, 1417 | 16, 1425 | | | 796, 797, 798 | 2234, 2235, 4258 | | 1307 | | | | | |
| Escherichia coli MS 175-1 (749544.3) | 75, 941, 1679 | 3303 | 3304 | 3305 | | 76, 3306, 3306 | | 756 | 764 | | | 1610, 1611, 1612 | 57 | | 102 | | | 1125 | 1125 | |
| Escherichia coli MS 182-1 (749545.3) | 4271, 4383 | 3966 | 3967 | 3968 | | 3969, 3969, 4277 | | 1678, 3366 | 1679, 3373 | | | 4052, 4053, 4054 | 521, 2116, 5029 | | 2423 | | | | | |
| Escherichia coli MS 185-1 (749546.3) | 915, 3765 | 4758 | 4757 | 4756 | | 3764, 3773, 4755, 4755 | | 2095, 2853 | 1418, 1419, 1830, 2088, 2854, 3978 | | | 2927, 2928, 2929 | 3014, 3015 | | 1611 | | | | | |
| Escherichia coli MS 187-1 (749547.3) | 111, 2562 | 1698 | 1697 | 1696 | | 1695, 1695, 3891 | | 1099, 3102 | 1092, 1100, 3101 | | | 2578, 2579, 2580, 2581 | 4285 | | 2663 | | | | | |
| Escherichia coli MS 196-1 (749548.3) | 2892, 3840 | 4665 | 4666 | 4667 | | 2893, 4668, 4668 | | 2793, 2946, 2947, 2948 | 2785, 2949 | | | 2223, 2224, 2225 | 1009, 1010, 2481 | | 13 | | | | | |
| Escherichia coli MS 198-1 (749549.3) | 772, 3619 | 2268 | 2269 | 2270 | | 773, 1597, 2271, 2271 | | 1966, 5003 | 1973, 5002 | | | 1884, 1885, 1886 | 4209, 4210 | | 1649 | | | | | |
| Escherichia coli MS 200-1 (749550.3) | 1620, 4502 | 3819 | 3818 | 3817 | | 1627, 3816, 3816, 3990 | | 1053, 1569 | 301, 1061, 1312, 1568, 2297, 2298 | | | 1791, 1792, 1793 | 994, 1823 | | 4439 | | | | | |
| Escherichia coli MS 21-1 (749527.3) | 1549, 4094 | 4321 | 4322 | 4323 | | 4324, 4324 | | 1495, 2779 | 1494, 2178, 2787, 3902, 3903 | | | 137, 138, 139 | 590 | | 5038 | | | | | |

| Organism | | | | | | | | | | | | | | | | | | | | |
|---|---|---|---|---|---|---|---|---|---|---|---|---|---|---|---|---|---|---|---|---|
| Escherichia coli MS 45-1 (749528.3) | 204, 3367 | 4488 | 4489 | 4490 | 4491, 4491 | | 1587, 4096 | 1586, 2505, 2936, 3037, 4103, 4217, 4313, 4314 | 2544, 2545, 2546 | | | | | | 1518 | | 3738 | | | |
| Escherichia coli MS 69-1 (749531.3) | 2105, 4145 | 1802 | 1801 | 1800 | 1799, 1799 | | 2055, 2554 | 2054, 2561 | 1998, 1999, 2000 | | | | | | 1309 | | 2382 | | | |
| Escherichia coli MS 78-1 (749532.3) | 1867, 3843 | 2626 | 2627 | 2628 | 337, 2629, 2629, 3842, 3844 | | 1625, 4066 | 1632, 4085 | 3427, 3428, 3429 | | | | | | 591, 3803, 4837 | | 4576 | | | |
| Escherichia coli MS 79-10 (796392.4) | 1370, 2205 | 2070 | 2071 | 2072 | 2073, 2073 | | 1421, 1897 | 1422, 1904 | 2133, 2134, 2135 | | | | | | 573, 1488 | | 4055 | 4678 | 4678 | |
| Escherichia coli MS 84-1 (749533.3) | 454, 2150 | 4702 | 4703 | 4704 | 4705, 4705 | | 419, 3490 | 418, 3483 | 362, 363, 364 | | | | | | 1457 | | 4385 | | | |
| Escherichia coli MS 85-1 (679202.3) | 2615, 3053 | 48 | 49 | 50 | 51, 51 | | 1012, 2581 | 1019, 2580 | 2525, 2526, 2527 | | | | | | 1669 | | 4077 | | | |
| Escherichia coli O157:H7 str. Sakai (386585.9) | 1838, 2979 | 904 | 905 | 906 | 907, 907, 2980 | | 3030, 4185 | 3031, 4193, 4548 | 3166, 3167, 3168 | | | | | | 4130 | | 5331 | | | |
| Escherichia coli SE11 (409438.11) | 1422, 2467 | 936 | 937 | 938 | 939, 939 | | 2516, 3585 | 2517, 3592 | 2568, 2569, 2570 | | | | | | 492, 1885, 3525 | | 4599 | | | |
| Escherichia coli SE15 (431946.3) | 1261, 2016 | 714 | 715 | 716 | 717, 717 | | 2063, 3102 | 2064, 3109, 3221, 3641, 3642 | 2120, 2121, 2122 | | | | | | 323 | | 4161 | | | |
| Escherichia coli str. K-12 substr. MG1655 (511145.12) | 1284, 2119 | 782 | 783 | 784 | 785, 785 | | 2172, 3226 | 2173, 3232 | 2231, 2232, 2233 | | | | | | 352, 1688, 3170 | | 4250 | | | |
| Escherichia coli UTI89 (364106.8) | 1483, 2342 | 841 | 842 | 843 | 844, 844 | | 1176, 2397, 3528 | 1174, 1175, 2398, 3535, 3647, 3921, 4110, 4111 | 2453, 2454, 2455 | | | | | | 482, 3480 | | 4594 | | | |
| Escherichia sp. 1_1_43 (457400.3) | | | | | | | | | | | | | | | 1032 | | 401 | | | |
| Escherichia sp. 3_2_53FAA (469598.5) | 2359, 2622 | 266 | 265 | 264 | 263, 263 | | 1082, 2678 | 962, 1075, 2679, 4279, 4486, 4487 | 2735, 2736, 2737 | | | | | | 627, 1130 | | 3703 | | | |
| Escherichia sp. 4_1B (457401.3) | 813, 2442 | 1765 | 1764 | 1763 | 1762, 1762, 2441, 2443 | | 2059, 3057 | 2058, 2554, 3049 | 2004, 2005, 2006 | | | | | | 2322, 3114 | | 3718 | | | |
| Eubacterium biforme DSM 3989 (518637.5) | | 1823 | 1822 | 1824 | 137, 965, 1821 | 1074 | 29 | | 2258, 2259, 2260 | 1075, 1077, 1078, 1079 | | | | | 128, 129 | | | | | |
| Eubacterium cylindroides T2-87 (717960.3) | | | | | 979 | 1442, 1443 | | | 1438, 1439, 1441 | | | | | | | | | | | |
| Eubacterium dolichum DSM 3991 (428127.7) | 740 | | | | 739, 1442 | | 1786 | | | | | | | | 32, 33 | | | | | |
| Eubacterium hallii DSM 3353 (411469.3) | | 1606 | 941 | 942 | 2223, 2636, 2646 | | | | 1597, 1598, 1599, 1605 | | | 1491 | | | 1490 | | | | | |
| Eubacterium rectale DSM 17629 (657318.4) | | 228 | 2177 | 2178 | 1361 | 2888 | | | 1510, 1511, 1512, 1516 | 2884, 2885, 2886 | 784 | 1447 | | | 1435, 1450, 1451, 2175 | 2082, 2522 | | 1521 | 1521 | |
| Eubacterium rectale M104/1 (657317.3) | | 3310 | 2440 | 2439 | 240 | 959 | | | 3190, 3194, 3195, 3196 | 955, 956, 957, 2443, 2444, 2445 | 2392 | 451 | | | 446, 449, 461, 1617, 2442 | 1381, 2494 | | 3185 | 3185 | |
| Eubacterium siraeum 70/3 (657319.3) | 98 | 1586 | 2194, 2598 | 1153, 2198 | 1828, 2008 | | | | | | | | | | 533, 870, 936, 2193 | 1652 | | 269, 302, 1825 | 269, 302, 1825 | Transporter is not found |
| Eubacterium siraeum DSM 15702 (428128.7) | 1888 | 509 | 218 | 1173 | 391, 2344 | | | | | | | | | | 808, 2023, 2199 | 486 | | 394, 1311, 1339 | 394, 1311, 1339 | Transporter is not found |
| Eubacterium siraeum V10Sc8a (717961.3) | 853 | 1963 | 977 | 2245 | 1758, 2403 | | | | | | | | | | 99, 1355, 1780 | 1853 | | 632, 663, 1761 | 632, 663, 1761 | Transporter is not found |
| Eubacterium sp. 3_1_31 (457402.3) | | | | | 2375 | 231, 238 | 1386 | | | 235, 236, 237, 240 | | | | | | | | | | |
| Eubacterium ventriosum ATCC 27560 (411463.4) | | 589 | 591 | 590 | 1775 | | | | 1203, 1207, 1208 | | | 714 | | | 713, 2143 | | | | | |
| Faecalibacterium cf. prausnitzii KLE1255 (748224.3) | 523 | 457 | 535, 2487 | 2486 | 131 | 2586 | | | 1979, 1980, 1981 | 2582, 2583, 2584 | 992 | | | | 228, 1108, 1490, 1569, 1832 | | | | | |
| Faecalibacterium prausnitzii A2-165 (411483.3) | 904 | 196 | 926, 1631 | 927 | 194, 593 | 1155 | | | 729, 730, 731 | 1151, 1152, 1153 | | | | | 145, 511, 955, 1662, 1999 | | | 1664, 2550 | 2550 | |
| Faecalibacterium prausnitzii L2-6 (718252.3) | 1720 | 159 | 3212, 3231 | 3230 | 158 | 113 | 1292 | 1297 | 1271, 1272, 1273 | 115, 116, 117, 3087, 3088, 3089 | 2750 | | | | 697, 1177, 1625, 2395, 2475, 3086 | | | | | |
| Faecalibacterium prausnitzii M21/2 (411485.10) | 1540 | 1032 | 1527, 2200 | 2201 | 1034 | 2375 | | | 2456, 2457, 2458 | 2377, 2378, 2379 | | | | | 28, 38, 356, 387, 1474, 1697, 2374 | | 1476 | | | |

| | | | | | | | | | | | | | | | | | |
|---|---|---|---|---|---|---|---|---|---|---|---|---|---|---|---|---|---|
| Faecalibacterium prausnitzii SL3/3 (657322.3) | 2284 | 821 | 2271, 2801 | 2800 | | 823, 1643 | 2621 | | | 747, 748, 749 | 2617, 2618, 2619 | 538, 1365, 1398, 2249, 2407, 2417, 2622 | | | 1613, 2051, 2252 | 1613 | |
| Fusobacterium gonidiaformans ATCC 25563 (469615.3) | 711 | | 450 | 451 | | 452 | | | | 468, 469, 470 | | | | | | | |
| Fusobacterium mortiferum ATCC 9817 (469616.3) | 652, 2274 | 2212 | 744 | 743 | | 742, 1069 | | 509 | 508, 741 | 636, 637, 638 | | 359 | 971 | | 969, 1756 | 969, 1756 | |
| Fusobacterium necrophorum subsp. funduliforme 1_1_36S (742814.3) | 215 | 1959 | 1236, 1237 | 1238 | | 364, 386, 1239 | | | | 679, 680, 681, 682 | | | | | | | |
| Fusobacterium sp. 11_3_2 (457403.3) | 942 | 2285 | 2439 | 2440 | | 1099, 2441 | | | | 307, 308, 309 | | | | | | | |
| Fusobacterium sp. 12_1B (457404.3) | 457 | 770 | 2313 | 2314 | | 633, 846, 2063, 2315 | | 3260 | 3261 | 2370, 2371, 2372 | | | | | | | |
| Fusobacterium sp. 1_1_41FAA (469621.3) | 952 | 1907 | 110 | 111 | | 112 | | | | 378, 379, 380 | | | | | | | |
| Fusobacterium sp. 21_1A (469601.3) | 1353 | 1617 | 1879 | 1878 | | 1450, 1877 | | | | 827, 828, 829 | | | | | | | |
| Fusobacterium sp. 2_1_31 (469599.3) | 553 | 1100 | 2025 | 2024 | | 2023 | | | | 771, 772, 773 | | | | | | | |
| Fusobacterium sp. 3_1_27 (469602.3) | 678 | 1589 | 290 | 291 | | 292, 1922 | | | | 794, 795, 796 | | | | | | | |
| Fusobacterium sp. 3_1_33 (469603.3) | 259 | 1075 | 1471 | 1473 | | 356, 1474 | | | | 657, 658, 659, 660 | | | | | | | |
| Fusobacterium sp. 3_1_36A2 (469604.3) | 1311 | 711 | 1880 | 1881 | | 138, 1882 | | | | 1201, 1202, 1203 | | | | | | | |
| Fusobacterium sp. 3_1_5R (469605.3) | 1122 | | 278 | 277 | | 276 | | | | 960, 961, 962 | | | | | | | |
| Fusobacterium sp. 4_1_13 (469606.3) | 327 | 870 | 2019 | 2018 | | 417, 2017 | | | | 1559, 1560, 1561 | | | | | | | |
| Fusobacterium sp. 7_1 (457405.3) | 1319 | 1995 | 2301 | 2300 | | 2299 | | | | 650, 651, 652 | | | | | | | |
| Fusobacterium sp. D11 (556264.3) | 1352 | 2236 | 1203 | 1201 | | 1200, 1693 | | | | 151, 152, 153 | | | | | | | |
| Fusobacterium sp. D12 (556263.3) | 926 | | 1442 | 1443 | | 1444 | | | | 4, 5, 6 | | | | | | | |
| Fusobacterium ulcerans ATCC 49185 (469617.3) | 346 | 78 | 1383 | 1382 | | 189, 1381, 1846 | | 722, 3039 | 721, 3038 | 1324, 1325, 1326 | | | | | | | |
| Fusobacterium varium ATCC 27725 (469618.3) | 122 | 1978 | 744 | 743 | | 179, 742, 1903, 2026, 2147 | | 1213 | 1212 | 686, 687, 688 | | | | | | | |
| Gordonibacter pamelaeae 7-10-1-b (657308.3) | 413 | | | | | 940 | | | | | | | | | | | |
| Hahia alvei ATCC 51873 (1002364.3) | 340, 2406 | 2565, 3934 | 3933 | 3932 | | 3931 | | 908 | 902, 2566 | 3809, 3810, 3811 | | 3875 | | | 912 | 912 | |
| Helicobacter bilis ATCC 43879 (613026.4) | 1128 | | | | | 997, 1010, 2121 | | | | | | | | | | | |
| Helicobacter canadensis MIT 98-5491 (Pi(30719) (537970.9) | 1443 | | | | | 151 | | | | | | | | | | | |
| Helicobacter cinaedi ATCC BAA-847 (1206745.3) | 450 | | | | | 326, 877, 1852 | | | | | | | | | | | |
| Helicobacter cinaedi CCUG 18818 (537971.5) | 1824 | | | | | 603, 1625, 1942, 1943 | | | | | | | | | | | |
| Helicobacter pullorum MIT 98-5489 (537972.5) | 827 | | | | | 570 | | | | | | | | | | | |
| Helicobacter winghamensis ATCC BAA-430 (556267.4) | 788 | | | | | 484 | | | | | | | | | | | |
| Holdemania filiformis DSM 12042 (545696.5) | | 2721 | 2722 | | 3459 | 2272, 2770 | | | | 3170, 3171, 3172 | | | 385 | | | | |
| Klebsiella pneumoniae 1162281 (1037908.3) | 2183, 4204 | 4718 | 4719 | 4720 | | | | | 4163 | 2257, 2258, 2259, 2619 | | 3709 | 296, 2536 | 385 | 4155 | 4155 | UDP-glucose 4-epimerase is not found |
| Klebsiella pneumoniae subsp. pneumoniae WGLW5 (1203547.3) | 2399, 2757 | 910 | 911 | 912 | | 913, 913 | | | | 651, 2831, 2832, 2833, 5167 | | 1727, 5270 | 572, 4767 | 4664 | 5346 | 5346 | |
| Klebsiella sp. 1_1_55 (469608.3) | 480, 1674 | 1852 | 1853 | 1854 | | 1855, 1855 | | 2810 | 2806, 4917 | 1598, 1599, 1600 | | 1285 | 2482, 4036 | 4136 | 4909 | 4909 | |
| Klebsiella sp. 4_1_44FAA (665944.3) | 225, 527 | 1638 | 1639 | 1640 | | 1641, 1641 | | | 4752 | 146, 147, 148, 1379 | | | 1217 | 4135 | 4760 | 4760 | |
| Lachnospiraceae bacterium 1_1_57FAA (658081.3) | | 132 | 122 | 160 | | 183 | 613 | 262 | 263 | 123, 128, 129, 131, 136 | 605, 606, 615, 616, 617, 620 | 547, 1620, 2560 | 460 | 653, 1616 | | | |
| Lachnospiraceae bacterium 1_4_56FAA (658655.3) | | 2460 | 492 | 493 | | 2325 | 359 | | | | 53, 54, 55, 56 | 397, 676, 1285, 2025, 2109, 2110, 2238, 2239, 2765, 2824, 2847 | 2269 | 1370 | | | Transporter is not found |
| Lachnospiraceae bacterium 2_1_46FAA (742723.3) | | 506 | 2071 | 2070 | | 1789 | 1802 | | | 1605, 1606, 1607 | | 458, 776, 1372, 1391, 1476, 1630, 2069 | | | | | |
| Lachnospiraceae bacterium 2_1_58FAA (658082.3) | | 3009 | 235 | 236 | | 846 | 1677 | 1925 | 1929 | | 1671, 1683, 2212, 2213, 2214 | 1257, 1825, 2006, 2181, 2182, 3016, 3344 | 2058 | | 2396, 2397, 3230 | 2396, 2397, 3230 | Transporter is not found |
| Lachnospiraceae bacterium 3_1_46FAA (665950.3) | | 180 | 170 | 207 | | 222 | 2413 | 306 | 307 | 171, 176, 177, 179, 184 | 2406, 2407, 2415, 2416, 2417, 2420 | 1102, 1202, 2352 | 2260 | 1196, 2438 | | | |

| Organism | C2 | C3 | C4 | C5 | C6 | C7 | C8 | C9 | C10 | C11 | C12 | C13 | C14 | C15 | C16 | C17 | C18 | C19 | Notes |
|---|---|---|---|---|---|---|---|---|---|---|---|---|---|---|---|---|---|---|---|
| Lachnospiraceae bacterium 3_1_57FAA_CT1 (658086.3) | 5123 | 1859, 3117 | 1857 | 1556, 2892, 2098, 6940, 7094 | 96 | | 818, 822, 823, 824 | 2099, 2100, 2101 | | | | | 2047, 2555, 3818, 3819, 3839, 3847, 4082, 4145, 4350, 4709, 4802, 4882, 5010, 5077, 5785, 5876, 6084, 6170, 6651, 7155 | 1306, 1326, 4715, 6401 | 1098, 3814, 3815, 4072, 4254, 4277, 5132, 5778, 6434 | 4706, 4804, 5177, 5325 | 328, 695, 3456, 3955, 4283, 4403, 4707, 6583 | 3456, 4403 | |
| Lachnospiraceae bacterium 4_1_37FAA (552395.3) | 1223 | 1950 | 1949 | 2523 | 2536 | | 1941, 1942, 1943, 1944 | 2527, 2528, 2529 | | | | | 577, 1299, 1300, 1308, 1387, 1900, 2444 | | | | 959, 2219, 2919 | 959, 2219, 2919 | |
| Lachnospiraceae bacterium 5_1_57FAA (658085.3) | 1480 | 2991 | 2990 | 234, 1697 | 243 | 786 | 789 | 904, 909, 910, 911 | 237, 238, 239, 1688, 1689, 1690 | | | | 1698, 1699 | 1695, 1696 | | | | | |
| Lachnospiraceae bacterium 5_1_63FAA (658089.3) | 191, 203, 1301 | 192 | 190 | 189, 1186, 2061 | | 1308 | | | 193 | | | | 2812, 2813 | 395 | | | 1747 | 1747 | |
| Lachnospiraceae bacterium 6_1_63FAA (658063.3) | 1952 | 84, 963, 83 | | 1604 | 642 | | 885, 889, 890, 892 | 646, 647, 1608, 1609, 1610 | | | | | 576, 696, 812, 1588 | | | | 1750 | 1750 | |
| Lachnospiraceae bacterium 7_1_58FAA (658087.3) | 3203 | 4682 | | 5251 | 5075 | | | | | | | | 831 | | | 5072 | | | |
| Lachnospiraceae bacterium 8_1_57FAA (665951.3) | 656 | 666 | 629 | 607 | 851 | 530 | 529 | | | | | | 914, 1782, 2033 | 1000 | | | 811, 2037 | | |
| Lachnospiraceae bacterium 9_1_43BFAA (658088.3) | 513 | 1703 | 1702 | 1622, 2216, 2228 | | | 1695, 1696, 1697, 1698 | 2220, 2221, 2222 | | | | | 39, 360, 435, 436, 444, 1546, 2158 | | | | 787, 2549 | 787, 2549 | |
| Lactobacillus acidophilus ATCC 4796 (525306.3) | 305, 1823 | 1149 | 1151 | 270, 1150, 1710 | 1161 | | | | 1156, 1157 | | | | 1160 | 1054, 1154, 1155 | | | 691, 691, 1132 | 691, 1132 | |
| Lactobacillus acidophilus NCFM (272621.13) | 597, 1640 | 1379 | 1381 | 560, 1380, 1754 | 1390 | | | | 1385 | | | | 1290, 1384 | | | | 1364 | 1364 | |
| Lactobacillus amylolyticus DSM 11664 (585524.3) | 1502 | 394 | 396 | 395, 519, 1533 | 398 | | | | | | | | | | | | | | Transporter is not found |
| Lactobacillus antri DSM 16041 (525309.3) | 1060 | 1463 | 1358 | 1359 | 1035 | | | | 1369, 757 | | | | 1498, 1499, 1500 | 395 | | | 1915 | 1915 | |
| Lactobacillus brevis ATCC 367 (387344.15) | 624 | 356, 357 | 1809 | 1807 | 1808 | | | | 1810 | | | | 2161 | | | | 1666 | 1666 | |
| Lactobacillus brevis subsp. gravesensis ATCC 27305 (525310.3) | 2524 | 234, 1419 | 2574 | 1967, 2572 | 2573 | | 1784 | | 1324, 68 | | | | 899, 900 | 465, 1325 | | | 276, 510, 514 | 276, 510, 514 | |
| Lactobacillus buchneri ATCC 11577 (525318.3) | 2410, 2488 | 1132 | 1771 | 1709, 1769 | 1770 | | 1293 | | 1072 | | | | 648, 649 | | | | 166 | 166 | |
| Lactobacillus casei ATCC 334 (321967.11) | 1065 | 672, 1771 | 668 | 670 | 669, 1981 | 682, 683 | 680, 2598 | 681, 2540 | | | 675, 676, 677 | | 404 | | | | 1010, 1010, 2043 | 2043 | |
| Lactobacillus casei BL23 (543734.4) | 1176 | 697, 1912 | 693 | 695 | 694, 2074 | 707, 708 | 705, 2694 | 386, 706, 2632 | 276 | | 700, 701, 702 | 277, 278, 279, 280 | | | | | 1118, 2148 | 1118, 2148 | |
| Lactobacillus crispatus 125-2-CHN (575595.3) | 971, 972, 1316, 1317, 1318, 1941 | 688 | 690 | 689, 1187, 1974 | 696 | | 1902 | | 692 | | | | 694, 695 | 691 | | | | | |
| Lactobacillus delbrueckii subsp. bulgaricus ATCC 11842 (390333.7) | 507 | | | | 1602 | | | | 1071 | | | | 1070 | | | | | | |
| Lactobacillus delbrueckii subsp. bulgaricus ATCC BAA-365 (321956.7) | 495 | | | | 1625 | | | | 1086 | | | | 1085 | | | | | | |
| Lactobacillus fermentum ATCC 14931 (525325.3) | 791 | 259 | 571 | 570 | 1708 | | | | 567 | | | | 1656 | | | | 568 | 568 | |
| Lactobacillus fermentum IFO 3956 (334390.5) | 371 | | 1974 | 1973 | 327 | | | | 855 | | | | | | | 856 | 1956 | 1956 | |
| Lactobacillus gasseri ATCC 33323 (324831.13) | 686 | 1309 | 1311 | 1310, 1698 | 1313, 1533 | 479, 480 | 147, 160 | 144 | 17 | | 482, 483, 484 | 116, 117, 118, 119, 124 | | | | 856 | 253 | 253 | |
| Lactobacillus helveticus DPC 4571 (405566.6) | 682, 1880 | 1578 | 1580 | 629, 1579 | 1635 | | | | 1626, 1627, 1628, 1629, 1630 | | | | 1633, 1634 | 1625 | | | 492, 493, 494, 495 | 492, 493, 494, 495 | |
| Lactobacillus helveticus DSM 20075 (585520.4) | 1174 | 1719, 1720 | 1722 | 1210, 1409, 1721 | 1300 | | | | 1292, 1293, 1294, 1295, 1296 | | | | 1298, 1299 | 1291 | | | 1167 | 1167 | |
| Lactobacillus hilgardii ATCC 8290 (525327.3) | 1052 | | 2064 | 1110, 2066 | 2065 | | 1543 | | 259 | | | | 1897, 1898 | | | | 417 | 417 | |
| Lactobacillus johnsonii NCC 533 (257314.6) | 1414 | 718 | 716 | 717, 1750 | 710, 1584 | | 155, 170 | 152 | 16 | 272, 714 | | 128, 129, 130, 131, 135 | 711, 712 | 593, 715 | | | 273 | 273 | |
| Lactobacillus paracasei subsp. paracasei 8700:2 (537973.8) | 389 | 820, 2604 | 2600 | 2602 | 2018, 2601 | 2614, 2615 | 1460, 2338, 2612 | 1099, 2613, 2333 | | | 2607, 2608, 2609 | 2334, 2335, 2336, 2337 | | | | | 444, 2088 | 444, 2088 | |
| Lactobacillus paracasei subsp. paracasei ATCC 25302 (525337.3) | 358 | 703, 2385 | 2381 | 2383 | 2191, 2382, 2733 | 2394, 2395 | 1849, 2382 | 1904, 2393 | | | 2386, 2389, 2390 | 1132, 1133, 1134 | | | | | 413, 1353 | 413, 1353 | |
| Lactobacillus plantarum 16 (1327988.3) | 614 | 674, 1379, 2806 | 2801 | 1444, 2799 | 573, 2800, 3137 | | 1691 | | 2805 | | | | 2802, 2803 | | | | 167, 168, 2804 | 167, 168, 2804 | |
| Lactobacillus plantarum subsp. plantarum ATCC 14917 (525338.3) | 1332 | 1401, 2331 | 523 | 521, 2268 | 522, 1292 | | 2027 | | 510, 527 | | | | 524, 525 | 511 | | | 526 | 526 | |

| | | | | | | | | | | | | | | | | | | | | | | |
|---|---|---|---|---|---|---|---|---|---|---|---|---|---|---|---|---|---|---|---|---|---|---|
| Lactobacillus plantarum WCFS1 (220668.9) | 638 | 701, 1460 | 2898 | 1522, 2896 | | 596, 1018, 2897 | | 1773 | | | | | | | 2885, 2902 | | | 2899, 2900 | 2886 | 2901 | 2901 | |
| Lactobacillus reuteri CF48-3A (525341.3) | 525 | 695 | 203 | 204 | | 557, 1243 | | | | | | | | | 1218, 13 | | | 437, 438 | 1219 | 1620 | 1620 | |
| Lactobacillus reuteri DSM 20016 (557436.4) | 382 | | 1794 | 1793 | | 348 | | | | | | | | | 1097, 1786 | | | 296, 297 | 1096 | 919 | 919 | |
| Lactobacillus reuteri F275 (299033.6) | 365 | | 1740 | 1739 | | 333 | | | | | | | | | 1067, 1731 | | | 282, 283 | 1066 | 896 | 896 | |
| Lactobacillus reuteri MM2-3 (585517.3) | 1201 | | 232 | 231 | | 1008 | | | | | | | | | 1509, 224 | | | 1059, 1060 | 1508 | 824 | 824 | |
| Lactobacillus reuteri MM4-1A (548485.3) | 1647 | | 64 | 65 | | 1676 | | | | | | | | | 72, 913 | | | 1727, 1728 | 914 | 1253 | 1253 | |
| Lactobacillus reuteri SD2112 (491077.3) | 1038 | 1552 | 29 | 30 | | 1007, 1631 | | | | | | | | | 1655, 37, 38 | | | 956, 957 | 1654 | 1817 | 1817 | |
| Lactobacillus rhamnosus ATCC 21052 (1002365.5) | 1177 | 1935, 2035 | 2031 | 2033 | | 2032, 2298 | 2047, 2048 | 433, 523, 2045, 2742 | 500, 2046, 2671 | 427, 428 | | | | | | | 429, 430, 431, 432 | | | 1524, 1824, 2369 | 2369 | |
| Lactobacillus rhamnosus GG (Pr) 40637) (568703.30) | 1054 | 660, 1851 | 656 | 658 | | 657, 2012 | 669, 670 | 343, 667, 2584 | 415, 668, 2584 | 336, 337 | | | | | | 662 | 338, 340, 341, 342 | | | 992, 2086 | 992, 2086 | |
| Lactobacillus rhamnosus LMS2-1 (525361.3) | 64 | 784, 2173 | 2169 | 2171 | | 2170, 2444 | 2184, 2185 | 1200, 1980, 2182 | 1264, 1953, 2183 | 1786 | | | | | | 2175, 2176, 2177 | 1787, 1788, 1789, 1791 | | | 961, 1155, 2372 | 961, 1155, 2372 | |
| Lactobacillus ruminis ATCC 25644 (525362.3) | 1770 | 213 | 1281 | 1280 | | 215 | | 520 | | | | | | | 258, 301, 736 | | | 259 | 735 | 300 | 300 | |
| Lactobacillus sakei subsp. sakei 23K (314315.12) | 1467 | 744 | 741 | 743 | | 742, 1244 | | | | | | | | | 177 | | | 1675, 1676 | | 1756 | 1756 | |
| Lactobacillus salivarius ATCC 11741 (525364.3) | 1843 | 1544, 1830 | 348 | 346, 989 | | 347, 922, 1540 | | | | | | | | | 350 | | | 352 | | 345, 964 | 345, 964 | |
| Lactobacillus salivarius UCC118 (362948.14) | 1342 | 1355, 1586 | 458 | 460, 1111 | | 459, 1039, 1590 | | 241 | | | | | | | 454, 456 | | | 452 | | 461, 462, 1084 | 461, 462, 1084 | |
| Lactobacillus sp. 7_1_47FAA (665945.3) | 217 | | | | | 1054 | 840, 841 | 1111 | 843 | 849 | | | | | | 845, 847, 848 | 850, 851, 852, 853 | | | | | |
| Lactobacillus ultunensis DSM 16047 (525365.3) | 680, 1633, 1639 | 1466 | 1468 | 632, 838, 1467 | | 1474 | | | | | | | | | 1470 | | | 1472, 1473 | 1469 | 391, 1460, 1482, 1483, 1723 | 391, 1460, 1482, 1483, 1723 | |
| Leuconostoc mesenteroides subsp. cremoris ATCC 19254 (586220.3) | 603 | 65 | 67, 747 | 746 | | 1451 | | | | | | | | | 68, 750, 752 | | | 70, 71 | | | | |
| Listeria grayi DSM 20601 (525367.9) | 86 | 1151, 2253 | 2264 | 2266 | | 160, 2265 | | | 645 | | | | | | | | | | | 2246, 2247 | 2246, 2247 | Transporter is not found |
| Listeria innocua ATCC 33091 (1002366.3) | 1661 | 2357 | | | | 2356 | | 527, 778 | 1141 | | | | | | | | | | | | | |
| Megamonas funiformis YIT 11815 (742816.3) | 1973 | 2279 | 981 | 2170, 2281 | | 2282 | | | | | | | | | 852, 854 | | | 2030 | 2027 | 458, 1210, 2278 | 458, 1210, 2278 | |
| Megamonas hypermegale ART12/1 (657316.3) | 25 | 2690 | 1204 | 621, 2692, 2693 | | 2694 | | | | | | | | | 1114, 1115, 1117, 1118 | | | 1560 | 1567, 1568, 1569 | 1598, 2601, 2602, 2688 | 1598, 2601, 2602, 2688 | |
| Methanobrevibacter smithii ATCC 35061 (420247.6) | 1568 | | | | | 318, 1656 | | | | | | | | | | | | | | | | |
| Methanosphaera stadtmanae DSM 3091 (339860.6) | 1054 | | | | | 57, 66, 520, 952, 1041 | | | | | | | | | | | | | | | | |
| Mitsuokella multacida DSM 20544 (500635.8) | 141, 1791 | 2223 | 1330 | 1332, 1889 | | 680, 1331 | | | | | | | | | 138 | | | 139 | 66 | 140, 1342 | 140, 1342 | |
| Mollicutes bacterium D7 (556270.3) | 3238 | 2209 | 2210 | 2208 | | 1804, 2207, 3187, 3188 | 1624 | | | | | | 1615, 1619, 1620, 1621, 1623 | | | | | 1749, 2190, 2191, 2217 | 1632 | 2225 | 2225 | Transporter is not found |
| Odoribacter laneus YIT 12061 (742817.3) | | 1462, 1802, 1800 | | | 2822 | 988, 2213, 3269 | | | | | | | | 1801 | | | | 212, 1210, 1225, 1704, 2239, 2428, 2751, 3086 | | 365, 1351 | 365, 1351 | |
| Oxalobacter formigenes HOxBLS (556268.6) | 139, 1547 | | | | | 1854, 2340 | | | | | | | | | | | | | | | | |
| Oxalobacter formigenes OXCC13 (556269.4) | 317, 1373 | | | | | 139, 601 | | | | | | | | | | | | | | | | |
| Paenibacillus sp. HGF5 (908341.3) | 1164, 5468 | | 3314 | 3312 | | 225, 1999, 3313 | | | 6203 | 5909 | | | | 1433, 1434, 6385 | 942, 1217, 2508, 2510, 3110, 3532, 3757, 5397 | 3672, 3993 | 3381, 4611, 5327 | 814, 2370, 2383 | 657, 3488, 4610 | 657, 4610 | | |
| Paenibacillus sp. HGF7 (944559.3) | 889, 1272, 1637, 5231 | | 4321 | 4319 | | 782, 4320, 5481 | | 3326 | 3325 | | | | | | 926, 3290, 4207 | 742 | | 4960, 5265 | | | | Transporter is not found |
| Paenibacillus sp. HGH0039 (1078505.3) | 863, 1363, 3417, 4131 | | 2440 | 2438 | | 1478, 1491, 2439, 3524, 3630 | | 4398 | 4399 | | | | | | 3380, 3207, 4434, 3566 | | | 94, 4281 | | | | Transporter is not found |
| Parabacteroides distasonis ATCC 8503 (435591.13) | | 1654, 3706 | 1656 | | 1552 | 24, 495, 3797 | | | | | | | | 1655 | | | | 219, 461, 966, 1407, 1962, 3063, 3080, 3081, 3257 | 1502 | 84, 436 | 84, 436 | |

| Organism | C2 | C3 | C4 | C5 | C6 | C7 | C8 | C9 | C10 | C11 | C12 | C13 | C14 | C15 | C16 | C17 |
|---|---|---|---|---|---|---|---|---|---|---|---|---|---|---|---|---|
| Parabacteroides johnsonii DSM 18315 (537006.5) | | 993, 3337 | 991 | | 1142 | 3149 | | | | | 1392 | | 345, 369, 403, 478, 995, 1556, 1576, 1995, 2448, 2455, 2526, 2527, 2583, 2663, 2690, 2795, 3413, 3534, 3535, 3544 | 1137 | 1953, 2387 | 1953 |
| Parabacteroides merdae ATCC 43184 (411477.4) | | 1590, 2341 | 2339 | | 2825 | 3366 | | | | | 2340 | 253 | 192, 215, 252, 258, 1217, 1239, 1662, 1719, 2196, 2412, 2916, 3108, 3339, 3345, 3404, 3405 | 264 | 1894 | 1894 |
| Parabacteroides sp. D13 (563193.3) | | 607, 1226 | 1224 | | 2682 | 2610, 4119 | | | | | 1225 | | 76, 885, 886, 1090, 1091, 1633, 2200, 2575, 3198, 3911 | 2737 | 2479, 2685, 4072 | 2479, 4072 |
| Parabacteroides sp. D25 (658661.3) | | 221, 2761 | 219 | | 3176 | 3103 | | | | | 220 | | 632, 781, 782, 1185, 1917, 3069, 3677, 4104 | 3227 | 3044, 4195 | 3044, 4195 |
| Paraprevotella clara YIT 11840 (762968.3) | | 1448, 2564, 2567 | 2566 | | 949 | 3124 | | | | | 2565 | | 124, 179, 565, 1392, 1642, 1647, 1895, 1901, 1902, 2275, 2410, 2737, 2975, 3221, 3431 | 1644, 3249 | 774, 775 | 774, 775 |
| Paraprevotella xylaniphila YIT 11841 (762982.3) | | 85, 88, 514 | 87 | | 173 | 1788 | | | | | 86 | | 241, 338, 774, 861, 997, 1159, 1168, 1169, 1417, 2112, 2547, 2576, 2724, 2792, 3109 | 2521 | 1283, 1284 | 1283, 1284 |
| Parvimonas micra ATCC 33270 (411465.10) | 610 | | | | | 282 | | | | | | | | | | |
| Pediococcus acidilactici 7_4 (563194.3) | 933 | 197 | 701, 1879 | 699 | | 700, 1434 | 1859, 1860 | 1853 | 1858 | 188 | 704 | | 1862, 1863, 1864 | 187, 189, 190, 191 | 1120 | 1120 |
| Pediococcus acidilactici DSM 20284 (862514.3) | 1665 | 1193 | 101, 1389 | 103 | | 102, 905 | 1408, 1409 | | 1410 | 1203 | 98 | | 1404, 1405, 1406 | 1199, 1200, 1201, 1202, 1204 | 99, 100 | |
| Phascolarctobacterium sp. YIT 12067 (626939.3) | 474 | | | | | 475, 1389 | | | | | | | | | | |
| Prevotella copri DSM 18205 (537011.5) | | 2542, 2545, 2554, 3164, 3167 | 3166 | | 4110, 568 | 2038, 5045 | | | | | 2543, 3165 | | 470, 946, 1605, 2251, 3853, 3951, 4209, 5267 | 1902, 2311, 3913, 5411 | 2211, 3813 | 2211, 3813 |
| Prevotella oralis HGA0225 (1203550.3) | | 251, 254 | 253 | | 2334 | 529, 657, 720 | | | | | 252 | | 221, 1708, 1799, 1984 | | 1366, 1757 | |

| Organism | | | | | | | | | | | | | | | | | | | | | Notes |
|---|---|---|---|---|---|---|---|---|---|---|---|---|---|---|---|---|---|---|---|---|---|
| Prevotella salivae DSM 15606 (888832.3) | | 1009, 2303 | 2301 | | 2668 | 2115 | | | | | 2302 | | 863 | | | 671, 926, 1054, 1136, 2629 | 1287 | | 2160 | | |
| Propionibacterium sp. 5_U_42AFAA (450748.3) | 389, 1027 | | 282, 388 | | | 556 | 572 | 2270 | | | 575, 576, 577 | | 283 | | | 460, 2034 | | | | | Galactose-1-phosphate uridylyltransferase is not found |
| Propionibacterium sp. HGH0353 (1203571.3) | 1043, 1675 | | 934, 1042 | | | 1220 | 2212 | 225 | | | | | 935 | | | 344 | | | 463 | 463 | Galactose-1-phosphate uridylyltransferase is not found |
| Proteus mirabilis WGLW6 (1125694.3) | 1436, 3617 | 1876 | 1877 | 1878 | | 1879 | | | 2026 | 2031 | | | 1874 | | | | | | | | |
| Proteus penneri ATCC 35198 (471881.3) | 3381 | 650, 651 | 652 | 653 | | 654 | | | 854, 855 | 863 | | | 646, 647 | | | | | | | | |
| Providencia alcalifaciens DSM 30120 (520999.6) | 1625 | 1007 | | | | 3332 | | | | | | | | | | | | | | | |
| Providencia rettgeri DSM 1131 (521000.6) | 1866 | 504, 534 | 533 | 532 | | 531, 1652 | | | 1200 | 1195 | | | 536 | | | | | | | | |
| Providencia rustigianii DSM 4541 (500637.6) | 3284, 3663 | 1426, 2201, 4929, 6165 | 1425, 4930 | 1424, 4931 | | 25, 26, 1423, 4932, 7056, 7057 | | | 397, 6693 | 391, 6699 | | | 1428, 4927 | | | | | | | | |
| Providencia stuartii ATCC 25827 (471874.6) | 1916, 4263 | 2441, 2442 | 2443 | 2444 | | 2445, 3228 | | | 3060 | 3066 | | | 2439 | | | | | | | | |
| Pseudomonas sp. 2_1_26 (665948.3) | 4577 | | | | | 144, 1095 | | | | | | | | | | | | | | | |
| Ralstonia sp. 5_7_47FAA (658664.3) | 245, 3153 | | | | | 312, 893, 4022, 4245 | | | 4146 | | | | | | | | | | | | |
| Roseburia intestinalis L1-82 (536231.5) | | 115, 2657 | 2807 | 2806 | | 883, 1334, 1348, 1381, 1986, 3390 | 3250 | | 390, 395, 396, 397 | 3251, 3252, 3253, 3255 | | 659 | | | | 652, 660, 761, 1247, 1252, 2809, 3145, 3275, 3888 | 642 | 2316, 2827 | 370, 637, 758, 1240, 1255 | 370, 758, 1240 | |
| Roseburia intestinalis M50/1 (657315.3) | | 583, 1137 | 1034 | 1035 | | 1499, 3291, 3719 | 3559 | | 822, 823, 824, 829 | 3553, 3554, 3556, 3557, 3558 | 1059 | 2078 | | | | 1031, 1333, 2077, 2086, 3165, 3166, 3821, 3822 | 2092 | 1011, 3602 | 571, 851, 2096, 3818 | 571, 851, 3818 | |
| Roseburia intestinalis XB6B4 (718255.3) | | 467, 3902 | 3797 | 3798 | | 855, 2207, 2363 | 2521 | | 2953, 2954, 2955, 2960 | 2522, 2523, 2524, 2526 | | 3357 | | | | 1306, 2080, 2263, 3365, 3794 | 975, 3371 | 2478 | 455, 2266, 2980, 2981, 3388 | 455, 2266, 2980, 2981 | |
| Roseburia inulinivorans DSM 16841 (622312.4) | | 2760, 3881 | 2082 | 2083 | | 2297, 2317, 2335, 3268 | 1688 | | 1759, 1760, 1761, 1766 | 1694, 1695, 1696 | | | | | | 527, 813, 1106, 1711, 3213, 3214, 3757 | | 2974 | 1102, 1973 | 1102, 1973 | |
| Ruminococcaceae bacterium D16 (552398.3) | | 1778 | 676 | 674 | | 339, 675 | 347 | | 485, 486, 487 | 343, 344, 345 | 227 | | | | | 2826 | | | | | |
| Ruminococcus bromii L2-63 (657321.5) | 657 | 574 | 577 | 575 | | 576, 1672, 1997 | | | | | | | 578 | | | | | | 1759 | 1759 | |
| Ruminococcus gnavus ATCC 29149 (411470.6) | | 1567 | 326 | 327 | | 1608, 1743 | 1054 | | 915 | 911 | | 1048, 1393, 1394, 1395 | | | | 1, 837, 1425, 1426, 1427, 1574, 2276, 3276 | | 707 | 3134, 3135 | 3134, 3135 | Transporter is not found |
| Ruminococcus lactaris ATCC 29176 (471875.6) | | 804 | 565 | 564 | | 443 | 1843 | | 353, 358, 359, 360 | 1845, 1846, 1847 | 1736 | 2411 | | | | 495, 512, 2410 | | 1236 | | | |
| Ruminococcus obeum A2-162 (657314.3) | | 2766 | 1102 | 365 | | 3361 | 2912 | | 1719 | 1715 | 3540, 3541, 3542 | 370 | | | | 890, 3541, 3543 | | 3544 | 1335 | 1335 | |
| Ruminococcus obeum ATCC 29174 (411459.7) | | 1761 | 2325 | 2202 | | 208, 2138, 3503 | 339 | | 3062 | 3053, 3066 | 342, 343, 344, 347 | 1674 | | | | 159, 1109, 3298 | 160, 3298 | | 2296 | 2296 | |
| Ruminococcus sp. 18P13 (213810.4) | | 2109 | 592 | | | 218, 2110, 2295 | | | | | | | | | | 600 | 230 | | 229, 993 | 229 | |
| Ruminococcus sp. 5_1_39BFAA (457412.4) | | 53 | 1568 | 320, 2608 | | 1928 | 611 | | | | 1422, 1423, 1424 | 2232 | 311 | | | 312, 1784, 2371 | | 1570, 1576, 3149 | 2374, 2380 | 2374, 2380 | |
| Ruminococcus sp. SR1/5 (657323.3) | | 2774 | 3679 | 467, 1270 | | 3536 | 1418 | | 3503 | 3507 | 1413, 1414, 1415 | 1819 | | | | 1872, 1873, 2121, 3141 | 1416 | 1798 | 1513, 1801, 3584 | 1513, 1801, 3584 | |
| Ruminococcus torques ATCC 27756 (411460.6) | | 804 | 814 | 777 | | 761 | 1844 | | 616 | 615 | 800, 805, 807, 808, 813 | 1837, 1840, 1841, 1842, 1850 | | | | 1580, 1891, 1925 | | 2024 | 1920, 2203 | | |
| Ruminococcus torques L2-14 (657313.3) | | 2576 | 1607 | 1606 | | 2568 | | | 554 | 558, 2221 | 2844 | 1900 | | | | 1899, 2382, 2400 | | | 1187, 2019, 2405 | 1187, 2019, 2405 | |
| Salmonella enterica subsp. enterica serovar Typhimurium str. LT2 (99287.12) | 1848, 2220 | 806 | 807 | 808 | | 809, 809 | | | 3456 | 3452, 4000 | 2315, 2316, 2317 | | | | | | | | 4520 | | |
| Staphylococcus sp. HGB0015 (1078083.3) | 1787 | 2023 | 2081 | 2083 | | 2082 | | | 782, 783 | 1673 | 784, 1674, 1675 | | | 1676, 1677, 1678, 1679 | | | | | | | |
| Streptococcus anginosus 1_2_62CV (742820.3) | 1358 | 1752 | 357 | 358 | | 763, 1452, 1574 | | | 120, 121 | 118, 1751, 1745 | | | 122, 123, 124 | 1746, 1747, 1748, 1749 | | | | | | | Tagatose-6-phosphate kinase and transporter are not found |
| Streptococcus equinus ATCC 9812 (525379.3) | 877 | 1641 | 1601 | 1600 | | 1599 | | | 297 | | 661, 662 | | 1694 | | | | | | | | |
| Streptococcus infantarius subsp. infantarius ATCC BAA-102 (471872.6) | 741 | 693 | 633 | 632 | | 631, 1260 | | | 655, 656 | 266, 653 | | | 1694 | | | | | | 642 | 642 | |
| Streptococcus sp. 2_1_36FAA (469609.3) | 175 | 63 | 882 | 881 | | 801, 802, 880 | | | 1477, 1478, 1484, 1485 | 703, 1476, 1483 | 57 | | | 1479, 1480, 61 | 58, 59, 60, 61 | 1448 | | | | | |
| Streptococcus sp. HPH0090 (1203590.3) | 1331 | 1440 | 10 | 11 | | 169 | | | 395, 396 | 193, 397 | 398 | 1446 | | 391, 392, 393 | 1442, 1443, 1444, 1445 | 802 | | | 144 | 144 | |

| Organism | | | | | | | | | | | | | | | | | | | Notes |
|---|---|---|---|---|---|---|---|---|---|---|---|---|---|---|---|---|---|---|---|
| Streptomyces sp. HGB0020 (1078086.3) | 4132, 6643 | 5612, 5631 | 6596 | 6598 | | 201, 1060, 1419, 5479, 6329, 6597, 7263, 8450 | 5311 | 3644 | | | | 6599 | 471, 484, 554, 1594, 1669, 1951, 2943, 3105, 5115 | 2886, 3757, 3873, 5605, 6085 | 2412 | 794 | 1397, 3092, 3099, 3294, 3980, 3984 | 1397, 3099, 3980 | |
| Streptomyces sp. HPH0547 (1203592.3) | 5781 | | 5833 | 5831 | | 2488, 4078, 4487, 5832 | 2140 | 4042 | | | | 5830 | 4101 | 4656 | | 94 | | | |
| Subdoligranulum sp. 4_3_54A2FAA (665956.3) | 3008 | 245 | 3820 | 2263 | | 440, 1605, 3219, 3715 | 447 | | | 3496, 3497, 3498 | 443, 444, 445 | | 869, 1286, 1852, 3987, 4421 | 1050, 1051, 2356 | 2330, 3334 | 141, 2332 | 141, 2332 | | |
| Subdoligranulum variabile DSM 15176 (411471.5) | | 1522, 4575 | 1067, 4240 | 1389, 4442 | | 82, 651, 937, 1840, 2171, 3779, 4824, 5061, 5111, 5694 | 2375, 3110 | | | 710, 711, 712, 3719, 3720, 3721 | 2377, 2378, 2379, 3112, 3113, 3114 | | 143, 845, 852, 964, 2155, 2276, 2321, 3012, 3056, 3578, 3585, 5088, 5099, 5632 | 1930, 5241 | | | 962, 1392, 4445, 5086 | 962, 1392, 4445, 5086 | |
| Succinatimonas hippei YIT 12066 (762983.3) | 1796 | 513 | 209 | 208 | | 207 | | | | 2008, 2009, 2013 | | 1596 | | | | | | | |
| Sutterella wadsworthensis 3_1_45B (742821.3) | 440 | | | | | 432 | | | | | | | | | | | | | |
| Sutterella wadsworthensis HGA0223 (1203554.3) | 2242 | | | | | 2252 | | | | | | | | | | | | | |
| Tannerella sp. 6_1_58FAA_CT1 (665949.3) | | 1878, 2419 | 2421 | | 583 | 1502, 1987 | | | | | | 2420 | 350, 358, 362, 553, 668, 912, 1014, 1779, 1882, 2292, 2308, 2318, 2597, 3211, 3212, 3319, 3358 | 493 | | | 735, 2320 | 2320 | |
| Turicibacter sp. HGF1 (910310.3) | 1275, 2005, 2590, 2591 | 1505 | 1503 | 1502 | | 1501, 2586, 2589 | 1099 | 35 | 1113 | | 1100, 1101, 1110, 1116 | 2078 | 1511, 1512, 1996 | 2706 | 2622 | 1518 | 1517, 1768 | 1517, 1768 | |
| Turicibacter sp. PC909 (702450.3) | 287, 496 | 1179 | 1177 | 1176 | | 1175, 1677, 1837 | 384 | 1430 | 370 | | 367, 373, 382, 383 | | 274, 1184, 1185 | 2541 | 119 | 1191 | 450, 1190 | 450, 1190 | Transporter is not found |
| Veillonella sp. 3_1_44 (457416.3) | 48 | | | 990 | | 57, 60, 777 | | | | | | | | | | | | | Galactose kinase and transporter are not found |
| Veillonella sp. 6_1_27 (450749.3) | 48 | | | 1046 | | 56, 59, 814 | | | | | | | | | | | | | Galactose kinase and transporter are not found |
| Weissella paramesenteroides ATCC 33313 (585506.3) | 1641 | 83, 1663 | 135 | 134 | | 826, 1565 | | | | | | 125, 136 | 1536 | | | | 124 | 124 | |
| Yokenella regensburgei ATCC 43003 (1002368.3) | 3447, 3929 | 2406 | 2407 | 2408 | | 2409, 2409, 3927 | | 1544 | 1537 | 4016, 4017, 4018 | | 3258 | 2170 | | | 3284 | | | |

**Table S8** Glycosylic bonds found in the mucin glycans and corresponding glycosyl hydrolases.

| 2nd monosaccharide / amino acid | 1st monosaccharide | | | | |
|---|---|---|---|---|---|
| | **Fuc** | **Gal** | **GalNAc** | **GlcNAc** | **Neu5Ac** |
| **Gal** | α(1-2), Alpha-L-fucosidase [EC 3.2.1.51] | α(1-3), Alpha-galactosidase [EC 3.2.1.22] | α(1-3), Alpha-N-acetylgalactosaminidase [EC 3.2.1.49]; β(1-4), Beta-N-hexosaminidase [EC 3.2.1.52] | α(1-4), Alpha-N-acetylglucosaminidase [EC 3.2.1.50]; β(1-3), Beta-N-hexosaminidase [EC 3.2.1.52] | α(2-3), Neuraminidase [EC 3.2.1.18]; α(2-6), Neuraminidase [EC 3.2.1.18] |
| **GalNAc** | - | β(1-3), Beta-galactosidase [EC 3.2.1.23] | α(1-6), Alpha-N-acetylgalactosaminidase [EC 3.2.1.49]; β(1-3), Beta-N-hexosaminidase [EC 3.2.1.52] | β(1-3), Beta-N-hexosaminidase [EC 3.2.1.52]; β(1-6), Beta-N-hexosaminidase [EC 3.2.1.52] | α(2-6), Neuraminidase [EC 3.2.1.18] |
| **GlcNAc** | α(1-2), Alpha-L-fucosidase [EC 3.2.1.51]; α(1-3), Alpha-L-fucosidase [EC 3.2.1.51]; α(1-4), Alpha-L-fucosidase [EC 3.2.1.51] | β(1-3), Beta-galactosidase [EC 3.2.1.23]; β(1-4), Beta-galactosidase [EC 3.2.1.23] | - | β(1-4), Beta-N-hexosaminidase [EC 3.2.1.52] | - |
| **O-Ser/Thr** | - | - | α(1), Endo-alpha-N-acetylgalactosaminidase [EC 3.2.1.97] | - | - |

**Legend**
Fuc - L-fucose
Gal - galactose
GalNAc - N-acetyl-D-galactosamine
GlcNAc - N-acetyl-D-glucosamine
Neu5Ac - N-acetylneuraminic acid
O-Ser/Thr - Oxygen atom of the side chains in serine or
threonine in mucin protein chain

**Table S9.** Search of non-orthologous replacement the GalT by gene occurrence analysis. Genome of interest and the final result are shown by bold and blue.

| 1. Analyzed genomes | | | 2. Results of the gene occurrence analysis from IMG/JGI | | | | |
|---|---|---|---|---|---|---|---|
| **Organism name** | **PubSEED ID** | **IMG ID** | **Gene ID** | **Locus Tag** | **Gene Name** | **Length** | **Percent Identity** |
| **GalK+ / GalT- genomes** | | | | | | | |
| **Akkermansia muciniphila ATCC BAA-835** | 349741.6 | 642555104 | **642612071** | **Amuc_0031** | **hypothetical protein** | **291** | **51.2** |
| Bacteroides fragilis 638R | 862962.3 | 650377910 | 642612034 | Amuc_0034 | prolyl-tRNA synthetase (EC 6.1.1.15) | 506 | 49.1 |
| Bacteroides fragilis NCTC 9343 | 272559.17 | 2623620861 | 642612078 | Amuc_0037 | amino acid/polyamine/organocation transporter, APC superfam | 494 | 59.2 |
| Bacteroides fragilis YCH46 | 295405.11 | 2623620898 | 642612080 | Amuc_0060 | Alpha-N-acetylglucosaminidase | 848 | 46.4 |
| Bacteroides thetaiotaomicron VPI-5482 | 226186.12 | 2623620899 | 642612150 | Amuc_0105 | RND efflux system, outer membrane lipoprotein, NodT family | 457 | 32.9 |
| Bacteroides vulgatus ATCC 8482 | 435590.9 | 2623620366 | 642612162 | Amuc_0117 | agmatine deiminase (EC 3.5.3.12) | 351 | 38.3 |
| Bacteroides xylanisolvens XB1A | 657309.4 | 650377911 | 642612193 | Amuc_0148 | carboxynorspermidine dehydrogenase | 409 | 61 |
| Parabacteroides distasonis ATCC 8503 | 435591.13 | 2623620331 | 642612252 | Amuc_0206 | succinate CoA transferase | 498 | 89.1 |
| | | | 642612344 | Amuc_0290 | glycoside hydrolase family 2 sugar binding | 986 | 44 |
| **GalK+ / GalT- genomes** | | | 642612420 | Amuc_0369 | Beta-N-acetylhexosaminidase | 665 | 38.3 |
| Bacillus subtilis subsp. subtilis str. 168 | 224308.113 | 646311909 | 642612443 | Amuc_0392 | coagulation factor 5/8 type domain protein | 709 | 44.1 |
| Bifidobacterium animalis subsp. lactis AD011 | 442563.4 | 643348515 | 642612504 | Amuc_0454 | protein of unknown function DUF303 acetylesterase putative | 512 | 39.4 |
| Bifidobacterium longum DJO10A | 205913.11 | 642555107 | 642612541 | Amuc_0491 | sulfatase | 500 | 38.8 |
| Bifidobacterium longum NCC2705 | 206672.9 | 637000031 | 642612658 | Amuc_0607 | Endonuclease/exonuclease/phosphatase | 253 | 51.7 |
| Bifidobacterium longum subsp. infantis 157F | 565040.3 | 649633017 | 642612671 | Amuc_0620 | NAD+ synthetase | 644 | 51.6 |
| Bifidobacterium longum subsp. infantis ATCC 15697 = JCM 1222 | 391904.8 | 651053007 | 642612753 | Amuc_0701 | transcriptional regulator, LysR family | 307 | 42.8 |
| Bifidobacterium longum subsp. longum F8 | 722911.3 | 650377912 | 642612756 | Amuc_0704 | hypothetical protein | 417 | 50.2 |
| Bifidobacterium longum subsp. longum JCM 1217 | 565042.3 | 649633019 | 642612824 | Amuc_0773 | acyltransferase 3 | 382 | 49.5 |
| Butyrivibrio fibrisolvens 16/4 | 657324.3 | 650377919 | 642612901 | Amuc_0846 | coagulation factor 5/8 type domain protein | 704 | 43.2 |
| Coprococcus catus GD/7 | 717962.3 | 650377925 | 642612923 | Amuc_0868 | Beta-N-acetylhexosaminidase | 549 | 39.2 |
| Coprococcus sp. ART55/1 | 751585.3 | 2565956545 | 642613016 | Amuc_0962 | 4Fe-4S ferredoxin iron-sulfur binding domain protein | 588 | 37.9 |
| Enterococcus sp. 7L76 | 657310.3 | 650377930 | 642613081 | Amuc_1032 | Beta-N-acetylhexosaminidase | 518 | 34.4 |
| Escherichia coli O157:H7 str. Sakai | 386585.9 | 2623620697 | 642613110 | Amuc_1059 | acriflavin resistance protein | 1052 | 36.9 |
| Escherichia coli SE11 | 409438.11 | 643348551 | 642613171 | Amuc_1120 | alpha-L-fucosidase 2 | 796 | 44.4 |
| Escherichia coli SE15 | 431946.3 | 646862327 | 642613182 | Amuc_1131 | MATE efflux family protein | 445 | 53.2 |
| Escherichia coli str. K-12 substr. MG1655 | 511145.12 | 646311926 | 642613236 | Amuc_1185 | LL-diaminopimelate aminotransferase apoenzyme (EC 2.6.1.83 | 531 | 56.5 |
| Escherichia coli UTI89 | 364106.8 | 2623620781 | 642613461 | Amuc_1410 | putative phosphate/sulphate permease | 743 | 43.8 |
| Eubacterium rectale DSM 17629 | 657318.4 | 650377936 | 642613521 | Amuc_1461 | protein of unknown function DUF558 | 245 | 34.5 |
| Eubacterium rectale M104/1 | 657317.3 | 650377937 | 642613623 | Amuc_1563 | protein of unknown function DUF6 transmembrane | 305 | 47.5 |
| Eubacterium siraeum 70/3 | 657319.3 | 650377938 | 642613665 | Amuc_1605 | NADH dehydrogenase subunit M (EC 1.6.5.3) | 474 | 33.6 |
| Eubacterium siraeum V10Sc8a | 717961.3 | 2565956546 | 642613681 | Amuc_1621 | 4-alpha-glucanotransferase | 623 | 35.2 |
| Faecalibacterium prausnitzii L2-6 | 718252.3 | 650377941 | 642613694 | Amuc_1634 | conserved hypothetical protein, membrane | 746 | 39.9 |
| Faecalibacterium prausnitzii SL3/3 | 657322.3 | 650377940 | 642613697 | Amuc_1637 | alpha amylase catalytic region | 574 | 38 |
| Fusobacterium sp. 3_1_27 | 469602.3 | 2563366552 | 642613704 | Amuc_1644 | transglutaminase domain protein | 379 | 38.1 |
| Fusobacterium sp. 3_1_36A2 | 469604.3 | 2554235746 | 642613714 | Amuc_1655 | sulfatase | 554 | 44.1 |
| Fusobacterium sp. 7_1 | 457405.3 | 646206252 | 642613716 | Amuc_1657 | carboxynorspermidine decarboxylase (EC 4.1.1.-) | 400 | 42.4 |
| Lactobacillus brevis ATCC 367 | 387344.15 | 639633027 | 642613727 | Amuc_1669 | Beta-N-acetylhexosaminidase | 547 | 37.2 |
| Lactobacillus casei ATCC 334 | 321967.11 | 639633028 | 642613762 | Amuc_1703 | Ferroxidase | 164 | 45.1 |
| Lactobacillus fermentum IFO 3956 | 334390.5 | 641522633 | 642613818 | Amuc_1755 | sulfatase | 562 | 47.5 |
| Lactobacillus gasseri ATCC 33323 | 324831.13 | 639633030 | 642613843 | Amuc_1780 | protein of unknown function DUF58 | 296 | 39.1 |
| Lactobacillus helveticus DPC 4571 | 405566.6 | 641228495 | 642613857 | Amuc_1793 | acyltransferase 3 | 393 | 59.2 |
| Lactobacillus johnsonii NCC 533 | 257314.6 | 2623620932 | 642613865 | Amuc_1801 | peptidase S15 | 359 | 81.6 |
| Lactobacillus paracasei subsp. paracasei 8700:2 | 537973.8 | 643886139 | 642613877 | Amuc_1813 | MATE efflux family protein | 461 | 57.6 |
| Lactobacillus plantarum 16 | 1327988.3 | 2563366583 | 642613897 | Amuc_1833 | L-fucose transporter | 644 | 44.2 |
| Lactobacillus plantarum WCFS1 | 220668.9 | 2623620707 | 642613899 | Amuc_1835 | Exo-alpha-sialidase | 674 | 43.2 |
| Lactobacillus reuteri F275 | 299033.6 | 640427118 | 642613930 | Amuc_1869 | (1->4)-alpha-D-glucan synthase (ADP-glucose) | 432 | 43 |
| Lactobacillus reuteri SD2112 | 491077.3 | 650716048 | 642613940 | Amuc_1879 | hypothetical protein | 373 | 35.9 |
| Lactobacillus rhamnosus GG | 568703.3 | 644736382 | 642613946 | Amuc_1885 | exsB protein | 223 | 49.3 |
| Lactobacillus sakei subsp. sakei 23K | 314315.12 | 637000142 | 642613950 | Amuc_1889 | Sec-independent protein translocase, TatC subunit | 348 | 38 |
| Lactobacillus salivarius UCC118 | 362948.14 | 2623620709 | 642613963 | Amuc_1902 | hypothetical protein | 230 | 80.7 |
| Megamonas hypermegale ART12/1 | 657316.3 | 650377958 | 642613964 | Amuc_1903 | hypothetical protein | 335 | 77.7 |
| Roseburia intestinalis M50/1 | 657315.3 | 650377964 | 642613984 | Amuc_1924 | Beta-N-acetylhexosaminidase | 664 | 40.7 |
| Roseburia intestinalis XB6B4 | 718255.3 | 650377965 | 642614044 | Amuc_1984 | heterodimeric methylmalonyl-CoA mutase small subunit | 684 | 39 |
| Ruminococcus bromii L2-63 | 657321.5 | 650377966 | 642614078 | Amuc_2018 | Beta-N-acetylhexosaminidase | 493 | 34.2 |
| Ruminococcus obeum A2-162 | 657314.3 | 650377967 | 642614079 | Amuc_2019 | Beta-N-acetylhexosaminidase | 504 | 42.6 |
| Ruminococcus sp. 18P13 | 213810.4 | 650377969 | 642614080 | Amuc_2020 | methylmalonyl-CoA mutase metallochaperone MeaB | 339 | 49.4 |
| Ruminococcus sp. SR1/5 | 657323.3 | 650377968 | 642614127 | Amuc_2065 | acyltransferase 3 | 384 | 55.3 |
| Ruminococcus torques L2-14 | 657313.3 | 650377970 | 642614139 | Amuc_2077 | polysaccharide export protein | 260 | 42.1 |
| Salmonella enterica subsp. enterica serovar Typhimurium str. LT2 | 99287.12 | 2623620604 | 642614180 | Amuc_2118 | Nitrilase/cyanide hydratase and apolipoprotein N-acyltransfera | 285 | 54.5 |
| | | | 642614209 | Amuc_2148 | Beta-N-acetylhexosaminidase | 549 | 40.3 |
| | | | 642923615 | Amuc_0123 | - | 427 | 40.4 |
| | | | 642923645 | Amuc_1980 | - | 47 | 53.3 |

**Table S9.** Distribution of the catabolic pathways (CPs) and glycosyl hydrolases (GHs) for the splitting of mucing glycans and utilization of derived monosaccharides. [1]Pattern demonstrates a presence of CPs and GHs in each genome, every letter corresponds to one monosaccharide, order is the same as in previous columns; x, both CPs and GHs are absent; P, only CP is present; H, only GH is present; B, both CPs and GHs are present. [2]Donor, organism can be involved in monosaccharide exchange only as a donor; Acceptor, organism can be involved in monosaccharide exchange only as an acceptor; Mixed, organism is a donor for some monosaccharides but is an acceptor for the others.

| Organism | Catabolic pathways (CPs) and glycosyl hydrolases (GHs) for individual monosaccharides | | | | | | | | | | Pattern[1] | Numbers of catabolyzed and/or cleavable monosaccharides | | | Mode[2] |
| | Fuc | | GalNAc | | GlcNAc | | Neu5Ac | | Gal | | | CPs | GHs | CPs+GHs | |
| | CP | GH | CP | GH | CP | GH | CP | GH | CP | GH | | | | | |
|---|---|---|---|---|---|---|---|---|---|---|---|---|---|---|---|
| Abiotrophia defectiva ATCC 49176 (592010.4) | 1 | 1 | 0 | 1 | 0 | 1 | 0 | 1 | 1 | 1 | BHHHB | 2 | 5 | 2 | Donor |
| Acidaminococcus sp. D21 (563191.3) | 0 | 0 | 0 | 0 | 0 | 0 | 0 | 0 | 0 | 0 | xxxxx | 0 | 0 | 0 | - |
| Acidaminococcus sp. HPA0509 (1203555.3) | 0 | 0 | 0 | 0 | 0 | 0 | 0 | 0 | 0 | 0 | xxxxx | 0 | 0 | 0 | - |
| Acinetobacter junii SH205 (575587.3) | 0 | 0 | 0 | 0 | 0 | 0 | 0 | 0 | 0 | 0 | xxxxx | 0 | 0 | 0 | - |
| Actinomyces odontolyticus ATCC 17982 (411466.7) | 1 | 1 | 0 | 0 | 1 | 0 | 1 | 1 | 1 | 1 | BxPBB | 4 | 3 | 3 | Acceptor |
| Actinomyces sp. HPA0247 (1203556.3) | 1 | 1 | 0 | 1 | 1 | 1 | 1 | 1 | 1 | 1 | BHBBB | 4 | 5 | 4 | Donor |
| Akkermansia muciniphila ATCC BAA-835 (349741.6) | 1 | 1 | 0 | 1 | 0 | 1 | 1 | 1 | 1 | 1 | BHHBB | 3 | 5 | 3 | Donor |
| Alistipes indistinctus YIT 12060 (742725.3) | 0 | 1 | 0 | 1 | 0 | 1 | 0 | 0 | 1 | 1 | HHHxB | 1 | 4 | 1 | Donor |
| Alistipes shahii WAL 8301 (717959.3) | 0 | 1 | 0 | 1 | 0 | 1 | 0 | 1 | 1 | 1 | HHHHB | 1 | 5 | 1 | Donor |
| Anaerobaculum hydrogeniformans ATCC BAA-1850 (592015.5) | 0 | 0 | 0 | 0 | 0 | 0 | 0 | 0 | 0 | 0 | xxxxx | 0 | 0 | 0 | - |
| Anaerococcus hydrogenalis DSM 7454 (561177.4) | 0 | 0 | 0 | 0 | 0 | 0 | 0 | 0 | 0 | 0 | xxxxx | 0 | 0 | 0 | - |
| Anaerofustis stercorihominis DSM 17244 (445971.6) | 0 | 0 | 0 | 0 | 0 | 0 | 0 | 0 | 0 | 0 | xxxxx | 0 | 0 | 0 | - |
| Anaerostipes caccae DSM 14662 (411490.6) | 0 | 0 | 0 | 1 | 1 | 1 | 0 | 0 | 1 | 1 | xHBxB | 2 | 3 | 2 | Donor |
| Anaerostipes hadrus DSM 3319 (649757.3) | 0 | 0 | 0 | 1 | 1 | 1 | 0 | 0 | 1 | 1 | xHBxB | 2 | 3 | 2 | Donor |
| Anaerostipes sp. 3_2_56FAA (665937.3) | 0 | 0 | 0 | 1 | 1 | 1 | 0 | 0 | 1 | 1 | xHBxB | 2 | 3 | 2 | Donor |
| Anaerotruncus colihominis DSM 17241 (445972.6) | 1 | 0 | 0 | 0 | 1 | 0 | 1 | 0 | 1 | 0 | PxPPP | 4 | 0 | 0 | Acceptor |
| Bacillus smithii 7_3_47FAA (665952.3) | 0 | 0 | 0 | 1 | 1 | 1 | 0 | 0 | 0 | 1 | xHBxH | 1 | 3 | 1 | Donor |
| Bacillus sp. 7_6_55CFAA_CT2 (665957.3) | 0 | 0 | 0 | 0 | 0 | 0 | 0 | 0 | 1 | 0 | xxxxP | 1 | 0 | 0 | Acceptor |
| Bacillus subtilis subsp. subtilis str. 168 (224308.113) | 0 | 0 | 0 | 1 | 1 | 1 | 0 | 0 | 1 | 1 | xHBxB | 2 | 3 | 2 | Donor |
| Bacteroides caccae ATCC 43185 (411901.7) | 1 | 1 | 0 | 1 | 0 | 1 | 1 | 1 | 1 | 1 | BHHBB | 3 | 5 | 3 | Donor |
| Bacteroides capillosus ATCC 29799 (411467.6) | 0 | 0 | 0 | 1 | 0 | 1 | 0 | 0 | 1 | 1 | xHHxB | 1 | 3 | 1 | Donor |
| Bacteroides cellulosilyticus DSM 14838 (537012.5) | 1 | 1 | 0 | 1 | 1 | 1 | 0 | 0 | 1 | 1 | BHBxB | 3 | 4 | 3 | Donor |
| Bacteroides clarus YIT 12056 (762984.3) | 0 | 1 | 0 | 1 | 1 | 1 | 1 | 1 | 1 | 1 | HHBBB | 3 | 5 | 3 | Donor |
| Bacteroides coprocola DSM 17136 (470145.6) | 0 | 1 | 0 | 1 | 0 | 1 | 0 | 1 | 1 | 1 | HHHHB | 1 | 5 | 1 | Donor |
| Bacteroides coprophilus DSM 18228 (547042.5) | 1 | 1 | 0 | 1 | 0 | 1 | 1 | 1 | 1 | 1 | BHHBB | 3 | 5 | 3 | Donor |
| Bacteroides dorei DSM 17855 (483217.6) | 0 | 1 | 0 | 1 | 0 | 1 | 1 | 1 | 1 | 1 | HHHBB | 2 | 5 | 2 | Donor |
| Bacteroides eggerthii 1_2_48FAA (665953.3) | 0 | 1 | 0 | 1 | 1 | 1 | 1 | 0 | 1 | 1 | HHBPB | 3 | 4 | 2 | Mixed |
| Bacteroides eggerthii DSM 20697 (483216.6) | 0 | 1 | 0 | 1 | 1 | 1 | 1 | 0 | 1 | 1 | HHBPB | 3 | 4 | 2 | Mixed |
| Bacteroides finegoldii DSM 17565 (483215.6) | 1 | 1 | 0 | 1 | 1 | 1 | 0 | 1 | 1 | 1 | BHBHB | 3 | 5 | 3 | Donor |
| Bacteroides fluxus YIT 12057 (763034.3) | 1 | 1 | 0 | 1 | 0 | 1 | 1 | 1 | 1 | 1 | BHHBB | 3 | 5 | 3 | Donor |
| Bacteroides fragilis 3_1_12 (457424.5) | 1 | 1 | 0 | 1 | 0 | 1 | 1 | 1 | 1 | 1 | BHHBB | 3 | 5 | 3 | Donor |
| Bacteroides fragilis 638R (862962.3) | 1 | 1 | 0 | 1 | 0 | 1 | 1 | 1 | 1 | 1 | BHHBB | 3 | 5 | 3 | Donor |
| Bacteroides fragilis NCTC 9343 (272559.17) | 1 | 1 | 0 | 1 | 0 | 1 | 1 | 1 | 1 | 1 | BHHBB | 3 | 5 | 3 | Donor |
| Bacteroides fragilis YCH46 (295405.11) | 1 | 1 | 0 | 1 | 0 | 1 | 1 | 1 | 1 | 1 | BHHBB | 3 | 5 | 3 | Donor |
| Bacteroides intestinalis DSM 17393 (471870.8) | 1 | 1 | 0 | 1 | 0 | 1 | 0 | 1 | 1 | 1 | BHHHB | 2 | 5 | 2 | Donor |
| Bacteroides ovatus 3_8_47FAA (665954.3) | 1 | 1 | 0 | 1 | 1 | 1 | 1 | 1 | 1 | 1 | BHBBB | 4 | 5 | 4 | Donor |
| Bacteroides ovatus ATCC 8483 (411476.11) | 1 | 1 | 0 | 1 | 1 | 1 | 1 | 1 | 1 | 1 | BHBBB | 4 | 5 | 4 | Donor |
| Bacteroides ovatus SD CC 2a (702444.3) | 1 | 1 | 0 | 1 | 1 | 1 | 0 | 1 | 1 | 1 | BHBHB | 3 | 5 | 3 | Donor |
| Bacteroides ovatus SD CMC 3f (702443.3) | 1 | 1 | 0 | 1 | 1 | 1 | 0 | 1 | 1 | 1 | BHBHB | 3 | 5 | 3 | Donor |
| Bacteroides pectinophilus ATCC 43243 (483218.5) | 0 | 0 | 0 | 0 | 0 | 0 | 0 | 0 | 0 | 0 | xxxxx | 0 | 0 | 0 | - |
| Bacteroides plebeius DSM 17135 (484018.6) | 1 | 1 | 0 | 1 | 0 | 1 | 1 | 1 | 1 | 1 | BHHBB | 3 | 5 | 3 | Donor |
| Bacteroides sp. 1_1_14 (469585.3) | 1 | 1 | 0 | 1 | 1 | 1 | 1 | 1 | 1 | 1 | BHBBB | 4 | 5 | 4 | Donor |
| Bacteroides sp. 1_1_30 (457387.3) | 1 | 1 | 0 | 1 | 1 | 1 | 1 | 1 | 1 | 1 | BHBBB | 4 | 5 | 4 | Donor |
| Bacteroides sp. 1_1_6 (469586.3) | 1 | 1 | 0 | 1 | 1 | 1 | 1 | 1 | 1 | 1 | BHBBB | 4 | 5 | 4 | Donor |
| Bacteroides sp. 2_1_16 (469587.3) | 1 | 1 | 0 | 1 | 0 | 1 | 1 | 1 | 1 | 1 | BHHBB | 3 | 5 | 3 | Donor |
| Bacteroides sp. 2_1_22 (469588.3) | 1 | 1 | 0 | 1 | 1 | 1 | 1 | 1 | 1 | 1 | BHBBB | 4 | 5 | 4 | Donor |
| Bacteroides sp. 2_1_33B (469589.3) | 0 | 1 | 0 | 1 | 0 | 1 | 1 | 1 | 1 | 1 | HHHBB | 2 | 5 | 2 | Donor |
| Bacteroides sp. 2_1_56FAA (665938.3) | 1 | 1 | 0 | 1 | 0 | 1 | 1 | 1 | 1 | 1 | BHHBB | 3 | 5 | 3 | Donor |
| Bacteroides sp. 2_1_7 (457388.5) | 0 | 1 | 0 | 1 | 0 | 1 | 1 | 1 | 1 | 1 | HHHBB | 2 | 5 | 2 | Donor |

| Organism | | | | | | | | | | | Code | | | | Status |
|---|---|---|---|---|---|---|---|---|---|---|---|---|---|---|---|
| Bacteroides sp. 2_2_4 (469590.5) | 1 | 1 | 0 | 1 | 1 | 1 | 1 | 1 | 1 | 1 | BHBBB | 4 | 5 | 4 | Donor |
| Bacteroides sp. 20_3 (469591.4) | 0 | 1 | 0 | 1 | 0 | 1 | 1 | 1 | 1 | 1 | HHHBB | 2 | 5 | 2 | Donor |
| Bacteroides sp. 3_1_19 (469592.4) | 0 | 1 | 0 | 1 | 0 | 1 | 1 | 1 | 1 | 1 | HHHBB | 2 | 5 | 2 | Donor |
| Bacteroides sp. 3_1_23 (457390.3) | 1 | 1 | 0 | 1 | 1 | 1 | 1 | 1 | 1 | 1 | BHBBB | 4 | 5 | 4 | Donor |
| Bacteroides sp. 3_1_33FAA (457391.3) | 0 | 1 | 0 | 1 | 0 | 1 | 1 | 1 | 1 | 1 | HHHBB | 2 | 5 | 2 | Donor |
| Bacteroides sp. 3_1_40A (469593.3) | 0 | 1 | 0 | 1 | 0 | 1 | 1 | 1 | 1 | 1 | HHHBB | 2 | 5 | 2 | Donor |
| Bacteroides sp. 3_2_5 (457392.3) | 1 | 1 | 0 | 1 | 0 | 1 | 1 | 1 | 1 | 1 | BHHBB | 3 | 5 | 3 | Donor |
| Bacteroides sp. 4_1_36 (457393.3) | 1 | 1 | 0 | 1 | 0 | 1 | 0 | 0 | 1 | 1 | BHHxB | 2 | 4 | 2 | Donor |
| Bacteroides sp. 4_3_47FAA (457394.3) | 0 | 1 | 0 | 1 | 0 | 1 | 1 | 1 | 1 | 1 | HHHBB | 2 | 5 | 2 | Donor |
| Bacteroides sp. 9_1_42FAA (457395.6) | 0 | 1 | 0 | 1 | 0 | 1 | 1 | 1 | 1 | 1 | HHHBB | 2 | 5 | 2 | Donor |
| Bacteroides sp. D1 (556258.5) | 1 | 1 | 0 | 1 | 1 | 1 | 1 | 1 | 1 | 1 | BHBBB | 4 | 5 | 4 | Donor |
| Bacteroides sp. D2 (556259.3) | 1 | 1 | 0 | 1 | 1 | 1 | 1 | 1 | 1 | 1 | BHBBB | 4 | 5 | 4 | Donor |
| Bacteroides sp. D20 (585543.3) | 1 | 1 | 0 | 1 | 0 | 1 | 1 | 1 | 1 | 1 | BHHBB | 3 | 5 | 3 | Donor |
| Bacteroides sp. D22 (585544.3) | 1 | 1 | 0 | 1 | 1 | 1 | 1 | 1 | 1 | 1 | BHBBB | 4 | 5 | 4 | Donor |
| Bacteroides sp. HPS0048 (1078089.3) | 1 | 1 | 0 | 1 | 0 | 1 | 0 | 1 | 1 | 1 | BHHHB | 2 | 5 | 2 | Donor |
| Bacteroides stercoris ATCC 43183 (449673.7) | 0 | 1 | 0 | 1 | 0 | 1 | 1 | 1 | 1 | 1 | HHHBB | 2 | 5 | 2 | Donor |
| Bacteroides thetaiotaomicron VPI-5482 (226186.12) | 1 | 1 | 0 | 1 | 1 | 1 | 0 | 1 | 1 | 1 | BHBHB | 3 | 5 | 3 | Donor |
| Bacteroides uniformis ATCC 8492 (411479.10) | 1 | 1 | 0 | 1 | 1 | 1 | 0 | 1 | 1 | 1 | BHBHB | 3 | 5 | 3 | Donor |
| Bacteroides vulgatus ATCC 8482 (435590.9) | 0 | 1 | 0 | 1 | 0 | 1 | 1 | 1 | 1 | 1 | HHHBB | 2 | 5 | 2 | Donor |
| Bacteroides vulgatus PC510 (702446.3) | 0 | 1 | 0 | 1 | 0 | 1 | 1 | 1 | 1 | 1 | HHHBB | 2 | 5 | 2 | Donor |
| Bacteroides xylanisolvens SD CC 1b (702447.3) | 1 | 1 | 0 | 1 | 1 | 1 | 1 | 1 | 1 | 1 | BHBBB | 4 | 5 | 4 | Donor |
| Bacteroides xylanisolvens XB1A (657309.4) | 1 | 1 | 0 | 1 | 1 | 1 | 1 | 1 | 1 | 1 | BHBBB | 4 | 5 | 4 | Donor |
| Barnesiella intestinihominis YIT 11860 (742726.3) | 0 | 1 | 0 | 1 | 0 | 1 | 0 | 1 | 0 | 1 | HHHHH | 0 | 5 | 0 | Donor |
| Bifidobacterium adolescentis L2-32 (411481.5) | 0 | 0 | 0 | 1 | 1 | 1 | 0 | 0 | 1 | 1 | xHBxB | 2 | 3 | 2 | Donor |
| Bifidobacterium angulatum DSM 20098 = JCM 7096 (518635.7) | 0 | 0 | 0 | 1 | 0 | 1 | 0 | 0 | 1 | 1 | xHHxB | 1 | 3 | 1 | Donor |
| Bifidobacterium animalis subsp. lactis AD011 (442563.4) | 0 | 0 | 0 | 1 | 0 | 1 | 0 | 0 | 1 | 1 | xHHxB | 1 | 3 | 1 | Donor |
| Bifidobacterium bifidum NCIMB 41171 (398513.5) | 0 | 0 | 0 | 1 | 1 | 1 | 0 | 1 | 1 | 1 | xHBHB | 2 | 4 | 2 | Donor |
| Bifidobacterium breve DSM 20213 = JCM 1192 (518634.7) | 1 | 0 | 0 | 1 | 1 | 1 | 1 | 1 | 1 | 1 | PHBBB | 4 | 4 | 3 | Mixed |
| Bifidobacterium breve HPH0326 (1203540.3) | 1 | 0 | 0 | 1 | 1 | 1 | 1 | 1 | 1 | 1 | PHBBB | 4 | 4 | 3 | Mixed |
| Bifidobacterium catenulatum DSM 16992 = JCM 1194 (566552.6) | 0 | 0 | 0 | 1 | 0 | 1 | 0 | 0 | 1 | 1 | xHHxB | 1 | 3 | 1 | Donor |
| Bifidobacterium dentium ATCC 27678 (473819.7) | 0 | 1 | 0 | 0 | 1 | 0 | 0 | 0 | 1 | 1 | HxPxB | 2 | 2 | 1 | Mixed |
| Bifidobacterium gallicum DSM 20093 (561180.4) | 0 | 0 | 0 | 1 | 0 | 1 | 1 | 0 | 1 | 1 | xHHPB | 2 | 3 | 1 | Mixed |
| Bifidobacterium longum DJO10A (Prj:321) (205913.11) | 0 | 0 | 0 | 1 | 1 | 1 | 0 | 0 | 1 | 1 | xHBxB | 2 | 3 | 2 | Donor |
| Bifidobacterium longum NCC2705 (206672.9) | 0 | 0 | 0 | 1 | 1 | 1 | 0 | 0 | 1 | 1 | xHBxB | 2 | 3 | 2 | Donor |
| Bifidobacterium longum subsp. infantis 157F (565040.3) | 0 | 0 | 0 | 1 | 1 | 1 | 0 | 0 | 1 | 1 | xHBxB | 2 | 3 | 2 | Donor |
| Bifidobacterium longum subsp. infantis ATCC 15697 = JCM 1222 (3 | 1 | 1 | 0 | 1 | 1 | 1 | 1 | 1 | 1 | 1 | BHBBB | 4 | 5 | 4 | Donor |
| Bifidobacterium longum subsp. infantis ATCC 55813 (548480.3) | 0 | 0 | 0 | 1 | 1 | 1 | 0 | 0 | 1 | 1 | xHBxB | 2 | 3 | 2 | Donor |
| Bifidobacterium longum subsp. infantis CCUG 52486 (537937.5) | 0 | 0 | 0 | 1 | 1 | 1 | 0 | 0 | 1 | 1 | xHBxB | 2 | 3 | 2 | Donor |
| Bifidobacterium longum subsp. longum 2-2B (1161745.3) | 0 | 0 | 0 | 1 | 1 | 1 | 0 | 0 | 1 | 1 | xHBxB | 2 | 3 | 2 | Donor |
| Bifidobacterium longum subsp. longum 44B (1161743.3) | 0 | 0 | 0 | 1 | 1 | 1 | 0 | 0 | 1 | 1 | xHBxB | 2 | 3 | 2 | Donor |
| Bifidobacterium longum subsp. longum F8 (722911.3) | 0 | 0 | 0 | 1 | 1 | 1 | 0 | 0 | 1 | 1 | xHBxB | 2 | 3 | 2 | Donor |
| Bifidobacterium longum subsp. longum JCM 1217 (565042.3) | 0 | 0 | 0 | 1 | 1 | 1 | 0 | 0 | 1 | 1 | xHBxB | 2 | 3 | 2 | Donor |
| Bifidobacterium pseudocatenulatum DSM 20438 = JCM 1200 (5470 | 1 | 0 | 0 | 1 | 1 | 1 | 0 | 0 | 1 | 1 | PHBxB | 3 | 3 | 2 | Mixed |
| Bifidobacterium sp. 12_1_47BFAA (469594.3) | 0 | 0 | 0 | 1 | 1 | 1 | 0 | 0 | 1 | 1 | xHBxB | 2 | 3 | 2 | Donor |
| Bilophila wadsworthia 3_1_6 (563192.3) | 0 | 0 | 0 | 1 | 0 | 1 | 0 | 0 | 0 | 0 | xHHxx | 0 | 2 | 0 | Donor |
| Blautia hansenii DSM 20583 (537007.6) | 0 | 1 | 0 | 0 | 1 | 0 | 1 | 1 | 1 | 1 | HxPBB | 3 | 3 | 2 | Mixed |
| Blautia hydrogenotrophica DSM 10507 (476272.5) | 1 | 0 | 0 | 1 | 0 | 1 | 1 | 0 | 1 | 1 | PHHPB | 3 | 3 | 1 | Mixed |
| Bryantella formatexigens DSM 14469 (478749.5) | 0 | 0 | 0 | 1 | 1 | 1 | 1 | 1 | 1 | 1 | xHBBB | 3 | 4 | 3 | Donor |
| Burkholderiales bacterium 1_1_47 (469610.4) | 0 | 0 | 0 | 0 | 0 | 0 | 0 | 0 | 0 | 0 | xxxxx | 0 | 0 | 0 | - |
| butyrate-producing bacterium SM4/1 (245012.3) | 0 | 0 | 0 | 1 | 0 | 1 | 0 | 0 | 0 | 1 | xHHxH | 0 | 3 | 0 | Donor |
| butyrate-producing bacterium SS3/4 (245014.3) | 0 | 0 | 0 | 0 | 0 | 0 | 0 | 0 | 1 | 0 | xxxxP | 1 | 0 | 0 | Acceptor |
| butyrate-producing bacterium SSC/2 (245018.3) | 0 | 0 | 0 | 1 | 1 | 1 | 0 | 0 | 1 | 0 | xHBxP | 2 | 2 | 1 | Mixed |
| Butyricicoccus pullicaecorum 1.2 (1203606.4) | 0 | 0 | 1 | 1 | 1 | 1 | 0 | 0 | 1 | 1 | xBBxB | 3 | 3 | 3 | - |
| Butyrivibrio crossotus DSM 2876 (511680.4) | 0 | 0 | 0 | 0 | 0 | 0 | 0 | 0 | 0 | 1 | xxxxH | 0 | 1 | 0 | Donor |
| Butyrivibrio fibrisolvens 16/4 (657324.3) | 0 | 0 | 0 | 1 | 0 | 1 | 0 | 0 | 1 | 1 | xHHxB | 1 | 3 | 1 | Donor |

| Organism | | | | | | | | | | | Code | | | | Type |
|---|---|---|---|---|---|---|---|---|---|---|---|---|---|---|---|
| Campylobacter coli JV20 (864566.3) | 0 | 0 | 0 | 0 | 0 | 0 | 0 | 0 | 0 | 0 | xxxxx | 0 | 0 | 0 | - |
| Campylobacter upsaliensis JV21 (888826.3) | 0 | 0 | 0 | 0 | 0 | 0 | 0 | 0 | 0 | 0 | xxxxx | 0 | 0 | 0 | - |
| Catenibacterium mitsuokai DSM 15897 (451640.5) | 0 | 0 | 0 | 1 | 1 | 1 | 1 | 0 | 1 | 1 | xHBxB | 2 | 3 | 2 | Donor |
| Cedecea davisae DSM 4568 (566551.4) | 0 | 0 | 0 | 1 | 1 | 1 | 1 | 0 | 1 | 1 | xHBxB | 2 | 3 | 2 | Donor |
| Citrobacter freundii 4_7_47CFAA (742730.3) | 1 | 0 | 0 | 1 | 1 | 1 | 1 | 1 | 1 | 1 | PxxPB | 3 | 1 | 1 | Acceptor |
| Citrobacter sp. 30_2 (469595.3) | 1 | 0 | 0 | 1 | 0 | 1 | 1 | 1 | 1 | 1 | PBBPB | 5 | 3 | 3 | Acceptor |
| Citrobacter youngae ATCC 29220 (500640.5) | 1 | 0 | 0 | 1 | 0 | 1 | 1 | 1 | 1 | 1 | PPxPB | 4 | 1 | 1 | Acceptor |
| Clostridiales bacterium 1_7_47FAA (457421.5) | 1 | 0 | 0 | 1 | 1 | 1 | 1 | 1 | 1 | 1 | PHBPB | 4 | 3 | 2 | Mixed |
| Clostridium asparagiforme DSM 15981 (518636.5) | 0 | 0 | 0 | 1 | 1 | 1 | 1 | 0 | 1 | 1 | xHBPB | 3 | 3 | 2 | Mixed |
| Clostridium bartlettii DSM 16795 (445973.7) | 0 | 0 | 0 | 1 | 0 | 1 | 0 | 0 | 0 | 0 | xHHxx | 0 | 2 | 0 | Donor |
| Clostridium bolteae 90A5 (997893.5) | 1 | 0 | 0 | 1 | 1 | 1 | 1 | 0 | 1 | 1 | PHBPB | 4 | 3 | 2 | Mixed |
| Clostridium bolteae ATCC BAA-613 (411902.9) | 1 | 0 | 0 | 1 | 1 | 1 | 1 | 0 | 1 | 1 | PHBPB | 4 | 3 | 2 | Mixed |
| Clostridium celatum DSM 1785 (545697.3) | 1 | 1 | 1 | 1 | 0 | 1 | 1 | 0 | 1 | 1 | BBBHB | 4 | 5 | 4 | Donor |
| Clostridium cf. saccharolyticum K10 (717608.3) | 0 | 0 | 0 | 1 | 1 | 1 | 1 | 0 | 1 | 1 | xHBPB | 3 | 3 | 2 | Mixed |
| Clostridium citroniae WAL-17108 (742733.3) | 0 | 0 | 0 | 1 | 1 | 1 | 1 | 0 | 1 | 1 | xHBPB | 3 | 3 | 2 | Mixed |
| Clostridium clostridioforme 2_1_49FAA (742735.4) | 1 | 0 | 0 | 1 | 1 | 1 | 1 | 0 | 1 | 1 | PHBPB | 4 | 3 | 2 | Mixed |
| Clostridium difficile CD196 (645462.3) | 0 | 0 | 0 | 1 | 0 | 1 | 1 | 1 | 1 | 0 | xPxBP | 3 | 1 | 1 | Acceptor |
| Clostridium difficile NAP07 (525258.3) | 0 | 0 | 0 | 1 | 0 | 1 | 1 | 1 | 1 | 1 | xPPBP | 4 | 1 | 1 | Acceptor |
| Clostridium difficile NAP08 (525259.3) | 0 | 0 | 0 | 1 | 0 | 1 | 1 | 1 | 1 | 1 | xPPBP | 4 | 1 | 1 | Acceptor |
| Clostridium hathewayi WAL-18680 (742737.3) | 0 | 0 | 0 | 1 | 1 | 1 | 1 | 1 | 1 | 1 | xHBBB | 3 | 4 | 3 | Donor |
| Clostridium hiranonis DSM 13275 (500633.7) | 0 | 0 | 0 | 1 | 0 | 1 | 0 | 0 | 1 | 1 | xHHPP | 2 | 2 | 0 | Mixed |
| Clostridium hylemonae DSM 15053 (553973.6) | 1 | 0 | 0 | 1 | 0 | 1 | 1 | 0 | 1 | 1 | PxxBB | 3 | 2 | 2 | Acceptor |
| Clostridium leptum DSM 753 (428125.8) | 1 | 1 | 0 | 1 | 0 | 1 | 1 | 0 | 1 | 1 | BHHxB | 2 | 4 | 2 | Donor |
| Clostridium methylpentosum DSM 5476 (537013.3) | 0 | 1 | 0 | 0 | 0 | 0 | 0 | 0 | 0 | 0 | Hxxxx | 0 | 1 | 0 | Donor |
| Clostridium nexile DSM 1787 (500632.7) | 0 | 1 | 0 | 1 | 1 | 1 | 1 | 1 | 1 | 1 | HHBHB | 2 | 5 | 2 | Donor |
| Clostridium perfringens WAL-14572 (742739.3) | 1 | 1 | 0 | 1 | 1 | 1 | 1 | 1 | 1 | 1 | BHBBB | 4 | 5 | 4 | Donor |
| Clostridium ramosum DSM 1402 (445974.6) | 0 | 0 | 0 | 1 | 1 | 1 | 1 | 0 | 1 | 1 | xHBPB | 2 | 4 | 2 | Donor |
| Clostridium scindens ATCC 35704 (411468.9) | 0 | 1 | 0 | 0 | 0 | 1 | 0 | 0 | 1 | 1 | HxxBB | 2 | 3 | 2 | Donor |
| Clostridium sp. 7_2_43FAA (457396.3) | 0 | 1 | 0 | 1 | 1 | 1 | 1 | 1 | 1 | 1 | xBBBB | 4 | 4 | 4 | - |
| Clostridium sp. 7_3_54FAA (665940.3) | 0 | 0 | 0 | 1 | 0 | 1 | 1 | 0 | 1 | 1 | xHHPB | 2 | 3 | 1 | Mixed |
| Clostridium sp. D5 (556261.3) | 1 | 1 | 0 | 1 | 0 | 1 | 1 | 0 | 1 | 1 | BxPPB | 4 | 2 | 2 | Acceptor |
| Clostridium sp. HGF2 (908340.3) | 0 | 0 | 0 | 0 | 0 | 0 | 0 | 0 | 0 | 1 | xxxxH | 0 | 1 | 0 | Donor |
| Clostridium sp. L2-50 (411489.7) | 0 | 0 | 0 | 1 | 0 | 1 | 1 | 0 | 1 | 1 | xHHxH | 0 | 3 | 0 | Donor |
| Clostridium sp. M62/1 (411486.3) | 0 | 0 | 0 | 1 | 1 | 1 | 1 | 1 | 1 | 1 | xHBPB | 3 | 3 | 2 | Mixed |
| Clostridium sp. SS2/1 (411484.7) | 0 | 0 | 0 | 1 | 1 | 1 | 1 | 0 | 1 | 1 | xHBxB | 2 | 3 | 2 | Donor |
| Clostridium spiroforme DSM 1552 (428126.7) | 0 | 1 | 1 | 1 | 1 | 1 | 1 | 0 | 0 | 1 | HBBxH | 2 | 4 | 2 | Donor |
| Clostridium sporogenes ATCC 15579 (471871.7) | 0 | 0 | 0 | 1 | 0 | 1 | 1 | 1 | 0 | 1 | xxxHx | 0 | 1 | 0 | Donor |
| Clostridium symbiosum WAL-14163 (742740.3) | 0 | 0 | 0 | 1 | 1 | 1 | 1 | 0 | 1 | 1 | xHHPB | 2 | 3 | 1 | Mixed |
| Clostridium symbiosum WAL-14673 (742741.3) | 0 | 0 | 0 | 1 | 1 | 1 | 1 | 0 | 1 | 1 | xHHPB | 2 | 3 | 1 | Mixed |
| Collinsella aerofaciens ATCC 25986 (411903.6) | 0 | 0 | 1 | 1 | 1 | 1 | 1 | 0 | 1 | 1 | xPHPB | 3 | 2 | 1 | Mixed |
| Collinsella intestinalis DSM 13280 (521003.7) | 0 | 1 | 0 | 1 | 1 | 1 | 1 | 0 | 1 | 1 | xBHBx | 2 | 3 | 2 | Donor |
| Collinsella stercoris DSM 13279 (445975.6) | 0 | 1 | 1 | 1 | 1 | 1 | 1 | 1 | 1 | 1 | HBBxB | 3 | 4 | 3 | Donor |
| Collinsella tanakaei YIT 12063 (742742.3) | 0 | 0 | 0 | 1 | 1 | 1 | 1 | 0 | 1 | 1 | PPxPP | 4 | 0 | 0 | Acceptor |
| Coprobacillus sp. 29_1 (469596.3) | 0 | 0 | 0 | 1 | 1 | 1 | 1 | 0 | 1 | 1 | xHBHB | 2 | 4 | 2 | Donor |
| Coprobacillus sp. 3_3_56FAA (665941.3) | 0 | 0 | 0 | 1 | 1 | 1 | 1 | 1 | 1 | 1 | xHBHB | 2 | 4 | 2 | Donor |
| Coprobacillus sp. 8_2_54BFAA (469597.4) | 0 | 0 | 0 | 1 | 1 | 1 | 1 | 1 | 1 | 1 | xHBHB | 2 | 4 | 2 | Donor |
| Coprobacillus sp. D6 (556262.3) | 0 | 0 | 0 | 1 | 1 | 1 | 1 | 0 | 1 | 1 | xHBxB | 2 | 3 | 2 | Donor |
| Coprococcus catus GD/7 (717962.3) | 0 | 0 | 0 | 1 | 1 | 1 | 1 | 0 | 1 | 1 | xHBxB | 2 | 3 | 2 | Donor |
| Coprococcus comes ATCC 27758 (470146.3) | 0 | 0 | 0 | 1 | 0 | 1 | 1 | 0 | 1 | 1 | xxPxB | 2 | 1 | 1 | Acceptor |
| Coprococcus eutactus ATCC 27759 (411474.6) | 0 | 0 | 0 | 1 | 0 | 1 | 0 | 0 | 1 | 1 | xHHxH | 0 | 3 | 0 | Donor |
| Coprococcus sp. ART55/1 (751585.3) | 0 | 0 | 0 | 1 | 0 | 1 | 0 | 0 | 1 | 1 | xHHxH | 0 | 3 | 0 | Donor |
| Coprococcus sp. HPP0048 (1078091.3) | 1 | 1 | 0 | 1 | 1 | 1 | 1 | 1 | 1 | 1 | BHBBB | 4 | 5 | 4 | Donor |
| Coprococcus sp. HPP0074 (1078090.3) | 1 | 1 | 0 | 1 | 0 | 1 | 1 | 0 | 1 | 1 | BHBHB | 3 | 5 | 3 | Donor |
| Corynebacterium ammoniagenes DSM 20306 (649754.3) | 1 | 1 | 0 | 1 | 1 | 1 | 1 | 1 | 1 | 0 | PHBPP | 4 | 2 | 1 | Mixed |
| Corynebacterium sp. HFH0082 (1078764.3) | 0 | 0 | 0 | 0 | 0 | 0 | 0 | 0 | 1 | 0 | xxxxP | 1 | 0 | 0 | Acceptor |

| Organism | | | | | | | | | | | Code | | | | Role |
|---|---|---|---|---|---|---|---|---|---|---|---|---|---|---|---|
| Dermabacter sp. HFH0086 (1203568.3) | 0 | 0 | 0 | 0 | 1 | 0 | 1 | 1 | 1 | 1 | xxPBB | 3 | 2 | 2 | Acceptor |
| Desulfitobacterium hafniense DP7 (537010.4) | 0 | 0 | 0 | 1 | 0 | 1 | 0 | 0 | 0 | 0 | xHHxx | 0 | 2 | 0 | Donor |
| Desulfovibrio piger ATCC 29098 (411464.8) | 0 | 0 | 0 | 0 | 0 | 0 | 0 | 0 | 0 | 0 | xxxxx | 0 | 0 | 0 | - |
| Desulfovibrio sp. 3_1_syn3 (457398.5) | 0 | 0 | 0 | 1 | 0 | 1 | 0 | 0 | 0 | 0 | xHHxx | 0 | 2 | 0 | Donor |
| Dorea formicigenerans 4_6_53AFAA (742765.5) | 0 | 1 | 0 | 0 | 1 | 0 | 0 | 1 | 1 | 1 | HxPHB | 2 | 3 | 1 | Mixed |
| Dorea formicigenerans ATCC 27755 (411461.4) | 0 | 0 | 0 | 1 | 1 | 1 | 1 | 1 | 1 | 1 | xHBBB | 3 | 4 | 3 | Donor |
| Dorea longicatena DSM 13814 (411462.6) | 0 | 0 | 0 | 0 | 0 | 0 | 1 | 0 | 1 | 1 | xxxPB | 2 | 1 | 1 | Acceptor |
| Dysgonomonas gadei ATCC BAA-286 (742766.3) | 1 | 1 | 0 | 1 | 0 | 1 | 0 | 0 | 1 | 1 | BHHxB | 2 | 4 | 2 | Donor |
| Dysgonomonas mossii DSM 22836 (742767.3) | 1 | 1 | 0 | 1 | 0 | 1 | 0 | 0 | 1 | 1 | BHHxB | 2 | 4 | 2 | Donor |
| Edwardsiella tarda ATCC 23685 (500638.3) | 1 | 0 | 1 | 1 | 1 | 0 | 1 | 0 | 1 | 1 | PBPPB | 5 | 3 | 3 | Acceptor |
| Eggerthella sp. 1_3_56FAA (665943.3) | 0 | 0 | 0 | 0 | 0 | 0 | 1 | 0 | 0 | 0 | xxxPx | 1 | 0 | 0 | Acceptor |
| Eggerthella sp. HGA1 (910311.3) | 0 | 0 | 0 | 0 | 0 | 0 | 1 | 0 | 0 | 0 | xxxPx | 1 | 0 | 0 | Acceptor |
| Enterobacter cancerogenus ATCC 35316 (500639.8) | 1 | 0 | 1 | 1 | 1 | 1 | 1 | 0 | 1 | 1 | PBBPB | 5 | 3 | 3 | Acceptor |
| Enterobacter cloacae subsp. cloacae NCTC 9394 (718254.4) | 1 | 0 | 1 | 1 | 1 | 1 | 1 | 0 | 1 | 1 | PBBPB | 5 | 3 | 3 | Acceptor |
| Enterobacteriaceae bacterium 9_2_54FAA (469613.3) | 1 | 0 | 0 | 1 | 1 | 1 | 1 | 0 | 1 | 1 | PHBPB | 4 | 3 | 2 | Mixed |
| Enterococcus faecalis PC1.1 (702448.3) | 0 | 0 | 0 | 0 | 0 | 0 | 0 | 0 | 1 | 1 | xxxxB | 1 | 1 | 1 | - |
| Enterococcus faecalis TX0104 (491074.3) | 0 | 0 | 0 | 1 | 1 | 1 | 0 | 0 | 1 | 1 | xHBxB | 2 | 3 | 2 | Donor |
| Enterococcus faecalis TX1302 (749513.3) | 0 | 0 | 0 | 1 | 1 | 1 | 0 | 0 | 1 | 1 | xHBxB | 2 | 3 | 2 | Donor |
| Enterococcus faecalis TX1322 (525278.3) | 0 | 0 | 1 | 1 | 1 | 1 | 0 | 0 | 1 | 1 | xBBxB | 3 | 3 | 3 | - |
| Enterococcus faecalis TX1341 (749514.3) | 0 | 0 | 1 | 1 | 1 | 1 | 0 | 0 | 1 | 1 | xBBxB | 3 | 3 | 3 | - |
| Enterococcus faecalis TX1342 (749515.3) | 0 | 0 | 1 | 0 | 0 | 0 | 0 | 0 | 1 | 1 | xPxxB | 2 | 1 | 1 | Acceptor |
| Enterococcus faecalis TX1346 (749516.3) | 0 | 0 | 1 | 1 | 1 | 1 | 0 | 0 | 1 | 1 | xBBxB | 3 | 3 | 3 | - |
| Enterococcus faecalis TX2134 (749518.3) | 0 | 0 | 1 | 1 | 1 | 1 | 0 | 0 | 1 | 1 | xBBxB | 3 | 3 | 3 | - |
| Enterococcus faecalis TX2137 (749491.3) | 0 | 0 | 1 | 1 | 1 | 1 | 0 | 0 | 1 | 1 | xBBxB | 3 | 3 | 3 | - |
| Enterococcus faecalis TX4244 (749494.3) | 0 | 0 | 1 | 1 | 1 | 1 | 0 | 0 | 1 | 1 | xBBxB | 3 | 3 | 3 | - |
| Enterococcus faecium PC4.1 (791161.5) | 0 | 0 | 0 | 0 | 0 | 0 | 0 | 0 | 1 | 1 | xxxxB | 1 | 1 | 1 | - |
| Enterococcus faecium TX1330 (525279.3) | 0 | 0 | 0 | 0 | 0 | 0 | 0 | 0 | 1 | 1 | xxxxB | 1 | 1 | 1 | - |
| Enterococcus saccharolyticus 30_1 (742813.4) | 0 | 1 | 0 | 1 | 0 | 1 | 0 | 0 | 1 | 1 | HHHxB | 1 | 4 | 1 | Donor |
| Enterococcus sp. 7L76 (657310.3) | 0 | 0 | 1 | 0 | 0 | 0 | 0 | 0 | 1 | 1 | xPxxB | 2 | 1 | 1 | Acceptor |
| Erysipelotrichaceae bacterium 2_2_44A (457422.3) | 0 | 0 | 0 | 0 | 0 | 0 | 0 | 0 | 0 | 1 | xxxxH | 0 | 1 | 0 | Donor |
| Erysipelotrichaceae bacterium 21_3 (658657.3) | 0 | 0 | 0 | 0 | 0 | 0 | 0 | 0 | 0 | 1 | xxxxH | 0 | 1 | 0 | Donor |
| Erysipelotrichaceae bacterium 3_1_53 (658659.3) | 0 | 0 | 0 | 0 | 0 | 0 | 0 | 0 | 0 | 0 | xxxxx | 0 | 0 | 0 | - |
| Erysipelotrichaceae bacterium 5_2_54FAA (552396.3) | 0 | 0 | 0 | 1 | 1 | 1 | 0 | 0 | 0 | 0 | xHBxx | 1 | 2 | 1 | Donor |
| Erysipelotrichaceae bacterium 6_1_45 (469614.3) | 0 | 0 | 0 | 0 | 0 | 0 | 0 | 0 | 0 | 1 | xxxxH | 0 | 1 | 0 | Donor |
| Escherichia coli 4_1_47FAA (1127356.4) | 1 | 0 | 1 | 0 | 1 | 0 | 1 | 0 | 1 | 1 | PPPPB | 5 | 1 | 1 | Acceptor |
| Escherichia coli MS 107-1 (679207.4) | 1 | 0 | 1 | 0 | 1 | 0 | 1 | 0 | 1 | 1 | PPPPB | 5 | 1 | 1 | Acceptor |
| Escherichia coli MS 115-1 (749537.3) | 1 | 0 | 1 | 0 | 0 | 0 | 1 | 0 | 1 | 1 | PPxPB | 4 | 1 | 1 | Acceptor |
| Escherichia coli MS 116-1 (749538.3) | 1 | 0 | 1 | 0 | 0 | 0 | 1 | 0 | 1 | 1 | PPxPB | 4 | 1 | 1 | Acceptor |
| Escherichia coli MS 119-7 (679206.4) | 1 | 0 | 1 | 0 | 1 | 0 | 1 | 0 | 1 | 1 | PPPPB | 5 | 1 | 1 | Acceptor |
| Escherichia coli MS 124-1 (679205.4) | 1 | 0 | 1 | 0 | 0 | 0 | 1 | 0 | 1 | 1 | PPxPB | 4 | 1 | 1 | Acceptor |
| Escherichia coli MS 145-7 (679204.3) | 1 | 0 | 1 | 0 | 1 | 0 | 1 | 0 | 1 | 1 | PPPPB | 5 | 1 | 1 | Acceptor |
| Escherichia coli MS 146-1 (749540.3) | 1 | 0 | 1 | 0 | 0 | 0 | 1 | 0 | 1 | 1 | PPxPB | 4 | 1 | 1 | Acceptor |
| Escherichia coli MS 175-1 (749544.3) | 0 | 0 | 1 | 0 | 0 | 1 | 0 | 0 | 1 | 1 | xPHxB | 2 | 2 | 1 | Mixed |
| Escherichia coli MS 182-1 (749545.3) | 1 | 0 | 1 | 0 | 1 | 0 | 1 | 0 | 1 | 1 | PPPPB | 5 | 1 | 1 | Acceptor |
| Escherichia coli MS 185-1 (749546.3) | 1 | 0 | 1 | 0 | 0 | 0 | 1 | 0 | 1 | 1 | PPxPB | 4 | 1 | 1 | Acceptor |
| Escherichia coli MS 187-1 (749547.3) | 1 | 0 | 1 | 0 | 0 | 0 | 1 | 0 | 1 | 1 | PPxPB | 4 | 1 | 1 | Acceptor |
| Escherichia coli MS 196-1 (749548.3) | 1 | 0 | 1 | 0 | 1 | 0 | 1 | 0 | 1 | 1 | PPPPB | 5 | 1 | 1 | Acceptor |
| Escherichia coli MS 198-1 (749549.3) | 1 | 0 | 1 | 0 | 1 | 0 | 1 | 0 | 1 | 1 | PPPPB | 5 | 1 | 1 | Acceptor |
| Escherichia coli MS 200-1 (749550.3) | 1 | 0 | 1 | 0 | 1 | 0 | 1 | 0 | 1 | 1 | PPPPB | 5 | 1 | 1 | Acceptor |
| Escherichia coli MS 21-1 (749527.3) | 1 | 0 | 1 | 0 | 1 | 0 | 1 | 0 | 1 | 1 | PPPPB | 5 | 1 | 1 | Acceptor |
| Escherichia coli MS 45-1 (749528.3) | 1 | 0 | 1 | 0 | 0 | 0 | 1 | 0 | 1 | 1 | PPxPB | 4 | 1 | 1 | Acceptor |
| Escherichia coli MS 69-1 (749531.3) | 1 | 0 | 1 | 0 | 1 | 0 | 1 | 0 | 1 | 1 | PPPPB | 5 | 1 | 1 | Acceptor |
| Escherichia coli MS 78-1 (749532.3) | 1 | 0 | 1 | 0 | 1 | 0 | 1 | 0 | 1 | 1 | PPPPB | 5 | 1 | 1 | Acceptor |
| Escherichia coli MS 79-10 (796392.4) | 1 | 0 | 1 | 0 | 1 | 0 | 1 | 0 | 1 | 1 | PPPPB | 5 | 1 | 1 | Acceptor |
| Escherichia coli MS 84-1 (749533.3) | 1 | 0 | 1 | 0 | 0 | 0 | 1 | 0 | 1 | 1 | PPxPB | 4 | 1 | 1 | Acceptor |

| Organism | | | | | | | | | | | Motif | | | | Class |
|---|---|---|---|---|---|---|---|---|---|---|---|---|---|---|---|
| Escherichia coli MS 85-1 (679202.3) | 1 | 0 | 1 | 0 | 0 | 0 | 1 | 0 | 1 | 1 | PPxPB | 4 | 1 | 1 | Acceptor |
| Escherichia coli O157:H7 str. Sakai (386585.9) | 1 | 0 | 1 | 0 | 1 | 0 | 1 | 0 | 1 | 1 | PPPPB | 5 | 1 | 1 | Acceptor |
| Escherichia coli SE11 (409438.11) | 1 | 0 | 1 | 0 | 1 | 0 | 1 | 0 | 1 | 1 | PPPPB | 5 | 1 | 1 | Acceptor |
| Escherichia coli SE15 (431946.3) | 1 | 0 | 1 | 0 | 0 | 0 | 1 | 0 | 1 | 1 | PPxPB | 4 | 1 | 1 | Acceptor |
| Escherichia coli str. K-12 substr. MG1655 (511145.12) | 1 | 0 | 1 | 0 | 0 | 0 | 1 | 0 | 1 | 1 | PPxPB | 4 | 1 | 1 | Acceptor |
| Escherichia coli UTI89 (364106.8) | 1 | 0 | 1 | 0 | 0 | 0 | 1 | 0 | 1 | 1 | PPxPB | 4 | 1 | 1 | Acceptor |
| Escherichia sp. 1_1_43 (457400.3) | 1 | 0 | 0 | 0 | 0 | 1 | 1 | 0 | 0 | 1 | PxHPH | 2 | 2 | 0 | Mixed |
| Escherichia sp. 3_2_53FAA (469598.5) | 1 | 0 | 1 | 0 | 0 | 0 | 1 | 0 | 1 | 1 | PPxPB | 4 | 1 | 1 | Acceptor |
| Escherichia sp. 4_1_40B (457401.3) | 1 | 0 | 1 | 0 | 0 | 0 | 1 | 0 | 1 | 1 | PPxPB | 4 | 1 | 1 | Acceptor |
| Eubacterium biforme DSM 3989 (518637.5) | 0 | 1 | 0 | 0 | 1 | 0 | 0 | 0 | 1 | 1 | HxPxB | 2 | 2 | 1 | Mixed |
| Eubacterium cylindroides T2-87 (717960.3) | 0 | 0 | 0 | 0 | 0 | 0 | 0 | 0 | 0 | 0 | xxxxx | 0 | 0 | 0 | - |
| Eubacterium dolichum DSM 3991 (428127.7) | 0 | 0 | 0 | 1 | 0 | 1 | 0 | 0 | 0 | 1 | xHHxH | 0 | 3 | 0 | Donor |
| Eubacterium hallii DSM 3353 (411469.3) | 0 | 0 | 0 | 1 | 0 | 1 | 0 | 0 | 1 | 1 | xHHxB | 1 | 3 | 1 | Donor |
| Eubacterium rectale DSM 17629 (657318.4) | 0 | 0 | 0 | 1 | 1 | 1 | 0 | 0 | 1 | 1 | xHBxB | 2 | 3 | 2 | Donor |
| Eubacterium rectale M104/1 (657317.3) | 0 | 0 | 0 | 1 | 1 | 1 | 0 | 0 | 1 | 1 | xHBxB | 2 | 3 | 2 | Donor |
| Eubacterium siraeum 70/3 (657319.3) | 1 | 1 | 0 | 0 | 0 | 0 | 0 | 0 | 1 | 1 | BxxxB | 2 | 2 | 2 | - |
| Eubacterium siraeum DSM 15702 (428128.7) | 1 | 1 | 0 | 1 | 0 | 1 | 0 | 0 | 1 | 1 | BHHxB | 2 | 4 | 2 | Donor |
| Eubacterium siraeum V10Sc8a (717961.3) | 1 | 1 | 0 | 1 | 0 | 1 | 0 | 0 | 1 | 1 | BHHxB | 2 | 4 | 2 | Donor |
| Eubacterium sp. 3_1_31 (457402.3) | 0 | 0 | 0 | 1 | 1 | 1 | 0 | 0 | 0 | 0 | xHBxx | 1 | 2 | 1 | Donor |
| Eubacterium ventriosum ATCC 27560 (411463.4) | 0 | 0 | 0 | 0 | 0 | 0 | 0 | 0 | 1 | 1 | xxxxB | 1 | 1 | 1 | - |
| Faecalibacterium cf. prausnitzii KLE1255 (748224.3) | 0 | 0 | 0 | 0 | 0 | 0 | 1 | 0 | 1 | 1 | xxxPB | 2 | 1 | 1 | Acceptor |
| Faecalibacterium prausnitzii A2-165 (411483.3) | 0 | 0 | 0 | 1 | 1 | 1 | 1 | 0 | 1 | 1 | xHBPB | 3 | 3 | 2 | Mixed |
| Faecalibacterium prausnitzii L2-6 (718252.3) | 0 | 0 | 1 | 0 | 0 | 0 | 1 | 0 | 1 | 1 | xPxPB | 3 | 1 | 1 | Acceptor |
| Faecalibacterium prausnitzii M21/2 (411485.10) | 0 | 0 | 0 | 0 | 1 | 0 | 1 | 0 | 1 | 1 | xxPPB | 3 | 1 | 1 | Acceptor |
| Faecalibacterium prausnitzii SL3/3 (657322.3) | 0 | 0 | 0 | 0 | 1 | 0 | 1 | 0 | 1 | 1 | xxPPB | 3 | 1 | 1 | Acceptor |
| Fusobacterium gonidiaformans ATCC 25563 (469615.3) | 0 | 0 | 0 | 0 | 1 | 0 | 1 | 0 | 1 | 0 | xxPPP | 3 | 0 | 0 | Acceptor |
| Fusobacterium mortiferum ATCC 9817 (469616.3) | 0 | 0 | 0 | 0 | 1 | 0 | 1 | 0 | 1 | 0 | xxPPP | 3 | 1 | 1 | Acceptor |
| Fusobacterium necrophorum subsp. funduliforme 1_1_36S (742814 | 0 | 0 | 0 | 0 | 1 | 0 | 1 | 0 | 1 | 0 | xxPPP | 3 | 0 | 0 | Acceptor |
| Fusobacterium sp. 1_1_41FAA (469621.3) | 0 | 0 | 0 | 0 | 1 | 0 | 1 | 0 | 1 | 0 | xxPPP | 3 | 0 | 0 | Acceptor |
| Fusobacterium sp. 11_3_2 (457403.3) | 0 | 0 | 0 | 0 | 1 | 0 | 1 | 0 | 1 | 0 | xxPPP | 3 | 0 | 0 | Acceptor |
| Fusobacterium sp. 12_1B (457404.3) | 0 | 0 | 0 | 0 | 1 | 0 | 1 | 0 | 1 | 0 | xxPPP | 3 | 0 | 0 | Acceptor |
| Fusobacterium sp. 2_1_31 (469599.3) | 0 | 0 | 0 | 0 | 1 | 0 | 0 | 0 | 1 | 0 | xxPxP | 2 | 0 | 0 | Acceptor |
| Fusobacterium sp. 21_1A (469601.3) | 0 | 0 | 0 | 0 | 1 | 0 | 0 | 0 | 1 | 0 | xxPxP | 2 | 0 | 0 | Acceptor |
| Fusobacterium sp. 3_1_27 (469602.3) | 0 | 0 | 0 | 0 | 1 | 0 | 1 | 0 | 1 | 0 | xxPPP | 3 | 0 | 0 | Acceptor |
| Fusobacterium sp. 3_1_33 (469603.3) | 0 | 0 | 0 | 0 | 1 | 0 | 0 | 0 | 1 | 0 | xxPxP | 2 | 0 | 0 | Acceptor |
| Fusobacterium sp. 3_1_36A2 (469604.3) | 0 | 0 | 0 | 0 | 1 | 0 | 1 | 0 | 1 | 0 | xxPPP | 3 | 0 | 0 | Acceptor |
| Fusobacterium sp. 3_1_5R (469605.3) | 0 | 0 | 0 | 0 | 1 | 0 | 1 | 0 | 1 | 0 | xxPPP | 3 | 0 | 0 | Acceptor |
| Fusobacterium sp. 4_1_13 (469606.3) | 0 | 0 | 0 | 0 | 1 | 0 | 1 | 0 | 1 | 0 | xxPPP | 3 | 0 | 0 | Acceptor |
| Fusobacterium sp. 7_1 (457405.3) | 0 | 0 | 0 | 0 | 1 | 0 | 0 | 0 | 1 | 0 | xxPxP | 2 | 0 | 0 | Acceptor |
| Fusobacterium sp. D11 (556264.3) | 0 | 0 | 0 | 0 | 1 | 0 | 1 | 0 | 1 | 0 | xxPPP | 3 | 0 | 0 | Acceptor |
| Fusobacterium sp. D12 (556263.3) | 0 | 0 | 0 | 0 | 1 | 0 | 1 | 0 | 1 | 0 | xxPPP | 3 | 0 | 0 | Acceptor |
| Fusobacterium ulcerans ATCC 49185 (469617.3) | 0 | 0 | 0 | 0 | 1 | 0 | 1 | 0 | 1 | 0 | xxPPP | 3 | 0 | 0 | Acceptor |
| Fusobacterium varium ATCC 27725 (469618.3) | 1 | 0 | 0 | 0 | 1 | 0 | 1 | 0 | 1 | 0 | PxPPP | 4 | 0 | 0 | Acceptor |
| Gordonibacter pamelaeae 7-10-1-b (657308.3) | 0 | 0 | 0 | 0 | 0 | 0 | 1 | 0 | 0 | 0 | xxxPx | 1 | 0 | 0 | Acceptor |
| Hafnia alvei 51873 (1002364.3) | 1 | 0 | 0 | 1 | 1 | 1 | 1 | 0 | 1 | 1 | PHBPB | 4 | 3 | 2 | Mixed |
| Helicobacter bilis ATCC 43879 (613026.4) | 0 | 0 | 0 | 0 | 0 | 1 | 0 | 0 | 0 | 0 | xxHxx | 0 | 1 | 0 | Donor |
| Helicobacter canadensis MIT 98-5491 (Prj:30719) (537970.9) | 0 | 0 | 0 | 0 | 0 | 0 | 0 | 0 | 0 | 0 | xxxxx | 0 | 0 | 0 | - |
| Helicobacter cinaedi ATCC BAA-847 (1206745.3) | 0 | 0 | 0 | 0 | 0 | 1 | 0 | 0 | 0 | 0 | xxHxx | 0 | 1 | 0 | Donor |
| Helicobacter cinaedi CCUG 18818 (537971.5) | 0 | 0 | 0 | 0 | 0 | 0 | 0 | 0 | 0 | 0 | xxxxx | 0 | 0 | 0 | - |
| Helicobacter pullorum MIT 98-5489 (537972.5) | 0 | 0 | 0 | 0 | 0 | 0 | 0 | 0 | 0 | 0 | xxxxx | 0 | 0 | 0 | - |
| Helicobacter winghamensis ATCC BAA-430 (556267.4) | 0 | 0 | 0 | 0 | 0 | 0 | 0 | 0 | 0 | 0 | xxxxx | 0 | 0 | 0 | - |
| Holdemania filiformis DSM 12042 (545696.5) | 1 | 1 | 0 | 1 | 1 | 1 | 0 | 0 | 1 | 1 | BHBxB | 3 | 4 | 3 | Donor |
| Klebsiella pneumoniae 1162281 (1037908.3) | 1 | 0 | 0 | 0 | 1 | 1 | 0 | 0 | 1 | 1 | PxxxB | 2 | 1 | 1 | Acceptor |
| Klebsiella pneumoniae subsp. pneumoniae WGLW5 (1203547.3) | 1 | 0 | 0 | 0 | 1 | 1 | 0 | 0 | 1 | 1 | PxxxB | 2 | 1 | 1 | Acceptor |
| Klebsiella sp. 1_1_55 (469608.3) | 1 | 0 | 0 | 0 | 1 | 1 | 0 | 0 | 1 | 1 | PxxxB | 2 | 1 | 1 | Acceptor |

| Organism | c1 | c2 | c3 | c4 | c5 | c6 | c7 | c8 | c9 | c10 | Motif | n1 | n2 | n3 | Type |
|---|---|---|---|---|---|---|---|---|---|---|---|---|---|---|---|
| Klebsiella sp. 4_1_44FAA (665944.3) | 1 | 0 | 0 | 0 | 0 | 0 | 0 | 0 | 1 | 1 | PxxxB | 2 | 1 | 1 | Acceptor |
| Lachnospiraceae bacterium 1_1_57FAA (658081.3) | 0 | 1 | 1 | 1 | 0 | 1 | 0 | 1 | 1 | 1 | HBHHB | 2 | 5 | 2 | Donor |
| Lachnospiraceae bacterium 1_4_56FAA (658655.3) | 1 | 1 | 0 | 1 | 1 | 1 | 0 | 0 | 1 | 1 | BHBxB | 3 | 4 | 3 | Donor |
| Lachnospiraceae bacterium 2_1_46FAA (742723.3) | 0 | 1 | 0 | 1 | 1 | 1 | 0 | 1 | 1 | 1 | HHBHB | 2 | 5 | 2 | Donor |
| Lachnospiraceae bacterium 2_1_58FAA (658082.3) | 1 | 1 | 0 | 1 | 1 | 1 | 0 | 0 | 1 | 1 | BHBxB | 3 | 4 | 3 | Donor |
| Lachnospiraceae bacterium 3_1_46FAA (665950.3) | 0 | 1 | 1 | 1 | 0 | 1 | 0 | 1 | 1 | 1 | HBHHB | 2 | 5 | 2 | Donor |
| Lachnospiraceae bacterium 3_1_57FAA_CT1 (658086.3) | 1 | 1 | 0 | 1 | 1 | 1 | 1 | 1 | 1 | 1 | BHBBB | 4 | 5 | 4 | Donor |
| Lachnospiraceae bacterium 4_1_37FAA (552395.3) | 1 | 1 | 0 | 1 | 1 | 1 | 1 | 1 | 1 | 1 | BHBBB | 4 | 5 | 4 | Donor |
| Lachnospiraceae bacterium 5_1_57FAA (658085.3) | 0 | 1 | 0 | 0 | 0 | 0 | 1 | 1 | 1 | 1 | HxxBB | 2 | 3 | 2 | Donor |
| Lachnospiraceae bacterium 5_1_63FAA (658089.3) | 0 | 0 | 0 | 1 | 1 | 1 | 0 | 0 | 1 | 1 | xHBxB | 2 | 3 | 2 | Donor |
| Lachnospiraceae bacterium 6_1_63FAA (658083.3) | 0 | 1 | 0 | 0 | 1 | 0 | 0 | 0 | 1 | 1 | HxPxB | 2 | 2 | 1 | Mixed |
| Lachnospiraceae bacterium 7_1_58FAA (658087.3) | 0 | 0 | 0 | 0 | 0 | 0 | 1 | 1 | 0 | 1 | xxxBH | 1 | 2 | 1 | Donor |
| Lachnospiraceae bacterium 8_1_57FAA (665951.3) | 0 | 1 | 1 | 1 | 0 | 1 | 0 | 1 | 1 | 1 | HBHHB | 2 | 5 | 2 | Donor |
| Lachnospiraceae bacterium 9_1_43BFAA (658088.3) | 1 | 1 | 0 | 1 | 1 | 1 | 0 | 1 | 1 | 1 | BHBHB | 3 | 5 | 3 | Donor |
| Lactobacillus acidophilus ATCC 4796 (525306.3) | 0 | 0 | 0 | 0 | 0 | 0 | 0 | 0 | 1 | 1 | xxxxB | 1 | 1 | 1 | - |
| Lactobacillus acidophilus NCFM (272621.13) | 0 | 0 | 0 | 0 | 1 | 0 | 0 | 0 | 1 | 1 | xxPxB | 2 | 1 | 1 | Acceptor |
| Lactobacillus amylolyticus DSM 11664 (585524.3) | 0 | 0 | 0 | 1 | 1 | 1 | 0 | 0 | 1 | 0 | xHBxP | 2 | 2 | 1 | Mixed |
| Lactobacillus antri DSM 16041 (525309.3) | 0 | 0 | 0 | 0 | 0 | 0 | 1 | 0 | 1 | 1 | xxxPB | 2 | 1 | 1 | Acceptor |
| Lactobacillus brevis ATCC 367 (387344.15) | 0 | 0 | 0 | 0 | 0 | 0 | 0 | 0 | 1 | 1 | xxxxB | 1 | 1 | 1 | - |
| Lactobacillus brevis subsp. gravesensis ATCC 27305 (525310.3) | 0 | 0 | 0 | 1 | 0 | 1 | 0 | 0 | 1 | 1 | xHHxB | 1 | 3 | 1 | Donor |
| Lactobacillus buchneri ATCC 11577 (525318.3) | 0 | 0 | 0 | 1 | 0 | 1 | 0 | 0 | 1 | 1 | xHHxB | 1 | 3 | 1 | Donor |
| Lactobacillus casei ATCC 334 (321967.11) | 0 | 1 | 0 | 1 | 0 | 1 | 0 | 0 | 1 | 1 | HHHxB | 1 | 4 | 1 | Donor |
| Lactobacillus casei BL23 (543734.4) | 0 | 1 | 1 | 1 | 0 | 1 | 0 | 0 | 1 | 1 | HBHxB | 2 | 4 | 2 | Donor |
| Lactobacillus crispatus 125-2-CHN (575595.3) | 0 | 0 | 0 | 0 | 1 | 0 | 0 | 0 | 1 | 1 | xxPxB | 2 | 1 | 1 | Acceptor |
| Lactobacillus delbrueckii subsp. bulgaricus ATCC 11842 (390333.7) | 0 | 0 | 0 | 0 | 0 | 0 | 0 | 0 | 0 | 1 | xxxxH | 0 | 0 | 1 | Donor |
| Lactobacillus delbrueckii subsp. bulgaricus ATCC BAA-365 (321956...) | 0 | 0 | 0 | 0 | 1 | 0 | 0 | 0 | 0 | 1 | xxPxH | 1 | 1 | 1 | Mixed |
| Lactobacillus fermentum ATCC 14931 (525325.3) | 0 | 0 | 0 | 0 | 0 | 0 | 0 | 0 | 1 | 1 | xxxxB | 1 | 1 | 1 | - |
| Lactobacillus fermentum IFO 3956 (334390.5) | 0 | 0 | 0 | 0 | 0 | 0 | 0 | 0 | 1 | 1 | xxxxB | 1 | 1 | 1 | - |
| Lactobacillus gasseri ATCC 33323 (324831.13) | 0 | 0 | 1 | 0 | 0 | 1 | 0 | 0 | 1 | 1 | xPHxB | 2 | 2 | 1 | Mixed |
| Lactobacillus helveticus DPC 4571 (405566.6) | 0 | 0 | 0 | 0 | 0 | 0 | 0 | 0 | 1 | 1 | xxxxB | 1 | 1 | 1 | - |
| Lactobacillus helveticus DSM 20075 (585520.4) | 0 | 0 | 0 | 0 | 0 | 0 | 0 | 0 | 1 | 1 | xxxxB | 1 | 1 | 1 | - |
| Lactobacillus hilgardii ATCC 8290 (525327.3) | 0 | 0 | 0 | 1 | 0 | 1 | 0 | 0 | 1 | 1 | xHHxB | 1 | 3 | 1 | Donor |
| Lactobacillus johnsonii NCC 533 (257314.6) | 0 | 0 | 1 | 0 | 0 | 1 | 0 | 0 | 1 | 1 | xPHxB | 2 | 2 | 1 | Mixed |
| Lactobacillus paracasei subsp. paracasei 8700:2 (537973.8) | 0 | 1 | 1 | 1 | 0 | 1 | 0 | 0 | 1 | 1 | HBHxB | 2 | 4 | 2 | Donor |
| Lactobacillus paracasei subsp. paracasei ATCC 25302 (525337.3) | 0 | 1 | 1 | 1 | 0 | 1 | 0 | 0 | 1 | 1 | HBHxB | 2 | 4 | 2 | Donor |
| Lactobacillus plantarum 16 (1327988.3) | 0 | 0 | 0 | 0 | 1 | 0 | 1 | 0 | 1 | 1 | xxPPB | 3 | 1 | 1 | Acceptor |
| Lactobacillus plantarum subsp. plantarum ATCC 14917 (525338.3) | 0 | 0 | 0 | 0 | 1 | 0 | 0 | 0 | 1 | 1 | xxPxB | 2 | 1 | 1 | Acceptor |
| Lactobacillus plantarum WCFS1 (220668.9) | 0 | 0 | 0 | 0 | 1 | 0 | 1 | 0 | 1 | 1 | xxPPB | 3 | 1 | 1 | Acceptor |
| Lactobacillus reuteri CF48-3A (525341.3) | 0 | 0 | 0 | 0 | 0 | 0 | 0 | 0 | 1 | 1 | xxxxB | 1 | 1 | 1 | - |
| Lactobacillus reuteri DSM 20016 (557436.4) | 0 | 0 | 0 | 0 | 0 | 0 | 0 | 0 | 1 | 1 | xxxxB | 1 | 1 | 1 | - |
| Lactobacillus reuteri F275 (299033.6) | 0 | 0 | 0 | 0 | 0 | 0 | 0 | 0 | 1 | 1 | xxxxB | 1 | 1 | 1 | - |
| Lactobacillus reuteri MM2-3 (585517.3) | 0 | 0 | 0 | 0 | 0 | 0 | 0 | 0 | 1 | 1 | xxxxB | 1 | 1 | 1 | - |
| Lactobacillus reuteri MM4-1A (548485.3) | 0 | 0 | 0 | 0 | 0 | 0 | 0 | 0 | 1 | 1 | xxxxB | 1 | 1 | 1 | - |
| Lactobacillus reuteri SD2112 (491077.3) | 0 | 0 | 0 | 0 | 0 | 0 | 0 | 0 | 1 | 1 | xxxxB | 1 | 1 | 1 | - |
| Lactobacillus rhamnosus ATCC 21052 (1002365.5) | 0 | 1 | 1 | 0 | 0 | 1 | 0 | 0 | 1 | 1 | HPHxB | 2 | 3 | 1 | Mixed |
| Lactobacillus rhamnosus GG (Prj:40637) (568703.30) | 1 | 1 | 1 | 1 | 0 | 1 | 0 | 0 | 1 | 1 | BBHxB | 3 | 4 | 3 | Donor |
| Lactobacillus rhamnosus LMS2-1 (525361.3) | 0 | 1 | 1 | 0 | 0 | 1 | 0 | 0 | 1 | 1 | HPHxB | 2 | 3 | 1 | Mixed |
| Lactobacillus ruminis ATCC 25644 (525362.3) | 0 | 0 | 0 | 1 | 0 | 1 | 0 | 0 | 1 | 1 | xHHxB | 1 | 3 | 1 | Donor |
| Lactobacillus sakei subsp. sakei 23K (314315.12) | 0 | 0 | 0 | 0 | 1 | 0 | 1 | 0 | 1 | 1 | xxPPB | 3 | 1 | 1 | Acceptor |
| Lactobacillus salivarius ATCC 11741 (525364.3) | 0 | 0 | 0 | 1 | 0 | 1 | 1 | 0 | 1 | 1 | xHHPB | 2 | 3 | 1 | Mixed |
| Lactobacillus salivarius UCC118 (362948.14) | 0 | 0 | 0 | 1 | 0 | 1 | 0 | 1 | 1 | 1 | xHHHB | 2 | 4 | 2 | Donor |
| Lactobacillus sp. 7_1_47FAA (665945.3) | 0 | 0 | 1 | 0 | 0 | 0 | 0 | 0 | 0 | 1 | xPxxH | 1 | 1 | 0 | Mixed |
| Lactobacillus ultunensis DSM 16047 (525365.3) | 0 | 0 | 0 | 0 | 1 | 0 | 0 | 0 | 1 | 1 | xxPxB | 2 | 1 | 1 | Acceptor |
| Leuconostoc mesenteroides subsp. cremoris ATCC 19254 (586220...) | 0 | 0 | 0 | 0 | 0 | 0 | 0 | 0 | 1 | 1 | xxxxB | 1 | 1 | 1 | - |
| Listeria grayi DSM 20601 (525367.9) | 0 | 0 | 0 | 0 | 0 | 1 | 0 | 0 | 1 | 1 | xxHxB | 1 | 2 | 1 | Donor |

| Organism | | | | | | | | | | | Pattern | | | | |
|---|---|---|---|---|---|---|---|---|---|---|---|---|---|---|---|
| Listeria innocua ATCC 33091 (1002366.3) | 0 | 0 | 0 | 0 | 1 | 1 | 0 | 0 | 1 | 1 | xxxxx | 0 | 0 | 0 | - |
| Megamonas funiformis YIT 11815 (742816.3) | 1 | 0 | 0 | 1 | 1 | 1 | 1 | 0 | 1 | 1 | PHBPB | 4 | 3 | 2 | Mixed |
| Megamonas hypermegale ART12/1 (657316.3) | 1 | 0 | 0 | 1 | 1 | 1 | 1 | 0 | 1 | 1 | PHBPB | 4 | 3 | 2 | Mixed |
| Methanobrevibacter smithii ATCC 35061 (420247.6) | 0 | 0 | 0 | 0 | 0 | 0 | 0 | 0 | 0 | 0 | xxxxx | 0 | 0 | 0 | - |
| Methanosphaera stadtmanae DSM 3091 (339860.6) | 0 | 0 | 0 | 0 | 0 | 0 | 0 | 0 | 0 | 0 | xxxxx | 0 | 0 | 0 | - |
| Mitsuokella multacida DSM 20544 (500635.8) | 0 | 0 | 0 | 1 | 0 | 1 | 1 | 0 | 1 | 1 | xHHxB | 1 | 3 | 1 | Donor |
| Mollicutes bacterium D7 (556270.3) | 0 | 0 | 0 | 1 | 1 | 1 | 0 | 1 | 1 | 1 | xHBHB | 2 | 4 | 2 | Donor |
| Odoribacter laneus YIT 12061 (742817.3) | 0 | 1 | 0 | 1 | 0 | 1 | 1 | 1 | 1 | 1 | HHHBB | 2 | 5 | 2 | Donor |
| Oxalobacter formigenes HOxBLS (556268.6) | 0 | 0 | 0 | 0 | 0 | 0 | 0 | 0 | 0 | 0 | xxxxx | 0 | 0 | 0 | - |
| Oxalobacter formigenes OXCC13 (556269.4) | 0 | 0 | 0 | 0 | 0 | 0 | 0 | 0 | 0 | 0 | xxxxx | 0 | 0 | 0 | - |
| Paenibacillus sp. HGF5 (908341.3) | 1 | 0 | 0 | 1 | 1 | 1 | 0 | 1 | 1 | 1 | BHBHB | 3 | 5 | 3 | Donor |
| Paenibacillus sp. HGF7 (944559.3) | 1 | 0 | 0 | 1 | 1 | 1 | 0 | 0 | 1 | 1 | BHBxB | 3 | 4 | 3 | Donor |
| Paenibacillus sp. HGH0039 (1078505.3) | 1 | 0 | 0 | 1 | 1 | 1 | 0 | 1 | 1 | 1 | BHBHB | 3 | 5 | 3 | Donor |
| Parabacteroides distasonis ATCC 8503 (435591.13) | 0 | 0 | 0 | 0 | 1 | 1 | 1 | 1 | 1 | 1 | HHHBB | 2 | 5 | 2 | Donor |
| Parabacteroides johnsonii DSM 18315 (537006.5) | 0 | 0 | 0 | 0 | 1 | 1 | 1 | 1 | 1 | 1 | HHHBB | 2 | 5 | 2 | Donor |
| Parabacteroides merdae ATCC 43184 (411477.4) | 0 | 0 | 0 | 0 | 1 | 0 | 1 | 1 | 1 | 1 | HHHBB | 1 | 5 | 1 | Donor |
| Parabacteroides sp. D13 (563193.3) | 0 | 0 | 0 | 0 | 1 | 1 | 1 | 1 | 1 | 1 | HHHBB | 2 | 5 | 2 | Donor |
| Parabacteroides sp. D25 (658661.3) | 0 | 0 | 0 | 0 | 1 | 1 | 1 | 1 | 1 | 1 | HHHBB | 2 | 5 | 2 | Donor |
| Paraprevotella clara YIT 11840 (762968.3) | 1 | 1 | 0 | 0 | 0 | 0 | 0 | 0 | 1 | 1 | BxxxB | 2 | 2 | 2 | - |
| Paraprevotella xylaniphila YIT 11841 (762982.3) | 1 | 1 | 0 | 0 | 0 | 0 | 0 | 0 | 1 | 1 | BxxxB | 2 | 2 | 2 | - |
| Parvimonas micra ATCC 33270 (411465.10) | 0 | 0 | 0 | 0 | 0 | 0 | 0 | 0 | 0 | 0 | xxxxx | 0 | 0 | 0 | - |
| Pediococcus acidilactici 7_4 (563194.3) | 1 | 1 | 1 | 0 | 1 | 1 | 0 | 0 | 1 | 1 | BPHPB | 4 | 3 | 2 | Mixed |
| Pediococcus acidilactici DSM 20284 (862514.3) | 0 | 1 | 1 | 0 | 1 | 1 | 0 | 0 | 1 | 1 | HPHPB | 3 | 4 | 2 | Mixed |
| Phascolarctobacterium sp. YIT 12067 (626939.3) | 0 | 0 | 0 | 0 | 0 | 0 | 0 | 0 | 0 | 0 | xxxxx | 0 | 0 | 0 | - |
| Prevotella copri DSM 18205 (537011.5) | 0 | 0 | 0 | 1 | 0 | 1 | 0 | 0 | 1 | 1 | xHHxB | 1 | 3 | 1 | Donor |
| Prevotella oralis HGA0225 (1203550.3) | 0 | 1 | 0 | 1 | 1 | 1 | 0 | 0 | 1 | 1 | HHHBB | 2 | 5 | 2 | Donor |
| Prevotella salivae DSM 15606 (888832.3) | 0 | 1 | 0 | 1 | 1 | 1 | 0 | 0 | 1 | 1 | HHHBB | 2 | 5 | 2 | Donor |
| Propionibacterium sp. 5_U_42AFAA (450748.3) | 1 | 1 | 0 | 1 | 1 | 1 | 1 | 0 | 1 | 1 | BHBBB | 4 | 5 | 4 | Donor |
| Propionibacterium sp. HGH0353 (1203571.3) | 1 | 0 | 0 | 1 | 0 | 1 | 0 | 0 | 1 | 1 | xHHPB | 2 | 3 | 1 | Mixed |
| Proteus mirabilis WGLW6 (1125694.3) | 0 | 0 | 1 | 0 | 1 | 0 | 1 | 0 | 1 | 0 | xPPPP | 4 | 0 | 0 | Acceptor |
| Proteus penneri ATCC 35198 (471881.3) | 0 | 0 | 1 | 0 | 1 | 0 | 1 | 0 | 1 | 0 | xPPPP | 4 | 0 | 0 | Acceptor |
| Providencia alcalifaciens DSM 30120 (520999.6) | 0 | 0 | 0 | 0 | 1 | 0 | 0 | 0 | 1 | 0 | xxPxx | 1 | 0 | 0 | Acceptor |
| Providencia rettgeri DSM 1131 (521000.6) | 0 | 0 | 1 | 1 | 1 | 0 | 1 | 0 | 1 | 1 | xBBPP | 4 | 2 | 2 | Acceptor |
| Providencia rustigianii DSM 4541 (500637.6) | 0 | 0 | 1 | 0 | 1 | 0 | 1 | 0 | 1 | 0 | xPPxP | 3 | 0 | 0 | Acceptor |
| Providencia stuartii ATCC 25827 (471874.6) | 0 | 0 | 1 | 1 | 1 | 0 | 1 | 0 | 1 | 1 | xBBxP | 3 | 2 | 2 | Acceptor |
| Pseudomonas sp. 2_1_26 (665948.3) | 0 | 0 | 0 | 0 | 1 | 0 | 0 | 0 | 0 | 0 | xxxxx | 0 | 0 | 0 | - |
| Ralstonia sp. 5_7_47FAA (658664.3) | 0 | 0 | 0 | 0 | 1 | 0 | 0 | 0 | 0 | 0 | xxxxx | 0 | 0 | 0 | - |
| Roseburia intestinalis L1-82 (536231.5) | 0 | 1 | 0 | 1 | 1 | 1 | 0 | 0 | 1 | 1 | HHBxB | 2 | 4 | 2 | Donor |
| Roseburia intestinalis M50/1 (657315.3) | 0 | 1 | 0 | 1 | 1 | 1 | 0 | 0 | 1 | 1 | HHBxB | 2 | 4 | 2 | Donor |
| Roseburia intestinalis XB6B4 (718255.3) | 0 | 1 | 0 | 1 | 1 | 1 | 0 | 0 | 1 | 1 | HHBxB | 2 | 4 | 2 | Donor |
| Roseburia inulinivorans DSM 16841 (622312.4) | 1 | 0 | 0 | 1 | 1 | 1 | 1 | 0 | 1 | 1 | BHBPB | 4 | 4 | 3 | Mixed |
| Ruminococcaceae bacterium D16 (552398.3) | 1 | 0 | 0 | 1 | 0 | 1 | 1 | 0 | 1 | 1 | xHHPB | 2 | 3 | 1 | Mixed |
| Ruminococcus bromii L2-63 (657321.5) | 0 | 0 | 0 | 1 | 0 | 1 | 0 | 0 | 1 | 1 | xHHxP | 1 | 2 | 0 | Mixed |
| Ruminococcus gnavus ATCC 29149 (411470.6) | 1 | 1 | 0 | 1 | 0 | 1 | 1 | 0 | 1 | 1 | BxPBB | 4 | 3 | 3 | Acceptor |
| Ruminococcus lactaris ATCC 29176 (471875.6) | 0 | 1 | 0 | 1 | 1 | 1 | 0 | 0 | 1 | 1 | HHHBB | 2 | 5 | 2 | Donor |
| Ruminococcus obeum A2-162 (657314.3) | 1 | 0 | 0 | 1 | 0 | 1 | 1 | 0 | 1 | 1 | PxPxB | 3 | 1 | 1 | Acceptor |
| Ruminococcus obeum ATCC 29174 (411459.7) | 1 | 0 | 0 | 1 | 0 | 1 | 0 | 0 | 1 | 1 | BxxxB | 2 | 2 | 2 | - |
| Ruminococcus sp. 18P13 (213810.4) | 0 | 0 | 0 | 0 | 1 | 0 | 0 | 0 | 0 | 1 | xxHxH | 0 | 2 | 0 | Donor |
| Ruminococcus sp. 5_1_39BFAA (457412.4) | 1 | 1 | 0 | 1 | 0 | 1 | 0 | 0 | 1 | 1 | BxxPB | 3 | 2 | 2 | Acceptor |
| Ruminococcus sp. SR1/5 (657323.3) | 1 | 1 | 0 | 1 | 0 | 1 | 0 | 0 | 1 | 1 | BxPxB | 3 | 2 | 2 | Acceptor |
| Ruminococcus torques ATCC 27756 (411460.6) | 0 | 1 | 1 | 1 | 1 | 0 | 1 | 0 | 1 | 1 | HBHHB | 2 | 5 | 2 | Donor |
| Ruminococcus torques L2-14 (657313.3) | 1 | 0 | 1 | 0 | 1 | 0 | 1 | 0 | 1 | 1 | BxPPB | 4 | 2 | 2 | Acceptor |
| Salmonella enterica subsp. enterica serovar Typhimurium str. LT2 (...) | 0 | 0 | 1 | 0 | 1 | 1 | 1 | 1 | 1 | 1 | PxxBB | 3 | 2 | 2 | Acceptor |
| Staphylococcus sp. HGB0015 (1078083.3) | 0 | 0 | 1 | 0 | 1 | 0 | 1 | 1 | 1 | 1 | xPHBB | 3 | 3 | 2 | Mixed |
| Streptococcus anginosus 1_2_62CV (742820.3) | 0 | 0 | 0 | 0 | 1 | 1 | 0 | 1 | 1 | 1 | xxHPB | 2 | 2 | 1 | Mixed |

| Organism | | | | | | | | | | | Code | | | | Type |
|---|---|---|---|---|---|---|---|---|---|---|---|---|---|---|---|
| Streptococcus equinus ATCC 9812 (525379.3) | 0 | 0 | 0 | 0 | 0 | 0 | 0 | 0 | 1 | 0 | xxxxP | 1 | 0 | 0 | Acceptor |
| Streptococcus infantarius subsp. infantarius ATCC BAA-102 (47187 | 0 | 0 | 0 | 0 | 0 | 0 | 0 | 0 | 1 | 1 | xxxxB | 1 | 1 | 1 | - |
| Streptococcus sp. 2_1_36FAA (469609.3) | 0 | 1 | 0 | 1 | 0 | 1 | 1 | 0 | 1 | 1 | HHHPB | 2 | 4 | 1 | Mixed |
| Streptococcus sp. HPH0090 (1203590.3) | 0 | 1 | 0 | 1 | 0 | 1 | 1 | 1 | 1 | 1 | HHHBB | 2 | 5 | 2 | Donor |
| Streptomyces sp. HGB0020 (1078086.3) | 1 | 1 | 0 | 1 | 1 | 1 | 1 | 1 | 1 | 1 | BHBBB | 4 | 5 | 4 | Donor |
| Streptomyces sp. HPH0547 (1203592.3) | 1 | 1 | 0 | 1 | 1 | 1 | 1 | 1 | 1 | 1 | BHBBB | 4 | 5 | 4 | Donor |
| Subdoligranulum sp. 4_3_54A2FAA (665956.3) | 0 | 1 | 0 | 1 | 0 | 1 | 1 | 1 | 1 | 1 | HHHBB | 2 | 5 | 2 | Donor |
| Subdoligranulum variabile DSM 15176 (411471.5) | 0 | 0 | 0 | 1 | 1 | 1 | 0 | 0 | 1 | 1 | xHBxB | 2 | 3 | 2 | Donor |
| Succinatimonas hippei YIT 12066 (762983.3) | 0 | 0 | 0 | 0 | 0 | 0 | 0 | 0 | 1 | 0 | xxxxP | 1 | 0 | 0 | Acceptor |
| Sutterella wadsworthensis 3_1_45B (742821.3) | 0 | 0 | 0 | 0 | 0 | 0 | 0 | 0 | 0 | 0 | xxxxx | 0 | 0 | 0 | - |
| Sutterella wadsworthensis HGA0223 (1203554.3) | 0 | 0 | 0 | 0 | 0 | 0 | 0 | 0 | 0 | 0 | xxxxx | 0 | 0 | 0 | - |
| Tannerella sp. 6_1_58FAA_CT1 (665949.3) | 0 | 1 | 0 | 1 | 0 | 1 | 0 | 0 | 1 | 1 | HHHxB | 1 | 4 | 1 | Donor |
| Turicibacter sp. HGF1 (910310.3) | 0 | 1 | 0 | 1 | 1 | 1 | 1 | 0 | 1 | 1 | HHBPB | 3 | 4 | 2 | Mixed |
| Turicibacter sp. PC909 (702450.3) | 0 | 1 | 0 | 1 | 1 | 1 | 1 | 0 | 1 | 1 | HHBPB | 3 | 4 | 2 | Mixed |
| Veillonella sp. 3_1_44 (457416.3) | 0 | 0 | 0 | 1 | 0 | 1 | 0 | 0 | 1 | 0 | xHHxP | 1 | 2 | 0 | Mixed |
| Veillonella sp. 6_1_27 (450749.3) | 0 | 0 | 0 | 1 | 0 | 1 | 0 | 0 | 1 | 0 | xHHxP | 1 | 2 | 0 | Mixed |
| Weissella paramesenteroides ATCC 33313 (585506.3) | 0 | 0 | 0 | 0 | 1 | 0 | 0 | 0 | 1 | 1 | xxPxB | 2 | 1 | 1 | Acceptor |
| Yokenella regensburgei ATCC 43003 (1002368.3) | 1 | 0 | 1 | 1 | 1 | 1 | 1 | 0 | 1 | 1 | PBBPB | 5 | 3 | 3 | Acceptor |

**Table S12.** Properties of known mucin glycans found in human intestine. [1]Numbers of glycan structures in agreement with the Supplementary Figure S9. [2]Bonds found in glycan structures: "+", presence of this type of bond; "-", absence of this type of bond. [3]Specific features of the glycan structures: aGal, contain alpha-bound galactose; aGalNAc, contain alpha-bound N-acetylgalactosamine; aGlcNAc, contain alpha-bound N-acetylglucosamine;

| Mucin No[1] | Bonds[2] | | | | | | | | | | | | | | | | Specific groups[3] | No of genomes[4] |
|---|---|---|---|---|---|---|---|---|---|---|---|---|---|---|---|---|---|---|
| | Fuc-alpha1,2-Gal | Fuc-alpha1,4-GlcNAc | Gal-alpha1,3-Gal | Gal-beta1,3-GalNAc | Gal-beta1,3-GlcNAc | Gal-beta1,4-GlcNAc | GalNAc-alpha1,3-Gal | GalNAc-alpha1,6-Gal | GlcNAc-alpha1,4-Gal | GlcNAc-beta1,3-Gal | GlcNAc-beta1,3-GalNAc | GlcNAc-beta1,6-Gal | GlcNAc-beta1,6-GalNAc | Neu5Ac-alpha2,3-Gal | Neu5Ac-alpha2,6-Gal | Neu5Ac-alpha2,6-GalNAc | | |
| 1 | - | - | - | - | - | - | - | - | - | - | - | - | - | - | - | + | Sial | 112 |
| 2 | - | - | - | - | - | - | - | - | - | - | + | - | - | - | - | + | Sial | 96 |
| 3 | - | - | - | - | + | - | - | - | - | - | + | - | - | - | - | + | Sial | 95 |
| 4 | - | - | - | - | - | - | - | - | - | + | + | - | - | - | - | + | Sial | 95 |
| 5 | - | - | - | - | - | - | - | - | - | + | + | + | - | - | - | + | Sial | 95 |
| 6 | + | - | - | - | - | + | + | - | - | + | + | - | - | - | - | + | aGalNAc, Fuc, Sial | 9 |
| 7 | + | - | - | + | + | + | + | - | - | + | + | - | - | + | - | + | aGalNAc, Fuc, Sial | 9 |
| 8 | - | - | - | + | + | + | + | - | - | - | + | - | - | + | - | + | aGalNAc, Sial | 9 |
| 9 | + | - | - | + | + | + | + | - | - | - | + | - | - | + | - | + | aGalNAc, Fuc, Sial | 9 |
| 10 | - | - | - | - | + | - | - | - | - | - | + | - | - | - | - | + | Sial | 95 |
| 11 | - | - | - | - | - | - | - | - | - | + | + | - | - | - | - | - | | 217 |
| 12 | - | - | - | - | + | - | - | - | - | - | + | - | - | - | - | - | | 189 |
| 13 | - | - | - | - | - | - | - | - | - | + | + | - | - | - | - | - | | 189 |
| 14 | - | - | - | - | - | - | - | - | - | + | + | - | - | - | - | - | | 189 |
| 15 | + | - | - | - | - | - | - | - | - | + | + | - | - | - | - | - | aGalNAc, Fuc | 11 |
| 16 | - | - | - | - | + | - | - | - | - | + | + | - | - | - | - | - | aGalNAc | 12 |
| 17 | + | - | - | - | - | - | - | - | - | + | + | + | - | - | - | - | aGalNAc | 11 |
| 18 | - | - | - | - | - | - | - | - | - | + | + | - | - | - | - | - | | 189 |
| 19 | + | - | - | - | - | + | - | - | - | - | + | - | - | - | - | - | aGalNAc | 12 |
| 20 | + | - | - | - | - | + | - | - | - | + | - | - | - | - | - | - | aGalNAc, Fuc | 11 |
| 21 | - | - | - | - | + | + | - | - | - | - | + | - | - | - | - | - | aGalNAc | 12 |
| 22 | + | - | - | + | - | - | - | - | - | - | - | - | + | - | - | - | Fuc | 124 |
| 23 | + | - | - | + | - | - | - | - | - | - | - | + | - | - | - | - | Fuc | 107 |
| 24 | + | - | - | - | + | - | - | - | - | - | - | - | + | - | - | - | aGalNAc, Fuc | 11 |
| 25 | + | - | - | + | - | - | - | - | - | - | - | - | + | - | - | - | Fuc | 107 |
| 26 | + | - | - | - | + | - | - | - | - | - | - | - | + | - | - | - | Fuc | 107 |
| 27 | + | - | - | + | - | - | - | - | - | - | - | - | + | - | - | - | Fuc | 107 |
| 28 | + | - | - | + | - | - | - | - | - | - | - | - | + | - | - | - | Fuc | 107 |
| 29 | + | - | - | - | - | + | - | - | - | - | - | - | + | - | - | - | aGalNAc, Fuc | 11 |
| 30 | + | - | - | + | - | - | - | - | - | - | - | - | + | - | - | - | Fuc | 107 |
| 31 | - | - | + | + | - | - | - | - | - | + | - | - | + | - | - | - | aGal, Fuc | 96 |
| 32 | + | - | - | + | - | - | - | - | - | + | - | - | + | - | - | - | Fuc | 107 |
| 33 | + | + | - | + | - | - | - | - | - | - | - | - | + | - | - | - | Fuc | 107 |
| 34 | + | - | + | + | - | - | - | - | - | - | - | - | + | - | - | - | aGal, Fuc | 96 |
| 35 | + | - | + | + | - | - | - | - | - | - | - | - | + | - | - | - | aGal, Fuc | 96 |
| 36 | + | - | + | + | - | + | - | - | - | - | - | - | + | - | - | - | aGal, Fuc | 96 |
| 37 | + | - | - | + | - | + | - | - | - | - | - | - | + | - | - | - | aGalNAc, Fuc | 11 |
| 38 | + | - | - | + | - | + | - | - | - | - | - | - | + | - | - | - | aGalNAc, Fuc | 11 |
| 39 | - | - | - | - | - | - | - | - | - | + | - | - | + | - | - | - | | 189 |
| 40 | + | - | - | + | - | + | - | - | - | + | - | - | + | - | - | - | Fuc | 107 |
| 41 | + | - | - | + | - | + | - | - | - | + | - | - | + | - | - | - | aGalNAc, Fuc | 11 |
| 42 | - | - | + | + | - | + | - | - | - | - | - | - | + | - | - | - | aGal, Fuc | 96 |
| 43 | - | - | - | - | - | - | - | - | - | + | - | - | + | - | - | - | | 285 |
| 44 | - | - | - | + | - | - | - | - | - | - | - | + | - | - | - | - | | 189 |
| 45 | - | - | - | + | + | - | - | - | - | + | - | - | - | - | - | - | | 189 |
| 46 | - | - | - | + | - | - | - | - | - | - | - | + | - | - | - | - | | 189 |
| 47 | + | - | - | - | + | - | - | - | + | - | - | - | - | - | - | - | aGlcNAc, Fuc | 55 |
| 48 | - | - | - | + | - | - | - | - | - | - | - | - | - | - | - | + | Sial | 103 |
| 49 | - | - | - | + | - | - | - | - | - | - | - | - | - | + | - | - | Sial | 103 |
| 50 | + | - | - | + | - | - | - | - | - | - | - | - | - | - | - | + | Fuc, Sial | 87 |
| 51 | - | - | - | + | - | - | - | - | - | - | - | + | - | - | + | - | Sial | 95 |
| 52 | - | - | - | + | - | - | - | - | - | - | - | - | - | - | + | + | Sial | 103 |
| 53 | - | - | - | + | - | - | - | - | - | - | - | + | - | + | - | - | Sial | 95 |
| 54 | - | - | - | + | - | - | - | - | - | - | - | + | - | + | - | - | Sial | 95 |
| 55 | + | - | - | + | - | + | - | - | - | + | - | + | - | - | - | - | Fuc | 107 |
| 56 | + | + | - | + | - | + | - | - | - | + | - | + | - | - | - | - | aGal, Fuc | 96 |

**Table S13.** Repertoire of GHs and hydrolysed mucin glycans in the analysed HGM genomes. [1]Numbers of glycan structures correspond to that in Supplementary Figure S9.

| Organism (PubSEED ID) | Glycosyl Hydrolases | | | | | | | Hydrolyzed glycans | | Comments |
|---|---|---|---|---|---|---|---|---|---|---|
| | Alpha-N-acetylgalactosaminidases | Alpha-galactosidases | Alpha-L-fucosidases | Alpha-N-acetylglucosaminidase | Beta-galactosidases | Beta-hexosaminidases | Sialidases | Number | Glycans[1] | |
| Abiotrophia defectiva ATCC 49176 (592010.4) | 0 | 1 | 1 | 0 | 1 | 1 | 1 | 40 | 1, 2, 3, 4, 5, 10, 11, 12, 13, 14, 18, 22, 23, 25, 26, 27, 28, 30, 31, 32, 33, 34, 35, 36, 39, 40, 42, 43, 44, 45, 46, 48, 49, 50, 51, 52, 53, 54, 55, 56 | Glycans with alpha-GalNAc or alpha-GlcNAc cannot be hydrolyzed. |
| Acidaminococcus sp. D21 (563191.3) | 0 | 0 | 0 | 0 | 0 | 0 | 0 | 0 | - | No glycans are hydrolyzed. |
| Acidaminococcus sp. HPA0509 (1203555.3) | 0 | 0 | 0 | 0 | 0 | 0 | 0 | 0 | - | No glycans are hydrolyzed. |
| Acinetobacter junii SH205 (575587.3) | 0 | 0 | 0 | 0 | 0 | 0 | 0 | 0 | - | No glycans are hydrolyzed. |
| Actinomyces odontolyticus ATCC 17982 (411466.7) | 0 | 1 | 1 | 0 | 0 | 0 | 1 | 1 | 1 | Only Syalyl-Tn-antigen can be hydrolyzed. |
| Actinomyces sp. HPA0247 (1203556.3) | 0 | 0 | 1 | 0 | 1 | 1 | 1 | 34 | 1, 2, 3, 4, 5, 10, 11, 12, 13, 14, 18, 22, 23, 25, 26, 27, 28, 30, 32, 33, 39, 40, 43, 44, 45, 46, 48, 49, 50, 51, 52, 53, 54, 55 | Glycans with alpha-Gal, alpha-GalNAc, or alpha-GlcNAc cannot be hydrolyzed. |
| Akkermansia muciniphila ATCC BAA-835 (349741.6) | 0 | 1 | 1 | 1 | 1 | 1 | 1 | 41 | 1, 2, 3, 4, 5, 10, 11, 12, 13, 14, 18, 22, 23, 25, 26, 27, 28, 30, 31, 32, 33, 34, 35, 36, 39, 40, 42, 43, 44, 45, 46, 47, 48, 49, 50, 51, 52, 53, 54, 55, 56 | Glycans with alpha-GalNAc cannot be hydrolyzed. |
| Alistipes indistinctus YIT 12060 (742725.3) | 0 | 1 | 1 | 0 | 1 | 1 | 0 | 27 | 11, 12, 13, 14, 18, 22, 23, 25, 26, 27, 28, 30, 31, 32, 33, 34, 35, 36, 39, 40, 42, 43, 44, 45, 46, 55, 56 | Glycans with alpha-Gal or alpha-GalNAc as well syalated glycans cannot be hydrolyzed. |
| Alistipes shahii WAL 8301 (717959.3) | 0 | 1 | 1 | 1 | 1 | 1 | 1 | 41 | 1, 2, 3, 4, 5, 10, 11, 12, 13, 14, 18, 22, 23, 25, 26, 27, 28, 30, 31, 32, 33, 34, 35, 36, 39, 40, 42, 43, 44, 45, 46, 47, 48, 49, 50, 51, 52, 53, 54, 55, 56 | Glycans with alpha-GalNAc cannot be hydrolyzed. |
| Anaerobaculum hydrogeniformans ATCC BAA-1850 | 0 | 0 | 0 | 0 | 0 | 0 | 0 | 0 | - | No glycans are hydrolyzed. |
| Anaerococcus hydrogenalis DSM 7454 (561177.4) | 0 | 0 | 0 | 0 | 0 | 0 | 0 | 0 | - | No glycans are hydrolyzed. |
| Anaerofustis stercorihominis DSM 17244 (445971.6) | 0 | 0 | 0 | 0 | 0 | 0 | 0 | 0 | - | No glycans are hydrolyzed. |
| Anaerostipes caccae DSM 14662 (411490.6) | 0 | 1 | 0 | 0 | 0 | 1 | 0 | 1 | 11 | Only Core 3 structure can be hydrolyzed. |
| Anaerostipes hadrus DSM 3319 (649757.3) | 0 | 0 | 0 | 0 | 1 | 1 | 0 | 9 | 11, 12, 13, 14, 39, 43, 44, 45, 46 | Only ploylactosamine structures can be hydrolyzed. |
| Anaerostipes sp. 3_2_56FAA (665937.3) | 0 | 1 | 0 | 0 | 0 | 1 | 0 | 1 | 11 | Only Core 3 structure can be hydrolyzed. |
| Anaerotruncus colihominis DSM 17241 (445972.6) | 0 | 0 | 0 | 0 | 0 | 0 | 0 | 0 | - | No glycans are hydrolyzed. |
| Bacillus smithii 7_3_47FAA (665952.3) | 0 | 1 | 0 | 0 | 0 | 1 | 0 | 1 | 11 | Only Core 3 structure can be hydrolyzed. |
| Bacillus sp. 7_6_55CFAA_CT2 (665957.3) | 0 | 0 | 0 | 0 | 0 | 0 | 0 | 0 | - | No glycans are hydrolyzed. |
| Bacillus subtilis subsp. subtilis str. 168 (224308.113) | 0 | 1 | 0 | 0 | 1 | 1 | 0 | 9 | 11, 12, 13, 14, 39, 43, 44, 45, 46 | Only ploylactosamine structures can be hydrolyzed. |
| Bacteroides caccae ATCC 43185 (411901.7) | 0 | 1 | 1 | 1 | 1 | 1 | 1 | 41 | 1, 2, 3, 4, 5, 10, 11, 12, 13, 14, 18, 22, 23, 25, 26, 27, 28, 30, 31, 32, 33, 34, 35, 36, 39, 40, 42, 43, 44, 45, 46, 47, 48, 49, 50, 51, 52, 53, 54, 55, 56 | Glycans with alpha-GalNAc cannot be hydrolyzed. |
| Bacteroides capillosus ATCC 29799 (411467.6) | 0 | 0 | 0 | 0 | 1 | 1 | 0 | 9 | 11, 12, 13, 14, 39, 43, 44, 45, 46 | Only ploylactosamine structures can be hydrolyzed. |
| Bacteroides cellulosilyticus DSM 14838 (537012.5) | 0 | 1 | 1 | 1 | 1 | 1 | 0 | 28 | 11, 12, 13, 14, 18, 22, 23, 25, 26, 27, 28, 30, 31, 32, 33, 34, 35, 36, 39, 40, 42, 43, 44, 45, 46, 47, 55, 56 | Glycans with alpha-GalNAc as well syalated glycans cannot be hydrolyzed. |
| Bacteroides clarus YIT 12056 (762984.3) | 0 | 1 | 1 | 0 | 1 | 1 | 1 | 40 | 1, 2, 3, 4, 5, 10, 11, 12, 13, 14, 18, 22, 23, 25, 26, 27, 28, 30, 31, 32, 33, 34, 35, 36, 39, 40, 42, 43, 44, 45, 46, 48, 49, 50, 51, 52, 53, 54, 55, 56 | Glycans with alpha-GalNAc or alpha-GlcNAc cannot be hydrolyzed. |
| Bacteroides coprocola DSM 17136 (470145.6) | 0 | 1 | 1 | 0 | 1 | 1 | 1 | 40 | 1, 2, 3, 4, 5, 10, 11, 12, 13, 14, 18, 22, 23, 25, 26, 27, 28, 30, 31, 32, 33, 34, 35, 36, 39, 40, 42, 43, 44, 45, 46, 48, 49, 50, 51, 52, 53, 54, 55, 56 | Glycans with alpha-GalNAc or alpha-GlcNAc cannot be hydrolyzed. |
| Bacteroides coprophilus DSM 18228 (547042.5) | 0 | 1 | 1 | 1 | 1 | 1 | 1 | 41 | 1, 2, 3, 4, 5, 10, 11, 12, 13, 14, 18, 22, 23, 25, 26, 27, 28, 30, 31, 32, 33, 34, 35, 36, 39, 40, 42, 43, 44, 45, 46, 47, 48, 49, 50, 51, 52, 53, 54, 55, 56 | Glycans with alpha-GalNAc cannot be hydrolyzed. |
| Bacteroides dorei DSM 17855 (483217.6) | 0 | 1 | 1 | 1 | 1 | 1 | 1 | 41 | 1, 2, 3, 4, 5, 10, 11, 12, 13, 14, 18, 22, 23, 25, 26, 27, 28, 30, 31, 32, 33, 34, 35, 36, 39, 40, 42, 43, 44, 45, 46, 47, 48, 49, 50, 51, 52, 53, 54, 55, 56 | Glycans with alpha-GalNAc cannot be hydrolyzed. |
| Bacteroides eggerthii 1_2_48FAA (665953.3) | 0 | 1 | 1 | 0 | 1 | 1 | 0 | 27 | 11, 12, 13, 14, 18, 22, 23, 25, 26, 27, 28, 30, 31, 32, 33, 34, 35, 36, 39, 40, 42, 43, 44, 45, 46, 55, 56 | Glycans with alpha-Gal or alpha-GalNAc as well syalated glycans cannot be hydrolyzed. |
| Bacteroides eggerthii DSM 20697 (483216.6) | 0 | 0 | 1 | 0 | 1 | 1 | 0 | 21 | 11, 12, 13, 14, 18, 22, 23, 25, 26, 27, 28, 30, 32, 33, 39, 40, 43, 44, 45, 46, 55 | Only (fucosylated) ploylactosamine structures can be hydrolyzed. |
| Bacteroides finegoldii DSM 17565 (483215.6) | 0 | 1 | 1 | 1 | 1 | 1 | 1 | 41 | 1, 2, 3, 4, 5, 10, 11, 12, 13, 14, 18, 22, 23, 25, 26, 27, 28, 30, 31, 32, 33, 34, 35, 36, 39, 40, 42, 43, 44, 45, 46, 47, 48, 49, 50, 51, 52, 53, 54, 55, 56 | Glycans with alpha-GalNAc cannot be hydrolyzed. |
| Bacteroides fluxus YIT 12057 (763034.3) | 0 | 1 | 1 | 1 | 1 | 1 | 1 | 41 | 1, 2, 3, 4, 5, 10, 11, 12, 13, 14, 18, 22, 23, 25, 26, 27, 28, 30, 31, 32, 33, 34, 35, 36, 39, 40, 42, 43, 44, 45, 46, 47, 48, 49, 50, 51, 52, 53, 54, 55, 56 | Glycans with alpha-GalNAc cannot be hydrolyzed. |

| Organism | | | | | | | | | Glycans | Notes |
|---|---|---|---|---|---|---|---|---|---|---|
| Bacteroides fragilis 3_1_12 (457424.5) | 0 | 1 | 1 | 1 | 1 | 1 | 1 | 41 | 1, 2, 3, 4, 5, 10, 11, 12, 13, 14, 18, 22, 23, 25, 26, 27, 28, 30, 31, 32, 33, 34, 35, 36, 39, 40, 42, 43, 44, 45, 46, 47, 48, 49, 50, 51, 52, 53, 54, 55, 56 | Glycans with alpha-GalNAc cannot be hydrolyzed. |
| Bacteroides fragilis 638R (862962.3) | 0 | 1 | 1 | 1 | 1 | 1 | 1 | 41 | 1, 2, 3, 4, 5, 10, 11, 12, 13, 14, 18, 22, 23, 25, 26, 27, 28, 30, 31, 32, 33, 34, 35, 36, 39, 40, 42, 43, 44, 45, 46, 47, 48, 49, 50, 51, 52, 53, 54, 55, 56 | Glycans with alpha-GalNAc cannot be hydrolyzed. |
| Bacteroides fragilis NCTC 9343 (272559.17) | 0 | 1 | 1 | 1 | 1 | 1 | 1 | 41 | 1, 2, 3, 4, 5, 10, 11, 12, 13, 14, 18, 22, 23, 25, 26, 27, 28, 30, 31, 32, 33, 34, 35, 36, 39, 40, 42, 43, 44, 45, 46, 47, 48, 49, 50, 51, 52, 53, 54, 55, 56 | Glycans with alpha-GalNAc cannot be hydrolyzed. |
| Bacteroides fragilis YCH46 (295405.11) | 0 | 1 | 1 | 1 | 1 | 1 | 1 | 41 | 1, 2, 3, 4, 5, 10, 11, 12, 13, 14, 18, 22, 23, 25, 26, 27, 28, 30, 31, 32, 33, 34, 35, 36, 39, 40, 42, 43, 44, 45, 46, 47, 48, 49, 50, 51, 52, 53, 54, 55, 56 | Glycans with alpha-GalNAc cannot be hydrolyzed. |
| Bacteroides intestinalis DSM 17393 (471870.8) | 0 | 1 | 1 | 1 | 1 | 1 | 1 | 41 | 1, 2, 3, 4, 5, 10, 11, 12, 13, 14, 18, 22, 23, 25, 26, 27, 28, 30, 31, 32, 33, 34, 35, 36, 39, 40, 42, 43, 44, 45, 46, 47, 48, 49, 50, 51, 52, 53, 54, 55, 56 | Glycans with alpha-GalNAc cannot be hydrolyzed. |
| Bacteroides ovatus 3_8_47FAA (665954.3) | 0 | 1 | 1 | 1 | 1 | 1 | 1 | 41 | 1, 2, 3, 4, 5, 10, 11, 12, 13, 14, 18, 22, 23, 25, 26, 27, 28, 30, 31, 32, 33, 34, 35, 36, 39, 40, 42, 43, 44, 45, 46, 47, 48, 49, 50, 51, 52, 53, 54, 55, 56 | Glycans with alpha-GalNAc cannot be hydrolyzed. |
| Bacteroides ovatus ATCC 8483 (411476.11) | 0 | 1 | 1 | 1 | 1 | 1 | 1 | 41 | 1, 2, 3, 4, 5, 10, 11, 12, 13, 14, 18, 22, 23, 25, 26, 27, 28, 30, 31, 32, 33, 34, 35, 36, 39, 40, 42, 43, 44, 45, 46, 47, 48, 49, 50, 51, 52, 53, 54, 55, 56 | Glycans with alpha-GalNAc cannot be hydrolyzed. |
| Bacteroides ovatus SD CC 2a (702444.3) | 0 | 1 | 1 | 1 | 1 | 1 | 1 | 41 | 1, 2, 3, 4, 5, 10, 11, 12, 13, 14, 18, 22, 23, 25, 26, 27, 28, 30, 31, 32, 33, 34, 35, 36, 39, 40, 42, 43, 44, 45, 46, 47, 48, 49, 50, 51, 52, 53, 54, 55, 56 | Glycans with alpha-GalNAc cannot be hydrolyzed. |
| Bacteroides ovatus SD CMC 3f (702443.3) | 1 | 1 | 1 | 1 | 1 | 1 | 1 | 56 | 1, 2, 3, 4, 5, 6, 7, 8, 9, 10, 11, 12, 13, 14, 15, 16, 17, 18, 19, 20, 21, 22, 23, 24, 25, 26, 27, 28, 29, 30, 31, 32, 33, 34, 35, 36, 37, 38, 39, 40, 41, 42, 43, 44, 45, 46, 47, 48, 49, 50, 51, 52, 53, 54, 55, 56 | All glycans can be hydrolyzed. |
| Bacteroides pectinophilus ATCC 43243 (483218.5) | 0 | 0 | 0 | 0 | 0 | 0 | 0 | 0 | . | No glycans are hydrolyzed. |
| Bacteroides plebeius DSM 17135 (484018.6) | 0 | 1 | 1 | 1 | 1 | 1 | 1 | 41 | 1, 2, 3, 4, 5, 10, 11, 12, 13, 14, 18, 22, 23, 25, 26, 27, 28, 30, 31, 32, 33, 34, 35, 36, 39, 40, 42, 43, 44, 45, 46, 47, 48, 49, 50, 51, 52, 53, 54, 55, 56 | Glycans with alpha-GalNAc cannot be hydrolyzed. |
| Bacteroides sp. 1_1_14 (469585.3) | 0 | 1 | 1 | 1 | 1 | 1 | 1 | 41 | 1, 2, 3, 4, 5, 10, 11, 12, 13, 14, 18, 22, 23, 25, 26, 27, 28, 30, 31, 32, 33, 34, 35, 36, 39, 40, 42, 43, 44, 45, 46, 47, 48, 49, 50, 51, 52, 53, 54, 55, 56 | Glycans with alpha-GalNAc cannot be hydrolyzed. |
| Bacteroides sp. 1_1_30 (457387.3) | 0 | 1 | 1 | 1 | 1 | 1 | 1 | 41 | 1, 2, 3, 4, 5, 10, 11, 12, 13, 14, 18, 22, 23, 25, 26, 27, 28, 30, 31, 32, 33, 34, 35, 36, 39, 40, 42, 43, 44, 45, 46, 47, 48, 49, 50, 51, 52, 53, 54, 55, 56 | Glycans with alpha-GalNAc cannot be hydrolyzed. |
| Bacteroides sp. 1_1_6 (469586.3) | 0 | 1 | 1 | 1 | 1 | 1 | 1 | 41 | 1, 2, 3, 4, 5, 10, 11, 12, 13, 14, 18, 22, 23, 25, 26, 27, 28, 30, 31, 32, 33, 34, 35, 36, 39, 40, 42, 43, 44, 45, 46, 47, 48, 49, 50, 51, 52, 53, 54, 55, 56 | Glycans with alpha-GalNAc cannot be hydrolyzed. |
| Bacteroides sp. 2_1_16 (469587.3) | 0 | 1 | 1 | 1 | 1 | 1 | 1 | 41 | 1, 2, 3, 4, 5, 10, 11, 12, 13, 14, 18, 22, 23, 25, 26, 27, 28, 30, 31, 32, 33, 34, 35, 36, 39, 40, 42, 43, 44, 45, 46, 47, 48, 49, 50, 51, 52, 53, 54, 55, 56 | Glycans with alpha-GalNAc cannot be hydrolyzed. |
| Bacteroides sp. 2_1_22 (469588.3) | 0 | 1 | 1 | 1 | 1 | 1 | 1 | 41 | 1, 2, 3, 4, 5, 10, 11, 12, 13, 14, 18, 22, 23, 25, 26, 27, 28, 30, 31, 32, 33, 34, 35, 36, 39, 40, 42, 43, 44, 45, 46, 47, 48, 49, 50, 51, 52, 53, 54, 55, 56 | Glycans with alpha-GalNAc cannot be hydrolyzed. |
| Bacteroides sp. 2_1_33B (469589.3) | 0 | 1 | 1 | 0 | 1 | 1 | 1 | 40 | 1, 2, 3, 4, 5, 10, 11, 12, 13, 14, 18, 22, 23, 25, 26, 27, 28, 30, 31, 32, 33, 34, 35, 36, 39, 40, 42, 43, 44, 45, 46, 48, 49, 50, 51, 52, 53, 54, 55, 56 | Glycans with alpha-GalNAc or alpha-GlcNAc cannot be hydrolyzed. |
| Bacteroides sp. 2_1_56FAA (665938.3) | 0 | 1 | 1 | 1 | 1 | 1 | 1 | 41 | 1, 2, 3, 4, 5, 10, 11, 12, 13, 14, 18, 22, 23, 25, 26, 27, 28, 30, 31, 32, 33, 34, 35, 36, 39, 40, 42, 43, 44, 45, 46, 47, 48, 49, 50, 51, 52, 53, 54, 55, 56 | Glycans with alpha-GalNAc cannot be hydrolyzed. |
| Bacteroides sp. 2_1_7 (457388.5) | 0 | 1 | 1 | 0 | 1 | 1 | 1 | 40 | 1, 2, 3, 4, 5, 10, 11, 12, 13, 14, 18, 22, 23, 25, 26, 27, 28, 30, 31, 32, 33, 34, 35, 36, 39, 40, 42, 43, 44, 45, 46, 48, 49, 50, 51, 52, 53, 54, 55, 56 | Glycans with alpha-GalNAc or alpha-GlcNAc cannot be hydrolyzed. |

| Name | 1 | 2 | 3 | 4 | 5 | 6 | 7 | # | List | Description |
|---|---|---|---|---|---|---|---|---|---|---|
| Bacteroides sp. 2_2_4 (469590.5) | 1 | 1 | 1 | 1 | 1 | 1 | 1 | 56 | 1, 2, 3, 4, 5, 6, 7, 8, 9, 10, 11, 12, 13, 14, 15, 16, 17, 18, 19, 20, 21, 22, 23, 24, 25, 26, 27, 28, 29, 30, 31, 32, 33, 34, 35, 36, 37, 38, 39, 40, 41, 42, 43, 44, 45, 46, 47, 48, 49, 50, 51, 52, 53, 54, 55, 56 | All glycans can be hydrolyzed. |
| Bacteroides sp. 20_3 (469591.4) | 0 | 1 | 1 | 0 | 1 | 1 | 1 | 40 | 1, 2, 3, 4, 5, 10, 11, 12, 13, 14, 18, 22, 23, 25, 26, 27, 28, 30, 31, 32, 33, 34, 35, 36, 39, 40, 42, 43, 44, 45, 46, 48, 49, 50, 51, 52, 53, 54, 55, 56 | Glycans with alpha-GalNAc or alpha-GlcNAc cannot be hydrolyzed. |
| Bacteroides sp. 3_1_19 (469592.4) | 0 | 1 | 1 | 0 | 1 | 1 | 1 | 40 | 1, 2, 3, 4, 5, 10, 11, 12, 13, 14, 18, 22, 23, 25, 26, 27, 28, 30, 31, 32, 33, 34, 35, 36, 39, 40, 42, 43, 44, 45, 46, 48, 49, 50, 51, 52, 53, 54, 55, 56 | Glycans with alpha-GalNAc or alpha-GlcNAc cannot be hydrolyzed. |
| Bacteroides sp. 3_1_23 (457390.3) | 1 | 1 | 1 | 1 | 1 | 1 | 1 | 56 | 1, 2, 3, 4, 5, 6, 7, 8, 9, 10, 11, 12, 13, 14, 15, 16, 17, 18, 19, 20, 21, 22, 23, 24, 25, 26, 27, 28, 29, 30, 31, 32, 33, 34, 35, 36, 37, 38, 39, 40, 41, 42, 43, 44, 45, 46, 47, 48, 49, 50, 51, 52, 53, 54, 55, 56 | All glycans can be hydrolyzed. |
| Bacteroides sp. 3_1_33FAA (457391.3) | 0 | 1 | 1 | 1 | 1 | 1 | 1 | 41 | 1, 2, 3, 4, 5, 10, 11, 12, 13, 14, 18, 22, 23, 25, 26, 27, 28, 30, 31, 32, 33, 34, 35, 36, 39, 40, 42, 43, 44, 45, 46, 47, 48, 49, 50, 51, 52, 53, 54, 55, 56 | Glycans with alpha-GalNAc cannot be hydrolyzed. |
| Bacteroides sp. 3_1_40A (469593.3) | 0 | 1 | 1 | 1 | 1 | 1 | 1 | 41 | 1, 2, 3, 4, 5, 10, 11, 12, 13, 14, 18, 22, 23, 25, 26, 27, 28, 30, 31, 32, 33, 34, 35, 36, 39, 40, 42, 43, 44, 45, 46, 47, 48, 49, 50, 51, 52, 53, 54, 55, 56 | Glycans with alpha-GalNAc cannot be hydrolyzed. |
| Bacteroides sp. 3_2_5 (457392.3) | 0 | 1 | 1 | 1 | 1 | 1 | 1 | 41 | 1, 2, 3, 4, 5, 10, 11, 12, 13, 14, 18, 22, 23, 25, 26, 27, 28, 30, 31, 32, 33, 34, 35, 36, 39, 40, 42, 43, 44, 45, 46, 47, 48, 49, 50, 51, 52, 53, 54, 55, 56 | Glycans with alpha-GalNAc cannot be hydrolyzed. |
| Bacteroides sp. 4_1_36 (457393.3) | 0 | 1 | 1 | 0 | 1 | 1 | 0 | 27 | 11, 12, 13, 14, 18, 22, 23, 25, 26, 27, 28, 30, 31, 32, 33, 34, 35, 36, 39, 40, 42, 43, 44, 45, 46, 55, 56 | Glycans with alpha-Gal or alpha-GalNAc as well syalated glycans cannot be hydrolyzed. |
| Bacteroides sp. 4_3_47FAA (457394.3) | 0 | 1 | 1 | 1 | 1 | 1 | 1 | 41 | 1, 2, 3, 4, 5, 10, 11, 12, 13, 14, 18, 22, 23, 25, 26, 27, 28, 30, 31, 32, 33, 34, 35, 36, 39, 40, 42, 43, 44, 45, 46, 47, 48, 49, 50, 51, 52, 53, 54, 55, 56 | Glycans with alpha-GalNAc cannot be hydrolyzed. |
| Bacteroides sp. 9_1_42FAA (457395.6) | 0 | 1 | 1 | 1 | 1 | 1 | 1 | 41 | 1, 2, 3, 4, 5, 10, 11, 12, 13, 14, 18, 22, 23, 25, 26, 27, 28, 30, 31, 32, 33, 34, 35, 36, 39, 40, 42, 43, 44, 45, 46, 47, 48, 49, 50, 51, 52, 53, 54, 55, 56 | Glycans with alpha-GalNAc cannot be hydrolyzed. |
| Bacteroides sp. D1 (556258.5) | 0 | 1 | 1 | 1 | 1 | 1 | 1 | 41 | 1, 2, 3, 4, 5, 10, 11, 12, 13, 14, 18, 22, 23, 25, 26, 27, 28, 30, 31, 32, 33, 34, 35, 36, 39, 40, 42, 43, 44, 45, 46, 47, 48, 49, 50, 51, 52, 53, 54, 55, 56 | Glycans with alpha-GalNAc cannot be hydrolyzed. |
| Bacteroides sp. D2 (556259.3) | 0 | 1 | 1 | 1 | 1 | 1 | 1 | 41 | 1, 2, 3, 4, 5, 10, 11, 12, 13, 14, 18, 22, 23, 25, 26, 27, 28, 30, 31, 32, 33, 34, 35, 36, 39, 40, 42, 43, 44, 45, 46, 47, 48, 49, 50, 51, 52, 53, 54, 55, 56 | Glycans with alpha-GalNAc cannot be hydrolyzed. |
| Bacteroides sp. D20 (585543.3) | 0 | 1 | 1 | 0 | 1 | 1 | 1 | 40 | 1, 2, 3, 4, 5, 10, 11, 12, 13, 14, 18, 22, 23, 25, 26, 27, 28, 30, 31, 32, 33, 34, 35, 36, 39, 40, 42, 43, 44, 45, 46, 48, 49, 50, 51, 52, 53, 54, 55, 56 | Glycans with alpha-GalNAc or alpha-GlcNAc cannot be hydrolyzed. |
| Bacteroides sp. D22 (585544.3) | 0 | 1 | 1 | 1 | 1 | 1 | 1 | 41 | 1, 2, 3, 4, 5, 10, 11, 12, 13, 14, 18, 22, 23, 25, 26, 27, 28, 30, 31, 32, 33, 34, 35, 36, 39, 40, 42, 43, 44, 45, 46, 47, 48, 49, 50, 51, 52, 53, 54, 55, 56 | Glycans with alpha-GalNAc cannot be hydrolyzed. |
| Bacteroides sp. HPS0048 (1078089.3) | 0 | 1 | 1 | 1 | 1 | 1 | 1 | 41 | 1, 2, 3, 4, 5, 10, 11, 12, 13, 14, 18, 22, 23, 25, 26, 27, 28, 30, 31, 32, 33, 34, 35, 36, 39, 40, 42, 43, 44, 45, 46, 47, 48, 49, 50, 51, 52, 53, 54, 55, 56 | Glycans with alpha-GalNAc cannot be hydrolyzed. |
| Bacteroides stercoris ATCC 43183 (449673.7) | 0 | 1 | 1 | 0 | 1 | 1 | 1 | 40 | 1, 2, 3, 4, 5, 10, 11, 12, 13, 14, 18, 22, 23, 25, 26, 27, 28, 30, 31, 32, 33, 34, 35, 36, 39, 40, 42, 43, 44, 45, 46, 48, 49, 50, 51, 52, 53, 54, 55, 56 | Glycans with alpha-GalNAc or alpha-GlcNAc cannot be hydrolyzed. |
| Bacteroides thetaiotaomicron VPI-5482 (226186.12) | 0 | 1 | 1 | 1 | 1 | 1 | 1 | 41 | 1, 2, 3, 4, 5, 10, 11, 12, 13, 14, 18, 22, 23, 25, 26, 27, 28, 30, 31, 32, 33, 34, 35, 36, 39, 40, 42, 43, 44, 45, 46, 47, 48, 49, 50, 51, 52, 53, 54, 55, 56 | Glycans with alpha-GalNAc cannot be hydrolyzed. |
| Bacteroides uniformis ATCC 8492 (411479.10) | 0 | 1 | 1 | 0 | 1 | 1 | 1 | 40 | 1, 2, 3, 4, 5, 10, 11, 12, 13, 14, 18, 22, 23, 25, 26, 27, 28, 30, 31, 32, 33, 34, 35, 36, 39, 40, 42, 43, 44, 45, 46, 48, 49, 50, 51, 52, 53, 54, 55, 56 | Glycans with alpha-GalNAc or alpha-GlcNAc cannot be hydrolyzed. |
| Bacteroides vulgatus ATCC 8482 (435590.9) | 0 | 1 | 1 | 1 | 1 | 1 | 1 | 41 | 1, 2, 3, 4, 5, 10, 11, 12, 13, 14, 18, 22, 23, 25, 26, 27, 28, 30, 31, 32, 33, 34, 35, 36, 39, 40, 42, 43, 44, 45, 46, 47, 48, 49, 50, 51, 52, 53, 54, 55, 56 | Glycans with alpha-GalNAc cannot be hydrolyzed. |
| Bacteroides vulgatus PC510 (702446.3) | 0 | 1 | 1 | 1 | 1 | 1 | 1 | 41 | 1, 2, 3, 4, 5, 10, 11, 12, 13, 14, 18, 22, 23, 25, 26, 27, 28, 30, 31, 32, 33, 34, 35, 36, 39, 40, 42, 43, 44, 45, 46, 47, 48, 49, 50, 51, 52, 53, 54, 55, 56 | Glycans with alpha-GalNAc cannot be hydrolyzed. |

| Organism | | | | | | | | # | Structures | Notes |
|---|---|---|---|---|---|---|---|---|---|---|
| Bacteroides xylanisolvens SD CC 1b (702447.3) | 0 | 1 | 1 | 1 | 1 | 1 | 1 | 41 | 1, 2, 3, 4, 5, 10, 11, 12, 13, 14, 18, 22, 23, 25, 26, 27, 28, 30, 31, 32, 33, 34, 35, 36, 39, 40, 42, 43, 44, 45, 46, 47, 48, 49, 50, 51, 52, 53, 54, 55, 56 | Glycans with alpha-GalNAc cannot be hydrolyzed. |
| Bacteroides xylanisolvens XB1A (657309.4) | 0 | 1 | 1 | 1 | 1 | 1 | 1 | 41 | 1, 2, 3, 4, 5, 10, 11, 12, 13, 14, 18, 22, 23, 25, 26, 27, 28, 30, 31, 32, 33, 34, 35, 36, 39, 40, 42, 43, 44, 45, 46, 47, 48, 49, 50, 51, 52, 53, 54, 55, 56 | Glycans with alpha-GalNAc cannot be hydrolyzed. |
| Barnesiella intestinihominis YIT 11860 (742726.3) | 0 | 1 | 1 | 1 | 1 | 1 | 1 | 41 | 1, 2, 3, 4, 5, 10, 11, 12, 13, 14, 18, 22, 23, 25, 26, 27, 28, 30, 31, 32, 33, 34, 35, 36, 39, 40, 42, 43, 44, 45, 46, 47, 48, 49, 50, 51, 52, 53, 54, 55, 56 | Glycans with alpha-GalNAc cannot be hydrolyzed. |
| Bifidobacterium adolescentis L2-32 (411481.5) | 0 | 1 | 0 | 0 | 1 | 1 | 0 | 9 | 11, 12, 13, 14, 39, 43, 44, 45, 46 | Only polyactosamine structures can be hydrolyzed. |
| Bifidobacterium angulatum DSM 20098 = JCM 7096 | 0 | 1 | 0 | 0 | 1 | 1 | 0 | 9 | 11, 12, 13, 14, 39, 43, 44, 45, 46 | Only polyactosamine structures can be hydrolyzed. |
| Bifidobacterium animalis subsp. lactis AD011 (442563... | 0 | 1 | 0 | 0 | 1 | 1 | 0 | 9 | 11, 12, 13, 14, 39, 43, 44, 45, 46 | Only polyactosamine structures can be hydrolyzed. |
| Bifidobacterium bifidum NCIMB 41171 (398513.5) | 0 | 1 | 0 | 1 | 1 | 1 | 1 | 22 | 1, 2, 3, 4, 5, 10, 11, 12, 13, 14, 18, 39, 43, 44, 45, 46, 48, 49, 51, 52, 53, 54 | Glycans with alpha-GalNAc as well syalated glycans cannot be hydrolyzed. |
| Bifidobacterium breve DSM 20213 = JCM 1192 (518... | 0 | 1 | 0 | 0 | 1 | 1 | 1 | 22 | 1, 2, 3, 4, 5, 10, 11, 12, 13, 14, 18, 39, 43, 44, 45, 46, 48, 49, 51, 52, 53, 54 | Glycans with alpha-GalNAc as well syalated glycans cannot be hydrolyzed. |
| Bifidobacterium breve HPH0326 (1203540.3) | 0 | 1 | 0 | 0 | 1 | 1 | 1 | 22 | 1, 2, 3, 4, 5, 10, 11, 12, 13, 14, 18, 39, 43, 44, 45, 46, 48, 49, 51, 52, 53, 54 | Glycans with alpha-GalNAc as well syalated glycans cannot be hydrolyzed. |
| Bifidobacterium catenulatum DSM 16992 = JCM 1194... | 0 | 1 | 0 | 0 | 1 | 1 | 0 | 9 | 11, 12, 13, 14, 39, 43, 44, 45, 46 | Only polyactosamine structures can be hydrolyzed. |
| Bifidobacterium dentium ATCC 27678 (473819.7) | 0 | 1 | 1 | 0 | 0 | 0 | 0 | 2 | 22, 43 | Only (fucosylated) Core 1 structure can be |
| Bifidobacterium gallicum DSM 20093 (561180.4) | 0 | 1 | 0 | 0 | 1 | 1 | 0 | 9 | 11, 12, 13, 14, 39, 43, 44, 45, 46 | Only polyactosamine structures can be hydrolyzed. |
| Bifidobacterium longum DJO10A (Pr;321) (205913.1... | 0 | 1 | 0 | 0 | 1 | 1 | 0 | 9 | 11, 12, 13, 14, 39, 43, 44, 45, 46 | Only polyactosamine structures can be hydrolyzed. |
| Bifidobacterium longum NCC2705 (206672.9) | 0 | 1 | 0 | 0 | 1 | 1 | 0 | 9 | 11, 12, 13, 14, 39, 43, 44, 45, 46 | Only polyactosamine structures can be hydrolyzed. |
| Bifidobacterium longum subsp. infantis 157F (565040... | 0 | 1 | 0 | 0 | 1 | 1 | 0 | 9 | 11, 12, 13, 14, 39, 43, 44, 45, 46 | Only polyactosamine structures can be hydrolyzed. |
| Bifidobacterium longum subsp. infantis ATCC 15697 | 0 | 1 | 1 | 0 | 1 | 1 | 1 | 40 | 1, 2, 3, 4, 5, 10, 11, 12, 13, 14, 18, 22, 23, 25, 26, 27, 28, 30, 31, 32, 33, 34, 35, 36, 39, 40, 42, 43, 44, 45, 46, 48, 49, 50, 51, 52, 53, 54, 55, 56 | Glycans with alpha-GalNAc or alpha-GlcNAc cannot be hydrolyzed. |
| Bifidobacterium longum subsp. infantis ATCC 55813 | 0 | 1 | 0 | 0 | 1 | 1 | 0 | 9 | 11, 12, 13, 14, 39, 43, 44, 45, 46 | Only polyactosamine structures can be hydrolyzed. |
| Bifidobacterium longum subsp. infantis CCUG 52486 | 0 | 1 | 0 | 0 | 1 | 1 | 0 | 9 | 11, 12, 13, 14, 39, 43, 44, 45, 46 | Only polyactosamine structures can be hydrolyzed. |
| Bifidobacterium longum subsp. longum 2-2B (116174... | 0 | 1 | 0 | 0 | 1 | 1 | 0 | 9 | 11, 12, 13, 14, 39, 43, 44, 45, 46 | Only polyactosamine structures can be hydrolyzed. |
| Bifidobacterium longum subsp. longum 44B (116174... | 0 | 1 | 0 | 0 | 1 | 1 | 0 | 9 | 11, 12, 13, 14, 39, 43, 44, 45, 46 | Only polyactosamine structures can be hydrolyzed. |
| Bifidobacterium longum subsp. longum F8 (722911.3... | 0 | 1 | 0 | 0 | 1 | 1 | 0 | 9 | 11, 12, 13, 14, 39, 43, 44, 45, 46 | Only polyactosamine structures can be hydrolyzed. |
| Bifidobacterium longum subsp. longum JCM 1217 (56... | 0 | 1 | 0 | 0 | 1 | 1 | 0 | 9 | 11, 12, 13, 14, 39, 43, 44, 45, 46 | Only polyactosamine structures can be hydrolyzed. |
| Bifidobacterium pseudocatenulatum DSM 20438 = JC... | 0 | 1 | 0 | 0 | 1 | 1 | 0 | 9 | 11, 12, 13, 14, 39, 43, 44, 45, 46 | Only polyactosamine structures can be hydrolyzed. |
| Bifidobacterium sp. 12_1_47BFAA (469594.3) | 0 | 1 | 0 | 0 | 1 | 1 | 0 | 9 | 11, 12, 13, 14, 39, 43, 44, 45, 46 | Only polyactosamine structures can be hydrolyzed. |
| Bilophila wadsworthia 3_1_6 (563192.3) | 0 | 0 | 0 | 0 | 0 | 1 | 0 | 1 | 11 | Only Core 3 structure can be hydrolyzed. |
| Blautia hansenii DSM 20583 (537007.6) | 0 | 1 | 1 | 0 | 1 | 0 | 1 | 6 | 1, 22, 43, 48, 49, 50, 52 | Only Syalyl-Tn-antigen and (fucosylated or syalated) Core 3 structure can be hydrolyzed. |
| Blautia hydrogenotrophica DSM 10507 (476272.5) | 0 | 0 | 0 | 0 | 1 | 1 | 0 | 9 | 11, 12, 13, 14, 39, 43, 44, 45, 46 | Only polyactosamine structures can be hydrolyzed. |
| Bryantella formatexigens DSM 14469 (478749.5) | 0 | 1 | 0 | 0 | 1 | 1 | 1 | 22 | 1, 2, 3, 4, 5, 10, 11, 12, 13, 14, 18, 39, 43, 44, 45, 46, 48, 49, 51, 52, 53, 54 | Glycans with alpha-GalNAc as well syalated glycans cannot be hydrolyzed. |
| Burkholderiales bacterium 1_1_47 (469610.4) | 0 | 0 | 0 | 0 | 0 | 0 | 0 | 0 | - | No glycans can be hydrolyzed. |
| butyrate-producing bacterium SM4/1 (245012.3) | 0 | 1 | 0 | 0 | 0 | 1 | 0 | 1 | 11 | Only Core 3 structure can be hydrolyzed. |
| butyrate-producing bacterium SS3/4 (245014.3) | 0 | 0 | 0 | 0 | 0 | 0 | 0 | 0 | - | No glycans can be hydrolyzed. |
| butyrate-producing bacterium SSC/2 (245018.3) | 0 | 0 | 0 | 0 | 0 | 1 | 0 | 1 | 11 | Only Core 3 structure can be hydrolyzed. |
| Butyricicoccus pullicaecorum 1.2 (1203606.4) | 0 | 1 | 0 | 0 | 1 | 1 | 0 | 9 | 11, 12, 13, 14, 39, 43, 44, 45, 46 | Only polyactosamine structures can be hydrolyzed. |
| Butyrivibrio crossotus DSM 2876 (511680.4) | 0 | 1 | 0 | 0 | 1 | 0 | 0 | 1 | 43 | Only polyactosamine structures can be hydrolyzed. |
| Butyrivibrio fibrisolvens 16/4 (657324.3) | 0 | 1 | 0 | 0 | 1 | 1 | 0 | 9 | 11, 12, 13, 14, 39, 43, 44, 45, 46 | Only polyactosamine structures can be hydrolyzed. |
| Campylobacter coli JV20 (864566.3) | 0 | 0 | 0 | 0 | 0 | 0 | 0 | 0 | - | No glycans are hydrolyzed. |
| Campylobacter upsaliensis JV21 (888826.3) | 0 | 0 | 0 | 0 | 0 | 0 | 0 | 0 | - | No glycans are hydrolyzed. |
| Catenibacterium mitsuokai DSM 15897 (451640.5) | 0 | 1 | 0 | 0 | 1 | 1 | 0 | 9 | 11, 12, 13, 14, 39, 43, 44, 45, 46 | Only polyactosamine structures can be hydrolyzed. |
| Cedecea davisae DSM 4568 (566551.4) | 1 | 0 | 0 | 0 | 0 | 0 | 0 | 0 | - | No glycans are hydrolyzed. |
| Citrobacter freundii 4_7_47CFAA (742730.3) | 0 | 1 | 0 | 0 | 1 | 0 | 0 | 1 | 43 | Only Core 1 structure can be hydrolyzed. |
| Citrobacter sp. 30_2 (469595.3) | 0 | 1 | 0 | 0 | 1 | 1 | 0 | 9 | 11, 12, 13, 14, 39, 43, 44, 45, 46 | Only polyactosamine structures can be hydrolyzed. |
| Citrobacter youngae ATCC 29220 (500640.5) | 0 | 1 | 0 | 0 | 1 | 0 | 0 | 1 | 43 | Only Core 1 structure can be hydrolyzed. |
| Clostridiales bacterium 1_7_47FAA (457421.5) | 0 | 1 | 0 | 0 | 1 | 1 | 0 | 9 | 11, 12, 13, 14, 39, 43, 44, 45, 46 | Only polyactosamine structures can be hydrolyzed. |
| Clostridium asparagiforme DSM 15981 (518636.5) | 0 | 1 | 0 | 0 | 1 | 1 | 0 | 9 | 11, 12, 13, 14, 39, 43, 44, 45, 46 | Only polyactosamine structures can be hydrolyzed. |
| Clostridium bartlettii DSM 16795 (445973.7) | 0 | 0 | 0 | 0 | 0 | 1 | 0 | 1 | 11 | Only Core 3 structure can be hydrolyzed. |
| Clostridium bolteae 90A5 (997893.5) | 0 | 1 | 0 | 0 | 1 | 1 | 0 | 9 | 11, 12, 13, 14, 39, 43, 44, 45, 46 | Only polyactosamine structures can be hydrolyzed. |
| Clostridium bolteae ATCC BAA-613 (411902.9) | 0 | 1 | 0 | 0 | 1 | 1 | 0 | 9 | 11, 12, 13, 14, 39, 43, 44, 45, 46 | Only polyactosamine structures can be hydrolyzed. |
| Clostridium celatum DSM 1785 (545697.3) | 1 | 0 | 1 | 1 | 1 | 1 | 1 | 50 | 1, 2, 3, 4, 5, 6, 7, 8, 9, 10, 11, 12, 13, 14, 15, 16, 17, 18, 19, 20, 21, 22, 23, 24, 25, 26, 27, 28, 29, 30, 32, 33, 37, 38, 39, 40, 41, 43, 44, 45, 46, 47, 48, 49, 50, 51, 52, 53, 54, 55 | Glycans with alpha-Gal cannot be hydrolyzed. |
| Clostridium cf. saccharolyticum K10 (717608.3) | 0 | 1 | 0 | 0 | 1 | 1 | 0 | 9 | 11, 12, 13, 14, 39, 43, 44, 45, 46 | Only polyactosamine structures can be hydrolyzed. |
| Clostridium citroniae WAL-17108 (742733.3) | 0 | 1 | 0 | 0 | 1 | 1 | 0 | 9 | 11, 12, 13, 14, 39, 43, 44, 45, 46 | Only polyactosamine structures can be hydrolyzed. |
| Clostridium clostridioforme 2_1_49FAA (742735.4) | 0 | 1 | 0 | 0 | 1 | 1 | 0 | 9 | 11, 12, 13, 14, 39, 43, 44, 45, 46 | Only polyactosamine structures can be hydrolyzed. |
| Clostridium difficile CD196 (645462.3) | 0 | 0 | 0 | 0 | 0 | 0 | 1 | 1 | 1 | Only Syalyl-Tn-antigen can be hydrolyzed. |
| Clostridium difficile NAP07 (525258.3) | 0 | 0 | 0 | 0 | 0 | 0 | 1 | 1 | 1 | Only Syalyl-Tn-antigen can be hydrolyzed. |
| Clostridium difficile NAP08 (525259.3) | 0 | 0 | 0 | 0 | 0 | 0 | 1 | 1 | 1 | Only Syalyl-Tn-antigen can be hydrolyzed. |
| Clostridium hathewayi WAL-18680 (742737.3) | 0 | 1 | 0 | 0 | 1 | 1 | 1 | 22 | 1, 2, 3, 4, 5, 10, 11, 12, 13, 14, 18, 39, 43, 44, 45, 46, 48, 49, 51, 52, 53, 54 | Glycans with alpha-GalNAc as well syalated glycans cannot be hydrolyzed. |
| Clostridium hiranonis DSM 13275 (500633.7) | 0 | 0 | 0 | 0 | 0 | 1 | 0 | 1 | 11 | Only Core 3 structure can be hydrolyzed. |

| Organism | | | | | | | | Total | Glycans | Notes |
|---|---|---|---|---|---|---|---|---|---|---|
| Clostridium hylemonae DSM 15053 (553973.6) | 0 | 1 | 0 | 0 | 1 | 0 | 1 | 5 | 1, 43, 48, 49, 52 | Only (sylated) Core 3 structure can be hydrolyzed. |
| Clostridium leptum DSM 753 (428125.8) | 0 | 1 | 1 | 0 | 1 | 1 | 0 | 27 | 11, 12, 13, 14, 18, 22, 23, 25, 26, 27, 28, 30, 31, 32, 33, 34, 35, 36, 39, 40, 42, 43, 44, 45, 46, 55, 56 | Glycans with alpha-Gal or alpha-GalNAc as well syalated glycans cannot be hydrolyzed. |
| Clostridium methylpentosum DSM 5476 (537013.3) | 0 | 0 | 1 | 0 | 0 | 0 | 0 | 0 | - | No glycans are hydrolyzed. |
| Clostridium nexile DSM 1787 (500632.7) | 0 | 1 | 1 | 0 | 1 | 1 | 1 | 40 | 1, 2, 3, 4, 5, 10, 11, 12, 13, 14, 18, 22, 23, 25, 26, 27, 28, 30, 31, 32, 33, 34, 35, 36, 39, 40, 42, 43, 44, 45, 46, 48, 49, 50, 51, 52, 53, 54, 55, 56 | Glycans with alpha-GalNAc or alpha-GlcNAc cannot be hydrolyzed. |
| Clostridium perfringens WAL-14572 (742739.3) | 1 | 1 | 1 | 1 | 1 | 1 | 1 | 56 | 1, 2, 3, 4, 5, 6, 7, 8, 9, 10, 11, 12, 13, 14, 15, 16, 17, 18, 19, 20, 21, 22, 23, 24, 25, 26, 27, 28, 29, 30, 31, 32, 33, 34, 35, 36, 37, 38, 39, 40, 41, 42, 43, 44, 45, 46, 47, 48, 49, 50, 51, 52, 53, 54, 55, 56 | All glycans can be hydrolyzed. |
| Clostridium ramosum DSM 1402 (445974.6) | 0 | 1 | 0 | 0 | 1 | 1 | 1 | 22 | 1, 2, 3, 4, 5, 10, 11, 12, 13, 14, 18, 39, 43, 44, 45, 46, 48, 49, 51, 52, 53, 54 | Glycans with alpha-GalNAc as well syalated glycans cannot be hydrolyzed. |
| Clostridium scindens ATCC 35704 (411468.9) | 0 | 0 | 1 | 0 | 1 | 0 | 1 | 6 | 1, 22, 43, 48, 49, 50, 52 | Only Syalyl-Tn-antigen and (fucosylated or sylated) Core 3 structure can be hydrolyzed. |
| Clostridium sp. 7_2_43FAA (457396.3) | 0 | 1 | 0 | 0 | 1 | 1 | 1 | 22 | 1, 2, 3, 4, 5, 10, 11, 12, 13, 14, 18, 39, 43, 44, 45, 46, 48, 49, 51, 52, 53, 54 | Glycans with alpha-GalNAc as well syalated glycans cannot be hydrolyzed. |
| Clostridium sp. 7_3_54FAA (665940.3) | 0 | 0 | 0 | 0 | 1 | 1 | 0 | 9 | 11, 12, 13, 14, 39, 43, 44, 45, 46 | Only ploylactosamine structures can be hydrolyzed. |
| Clostridium sp. D5 (556261.3) | 0 | 1 | 1 | 0 | 1 | 0 | 0 | 2 | 22, 43 | Only (fucosylated) Core 1 structure can be |
| Clostridium sp. HGF2 (908340.3) | 0 | 1 | 0 | 0 | 0 | 0 | 0 | 0 | - | No glycans are hydrolyzed. |
| Clostridium sp. L2-50 (411489.7) | 0 | 1 | 0 | 0 | 1 | 1 | 0 | 9 | 11, 12, 13, 14, 39, 43, 44, 45, 46 | Only ploylactosamine structures can be hydrolyzed. |
| Clostridium sp. M62/1 (411486.3) | 0 | 1 | 0 | 0 | 0 | 1 | 0 | 1 | 11 | Only Core 3 structure can be hydrolyzed. |
| Clostridium sp. SS2/1 (411484.7) | 0 | 0 | 0 | 0 | 1 | 1 | 0 | 9 | 11, 12, 13, 14, 39, 43, 44, 45, 46 | Only ploylactosamine structures can be hydrolyzed. |
| Clostridium spiroforme DSM 1552 (428126.7) | 1 | 0 | 1 | 1 | 1 | 1 | 1 | 33 | 11, 12, 13, 14, 15, 16, 17, 18, 19, 20, 21, 22, 23, 24, 25, 26, 27, 28, 29, 30, 32, 33, 37, 38, 39, 40, 41, 43, 44, 45, 46, 47, 55 | Glycans with alpha-Gal as well syalated glycans cannot be hydrolyzed. |
| Clostridium sporogenes ATCC 15579 (471871.7) | 0 | 0 | 0 | 0 | 0 | 0 | 1 | 1 | 1 | Only Syalyl-Tn-antigen can be hydrolyzed. |
| Clostridium symbiosum WAL-14163 (742740.3) | 0 | 0 | 0 | 0 | 1 | 1 | 0 | 9 | 11, 12, 13, 14, 39, 43, 44, 45, 46 | Only ploylactosamine structures can be hydrolyzed. |
| Clostridium symbiosum WAL-14673 (742741.3) | 0 | 0 | 0 | 0 | 1 | 1 | 0 | 9 | 11, 12, 13, 14, 39, 43, 44, 45, 46 | Only ploylactosamine structures can be hydrolyzed. |
| Collinsella aerofaciens ATCC 25986 (411903.6) | 0 | 0 | 0 | 0 | 0 | 0 | 0 | 0 | - | No glycans are hydrolyzed. |
| Collinsella intestinalis DSM 13280 (521003.7) | 0 | 0 | 0 | 0 | 0 | 1 | 1 | 3 | 1, 2, 11 | Only Syalyl-Tn-antigen and (sylated) Core 3 structure can be hydrolyzed. |
| Collinsella stercoris DSM 13279 (445975.6) | 1 | 0 | 1 | 1 | 1 | 1 | 0 | 33 | 11, 12, 13, 14, 15, 16, 17, 18, 19, 20, 21, 22, 23, 24, 25, 26, 27, 28, 29, 30, 32, 33, 37, 38, 39, 40, 41, 43, 44, 45, 46, 47, 55 | Glycans with alpha-Gal as well syalated glycans cannot be hydrolyzed. |
| Collinsella tanakaei YIT 12063 (742742.3) | 0 | 0 | 0 | 0 | 0 | 0 | 0 | 0 | - | No glycans are hydrolyzed. |
| Coprobacillus sp. 29_1 (469596.3) | 0 | 0 | 0 | 0 | 1 | 1 | 1 | 22 | 1, 2, 3, 4, 5, 10, 11, 12, 13, 14, 18, 39, 43, 44, 45, 46, 48, 49, 51, 52, 53, 54 | Glycans with alpha-GalNAc as well syalated glycans cannot be hydrolyzed. |
| Coprobacillus sp. 3_3_56FAA (665941.3) | 0 | 1 | 0 | 0 | 1 | 1 | 1 | 22 | 1, 2, 3, 4, 5, 10, 11, 12, 13, 14, 18, 39, 43, 44, 45, 46, 48, 49, 51, 52, 53, 54 | Glycans with alpha-GalNAc as well syalated glycans cannot be hydrolyzed. |
| Coprobacillus sp. 8_2_54BFAA (469597.4) | 0 | 1 | 0 | 0 | 1 | 1 | 1 | 22 | 1, 2, 3, 4, 5, 10, 11, 12, 13, 14, 18, 39, 43, 44, 45, 46, 48, 49, 51, 52, 53, 54 | Glycans with alpha-GalNAc as well syalated glycans cannot be hydrolyzed. |
| Coprobacillus sp. D6 (556262.3) | 0 | 0 | 0 | 0 | 1 | 1 | 1 | 22 | 1, 2, 3, 4, 5, 10, 11, 12, 13, 14, 18, 39, 43, 44, 45, 46, 48, 49, 51, 52, 53, 54 | Glycans with alpha-GalNAc as well syalated glycans cannot be hydrolyzed. |
| Coprococcus catus GD/7 (717962.3) | 0 | 0 | 0 | 0 | 1 | 1 | 0 | 9 | 11, 12, 13, 14, 39, 43, 44, 45, 46 | Only ploylactosamine structures can be hydrolyzed. |
| Coprococcus comes ATCC 27758 (470146.3) | 0 | 1 | 0 | 0 | 0 | 0 | 0 | 1 | 43 | Only Core 1 structure can be hydrolyzed. |
| Coprococcus eutactus ATCC 27759 (411474.6) | 0 | 1 | 0 | 0 | 1 | 1 | 0 | 9 | 11, 12, 13, 14, 39, 43, 44, 45, 46 | Only ploylactosamine structures can be hydrolyzed. |
| Coprococcus sp. ART55/1 (751585.3) | 0 | 1 | 0 | 0 | 1 | 1 | 0 | 9 | 11, 12, 13, 14, 39, 43, 44, 45, 46 | Only ploylactosamine structures can be hydrolyzed. |
| Coprococcus sp. HPP0048 (1078091.3) | 0 | 1 | 1 | 0 | 1 | 1 | 1 | 40 | 1, 2, 3, 4, 5, 10, 11, 12, 13, 14, 18, 22, 23, 25, 26, 27, 28, 30, 31, 32, 33, 34, 35, 36, 39, 40, 42, 43, 44, 45, 46, 48, 49, 50, 51, 52, 53, 54, 55, 56 | Glycans with alpha-GalNAc or alpha-GlcNAc cannot be hydrolyzed. |
| Coprococcus sp. HPP0074 (1078090.3) | 0 | 1 | 1 | 0 | 1 | 1 | 1 | 40 | 1, 2, 3, 4, 5, 10, 11, 12, 13, 14, 18, 22, 23, 25, 26, 27, 28, 30, 31, 32, 33, 34, 35, 36, 39, 40, 42, 43, 44, 45, 46, 48, 49, 50, 51, 52, 53, 54, 55, 56 | Glycans with alpha-GalNAc or alpha-GlcNAc cannot be hydrolyzed. |
| Corynebacterium ammoniagenes DSM 20306 (64975...) | 0 | 0 | 0 | 0 | 0 | 1 | 0 | 1 | 11 | Only Core 3 structure can be hydrolyzed. |
| Corynebacterium sp. HFH0082 (1078764.3) | 0 | 0 | 0 | 0 | 0 | 0 | 0 | 0 | - | No glycans are hydrolyzed. |
| Dermabacter sp. HFH0086 (1203568.3) | 0 | 1 | 0 | 0 | 0 | 0 | 1 | 1 | 1 | Only Syalyl-Tn-antigen can be hydrolyzed. |
| Desulfitobacterium hafniense DP7 (537010.4) | 0 | 0 | 0 | 0 | 0 | 1 | 0 | 1 | 11 | Only Core 3 structure can be hydrolyzed. |
| Desulfovibrio piger ATCC 29098 (411464.8) | 0 | 0 | 0 | 0 | 0 | 0 | 0 | 0 | - | No glycans are hydrolyzed. |
| Desulfovibrio sp. 3_1_syn3 (457398.5) | 0 | 0 | 0 | 0 | 0 | 1 | 0 | 1 | 11 | Only Core 3 structure can be hydrolyzed. |
| Dorea formicigenerans 4_6_53AFAA (742765.5) | 0 | 0 | 1 | 0 | 1 | 0 | 1 | 6 | 1, 22, 43, 48, 49, 50, 52 | Only Syalyl-Tn-antigen and (fucosylated or sylated) Core 3 structure can be hydrolyzed. |
| Dorea formicigenerans ATCC 27755 (411461.4) | 0 | 1 | 0 | 0 | 1 | 1 | 1 | 22 | 1, 2, 3, 4, 5, 10, 11, 12, 13, 14, 18, 39, 43, 44, 45, 46, 48, 49, 51, 52, 53, 54 | Glycans with alpha-GalNAc as well syalated glycans cannot be hydrolyzed. |
| Dorea longicatena DSM 13814 (411462.6) | 0 | 1 | 0 | 0 | 0 | 0 | 0 | 1 | 43 | Only Core 1 structure can be hydrolyzed. |
| Dysgonomonas gadei ATCC BAA-286 (742766.3) | 0 | 1 | 1 | 0 | 1 | 1 | 0 | 27 | 11, 12, 13, 14, 18, 22, 23, 25, 26, 27, 28, 30, 31, 32, 33, 34, 35, 36, 39, 40, 42, 43, 44, 45, 46, 55, 56 | Glycans with alpha-Gal or alpha-GalNAc as well syalated glycans cannot be hydrolyzed. |
| Dysgonomonas mossii DSM 22836 (742767.3) | 0 | 1 | 1 | 0 | 1 | 1 | 0 | 27 | 11, 12, 13, 14, 18, 22, 23, 25, 26, 27, 28, 30, 31, 32, 33, 34, 35, 36, 39, 40, 42, 43, 44, 45, 46, 55, 56 | Glycans with alpha-Gal or alpha-GalNAc as well syalated glycans cannot be hydrolyzed. |
| Edwardsiella tarda ATCC 23685 (500638.3) | 0 | 1 | 0 | 0 | 0 | 1 | 0 | 1 | 11 | Only Core 3 structure can be hydrolyzed. |
| Eggerthella sp. 1_3_56FAA (665943.3) | 0 | 0 | 0 | 0 | 0 | 0 | 0 | 0 | - | No glycans are hydrolyzed. |
| Eggerthella sp. HGA1 (910311.3) | 0 | 0 | 0 | 0 | 0 | 0 | 0 | 0 | - | No glycans are hydrolyzed. |

| Organism | 1 | 2 | 3 | 4 | 5 | 6 | 7 | 8 | n | Structures | Description |
|---|---|---|---|---|---|---|---|---|---|---|---|
| Enterobacter cancerogenus ATCC 35316 (500639.8) | 0 | 0 | 0 | 0 | 1 | | | 0 | 9 | 11, 12, 13, 14, 39, 43, 44, 45, 46 | Only ploylactosamine structures can be hydrolyzed. |
| Enterobacter cloacae subsp. cloacae NCTC 9394 (7... | 0 | 1 | 0 | 0 | 1 | 1 | | 0 | 9 | 11, 12, 13, 14, 39, 43, 44, 45, 46 | Only ploylactosamine structures can be hydrolyzed. |
| Enterobacteriaceae bacterium 9_2_54FAA (469613.3 | 0 | 0 | 0 | 0 | 1 | 1 | | 0 | 9 | 11, 12, 13, 14, 39, 43, 44, 45, 46 | Only ploylactosamine structures can be hydrolyzed. |
| Enterococcus faecalis PC1.1 (702448.3) | 0 | 1 | 0 | 0 | 1 | 0 | | 0 | 1 | 43 | Only Core 1 structure can be hydrolyzed. |
| Enterococcus faecalis TX0104 (491074.3) | 0 | 0 | 0 | 0 | 1 | | | 0 | 9 | 11, 12, 13, 14, 39, 43, 44, 45, 46 | Only ploylactosamine structures can be hydrolyzed. |
| Enterococcus faecalis TX1302 (749513.3) | 0 | 0 | 0 | 0 | 1 | 1 | | 0 | 9 | 11, 12, 13, 14, 39, 43, 44, 45, 46 | Only ploylactosamine structures can be hydrolyzed. |
| Enterococcus faecalis TX1322 (525278.3) | 0 | 0 | 0 | 0 | 1 | 1 | | 0 | 9 | 11, 12, 13, 14, 39, 43, 44, 45, 46 | Only ploylactosamine structures can be hydrolyzed. |
| Enterococcus faecalis TX1341 (749514.3) | 0 | 0 | 0 | 0 | 1 | | | 0 | 9 | 11, 12, 13, 14, 39, 43, 44, 45, 46 | Only ploylactosamine structures can be hydrolyzed. |
| Enterococcus faecalis TX1342 (749515.3) | 0 | 0 | 0 | 0 | 1 | | | 0 | 1 | 43 | Only Core 1 structure can be hydrolyzed. |
| Enterococcus faecalis TX1346 (749516.3) | 0 | 0 | 0 | 0 | | 0 | 1 | 0 | 1 | 11 | Only Core 3 structure can be hydrolyzed. |
| Enterococcus faecalis TX2134 (749518.3) | 0 | 0 | 0 | 0 | 1 | 1 | | 0 | 9 | 11, 12, 13, 14, 39, 43, 44, 45, 46 | Only ploylactosamine structures can be hydrolyzed. |
| Enterococcus faecalis TX2137 (749491.3) | 0 | 0 | 0 | 0 | 1 | 1 | | 0 | 9 | 11, 12, 13, 14, 39, 43, 44, 45, 46 | Only ploylactosamine structures can be hydrolyzed. |
| Enterococcus faecalis TX4244 (749494.3) | 0 | 0 | 0 | 0 | 1 | | | 0 | 9 | 11, 12, 13, 14, 39, 43, 44, 45, 46 | Only ploylactosamine structures can be hydrolyzed. |
| Enterococcus faecium PC4.1 (791161.5) | 0 | 1 | 0 | 0 | 1 | 0 | | 0 | 1 | 43 | Only Core 1 structure can be hydrolyzed. |
| Enterococcus faecium TX1330 (525279.3) | 0 | 1 | 0 | 0 | 1 | 0 | | 0 | 1 | 43 | Only Core 1 structure can be hydrolyzed. |
| Enterococcus saccharolyticus 30_1 (742813.4) | 0 | 1 | 1 | 0 | 1 | 1 | | 0 | 27 | 11, 12, 13, 14, 18, 22, 23, 25, 26, 27, 28, 30, 31, 32, 33, 34, 35, 36, 39, 40, 42, 43, 44, 45, 46, 55, 56 | Glycans with alpha-Gal or alpha-GalNAc as well syalated glycans cannot be hydrolyzed. |
| Enterococcus sp. 7L76 (657310.3) | 0 | 0 | 0 | 0 | 1 | 0 | | 0 | 1 | 43 | Only Core 1 structure can be hydrolyzed. |
| Erysipelotrichaceae bacterium 2_2_44A (457422.3) | 0 | 1 | 0 | 0 | 0 | 0 | 0 | 0 | 0 | | No glycans are hydrolyzed. |
| Erysipelotrichaceae bacterium 21_3 (658657.3) | 0 | 1 | 0 | 0 | 0 | 0 | 0 | 0 | 0 | | No glycans are hydrolyzed. |
| Erysipelotrichaceae bacterium 3_1_53 (658659.3) | 0 | 0 | 0 | 0 | 0 | 0 | 0 | 0 | 0 | - | |
| Erysipelotrichaceae bacterium 5_2_54FAA (552396.3 | 0 | 0 | 0 | 1 | 0 | 1 | 0 | 0 | 1 | 11 | Only Core 3 structure can be hydrolyzed. |
| Erysipelotrichaceae bacterium 6_1_45 (469614.3) | 0 | 1 | 0 | 0 | 0 | 0 | 0 | 0 | 0 | | No glycans are hydrolyzed. |
| Escherichia coli 4_1_47FAA (1127356.4) | 0 | 1 | 0 | 0 | 1 | 0 | 0 | 0 | 1 | 43 | Only Core 1 structure can be hydrolyzed. |
| Escherichia coli MS 107-1 (679207.4) | 0 | 1 | 0 | 0 | 1 | 0 | 0 | 0 | 1 | 43 | Only Core 1 structure can be hydrolyzed. |
| Escherichia coli MS 115-1 (749537.3) | 0 | 1 | 0 | 0 | 1 | 0 | 0 | 0 | 1 | 43 | Only Core 1 structure can be hydrolyzed. |
| Escherichia coli MS 116-1 (749538.3) | 0 | 1 | 0 | 0 | 1 | 0 | 0 | 0 | 1 | 43 | Only Core 1 structure can be hydrolyzed. |
| Escherichia coli MS 119-7 (679206.4) | 0 | 1 | 0 | 0 | 1 | 0 | 0 | 0 | 1 | 43 | Only Core 1 structure can be hydrolyzed. |
| Escherichia coli MS 124-1 (679205.4) | 0 | 1 | 0 | 0 | 1 | 0 | 0 | 0 | 1 | 43 | Only Core 1 structure can be hydrolyzed. |
| Escherichia coli MS 145-7 (702404.3) | 0 | 1 | 0 | 0 | 1 | 0 | 0 | 0 | 1 | 43 | Only Core 1 structure can be hydrolyzed. |
| Escherichia coli MS 146-1 (749540.3) | 0 | 1 | 0 | 0 | 1 | 0 | 0 | 0 | 1 | 43 | Only Core 1 structure can be hydrolyzed. |
| Escherichia coli MS 175-1 (749544.3) | 0 | 1 | 0 | 0 | 1 | 0 | 0 | 0 | 1 | 43 | Only Core 1 structure can be hydrolyzed. |
| Escherichia coli MS 182-1 (749545.3) | 0 | 1 | 0 | 0 | 1 | 0 | 0 | 0 | 1 | 43 | Only Core 1 structure can be hydrolyzed. |
| Escherichia coli MS 185-1 (749546.3) | 0 | 1 | 0 | 0 | 1 | 0 | 0 | 0 | 1 | 43 | Only Core 1 structure can be hydrolyzed. |
| Escherichia coli MS 187-1 (749547.3) | 0 | 1 | 0 | 0 | 1 | 0 | 0 | 0 | 1 | 43 | Only Core 1 structure can be hydrolyzed. |
| Escherichia coli MS 196-1 (749548.3) | 0 | 1 | 0 | 0 | 1 | 0 | 0 | 0 | 1 | 43 | Only Core 1 structure can be hydrolyzed. |
| Escherichia coli MS 198-1 (749549.3) | 0 | 1 | 0 | 0 | 1 | 0 | 0 | 0 | 1 | 43 | Only Core 1 structure can be hydrolyzed. |
| Escherichia coli MS 200-1 (749550.3) | 0 | 1 | 0 | 0 | 1 | 0 | 0 | 0 | 1 | 43 | Only Core 1 structure can be hydrolyzed. |
| Escherichia coli MS 21-1 (749527.3) | 0 | 1 | 0 | 0 | 1 | 0 | 0 | 0 | 1 | 43 | Only Core 1 structure can be hydrolyzed. |
| Escherichia coli MS 45-1 (749528.3) | 0 | 1 | 0 | 0 | 1 | 0 | 0 | 0 | 1 | 43 | Only Core 1 structure can be hydrolyzed. |
| Escherichia coli MS 69-1 (749531.3) | 0 | 1 | 0 | 0 | 1 | 0 | 0 | 0 | 1 | 43 | Only Core 1 structure can be hydrolyzed. |
| Escherichia coli MS 78-1 (749532.3) | 0 | 1 | 0 | 0 | 1 | 0 | 0 | 0 | 1 | 43 | Only Core 1 structure can be hydrolyzed. |
| Escherichia coli MS 79-10 (796392.4) | 0 | 1 | 0 | 0 | 1 | 0 | 0 | 0 | 1 | 43 | Only Core 1 structure can be hydrolyzed. |
| Escherichia coli MS 84-1 (749533.3) | 0 | 1 | 0 | 0 | 1 | 0 | 0 | 0 | 1 | 43 | Only Core 1 structure can be hydrolyzed. |
| Escherichia coli MS 85-1 (679202.3) | 0 | 1 | 0 | 0 | 1 | 0 | 0 | 0 | 1 | 43 | Only Core 1 structure can be hydrolyzed. |
| Escherichia coli O157:H7 str. Sakai (386585.9) | 0 | 1 | 0 | 0 | 1 | 0 | 0 | 0 | 1 | 43 | Only Core 1 structure can be hydrolyzed. |
| Escherichia coli SE11 (409438.11) | 0 | 1 | 0 | 0 | 1 | 0 | 0 | 0 | 1 | 43 | Only Core 1 structure can be hydrolyzed. |
| Escherichia coli SE15 (431946.3) | 0 | 1 | 0 | 0 | 1 | 0 | 0 | 0 | 1 | 43 | Only Core 1 structure can be hydrolyzed. |
| Escherichia coli str. K-12 substr. MG1655 (511145.12 | 0 | 1 | 0 | 0 | 1 | 0 | 0 | 0 | 1 | 43 | Only Core 1 structure can be hydrolyzed. |
| Escherichia coli UTI89 (364106.8) | 0 | 1 | 0 | 0 | 1 | 0 | 0 | 0 | 1 | 43 | Only Core 1 structure can be hydrolyzed. |
| Escherichia sp. 1_1_43 (457400.3) | 0 | 1 | 0 | 0 | 1 | 0 | 0 | 0 | 1 | 43 | Only Core 1 structure can be hydrolyzed. |
| Escherichia sp. 3_2_53FAA (469598.5) | 0 | 1 | 0 | 0 | 1 | 0 | 0 | 0 | 1 | 43 | Only Core 1 structure can be hydrolyzed. |
| Escherichia sp. 4_1_40B (457401.3) | 0 | 1 | 0 | 0 | 1 | 0 | 0 | 0 | 1 | 43 | Only Core 1 structure can be hydrolyzed. |
| Eubacterium biforme DSM 3989 (518637.5) | 0 | 0 | 1 | 0 | 1 | 0 | 0 | 0 | 2 | 22, 43 | Only Core 1 structure can be |
| Eubacterium cylindroides T2-87 (717960.3) | 0 | 0 | 0 | 0 | 0 | 0 | 0 | 0 | 0 | - | No glycans are hydrolyzed. |
| Eubacterium dolichum DSM 3991 (428127.7) | 1 | 0 | 0 | 1 | | 1 | | 0 | 13 | 11, 12, 13, 14, 16, 18, 19, 21, 39, 43, 44, 45, 46 | Only (sylated) ploylactosamine structures can be |
| Eubacterium hallii DSM 3353 (411469.3) | 0 | 0 | 0 | 0 | 1 | 1 | | 0 | 9 | 11, 12, 13, 14, 39, 43, 44, 45, 46 | Only ploylactosamine structures can be hydrolyzed. |
| Eubacterium rectale DSM 17629 (657318.4) | 0 | 0 | 0 | 0 | 1 | 1 | | 0 | 9 | 11, 12, 13, 14, 39, 43, 44, 45, 46 | Only ploylactosamine structures can be hydrolyzed. |
| Eubacterium rectale M104/1 (657317.3) | 0 | 0 | 0 | 0 | 1 | 1 | | 0 | 9 | 11, 12, 13, 14, 39, 43, 44, 45, 46 | Only ploylactosamine structures can be hydrolyzed. |
| Eubacterium siraeum 70/3 (657319.3) | 0 | 1 | 1 | 0 | 1 | 0 | | 0 | 2 | 22, 43 | Only (fucosylated) Core 1 structure can be |
| Eubacterium siraeum DSM 15702 (428128.7) | 0 | 1 | 1 | 0 | 1 | 1 | | 0 | 27 | 11, 12, 13, 14, 18, 22, 23, 25, 26, 27, 28, 30, 31, 32, 33, 34, 35, 36, 39, 40, 42, 43, 44, 45, 46, 55, 56 | Glycans with alpha-Gal or alpha-GalNAc as well syalated glycans cannot be hydrolyzed. |
| Eubacterium siraeum V10Sc8a (717961.3) | 0 | 1 | 1 | 0 | 1 | 1 | | 0 | 27 | 11, 12, 13, 14, 18, 22, 23, 25, 26, 27, 28, 30, 31, 32, 33, 34, 35, 36, 39, 40, 42, 43, 44, 45, 46, 55, 56 | Glycans with alpha-Gal or alpha-GalNAc as well syalated glycans cannot be hydrolyzed. |
| Eubacterium sp. 3_1_31 (457402.3) | 0 | 0 | 0 | 0 | 0 | 1 | | 0 | 1 | 11 | Only Core 3 structure can be hydrolyzed. |
| Eubacterium ventriosum ATCC 27560 (411463.4) | 0 | 0 | 0 | 0 | 1 | 0 | 0 | 0 | 1 | 43 | Only Core 1 structure can be hydrolyzed. |
| Faecalibacterium cf. prausnitzii KLE1255 (748224.3) | 0 | 0 | 0 | 0 | 1 | 0 | 0 | 0 | 1 | 43 | Only Core 1 structure can be hydrolyzed. |
| Faecalibacterium prausnitzii A2-165 (411483.3) | 0 | 1 | 0 | 0 | 1 | 1 | | 0 | 9 | 11, 12, 13, 14, 39, 43, 44, 45, 46 | Only ploylactosamine structures can be hydrolyzed. |
| Faecalibacterium prausnitzii L2-6 (718252.3) | 0 | 0 | 0 | 0 | 1 | 0 | 0 | 0 | 1 | 43 | Only Core 1 structure can be hydrolyzed. |
| Faecalibacterium prausnitzii M21/2 (411485.10) | 0 | 1 | 0 | 0 | 1 | 0 | 0 | 0 | 1 | 43 | Only Core 1 structure can be hydrolyzed. |
| Faecalibacterium prausnitzii SL3/3 (657322.3) | 0 | 0 | 0 | 0 | 1 | 0 | 0 | 0 | 1 | 43 | Only Core 1 structure can be hydrolyzed. |
| Fusobacterium gonidiaformans ATCC 25563 (469615 | 0 | 0 | 0 | 0 | 0 | 0 | 0 | 0 | 0 | - | No glycans are hydrolyzed. |
| Fusobacterium mortiferum ATCC 9817 (469616.3) | 0 | 1 | 0 | 0 | 1 | 0 | 0 | 0 | 1 | 43 | Only Core 1 structure can be hydrolyzed. |

| Organism | C1 | C2 | C3 | C4 | C5 | C6 | C7 | C8 | # | Glycans | Notes |
|---|---|---|---|---|---|---|---|---|---|---|---|
| Fusobacterium necrophorum subsp. funduliforme 1_1 | 0 | 0 | 0 | 0 | 0 | 0 | 0 | 0 | | - | No glycans are hydrolyzed. |
| Fusobacterium sp. 1_1_41FAA (469621.3) | 0 | 0 | 0 | 0 | 0 | 0 | 0 | 0 | | - | No glycans are hydrolyzed. |
| Fusobacterium sp. 11_3_2 (457403.3) | 0 | 0 | 0 | 0 | 0 | 0 | 0 | 0 | | - | No glycans are hydrolyzed. |
| Fusobacterium sp. 12_1B (457404.3) | 0 | 0 | 0 | 0 | 0 | 0 | 0 | 0 | | - | No glycans are hydrolyzed. |
| Fusobacterium sp. 2_1_31 (469599.3) | 0 | 0 | 0 | 0 | 0 | 0 | 0 | 0 | | - | No glycans are hydrolyzed. |
| Fusobacterium sp. 21_1A (469601.3) | 0 | 0 | 0 | 0 | 0 | 0 | 0 | 0 | | - | No glycans are hydrolyzed. |
| Fusobacterium sp. 3_1_27 (469602.3) | 0 | 0 | 0 | 0 | 0 | 0 | 0 | 0 | | - | No glycans are hydrolyzed. |
| Fusobacterium sp. 3_1_33 (469603.3) | 0 | 0 | 0 | 0 | 0 | 0 | 0 | 0 | | - | No glycans are hydrolyzed. |
| Fusobacterium sp. 3_1_36A2 (469604.3) | 0 | 0 | 0 | 0 | 0 | 0 | 0 | 0 | | - | No glycans are hydrolyzed. |
| Fusobacterium sp. 3_1_5R (469605.3) | 0 | 0 | 0 | 0 | 0 | 0 | 0 | 0 | | - | No glycans are hydrolyzed. |
| Fusobacterium sp. 4_1_13 (469606.3) | 0 | 0 | 0 | 0 | 0 | 0 | 0 | 0 | | - | No glycans are hydrolyzed. |
| Fusobacterium sp. 7_1 (457405.3) | 0 | 0 | 0 | 0 | 0 | 0 | 0 | 0 | | - | No glycans are hydrolyzed. |
| Fusobacterium sp. D11 (556264.3) | 0 | 0 | 0 | 0 | 0 | 0 | 0 | 0 | | - | No glycans are hydrolyzed. |
| Fusobacterium sp. D12 (556263.3) | 0 | 0 | 0 | 0 | 0 | 0 | 0 | 0 | | - | No glycans are hydrolyzed. |
| Fusobacterium ulcerans ATCC 49185 (469617.3) | 0 | 0 | 0 | 0 | 0 | 0 | 0 | 0 | | - | No glycans are hydrolyzed. |
| Fusobacterium varium ATCC 27725 (469618.3) | 0 | 0 | 0 | 0 | 0 | 0 | 0 | 0 | | - | No glycans are hydrolyzed. |
| Gordonibacter pamelaeae 7-10-1-b (657308.3) | 0 | 0 | 0 | 0 | 0 | 0 | 0 | 0 | | - | No glycans are hydrolyzed. |
| Hafnia alvei ATCC 51873 (1002364.3) | 0 | 1 | 0 | 0 | 1 | 1 | 0 | 9 | | 11, 12, 13, 14, 39, 43, 44, 45, 46 | Only ploylactosamine structures can be hydrolyzed. |
| Helicobacter bilis ATCC 43879 (613026.4) | 0 | 0 | 0 | 0 | 0 | 0 | 0 | 0 | | - | No glycans are hydrolyzed. |
| Helicobacter canadensis MIT 98-5491 (Prj:30719) (53 | 0 | 0 | 0 | 0 | 0 | 0 | 0 | 0 | | - | No glycans are hydrolyzed. |
| Helicobacter cinaedi ATCC BAA-847 (1206745.3) | 0 | 0 | 0 | 0 | 0 | 0 | 0 | 0 | | - | No glycans are hydrolyzed. |
| Helicobacter cinaedi CCUG 18818 (537971.5) | 0 | 0 | 0 | 0 | 0 | 0 | 0 | 0 | | - | No glycans are hydrolyzed. |
| Helicobacter pullorum MIT 98-5489 (537972.5) | 0 | 0 | 0 | 0 | 0 | 0 | 0 | 0 | | - | No glycans are hydrolyzed. |
| Helicobacter winghamensis ATCC BAA-430 (556267 | 0 | 0 | 0 | 0 | 0 | 0 | 0 | 0 | | - | No glycans are hydrolyzed. |
| Holdemania filiformis DSM 12042 (545696.5) | 0 | 0 | 1 | 0 | 1 | 1 | 0 | 21 | | 11, 12, 13, 14, 18, 22, 23, 25, 26, 27, 28, 30, 32, 33, 39, 40, 43, 44, 45, 46, 55 | Only (fucosylated) ploylactosamine structures can be hydrolyzed. |
| Klebsiella pneumoniae 1162281 (1037908.3) | 0 | 1 | 0 | 0 | 1 | 0 | 0 | 1 | | 43 | Only Core 1 structure can be hydrolyzed. |
| Klebsiella pneumoniae subsp. pneumoniae WGLW5 | 0 | 1 | 0 | 0 | 1 | 0 | 0 | 1 | | 43 | Only Core 1 structure can be hydrolyzed. |
| Klebsiella sp. 1_1_55 (469608.3) | 0 | 1 | 0 | 0 | 1 | 0 | 0 | 1 | | 43 | Only Core 1 structure can be hydrolyzed. |
| Klebsiella sp. 4_1_44FAA (665944.3) | 0 | 1 | 0 | 0 | 1 | 0 | 0 | 1 | | 43 | Only Core 1 structure can be hydrolyzed. |
| Lachnospiraceae bacterium 1_1_57FAA (658081.3) | 1 | 1 | 1 | 1 | 1 | 1 | 1 | 56 | | 1, 2, 3, 4, 5, 6, 7, 8, 9, 10, 11, 12, 13, 14, 15, 16, 17, 18, 19, 20, 21, 22, 23, 24, 25, 26, 27, 28, 29, 30, 31, 32, 33, 34, 35, 36, 37, 38, 39, 40, 41, 42, 43, 44, 45, 46, 47, 48, 49, 50, 51, 52, 53, 54, 55, 56 | All glycans can be hydrolyzed. |
| Lachnospiraceae bacterium 1_4_56FAA (658655.3) | 0 | 1 | 1 | 0 | 1 | 1 | 0 | 27 | | 11, 12, 13, 14, 18, 22, 23, 25, 26, 27, 28, 30, 31, 32, 33, 34, 35, 36, 39, 40, 42, 43, 44, 45, 46, 55, 56 | Glycans with alpha-Gal or alpha-GalNAc as well syalated glycans cannot be hydrolyzed. |
| Lachnospiraceae bacterium 2_1_46FAA (742723.3) | 0 | 0 | 1 | 1 | 1 | 1 | 1 | 35 | | 1, 2, 3, 4, 5, 10, 11, 12, 13, 14, 18, 22, 23, 25, 26, 27, 28, 30, 32, 33, 39, 40, 43, 44, 45, 46, 47, 48, 49, 50, 51, 52, 53, 54, 55 | Glycans with alpha-Gal or alpha-GalNAc cannot be hydrolyzed. |
| Lachnospiraceae bacterium 2_1_58FAA (658082.3) | 0 | 1 | 1 | 0 | 1 | 1 | 0 | 27 | | 11, 12, 13, 14, 18, 22, 23, 25, 26, 27, 28, 30, 31, 32, 33, 34, 35, 36, 39, 40, 42, 43, 44, 45, 46, 55, 56 | Glycans with alpha-Gal or alpha-GalNAc as well syalated glycans cannot be hydrolyzed. |
| Lachnospiraceae bacterium 3_1_46FAA (665950.3) | 1 | 1 | 1 | 1 | 1 | 1 | 1 | 56 | | 1, 2, 3, 4, 5, 6, 7, 8, 9, 10, 11, 12, 13, 14, 15, 16, 17, 18, 19, 20, 21, 22, 23, 24, 25, 26, 27, 28, 29, 30, 31, 32, 33, 34, 35, 36, 37, 38, 39, 40, 41, 42, 43, 44, 45, 46, 47, 48, 49, 50, 51, 52, 53, 54, 55, 56 | All glycans can be hydrolyzed. |
| Lachnospiraceae bacterium 3_1_57FAA_CT1 (65808 | 0 | 1 | 1 | 0 | 1 | 1 | 1 | 40 | | 1, 2, 3, 4, 5, 10, 11, 12, 13, 14, 18, 22, 23, 25, 26, 27, 28, 30, 31, 32, 33, 34, 35, 36, 39, 40, 42, 43, 44, 45, 46, 48, 49, 50, 51, 52, 53, 54, 55, 56 | Glycans with alpha-GalNAc or alpha-GlcNAc cannot be hydrolyzed. |
| Lachnospiraceae bacterium 4_1_37FAA (552395.3) | 0 | 1 | 1 | 0 | 1 | 1 | 1 | 40 | | 1, 2, 3, 4, 5, 10, 11, 12, 13, 14, 18, 22, 23, 25, 26, 27, 28, 30, 31, 32, 33, 34, 35, 36, 39, 40, 42, 43, 44, 45, 46, 48, 49, 50, 51, 52, 53, 54, 55, 56 | Glycans with alpha-GalNAc or alpha-GlcNAc cannot be hydrolyzed. |
| Lachnospiraceae bacterium 5_1_57FAA (658085.3) | 0 | 0 | 1 | 0 | 1 | 0 | 1 | 6 | | 1, 22, 43, 48, 49, 50, 52 | Only Syalyl-Tn-antigen and (fucosylated or syalated) Core 3 structure can be hydrolyzed. |
| Lachnospiraceae bacterium 5_1_63FAA (658089.3) | 0 | 1 | 0 | 0 | 1 | 1 | 0 | 9 | | 11, 12, 13, 14, 39, 43, 44, 45, 46 | Only ploylactosamine structures can be hydrolyzed. |
| Lachnospiraceae bacterium 6_1_63FAA (658083.3) | 0 | 1 | 1 | 0 | 1 | 0 | 0 | 2 | | 22, 43 | Only (fucosylated) Core 1 structure can be hydrolyzed |
| Lachnospiraceae bacterium 7_1_58FAA (658087.3) | 0 | 0 | 1 | 0 | 1 | 0 | 1 | 5 | | 1, 43, 48, 49, 52 | Only (sylated) Core 3 structure can be hydrolyzed. |
| Lachnospiraceae bacterium 8_1_57FAA (665951.3) | 1 | 1 | 1 | 1 | 1 | 1 | 1 | 56 | | 1, 2, 3, 4, 5, 6, 7, 8, 9, 10, 11, 12, 13, 14, 15, 16, 17, 18, 19, 20, 21, 22, 23, 24, 25, 26, 27, 28, 29, 30, 31, 32, 33, 34, 35, 36, 37, 38, 39, 40, 41, 42, 43, 44, 45, 46, 47, 48, 49, 50, 51, 52, 53, 54, 55, 56 | All glycans can be hydrolyzed. |
| Lachnospiraceae bacterium 9_1_43BFAA (658088.3) | 0 | 1 | 1 | 0 | 1 | 1 | 1 | 40 | | 1, 2, 3, 4, 5, 10, 11, 12, 13, 14, 18, 22, 23, 25, 26, 27, 28, 30, 31, 32, 33, 34, 35, 36, 39, 40, 42, 43, 44, 45, 46, 48, 49, 50, 51, 52, 53, 54, 55, 56 | Glycans with alpha-GalNAc or alpha-GlcNAc cannot be hydrolyzed. |
| Lactobacillus acidophilus ATCC 4796 (525306.3) | 0 | 1 | 0 | 0 | 1 | 0 | 0 | 1 | | 43 | Only Core 1 structure can be hydrolyzed. |
| Lactobacillus acidophilus NCFM (272621.13) | 0 | 1 | 0 | 0 | 1 | 0 | 0 | 1 | | 43 | Only Core 1 structure can be hydrolyzed. |
| Lactobacillus amylolyticus DSM 11664 (585524.3) | 0 | 0 | 0 | 0 | 0 | 1 | 0 | 1 | | 11 | Only Core 3 structure can be hydrolyzed. |
| Lactobacillus antri DSM 16041 (525309.3) | 0 | 1 | 0 | 0 | 1 | 0 | 0 | 1 | | 43 | Only Core 1 structure can be hydrolyzed. |
| Lactobacillus brevis ATCC 367 (387344.15) | 0 | 1 | 0 | 0 | 1 | 0 | 0 | 1 | | 43 | Only Core 1 structure can be hydrolyzed. |
| Lactobacillus brevis subsp. gravesensis ATCC 27305 | 0 | 1 | 0 | 0 | 1 | 1 | 0 | 9 | | 11, 12, 13, 14, 39, 43, 44, 45, 46 | Only ploylactosamine structures can be hydrolyzed. |
| Lactobacillus buchneri ATCC 11577 (525318.3) | 0 | 1 | 0 | 0 | 1 | 1 | 0 | 9 | | 11, 12, 13, 14, 39, 43, 44, 45, 46 | Only ploylactosamine structures can be hydrolyzed. |

| Organism | | | | | | | | N | Elements | Notes |
|---|---|---|---|---|---|---|---|---|---|---|
| Lactobacillus casei ATCC 334 (321967.11) | 0 | 1 | 1 | 0 | 1 | 1 | 0 | 27 | 11, 12, 13, 14, 18, 22, 23, 25, 26, 27, 28, 30, 31, 32, 33, 34, 35, 36, 39, 40, 42, 43, 44, 45, 46, 55, 56 | Glycans with alpha-Gal or alpha-GalNAc as well sylated glycans cannot be hydrolyzed. |
| Lactobacillus casei BL23 (543734.4) | 0 | 1 | 1 | 0 | 0 | 1 | 0 | 1 | 11 | Only Core 3 structure can be hydrolyzed. |
| Lactobacillus crispatus 125-2-CHN (575595.3) | 0 | 0 | 0 | 0 | 1 | 0 | 0 | 1 | 43 | Only Core 1 structure can be hydrolyzed. |
| Lactobacillus delbrueckii subsp. bulgaricus ATCC 118 | 0 | 0 | 0 | 0 | 1 | 0 | 0 | 1 | 43 | Only Core 1 structure can be hydrolyzed. |
| Lactobacillus delbrueckii subsp. bulgaricus ATCC BA | 0 | 0 | 0 | 0 | 1 | 0 | 0 | 1 | 43 | Only Core 1 structure can be hydrolyzed. |
| Lactobacillus fermentum ATCC 14931 (525325.3) | 0 | 1 | 0 | 0 | 1 | 0 | 0 | 1 | 43 | Only Core 1 structure can be hydrolyzed. |
| Lactobacillus fermentum IFO 3956 (334390.5) | 0 | 1 | 0 | 0 | 1 | 0 | 0 | 1 | 43 | Only Core 1 structure can be hydrolyzed. |
| Lactobacillus gasseri ATCC 33323 (324831.13) | 0 | 0 | 0 | 0 | 0 | 0 | 0 | 0 | | No glycans are hydrolyzed. |
| Lactobacillus helveticus DPC 4571 (405566.6) | 0 | 1 | 0 | 0 | 1 | 0 | 0 | 1 | 43 | Only Core 1 structure can be hydrolyzed. |
| Lactobacillus helveticus DSM 20075 (585520.4) | 0 | 1 | 0 | 0 | 1 | 0 | 0 | 1 | 43 | Only Core 1 structure can be hydrolyzed. |
| Lactobacillus hilgardii ATCC 8290 (525327.3) | 0 | 1 | 0 | 0 | 1 | 1 | 0 | 9 | 11, 12, 13, 14, 39, 43, 44, 45, 46 | Only ploylactosamine structures can be hydrolyzed. |
| Lactobacillus johnsonii NCC 533 (257314.6) | 0 | 1 | 0 | 0 | 1 | 0 | 0 | 1 | 43 | Only Core 1 structure can be hydrolyzed. |
| Lactobacillus paracasei subsp. paracasei 8700:2 (53 | 0 | 1 | 1 | 0 | 0 | 1 | 0 | 1 | 11 | Only Core 3 structure can be hydrolyzed. |
| Lactobacillus paracasei subsp. paracasei ATCC 2530 | 0 | 1 | 1 | 0 | 0 | 1 | 0 | 1 | 11 | Only Core 3 structure can be hydrolyzed. |
| Lactobacillus plantarum 16 (1327988.3) | 0 | 1 | 0 | 0 | 1 | 0 | 0 | 1 | 43 | Only Core 1 structure can be hydrolyzed. |
| Lactobacillus plantarum subsp. plantarum ATCC 149 | 0 | 1 | 0 | 0 | 1 | 0 | 0 | 1 | 43 | Only Core 1 structure can be hydrolyzed. |
| Lactobacillus plantarum WCFS1 (220668.9) | 0 | 1 | 0 | 0 | 1 | 0 | 0 | 1 | 43 | Only Core 1 structure can be hydrolyzed. |
| Lactobacillus reuteri CF48-3A (525341.3) | 0 | 1 | 0 | 0 | 1 | 0 | 0 | 1 | 43 | Only Core 1 structure can be hydrolyzed. |
| Lactobacillus reuteri DSM 20016 (557436.4) | 0 | 1 | 0 | 0 | 1 | 0 | 0 | 1 | 43 | Only Core 1 structure can be hydrolyzed. |
| Lactobacillus reuteri F275 (299033.6) | 0 | 1 | 0 | 0 | 1 | 0 | 0 | 1 | 43 | Only Core 1 structure can be hydrolyzed. |
| Lactobacillus reuteri MM2-3 (585517.3) | 0 | 1 | 0 | 0 | 1 | 0 | 0 | 1 | 43 | Only Core 1 structure can be hydrolyzed. |
| Lactobacillus reuteri MM4-1A (548485.3) | 0 | 1 | 0 | 0 | 1 | 0 | 0 | 1 | 43 | Only Core 1 structure can be hydrolyzed. |
| Lactobacillus reuteri SD2112 (491077.3) | 0 | 1 | 0 | 0 | 1 | 0 | 0 | 1 | 43 | Only Core 1 structure can be hydrolyzed. |
| Lactobacillus rhamnosus ATCC 21052 (1002365.5) | 0 | 1 | 1 | 0 | 0 | 0 | 0 | 0 | | No glycans are hydrolyzed. |
| Lactobacillus rhamnosus GG (Prj:40637) (568703.30 | 0 | 1 | 1 | 0 | 0 | 1 | 0 | 1 | 11 | Only Core 3 structure can be hydrolyzed. |
| Lactobacillus rhamnosus LMS2-1 (525361.3) | 0 | 1 | 1 | 0 | 0 | 0 | 0 | 0 | | No glycans are hydrolyzed. |
| Lactobacillus ruminis ATCC 25644 (525362.3) | 0 | 1 | 0 | 0 | 1 | 1 | 0 | 9 | 11, 12, 13, 14, 39, 43, 44, 45, 46 | Only ploylactosamine structures can be hydrolyzed. |
| Lactobacillus sakei subsp. sakei 23K (314315.12) | 0 | 1 | 0 | 0 | 1 | 0 | 0 | 1 | 43 | Only Core 1 structure can be hydrolyzed. |
| Lactobacillus salivarius ATCC 11741 (525364.3) | 0 | 1 | 0 | 0 | 1 | 1 | 0 | 9 | 11, 12, 13, 14, 39, 43, 44, 45, 46 | Only ploylactosamine structures can be hydrolyzed. |
| Lactobacillus salivarius UCC118 (362948.14) | 0 | 1 | 0 | 0 | 1 | 1 | 1 | 22 | 1, 2, 3, 4, 5, 10, 11, 12, 13, 14, 18, 39, 43, 44, 45, 46, 48, 49, 51, 52, 53, 54 | Glycans with alpha-GalNAc as well sylated glycans cannot be hydrolyzed. |
| Lactobacillus sp. 7_1_47FAA (665945.3) | 0 | 0 | 0 | 0 | 0 | 0 | 0 | 0 | | No glycans are hydrolyzed. |
| Lactobacillus ultunensis DSM 16047 (525365.3) | 0 | 1 | 0 | 0 | 1 | 0 | 0 | 1 | 43 | Only Core 1 structure can be hydrolyzed. |
| Leuconostoc mesenteroides subsp. cremoris ATCC | 0 | 1 | 0 | 0 | 1 | 0 | 0 | 1 | 43 | Only Core 1 structure can be hydrolyzed. |
| Listeria grayi DSM 20601 (525367.9) | 0 | 1 | 0 | 0 | 0 | 0 | 0 | 0 | | No glycans are hydrolyzed. |
| Listeria innocua ATCC 33091 (1002366.3) | 0 | 0 | 0 | 0 | 0 | 0 | 0 | 0 | - | No glycans are hydrolyzed. |
| Megamonas funiformis YIT 11815 (742816.3) | 0 | 1 | 0 | 0 | 1 | 1 | 0 | 9 | 11, 12, 13, 14, 39, 43, 44, 45, 46 | Only ploylactosamine structures can be hydrolyzed. |
| Megamonas hypermegale ART12/1 (657316.3) | 0 | 1 | 0 | 0 | 1 | 1 | 0 | 9 | 11, 12, 13, 14, 39, 43, 44, 45, 46 | Only ploylactosamine structures can be hydrolyzed. |
| Methanobrevibacter smithii ATCC 35061 (420247.6) | 0 | 0 | 0 | 0 | 0 | 0 | 0 | 0 | - | No glycans are hydrolyzed. |
| Methanosphaera stadtmanae DSM 3091 (339860.6) | 0 | 0 | 0 | 0 | 0 | 0 | 0 | 0 | | No glycans are hydrolyzed. |
| Mitsuokella multacida DSM 20544 (500635.8) | 0 | 1 | 0 | 0 | 1 | 1 | 0 | 9 | 11, 12, 13, 14, 39, 43, 44, 45, 46 | Only ploylactosamine structures can be hydrolyzed. |
| Mollicutes bacterium D7 (556270.3) | 0 | 1 | 0 | 0 | 1 | 1 | 1 | 22 | 1, 2, 3, 4, 5, 10, 11, 12, 13, 14, 18, 39, 43, 44, 45, 46, 48, 49, 51, 52, 53, 54 | Glycans with alpha-GalNAc as well sylated glycans cannot be hydrolyzed. |
| Odoribacter laneus YIT 12061 (742817.3) | 0 | 1 | 0 | 0 | 1 | 1 | 1 | 40 | 1, 2, 3, 4, 5, 10, 11, 12, 13, 14, 18, 22, 23, 25, 26, 27, 28, 30, 31, 32, 33, 34, 35, 36, 39, 40, 42, 43, 44, 45, 46, 48, 49, 50, 51, 52, 53, 54, 55, 56 | Glycans with alpha-GalNAc or alpha-GlcNAc cannot be hydrolyzed. |
| Oxalobacter formigenes HOxBLS (556268.6) | 0 | 0 | 0 | 0 | 0 | 0 | 0 | 0 | - | No glycans are hydrolyzed. |
| Oxalobacter formigenes OXCC13 (556269.4) | 0 | 0 | 0 | 0 | 0 | 0 | 0 | 0 | - | No glycans are hydrolyzed. |
| Paenibacillus sp. HGF5 (908341.3) | 0 | 1 | 0 | 0 | 1 | 1 | 1 | 40 | 1, 2, 3, 4, 5, 10, 11, 12, 13, 14, 18, 22, 23, 25, 26, 27, 28, 30, 31, 32, 33, 34, 35, 36, 39, 40, 42, 43, 44, 45, 46, 48, 49, 50, 51, 52, 53, 54, 55, 56 | Glycans with alpha-GalNAc or alpha-GlcNAc cannot be hydrolyzed. |
| Paenibacillus sp. HGF7 (944559.3) | 0 | 1 | 0 | 0 | 1 | 1 | 0 | 27 | 11, 12, 13, 14, 18, 22, 23, 25, 26, 27, 28, 30, 31, 32, 33, 34, 35, 36, 39, 40, 42, 43, 44, 45, 46, 55, 56 | Glycans with alpha-Gal or alpha-GalNAc as well sylated glycans cannot be hydrolyzed. |
| Paenibacillus sp. HGH0039 (1078505.3) | 0 | 1 | 0 | 0 | 1 | 1 | 1 | 40 | 1, 2, 3, 4, 5, 10, 11, 12, 13, 14, 18, 22, 23, 25, 26, 27, 28, 30, 31, 32, 33, 34, 35, 36, 39, 40, 42, 43, 44, 45, 46, 48, 49, 50, 51, 52, 53, 54, 55, 56 | Glycans with alpha-GalNAc or alpha-GlcNAc cannot be hydrolyzed. |
| Parabacteroides distasonis ATCC 8503 (435591.13) | 0 | 1 | 0 | 0 | 1 | 1 | 1 | 40 | 1, 2, 3, 4, 5, 10, 11, 12, 13, 14, 18, 22, 23, 25, 26, 27, 28, 30, 31, 32, 33, 34, 35, 36, 39, 40, 42, 43, 44, 45, 46, 48, 49, 50, 51, 52, 53, 54, 55, 56 | Glycans with alpha-GalNAc or alpha-GlcNAc cannot be hydrolyzed. |
| Parabacteroides johnsonii DSM 18315 (537006.5) | 0 | 1 | 0 | 1 | 1 | 1 | 1 | 41 | 1, 2, 3, 4, 5, 10, 11, 12, 13, 14, 18, 22, 23, 25, 26, 27, 28, 30, 31, 32, 33, 34, 35, 36, 39, 40, 42, 43, 44, 45, 46, 47, 48, 49, 50, 51, 52, 53, 54, 55, 56 | Glycans with alpha-GalNAc cannot be hydrolyzed. |
| Parabacteroides merdae ATCC 43184 (411477.4) | 0 | 1 | 0 | 1 | 1 | 1 | 1 | 41 | 1, 2, 3, 4, 5, 10, 11, 12, 13, 14, 18, 22, 23, 25, 26, 27, 28, 30, 31, 32, 33, 34, 35, 36, 39, 40, 42, 43, 44, 45, 46, 47, 48, 49, 50, 51, 52, 53, 54, 55, 56 | Glycans with alpha-GalNAc cannot be hydrolyzed. |
| Parabacteroides sp. D13 (563193.3) | 0 | 1 | 1 | 0 | 1 | 1 | 1 | 40 | 1, 2, 3, 4, 5, 10, 11, 12, 13, 14, 18, 22, 23, 25, 26, 27, 28, 30, 31, 32, 33, 34, 35, 36, 39, 40, 42, 43, 44, 45, 46, 48, 49, 50, 51, 52, 53, 54, 55, 56 | Glycans with alpha-GalNAc or alpha-GlcNAc cannot be hydrolyzed. |
| Parabacteroides sp. D25 (658661.3) | 0 | 1 | 1 | 0 | 1 | 1 | 1 | 40 | 1, 2, 3, 4, 5, 10, 11, 12, 13, 14, 18, 22, 23, 25, 26, 27, 28, 30, 31, 32, 33, 34, 35, 36, 39, 40, 42, 43, 44, 45, 46, 48, 49, 50, 51, 52, 53, 54, 55, 56 | Glycans with alpha-GalNAc or alpha-GlcNAc cannot be hydrolyzed. |

| Organism | | | | | | | | | | Notes |
|---|---|---|---|---|---|---|---|---|---|---|
| Paraprevotella clara YIT 11840 (762968.3) | 0 | 1 | 1 | 0 | 1 | 0 | 0 | 2 | 22, 43 | Only (fucosylated) Core 1 structure can be |
| Paraprevotella xylaniphila YIT 11841 (762982.3) | 0 | 1 | 1 | 0 | 1 | 0 | 0 | 2 | 22, 43 | Only (fucosylated) Core 1 structure can be |
| Parvimonas micra ATCC 33270 (411465.10) | 0 | 0 | 0 | 0 | 0 | 0 | 0 | 0 | - | No glycans are hydrolyzed. |
| Pediococcus acidilactici 7_4 (563194.3) | 0 | 1 | 1 | 0 | 0 | 0 | 0 | 0 | - | No glycans are hydrolyzed. |
| Pediococcus acidilactici DSM 20284 (862514.3) | 0 | 0 | 1 | 0 | 1 | 0 | 1 | 6 | 1, 22, 43, 48, 49, 50, 52 | Only Sialyl-Tn-antigen and (fucosylated or sialated) |
| Phascolarctobacterium sp. YIT 12067 (626939.3) | 0 | 0 | 0 | 0 | 0 | 0 | 0 | 0 | - | No glycans are hydrolyzed. |
| Prevotella copri DSM 18205 (537011.5) | 0 | 1 | 0 | 0 | 1 | 1 | 0 | 9 | 11, 12, 13, 14, 39, 43, 44, 45, 46 | Only ployplactosamine structures can be hydrolyzed. |
| Prevotella oralis HGA0225 (1203550.3) | 0 | 1 | 1 | 1 | 1 | 1 | 1 | 41 | 1, 2, 3, 4, 5, 10, 11, 12, 13, 14, 18, 22, 23, 25, 26, 27, 28, 30, 31, 32, 33, 34, 35, 36, 39, 40, 42, 43, 44, 45, 46, 47, 48, 49, 50, 51, 52, 53, 54, 55, 56 | Glycans with alpha-GalNAc cannot be hydrolyzed. |
| Prevotella saliviae DSM 15606 (888832.3) | 0 | 1 | 1 | 0 | 1 | 1 | 1 | 40 | 1, 2, 3, 4, 5, 10, 11, 12, 13, 14, 18, 22, 23, 25, 26, 27, 28, 30, 31, 32, 33, 34, 35, 36, 39, 40, 42, 43, 44, 45, 46, 48, 49, 50, 51, 52, 53, 54, 55, 56 | Glycans with alpha-GalNAc or alpha-GlcNAc cannot be hydrolyzed. |
| Propionibacterium sp. 5_U_42AFAA (450748.3) | 0 | 0 | 1 | 0 | 1 | 1 | 1 | 34 | 1, 2, 3, 4, 5, 10, 11, 12, 13, 14, 18, 22, 23, 25, 26, 27, 28, 30, 32, 33, 39, 40, 43, 44, 45, 46, 48, 49, 50, 51, 52, 53, 54, 55 | Glycans with alpha-Gal, alpha-GalNAc, or alpha-GlcNAc cannot be hydrolyzed. |
| Propionibacterium sp. HGH0353 (1203571.3) | 0 | 1 | 0 | 0 | 1 | 1 | 0 | 9 | 11, 12, 13, 14, 39, 43, 44, 45, 46 | Only ployplactosamine structures can be hydrolyzed. |
| Proteus mirabilis WGLW6 (1125694.3) | 0 | 0 | 0 | 0 | 0 | 0 | 0 | 0 | - | No glycans are hydrolyzed. |
| Proteus penneri ATCC 35198 (471881.3) | 0 | 0 | 0 | 0 | 0 | 0 | 0 | 0 | - | No glycans are hydrolyzed. |
| Providencia alcalifaciens DSM 30120 (520999.6) | 0 | 0 | 0 | 0 | 0 | 0 | 0 | 0 | - | No glycans are hydrolyzed. |
| Providencia rettgeri DSM 1131 (521000.6) | 0 | 0 | 0 | 0 | 0 | 1 | 0 | 1 | 11 | Only Core 3 structure can be hydrolyzed. |
| Providencia rustigianii DSM 4541 (500637.6) | 0 | 0 | 0 | 0 | 0 | 0 | 0 | 0 | - | No glycans are hydrolyzed. |
| Providencia stuartii ATCC 25827 (471874.6) | 0 | 0 | 0 | 0 | 0 | 1 | 0 | 1 | 11 | Only Core 3 structure can be hydrolyzed. |
| Pseudomonas sp. 2_1_26 (665948.3) | 0 | 0 | 0 | 0 | 0 | 0 | 0 | 0 | - | No glycans are hydrolyzed. |
| Ralstonia sp. 5_7_47FAA (658664.3) | 0 | 0 | 0 | 0 | 0 | 0 | 0 | 0 | - | No glycans are hydrolyzed. |
| Roseburia intestinalis L1-82 (536231.5) | 0 | 1 | 1 | 0 | 1 | 1 | 0 | 27 | 11, 12, 13, 14, 18, 22, 23, 25, 26, 27, 28, 30, 31, 32, 33, 34, 35, 36, 39, 40, 42, 43, 44, 45, 46, 55, 56 | Glycans with alpha-Gal or alpha-GalNAc as well syalated glycans cannot be hydrolyzed. |
| Roseburia intestinalis M50/1 (657315.3) | 0 | 1 | 1 | 0 | 1 | 1 | 0 | 27 | 11, 12, 13, 14, 18, 22, 23, 25, 26, 27, 28, 30, 31, 32, 33, 34, 35, 36, 39, 40, 42, 43, 44, 45, 46, 55, 56 | Glycans with alpha-Gal or alpha-GalNAc as well syalated glycans cannot be hydrolyzed. |
| Roseburia intestinalis XB6B4 (718255.3) | 0 | 1 | 1 | 0 | 1 | 1 | 0 | 27 | 11, 12, 13, 14, 18, 22, 23, 25, 26, 27, 28, 30, 31, 32, 33, 34, 35, 36, 39, 40, 42, 43, 44, 45, 46, 55, 56 | Glycans with alpha-Gal or alpha-GalNAc as well syalated glycans cannot be hydrolyzed. |
| Roseburia inulinivorans DSM 16841 (622312.4) | 0 | 1 | 1 | 0 | 1 | 1 | 0 | 27 | 11, 12, 13, 14, 18, 22, 23, 25, 26, 27, 28, 30, 31, 32, 33, 34, 35, 36, 39, 40, 42, 43, 44, 45, 46, 55, 56 | Glycans with alpha-Gal or alpha-GalNAc as well syalated glycans cannot be hydrolyzed. |
| Ruminococcaceae bacterium D16 (552398.3) | 0 | 1 | 0 | 0 | 1 | 1 | 0 | 9 | 11, 12, 13, 14, 39, 43, 44, 45, 46 | Only ployplactosamine structures can be hydrolyzed. |
| Ruminococcus bromii L2-63 (657321.5) | 0 | 0 | 0 | 0 | 0 | 1 | 0 | 1 | 11 | Only Core 3 structure can be hydrolyzed. |
| Ruminococcus gnavus ATCC 29149 (411470.6) | 0 | 1 | 1 | 0 | 1 | 0 | 1 | 6 | 1, 22, 43, 48, 49, 50, 52 | Only Sialyl-Tn-antigen and (fucosylated or syalated) Core 3 structure can be hydrolyzed. |
| Ruminococcus lactaris ATCC 29176 (471875.6) | 0 | 0 | 1 | 1 | 1 | 1 | 1 | 35 | 1, 2, 3, 4, 5, 10, 11, 12, 13, 14, 18, 22, 23, 25, 26, 27, 28, 30, 32, 33, 39, 40, 43, 44, 45, 46, 47, 48, 49, 50, 51, 52, 53, 54, 55 | Glycans with alpha-Gal or alpha-GalNAc cannot be hydrolyzed. |
| Ruminococcus obeum A2-162 (657314.3) | 0 | 0 | 0 | 0 | 1 | 0 | 0 | 1 | 43 | Only Core 1 structure can be hydrolyzed. |
| Ruminococcus obeum ATCC 29174 (411459.7) | 0 | 1 | 1 | 0 | 1 | 0 | 0 | 2 | 22, 43 | Only (fucosylated) Core 1 structure can be |
| Ruminococcus sp. 18P13 (213810.4) | 0 | 1 | 0 | 0 | 1 | 0 | 0 | 1 | 43 | Only Core 1 structure can be hydrolyzed. |
| Ruminococcus sp. 5_1_39BFAA (457412.4) | 0 | 1 | 1 | 0 | 1 | 0 | 0 | 2 | 22, 43 | Only (fucosylated) Core 1 structure can be |
| Ruminococcus sp. SR1/5 (657323.3) | 0 | 1 | 1 | 0 | 1 | 0 | 0 | 2 | 22, 43 | Only (fucosylated) Core 1 structure can be |
| Ruminococcus torques ATCC 27756 (411460.6) | 1 | 1 | 1 | 1 | 1 | 1 | 1 | 56 | 1, 2, 3, 4, 5, 6, 7, 8, 9, 10, 11, 12, 13, 14, 15, 16, 17, 18, 19, 20, 21, 22, 23, 24, 25, 26, 27, 28, 29, 30, 31, 32, 33, 34, 35, 36, 37, 38, 39, 40, 41, 42, 43, 44, 45, 46, 47, 48, 49, 50, 51, 52, 53, 54, 55, 56 | All glycans can be hydrolyzed. |
| Ruminococcus torques L2-14 (657313.3) | 0 | 1 | 1 | 0 | 1 | 0 | 0 | 2 | 22, 43 | Only (fucosylated) Core 1 structure can be |
| Salmonella enterica subsp. enterica serovar Typhimu | 0 | 0 | 0 | 0 | 0 | 0 | 1 | 1 | 1 | Only Sialyl-Tn-antigen can be hydrolyzed. |
| Staphylococcus sp. HGB0015 (1078083.3) | 0 | 0 | 0 | 0 | 0 | 0 | 1 | 1 | 1 | Only Sialyl-Tn-antigen can be hydrolyzed. |
| Streptococcus anginosus 1_2_62CV (742820.3) | 0 | 0 | 0 | 0 | 0 | 0 | 0 | 0 | - | No glycans are hydrolyzed. |
| Streptococcus equinus ATCC 9812 (525379.3) | 0 | 0 | 0 | 0 | 0 | 0 | 0 | 0 | - | No glycans are hydrolyzed. |
| Streptococcus infantarius subsp. infantarius ATCC B | 0 | 1 | 0 | 0 | 0 | 0 | 0 | 0 | - | No glycans are hydrolyzed. |
| Streptococcus sp. 2_1_36FAA (469609.3) | 0 | 0 | 1 | 0 | 1 | 1 | 0 | 21 | 11, 12, 13, 14, 18, 22, 23, 25, 26, 27, 28, 30, 32, 33, 39, 40, 43, 44, 45, 46, 55 | Only (fucosylated) ployplactosamine structures can be hydrolyzed. |
| Streptococcus sp. HPH0090 (1203590.3) | 0 | 0 | 1 | 0 | 1 | 1 | 1 | 34 | 1, 2, 3, 4, 5, 10, 11, 12, 13, 14, 18, 22, 23, 25, 26, 27, 28, 30, 31, 32, 33, 39, 40, 43, 44, 45, 46, 48, 49, 50, 51, 52, 53, 54, 55 | Glycans with alpha-Gal, alpha-GalNAc, or alpha-GlcNAc cannot be hydrolyzed. |
| Streptomyces sp. HGB0020 (1078086.3) | 0 | 1 | 1 | 0 | 1 | 1 | 1 | 40 | 1, 2, 3, 4, 5, 10, 11, 12, 13, 14, 18, 22, 23, 25, 26, 27, 28, 30, 31, 32, 33, 34, 35, 36, 39, 40, 42, 43, 44, 45, 46, 48, 49, 50, 51, 52, 53, 54, 55, 56 | Glycans with alpha-GalNAc or alpha-GlcNAc cannot be hydrolyzed. |
| Streptomyces sp. HPH0547 (1203592.3) | 0 | 1 | 1 | 1 | 1 | 1 | 1 | 41 | 1, 2, 3, 4, 5, 10, 11, 12, 13, 14, 18, 22, 23, 25, 26, 27, 28, 30, 31, 32, 33, 34, 35, 36, 39, 40, 42, 43, 44, 45, 46, 47, 48, 49, 50, 51, 52, 53, 54, 55, 56 | Glycans with alpha-GalNAc cannot be hydrolyzed. |

| Species | | | | | | | | Number | Hydrolyzed glycans | Comments |
|---|---|---|---|---|---|---|---|---|---|---|
| Subdoligranulum sp. 4_3_54A2FAA (665956.3) | 0 | 1 | 1 | 0 | 1 | 1 | 1 | 40 | 1, 2, 3, 4, 5, 10, 11, 12, 13, 14, 18, 22, 23, 25, 26, 27, 28, 30, 31, 32, 33, 34, 35, 36, 39, 40, 42, 43, 44, 45, 46, 48, 49, 50, 51, 52, 53, 54, 55, 56 | Glycans with alpha-GalNAc or alpha-GlcNAc cannot be hydrolyzed. |
| Subdoligranulum variabile DSM 15176 (411471.5) | 0 | 1 | 0 | 0 | 1 | 1 | 0 | 9 | 11, 12, 13, 14, 39, 43, 44, 45, 46 | Only ploylactosamine structures can be hydrolyzed. |
| Succinatimonas hippei YIT 12066 (762983.3) | 0 | 0 | 0 | 0 | 0 | 0 | 0 | 0 | - | No glycans are hydrolyzed. |
| Sutterella wadsworthensis 3_1_45B (742821.3) | 0 | 0 | 0 | 0 | 0 | 0 | 0 | 0 | - | No glycans are hydrolyzed. |
| Sutterella wadsworthensis HGA0223 (1203554.3) | 0 | 0 | 0 | 0 | 0 | 0 | 0 | 0 | - | No glycans are hydrolyzed. |
| Tannerella sp. 6_1_58FAA_CT1 (665949.3) | 0 | 1 | 1 | 0 | 1 | 1 | 0 | 27 | 11, 12, 13, 14, 18, 22, 23, 25, 26, 27, 28, 30, 31, 32, 33, 34, 35, 36, 39, 40, 42, 43, 44, 45, 46, 55, 56 | Glycans with alpha-Gal or alpha-GalNAc as well syalated glycans cannot be hydrolyzed. |
| Turicibacter sp. HGF1 (910310.3) | 0 | 1 | 1 | 0 | 1 | 1 | 0 | 27 | 11, 12, 13, 14, 18, 22, 23, 25, 26, 27, 28, 30, 31, 32, 33, 34, 35, 36, 39, 40, 42, 43, 44, 45, 46, 55, 56 | Glycans with alpha-Gal or alpha-GalNAc as well syalated glycans cannot be hydrolyzed. |
| Turicibacter sp. PC909 (702450.3) | 0 | 1 | 1 | 0 | 1 | 1 | 0 | 27 | 11, 12, 13, 14, 18, 22, 23, 25, 26, 27, 28, 30, 31, 32, 33, 34, 35, 36, 39, 40, 42, 43, 44, 45, 46, 55, 56 | Glycans with alpha-Gal or alpha-GalNAc as well syalated glycans cannot be hydrolyzed. |
| Veillonella sp. 3_1_44 (457416.3) | 0 | 0 | 0 | 0 | 0 | 1 | 0 | 1 | 11 | Only Core 3 structure can be hydrolyzed. |
| Veillonella sp. 6_1_27 (450749.3) | 0 | 0 | 0 | 0 | 0 | 1 | 0 | 1 | 11 | Only Core 3 structure can be hydrolyzed. |
| Weissella paramesenteroides ATCC 33313 (585506.) | 0 | 1 | 0 | 0 | 1 | 0 | 0 | 1 | 43 | Only Core 1 structure can be hydrolyzed. |
| Yokenella regensburgei ATCC 43003 (1002368.3) | 0 | 1 | 0 | 0 | 1 | 1 | 0 | 9 | 11, 12, 13, 14, 39, 43, 44, 45, 46 | Only ploylactosamine structures can be hydrolyzed. |

**Summary**

| Number of | Number | Hydrolyzed glycans | Comments |
|---|---|---|---|
| 0 | 76 | - | No glycans are hydrolyzed. |
| 1 | 8 | | Only Syalyl-Tn-antigen can be hydrolyzed. |
| 1 | 27 | 11 | Only Core 3 structure can be hydrolyzed. |
| 1 | 77 | 43 | Only Core 1 structure can be hydrolyzed. |
| 2 | 11 | 22, 43 | Only (fucosylated) Core 1 structure can be hydrolyzed. |
| 3 | 1 | 1, 2, 11 | Only Syalyl-Tn-antigen and (syalated) Core 3 structure can be hydrolyzed. |
| 5 | 2 | 1, 43, 48, 49, 52 | Only (syalated) Core 3 structure can be hydrolyzed. |
| 7 | 6 | 1, 22, 43, 48, 49, 50, 52 | Only Syalyl-Tn-antigen and (fucosylated or syalated) Core 3 structure can be hydrolyzed. |
| 9 | 56 | 11, 12, 13, 14, 39, 43, 44, 45, 46 | Only ploylactosamine structures can be hydrolyzed. |
| 13 | 1 | 11, 12, 13, 14, 16, 18, 19, 21, 39, 43, 44, 45, 46 | Only (syalated) ploylactosamine structures can be hydrolyzed. |
| 21 | 3 | 11, 12, 13, 14, 18, 22, 23, 25, 26, 27, 28, 30, 32, 33, 39, 40, 43, 44, 45, 46, 55 | Only (fucosylated) ploylactosamine structures can be hydrolyzed. |
| 22 | 14 | 1, 2, 3, 4, 5, 10, 11, 12, 13, 14, 18, 39, 43, 44, 45, 46, 48, 49, 51, 52, 53, 54 | Glycans with alpha-GalNAc as well syalated glycans cannot be hydrolyzed. |
| 27 | 20 | 11, 12, 13, 14, 18, 22, 23, 25, 26, 27, 28, 30, 31, 32, 33, 34, 35, 36, 39, 40, 42, 43, 44, 45, 46, 55, 56 | Glycans with alpha-Gal or alpha-GalNAc as well syalated glycans cannot be hydrolyzed. |
| 28 | 1 | 11, 12, 13, 14, 18, 22, 23, 25, 26, 27, 28, 30, 31, 32, 33, 34, 35, 36, 39, 40, 42, 43, 44, 45, 46, 47, 55, 56 | Glycans with alpha-GalNAc as well syalated glycans cannot be hydrolyzed. |
| 33 | 2 | 11, 12, 13, 14, 15, 16, 17, 18, 19, 20, 21, 22, 23, 24, 25, 26, 27, 28, 29, 30, 32, 33, 37, 38, 39, 40, 41, 43, 44, 45, 46, 47, 55 | Glycans with alpha-Gal as well syalated glycans cannot be hydrolyzed. |
| 34 | 3 | 1, 2, 3, 4, 5, 10, 11, 12, 13, 14, 18, 22, 23, 25, 26, 27, 28, 30, 32, 33, 39, 40, 43, 44, 45, 46, 48, 49, 50, 51, 52, 53, 54, 55 | Glycans with alpha-Gal, alpha-GalNAc, or alpha-GlcNAc cannot be hydrolyzed. |
| 35 | 2 | 1, 2, 3, 4, 5, 10, 11, 12, 13, 14, 18, 22, 23, 25, 26, 27, 28, 30, 32, 33, 39, 40, 43, 44, 45, 46, 47, 48, 49, 50, 51, 52, 53, 54, 55 | Glycans with alpha-Gal or alpha-GalNAc cannot be hydrolyzed. |
| 40 | 26 | 1, 2, 3, 4, 5, 10, 11, 12, 13, 14, 18, 22, 23, 25, 26, 27, 28, 30, 31, 32, 33, 34, 35, 36, 39, 40, 42, 43, 44, 45, 46, 48, 49, 50, 51, 52, 53, 54, 55, 56 | Glycans with alpha-GalNAc or alpha-GlcNAc cannot be hydrolyzed. |
| 41 | 41 | 1, 2, 3, 4, 5, 10, 11, 12, 13, 14, 18, 22, 23, 25, 26, 27, 28, 30, 31, 32, 33, 34, 35, 36, 39, 40, 42, 43, 44, 45, 46, 47, 48, 49, 50, 51, 52, 53, 54, 55, 56 | Glycans with alpha-GalNAc cannot be hydrolyzed. |
| 50 | 1 | 1, 2, 3, 4, 5, 6, 7, 8, 9, 10, 11, 12, 13, 14, 15, 16, 17, 18, 19, 20, 21, 22, 23, 24, 25, 26, 27, 28, 29, 30, 32, 33, 37, 38, 39, 40, 41, 43, 44, 45, 46, 47, 48, 49, 50, 51, 52, 53, 54, 55 | Glycans with alpha-Gal cannot be hydrolyzed. |
| 56 | 8 | 1, 2, 3, 4, 5, 6, 7, 8, 9, 10, 11, 12, 13, 14, 15, 16, 17, 18, 19, 20, 21, 22, 23, 24, 25, 26, 27, 28, 29, 30, 31, 32, 33, 34, 35, 36, 37, 38, 39, 40, 41, 42, 43, 44, 45, 46, 47, 48, 49, 50, 51, 52, 53, 54, 55, 56 | All glycans can be hydrolyzed. |

Table S16. Mycotoxin gene clusters and their distribution across the fungal genomes analyzed in this study. Presence or absence of a gene cluster is represented by 1 (present) or 0 (absent) for each column of the gene cluster analyzed.